\documentclass[11pt,twoside]{ILCTDR}

\usepackage{multirow} 

\usepackage{savesym}
\savesymbol{iint}
\savesymbol{iiint}
\usepackage{amsmath}
\restoresymbol{}{iint}
\restoresymbol{}{iiint}
\usepackage{authblk}

\usepackage{amssymb}
\usepackage{enumitem}
\usepackage{booktabs}

\usepackage{hepnames}
\usepackage{hepunits}
\usepackage{graphicx}
\usepackage[section]{placeins}
\usepackage{adjustbox}
\usepackage{anyfontsize}

\usepackage{datatool-base}

\graphicspath{
   {./Introduction/figs/}
   {./Detector/figs/}
   {./figs/}
   {./ILC/figs/}
   {./ILDfigs/}
   {./Integration/figs/}
   {./Modelling/figs/}
   {./Performance/figs/}
   {./References/figs/}
	 {./}
}

\newcommand{\fix}[1]{\textcolor{red}{\texttt{#1}}} 

\newcommand\writer[2]{{\marginpar{\small \vspace{-2cm}\hspace{-4cm}
\colorbox{green}{\begin{minipage}{4cm}{{#1 \\#2 pages}}\end{minipage}}}}}
\renewcommand\writer[2]{{\ifx\xxyy\undefined {} \else #1 #2 \fi}}

\newcommand{\DBDnumbers}{}

\makenomenclature
\begin{document}
\setcounter{chapter}{0}
\setcounter{part}{1}
%
%
\setcounter{chapter}{99}
%
%
\frontmatter
%
%
%
%
\begin{titlepage}
\begin{center}
\rightline{DESY 20-034}
\rightline{KEK Preprint 2019-57}
 ~\vskip 0.5cm

    {\huge  International} 
    {\huge  Large} 
    {\huge  Detector}
\vskip 3cm

\noindent\makebox[\linewidth]{\rule{\textwidth}{0.4pt}}
\begin{center}
\includegraphics[height=3.3cm]{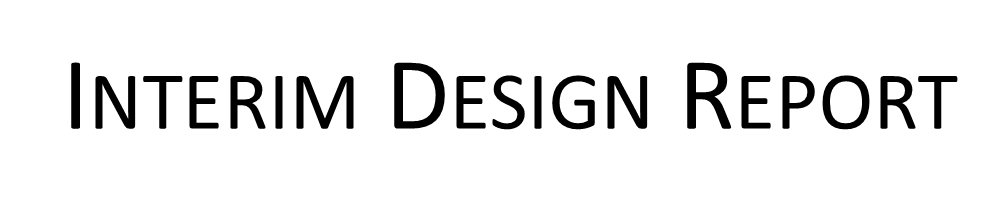}
\end{center}

\noindent\makebox[\linewidth]{\rule{\textwidth}{0.4pt}}
\vskip 1.2cm





{\Large The ILD Concept Group}


\vskip 3cm
\end{center}
\begin{center}
\includegraphics[height=5cm]{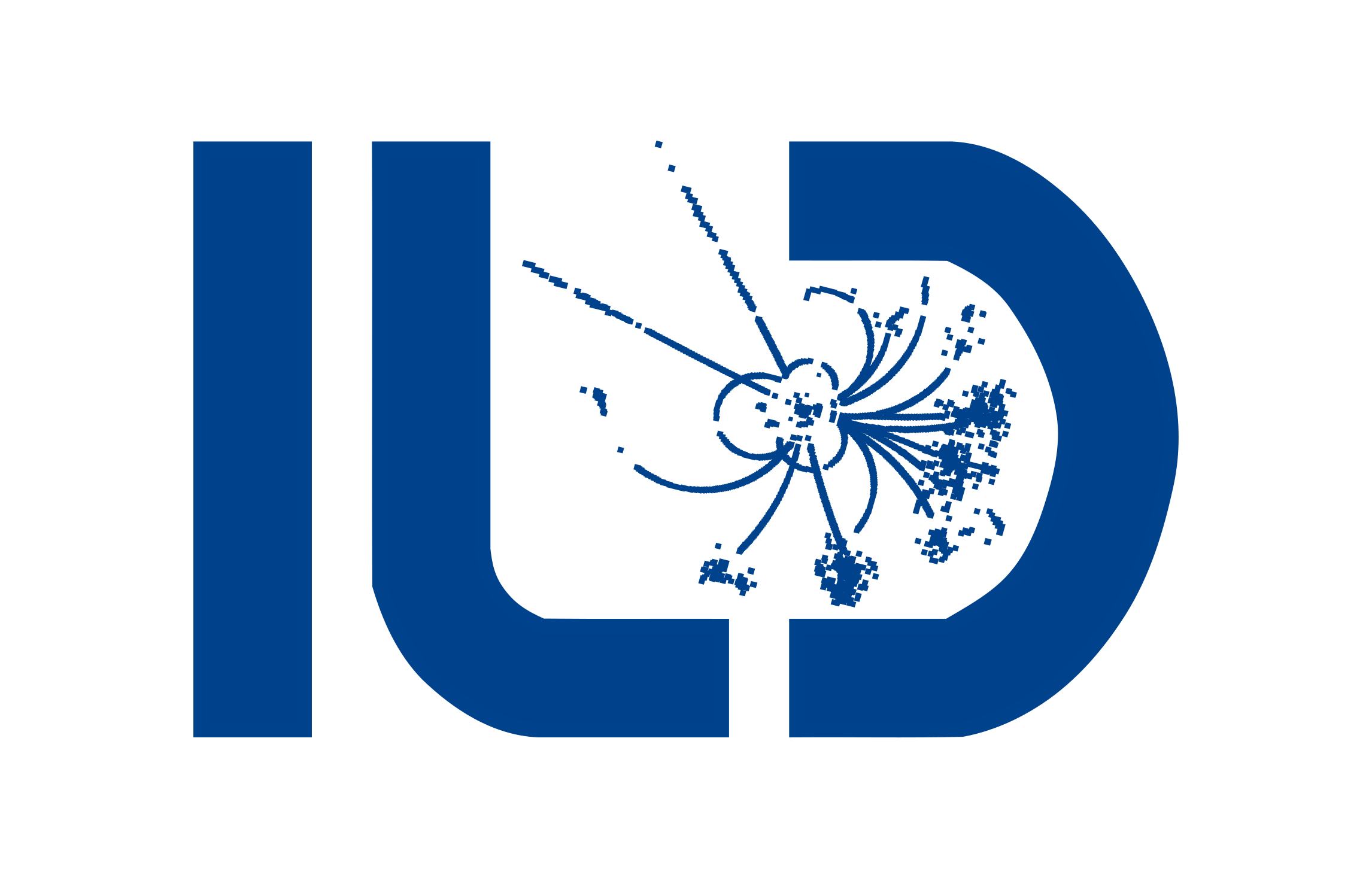}
\end{center}

  \vfill\ 
    
    \centering{\large \today}
\end{titlepage}

\newpage\thispagestyle{empty}
\cleardoublepage
\newcommand{\IDRinst}[2]{\noindent  $^{\;\;\:\:#1}$\begin{minipage}[t]{0.95\textwidth}{#2}\end{minipage}\\}
\newcommand{\IDRauth}[2]{\noindent {#2}$^{#1}$,}
\newcommand{\IDRauthl}[2]{\noindent {#2}$^{#1}$}
\chapter*{IDR Editors}

Ties Behnke$^1$, Karsten Buesser$^1$, Keisuke Fujii$^2$, Frank Gaede$^1$, Kiyotomo Kawagoe$^3$, Jenny List$^1$, Akiya Miyamoto$^2$, Claude Vall\'ee$^4$, Henri Videau$^5$ \\

\vspace{5mm}
\IDRinst{1}{Deutsches Elektronen-Synchrotron, Hamburg and Zeuthen, Germany}
\IDRinst{2}{High Energy Accelerator Research Organisation, KEK, Tsukuba, Ibaraki, Japan}
\IDRinst{3}{Kyushu University, Department of Physics, Research Center for Advanced Particle Physics, Fukuoka, Japan}
\IDRinst{4}{Aix Marseille Univ, CNRS/IN2P3, CPPM, Marseille, France}
\IDRinst{5}{Laboratoire Leprince-Ringuet, Institut Polytechnique de Paris, Palaiseau, France}

\chapter*{ ILD authors}


\begin{center}
\IDRauth{52}{\DTLinitials{Halina}~Abramowicz}
\IDRauth{64}{\DTLinitials{Tatjana}~Agatonovic Jovin}
\IDRauth{55}{\DTLinitials{Oscar}~Alonso}
\IDRauth{56}{\DTLinitials{Mohammad Sohail}~Amjad}
\IDRauth{21}{\DTLinitials{Fenfen}~An}
\IDRauth{16}{\DTLinitials{Ladislav}~Andricek}
\IDRauth{35}{\DTLinitials{Marc}~Anduze}
\IDRauth{29}{\DTLinitials{Justin}~Anguiano}
\IDRauth{37}{\DTLinitials{Evgeny}~Antonov}
\IDRauth{30}{\DTLinitials{Yumi}~Aoki}
\IDRauth{27}{\DTLinitials{Fernando}~Arteche}
\IDRauth{8}{\DTLinitials{David}~Atti\'e}
\IDRauth{60}{\DTLinitials{Volker}~B\"uscher}
\IDRauth{12}{\DTLinitials{Ole}~Bach}
\IDRauth{35}{\DTLinitials{Vladislav}~Balagura}
\IDRauth{25}{\DTLinitials{J\'erome}~Baudot}
\IDRauth{17}{\DTLinitials{Vadim}~Begunov}
\IDRauth{22}{\DTLinitials{Subhasish}~Behera}
\IDRauth{12}{\DTLinitials{Ties}~Behnke}
\IDRauth{7}{\DTLinitials{Alain}~Bellerive}
\IDRauth{9}{\DTLinitials{Daniel}~Belver}
\IDRauth{52}{\DTLinitials{Yan}~Benhammou}
\IDRauth{12}{\DTLinitials{Mikael}~Berggren}
\IDRauth{8}{\DTLinitials{Christophe}~Berriaud}
\IDRauth{25}{\DTLinitials{Gregory}~Bertolone}
\IDRauth{8}{\DTLinitials{Marc}~Besan\c con}
\IDRauth{25}{\DTLinitials{Auguste}~Besson}
\IDRauth{12}{\DTLinitials{Jakob}~Beyer}
\IDRauth{54}{\DTLinitials{Oleg}~Bezshyyko}
\IDRauth{49}{\DTLinitials{Deb Sankar}~Bhattacharya}
\IDRauth{49}{\DTLinitials{Purba}~Bhattacharya}
\IDRauth{45}{\DTLinitials{Gerald}~Blazey}
\IDRauth{12}{\DTLinitials{Vladimir}~Bocharnikov}
\IDRauth{19}{\DTLinitials{Marca}~Boronat}
\IDRauth{12}{\DTLinitials{Oleksandr}~Borysov}
\IDRauth{3}{\DTLinitials{Robert}~Bosley}
\IDRauth{35}{\DTLinitials{Vincent}~Boudry}
\IDRauth{36}{\DTLinitials{Djamel}~Boumediene}
\IDRauth{34}{\DTLinitials{Christian}~Bourgeois}
\IDRauth{64}{\DTLinitials{Ivanka}~Bozovic Jelisavcic}
\IDRauth{34}{\DTLinitials{Dominique}~Breton}
\IDRauth{12}{\DTLinitials{Eldwan}~Brianne}
\IDRauth{35}{\DTLinitials{Jean-Claude}~Brient}
\IDRauth{58}{\DTLinitials{Konrad}~Briggl}
\IDRauth{12}{\DTLinitials{Karsten}~Buesser}
\IDRauth{46}{\DTLinitials{Stephane}~Callier}
\IDRauth{9}{\DTLinitials{Enrique}~Calvo Alamillo}
\IDRauth{9}{\DTLinitials{Camilo}~Carrillo}
\IDRauth{19}{\DTLinitials{Ana}~Catal\'an}
\IDRauth{37}{\DTLinitials{Marina}~Chadeeva}
\IDRauth{60}{\DTLinitials{Phi}~Chau}
\IDRauth{12,a}{\DTLinitials{Madalina}~Chera}
\IDRauth{17}{\DTLinitials{Boris}~Chetverushkin}
\IDRauth{20}{\DTLinitials{Marcin}~Chrzaszcz}
\IDRauth{25}{\DTLinitials{Gilles}~Claus}
\IDRauth{8}{\DTLinitials{Paul}~Colas}
\IDRauth{25}{\DTLinitials{Claude}~Colledani}
\IDRauth{26}{\DTLinitials{Christophe}~Combaret}
\IDRauth{38}{\DTLinitials{R\'emi}~Cornat}
\IDRauth{40}{\DTLinitials{Francois}~Corriveau}
\IDRauth{41}{\DTLinitials{Mikhail}~Danilov}
\IDRauth{46}{\DTLinitials{Christophe}~de la Taille}
\IDRauth{33}{\DTLinitials{Yuto}~Deguchi}
\IDRauth{5}{\DTLinitials{Klaus}~Desch}
\IDRauth{55}{\DTLinitials{Angel}~Dieguez}
\IDRauth{12}{\DTLinitials{Ralf}~Diener}
\IDRauth{7}{\DTLinitials{Madhu}~Dixit}
\IDRauth{21}{\DTLinitials{Mingyi}~Dong}
\IDRauth{25}{\DTLinitials{Andrei}~Dorokhov}
\IDRauth{25}{\DTLinitials{Guy}~Dozi\`ere}
\IDRauth{37}{\DTLinitials{Alexey}~Drutskoy}
\IDRauth{46}{\DTLinitials{Frederic}~Dulucq}
\IDRauth{45}{\DTLinitials{Alexander}~Dyshkant}
\IDRauth{35}{\DTLinitials{Evelyne}~Edy}
\IDRauth{2}{\DTLinitials{Gerald}~Eigen}
\IDRauth{12}{\DTLinitials{Ulrich}~Einhaus}
\IDRauth{25}{\DTLinitials{Ziad}~El Bitar}
\IDRauth{6}{\DTLinitials{Amine}~Elkhalii}
\IDRauth{42}{\DTLinitials{Lorenz}~Emberger}
\IDRauth{19}{\DTLinitials{Danniel}~Esperante}
\IDRauth{12}{\DTLinitials{R\'emi}~Et\'e}
\IDRauth{21}{\DTLinitials{Yaquan}~Fang}
\IDRauth{12}{\DTLinitials{Oleksiy}~Fedorchuk}
\IDRauth{1}{\DTLinitials{Miroslaw}~Firlej}
\IDRauth{1}{\DTLinitials{Tomasz}~Fiutowski}
\IDRauth{51}{\DTLinitials{Ivor}~Fleck}
\IDRauth{8}{\DTLinitials{Nicolas}~Fourches}
\IDRauth{9}{\DTLinitials{Mar\'ia Cruz}~Fouz}
\IDRauth{45}{\DTLinitials{Kurt}~Francis}
\IDRauth{18}{\DTLinitials{Kazuki}~Fujii}
\IDRauth{30}{\DTLinitials{Keisuke}~Fujii}
\IDRauth{19}{\DTLinitials{Esteban}~Fullana}
\IDRauth{48}{\DTLinitials{Takahiro}~Fusayasu}
\IDRauth{19}{\DTLinitials{Juan}~Fuster}
\IDRauth{12}{\DTLinitials{Peter}~G\"ottlicher}
\IDRauth{12}{\DTLinitials{Karsten}~Gadow}
\IDRauth{12}{\DTLinitials{Frank}~Gaede}
\IDRauth{34}{\DTLinitials{Alexandre}~Gallas}
\IDRauth{8}{\DTLinitials{Serguei}~Ganjour}
\IDRauth{19}{\DTLinitials{Ignacio}~Garc\'ia}
\IDRauth{9}{\DTLinitials{Hector}~Garc\'ia Cabrera}
\IDRauth{26}{\DTLinitials{Guillaume}~Garillot}
\IDRauth{14}{\DTLinitials{Erika}~Garutti}
\IDRauth{35}{\DTLinitials{Franck}~Gastaldi}
\IDRauth{38}{\DTLinitials{Patrick}~Ghislain}
\IDRauth{25}{\DTLinitials{Mathieu}~Goffe}
\IDRauth{19}{\DTLinitials{Pablo}~Gomis}
\IDRauth{21}{\DTLinitials{Wenxuan}~Gong}
\IDRauth{34}{\DTLinitials{Alexandre}~Gonnin}
\IDRauth{22}{\DTLinitials{Deepanjali}~Goswami}
\IDRauth{33}{\DTLinitials{Kiichi}~Goto}
\IDRauth{42}{\DTLinitials{Christian}~Graf}
\IDRauth{12}{\DTLinitials{Ingrid-Maria}~Gregor}
\IDRauth{26}{\DTLinitials{Gerald}~Grenier}
\IDRauth{35}{\DTLinitials{R\'ei}~Guillaumat}
\IDRauth{12}{\DTLinitials{Moritz}~Habermehl}
\IDRauth{12}{\DTLinitials{Lars}~Hagge}
\IDRauth{12,b}{\DTLinitials{Oskar}~Hartbrich}
\IDRauth{44}{\DTLinitials{Fred}~Hartjes}
\IDRauth{12}{\DTLinitials{Hans}~Henschel}
\IDRauth{12}{\DTLinitials{Daniel}~Heuchel}
\IDRauth{23}{\DTLinitials{Salvador}~Hidalgo}
\IDRauth{25}{\DTLinitials{Abdelkader}~Himmi}
\IDRauth{21}{\DTLinitials{Tao}~Hu}
\IDRauth{25}{\DTLinitials{Christine}~Hu-Guo}
\IDRauth{26}{\DTLinitials{Jean-Christophe Tibor}~Ianigro}
\IDRauth{1}{\DTLinitials{Marek}~Idzik}
\IDRauth{34}{\DTLinitials{Adrian}~Irles}
\IDRauth{50}{\DTLinitials{Hiroki}~Ishihara}
\IDRauth{30}{\DTLinitials{Akimasa}~Ishikawa}
\IDRauth{39}{\DTLinitials{Leif}~J\"onsson}
\IDRauth{25}{\DTLinitials{Kimmo}~Jaaskelainen}
\IDRauth{30}{\DTLinitials{Daniel}~Jeans}
\IDRauth{34}{\DTLinitials{Jimmy}~Jeglot}
\IDRauth{64}{\DTLinitials{Goran}~Kacarevic}
\IDRauth{25}{\DTLinitials{Maciej}~Kachel}
\IDRauth{18}{\DTLinitials{Shogo}~Kajiwara}
\IDRauth{62}{\DTLinitials{Jan}~Kalinowski}
\IDRauth{5}{\DTLinitials{Jochen}~Kaminski}
\IDRauth{18}{\DTLinitials{Yoshio}~Kamiya}
\IDRauth{12}{\DTLinitials{Robert}~Karl}
\IDRauth{18}{\DTLinitials{Yu}~Kato}
\IDRauth{31}{\DTLinitials{Yukihiro}~Kato}
\IDRauth{12}{\DTLinitials{Shin-ichi}~Kawada}
\IDRauth{33}{\DTLinitials{Kiyotomo}~Kawagoe}
\IDRauth{13}{\DTLinitials{Sameen A}~Khan}
\IDRauth{12}{\DTLinitials{Claus}~Kleinwort}
\IDRauth{44}{\DTLinitials{Peter}~Kluit}
\IDRauth{30}{\DTLinitials{Makoto}~Kobayashi}
\IDRauth{16}{\DTLinitials{Christian}~Koffmane}
\IDRauth{61,d}{\DTLinitials{Sachio}~Komamiya}
\IDRauth{37}{\DTLinitials{Sergey}~Korpachev}
\IDRauth{50}{\DTLinitials{Katsushige}~Kotera}
\IDRauth{12}{\DTLinitials{Uwe}~Kr\"amer}
\IDRauth{12}{\DTLinitials{Katja}~Kr\"uger}
\IDRauth{35}{\DTLinitials{Jonas}~Kunath}
\IDRauth{30}{\DTLinitials{Masakazu}~Kurata}
\IDRauth{26}{\DTLinitials{Tibor}~Kurca}
\IDRauth{24,c}{\DTLinitials{Jir\'i}~Kvasnicka}
\IDRauth{38}{\DTLinitials{Didier}~Lacour}
\IDRauth{26}{\DTLinitials{Imad}~Laktineh}
\IDRauth{12}{\DTLinitials{Wolfgang}~Lange}
\IDRauth{12}{\DTLinitials{Suvi-Leena}~Lehtinen}
\IDRauth{20}{\DTLinitials{Tadeusz}~Lesiak}
\IDRauth{52}{\DTLinitials{Aharon}~Levy}
\IDRauth{52}{\DTLinitials{Itamar}~Levy}
\IDRauth{26}{\DTLinitials{Bo}~Li}
\IDRauth{21}{\DTLinitials{Gang}~Li}
\IDRauth{44}{\DTLinitials{Cornelis}~Ligtenberg}
\IDRauth{12}{\DTLinitials{Benno}~List}
\IDRauth{12}{\DTLinitials{Jenny}~List}
\IDRauth{18}{\DTLinitials{Linghui}~Liu}
\IDRauth{21}{\DTLinitials{Yong}~Liu}
\IDRauth{21}{\DTLinitials{Zhenan}~Liu}
\IDRauth{12}{\DTLinitials{Wolfgang}~Lohmann}
\IDRauth{35}{\DTLinitials{Marc}~Louzir}
\IDRauth{12}{\DTLinitials{Shaojun}~Lu}
\IDRauth{39}{\DTLinitials{Bjoern}~Lundberg}
\IDRauth{34}{\DTLinitials{Jihane}~Maalmi}
\IDRauth{35}{\DTLinitials{Fr\'ed\'eric}~Magniette}
\IDRauth{49}{\DTLinitials{Nayana}~Majumdar}
\IDRauth{30}{\DTLinitials{Yasuhiro}~Makida}
\IDRauth{12}{\DTLinitials{Paul}~Malek}
\IDRauth{9}{\DTLinitials{Jes\'us}~Mar\'in}
\IDRauth{63}{\DTLinitials{John}~Marshall}
\IDRauth{14}{\DTLinitials{Stephan}~Martens}
\IDRauth{46}{\DTLinitials{Gisele}~Martin-Chassard}
\IDRauth{60}{\DTLinitials{Lucia}~Masetti}
\IDRauth{18}{\DTLinitials{Ryunosuke}~Masuda}
\IDRauth{26}{\DTLinitials{Herve}~Mathez}
\IDRauth{30}{\DTLinitials{Takeshi}~Matsuda}
\IDRauth{47}{\DTLinitials{Kirk T}~McDonald}
\IDRauth{17}{\DTLinitials{Dmitry}~Mikhaylov}
\IDRauth{26}{\DTLinitials{Laurent}~Mirabito}
\IDRauth{37}{\DTLinitials{Sergey}~Miroshin}
\IDRauth{15}{\DTLinitials{Winfried}~Mitaroff}
\IDRauth{30}{\DTLinitials{Akiya}~Miyamoto}
\IDRauth{30}{\DTLinitials{Takahiro}~Mizuno}
\IDRauth{39}{\DTLinitials{Ulf}~Mj\"ornmark}
\IDRauth{18}{\DTLinitials{Takanori}~Mogi}
\IDRauth{12}{\DTLinitials{Gudrid}~Moortgat-Pick}
\IDRauth{25}{\DTLinitials{Fr\'ed\'eric}~Morel}
\IDRauth{55}{\DTLinitials{Sergio}~Moreno}
\IDRauth{18}{\DTLinitials{Toshinori}~Mori}
\IDRauth{1}{\DTLinitials{Jakub}~Moron}
\IDRauth{11}{\DTLinitials{David}~Moya}
\IDRauth{49}{\DTLinitials{Supratik}~Mukhopadhyay}
\IDRauth{58}{\DTLinitials{Yonathan}~Munwes}
\IDRauth{53}{\DTLinitials{Tadashi}~Nagamine}
\IDRauth{35}{\DTLinitials{J\'erome}~Nanni}
\IDRauth{8}{\DTLinitials{Olivier}~Napoly}
\IDRauth{28}{\DTLinitials{Shinya}~Narita}
\IDRauth{9}{\DTLinitials{Jose Javier}~Navarrete}
\IDRauth{28}{\DTLinitials{Kentaro}~Negishi}
\IDRauth{16}{\DTLinitials{Jelena}~Ninkovic}
\IDRauth{51}{\DTLinitials{Amir}~Noori Shirazi}
\IDRauth{30}{\DTLinitials{Tomohisa}~Ogawa}
\IDRauth{30}{\DTLinitials{Takahiro}~Okamura}
\IDRauth{30}{\DTLinitials{Tsunehiko}~Omori}
\IDRauth{43}{\DTLinitials{Hiroaki}~Ono}
\IDRauth{18}{\DTLinitials{Wataru}~Ootani}
\IDRauth{39}{\DTLinitials{Anders}~Oskarsson}
\IDRauth{39}{\DTLinitials{Lennart}~\"Ostermann}
\IDRauth{21}{\DTLinitials{Qun}~Ouyang}
\IDRauth{34}{\DTLinitials{Roman}~P\"oschl}
\IDRauth{38}{\DTLinitials{Jean-Marc}~Parraud}
\IDRauth{20}{\DTLinitials{Bogdan}~Pawlik}
\IDRauth{23}{\DTLinitials{Giulio}~Pellegrini}
\IDRauth{19}{\DTLinitials{Martin}~Perello}
\IDRauth{25}{\DTLinitials{Alejandro}~Perez}
\IDRauth{25}{\DTLinitials{Hung}~Pham}
\IDRauth{27}{\DTLinitials{Javier}~Piedrafita}
\IDRauth{35}{\DTLinitials{Thomas}~Pierre-Emile}
\IDRauth{57}{\DTLinitials{Antoine}~Pingault}
\IDRauth{12}{\DTLinitials{Olin Lyod}~Pinto}
\IDRauth{24}{\DTLinitials{Ivo}~Pol\'ak}
\IDRauth{41}{\DTLinitials{Elena}~Popova}
\IDRauth{22}{\DTLinitials{Poulose}~Poulose}
\IDRauth{27}{\DTLinitials{Alvaro}~Pradas}
\IDRauth{12}{\DTLinitials{Volker}~Prahl}
\IDRauth{3}{\DTLinitials{Tony}~Price}
\IDRauth{12}{\DTLinitials{Ambra}~Provenza}
\IDRauth{9}{\DTLinitials{Jes\'us}~Puerta Pelayo}
\IDRauth{21}{\DTLinitials{Huirong}~Qi}
\IDRauth{12}{\DTLinitials{Yasser}~Radkhorrami}
\IDRauth{46}{\DTLinitials{Ludovic}~Raux}
\IDRauth{44}{\DTLinitials{Gerhard}~Raven}
\IDRauth{12}{\DTLinitials{Mathias}~Reinecke}
\IDRauth{34}{\DTLinitials{Francois}~Richard}
\IDRauth{16}{\DTLinitials{Rainer}~Richter}
\IDRauth{12}{\DTLinitials{Sabine}~Riemann}
\IDRauth{60}{\DTLinitials{Maria Soledad}~Robles Manzano}
\IDRauth{29}{\DTLinitials{Christopher}~Rogan}
\IDRauth{14}{\DTLinitials{Jack}~Rolph}
\IDRauth{19}{\DTLinitials{Eduardo}~Ros}
\IDRauth{60}{\DTLinitials{Anna}~Rosmanitz}
\IDRauth{29}{\DTLinitials{Christophe}~Royon}
\IDRauth{21}{\DTLinitials{Manqi}~Ruan}
\IDRauth{11}{\DTLinitials{Alberto}~Ruiz-Jimeno}
\IDRauth{53}{\DTLinitials{Tomoyuki}~Sanuki}
\IDRauth{12}{\DTLinitials{Swathi Kollassery}~Sasikumar}
\IDRauth{53}{\DTLinitials{Yo}~Sato}
\IDRauth{17}{\DTLinitials{Andrey}~Saveliev}
\IDRauth{17}{\DTLinitials{Valery}~Saveliev}
\IDRauth{12}{\DTLinitials{Oliver}~Sch\"afer}
\IDRauth{60}{\DTLinitials{Christian}~Schmitt}
\IDRauth{12}{\DTLinitials{Uwe}~Schneekloth}
\IDRauth{12}{\DTLinitials{Thomas}~Sch\"orner-Sadenius}
\IDRauth{58}{\DTLinitials{Hans-Christian}~Schultz-Coulon}
\IDRauth{12}{\DTLinitials{Sergej}~Schuwalow}
\IDRauth{12}{\DTLinitials{Felix}~Sefkow}
\IDRauth{46}{\DTLinitials{Nathalie}~Seguin-Moreau}
\IDRauth{33}{\DTLinitials{Izumi}~Sekiya}
\IDRauth{42}{\DTLinitials{Ronald}~Settles}
\IDRauth{4}{\DTLinitials{L}~Shekhtman}
\IDRauth{58}{\DTLinitials{Wei}~Shen}
\IDRauth{50}{\DTLinitials{Ryousuke}~Shiraz}
\IDRauth{28}{\DTLinitials{Aiko}~Shoji}
\IDRauth{42}{\DTLinitials{Frank}~Simon}
\IDRauth{12}{\DTLinitials{Klaus}~Sinram}
\IDRauth{64}{\DTLinitials{Ivan}~Smiljanic}
\IDRauth{25}{\DTLinitials{Matthieu}~Specht}
\IDRauth{12}{\DTLinitials{Richard}~Stromhagen}
\IDRauth{12}{\DTLinitials{Yuji}~Sudo}
\IDRauth{33}{\DTLinitials{Taikan}~Suehara}
\IDRauth{30}{\DTLinitials{Yasuhiro}~Sugimoto}
\IDRauth{48}{\DTLinitials{Akira}~Sugiyama}
\IDRauth{1}{\DTLinitials{Krzysztof}~Swientek}
\IDRauth{59}{\DTLinitials{Tohru}~Takahashi}
\IDRauth{50}{\DTLinitials{Tohru}~Takeshita}
\IDRauth{50}{\DTLinitials{Yukinaru}~Tamaya}
\IDRauth{18}{\DTLinitials{Tomohiko}~Tanabe}
\IDRauth{30}{\DTLinitials{Toshiaki}~Tauchi}
\IDRauth{4}{\DTLinitials{Valery}~Telnov}
\IDRauth{1}{\DTLinitials{Pzremyslaw}~Terlecki}
\IDRauth{34}{\DTLinitials{Alice}~Thiebault}
\IDRauth{18}{\DTLinitials{Junping}~Tian}
\IDRauth{44}{\DTLinitials{Jan}~Timmermans}
\IDRauth{8}{\DTLinitials{Maxim}~Titov}
\IDRauth{12}{\DTLinitials{Huong Lan}~Tran}
\IDRauth{50}{\DTLinitials{Reima}~Tread}
\IDRauth{12}{\DTLinitials{Dimitra}~Tsionou}
\IDRauth{18}{\DTLinitials{Naoki}~Tsuji}
\IDRauth{8}{\DTLinitials{Boris}~Tuchming}
\IDRauth{57}{\DTLinitials{Michael}~Tytgat}
\IDRauth{53}{\DTLinitials{Takayuki}~Ueno}
\IDRauth{33}{\DTLinitials{Yuto}~Uesugi}
\IDRauth{50}{\DTLinitials{Satoru}~Uozumi}
\IDRauth{25}{\DTLinitials{Isabelle}~Valin}
\IDRauth{10}{\DTLinitials{Claude}~Vall\'ee}
\IDRauth{44}{\DTLinitials{Harry}~van der Graaf}
\IDRauth{44}{\DTLinitials{Naomi}~van der Kolk}
\IDRauth{29}{\DTLinitials{Brian}~van Doren}
\IDRauth{9}{\DTLinitials{Antonio}~Verdugo de Osa}
\IDRauth{19}{\DTLinitials{Guillem}~Vidal}
\IDRauth{35}{\DTLinitials{Henri}~Videau}
\IDRauth{11}{\DTLinitials{Iv\'an}~Vila}
\IDRauth{19}{\DTLinitials{Miguel Angel}~Villarrejo}
\IDRauth{17}{\DTLinitials{Denis}~Volkov}
\IDRauth{19}{\DTLinitials{Marcel}~Vos}
\IDRauth{64}{\DTLinitials{Natasa}~Vukasinovic}
\IDRauth{12}{\DTLinitials{Yan}~Wang}
\IDRauth{32}{\DTLinitials{Takashi}~Watanabe}
\IDRauth{3}{\DTLinitials{Nigel}~Watson}
\IDRauth{51}{\DTLinitials{Ulrich}~Werthenbach}
\IDRauth{29}{\DTLinitials{Graham W}~Wilson}
\IDRauth{56,c}{\DTLinitials{Matthew}~Wing}
\IDRauth{3}{\DTLinitials{Alasdair}~Winter}
\IDRauth{25}{\DTLinitials{Marc}~Winter}
\IDRauth{20}{\DTLinitials{Tomasz}~Wojto\'n}
\IDRauth{53}{\DTLinitials{Hitoshi}~Yamamoto}
\IDRauth{18}{\DTLinitials{Satoru}~Yamashita}
\IDRauth{53}{\DTLinitials{Ryo}~Yonamine}
\IDRauth{33}{\DTLinitials{Tamaki}~Yoshioka}
\IDRauth{21}{\DTLinitials{Boxiang}~Yu}
\IDRauth{21}{\DTLinitials{Dan}~Yu}
\IDRauth{58}{\DTLinitials{Zhenxiong}~Yuan}
\IDRauth{30}{\DTLinitials{Keita}~Yumino}
\IDRauth{62}{\DTLinitials{Aleksander Filip}~Zarnecki}
\IDRauth{6}{\DTLinitials{Christian}~Zeitnitz}
\IDRauth{34}{\DTLinitials{Dirk}~Zerwas}
\IDRauth{21}{\DTLinitials{Hang}~Zhao}
\IDRauth{21}{\DTLinitials{Jingzhou}~Zhao}
\IDRauth{25}{\DTLinitials{Ruiguang}~Zhao}
\IDRauth{25}{\DTLinitials{Y\"ue}~Zhao}
\IDRauth{21}{\DTLinitials{Hongbo}~Zhu}
\IDRauthl{45}{\DTLinitials{Vishnu}~Zutshi}
\vspace{3mm}
\IDRinst{1}{AGH University of Science and Technology, Faculty of Physics and Applied Computer Science, Krakow, Poland}
\IDRinst{2}{University of Bergen, Institute for Physics and Technology, Bergen, Norway}
\IDRinst{3}{University of Birmingham, School of Physics and Astronomy, Edgbaston, UK}
\IDRinst{4}{Budker Institute of Nuclear Physics, Siberian Branch Russian Academy of Sciences and Novosibirsk State University, Novosibirsk, Russia}
\IDRinst{5}{University of Bonn, Physikalisches Institut, Bonn, Germany}
\IDRinst{6}{Bergische Universit\"at Wuppertal, Fakult\"at 4/ Physik, Wuppertal, Germany}
\IDRinst{7}{Carleton University, Ottawa, Canada}
\IDRinst{8}{Irfu, CEA, Universit\'e Paris Saclay, Gif sur Yvette, France}
\IDRinst{9}{CIEMAT - Centro de Investigaciones Energ\'eticas, Medioambientales y Tecnol\'ogicas, Madrid, Spain}
\IDRinst{10}{Aix Marseille Univ, CNRS/IN2P3, CPPM, Marseille, France}
\IDRinst{11}{CSIC - University of Cantabria, Instituto de F\'isica de Cantabria, Santander, Spain}
\IDRinst{12}{Deutsches Elektronen-Synchrotron, Hamburg, Germany}
\IDRinst{13}{Dhofar University, Salalah, Oman}
\IDRinst{14}{University of Hamburg, Faculty of Mathematics, Informatics and Natural Sciences, Hamburg, Germany}
\IDRinst{15}{Institute of High Energy Physics, Austrian Academy of Science, Wien, Austria}
\IDRinst{16}{Halbeiterlabor der Max-Planck-Gesellschaft, M\"unchen, Germany}
\IDRinst{17}{Keldysh Institute of Applied Mathematics, Russian Academy of Sciences, Moscow, Russia}
\IDRinst{18}{International Center for Elementary Particle Physics (ICEPP), The University of Tokyo, Tokyo, Japan}
\IDRinst{19}{CSIC-University of Valencia, Instituto de F\'isica Corpuscular, Paterna, Spain}
\IDRinst{20}{H.Niewodniczanski Institute of Nuclear Physics, Polish Academy of Sciences (IFJ PAN), Krak\'ow, Poland}
\IDRinst{21}{Institute of High Energy Physics, Chinese Academy of Science, Beijing, China}
\IDRinst{22}{Indian Institute of Technology Guwahati, Assam, India}
\IDRinst{23}{Centro Nacional de Microelectr\'onica (IMB-CNM-CSIC), Barcelona, Spain}
\IDRinst{24}{Institute of Physics of the Czech Academy of Sciences, Prague 8, Czech Republic}
\IDRinst{25}{Institut Pluridisciplinaire Hubert Curien, Strasbourg, France}
\IDRinst{26}{Institut de Physique des deux infinis de Lyon, Villeurbanne, France}
\IDRinst{27}{Instituto Tecnol\'ogico de Arag\`on, Zaragoza, Spain}
\IDRinst{28}{Iwate University, Morioka, Japan}
\IDRinst{29}{University of Kansas, Department of Physics and Astronomy, Lawrence, KS, USA}
\IDRinst{30}{High Energy Accelerator Research Organisation, KEK, Tsukuba, Ibaraki, Japan}
\IDRinst{31}{Kindai University, Department of Physics, Higashi Osaka, Japan}
\IDRinst{32}{Kogakuin University, Shinjuku-ku, Tokyo, Japan}
\IDRinst{33}{Kyushu University, Department of Physics, Research Center for Advanced Particle Physics, Fukuoka, Japan}
\IDRinst{34}{Universit\'e Paris-Saclay, CNRS/IN2P3, IJCLab, Orsay, France}
\IDRinst{35}{Laboratoire Leprince-Ringuet, Institut Polytechnique de Paris, Palaiseau, France}
\IDRinst{36}{Laboratoire de Physique de Clermont, CNRS/IN2P3, Aubi\`ere, France}
\IDRinst{37}{P.N. Lebedev Physical Institute of the Russian Academy of Sciences (LPI), Moscow, Russia}
\IDRinst{38}{Laboratoire de Physique Nucl\'eaire et de Hautes Energies (LPNHE), Sorbonne Universit\'e, Paris-Diderot Sorbonne Paris Cit\'e, CNRS/IN2P3, Paris, France}
\IDRinst{39}{Lund University, Physics Department, Lund, Sweden}
\IDRinst{40}{McGill University, Department of Physics, Montreal, Quebec, Canada}
\IDRinst{41}{National Research Nuclear University, Moscow, Russia}
\IDRinst{42}{Max-Planck-Institut {f\"u}r Physik, M\"unchen, Germany}
\IDRinst{43}{Nippon Dental University School of Life Dentistry at Niigata, Niigata, Japan}
\IDRinst{44}{Nikhef, National Institute for Subatomic Physics, Amsterdam, Netherlands}
\IDRinst{45}{Northern Illinois University, Department of Physics, DeKalb, USA}
\IDRinst{46}{Centre de micro\'electronique OMEGA, Institut Polytechnique de Paris, Palaiseau, France}
\IDRinst{47}{Princeton University, Department of Physics, Princeton, NJ, USA}
\IDRinst{48}{Saga University, Department of Physics, Saga, Japan}
\IDRinst{49}{Saha Institute of Nuclear Physics, Kolkata, India}
\IDRinst{50}{Shinshu University, Department of Physics, Matsumoto, Japan}
\IDRinst{51}{Universit\"at Siegen, Fakult\"at IV, Department of Physik, Siegen, Germany}
\IDRinst{52}{Tel Aviv University, Raymond \& Beverly Sackler School of Physics \& Astronomy, Tel Aviv, Israel}
\IDRinst{53}{Tohoku University, Sendai, Japan}
\IDRinst{54}{Taras Shevchenko National University of Kyiv (TSNUK), Kyiv, Ukraine}
\IDRinst{55}{University of Barcelona, Barcelona, Spain}
\IDRinst{56}{University College London, London, UK}
\IDRinst{57}{University of Ghent, Dept. of Physics and Astronomy, Gent, Belgium}
\IDRinst{58}{Universit\"at Heidelberg, Kirchhoff-Institute f\"ur Physik, Heidelberg, Germany}
\IDRinst{59}{Hiroshima University, Higashi Hiroshima, Japan}
\IDRinst{60}{Johannes Gutenberg-Universität Mainz, Institute of Physics, Mainz, Germany}
\IDRinst{61}{The University of Tokyo, Graduate School of Science, Tokyo, Japan}
\IDRinst{62}{Faculty of Physics, University of Warsaw, Warszawa, Poland}
\IDRinst{63}{University of Warwick, Coventry, United Kingdom}
\IDRinst{64}{VINCA Institute of Nuclear Sciences, University of Belgrade, Belgrade, Serbia}

\vspace{3mm}
\IDRinst{a}{{\it {now at} Aachen Institute for Advanced Study in Computational Engineering Science (AICES),
RWTH Aachen University, Aachen, Germany}}
\IDRinst{b}{\it{also at} University of Hawai at Menoa, Department of Physics and Astronomy, Honolulu, Hawaii 96822}
\IDRinst{c}{ \it{also at} Deutsches Elektronen Synchrotron, Hamburg and Zeuthen, Germany}
\IDRinst{d}{\it{now at} Waseda Research Institute for Science and Engineering, Tokyo, Japan\vspace{3 mm}}

\end{center}

\chapter*{Acknowledgements}
We acknowledge the support by: \\
the Natural Sciences and Engineering Research Council (NSERC) (Canada);\\ 
DRF (Direction de la Recherche fondamentale), CEA/DRF/IRFU, Universite Paris Saclay, 
and CNRS/IN2P3 (France);\\
Deutsche Forschungsgemeinschaft (DFG) under Germany{'}s Excellence Strategy -- EXC 2121 Quantum Universe -- 390833306,
the German Federal Ministry of Education and Research (BMBF),
the Helmholtz Association (Germany);\\
the Israel Science Foundation (ISF), Israel German Foundation (GIF), the I-CORE program of the Israel Planning and Budgeting Committee, Israel Academy of Sciences and Humanities (Israel); \\
MEXT and JSPS under Grants-in-Aid for Science Research 23000002, 15H02083, 16H02173, 16H02176, and 17H02882 (Japan); \\
the National Science Centre, Poland, within the OPUS project 
under contract UMO-2017/25/B/ST2/00496 (2018-2021) and the HARMONIA 
project under contract UMO-2015/18/M/ST2/00518 (2016-2020) (Poland);\\
the Russian government under Grant RFBR 20-52-12056 (Russia);\\
the Spanish government  MINECO/AEI through the Particle Physics program and the Unidad de Excelencia Maria de Maeztu, ref. MDM-2017-0765 (Spain); \\
MPNTR through the national project OI171012 (Serbia); \\
the Science and Technology Facilities Council, UK; \\
the Department of Energy (DOE) and the National Science Foundation (NSF), USA;\\
and the European H2020 project AIDA-2020, GA no. 654168.

We would like to thank the LCC generator working group for providing the generator files used in the Monte Carlo productions for the different analyses. This work has greatly benefited
from computing services provided by the ILC Virtual Organization, supported by the national resource
providers of the EGI Federation and the Open Science GRID, and of those of the German National Analysis Facility (NAF).

We like to particularly acknowledge the very fruitful discussions of the document during the final stages with Katsuo Tokushuku (KEK), Lucie Linssen (CERN) and Paul Grannis (State University of New York at Stony Brook).

%


%
%
%

%
%
%
\cleardoublepage\setcounter{page}{1}
%
%
%
%
%
%
\tableofcontents 
\addcontentsline{toc}{chapter}{Contents}
%
\setcounter{secnumdepth}{0}
%
%
%
\mainmatter
%
%
\setcounter{page}{1}
%
%
\setcounter{chapter}{0}
\setcounter{secnumdepth}{3} \setcounter{subsection}{0}
\setcounter{subsubsection}{0}
\usecounter{subsubsection}

\renewcommand{\DBDnumbers}{-numbers}

\chapter{Introduction}
\label{chap:introduction}
The ILD detector is proposed for an electron-positron collider with collision centre-of-mass energies from 90~\GeV~to about 1~\TeV. It has been developed over the last 10 years by an international team of scientists with the goal to design and eventually propose a fully integrated detector, primarily for the International Linear Collider, ILC.

The fundamental ideas and concepts behind the ILD detector have been discussed in two previous documents, the letter of intent \cite{ild:bib:ILDloi} and the detailed baseline document, DBD \cite{ild:bib:ilddbd}. 
The ILD concept has been scrutinized by international groups at different occasions. After the publication of the letter of intent \cite{ild:bib:ILDloi} an international expert team reviewed ILD alongside two other detector concepts, SiD \cite{bib:sid:loi} and the fourth detector concept \cite{bib:4th:loi}. SiD and ILD were both validated as potential candidate experiments at the ILC. 

ILD has been designed as a multi-purpose detector. It should deliver outstanding physics performance for collision energies between 90~\GeV~and 1~TeV. The detector has been optimized to perform excellently at the initial  energy of 250~\GeV, while maintaining full physics capabilities at higher energies. 

 A central element of the design is the capability of the detector to reconstruct precisely complex hadronic final states as well as events with leptons or missing energy in the final state. To achieve these goals, precision detector elements such as vertex detectors are combined with a large volume time projection chamber for excellent tracking efficiency and with a highly granular calorimeter, in an overall design philosophy called particle flow, developed for optimal global event reconstruction. 

In this document the current state of the design of the ILD detector is summarised. The technologies which are proposed for the different parts of the detector are introduced. An extensive benchmarking has been performed to demonstrate the physics performance of ILD. In order to ensure a detector adequacy for the whole ILC program, including possible future energy upgrades, benchmarking has been mostly done at a collision c.m.s energy of 500 GeV instead of the initial 250 GeV energy. Two detector configurations, a large and a smaller one, have been simulated in detail to provide guidelines towards an optimal balance between performance and cost. 

A lot of the work presented in this report is based on detector R\&D work which has taken place over the past decade to develop the necessary technologies. 
This work has been typically conducted within dedicated R\&D collaborations, which are independent but maintain very close connections to ILD. All technologies selected by ILD for its subsystems have been proven experimentally to meet the performance goals, or to come very close. Developing a very powerful detector concept over a long period of time requires balancing cutting edge technologies, which might become available while the concept is being developed, with safe and sound solutions. ILD in many cases is pursuing more than one technological option, to remain flexible and to be able to adapt to new developments. The concept group  wants to remain open and flexible to be prepared to select the most modern and most powerful technology once it is necessary. 

The ILD concept group has currently 64 member institutes from all around the world. The group has evolved into a proto-collaboration, which is positioning itself to move forward with a proposal for a detector at the ILC or other proposed electron-positron facilities.

The document starts with a short review of the science goals of the ILC, and how the goals can be achieved today with the detector technologies at hand. After a discussion of the ILC and the environment in which the experiment will take place, the detector is described in more detail, including the status of the development of the technologies foreseen for each subdetector. The integration of the different sub-systems into an integrated detector is discussed, as is the interface between the detector and the collider. This is followed by a concise summary of the benchmarking which has been performed in order to find an optimal balance between performance and cost. To the end the costing methodology used by ILD is presented, and an updated cost estimate for the detector is presented. The report closes with a summary of the current status and of planned future actions.

\chapter{Science with ILC}
\label{chap:science}

The ultimate goal of fundamental physics is to achieve a unified understanding of nature, including matter, forces and space-time. Within the Big Bang model, the early universe before formation of hydrogen atoms is characterised by a hot dense state which was opaque to light and cannot be directly observed by optical telescopes. Its behaviour is governed by microscopic physics of fundamental particles and forces. Energy frontier collider experiments provide a unique opportunity to investigate the physics of this early universe by reproducing in a controlled manner reactions that happen in very hot dense states. Our current understanding of microscopic physics is summarised in the Standard Model (SM) of particle physics. The SM consists of matter fermions (quarks and leptons), force carrying vector bosons (gauge bosons), and a scalar boson (Higgs boson) designed specifically to give masses, where needed, to otherwise massless SM particles by breaking the electroweak symmetry through Higgs condensation in the vacuum. The 2012 discovery of the Higgs boson at the LHC has completed the SM particle spectrum. 

Up to now, the SM has survived all the intense scrutiny through searches for new particles at energy frontier colliders such as LEP, HERA, Tevatron, and, most recently, the LHC, as well as through precision measurements of electroweak observables. However, while being extremely successful, the SM leaves many open questions such as: What is the nature of dark matter and dark energy?  Why does matter dominate over antimatter in the universe? What is the origin of neutrino masses and mixings? And why did the Higgs field fill the entire universe and why at the electroweak scale? These questions call for physics beyond the Standard Model (BSM). The Higgs boson, through its central role at the heart of the Standard Model and its very specific scalar nature, opens new avenues to address these issues. The Higgs boson discovery hence marked the start of a new voyage towards a better understanding of the physics of the early universe.

The ILC is going to explore uncharted waters at and beyond the electroweak scale, corresponding to about one trillionth of a second after the Big Bang, in order to directly address the question: why did the Higgs field fill the entire universe and why did it do so at the electroweak scale? Precision Higgs studies are of utmost importance to answer this question. The ILC will fully exploit the Higgs boson as a new tool to discover BSM physics responsible for the mystery of the electroweak scale.

A key observable is precision measurements of Higgs couplings to various SM particles. In the SM the Higgs boson's couplings to various SM particles are proportional to their masses. BSM physics modifies this proportionality, leaving its nature imprinted in the deviation pattern from the SM, which would allow to discriminate possible new phenomena such as existence of another dimension, a deeper stratum of matter, or a "multi-verse".
%
%

The ATLAS and CMS experiments at the LHC have measured the Higgs couplings to the vector bosons and the third generation matter fermions and found them consistent with the SM predictions within errors at the 10\,\% level~\cite{Cepeda:2019klc}. The high-luminosity LHC (HL-LHC) is expected to significantly improve the measurement accuracy to a 2-4\,\% level~\cite{Cepeda:2019klc}. These accuracies would be, however, insufficient, since deviations are typically expected at a level of 5\,\% or smaller due to the famous decoupling theorem~\cite{Ref:Decoupling}. To see the deviations and their pattern and to decide the future direction of particle physics, a percent level precision for various Higgs couplings is needed.

The primary goal of the ILD experiment is to measure various Higgs couplings to a \% level or better, so as to make a decisive step in deciphering the physics beyond the Standard Model, and to help to 
decide the future direction of particle physics from their deviation pattern. ILD is also to measure properties of the $W$ and $Z$ bosons and fermion pair production with unprecedented precision, while searching directly for new particles with unprecedented sensitivity, in order to further elucidate new physics that lies beyond the electroweak scale.

To achieve this goal, the ILD experiment has to reach a new level of precision in the reconstruction of final states. It aims at reconstructing every event at the level of quarks, leptons, and fundamental bosons including gauge bosons and the Higgs boson, so as to see events as if viewing a Feynman diagram. For this purpose, the ILD design has been optimized for Particle Flow Analysis (PFA), enforced by precision vertexing to tag heavy flavors and hermeticity to indirectly detect invisible particles such as neutrinos and dark matter particles. 

To benefit maximally from the energy upgradability of the ILC machine, ILD has been designed to be an experiment at a collider providing electron-positron collisions at centre-of-mass energies between 90\,GeV and about 1\,TeV, which allows a broad and long-term physics program that will evolve depending on the results from earlier stages. New particle searches at higher energies guided by a specific deviation pattern of Higgs couplings found at 250\,GeV are a typical case of such evolution. There is, however, a set of guaranteed physics of crucial importance at higher energy stages. The top quark, which is the heaviest in the SM and hence expected to be tightly coupled to the electroweak symmetry breaking sector, will enter our physics program at $\sqrt{s} \gtrsim 350$\,GeV. The 500\,GeV stage allows to directly access the top Yukawa coupling through $t\bar{t}H$ production and the triple Higgs coupling through $ZHH$ production. The measurement precision for the top Yukawa and the triple Higgs couplings will be significantly improved at 1\,TeV. An up-to-date presentation of the physics case of the ILC project can be found 
 in \cite{Ref:ILCESU1} and in an extended version of this paper \cite{Bambade:2019fyw}.

In the context of the ILD group a broad range of studies have been undertaken to understand the potential of the ILD experiment at the ILC up to 1\,TeV. 
It is important to point out that these studies are based on fully simulated events, using a realistic detector model and advanced reconstruction software, and in many cases include estimates of major systematic effects. This is particularly important when estimating the physics reach of ILC and ILD for specific measurements. Determining, for example, the branching ratios of the Higgs boson at the percent level depends critically on the detector performance, and thus on the quality of the event simulation and reconstruction.

It should also be emphasized that, in many cases, the performance used in the physics analyses has been tested against prototype experiments. The performance numbers decisive for vertexing, tracking, and calorimetry are all based on results from test beam experiments. These test beam results include those regarding single particle resolution for neutral and charged particles, particle separation in jets, one-to-one linking power between charged tracks and calorimeter clusters, and many aspects of detailed shower analyses. The PFA performance, a crucial element that decides the ILD physics reach, has been corroborated by these test beam results, though its full demonstration has to wait for a larger scale test beam experiment that combines these major detector aspects. 
A very brief summary of the main results from the ILD full simulation studies is given in the next sections of this chapter.

\section{Higgs Physics}
\begin{figure}[htbp]
\begin{center}
 \includegraphics[width=0.48\textwidth]{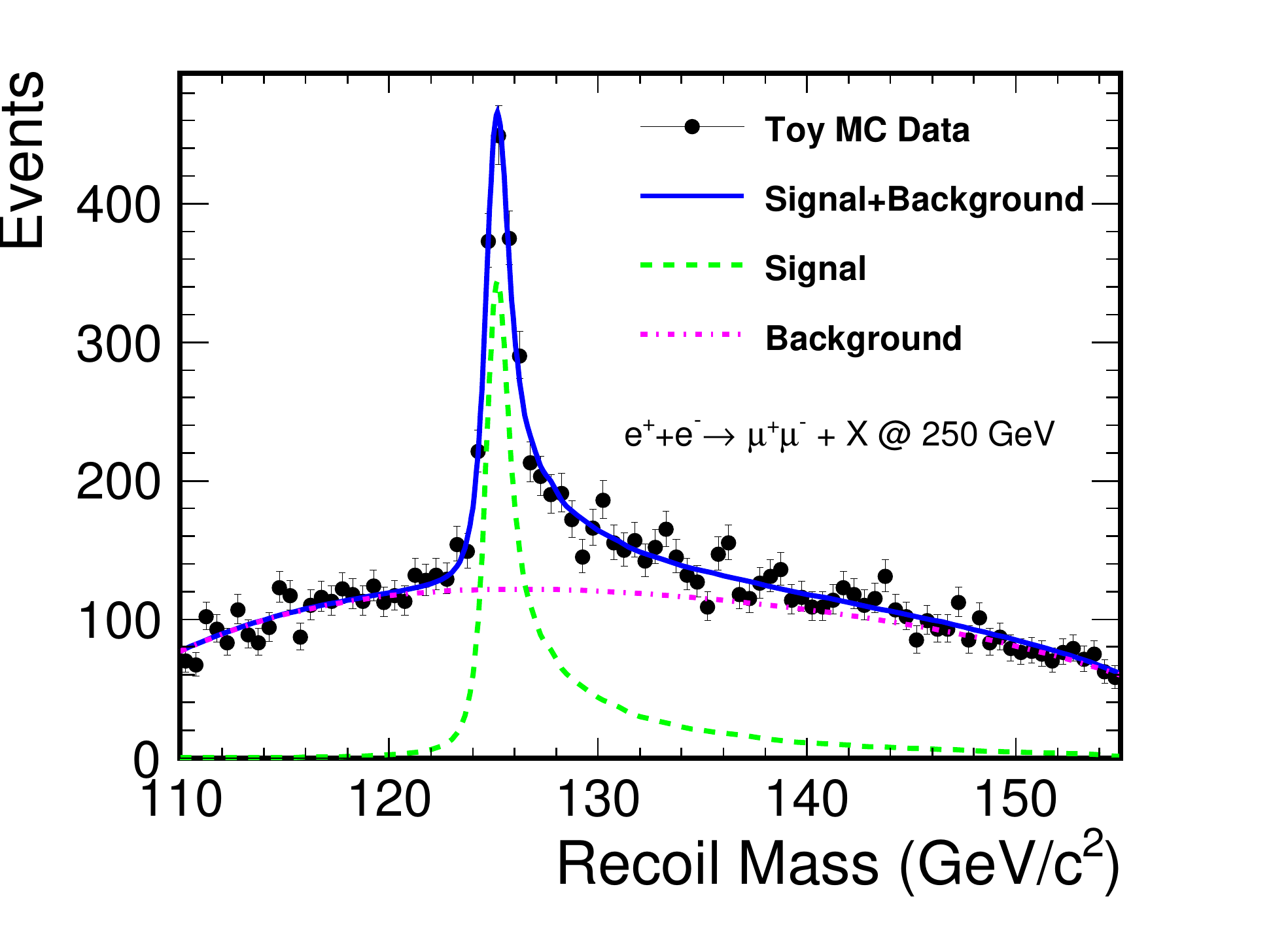}\hspace{2mm}
 \includegraphics[width=0.48\textwidth]{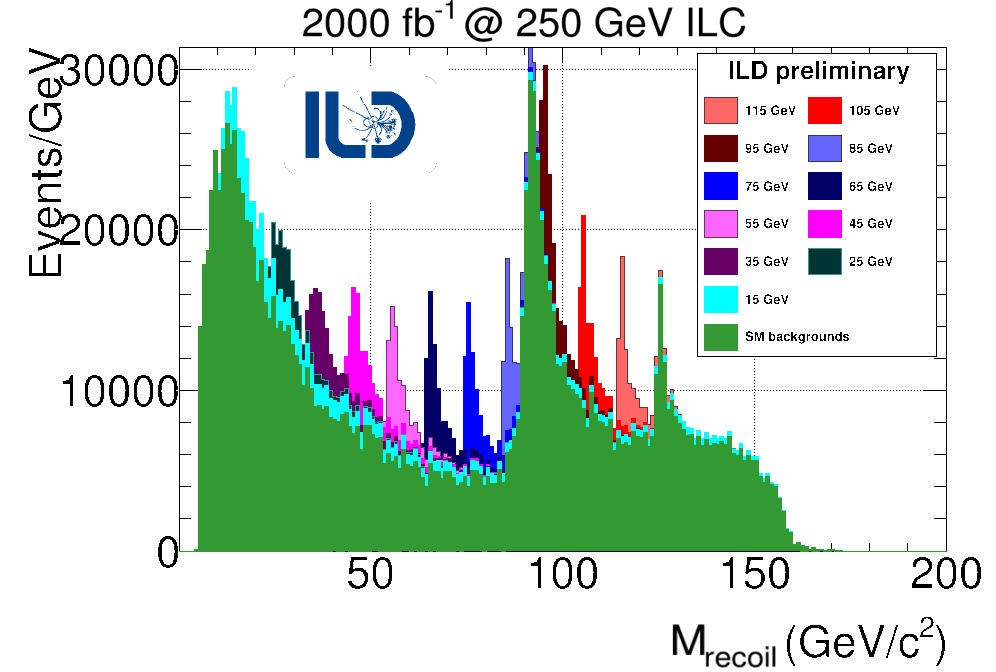}
\end{center}
\caption{Recoil mass distribution for $e^+e^- \to ZH$ followed by a $Z \to \mu^+\mu^-$ decay at $\sqrt{s}=250\,$GeV (left). A plot similar to the left figure but with an additional scalar particle recoiling against the Z boson (right). 
}
\label{fig:Mhrecoilmm}
\end{figure}

One of the most important Higgs measurements at the ILC is that of the $e^+e^- \to ZH$ process with the recoil mass technique, which allows to measure the total $ZH$ production cross section ($\sigma_{ZH}$) independently of the decay modes, and hence to absolutely normalise Higgs couplings. This is in contrast to measurements at the LHC, where initial state 4-momenta of colliding partons are unknown and hence the recoil mass technique cannot be applied. Figure \ref{fig:Mhrecoilmm} (left) shows the mass distribution of the system recoiling against the $\mu^+\mu^-$ pair from a $Z$ decay\cite{Yan:2016xyx}. A clear Higgs peak sticks out from the background, independently of the decay products of the Higgs boson. This will allow to determine the Higgs boson mass to $14$\,MeV with $2\,{\rm ab}^{-1}$ at $\sqrt{s}=250$\,GeV.
With the same integrated luminosity, by combining the $Z \to e^+e^-$ and $Z \to q\bar{q}$ channels, the absolutely normalised $\sigma_{ZH}$ can be measured to 2.0\,\% for both $e^-_Le^+_R$ and $e^-_Re^+_L$ beam polarisations.  

The same technique can be used to search for a new scalar boson or the Higgs to dark matter invisible decays. 
Figure\,\ref{fig:Mhrecoilmm} (right) shows expected recoil-mass peaks of additional scalar bosons with various different masses, assuming, here for display purpose, that they have the same coupling to $Z$ as the SM Higgs boson. 
Notice that the high momentum resolution from the ILD tracking is essential to see these sharp peaks with a width dominated by the beam energy spread instead of detector resolution.
The ILD will allow to put a 95\,\% C.L. upper limit of 1\,\% or better on the cross section normalized to the SM value for such extra scalar bosons with masses below 120\,GeV.
The search for the invisible Higgs decays can take advantage of the higher branching ratio for hadronic $Z$ decays. With the 2\,ab$^{-1}$ at $\sqrt{s}=250$\,GeV, ILD will be able to put a 95\,\% C.L. upper limit of 0.3\,\% on $BR(H\to \mathrm{invisible})$\cite{Ref:Hinvisible}.
%

The $\sigma_{ZH}$ measured with the recoil mass technique is used to extract the branching ratio of the Higgs boson to a pair of SM particles ($X$) from its corresponding $\sigma_{ZH} \times BR(H \to XX)$ measurement. Here the ILC's clean environment and ILD's excellent flavor tagging capability play a central role to separate $H \to c\bar{c}$ and $H \to gg$ decays, not to mention the dominant $H \to b\bar{b}$ decay.
While the direct detection of the $H \to c\bar{c}$ and $H \to gg$ decays is challenging at the LHC, the LHC can measure the ratios of branching ratios to high precision for decay modes with small branching ratios as long as they have clean signatures. For a Higgs decay such as $H \to \gamma\gamma$ which has a branching ratio of a per-mille level, its measurement at the ILC alone will be statistics-limited. The combination of the measurement of the ratio, $BR(H \to \gamma\gamma)/BR(H \to ZZ^*)$ at the LHC and that of  $BR(H \to ZZ^*)$ at the ILC allows the measurement of the $H\to\gamma\gamma$ coupling to 1\,\%, which is a typical example of LHC-ILC synergy.

\begin{figure}[htbp]
\begin{center}
 \includegraphics[width=0.5\textwidth]{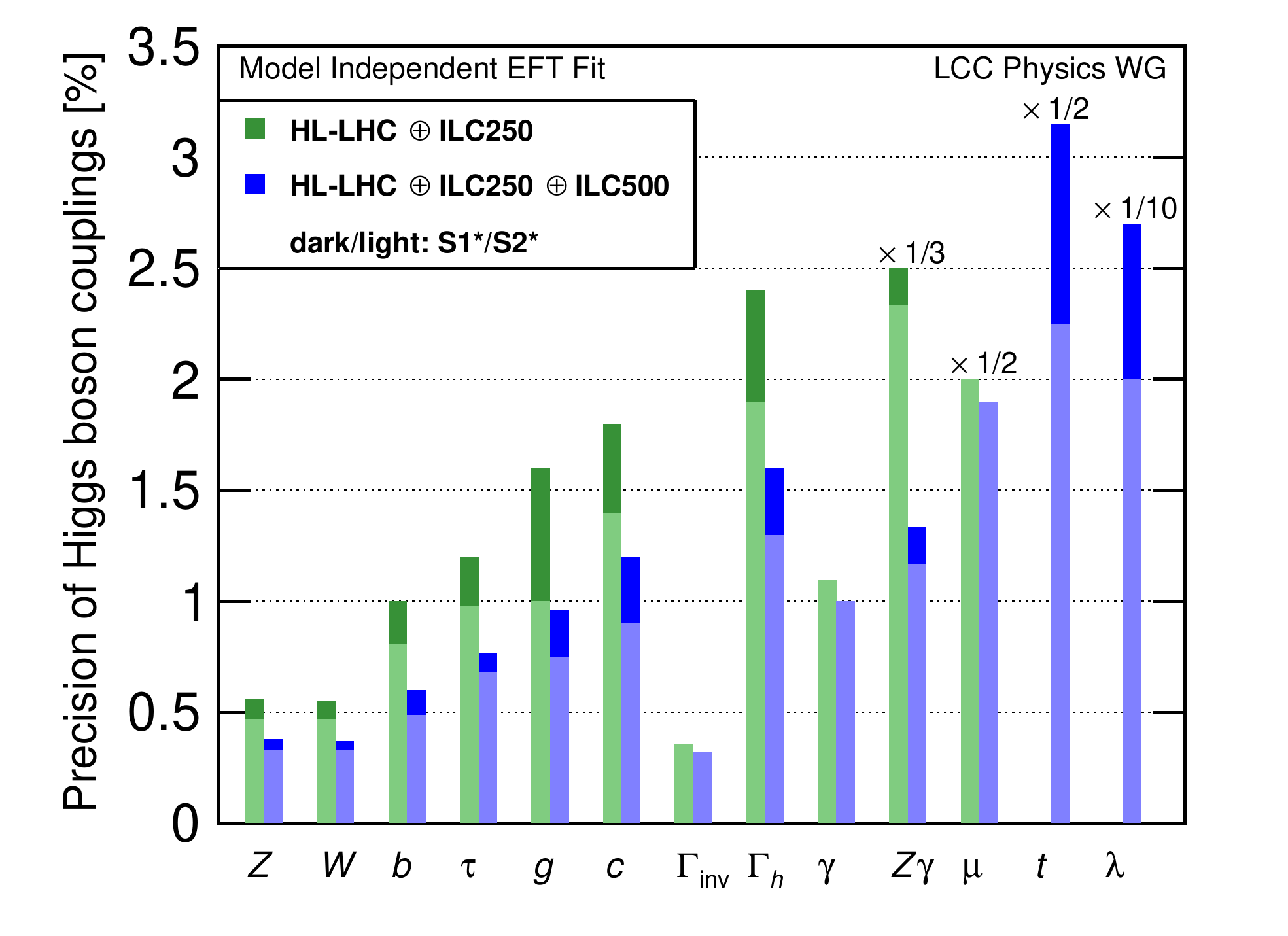}
\end{center}
\caption{Projected precisions of Higgs boson couplings from the SMEFT fit for the 250\,GeV ILC (green) and its upgrade to 500\,GeV (blue), where the lighter bands assume some future improvements in ILC measurements\cite{Ref:ILCESU1}.
}
\label{fig:Manhattan}
\end{figure}

In order to extract an absolutely normalized Higgs coupling, $g_{HXX}$, from the corresponding measured branching ratio, $BR(H \to XX)$, the total width, $\Gamma_H$, of the Higgs boson needs to be known. The total width is, however, only 4\,MeV in the SM, which is too small for a direct measurement. The most recent method to overcome this difficulty and to determine the Higgs couplings is to perform a global fit in the framework of the SM Effective Field Theory (SMEFT)\cite{Barklow:2017suo,Barklow:2017awn}. The SMEFT framework links observables directly involving the Higgs boson to those without the Higgs boson, through the $SU(2) \times U(1)$ gauge symmetry. This allows to make full use of all the measurements with ILD, not only those directly involving the Higgs boson, but all the others regarding precision electroweak observables or processes without the Higgs boson such as $e^+e^- \to W^+W^-$\cite{Karl:2019hes}. The beam polarizations play a crucial role here to lift degeneracies between different EFT operators and to control systematic uncertainties. Figure\,\ref{fig:Manhattan} shows the projected precision of various Higgs couplings from the SMEFT fit for the 250\,GeV ILC (green) and its upgrade to 500\,GeV (blue), where the lighter bands assume some future improvements in ILC measurements\cite{Ref:ILCESU1}.
It can be seen that, already at 250\,GeV, the precisions reach the target level of 1\,\% for most of the major couplings, which will be significantly improved at 500\,GeV.

As already mentioned above the 500\,GeV stage of the ILC will provide direct access to the triple Higgs coupling, through the double Higgs-strahlung process $e^+e^- \to ZHH$, which has its cross section maximum at around 500\,GeV. The measurement of the triple Higgs coupling is not only important for determining the shape of the Higgs potential but also crucial for testing the idea of electroweak baryogenesis. Models of this type require the electroweak phase transition to be strongly of first order, and predict an upward deviation, as large as 100\,\% from the SM prediction\cite{Ref:EWBG1,Ref:EWBG2,Ref:EWBG3}. With 4\,ab$^{-1}$ at 500\,GeV, the ILC can measure the triple Higgs coupling to 27\,\% for the SM case\cite{Ref:Claude}. The precision would be improved to 15\,\% if the upward deviation is +100\,\%, being accurate enough to test such models. 
%
%
\section{BSM Physics}
 The previous section has shown how the ILD experiment will allow to achieve the ILC's primary goal as a Higgs factory. The ILC is, however, not only a Higgs factory but also a new particle discovery machine. In general, lepton colliders are complementary to hadron colliders because of their high sensitivities to regions with small cross sections and compressed mass spectra, which are challenging at hadron colliders such as the LHC. In many cases the current best limits in such regions are still those from the LEP experiments. Though the ILC's initial collision energy, 250\,GeV, is not very much higher than the highest energy of LEP2, there are four reasons why the ILC will significantly enhance the sensitivities to these difficult regions. First, the ILC's integrated luminosity at 250\,GeV is $10^3$ times larger than that collected by the four LEP experiments together at the highest energies. Second, the ILD detector is much more advanced than LEP detectors thanks to the progress in detector technologies since the LEP time. Third, the ILC provides polarised beams, a very powerful tool to control signal and background processes. Finally, the ILC's beam structure allows trigger-less data taking. 
 
\begin{figure}[htbp]
\begin{center}
 \includegraphics[width=0.5\textwidth]{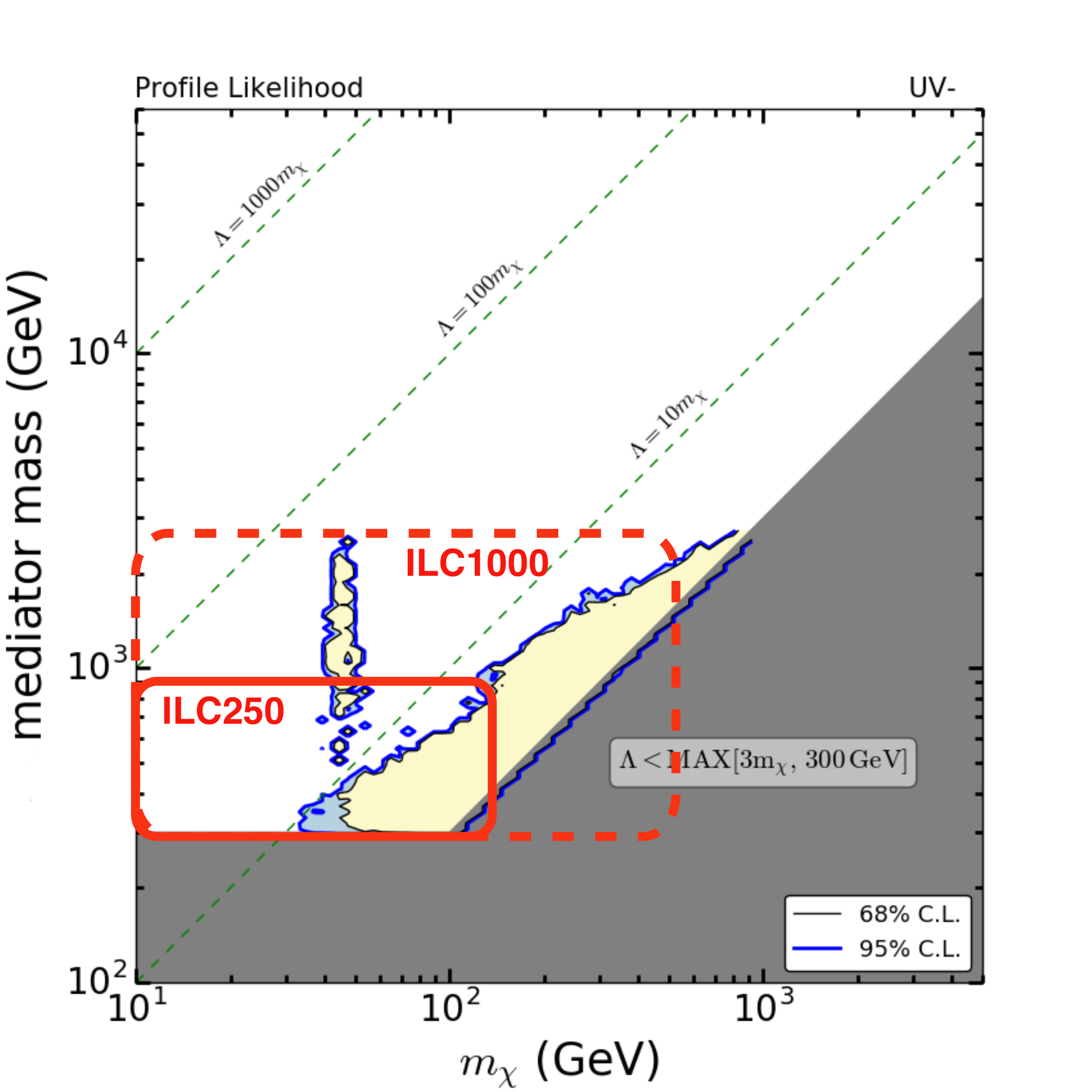}
\end{center}
\caption{Portion of WIMP parameter space (yellow area) expected to survive searches at the (HL-)LHC as well as current and future direct searches. The solid and dashed red boxes indicate, respectively, the regions the 250\,GeV ILC and its 1\,TeV upgrade will be sensitive to\cite{Habermehl:2018yul}.
}
\label{fig:ilcWIMP}\label{fig:WIMPleft}
\end{figure}

 The search for weakly interacting massive particles (WIMP) pair production is one of the most important targets of new particle searches both at the LHC and the ILC. The LHC has a mass reach much higher than the ILC. Nevertheless, there will be a significant fraction of WIMP parameter space left unexplored even after the HL-LHC. The yellow area in Fig.\,\ref{fig:WIMPleft} shows the remaining allowed regions after searches at the (HL-)LHC as well as current or future direct detection experiments\cite{Habermehl:2017dxh}.
The red solid box indicates the region the 250\,GeV ILC will probe, which is already substantial. The 1\,TeV upgrade will expand the ILC's sensitivity to the red dashed box covering a large fraction of the remaining WIMP parameter space\cite{Habermehl:2017dxh}. In addition to the beam polarisations, the excellent calorimetric coverage of the ILD detector, hermetic down to about 6\,mrad to the beam axes, is essential to veto Bhabha background so as to achieve this high sensitivity. 

A search for higgsinos is another example for which detection is challenging at the LHC because of their compressed mass spectrum. Such higgsinos are expected to be light and their mass differences typically below 20\,GeV in the natural SUSY models, while the other SUSY particles may be beyond the reach of the HL-LHC. As demonstrated in \cite{Baer:2016new}, ILD can not only discover such higgsinos but also measure their masses and production cross sections, which can be used, together with the Higgs-related measurements described above, in a global fit to extract underlying SUSY model parameters. This in turn provides an opportunity to test high scale physics such as gaugino mass unification and various SUSY breaking scenarios. Here again, the ILD's high sensitivity to higgsinos with small mass differences is due to the hermeticity of the ILD detector and its excellent tracking capability over a wide momentum range.

\section{Top Quark Physics}
As already mentioned, the top quark, being the heaviest in the SM, might hold the key to the mystery of the electroweak symmetry breaking. Its measurement starts around the $t\bar{t}$ threshold in the 350\,GeV stage of the ILC. The $t\bar{t}$ threshold provides an ideal laboratory to measure a short-distance top quark mass such as $m_t(\overline{\mathrm MS})$. ILD can measure $m_t(\overline{\mathrm MS})$ to 50\,MeV~\cite{Horiguchi:2013wra, Vos:2016til} and, together with the aforementioned Higgs mass measurement, test stability of the SM vacuum. 

\begin{figure}[htbp]
\begin{center}
 \includegraphics[width=0.6\textwidth]{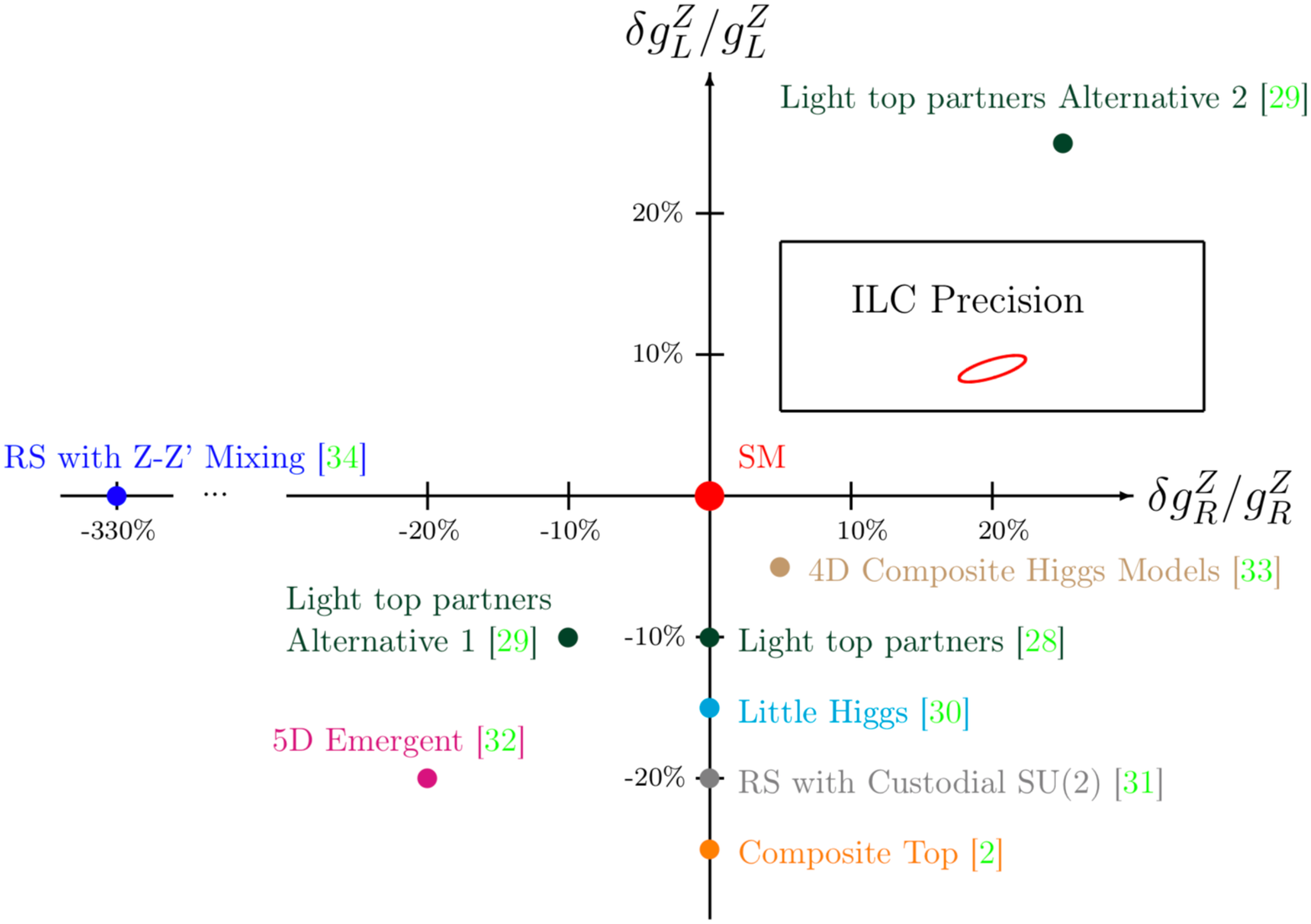}
\end{center}
\caption{Expected deviations of the $t_L$ and $t_R$ couplings to the $Z$ boson for various BSM models compared with the expected ILC precision for 500\,fb$^{-1}$ at 500\,GeV\cite{Amjad:2015mma}.
}
\label{fig:ttZ_gLgR}
\end{figure}

In the 500\,GeV stage, our main focus of top quark physics will shift to form factor measurements for the top quark couplings to the photon and the $Z$ boson. Partially composite top quarks, which often accrue from composite Higgs models, are to modify the $t\bar{t}Z$ form factors and cause significant deviations from the SM expectations with characteristic deviation pattern for couplings to $t_L$ and $t_R$. ILD can determine these form factors separately to accuracies better than 0.3\%
by measuring the cross section and forward-backward asymmetry for the $t\bar{t}$ pair production with 4\,ab$^{-1}$ at 500\,GeV\cite{Amjad:2015mma}. 
Figure\,\ref{fig:ttZ_gLgR} compares the expected ILD precision for 500\,fb$^{-1}$ with various BSM models.
The form factor measurements will hence provide another handle to elucidate BSM physics responsible for the electroweak symmetry breaking, in addition to the precision Higgs measurements described above. It should be emphasized that combination of vertex charge and the Kaon ID using $dE/dx$ information from the ILD TPC plays a substantial role in the forward-backward asymmetry measurement of the $t\bar{t}$ production\cite{Ref:bilokin2017}.

\section{Benchmarking Studies}
While the above physics case studies are based on the version of the ILD detector presented in the DBD~\cite{ild:bib:ilddbd}, ILD has recently initiated a systematic benchmarking effort to study the performance of the latest updated ILD concept, and in particular to determine the correlations between science objectives and detector performance. The list of benchmark analyses is given in table \ref{tab-benchmark} and their results presented in Chapter 8. 

It is important to note that, although the ILC will start operation at the centre-of-mass energy of 250\,GeV, the ILD detector is being designed to meet more challenging requirements at higher centre-of-mass energies, since major parts of the detector, e.g.\ the coil, the yoke and the main calorimeters are not expected to be replaced when upgrading the accelerator. Therefore, most of the detector benchmark analyses are performed at the centre-of-mass energy of 500\,GeV, and one benchmark even at 1\,TeV. The assumed integrated luminosities and beam polarisation settings follow the canonical running scenario~\cite{Barklow:2015tja}.

In addition to the well-established performance aspects of the ILD detector, the potential of new features not yet incorporated in the existing detector prototypes, e.g.\ time-of-flight information, has also started to be evaluated.

\begin{table}[thb]
  \setlength\extrarowheight{5pt}
    \centering
    \begin{tabular}{ p{4cm} p{5cm} p{5cm}}
 \hline
{\bf    Measurement}     & {\bf Main physics question} & {\bf main issue addressed} \\
\hline
Higgs mass in $H\rightarrow b {\bar b}$         &  Precision Higgs mass determination &Flavour tag, jet energy resolution, lepton momentum resolution  \\
\hline
Branching ratio $H \rightarrow \mu^+\mu^-$ & Rare decay, Higgs Yukawa coupling to muons & High-momentum $p_t$ resolution, $\mu$ identification \\
\hline
Limit on $H \rightarrow$ invisible & Hidden sector / Higgs portal & Jet energy resolution, $Z$ or recoil mass resolution, hermeticity\\
\hline
Coupling between Z and left-handed $\tau$ & Contact interactions, new physics related to 3rd generation & Highly boosted topologies, $\tau$ reconstruction, $\pi^0$ reconstruction \\
\hline
Cross section of ${e^+e^- \rightarrow \nu \nu qqqq}$ & Vector Bosons Scattering, test validity of SM at high energies&  $W/Z$ separation, jet energy resolution, hermeticity\\
\hline
Left-Right asymmetry in $e^+e^- \rightarrow \gamma Z$ & Full dim-6 EFT interpretation of Higgs measurements &  Jet energy scale calibration, lepton and photon reconstruction \\
\hline
Hadronic branching ratios for $H\rightarrow b \bar b $ and $c \bar c$ & New physics modifying the Higgs couplings &  Flavour tag, jet energy resolution\\

\hline
$A_{FB}, A_{LR}$ from ${e^+e^- \to b\bar{b}}$ and $t \bar t \rightarrow b\bar{b} qqqq / b \bar{b} qql\nu$ & Form factors, electroweak coupling &  Flavour tag, PID, (multi-)jet final states with jet and vertex charge\\
\hline

Discovery range for low $\Delta M$ Higgsinos & Testing SUSY in an area inaccessible for the LHC& Tracks with very low $p_t$, ISR photon identification, finding multiple vertices\\
\hline
Discovery range for WIMP's in mono-photon channel & Invisible particles, Dark sector & Photon detection at all angles, tagging power in the very forward calorimeters\\
\hline
Discovery range for extra Higgs bosons in $e^+e^- \rightarrow Zh$ & Additional scalars with reduced couplings to the $Z$ & Isolated muon finding, ISR photon identification.\\
\hline
\end{tabular}
    \caption{Table of benchmark reactions which are used by ILD to optimize the detector performance. The analyses are mostly conducted at 500\,GeV centre-of-mass energy, to optimally study the detector sensitivty. The channel, the physics motivation, and the main detector performance parameters are given.}
    \label{tab-benchmark}
\end{table}

\chapter{The ILC Environment}
\label{chap:ilc}
\label{ild:sec:ilc}

In this chapter the latest status of the International Linear Collider layout and design parameters are summarized, as well as the corresponding experimental conditions for the ILD detector.

\section{The International Linear Collider Project}
The ILC is a high-luminosity linear electron-positron collider based on the 1.3~GHz superconducting RF accelerating technology. The ILC has been initially proposed for an energy range of 200-500~\GeV, upgradable to 1\,TeV~\cite{Behnke:2013xla}. After the discovery of the Higgs Boson at 125~\GeV~mass, the baseline of the ILC has been re-configured with an initial stage at 250~\GeV~ center-of-mass energy~\cite{Bambade:2019fyw}, which provides a rich harvest for a precision Standard Model and Higgs physics programme. The baseline ILC can be extended to higher energies and luminosities in well defined upgrade scenarios. The basic beam parameters for the baseline and potential upgrades are given in table~\ref{tab:ilc-params}.

\begin{table}[tbhp]
\begin{adjustbox}{width=1.\textwidth,center=\textwidth}
\begin{tabular}{lccccccc}
Quantity & Symbol & Unit & Initial & ${\mathcal{L}}$ Upgrade & TDR &  \multicolumn{2}{c}{Upgrades} \\
\hline
Centre of mass energy & $\sqrt{s}$ & ${\mathrm{\GeV}}$ & $250$ & $250$ & $250$ & $500$ & $1000$ \\
Luminosity & \multicolumn{2}{c}{${\mathcal{L}}$ ~~~~$10^{34}{\mathrm{cm^{-2}s^{-1}}}$} & $1.35$ & $2.7$ & $0.82$ & $1.8 / 3.6$ & $4.9$ \\
Polarisation for $e^- (e^+)$ & $P\sub{-} (P\sub{+})$ & & ~$80\,\% (30\,\%)$~ &  ~$80\,\% (30\,\%)$~ &  ~$80\,\% (30\,\%)$~ &~$80\,\% (30\,\%)$~ &  ~$80\,\% (20\,\%)$~  \\
Repetition frequency &$f\sub{{rep}}$ & ${\mathrm{Hz}}$  & $5$ & $5$ & $5$ & $5$ & $4$ \\
Bunches per pulse  &$n\sub{{bunch}}$ & 1  & $1312$ & $2625$ & $1312$ & $1312 / 2625$ & $2450$ \\
Bunch population  &$N\sub{{e}}$ & $10^{10}$ & $2$ &  $2$ & $2$ & $2$ & $1.74$ \\
Linac bunch interval & $\Delta t\sub{{b}}$ & ${\mathrm{ns}}$ & $554$ & $366$ & $554$ & $554 / 366$ & $366$ \\
Beam current in pulse & $I\sub{{pulse}}$ & ${\mathrm{mA}}$& $5.8$ & $5.8$& $8.8$ & $5.8$ & $7.6$  \\
Beam pulse duration  & $t\sub{{pulse}}$ & ${\mathrm{\mu s}}$ & $727$ & $961$ & $727$ & $727 / 961$ & $897$ \\
Average beam power  & $P\sub{{ave}}$   & ${\mathrm{MW}}$ & $5.3$ & $10.5$ & $10.5$ & $10.5 / 21$  & $27.2$ \\  
Norm. hor. emitt. at IP & $\gamma\epsilon\sub{{x}}$ & ${\mathrm{\mu m}}$& $5$ & $5$ & $10$ & $10$ & $10$  \\ 
Norm. vert. emitt. at IP & $\gamma\epsilon\sub{{y}}$ & ${\mathrm{nm}}$ & $35$ & $35$ & $35$ & $35$ & $30$ \\ 
RMS hor. beam size at IP  & $\sigma^*\sub{{x}}$ & ${\mathrm{nm}}$  & $516$ & $516$ & $729$ & $474$ & $335$ \\
RMS vert. beam size at IP &$\sigma^*\sub{{y}}$ & ${\mathrm{nm}}$ & $7.7$  & $7.7$  & $7.7$  & $5.9$ & $2.7$ \\
Luminosity in top $1\,\%$ & ${\mathcal{L}}\sub{0.01} / {\mathcal{L}}$ &  & $73\,\%$  &  $73\,\%$ & $87.1\,\%$  & $58.3\,\%$ & $44.5\,\%$\\
Energy loss from beamstrahlung  & $\delta\sub{BS}$ &  & $2.6\,\%$  & $2.6\,\%$  & $0.97\,\%$  & $4.5\,\%$ & $10.5\,\%$ \\
Site AC power  & $P\sub{{site}}$ &  ${\mathrm{MW}}$ & $129$ & & $122$ & $163$ & $300$ \\
Site length & $L\sub{{site}}$ &  ${\mathrm{km}}$ & $20.5$ & $20.5$ & $31$ & $31$ & $40$ \\
\end{tabular}
\caption{Summary table of the ILC accelerator parameters in the initial 250~\GeV~staged configuration (with TDR parameters at 250~\GeV~given for comparison) and possible upgrades~\cite{Bambade:2019fyw}. A 500~\GeV~machine could also be operated at 250~\GeV~with 10~Hz repetition rate, bringing the maximum luminosity to 5.4$\cdot$10$^{34}$cm$^{-2}$s$^{-1}$~\cite{Harrison:2013nva}.
\label{tab:ilc-params}}
\end{adjustbox}
\end{table}

Recently, options for running the ILC-250 machine at the Z pole have been re-evaluated~\cite{Yokoya:2019rhx}. Luminosities in the order of 2.5 $\times$ 10$^{33}$ cm$^{-2}$s$^{-1}$ are in reach with moderate modifications of the ILC beam parameters. Increasing this value to $\approx$ 5 $\times$ 10$^{33}$ cm$^{-2}$s$^{-1}$ can be realised when the number of bunches per pulse is doubled, as currently suggested for the luminosity upgrade scenarios at higher energies~(c.f.~table~\ref{tab:ilc-params}).

\begin{figure}[!ht]
\centering
\includegraphics[width=0.9\hsize]{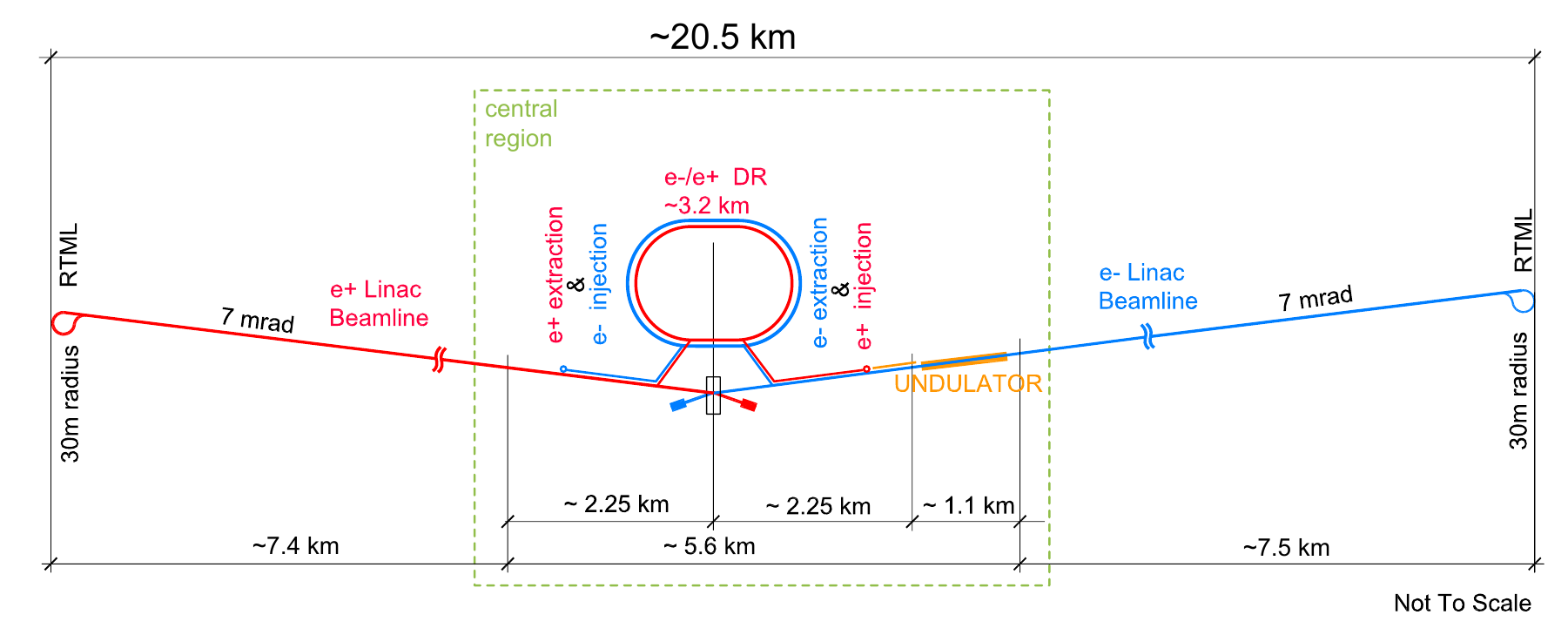}

\caption{\label{ild:fig:ilc_footprint}Layout of the ILC in the 250~\GeV~baseline configuration~\cite{Bambade:2019fyw}.}
\end{figure}

\begin{figure}[htbp]
\centering
\includegraphics[width=0.9\hsize]{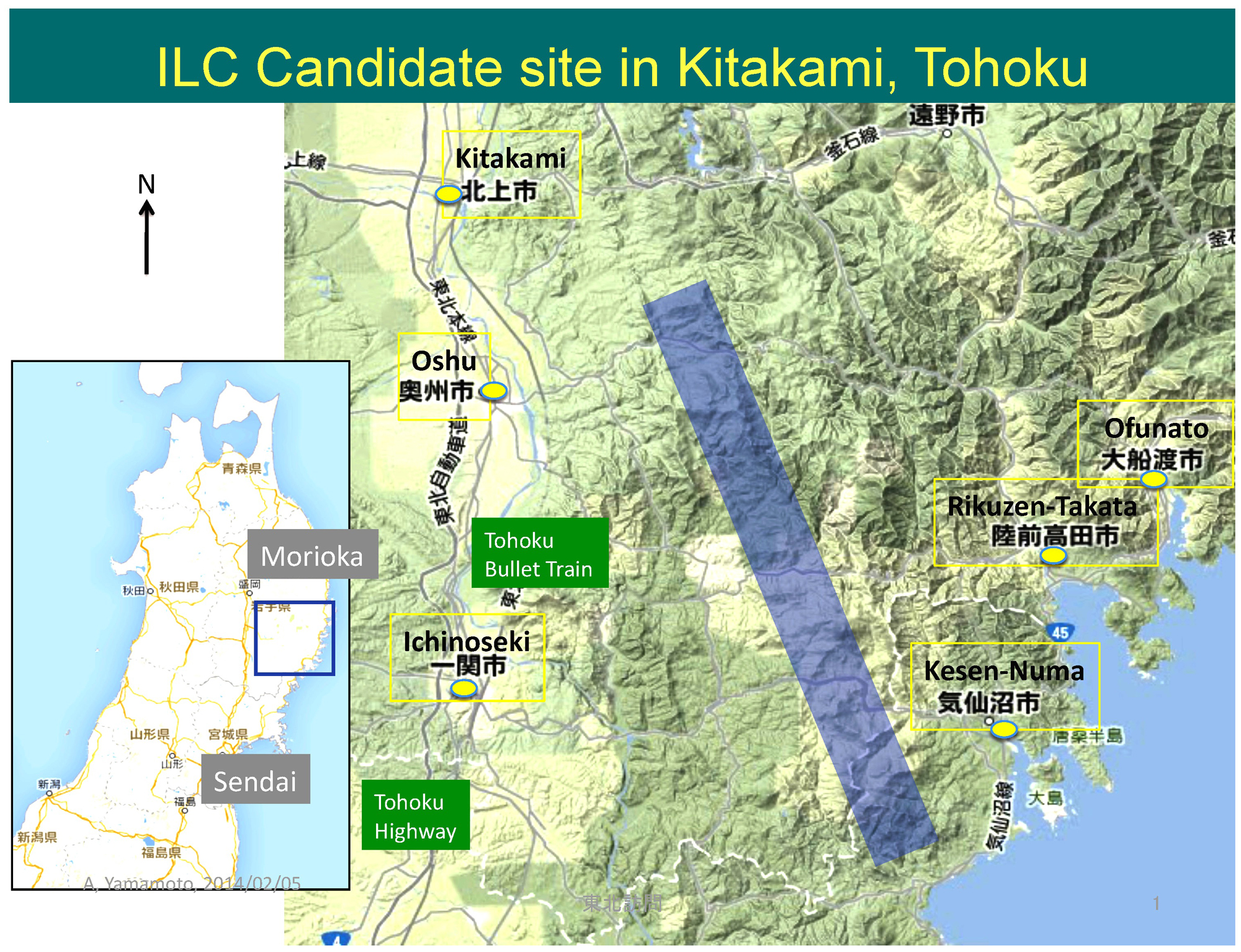}
\caption{\label{ild:fig:ilc_site}Location of the ILC candidate site in the Kitakami mountains of Tohoku, Japan~\cite{ild:bib:Newsline_Kitakami}.}
\end{figure}

\begin{figure}[htbp]
\centering
\includegraphics[width=0.8\hsize]{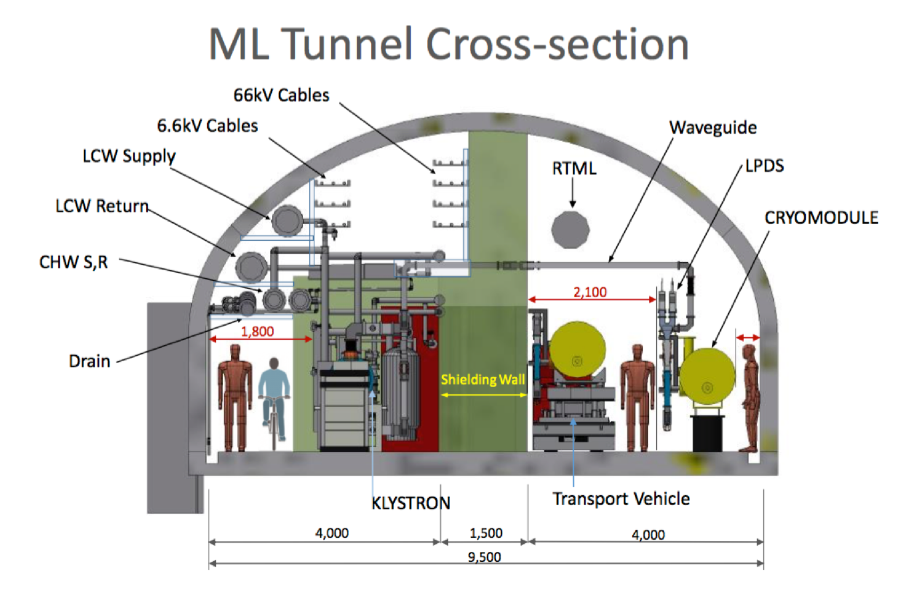}
\caption{\label{ild:fig:ilc_tunnel}Cross section of the ILC main linac tunnel~\cite{Bambade:2019fyw}.}
\end{figure}

The layout of the baseline 250~\GeV~ILC facility is shown in figure~\ref{ild:fig:ilc_footprint}. The total linear accelerator length is about 20.5~km with an experimental area hosting two detectors in push-pull configuration in the interaction region located at the center. The ILC design foresees a crossing angle of 14~mrad between the linac arms.
The Japan Association of High Energy Physicists (JAHEP) has proposed that Japan hosts the ILC as a staged project~\cite{ild:bib:JAHEP}. A possible site for the construction of the ILC has been idenfitied in the Kitakami mountains in the Tohoku area in the north of Honshu main island, about 400~km north of Tokyo. Figure~\ref{ild:fig:ilc_site} shows the location that allows for a total linac length of about 50~km, and therefore extension space for upgrade scenarios, in good geological conditions. Figure~\ref{ild:fig:ilc_tunnel} shows the cross section of the ILC main linac tunnel with the cryomodules in the right section and the klystrons and RF distribution in the left section.
The specific location of the ILC site and the local conditions, e.g. in terms of street access and topography, has implications on the design, assembly and operations of ILD, as discussed in section~\ref{ild:sec:access}.

\section{Integration of ILD into the experimental environment}
ILD is designed to operate in a push-pull arrangement with another detector, sharing one common ILC interaction region. In this scheme ILD sits on a movable platform in the underground experimental hall. This platform allows for a roll-in of ILD from the parking position into the beam and vice versa within a few hours. The detector can be fully opened and maintained in the parking position.

The current mechanical design of ILD assumes an initial assembly of the detector on the surface, similar to the construction of CMS at the LHC. A vertical shaft from the surface into the underground experimental cavern allows ILD to be lowered in five large segments, corresponding to the five yoke rings.

ILD is designed to fully cope with the ILC beam conditions~(c.f.~Section~\ref{ild:sec:beam_backgrounds}). The expected levels of beam induced backgrounds have been simulated and are seen to be at tolerable levels, {\it e.g.} for the vertex detectors. Judiciously placed shielding keeps scattered backgrounds under control. Regarding the collider, the design of the interaction region and the collimation system has been defined so as to keep the external background sources at levels below the detector requirements.

ILD is self-shielding with respect to radiation and magnetic fields to enable the operation and maintenance of equipment surrounding the detector, {\it e.g.} cryogenics. Of paramount importance is the possibility to operate and maintain the second ILC push-pull detector in the underground cavern during ILC operation~(c.f.~Section~\ref{ild:sec:external_integration}).

\section{Experimental Conditions}
 
\subsection{Beam Conditions} \label{sec:beam:conditions}
\begin{figure}[htbp]
\begin{subfigure}{0.49\hsize} \includegraphics[width=\textwidth]{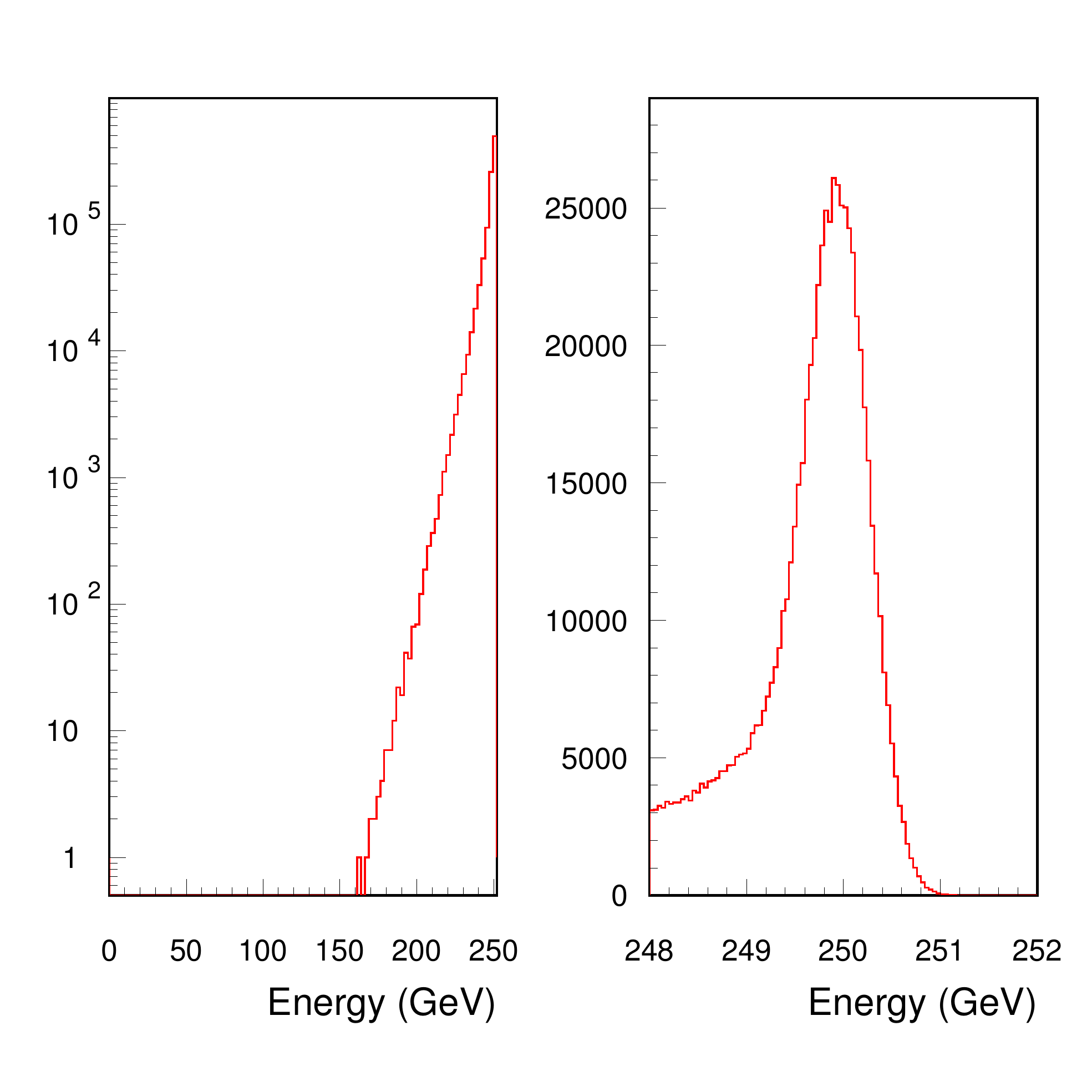}
 \caption{ \label{fig:ilc:ecmspect-250}}
 \end{subfigure}
\begin{subfigure}{0.49\hsize} \includegraphics[width=\textwidth]{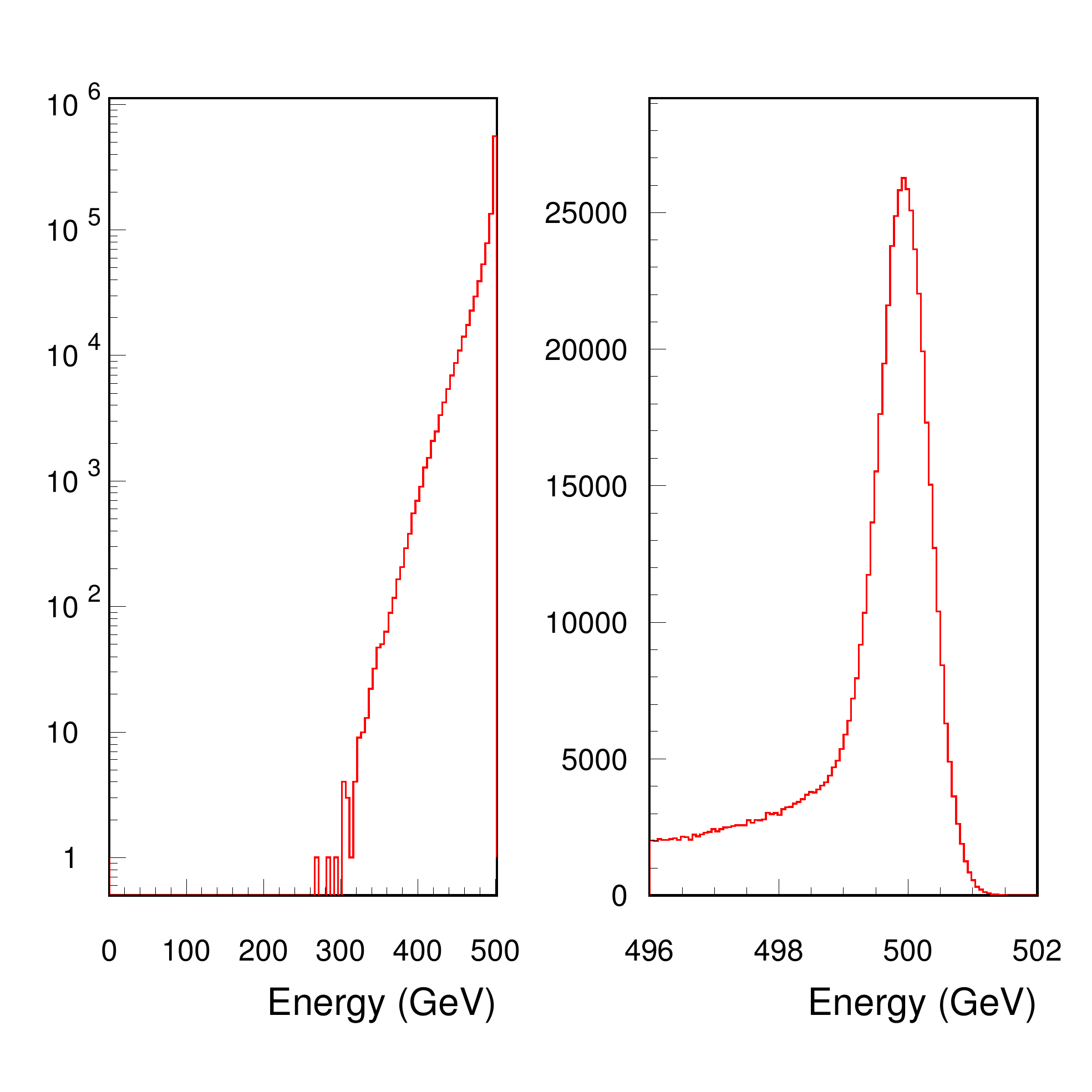}
 \caption{  \label{fig:ilc:ecmspect-500}}
 \end{subfigure}
\caption{
(a) Luminosity spectrum for ILC-250.
(b) Luminosity spectrum for the 500~\GeV~upgrade machine.
Spectra are generated with Guinea Pig (see section~\ref{sec:generator}).
}
\label{fig:ilc:ecmspect}
\end{figure}

The ILC beams have specific properties that are optimised for the physics output of the experimental programme:
\begin{itemize}
\item High instantaneous luminosity: 1.35$\cdot$10$^{34}$cm$^{-2}$s$^{-1}$ at $\sqrt{s}=$250~\GeV~cms;
\item Longitudinal polarisation for the electron (80\%) and positron (30\%) beams;
\item Moderate energy losses from beamstrahlung;
\item A pulse structure with pulse lengths of $\approx$ 1~ms and low repetition rates of 5-10~Hz which enables the use of power-pulsed readout schemes in the detector.
\item A beam crossing angle of 14 mrad at the collision point.
\end{itemize}
To reach high luminosities, the ILC bunches are focused to very small cross sections in the nanometre range at the interaction point (see Table~\ref{tab:ilc-params}). The new baseline ILC-250 was optimised for maximum luminosity with an increased beam focusing compared to the TDR, resulting in a slightly increased beam energy spread for interactions. The luminosity spectrum describes the distribution of the luminosity versus the centre-of-mass energy. Figure~\ref{fig:ilc:ecmspect-250} shows the luminosity spectrum for ILC-250, Figure~\ref{fig:ilc:ecmspect-500} for the 500~\GeV~upgrade.

The ILC beams produce three main sources of background in the detectors. Upstream of the collision region, the interaction of the electron and positron bunches with beam-line elements such as collimators produce high-energy highly penetrating muons parallel to the beam. Recent work has shown that the level of muon background can be reduced to a level tolerable for the detector \cite{Keller:2019aak}, with a hit occupancy well below the critical limits. 

In the collision region, the strongly focused beams emit beamstrahlung photons in the very forward directions which leave the detector towards the main beam dumps. Secondary electron and positron pairs stemming from photon conversions and interactions are a source of backgrounds especially for detectors close to the beampipe. Finally, the neutrons produced in the main beam dumps 300\,m downstream of the interaction region can travel back into the detector. The three sources have been studied in detail and are under control~(see~section~\ref{ild:sec:beam_backgrounds}).

\subsection{Machine Detector Interface}
The requirements of a linear collider have technical implications for the ILD detector. All those aspects are summarised in the Machine-Detector Interface that has been specified and designed in close collaboration with the ILC machine groups and SiD~\cite{bib:sid:loi}, the other currently proposed detector for the ILC.

\subsubsection{Push-Pull}
\begin{figure}[h!]
\centering
\includegraphics[width=0.8\hsize]{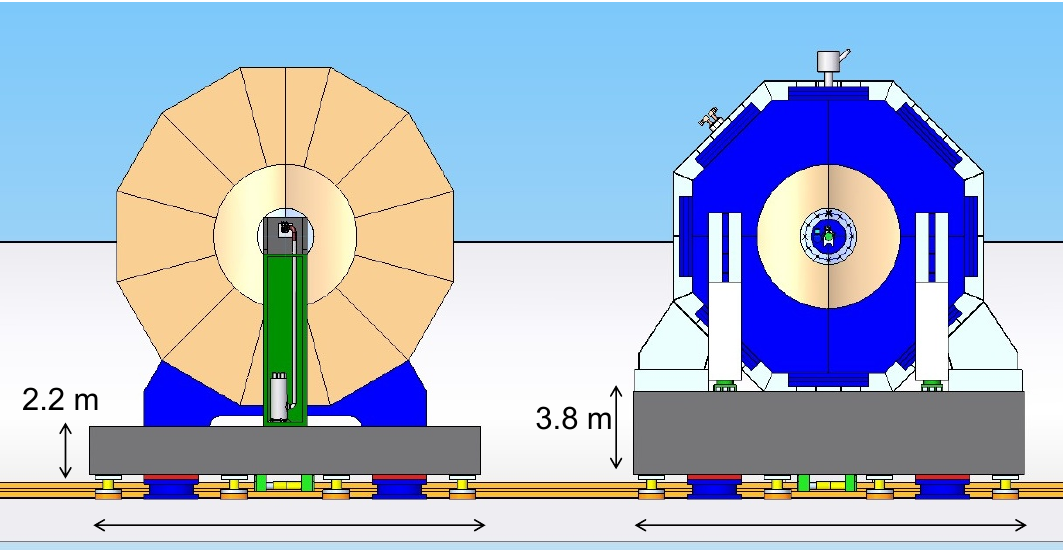}
\caption{\label{ild:fig:push_pull}Conceptual design of the push-pull system for ILD (left) and SiD (right). Both detectors are mounted on movable concrete platforms~\cite{Behnke:2013xla}.}
\end{figure}

The ILC is proposed to have a common Beam Delivery System that serves one interaction point shared between two detectors, ILD and SiD, that operate in a push-pull scheme~\cite{Behnke:2013xla}. In such a scheme, one detector is taking data on the ILC beam, while the other one is parked close by and waits for its turn to move in. Figure~\ref{ild:fig:push_pull} shows a conceptual design of ILD and SiD in such a push-pull configuration where both detectors are mounted on movable concrete platforms. The system is designed for short turn-around times of one day or less.
The requirements for such a push-pull operation scheme have been defined between both detector collaborations~\cite{Parker:2009zz}. The impact of these requirements on the ILD design is discussed in more detail in section~\ref{ild:sec:external_integration}.

\subsubsection{Accelerator Elements}
\begin{figure}[h!]
\centering
\includegraphics[width=0.8\hsize]{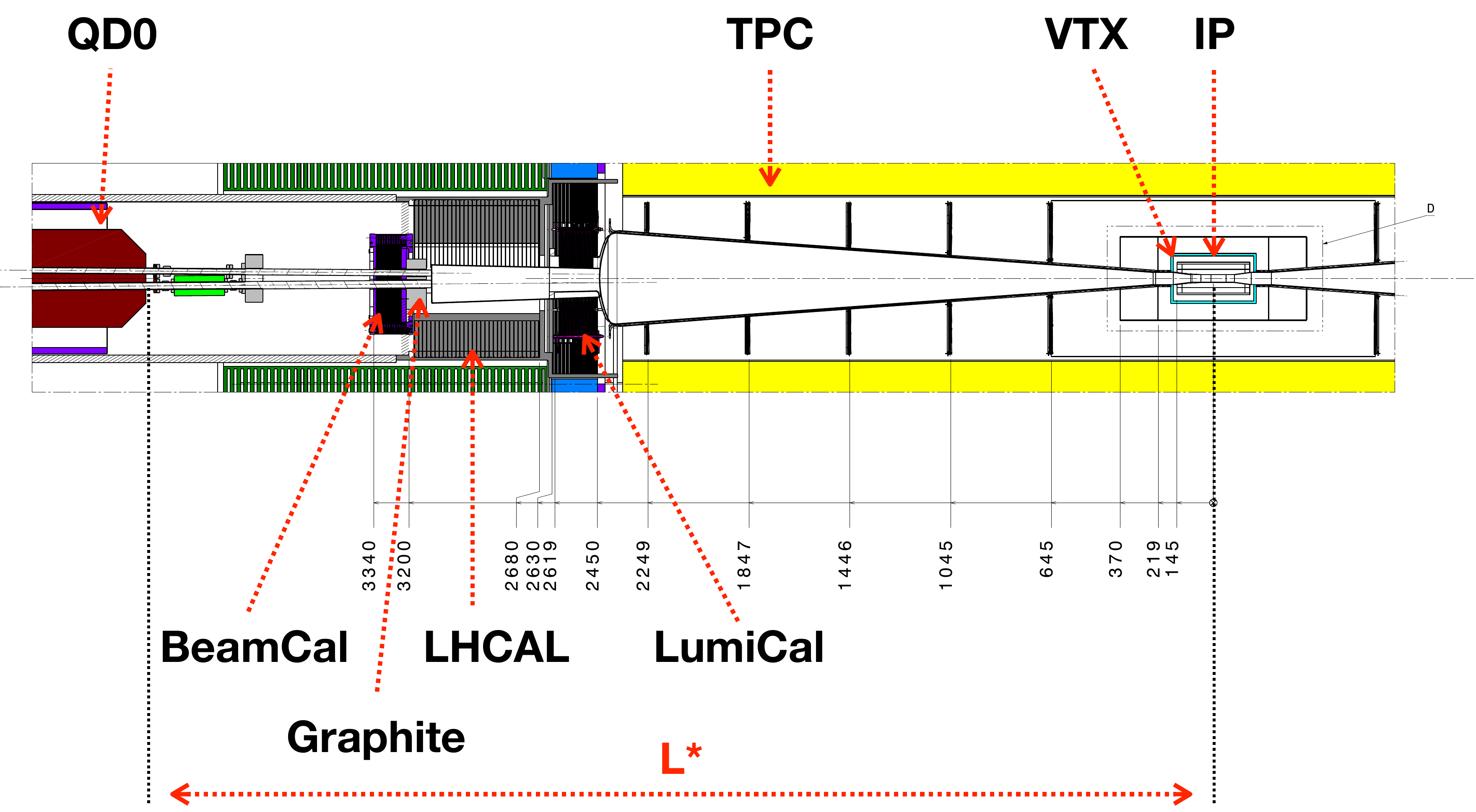}
\caption{\label{ild:fig:Forward_QD0}ILD forward region. The focal length $L^*$ in this design is 4.1 m. The interaction point IP is on the right of the picture, the main elements of the forward system are indicated in the drawing.}
\end{figure}

The final focus magnets of the ILC are at a close focal length so that they have to be accommodated by the detectors. The closest magnet to the interaction point, the QD0 quadrupole, is an integral part of ILD, as can be seen in figures~\ref{ild:fig:Forward_QD0} and~\ref{fig:det:quad}. The QD0 magnet packages move together with the detector in case of push-pull operations, i.e. SiD and ILD both have their own set of magnets.
In the ILC TDR, two focal lengths of the QD0 magnets were foreseen, 4.4~m for ILD and 3.5~m for SiD~\cite{Behnke:2013xla}. Later, the design of the ILC has been changed and the focal lengths were harmonised at 4.1~m~\cite{ild:bib:lstar_cr}. For ILD, this required a re-design of the forward region, since the QD0 magnet has moved closer to the interaction point, while the detector dimension remained unchanged. A vacuum pump located close to the interaction point had to be removed to provide the required space for the change. The pump removal was expected to deteriorate the vacuum levels in the central beam pipe of ILD. However, a study of the possible impact on backgrounds from beam-gas reactions showed that the estimated levels are still negligible~\cite{ild:bib:beam_gas}.

\chapter{The ILD detector concept}
\label{chap:ILD}
In this chapter, the concept of the ILD detector is presented. The basic idea has not changed since the ILD-DBD and is shortly repeated below. Recently, a re-optimisation process of the ILD configuration has been initiated in order to find an adequate balance between performance and cost of the detector. The rationale of this optimisation is presented as well as the two detector models considered for the rest of this document for performance and cost evaluation. 

\section{The overall ILD concept}

\begin{figure}[t!]
\centering
\includegraphics[width=1.0\hsize]{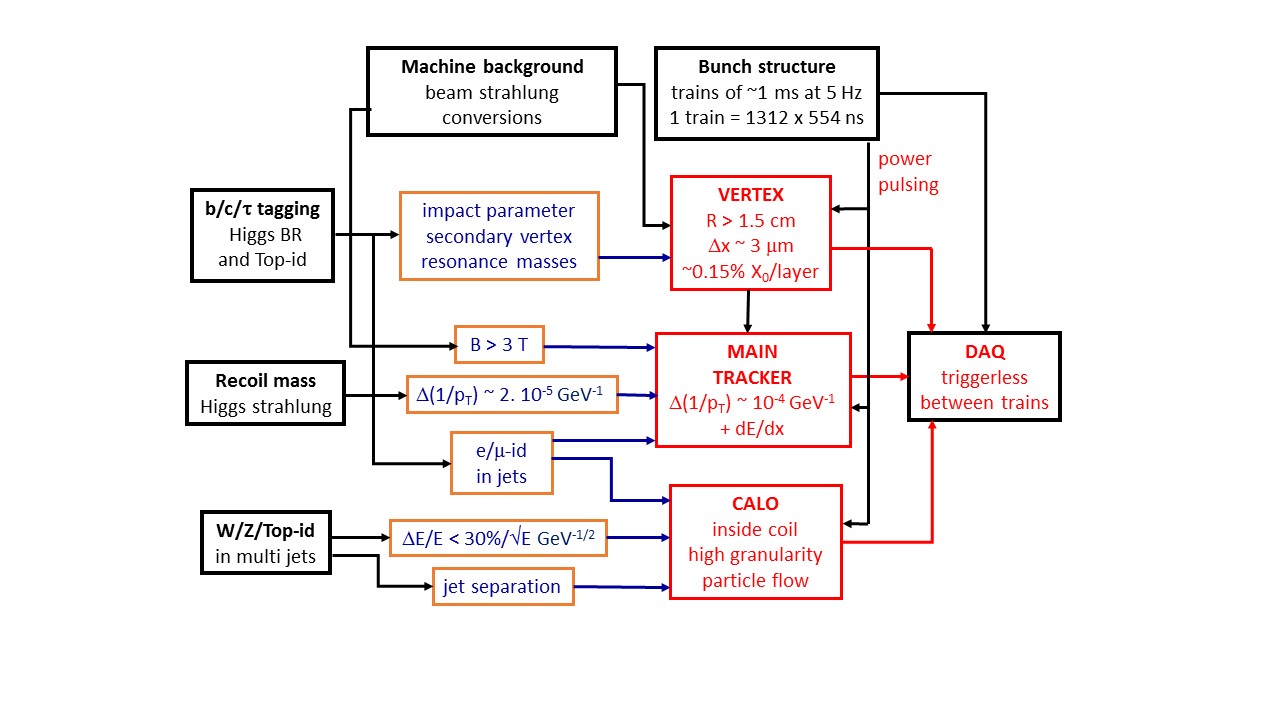}
\caption{Interplay between ILC machine characteristics, physics requirements and detector specifications.}
\label{fig:ILD:specifications}
\end{figure}

The overall ILD concept is based on the ideas presented in the DBD~\cite{ild:bib:ilddbd}. The global detector layout and the performance of the subdetectors are tightly linked to the accelerator characteristics and the physics requirements, as summarised in figure ~\ref{fig:ILD:specifications}. 

The high beamstrahlung background at the collision point requires a magnetic field higher than 3\,T to confine most of the low-energy electron pairs  within the beam pipe, and sets a minimum of $\approx$1.5~cm for the closest distance of approach of the vertex detector inner layer from the beamline. On the other hand the bunch structure of short trains separated by long idle periods sets rather relaxed conditions on the data acquisition, with the possibility to avoid a hardware trigger system. This in addition allows to power the front-end electronics only during active bunch trains (so-called "power-pulsing" mode), which minimises the subdetector cooling requirements and associated material budgets.  

The subdetector specifications are tightly linked to the physics requirements from precision Higgs and electroweak physics. The dominant Higgs strahlung process, which at an $e^+e^-$ collider provides the unique opportunity to tag Higgs production independently of Higgs decay mode, requires a very high-precision momentum measurement of isolated particles from Z decays and hence a high precision main tracker. The efficient tagging of quark and lepton flavours to disentangle Higgs couplings requires a very high precision and low-material vertex detector, which also improves on the particle momentum measurements, as well as a high calorimeter granularity to identify leptons in jets. Similarly an efficient identification of W, Z and top hadronic decays in a crowded multijet environment needs a high jet energy resolution, twice better than currently realized at LHC, as well as an efficient spatial jet separation. ILD considers that the best concept to meet these requirements altogether is particle flow, where the charged and neutral particle contents of the jets are measured with the high performance trackers and the high granularity calorimeters, respectively. Within this scheme, an efficient match between the trackers and the calorimeters requires the calorimeters to be positioned inside the coil.     

\section{Optimizing ILD}

The baseline ILD layout of the DBD~\cite{ild:bib:ilddbd} had intentionally large dimensions in order to maximize the tracking performance and the particle flow capabilities of the calorimeters. The main cost drivers of the DBD detector were the electromagnetic calorimeter and the coil/yoke system, for which specific options are considered to reduce their costs (chapters 6.4 and 9). In the past years a re-optimization process of the detector global dimensions has been launched to identify an optimal point in the cost-performance space.

In a first step, a parametric study \cite{Ref:bib:TPCOPT} of the dependence of cost and performance as function of the outer radius and length of the main tracker (the TPC in ILD) has been performed. 
A simple model has been constructed, based on the cost estimate published as part of the ILD DBD~\cite{ild:bib:ilddbd}. In this model the cost of each subdetector is scaled as a function of the size based on simple scaling laws. Sensitive detector elements like Silicon planes are scaled with the total area, while mechanical elements - for example, the absorber in a calorimeter - scale with the volume. The reference is always the DBD cost estimate. To study the effect of changing the TPC radius and length, all other dimensions outside of the TPC are tied to the TPC radius and TPC length. Clearances between detectors are kept constant, and do not scale. In this way, an overall cost scaling of the ILD detector can be computed. Comparison with the more detailed updated costing presented in chapter 9 shows that this parametric scaling if correct at the level of 20-30\%. 

The performance of the detector is measured by a combined performance estimator, based on a few observables mostly from Higgs physics. Essentially, these are the tracking performance (momentum resolution and impact parameter resolution), the Higgs mass precision (with and without beam-strahlung background), the inverse of the significance of b-tagging, and the minimum transverse momentum to reach the last layer of the vertex detector. All numbers are normalised to the performance of the DBD detector. For a more complete description of the method and the definitions, see \cite{Ref:bib:TPCOPT}. 

\begin{figure}[t!]
\centering
\includegraphics[width=0.5\hsize]{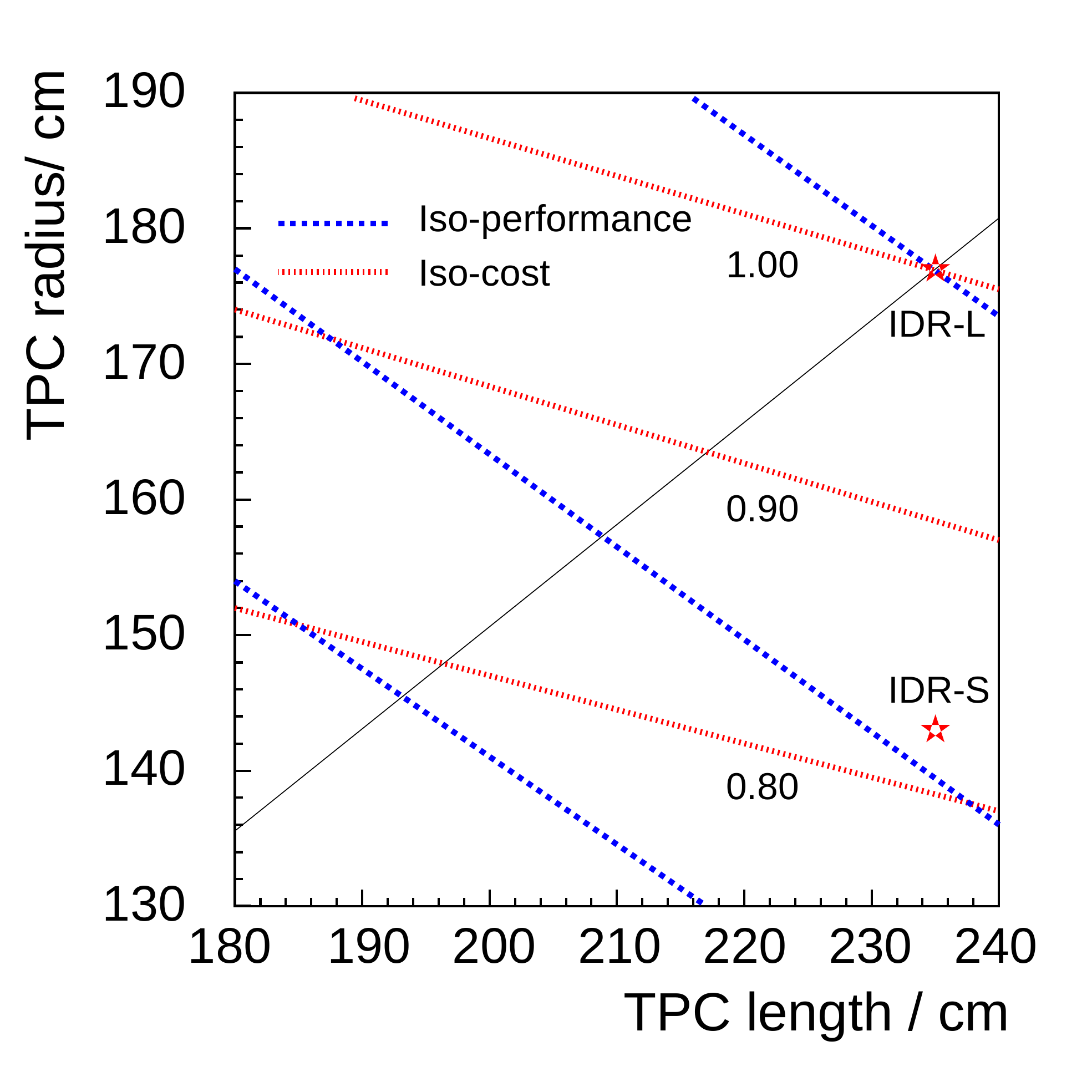}
\caption{Iso-performance and iso-cost curves resulting from a parametric evaluation of the ILD detector in the tracker radius-length parameter space. The stars indicate the location of the different ILD models in this parameter space. The black diagonal line indicates dimension changes at a constant r/z aspect ratio.}
\label{fig:ILD:aspect_ratio}
\end{figure}

Two types of iso-curves are then defined in the space opened by the TPC radius and length: Equal performance, and equal cost. In figure ~\ref{fig:ILD:aspect_ratio} iso-cost and iso-performance curves are shown. The three red lines correspond to costs relative to the DBD of (from top to bottom) 100\%, 90\% and 80\%. The blue lines indicate equal-performance lines. From the plot it can be seen that the dependence of the cost on the detector radius is steeper than on the length, while the performance scales roughly the same for both. This shows that targeting a given cost reduction maintains a higher performance when reducing the radius while keeping the overall length unchanged, instead of keeping the aspect ratio r/z unchanged.  

Based on this study, detector models with different sizes were defined with the following guidelines:

\begin{itemize}
    
\item The number of detector models is limited to two to maintain simulations and analyses at a manageable level.

\item One of the models ("IDR-L") has dimensions similar to those of the DBD model, in order to have a well understood reference in the studies. The only changes compared to the DBD are associated to the collider parameter evolution (e.g. the new L* optics, see chapter 3), and to better understanding of the subdetector technology constraints. This resulted in small changes to the TPC outer radius and length. 

\item The second model ("IDR-S") has a reduced outer radius of the main tracker while keeping its length unchanged. The smaller radius has to be far enough from the IDR-L radius to provide a significant lever-arm for the comparison. The chosen value is equal to that of the new CLIC detector model CLICdp~\cite{Arominski:2018uuz}, half way of the even smaller radius of the SiD detector~\cite{ild:bib:ilddbd}. With this choice IDR-S has similar outer tracker dimensions to CLICdp for both radius and length. This offers the possibility to compare the performance of the TPC option to the all-silicon option favored by CLICdp. 

\item All other components of IDR-S are similar to IDR-L. The inner tracking and very forward detectors are identical. The calorimeter depths and cell sizes are also kept unchanged, and the number of cells is reduced only as function of the calorimeter radii. All external systems such as coil, yoke and endcaps have their radial dimensions reduced accordingly.

\item In order to compensate for the smaller tracking lever-arm, the nominal magnetic field of IDR-S is increased from 3.5\,T to 4\,T.

\end{itemize}

\begin{figure}[t!]
\centering
\includegraphics[width=1.0\hsize]{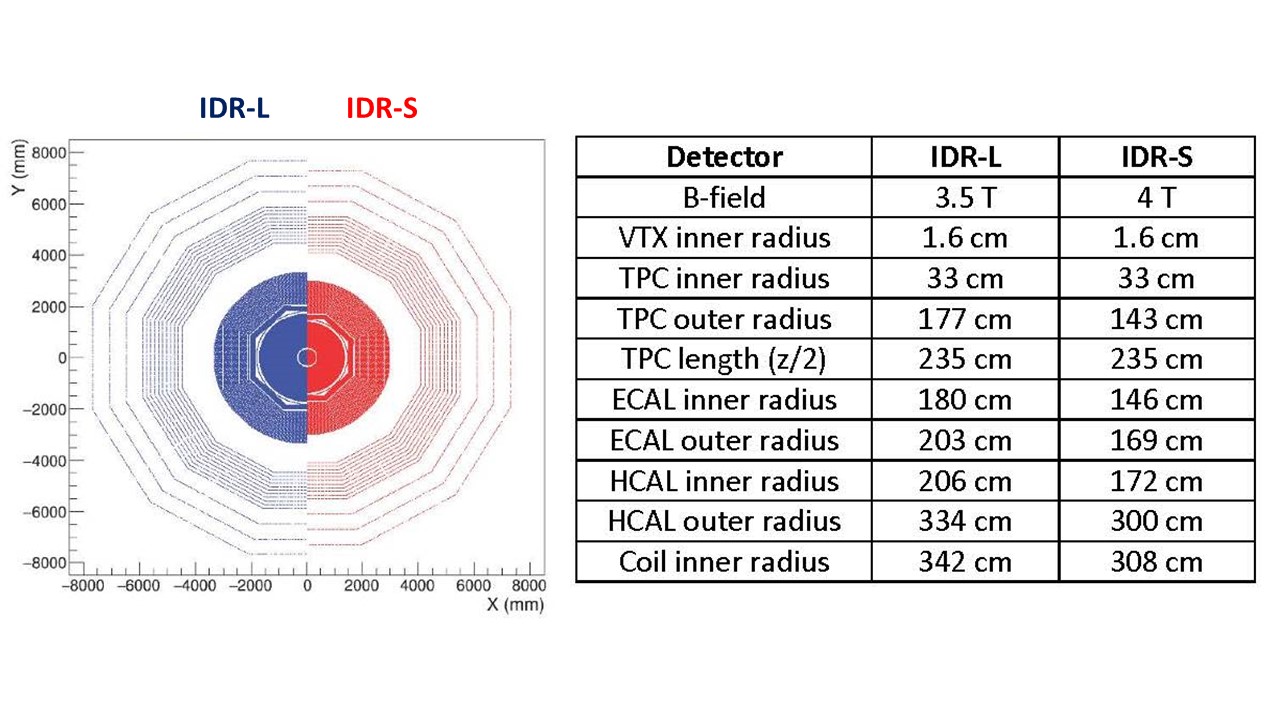}
\caption{The large ("IDR-L") and small ("IDR-S") models used for the ILD optimization: R-$\phi$ view (left) and main subdetector dimensions (right).}
\label{fig:ILD:sizes}
\end{figure}

The resulting IDR-L and IDR-S dimensions are summarised in figure ~\ref{fig:ILD:sizes}. Both models were used for detailed simulations of physics benchmark samples. The simulation framework is described in chapter~\ref{chap:modelling}. The detector and physics performance of both models are compared in chapter~\ref{chap:performance} and their costing estimated in chapter~\ref{chap:costing}. \vfill

\newcommand{\ilds}[1]{{\em#1}$^s$} 

\chapter{Detector Layout and Technologies}
\label{chap:technologies}

In the first section of this chapter the overall layout of ILD and the conceptual design of the subdetectors is summarised, with a focus on the evolution since the DBD. In the second section the results of the intense R\&D performed since the DBD on the subdetector technologies is presented, and possible future steps towards a final ILD design are discussed. This chapter is intended to give a snapshot of the current state of the art and a view of the trends of future studies for the coming years.      
\section{Overall Structure of the Detector}

The geometrical structure of the ILD detector and the individual layouts of subdetectors were described in detail in the ILD LOI~\cite{ild:bib:ILDloi} and the ILD DBD~\cite{ild:bib:ilddbd}. In this section the main characteristics with a special emphasis on recent developments and open options is given. The main design changes implemented since the DBD take into account the continuous progress in detection technologies and the new optics of the ILC interaction region (see chapter~\ref{chap:ilc}). 


\subsection{Global structure and parameters}

\begin{figure}[th!]
\centering
\includegraphics[width=0.85\hsize]{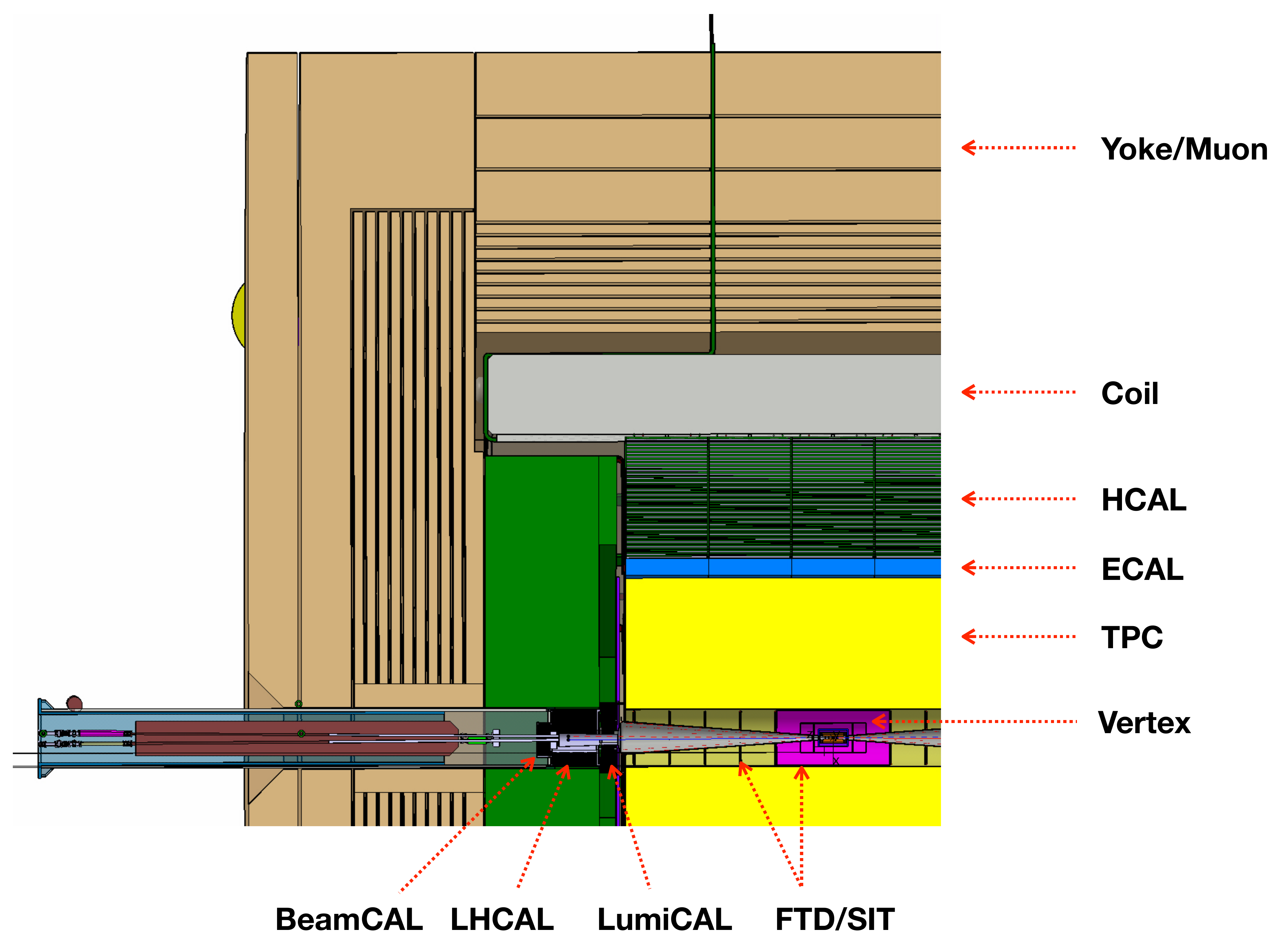}
\caption{r-z view of an ILD quadrant. The interaction point is on the lower right of the picture.}
\label{fig:det:quad}
\end{figure}

The overall ILD detector structure is shown in figure ~\ref{fig:det:quad}: a high precision vertex detector positioned very close to the interaction point is followed by a hybrid tracking layout, realised as a combination of silicon tracking with a time projection chamber, and by a calorimeter system. The complete system is located inside a large solenoid providing a nominal magnetic field of 3.5T (IDR-L) or 4T (IDR-S). On the outside of the coil, the iron return yoke is instrumented as a muon system and as a tail catcher calorimeter. 
The main geometrical parameters are summarised in table~\ref{ild:tab:barrelpara} and table~\ref{ild:tab:endcappara}.

%
%
\begin{table}\hspace*{-0cm}\small
\begin{tabular}{ l p{0.05\hsize}p{0.04\hsize}p{0.04\hsize} p{0.20\hsize}p{0.20\hsize}p{0.20\hsize} }
\toprule
\multicolumn{7}{l}{{\bf Barrel system}}\\
\midrule
System & $r_{\rm{in}}$ & $r_{\rm{out}}$ & $z_{\rm{max}}$  & technology & \multicolumn{2}{l}{comments}\\
       & \multicolumn{3}{c}{[mm]}   &&&\\
\midrule
VTX    & 16         & 60        & 125       & silicon pixel sensors  & 3 double layers at                         & $ r_{0} = \unit{16, 37, 58}{\mm}$  \\
       &            &           &           &                        & $ \sigma_{r\phi,z}  = \unit{3.0}{\micron} $ & (layers 1-6)  \\
       &            &           &           &                        & $ \sigma_{t}  = \unit{2-4}{\mu}s $ &   \\    
       &            &           & &&&   \\
SIT    & 153        & 303       & 644       & silicon pixel sensors  & 2 double layers at                         & $ r = \unit{155, 301}{\mm}$  \\
       &            &           &           &                        & $ \sigma_{r\phi,z}  = \unit{5.0}{\micron} $ & (layers 1-4)  \\
       &            &           &           &                        & $ \sigma_{t}  = \unit{0.5-1}{\mu}s $ &   \\    
       &            &           & &&&   \\
TPC    & 329        & 1770      & 2350      & MPGD readout           &  220 (\ilds{163}) layers                                & $\sigma_{r\phi} \approx \unit{60-100}{\micron}$ \\
       &            & \ilds{1427} &           &                        &  \unit{$1\times 6$}{\mm^2} pads                  &   \\
       &            &           & &&& \\
SET    & 1773       & 1776      & 2300      & silicon strip sensors  &  1 double layer at                         & $ r = \unit{1774}{\mm}$ \\
       &\ilds{1430} &\ilds{1433}&           &                        &  $ \sigma_{r\phi}  = \unit{7.0}{\micron} $  & $\phi_{\mathrm{stereo}}=\unit{7}{^\circ}$  \\
\midrule
ECAL    & 1805      & 2028      & 2350      & W absorber             &   30 layers                          &   \\ 
        &\ilds{1462}&\ilds{1685}& &&& \\
        &           &           &           & silicon sensor         &   \unit{$5\times 5$}{\mm^2} cells    & SiECAL \\
        &           &           &           & scintilator sensor     &   \unit{$5\times 45$}{\mm^2} strips  & ScECAL \\
        &           &           & &&& \\
HCAL    &  2058     & 3345      & 2350      & Fe absorber                    &   48 layers                         &   \\ 
        &\ilds{1715}&\ilds{3002}& &&& \\
        &           &           &           & scintilator sensor, analogue   &   \unit{$3\times 3$}{\cm^2} cells   & AHCAL \\
        &           &           &           & RPC gas sensor,  semi-digital  &   \unit{$1\times 1$}{\cm^2} cells   & SDHCAL \\
        &           &           & &&& \\
\midrule
Coil    & 3425      & 4175      & 3872      &                                & 3.5 T field                         &   int.lengths = $2 \lambda $  \\
        &\ilds{3082}&\ilds{3832}& & &\ilds{4.0 T field} & \\
Muon    & 4450      & 7755      & 4047      &  scintillator  sensor          & 14  layers                          &   \\
        &\ilds{4107}&\ilds{7412}&           &                                & \unit{$3\times 3$}{\cm^2} cells     & \\
\bottomrule
\end{tabular}
\caption{\label{ild:tab:barrelpara}List of the main parameters of the large and small ILD detector models for the barrel part. Numbers correspond to the sensitive parts of the subdetectors.
  The numbers of the inner and outer radii refer to the distance from the IP at orthogonal impact
  of the corresponding detector plane. The parameters of the small model ($\delta r=343$~mm) are labeled with superscript$^s$.}
\end{table}

\begin{figure}[th!]

\begin{subfigure}{0.45\textwidth}
\includegraphics[width=\textwidth]{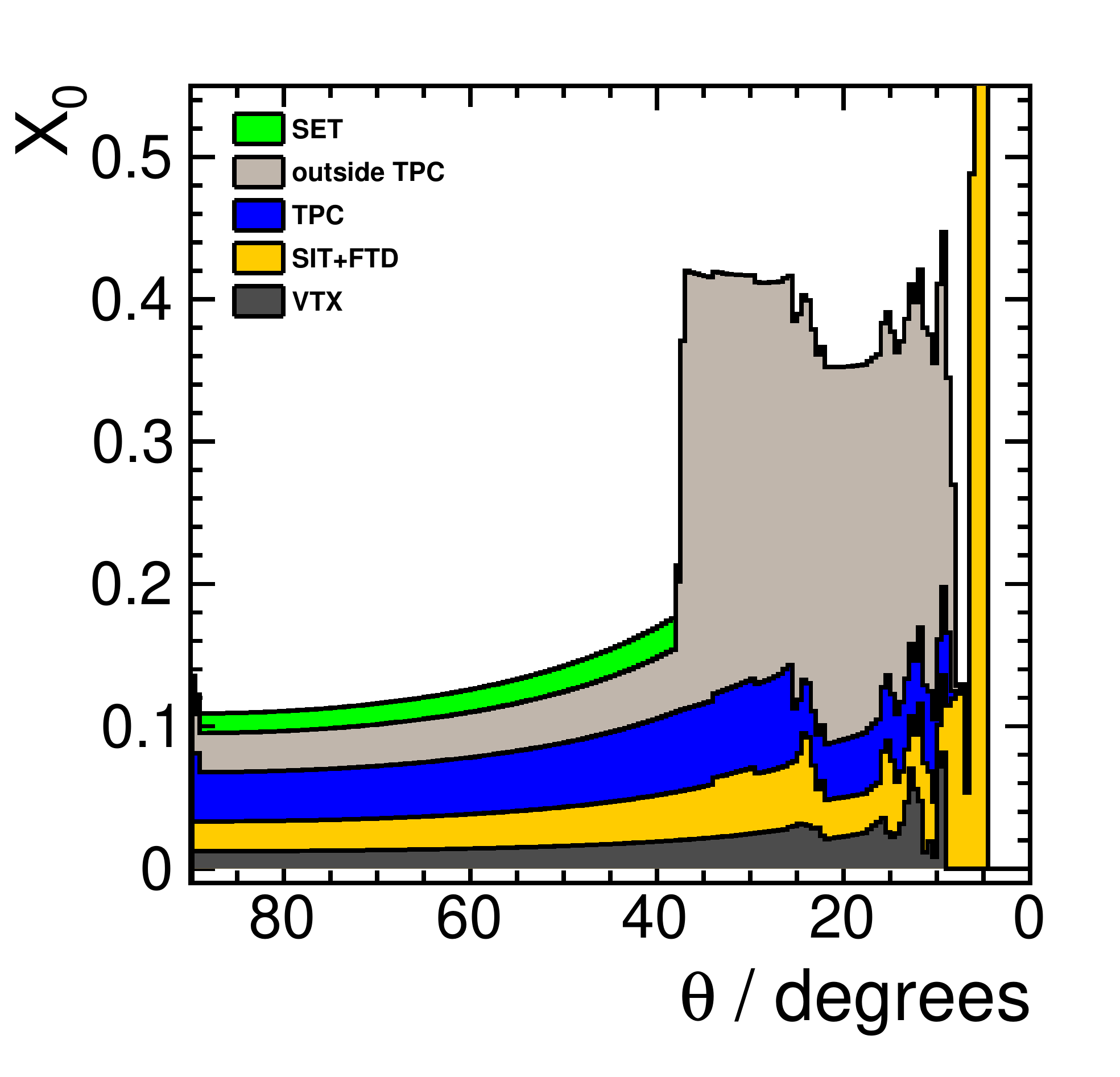}
 \caption{ \label{fig:det:material:x0}}
\end{subfigure}
\begin{subfigure}{0.45\textwidth}
\includegraphics[width=\textwidth]{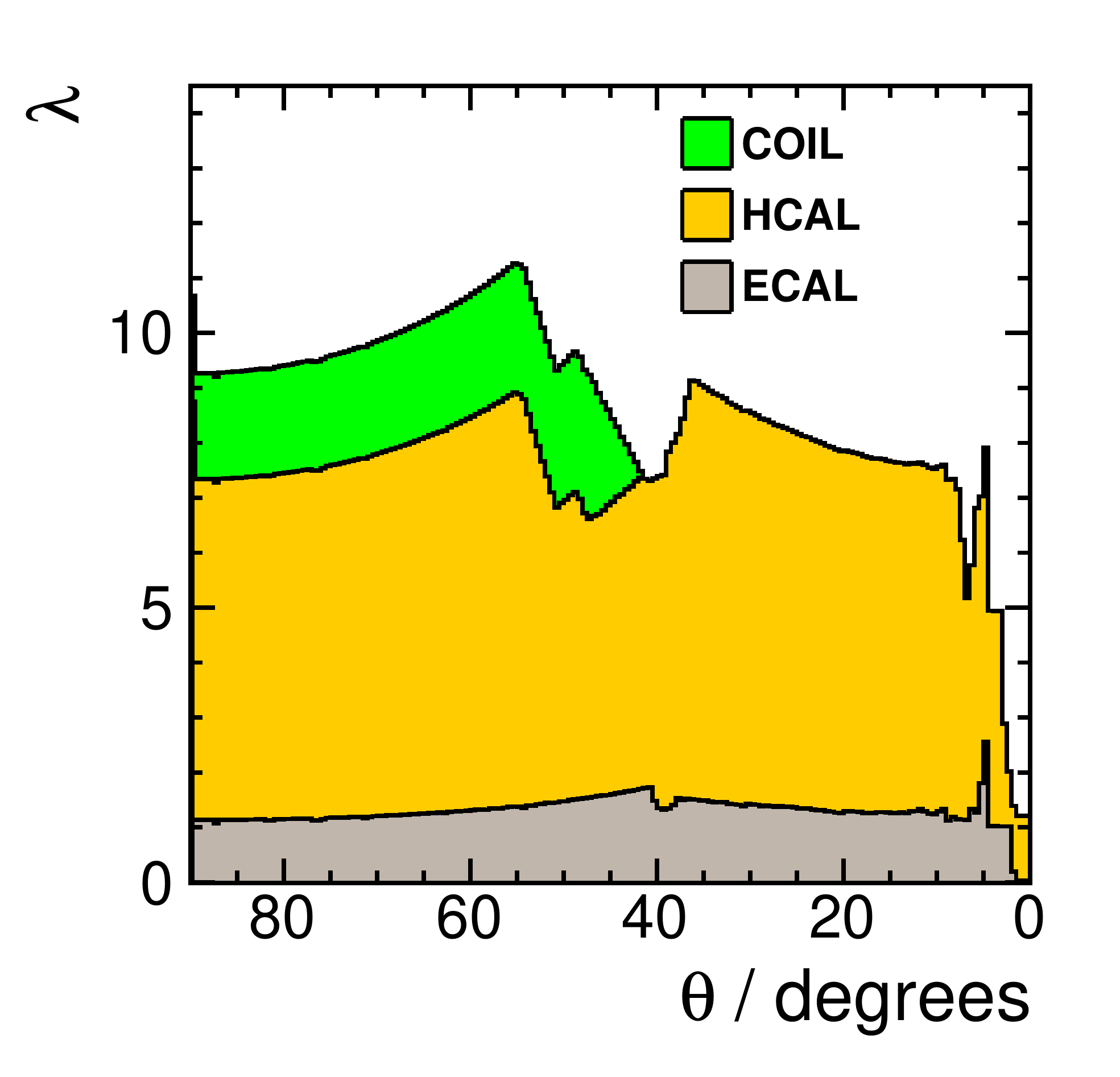}
 \caption{ \label{fig:det:material:lambda}}
\end{subfigure}
\caption[Material in the ILD detector]{\label{fig:det:material} (a) Average total radiation length of the tracker material as a function of polar angle. (b) Total interaction length seen up to the end of the electromagnetic calorimeter, the hadronic calorimeter and the solenoid coil, respectively.}
\end{figure}

\begin{table}\hspace*{-0cm}\small
\begin{tabular}{ l p{0.04\hsize}p{0.04\hsize}p{0.04\hsize}p{0.04\hsize}  p{0.18\hsize}p{0.18\hsize}p{0.18\hsize} }

\toprule
\multicolumn{8}{ l }{{\bf End cap system}}\\
\midrule
System  & $z_{\rm{min}}$  & $z_{\rm{max}}$  &  $r_{\rm{in}}$ &  $r_{\rm{out}}$ & technology  & \multicolumn{2}{l}{comments}\\
        & \multicolumn{4}{c}{[mm]}   &&&\\
\midrule

FTD     & 220         & 371        &         & 153     & silicon pixel sensors  & 2 discs         &   $ \sigma_{r\phi,z}  = \unit{3.0}{\micron} $ \\
        & 645         & 2212       &         & 300     & silicon strip sensors  & 5 double discs  &   $ \sigma_{r\phi }  = \unit{7.0}{\micron} $ \\
        &             &            &         &         &                        &                 &      $\phi_{\mathrm{stereo}}=\unit{7}{^\circ}$ \\
\midrule
ECAL    & 2411        & 2635       &  250    &   2096      & W absorber      &   30 layers                          & incl. EcalPlug\\
        &             &            &         & \ilds{1718} &                 &                                      & \\
        &             &            &         &             & silicon sensor       &   \unit{$5\times 5$}{\mm^2} cells    & SiECAL \\
        &             &            &         &             & scintillator sensor   &   \unit{$5\times 45$}{\mm^2} strips  & ScECAL \\
        &             &            & &&&& \\
HCAL    & 2650        & 3937       & 350     & 3226        & Fe absorber          &   48 layers                          &   \\
        &             &            &         &\ilds{2876}  &                      &                                      & \\
        &             &            &           &           & scintilator sensor, analogue   &   \unit{$3\times 3$}{\cm^2} cells   & AHCAL \\
        &             &            &           &           & RPC gas sensor,  semi-digital  &   \unit{$1\times 1$}{\cm^2} cells   & SDHCAL \\
        &             &            & &&&& \\
Muon    & 4072       &  6712       & 350     & 7716        & scintillator sensor & 12  layers   & \\
        &            &             &         &\ilds{7366}  &                     & \unit{$3\times 3$}{\cm^2} cells & \\
\midrule                                    
BeamCAL & 3115        & 3315       &  18     & 140    & W absorber      & 30 layers & \\
        &             &            &         &        & GaAs readout    &           & \\
        &             &            & &&&& \\
LumiCAL & 2412        & 2541       &  84     & 194    & W absorber      & 30 layers & \\
        &             &            &         &        & silicon sensor  &           & \\
        &             &            & &&&& \\
LHCAL   & 2680        & 3160       &  130     & 315   & W absorber      &                   &\\
\bottomrule
\end{tabular}
\caption{\label{ild:tab:endcappara}List of the main parameters of the large and small ILD detector models for the  end cap part. Numbers correspond to the sensitive parts of the subdetectors.
  The numbers of the inner and outer radii refer to the distance from the IP at orthogonal impact
  of the corresponding detector plane. The parameters of the small model are labeled with superscript$^s$.}
\end{table}
%
%
A key feature of the detector is the amount of material crossed by the particles: particle flow requires a thin tracker to minimise interactions before the calorimeter, and a thick calorimeter to fully absorb the showers and measure neutral hadrons. Figure~\ref{fig:det:material} shows the amount of radiation lengths of the tracker material and the total interaction lengths including the calorimeter system. 

\subsection{Subdetector layouts}
\label{ref:subsec:subdetectors}

The current design of subdetectors is presented including open options and critical aspects, as well as prospects for enhanced capabilities in the future. The most recent progress and status of each detection technology will be summarised in section~\ref{sec:subdetectors}.

\subsubsection{Vertex detector (VTX)}

\begin{figure}[t!]
\centering
\includegraphics[width=0.9\hsize]{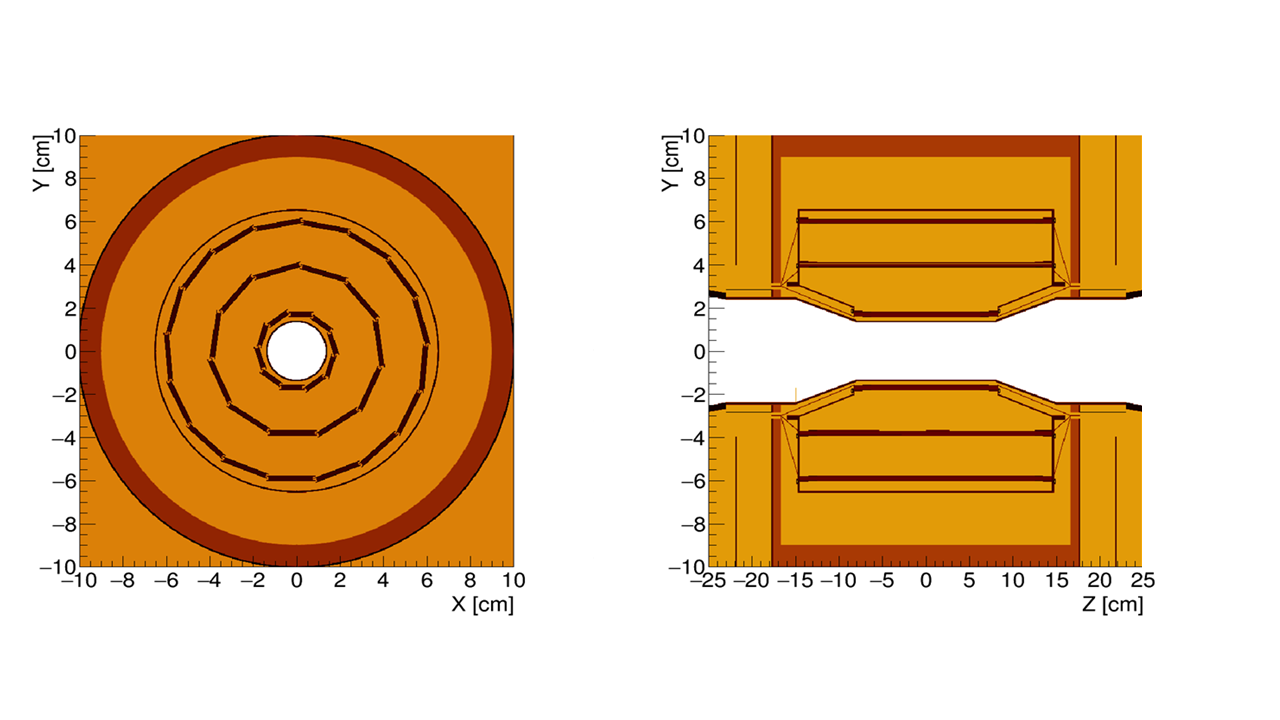}
\caption{Layout of the vertex detector. The inner layer is twice shorter than the others to minimize occupancy. The surrounding layer corresponds to the Faraday cage of the sensors, also acting as a cryostat for the FPCCD technology option.}
\label{fig:det:vertex}
\end{figure}

The vertex detector is realised as a multi-layer pixel detector with three double-layers (Figure~\ref{fig:det:vertex}). The detector has a pure barrel geometry. To minimise the occupancy from background hits,
the double-layer closest to the beam is twice shorter than the other two. 

Critical parameters of the VTX optimisation are the point resolution for secondary vertex tagging, required to be better than 3$\mu m$, and the material thickness which should not exceed $\simeq 0.15\%$ $X_0$ per layer to minimise multiple scattering. Three main technologies are under consideration to achieve the required goals:
\begin{itemize}
    \item {\bf CMOS pixels:} this well-established technology offers the advantages of high granularity with fully monolithic pixel digital electronics available from industrial processes. The most critical points of focus of current R\&D~\cite{Besson:2016ivb} are the readout speed, aimed to provide single bunch tagging capacity while keeping the power consumption low enough (with or without power pulsing), and the overall material budget of the layers.
    \item {\bf DEPFET pixels:} the DEPleted Field Effect Transistor (DEPFET) concept implements a first amplification stage in a FET in the high-resistivity, depleted silicon detector. Such active pixel detectors with a rolling shutter read-out can meet the stringent requirements of the ILD vertex detector~\cite{Alonso:2012ss,Richter:2003dn}. The technology offers the advantage of high granularity with a small layer material thickness, the digital electronics being shifted at the end of the ladders: the all-silicon ladder design~\cite{Andricek:2004cj}, that is fully self-supporting and requires no external support, minimises the material in the active area to approximately 0.12\% $X_{0}$. Critical aspects of DEPFETs are the industrialisation of the fabrication process and the integration of large detector surfaces.
    \item{\bf Fine Pixel CCD (FPCCD):} fine pixel CCD's ~\cite{Paredes:2014kda} offer the prospects for the highest granularities associated with low power consumption. Another advantage is the minimal material budget of the detector layer, however counter-balanced by the need of a cryostat to ensure low-temperature operation. Critical aspects under study are the readout speed and the resistance to radiation.  
\end{itemize}

\vspace{0.5cm}
The CMOS and DEPFET pixels have typical sizes of 20 microns and are readout in a continuous mode during bunch trains. The readout speed determines the capability to resolve individual bunches. The FPCCD pixels accumulate hits during one bunch train before readout and reset in-between trains. Their occupancy is kept acceptable thanks to a small pixel size of order 5 microns.  

\subsubsection{Silicon trackers (SIT, SET, FTD)}

\begin{figure}[t!]
\centering
\includegraphics[width=0.75\hsize]{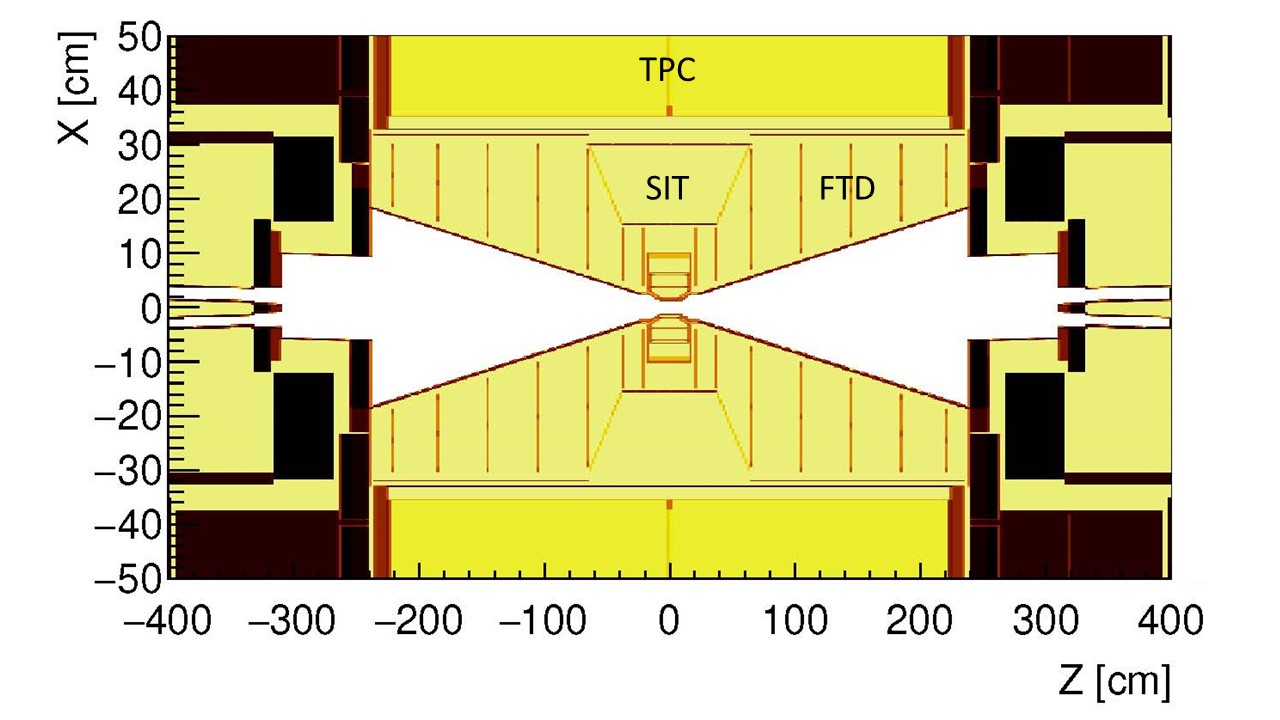}
\caption{Layout of the inner Silicon (SIT) and forward Silicon (FTD) trackers surrounding the vertex detector.}
\label{fig:det:silicon}
\end{figure}

A system of silicon trackers surrounds the VTX detector (Figure~\ref{fig:det:silicon}). 

In the barrel, two layers of silicon sensors are arranged as a silicon internal tracker (SIT) to bridge the gap between the VTX and the TPC, and a one-layer silicon external tracker (SET) is foreseen in-between the TPC and the ECAL. An external tracker similar to the SET, located between the TPC flange and ECAL endcap and named ETD in the DBD, is now no longer part of the ILD design. The baseline technology for the large area trackers of the DBD was silicon strips, but the progress made since then with CMOS detectors would now allow to equip the SIT with pixels instead of strips. The design of the SET is still open including the option to implement it as the first layer of the ECAL Calorimeter. The use of sensors with a high timing resolution of O(10ps) is considered in order to provide a time of flight (TOF) functionality  for particle identification. 

\begin{figure}[t!]
\centering
\includegraphics[width=0.75\hsize]{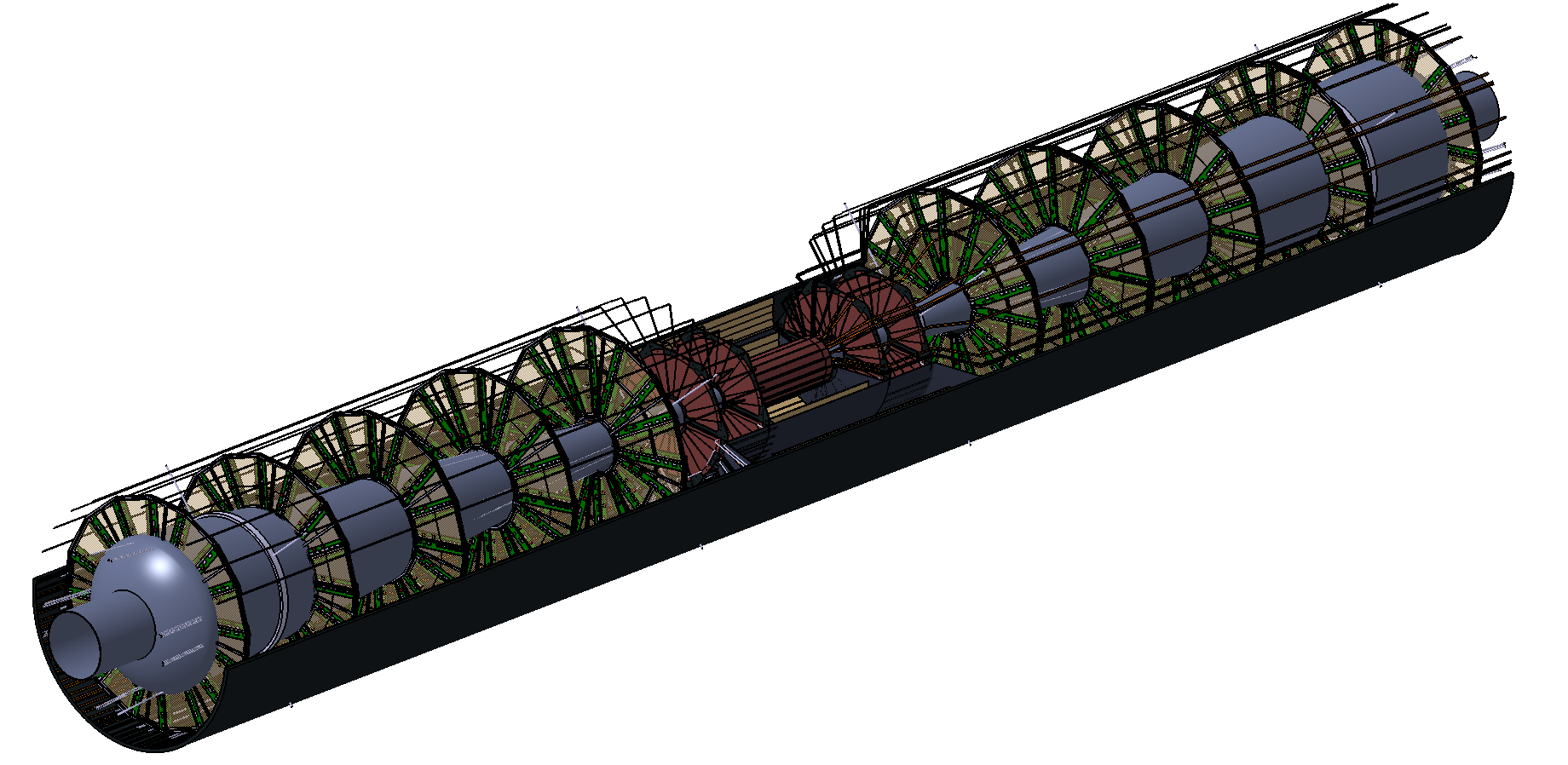}
\caption{Structure of the Forward Tracking Detector.}
\label{fig:det:FTD}
\end{figure}

The forward tracking detector (FTD) completes the coverage of the ILD experiment for charged particles emitted at shallow angles with respect to the beam (Figure~\ref{fig:det:FTD}). The FTD acceptance starts at 4.8 degrees, with at least one hit for tracks with polar angles below 35 degrees ($|\cos \theta| < $ 0.82) and nearly standalone tracking for tracks below 16 degrees ($|\cos \theta| <$ 0.96).

The first two disks are installed close to the vertex detector and are equipped with highly granular pixel detectors that provide precise 3D space points with a resolution of approximately 3-5 $\mu\mathrm{m}$. The technology candidates for these disks are similar to those of the vertex detector: CMOS, DEPFET, FPCCD, SOI, and future 3D-integrated devices. 

The remaining five disks extend out to the inner envelope of the TPC at $r=$ 30 cm. While the requirement on the resolution of the $r\phi$ measurement remains stringent, the cell dimension aligned with the $r$-coordinate is less constrained. These requirements can be met with micro-strip detectors: two single-sided detectors mounted under a small stereo angle may provide the required resolution. A solution based on CMOS monolithic pixel detectors with elongated pixels is likely to be quite competitive. Also the possibility of a TOF measurement in the outermost disks merits further study.

\subsubsection{Time projection chamber (TPC)}

\begin{figure}[t!]
\begin{subfigure}{0.45\textwidth}
\includegraphics[width=1\hsize]{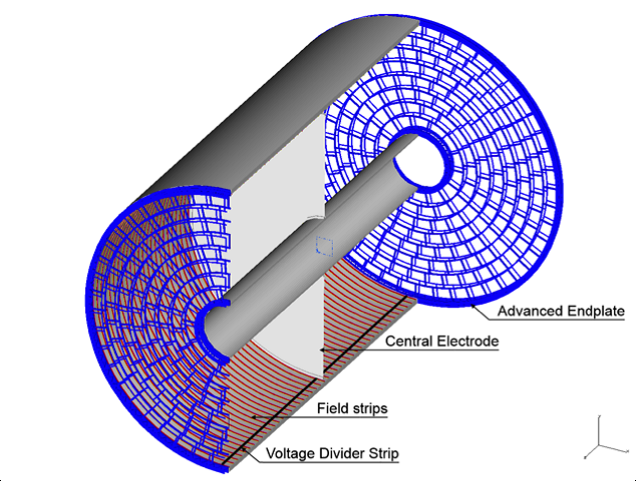}
\caption{}
\end{subfigure}
\begin{subfigure}{0.45\textwidth}
\includegraphics[width=1\hsize]{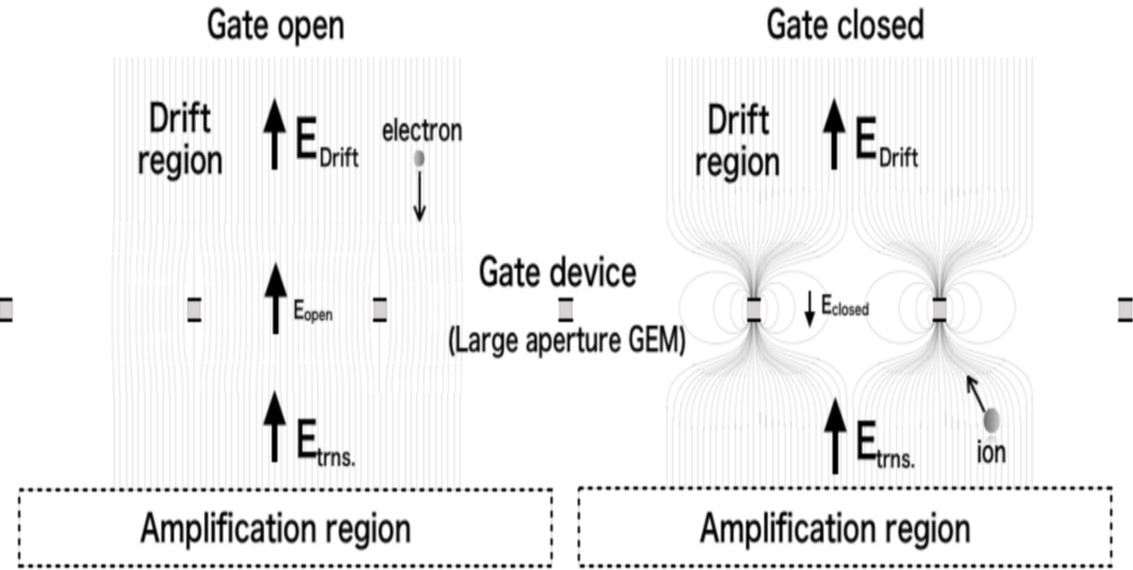}
\caption{}
\end{subfigure}
\caption[TPC layout]{(a) Conceptional layout of the TPC. (b) Principle of the ion GEM gating scheme showing the two electric field configurations within (left) and outside (right) bunch trains.}
\label{fig:det:TPC}
\end{figure}

A distinct feature of ILD is a large volume time projection chamber (Figure~\ref{fig:det:TPC} left). The TPC allows a continuous 3-dimensional tracking, dE/dx-based particle identification and minimum material. The required performance of the TPC as a standalone tracker is a momentum resolution $\sigma(1/p_T)$ better than $10^{-4}$GeV$^{-1}$, corresponding to a single point resolution of $100\mu$ over about 200 points, and a dE/dx resolution better than 5\%. One critical issue concerns potential field distortions due to ion accumulation within the chamber. At ILC this can be mitigated by implementing an ion gating between bunch trains, using large aperture GEM foils as shown in Figure~\ref{fig:det:TPC} right: during bunch trains, the voltage difference between the GEM sides is configured to allow drift electrons cross the GEM and reach the amplification region. Outside bunch trains, the voltage difference is reversed so that ions produced in the gas amplification region stay confined and are guided to the GEM surface where they are absorbed. 

\begin{figure}[t!]
\begin{subfigure}{0.49 \textwidth}
\includegraphics[width=1\hsize,viewport={0 -10 600 500},clip]{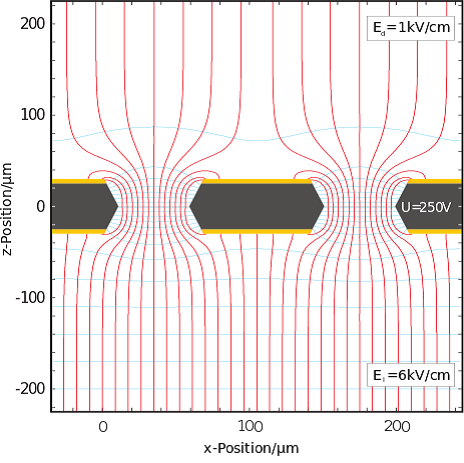} 
\caption{}
\end{subfigure}
\begin{subfigure}{0.49\textwidth}
\includegraphics[width=1.0\hsize]{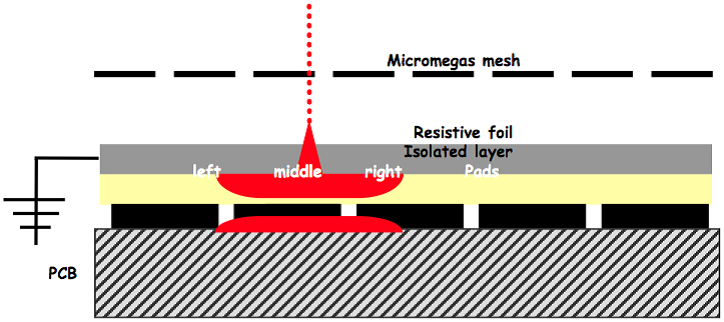}
\caption{}
\end{subfigure}
\caption[TPC readout]{(a) Electrical field map showing the amplification region in a GEM. (b) Schematic view of a micromegas readout.}
\label{fig:det:TPC_readout}
\end{figure}

Three options are under consideration for the ionisation signal amplification and readout:
\begin{itemize}
    \item GEM readout (Figure~\ref{fig:det:TPC_readout} left): the ionisation signal is amplified by passing through a GEM foil and is collected on pads.
    \item Micromegas readout (Figure~\ref{fig:det:TPC_readout} right): the ionisation signal is amplified between a mesh and the pad array where it is collected.
    \item GridPix: the ionisation signal is amplified as for the Micromegas case but collected on a fine array of silicon pixels providing individual pixel timing.
\end{itemize}
\vspace{0.5cm}
For the GEM and Micromegas options, the typical pad sizes are a few mm$^2$ (table~\ref{ild:tab:barrelpara}) and spatial resolution is improved by combining the track signals of several adjacent pads. For the GridPix option the pixel size of $\approx$50 microns matches the size of the mesh, providing pixel sensitivity to single ionisation electrons. The spatial resolution is improved and the dE/dx signal is measured by counting pixels or clusters. 

\subsubsection{Electromagnetic calorimeter (ECAL)}

\begin{figure}[t!]
\centering
\includegraphics[width=1.0\hsize]{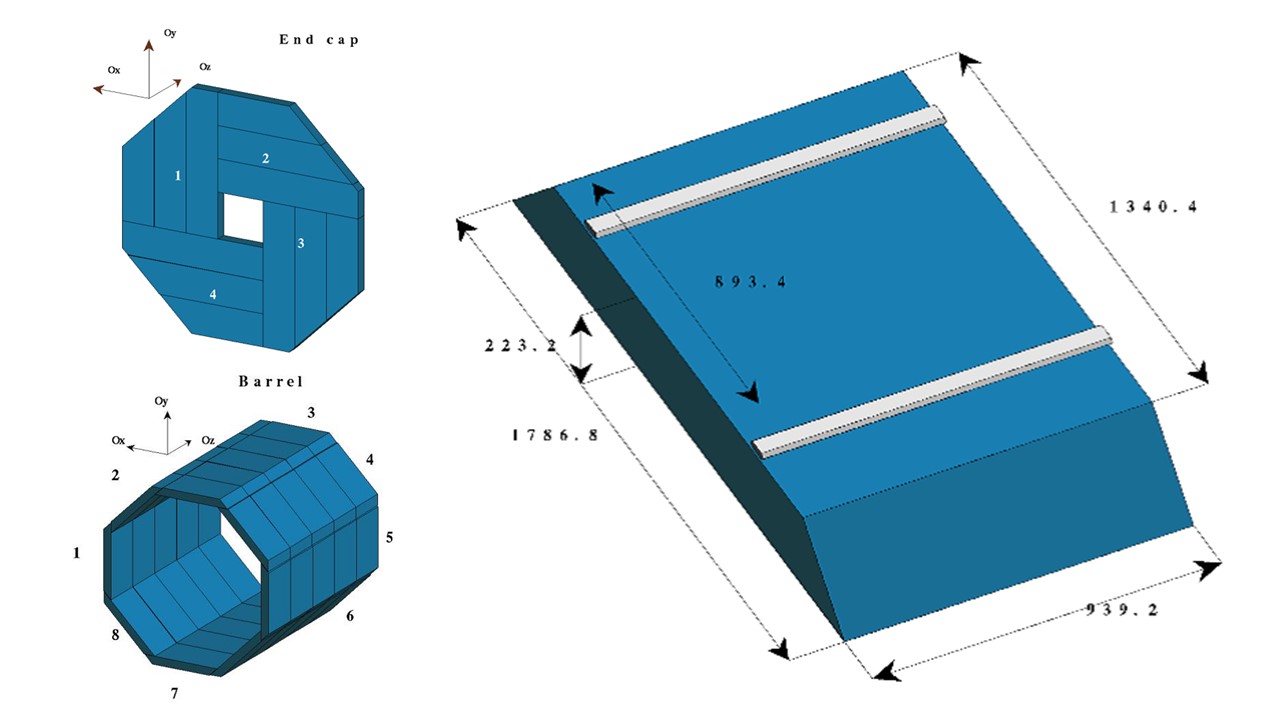}
\caption{Mechanical structure of the electromagnetic calorimeter: left: end-cap (top) and barrel (bottom); right: individual barrel module.}
\label{fig:det:ECAL}
\end{figure}

Electromagnetic showers are measured with a compact highly-segmented calorimeter (Figure~\ref{fig:det:ECAL}) with absorber plates made of tungsten. The ECAL barrel shape is octagonal with individual stacks laid such as to avoid projective dead zones in azimuth. The baseline number of layers is 30, with options to reduce the number to 26 or even 22, keeping the amount of radiation lengths identical and increasing the thickness of the sensitive medium to maintain a similar energy resolution for single particles. 

\begin{figure}[t!]
\centering
\includegraphics[width=0.8\hsize]{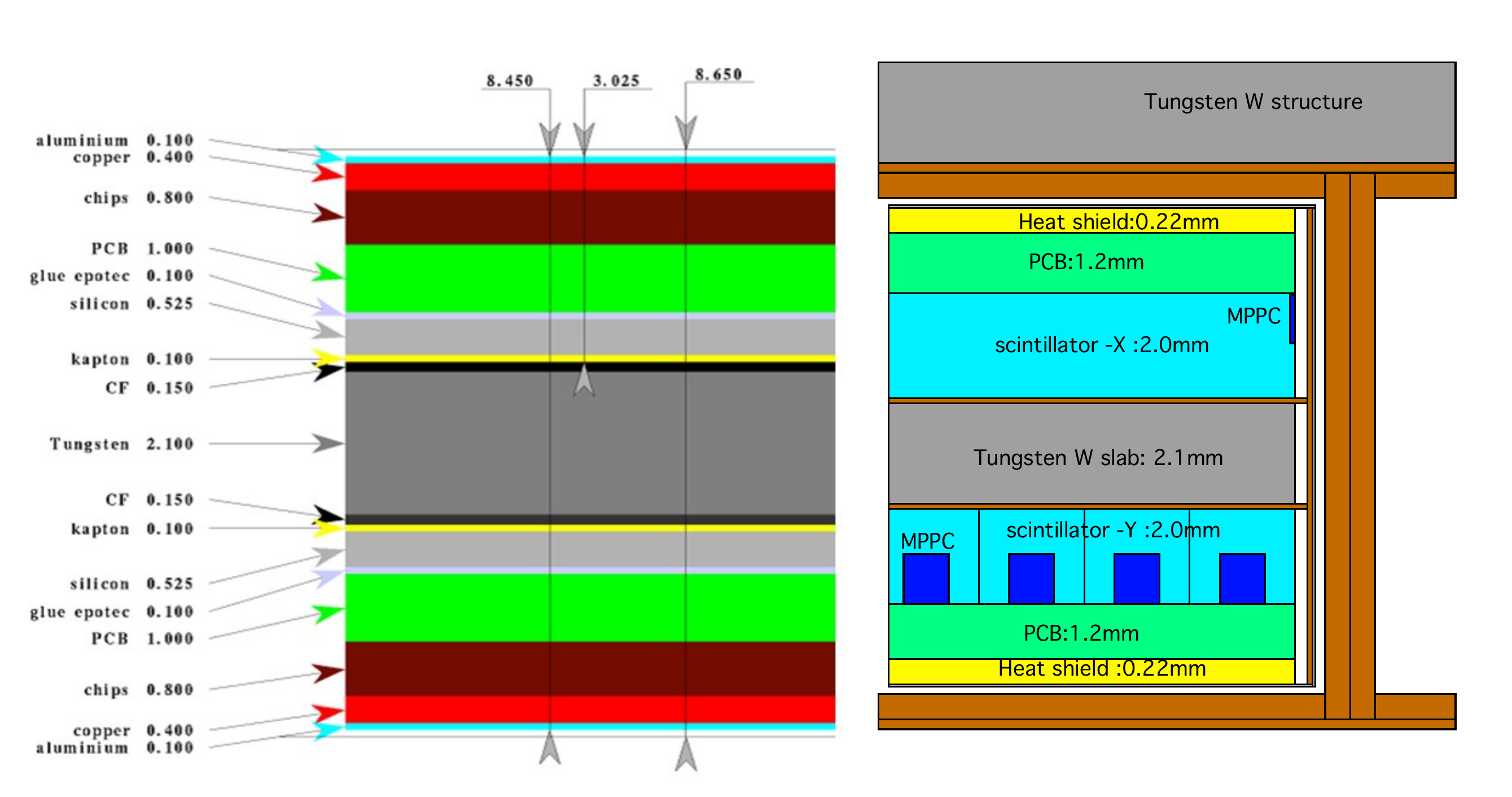}
\caption{ECAL sensitive layer: left: silicon option; right: scintillator option.}
\label{fig:det:ECAL_readout}
\end{figure}

The sensitive medium consists of silicon sensors with about 5x5 mm$^2$ pads bonded on a PCB equipped with front-end readout ASICs (Figure~\ref{fig:det:ECAL_readout} left). In order to reduce the costs it is considered to equip part of the sensitive layers with scintillator sensors readout through SiPMs (Figure~\ref{fig:det:ECAL_readout} right). In that case the scintillator strips would have a larger dimension of 45x5mm$^2$ with alternate orthogonal orientation. The option to equip the first layer with high-resolution timing sensors is also under consideration to provide a time-of-flight (TOF) functionality.

\subsubsection{Hadronic calorimeter (HCAL)}
\label{ild:sec:hcal}

\begin{figure}[t!]
\centering
\includegraphics[width=1.0\hsize]{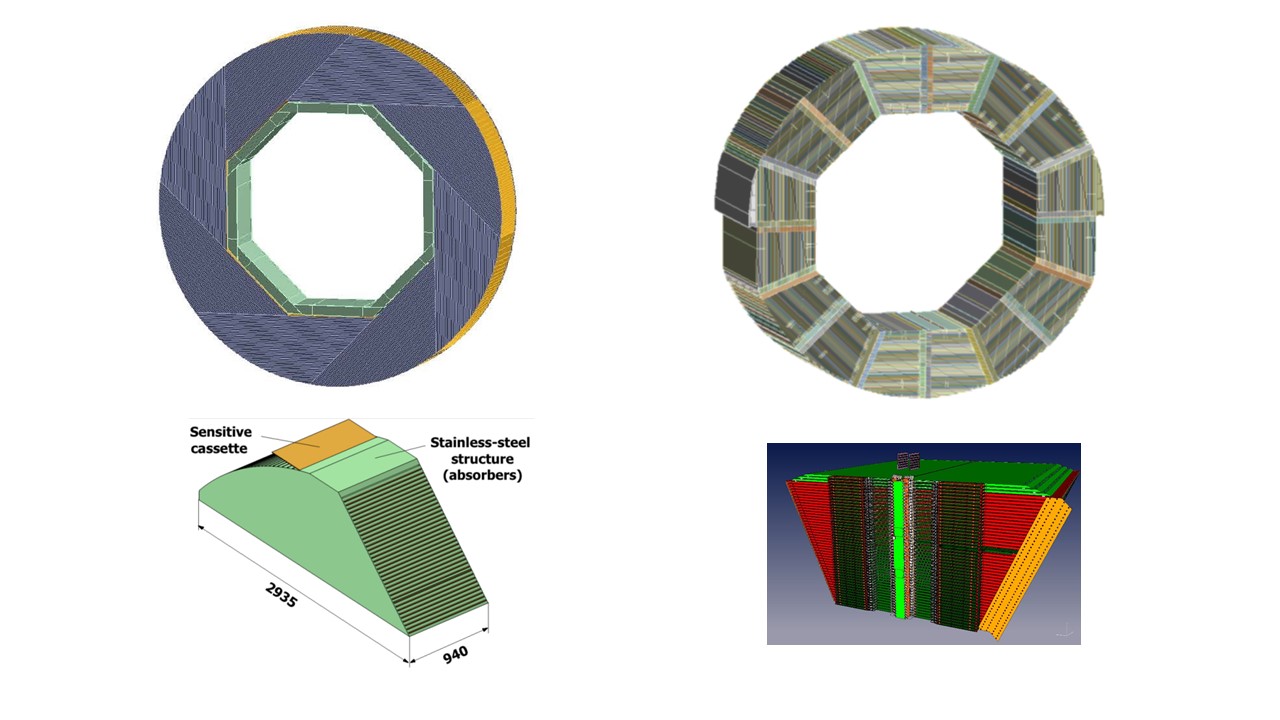}
\caption{HCAL barrel mechanical structure: full wheel (top) and individual stack (bottom) for the two configurations under consideration: the "Videau" option (left) and the "TESLA" option (right).}
\label{fig:det:HCAL}
\end{figure}

The hadronic calorimeter consists of 48 longitudinal samples with steel absorber plates. Two options are currently considered for the mechanical structure, differing mainly in the barrel region (Figure~\ref{fig:det:HCAL}): the "TESLA" barrel made of 2 wheels with signals extracted longitudinally in the gaps between the barrel and endcaps, and the "VIDEAU" barrel made of 3 or 5 wheels with signals extracted at the periphery between the HCAL and coil cryostat. The TESLA wheels are assembled from iron plates screwed to each other whereas the VIDEAU wheels are self sustainable welded structures. The VIDEAU configuration presents no projective dead zone in azimuth nor at $90^o$ polar angle, and offers a better mechanical stiffness, at the cost of a reduced accessibility of the interfaces for data concentration and power supply.

\begin{figure}[t!]
\centering
\includegraphics[width=1.0\hsize]{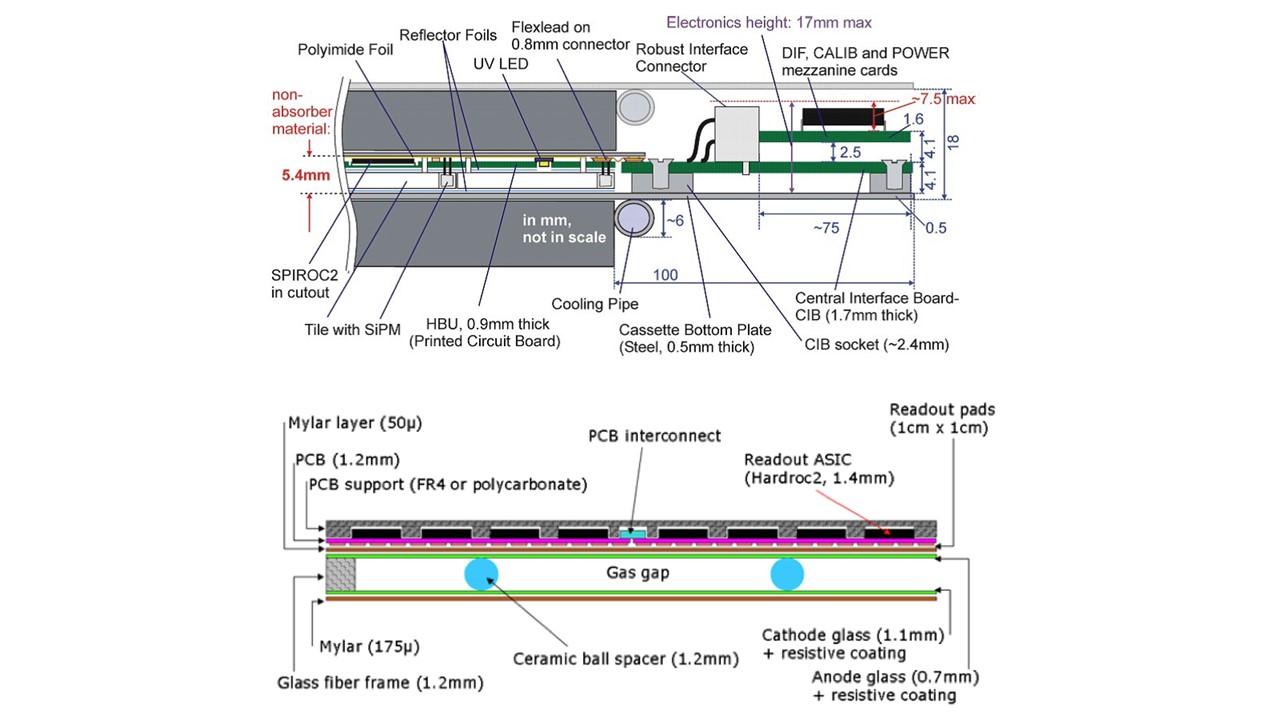}
\caption{HCAL sensitive layer; top: scintillator option; bottom: RPC option.}
\label{fig:det:HCAL_readout}
\end{figure}

The HCAL layers are instrumented with high granularity for an efficient separation of charged and neutral hadronic showers, necessary for particle flow, as well as for a good muon identification for flavor jet tagging. Two sensor options are under consideration (Figure~\ref{fig:det:HCAL_readout}): scintillator tiles of 3x3 cm$^2$ with analog readout through SiPMs, and RPCs with pads of 1 cm$^2$ readout with a semi-digital resolution of 2 bits. A timing functionality is implemented in the scintillator option and under consideration for RPCs: a resolution of O(1ns) already achieved with the scintillator option allows to tag late neutral components of the showers for particle flow, whereas a resolution of a few 10 ps, achievable with multigap RPCs, would provide TOF measurements. 


\subsubsection{Very forward detectors (LumiCAL, LHCAL, BeamCAL, ECAL ring)}

\begin{figure}[t!]
\centering
\includegraphics[width=0.4\hsize]{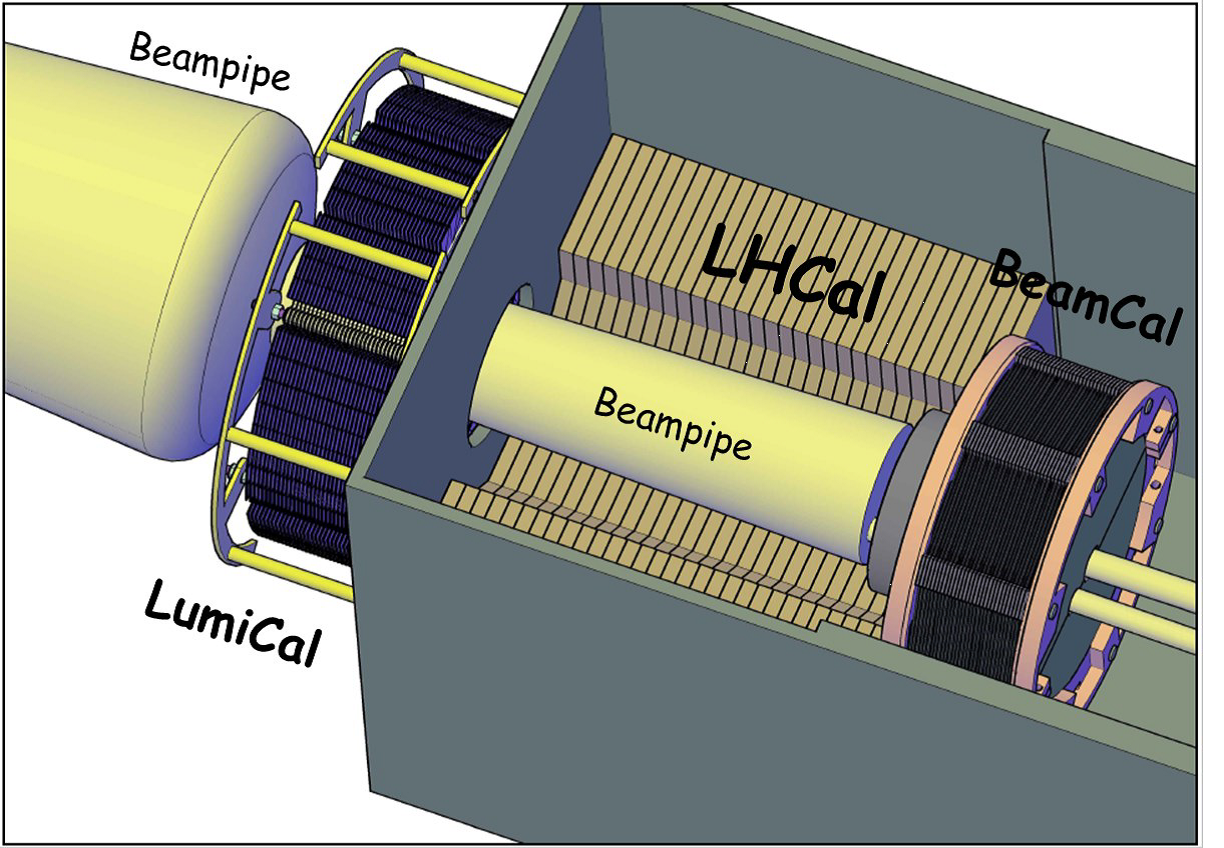}
\caption{Overal layout of the very forward detectors.}
\label{fig:det:VFS}
\end{figure}

The very forward region in the ILD detector is equipped with dedicated detectors to perform:
\begin{itemize}
\item a precise determination of the luminosity from Bhabha scattering electron pairs (LumiCAL);
\item an extension of the hadronic calorimetric coverage in the forward region (LHCAL);
\item calorimetric hermeticity down to 6~mrad from the beam pipe and fast monitoring of beam conditions (BeamCAL).
\item transition between the LumiCAL and the ECAL endcap (ECAL ring).
\end{itemize}
The longitudinal configuration of the LumiCAL+LHCAL+BeamCAL system has been adapted to the new beam optics as discussed in Chapter 3 (Figure~\ref{ild:fig:Forward_QD0}). The current overall layout of these detectors is shown in Figure~\ref{fig:det:VFS}.  

\begin{figure}[t!]
\centering
\includegraphics[width=0.8\hsize]{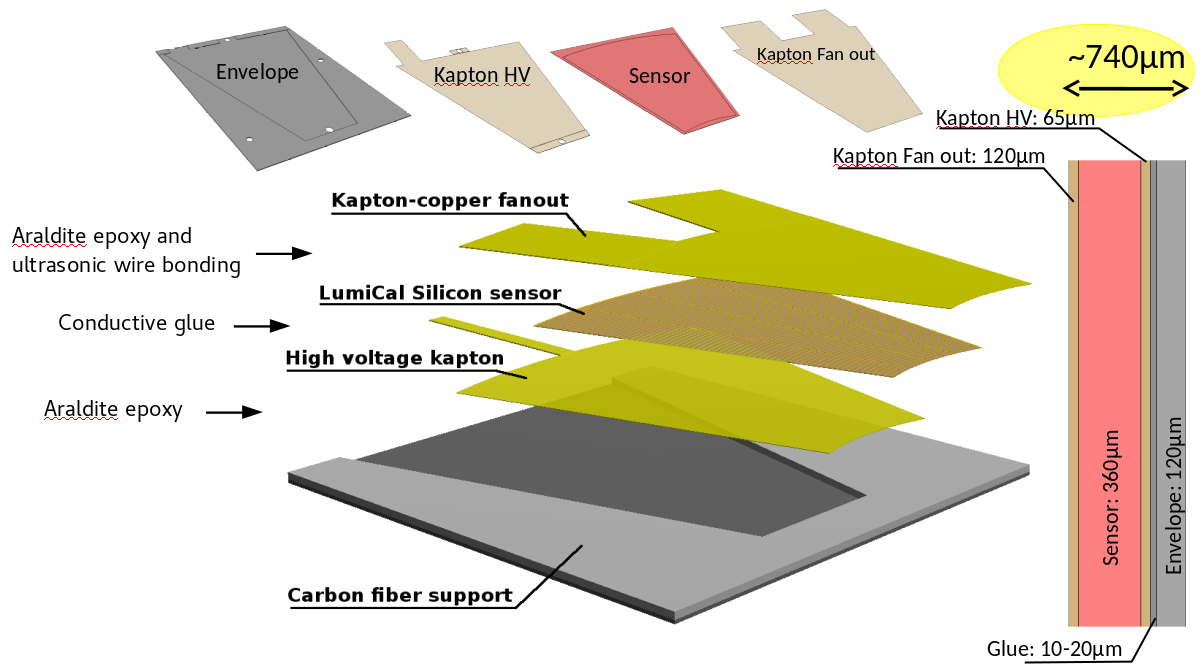}
\caption{Structure of a sensitive layer of the LumiCAL calorimeter.}
\label{fig:det:lumical}
\end{figure}

The very forward detectors are based on similar technologies as the ILD electromagnetic calorimeter, taking into account the specific conditions of the forward region such as the harder radiation environment or the need for an improved compactness to identify electromagnetic showers in a high occupancy environment. For the detectors positioned closest to the beam pipe new sensors such as saphire are considered. The LumiCAL will be based on silicon sensors similar to the ECAL (Figure~\ref{fig:det:lumical}), with thinner instrumentation layers to minimize the lateral spread of the showers.     

\subsubsection{Iron yoke instrumentation}

\begin{figure}[t!]
\centering
\includegraphics[width=1.0\hsize]{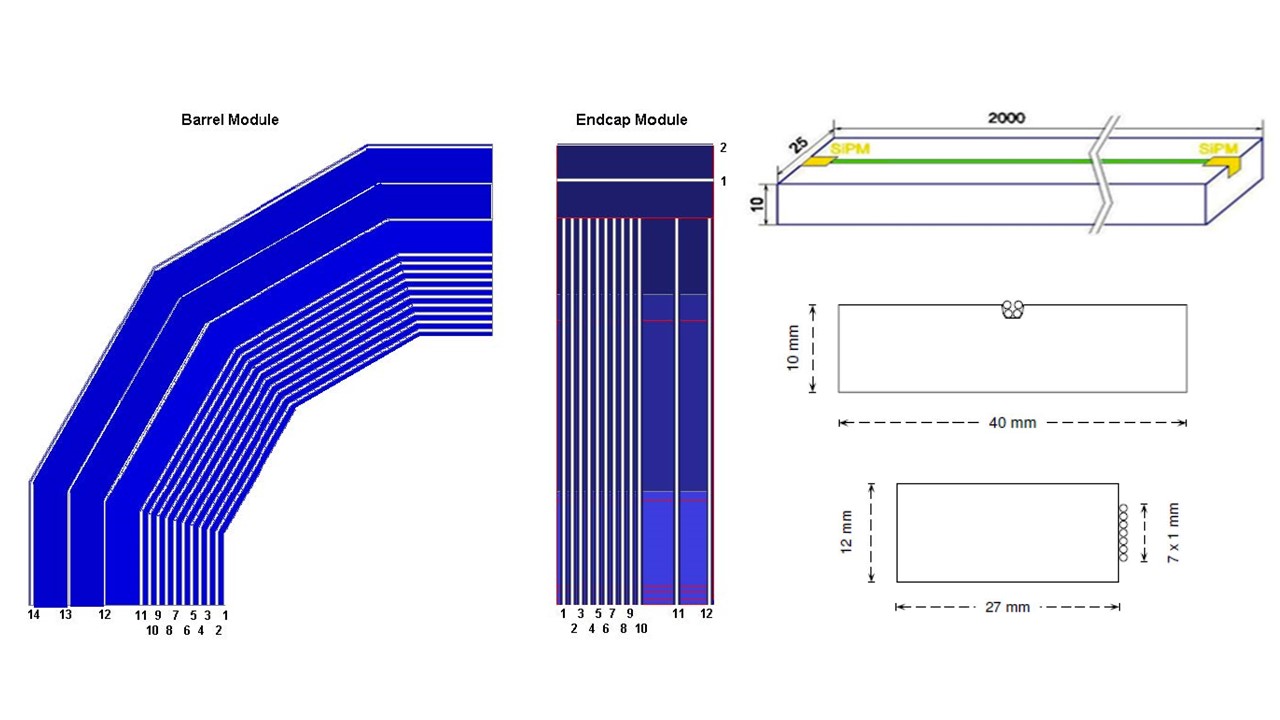}
\caption{Iron yoke baseline structure (left) and scintillator/SiPM instrumentation options (right).}
\label{fig:det:yoke}
\end{figure}

A large volume superconducting coil surrounds the calorimeters, creating an axial $B$-field of nominally 3.5 Tesla (IDR-L) or 4 Tesla (IDR-S). An iron yoke returns the magnetic flux of the solenoid, and, at the same time, serves as a muon filter, muon detector and tail catcher calorimeter. The baseline structure of the yoke is shown in Figure~\ref{fig:det:yoke} left for a configuration accounting for the maximum stray fields allowed in the push-pull configuration (chapter \ref{chap:ilc}). Optimization of the yoke size is ongoing to minimize the overal cost driven by the amount of iron (section 6.4.2). A number of iron yoke gaps will be instrumented for muon tracking and measurement of the tails of the hadronic showers. The required muon tracking precision is of the order of 1 cm in azimuth and of a few cm longitudinally. The instrumentation is expected to consist of scintillator bars but RPCs are also considered. In the scintillator option the light will be collected by wave length shifting (WLS) fiber readout with SiPMs at both ends~\cite{Denisov:2015jjl} (Figure~\ref{fig:det:yoke} right top). The WLS fibers will be positioned either in an extruded groove at the center of the large side of the bar (Figure~\ref{fig:det:yoke} right middle), or along the small side of the bar (Figure~\ref{fig:det:yoke} right bottom).  


\newpage
\section{Subdetector Technology Status}
\label{sec:subdetectors}


All ILD detector technologies under consideration have benefited from substantial developments since the DBD publication. Many activities are coordinated within worldwide R\&D Collaborations such as LCTPC~\cite{ild:bib:TPC_lctpc}, CALICE~\cite{ild:bib:CALICE} or FCAL~\cite{ild:bib:FCAL}. Compared to the DBD studies, which were still focused on intrinsic physics response and performance, many technologies have now developed operational implementations with technological prototypes which are mature for extrapolation to a full detector. Applications have indeed already started with many spin-offs to existing experiments such as the high-luminosity LHC detector upgrades. The experience gained with these projects will be a strong asset to the final design and construction of ILD.  

\subsection{Vertex Detector}

The vertex detector is a high-precision small device which is expected to be one of the latest subdetectors to be built and inserted within ILD. The development of optimal technologies can therefore proceed until a few years before the start of ILC. There has been much progress in this direction in the past 5 years for the three main options under consideration: CMOS, DEPFET and FPCCD sensors.

\subsubsection{CMOS sensors}

\begin{figure}[t!]
\centering
\includegraphics[width=0.6\hsize]{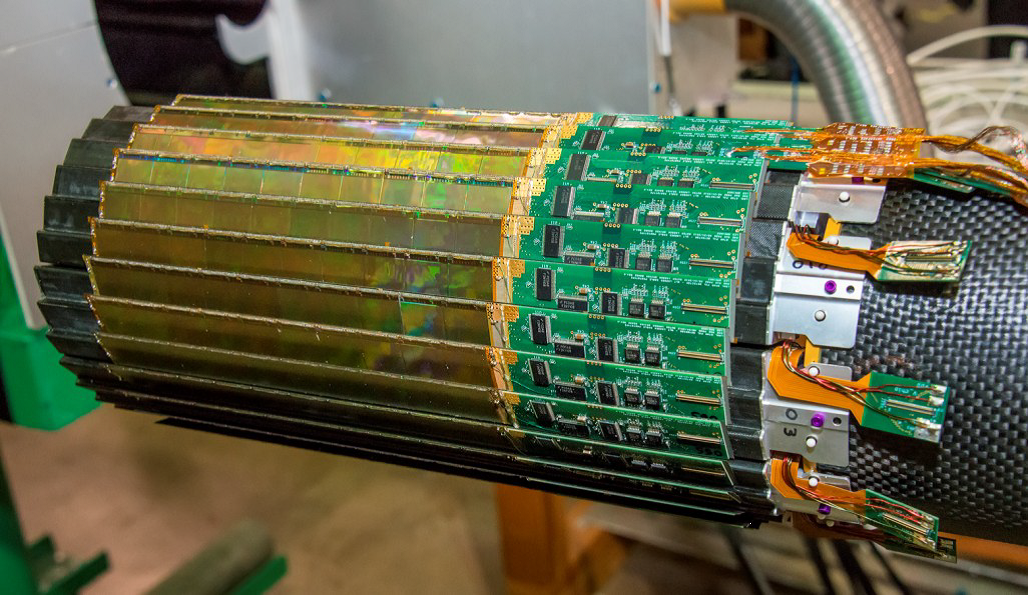}
\caption{The vertex detector of the STAR experiment based on the ULTIMATE CMOS pixel sensor~\cite{Contin:2017mck}.}
\label{fig:det:VTX_STAR}
\end{figure}

The use of CMOS sensors for particle physics has benefited a lot from the development in the past two decades of the MIMOSA chip series by the IPHC Institute. A first full scale particle physics detector application has been realized with the STAR vertex detector~\cite{Contin:2017mck} (Figure~\ref{fig:det:VTX_STAR}) on the RHIC hadron collider. Since then the technology has further developed as a widespread standard for pixel detectors, including many applications to e.g. LHC upgrades or new experiments. 

The general trend of performance improvements towards ILD specifications is summarized in table~\ref{ild:tab:CMOSdev}. Compared to STAR the new applications for the ALICE upgrade~\cite{AglieriRinella:2017lym} and CBM at FAIR~\cite{Koziel:2017loo} have moved to a technology with a smaller pattern, have implemented a new data driven readout scheme, and have improved the time resolution and power consumption to values close to ILD needs. The ALICE detector also concerns a very large area of more than 10 m$^2$, which qualifies the technology for the inner layers of a central tracker.

With these applications more attention is given to integration aspects of the technology. The chip intrinsic power consumption is now close to the ILD specification and could still be reduced by a factor 10-20 with power pulsing. To this respect a trade-off will have to be made between readout speed (related to time resolution) and power. With the expected heat production air cooling as done at STAR could be sufficient, but ILD has stronger constraints on the possible air flow due to a more forward instrumentation than STAR. This critical issue requires further studies. Low material ladder supports have been developed with the PLUME concept~\cite{Nomerotski:2011zz}, consisting of a thin foam layer carrying pixel chips on both sides as a double layer. First PLUME ladders have been built and a second version has been successfully operated for the BELLE II beam commissioning (Figure~\ref{fig:det:VTX_BELLE2} top).

\begin{table}\hspace*{-0cm}\small
\begin{tabular}{ l c c c c }
\toprule
DETECTOR: & STAR-PXL & ALICE-ITS & CBM-MVD & ILD-VXD \\
& (ULTIMATE) & (ALPIDE) & (MIMOSIS) & (PSIRA) \\
& 2014-16 & 2021-22 & 2021-22   & 2030 \\
\midrule
Technology (AMS): & 0.35 $\mu$m & 0.18 $\mu$m & 0.18 $\mu$m & $<$ 0.18 $\mu$m \\
Pixel size ($\mu$m$^2)$: & 20.7 x 20.7  & 27 x 29 & 22 x 33 & 22 x 22 or 18 x 18 \\
Readout mode:   & rolling shutter & data driven & data driven & data driven \\
Time resolution ($\mu$s):   & 135 & 5-10 & 5 & 1-4 \\
Power (mW/cm$^2$):   & 150 & 35  & 200  & 50-100 \\
Material ($X_0$/layer):   & 0.39\% & 0.3\%  & - & 0.15\% \\
\bottomrule
\end{tabular}
\caption{\label{ild:tab:CMOSdev}Development path of CMOS pixel sensors towards ILD.}
\end{table}

\subsubsection{DEPFET sensors}

\begin{figure}[t!]
\centering
\includegraphics[width=1.0\hsize]{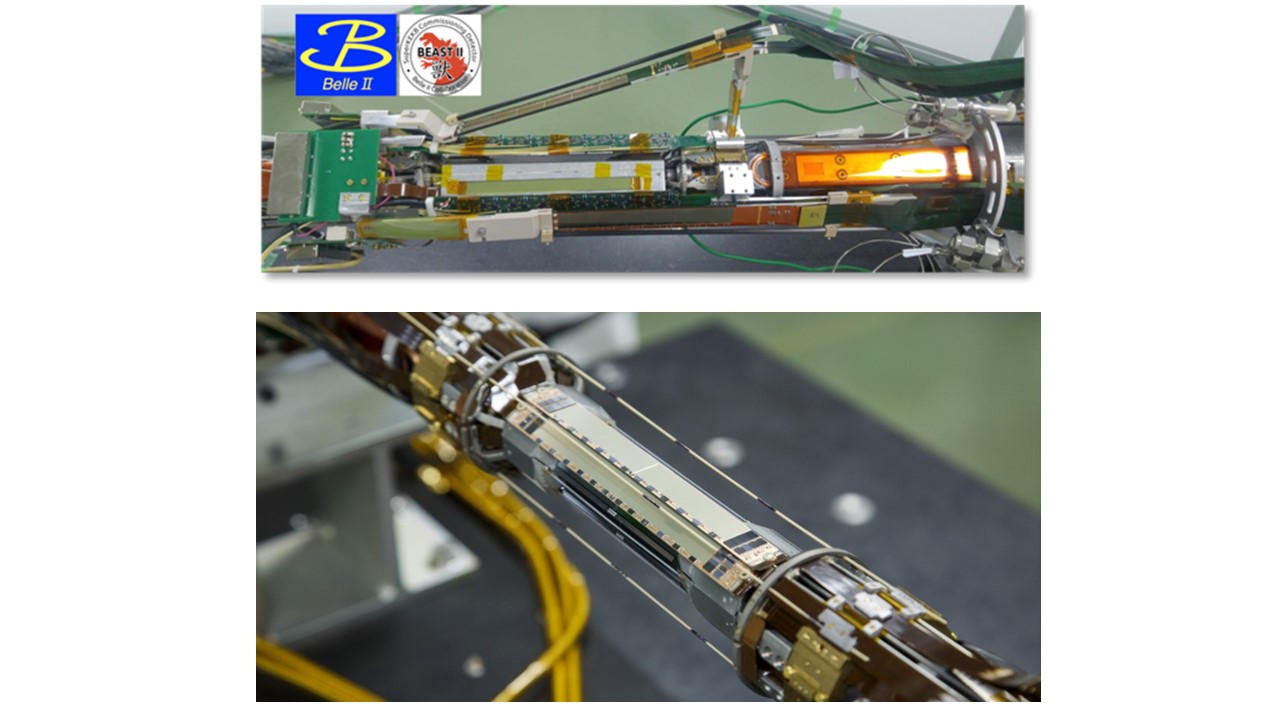}
\caption{Pixel detectors of the BELLE II experiment. Top: beam commissioning with PLUME CMOS (inclined sensors) and DEPFET (barrel) ladders. Bottom: the BELLE II DEPFET vertex detector.}
\label{fig:det:VTX_BELLE2}
\end{figure}

The development of DEPFET sensors in particle physics is reaching maturity. Following the demonstration of small
prototypes~\cite{Andricek:2011zza,Velthuis:2008zza} and first operational ladders five years ago the technology was 
chosen as the baseline for the vertex detector~\cite{Marinas:2011zz} of the Belle II experiment~\cite{Abe:2010gxa}. 
As many requirements of Belle II are similar to those
of the ILC, this can be seen as a 30\% prototype of the ILC vertex detector. DEPFET ladders have been successfully used 
in the BELLE II beam commissioning detector (BEAST 2). The first layer of the pixel vertex detector was installed in 2018
and is now in operation in the experiment (Figure~\ref{fig:det:VTX_BELLE2} bottom). Due to a low yield of the module assembly 
process, the second layer vertex detector is expected to be completed in 2020. While the experiment is 
taking data, studies towards future BELLE II upgrades based on advanced DEPFET technology are starting. The development of advanced DEPFET solutions for the ILD vertex detector is synergetic with this effort.

\subsubsection{FPCCD sensors}

\begin{figure}[t!]
\centering
\includegraphics[width=0.7\hsize]{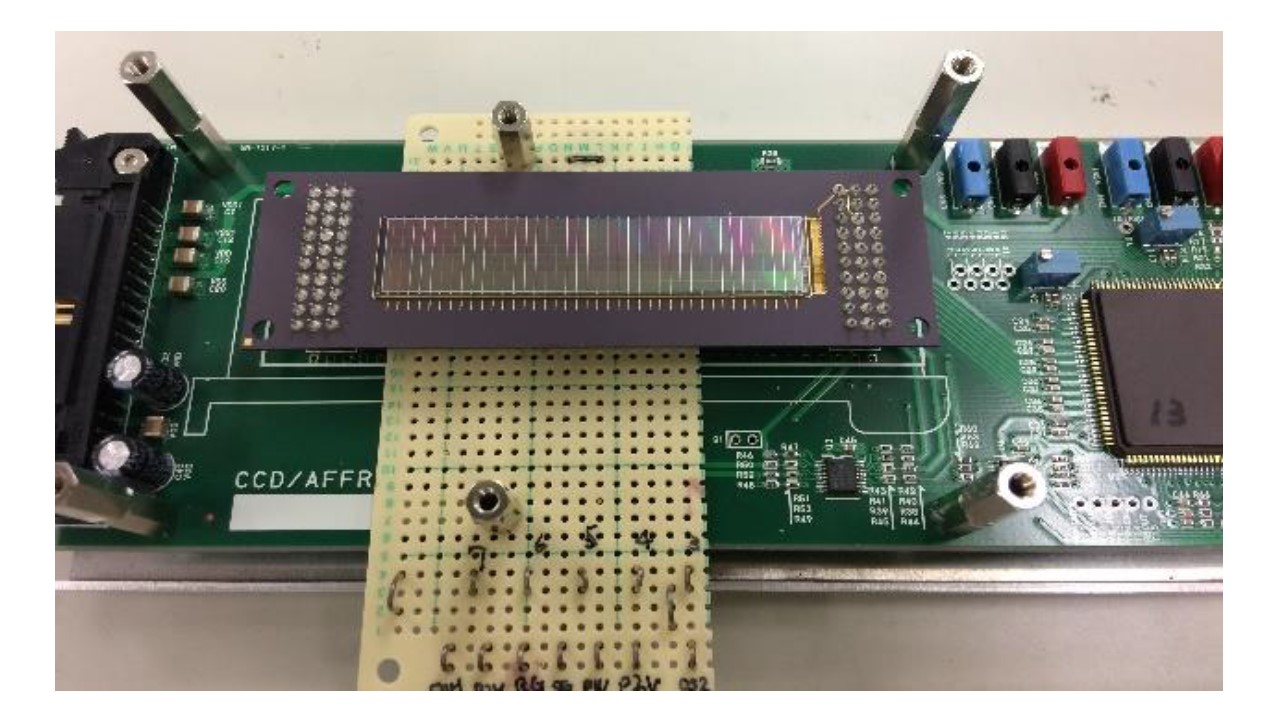}
\caption{First full size FPCCD ladder on its test bed.}
\label{fig:det:VTX_FPCCD}
\end{figure}

The FPCCD technology is not yet used in full size detector applications but a first large prototype has been built~\cite{ild:bib:FPCCD} with a sufficient size to cover the inner layer of the ILD vertex detector. The prototype is currently undergoing detailed characterization with e.g. radioactive source signals (Figure~\ref{fig:det:VTX_FPCCD}). Irradiation tests of FPCCDs are also being performed since radiation hardness is a critical aspect of this technology.   

\subsection{Silicon Trackers}

The silicon sensors surrounding the vertex detector will directly benefit from the pixel detector R\&D reported above as regards their high spatial resolution components. In the past years the specific silicon R\&D has focused around two lines: a generic development of high-resolution timing sensors, and the prototyping of mechanical support structures of the Forward Tracking Detector.

\subsubsection{iLGADs for precise tracking and time stamping (4D-tracking)}

The Low Gain Avalanche Detector (LGAD) is the baseline sensing technology of the recently proposed Minimum Ionizing Particle (MIP) end-cap timing detectors (MTD) at the ATLAS and CMS experiments. LGADs are n-on-p silicon detectors with an internal gain. To obtain this gain, an extra, highly doped, layer is added just below the p-n junction of a PIN diode. This highly doped region creates a very high electric field region. This electric field induces an avalanche multiplication of the electrons and thus create additional electron-hole pairs\cite{PELLEGRINI201412}. 

\begin{figure}[!htbp]
\centering
\includegraphics[width=1.0\hsize]{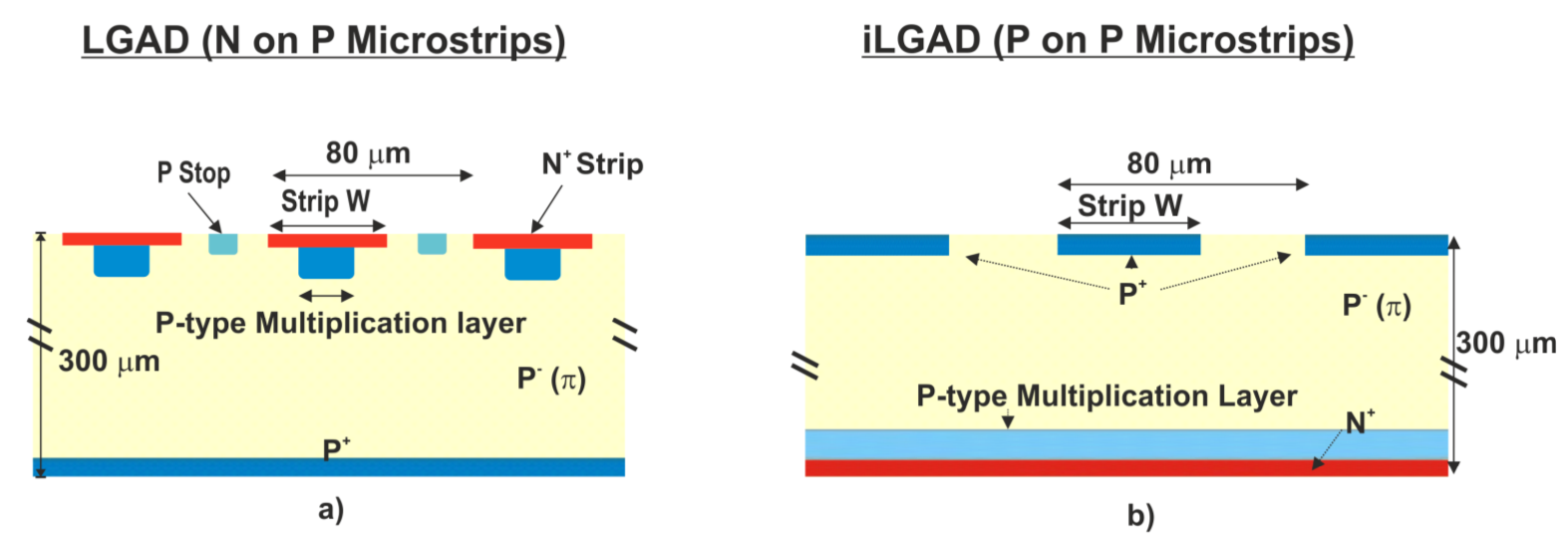}
\caption{Cross-sections of the core layout of LGAD (a) and iLGAD (b) micro-strip designs.}
\label{fig:det:cross-section}
\end{figure}

The current MTD sensor is designed as a multi-pad matrix detector delivering a poor position resolution, due to the relatively large pad area, around 1 mm$^2$; and a good timing resolution, around 20-30 ps. In its current technological implementation, the signal of MIP particles hitting the inter-pad region is visible but collected with a reduced amplification which severely degrades the timing resolution. This limitation is named as the LGAD fill-factor problem. For ILD, a true 4D tracking  must overcome the poor spatial resolution and fill factor limitations. A new p-in-p LGAD architecture named as inverse LGAD (iLGAD) tackles both issues~\cite{Carulla_2016}. Contrary to the conventional LGAD design, the iLGAD has a non-segmented multiplication layer, and it should ideally present a constant gain value over all the sensitive region of the device without gain drops between the signal collecting electrodes, see figure\,\ref{fig:det:cross-section}. This feature has been experimentally confirmed on a strip-like segmented iLGAD and compared against a conventional strip-like LGAD and PIN devices~\cite{Curras2019}. 

\begin{figure}[t]
\centering
\begin{subfigure}{0.48\textwidth}
\includegraphics[width=7cm]{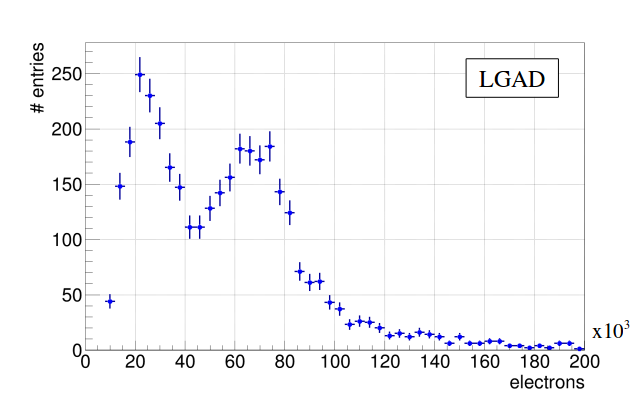}
\caption{}
\end{subfigure}
\begin{subfigure}{0.48\textwidth}
\includegraphics[width=7cm]{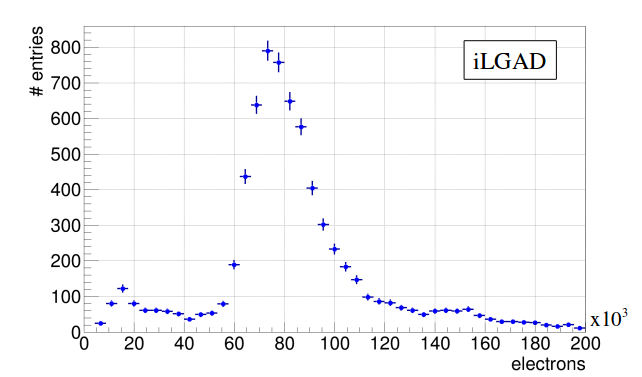}
\caption{}
\end{subfigure}
\caption{Charge distribution measured during the test beam for one LGAD strip detector (a) and one iLGAD strip detector (b). The fill-factor problem visible for the LGAD case is not present in the iLGAD structure.} 
\label{fig:det:fill-factor}
\centering
\end{figure}

The tracking performances of one LGAD and one iLGAD strip detector were studied in a test beam at CERN-SPS and compared with a standard PIN strip detector. These three strip detectors were unirradiated and consisted of 45\,strips with a 160\,$\mu$m pitch. 
The big advantage of the iLGAD technology was confirmed during the test beam. It was proven that while in the LGAD strip detector the signal is severely degraded in the inter-pad region, the iLGAD presents a very constant gain value over all the sensitive region of the device. These results are shown on figure\,\ref{fig:det:fill-factor}. The charge distribution of the LGAD measured during the beam tests presents two peaks: one around 24\,ke, corresponding to the MIP particles that cross the interstrip region where the generated signal is not amplified (same charge measure in the PIN strip); and one around 77\,ke, corresponding to the particles that cross the region where the signal is amplified. On the other hand, the same plot produced for the iLGAD detector presents only one peak in the charge distribution around 75\,ke. In this case,  the signals produced for all MIP particles that cross the sensitive region of the device are amplified, resulting in a much better and uniform response along the sensitive region. The tracking and timing performances were also quantified during the test beam: the position precision reaches a few tens of microns and the timing resolution stands between 20 and 40 picoseconds depending on the amplification step features.

\subsubsection{Thermo-mechanical studies of the FTD support structure}

\begin{figure}[t!]
\centering
\includegraphics[width=0.6\hsize]{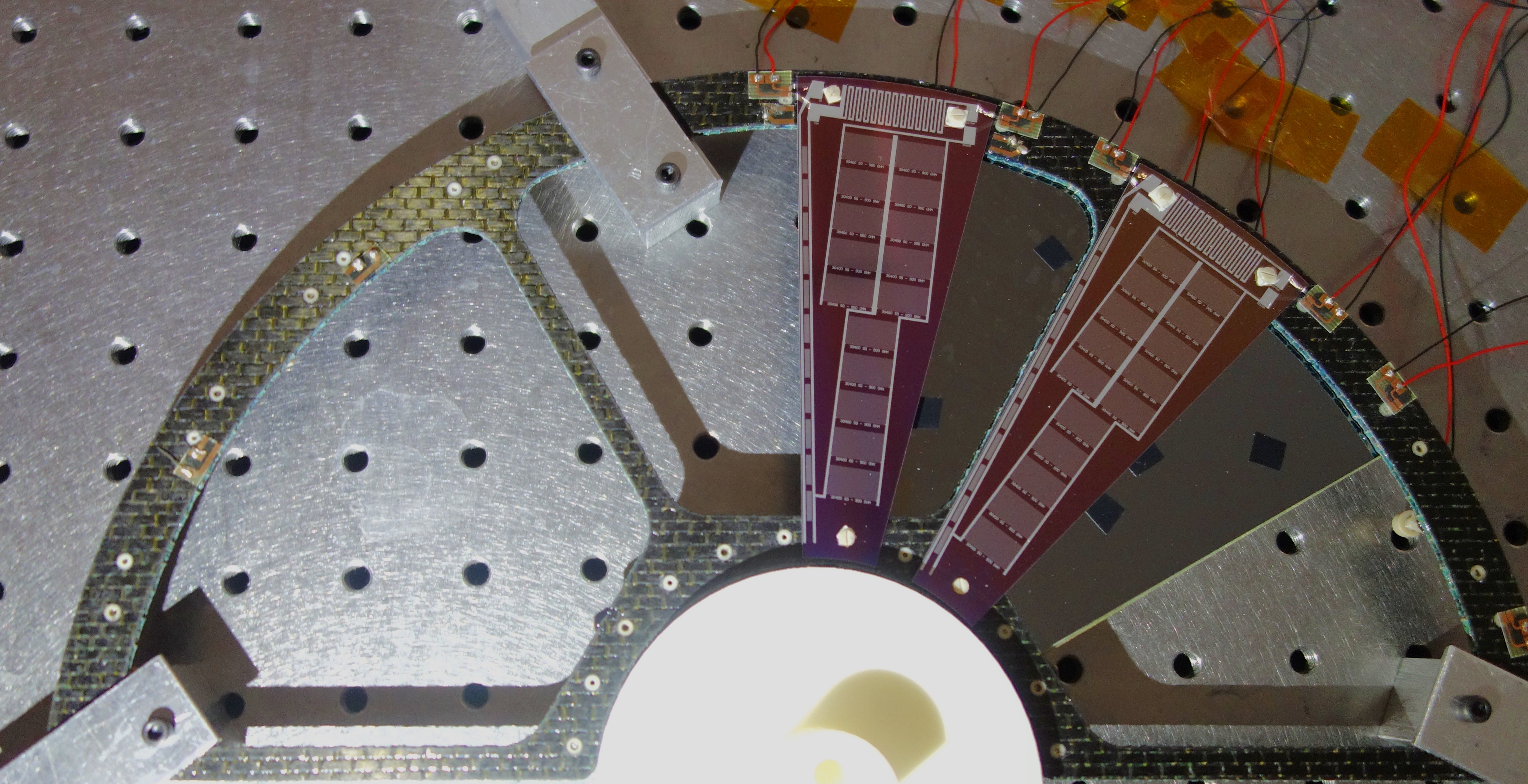}
\caption{FTD thermo-mechanical mockup for the 2 inner disks.}
\label{fig:det:FTD_mockup}
\end{figure}

The combination of the requirements of minimal material and a mechanical stability to 
the level of several $\mu\mathrm{m}$ represents quite a challenge. A realistic, 
full-scale mechanical prototype of the disks has been developed to characterize 
their thermo-mechanical performance 
in conditions that resemble those of the experiment (Figure~\ref{fig:det:FTD_mockup}). 
This {\em mock-up} is based on a carbon-fibre support disk
and 50~$\mu\mathrm{m}$ thick silicon petals. 

The carbon-fibre disk consists of
a 1 mm thick rohacell core covered on both sides by three-layer carbon-fibre skins. The resulting structure adds less than 0.04 \% of a radiation length ($X_0$) to the material budget, when averaged over the area of the disk. The mounting points for the Silicon sensors are formed by precisely machined 
PEEK inserts that are glued into the carbon-fibre structure. The gluing procedure controls the relative position of the mounting points to better than 50 $\mu \mathrm{m}$ with a custom jig. 

The Silicon petals were produced using the Silicon-on-Oxide process that is at the heart of the
all-silicon-ladder concept~\cite{Andricek:2004cj}. The 50 $\mu \mathrm{m}$ thick sensor area is supported by a 500 $\mu \mathrm{m}$ thick rim.

The contribution to the material budget of the sensors and support disks is 
summarized in Table~\ref{tab:ftd_disk_material_budget}. The Silicon sensors clearly
dominate the total contribution.

\begin{table}[h]
    \centering
    \begin{tabular}{lcc}
    \toprule
    Component                      & material (\% $X_0$) \\ \midrule
    Silicon petals (active area)  &         0.0500 \\
    Carbon fibre (incl. cyano-ester resin)     &   0.0380 \\
    Honeycomb core (Aramide)      &   0.0006 \\
    PEEK inserts                  &   0.0019 \\
    PEEK screws                   &   0.0014 \\
    glue                          &   0.0006  \\ \midrule
    total                         &     0.093      \\ \bottomrule
    \end{tabular}
    \caption{Contributions to the material budget of one disk of the forward tracking detector. The contributions are determined for perpendicular incidence and averaged over the area
    of the disk.}
    \label{tab:ftd_disk_material_budget}
\end{table}

The thermo-mechanical performance of the loaded disk has been tested extensively. The 
support disk is found to have a planarity of 200 $\mu \mathrm{m}$ (RMS). Despite the 
minimal material it is very stiff, with an eigenfrequency greater than 1~kHz. The
silicon petals are mounted kinematically, such that they are free to expand in response
to a thermal load, while distortions of the sensors out of the nominal plane remain
very tightly constrained. The torque applied to the screws must be carefully
chosen: a torque of 3~mN$\times$m is found to be optimal. 
With this choice, the first eigenfrequency of a free petal (167 Hz) is nearly
doubled (to $\sim$ 270 Hz) when the sensor is clamped to the disk. 

The impact of air cooling on the mechanical stability is studied with a local, 
laminar air flow. The power consumption pattern mimics that of a DEPFET
active pixel detector, assuming that the application of power pulsing reduces
the average power consumption by a factor 20.
In these conditions, a gentle, laminar flow of 1 $\mathrm{m/s}$ is found to be sufficient
to keep the temperature gradient over the sensor to within 10$^{\circ}$C, 
Vibrations due to air flow have an  amplitude of less than 1~$\mu \mathrm{m}$ 
for laminar air flow with a velocity up to 4 $\mathrm{m/s}$. 

These results indicate that an aggressive design based on a thin carbon-fibre
support disk and ultra-thin self-supporting Silicon petals can meet the 
stringent requirements on mechanical stability of the ILD experiment.
\subsection{Time Projection Chamber}
\label{chap:technologies:tpc}

The ILD TPC R\&D is being conducted mainly within the LCTPC Collaboration \cite{ild:bib:TPC_lctpc}. The history of these developments is described in the R\&D liaison report of the Linear Collider Collaboration~\cite{ild:bib:TPC_liaison} which contains many references (pp 36-60).

\begin{figure}[t!]
\centering
\includegraphics[width=1.0\hsize]{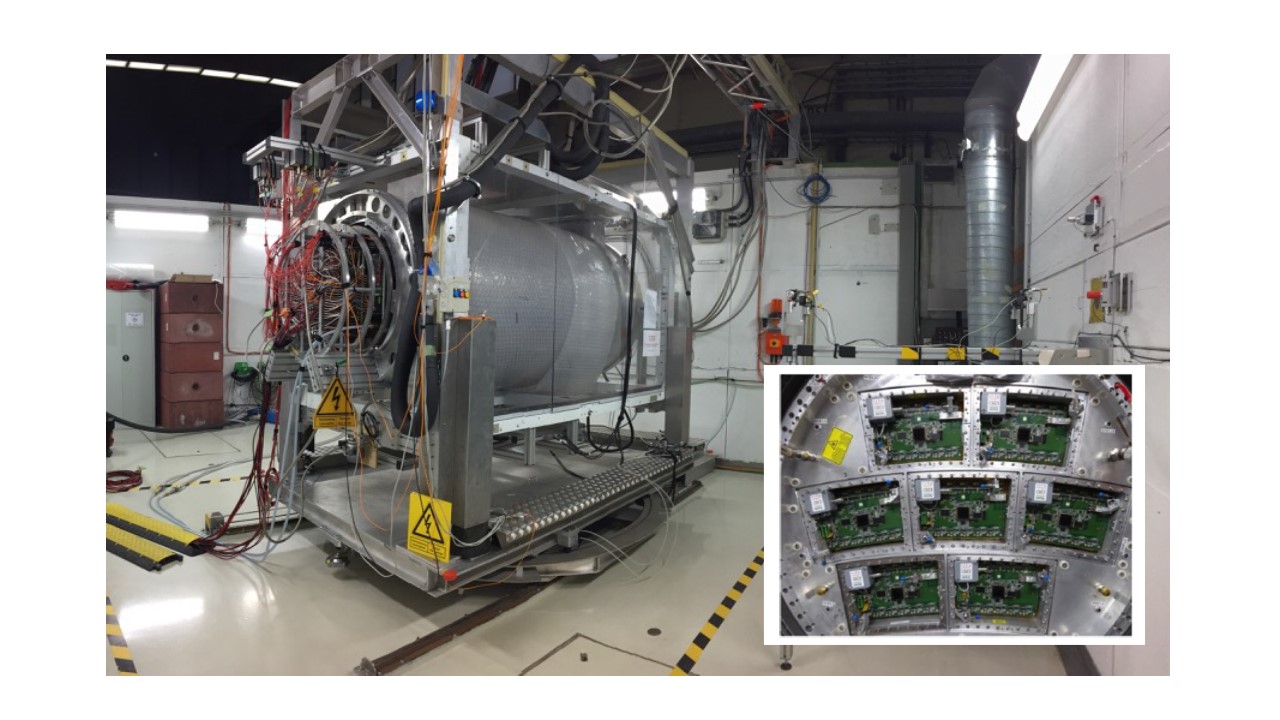}
\caption{The TPC test setup at DESY. The insert shows the geometrical structure of the TPC end cap which can host prototypes of detection planes.}
\label{fig:det:TPC_test_setup}
\end{figure}

The workhorse for validation of detector prototypes and operational conditions is the TPC test set-up installed permanently in the DESY test beam~\cite{ild:bib:TPC_desytb} (Figure~\ref{fig:det:TPC_test_setup}). The TPC is situated in a superconducting magnet providing a magnetic field of 1 Tesla, and the beam line is equipped with precise incident and outgoing particle beam telescopes allowing to quantify the TPC reconstruction precision as function of the particle parameters. The beam test set up is currently being upgraded with the high precision LYCORIS silicon telescope~\cite{ild:bib:TPC_lycoris}, and a new TPC field cage with reduced field distortion is being assembled.

\begin{figure}[t!]
\centering
\includegraphics[width=1.0\hsize]{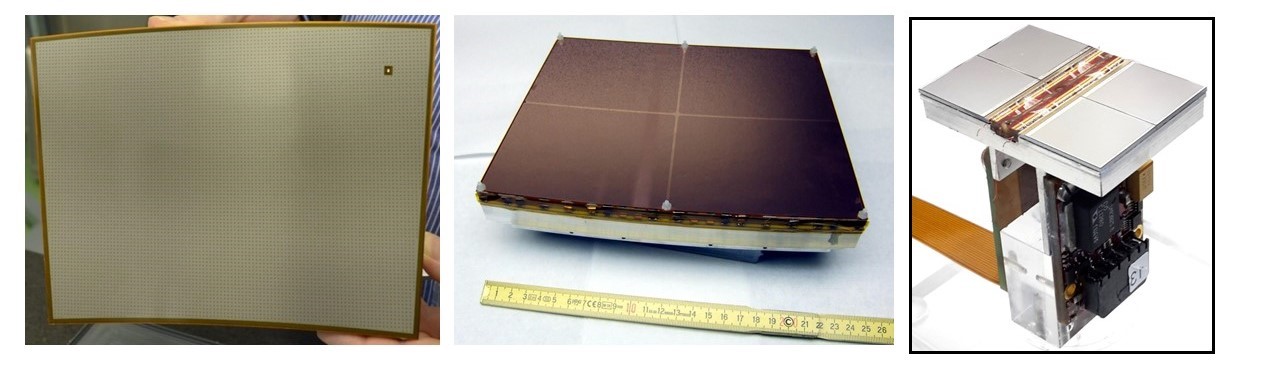}
\caption{TPC prototype detection modules for the three baseline technologies under consideration: Micromegas module (left), GEM module (middle) and GridPix QUAD module (right).}
\label{fig:det:TPC_prototypes}
\end{figure}

Significant progress has been seen in the manufacturing process of detection modules for each of the readout options. A new Micromegas layout with resistive anodes has been shown to exhibit reduced boundary distortions~\cite{ild:bib:TPC_distortions}. The flatness of the GEM modules has been improved significantly, increasing the gain uniformity by a factor 2~\cite{Malek:2017xol}. Operational GridPix "QUAD" modules have been built based on the TimePix3 pixel chip~\cite{ild:bib:TPC_quad}. Recent prototypes of the three types of detection modules are shown in Figure~\ref{fig:det:TPC_prototypes}.  

\begin{figure}[t!]
\centering
\begin{subfigure}{0.48\textwidth}
\includegraphics[width=1.0\hsize]{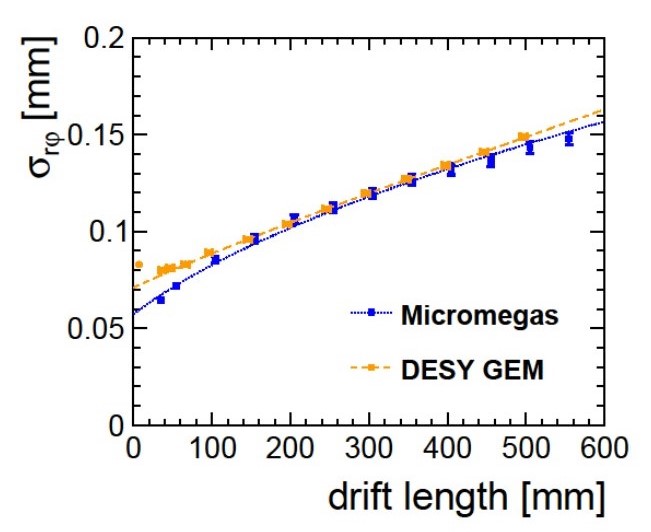}
\caption{}
\end{subfigure}
\begin{subfigure}{0.48\textwidth}
\includegraphics[width=1.0\hsize]{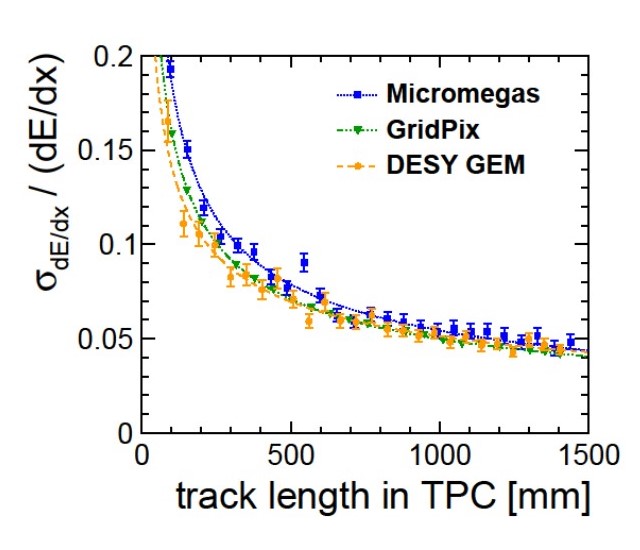}
\caption{}
\end{subfigure}
\caption{(a) TPC resolution on the track position in r$\phi$ as function of the drift length and (b) resolution on the ionisation loss dE/dx as function of the track length, for the three readout options under consideration.}
\label{fig:det:TPC_performances}
\end{figure}

The performance of the three technologies has been measured in beam tests. Figure~\ref{fig:det:TPC_performances} shows the measured point resolution in 1~T magnetic field for drift distances from 0 to 0.6 m. This can be safely extrapolated to $\sim~100$ $\mu$m in a field of 3.5~T at a drift length of 2.3~m: the higher magnetic field reduces the transverse diffusion constant in the selected gas from 91~$\mu \mathrm{m} / \sqrt{\mathrm{cm}}$ to 30 $\mu \mathrm{m} / \sqrt{\mathrm{cm}}$, which compensates for the longer drift length compared to the prototype set-up. The dE/dx resolution determined by the truncated mean method has been measured to be $4.6\%, 4.5\%$ and $4.2\%$, respectively, for Micromegas, GEM and GridPix technologies. It improves to $3.8\%$ for GridPix using a cluster counting method. In conclusion the target requirement of a spatial resolution of $100 \mu$m in the transverse plane, and a dE/dx resolution better than 5\% have been reached in all options.   

\begin{figure}[t!]
\centering
\includegraphics[width=0.55\hsize]{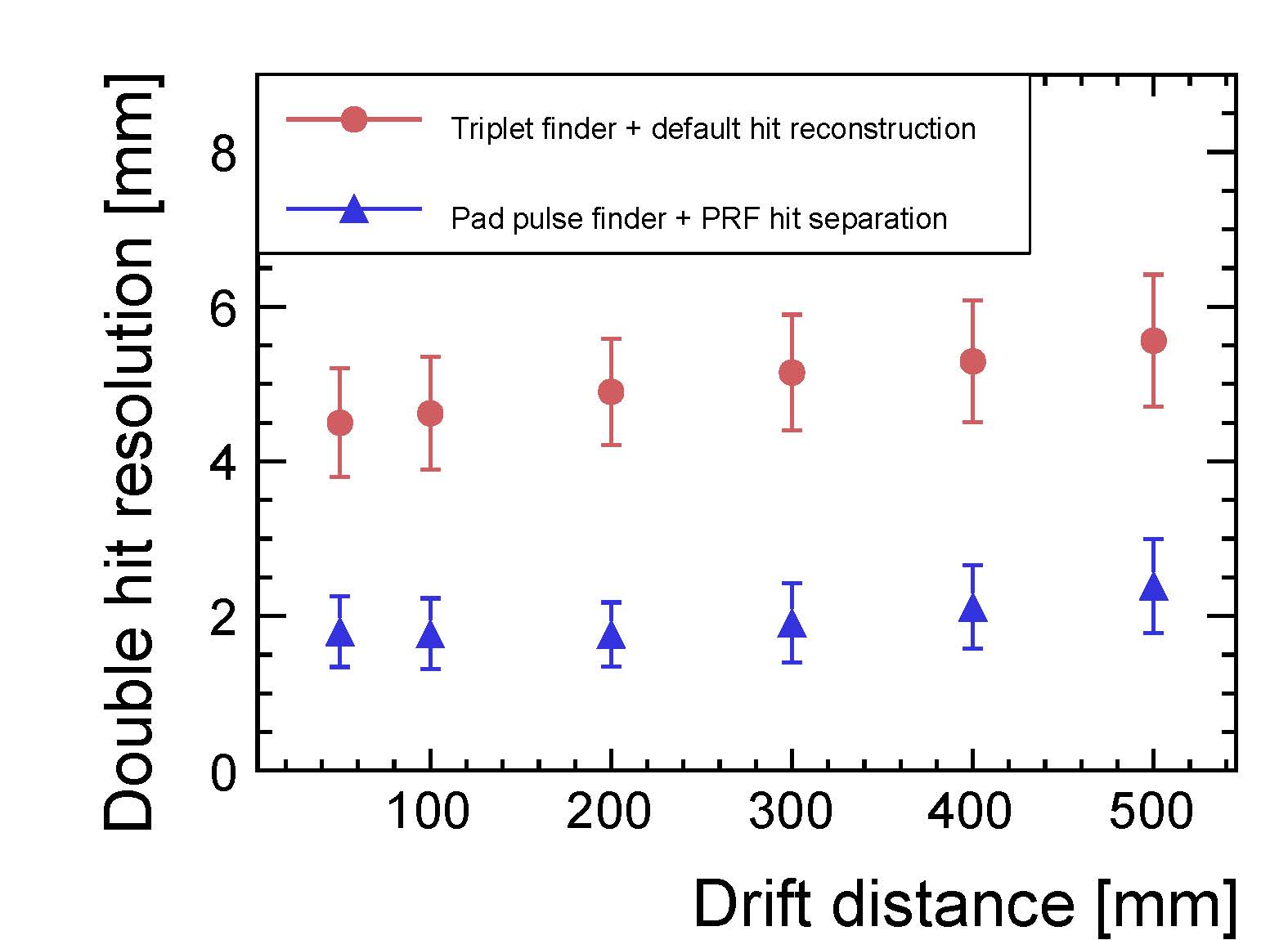}
\caption{TPC 2-hit separation as a function of the drift distance from beam test. Red dots are the results for standard hit reconstruction, blue triangles show the results of the improved algorithm. The data points shown correspond to the point where the separation efficiency has dropped to $50\%$, the bar describes the width of the error function, which is used to fit the transition region of the efficiency.}
\label{fig:det:TPC_separation}
\end{figure}

Two-track separation has also been investigated. A $47\%~\rm{X_0}$ steel target was introduced near the TPC wall to produce multi-track events suitable for this study.
From these events a 2-hit separation distance of 4 to 6~mm was measured depending on the drift distance. An algorithm based on fitting the double-hit charge deposition with expected pad response function width allowed this separation distance to be reduced to 2~mm, with 1.3~mm pads (Figure~\ref{fig:det:TPC_separation}).

A challenging aspect of the TPC operation is the cooling of the readout end-caps, which must be realised with minimal dead material.  For this a double phase $\mathrm{CO}_2$ cooling system with thin low-material fluid pipes has been developed and is shown to perform adequately in test-beam experiments. 

\begin{figure}[t!]
\centering
\begin{subfigure}{0.40\textwidth}
\includegraphics[width=1.0\hsize]{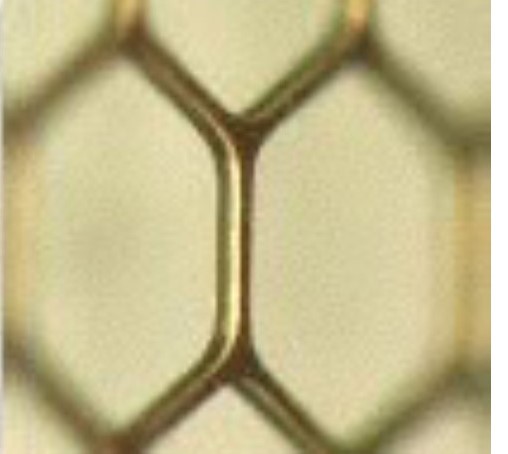}
\caption{}
\end{subfigure}
\begin{subfigure}{0.48\textwidth}
\includegraphics[width=1.0\hsize]{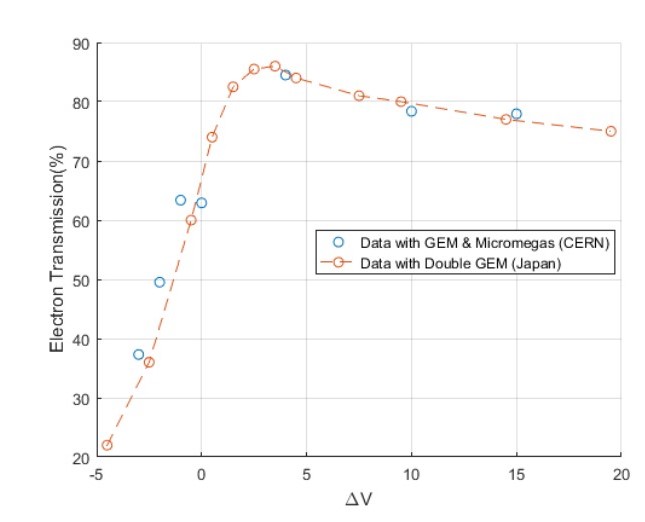}
\caption{}
\end{subfigure}
\caption{TPC gating: (a) detail of a GEM gating grid and (b) signal electron transparency with GEM gating as a function of the voltage difference (in Volts) across the two sides of the GEM plane in gate open mode.} 
\label{fig:det:TPC_gating}
\end{figure}

Also critical for ultimate performance is the mitigation of the drift field distortions which may develop from the accumulation of ions migrating from the amplification region into the drift volume. For this an ion gating scheme based on a large aperture GEM foil (Figure~\ref{fig:det:TPC_gating} left) has been implemented and beam tested~\cite{ild:bib:TPC_gatinginbeam}. To prevent the positive ions created by the avalanches in the amplification device from flowing back to the drift space, a counter-field is created by applying a polarisation $\Delta V$ of -20 V between the faces of the gating GEM foil in between beam train crossings. During train crossings the voltage difference is reversed to +3 V, opening the way to the incoming electrons to be multiplied. 

Results from test bench measurements show that a good transparency for drift electron signals can be maintained while reducing the accumulation of ions in the drift volume by one order of magnitude~\cite{ild:bib:TPC_gatingpaper} (Figure~\ref{fig:det:TPC_gating} right).

In conclusion all basic aspects of the TPC operation have shown to meet the requirements for an experiment at the ILC. 
\subsection{Electromagnetic Calorimeter}
\label{ild:sec:ECAL}
The technological status of both the silicon based and the scintillator based option considered for the highly granular multi-layer sampling electromagnetic calorimeter is described in this secion. 

\subsubsection{Silicon option (SiECAL)}

In the past five years the developments of the silicon option of the electromagnetic calorimeter has mostly focused on the design and construction of technological prototypes of the detector, and on performing beam tests. 

So far square wafers with thicknesses between $325\,\upmu m$ and $650\,\upmu m$ processed from 6" silicon ingots have been used. 
An increased thickness of $725\,\upmu m$, already available from 6" ingots, would be conducive in case wafers from 8" ingots can be reliably produced. It would allow to reduce the number of ECAL layers to 26, without significant degradation of the energy resolution compared to the current baseline design of 30 layers (section~\ref{ref:subsec:subdetectors}). For the final detector, the ongoing technological developments allow to foresee to use diode matrices with a standard thickness of $725\,\upmu m$ produced from 20\,cm wafers, but the final choice of the ingot size will depend on the silicon producers chosen at the time of building the calorimeter.



\begin{figure}[t!]
\centering
\begin{subfigure}{0.45\textwidth}
\includegraphics[width=0.9\hsize]{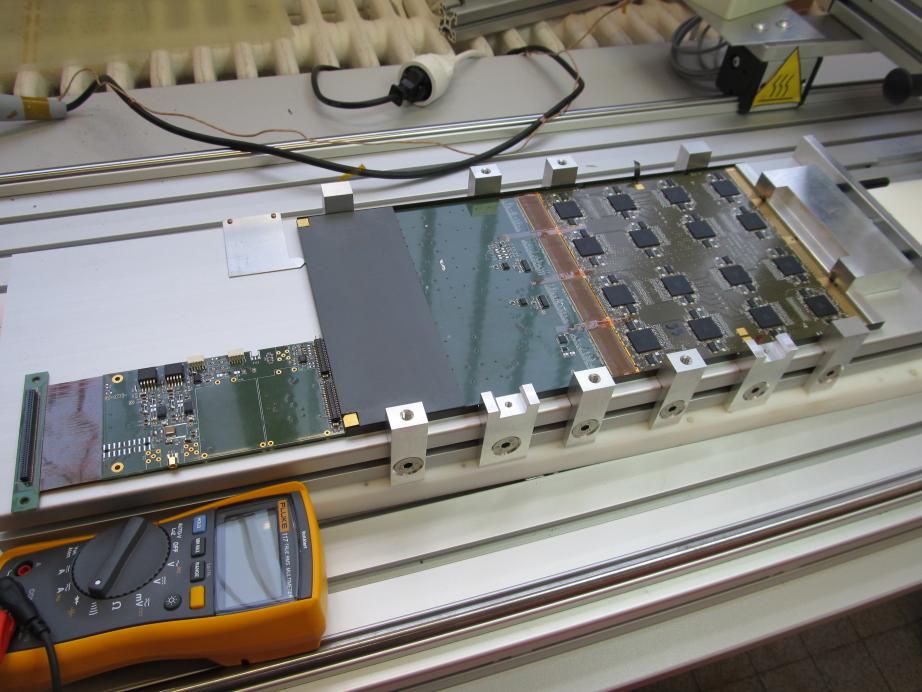}
\end{subfigure}
\begin{subfigure}{0.45\textwidth}
\includegraphics[width=1.1\hsize]{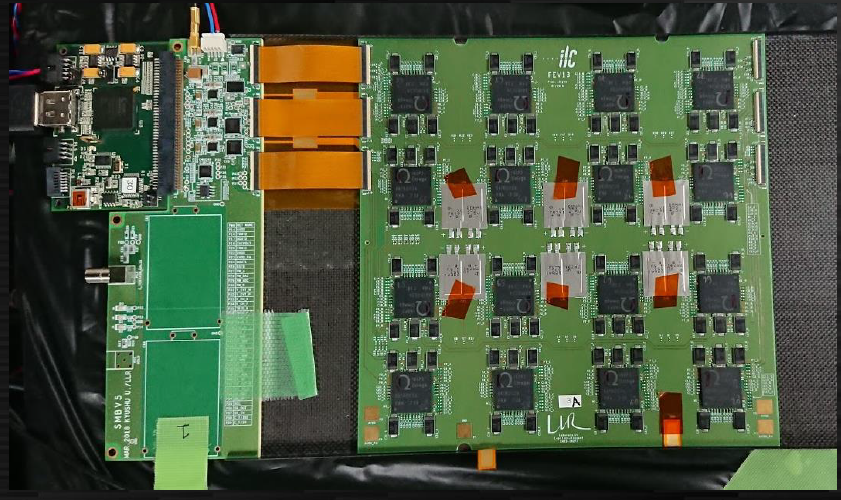}
\end{subfigure} \\ 
\includegraphics[width=0.50\hsize]{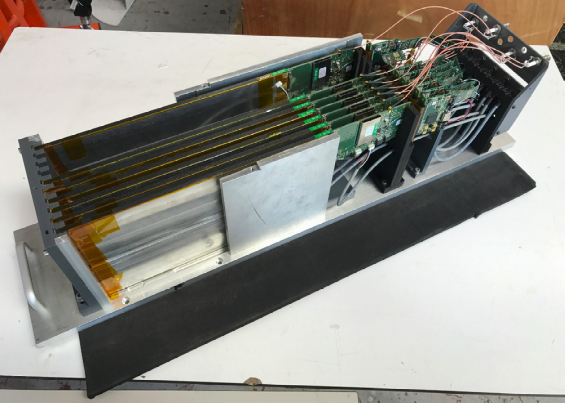}
\caption{Technological prototype of the SiECAL. Top: integrated ASUs (left: version \#11, right: version \#13); bottom: mechanical housing with integrated layers of the SiECAL prototype used in beam tests at DESY and CERN.}
\label{fig:det:SiWECAL_proto}
\end{figure}

Fully integrated layouts of the silicon active sensor units (ASU) have been designed with the required dimension of $18 \times 18\,{\rm cm^2}$ comprising 1024 channels. One ASU hosts 16 SKIROC ASICs~\cite{Callier:2011zz,Suehara:2018mqk} to process 64 channels each. The top part of Figure~\ref{fig:det:SiWECAL_proto} shows two recent versions of the ASU, both assembled with standard ball grid array (BGA) packaging of the ASICs. 
The most recent version (\#13) features a separation between the power supply of the preamplifier and other power lines of the SKIROC ASIC. 
Big capacitances are integrated into the board design to avoid large current peaks during power pulsing. 
This as well as earlier versions of the ASU have been operated in power pulsed mode since 2013. The bottom part of Figure~\ref{fig:det:SiWECAL_proto} shows the mechanical housing that can hold up to ten ASUs used in beam tests at CERN and DESY since 2012~\cite{Boudry:2014bxa,Poeschl:2015jma}. 

\begin{figure}[t!]
\centering
\begin{subfigure}{0.48\textwidth}
\includegraphics[width=1.0\hsize]{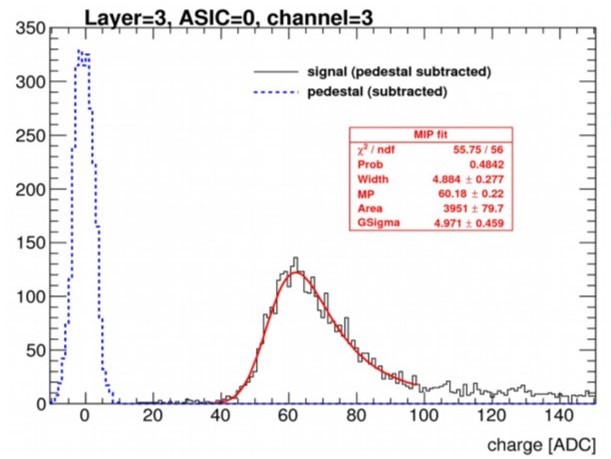}
\caption{}
\end{subfigure}
\begin{subfigure}{0.48\textwidth}
\includegraphics[width=1.0\hsize]{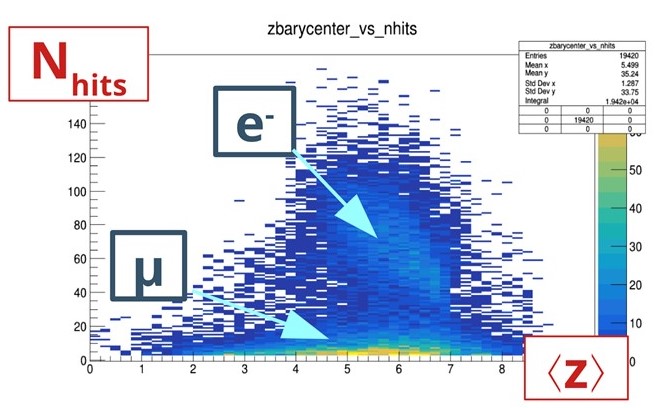}
\caption{}
\end{subfigure}
\caption{Particle response of the SiECAL prototype: (a) MIP response of a single pad. (b) Shower profile (number of pad hits as a function of longitudinal layer number) of muons and 80 GeV electrons.}
\label{fig:det:SiWECAL_signals}
\end{figure}

The response of the SiECAL technological prototype to particles is as expected~\cite{Kawagoe:2019dzh}. The signal-over-noise ratio relevant for the internal ASIC trigger has been evaluated to be about 12.8 for a wafer thickness of $325\,\upmu m$. This value allows for setting the trigger threshold below the single MIP level with high efficiency. 
The signal-over-noise at the level of charge measurement for triggered channels is 20.4 (Figure~\ref{fig:det:SiWECAL_signals} left), or better, depending on the wafer thickness. Therefore a threshold cut in the ASIC at 0.5 MIP can be applied without noise problems and with good efficiency.
 Response to high energy electrons has been measured at CERN in 2018 in a combined test with the SDHCAL (next section) as shown in Figure~\ref{fig:det:SiWECAL_signals} right.


\begin{figure}[t!]
\centering
\includegraphics[width=0.32\hsize,angle=90]{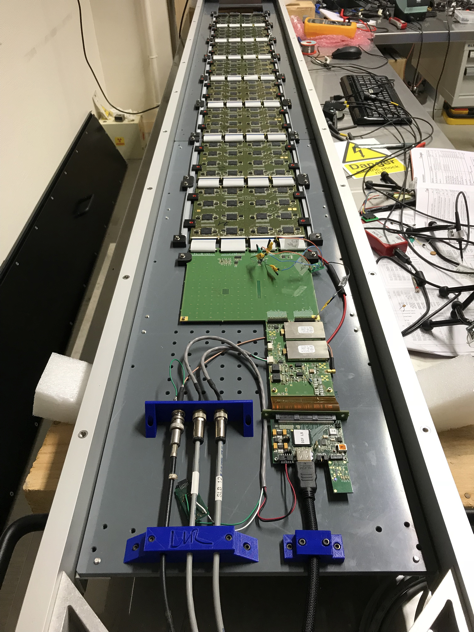}
\caption{The first prototype of a long slab of the SiECAL.}
\label{fig:det:siw-longslab}
\end{figure}

A long detector slab featuring a chain of eight ASUs has been built (Figure~\ref{fig:det:siw-longslab}) and tested with MIPs~\cite{Magniette:2019nyg}. 
A 10\% signal drop has been observed along the full length of the slab and could be attributed to power voltage drops and clock reflections. An improved long slab is under construction to correct these effects and validate the system. 

\begin{figure}[t!]
\centering
\begin{subfigure}{0.48\textwidth}
\includegraphics[width=1\hsize]{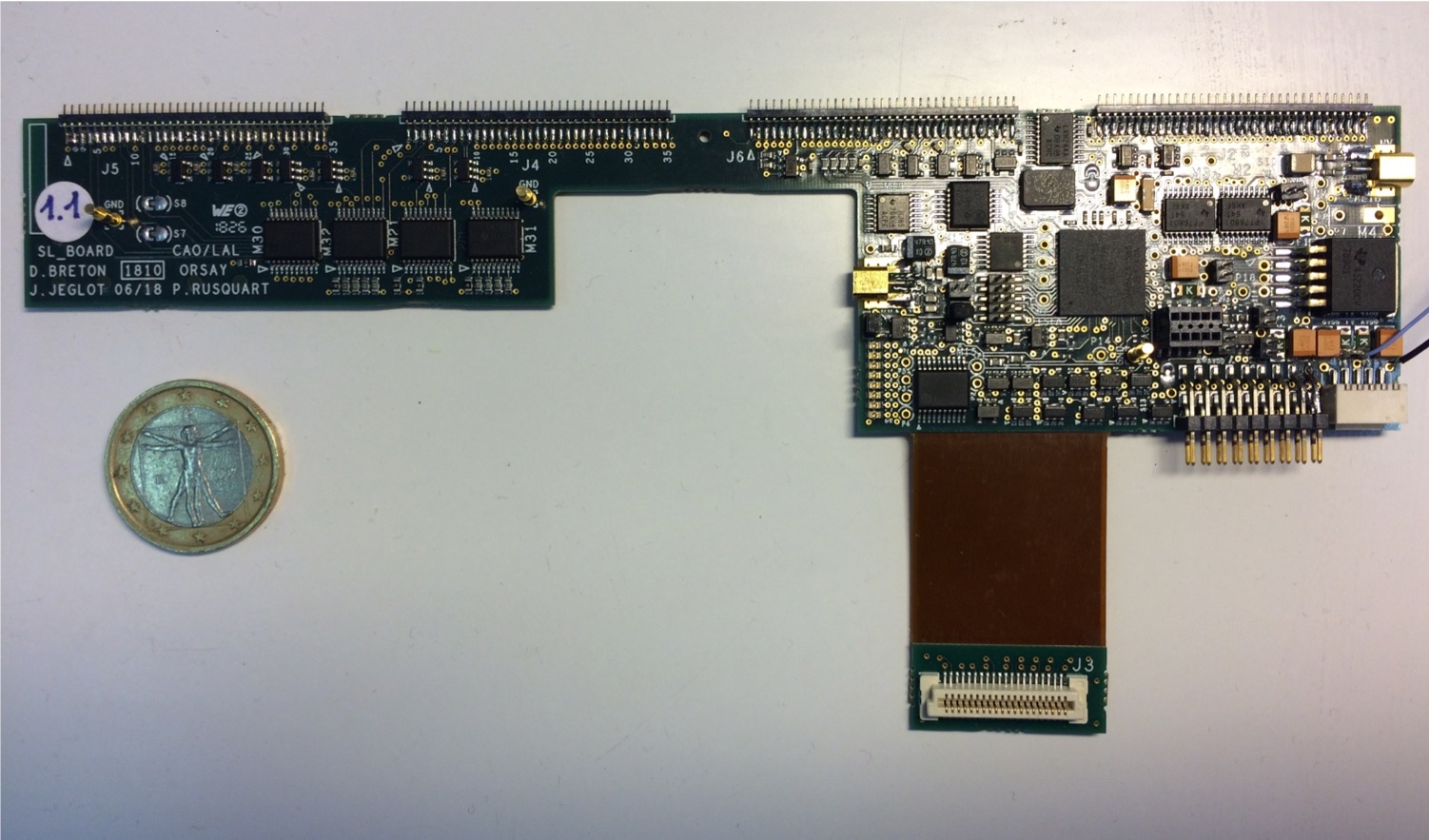}
\caption{}
\end{subfigure}
\begin{subfigure}{0.48\textwidth}
\includegraphics[width=1\hsize]{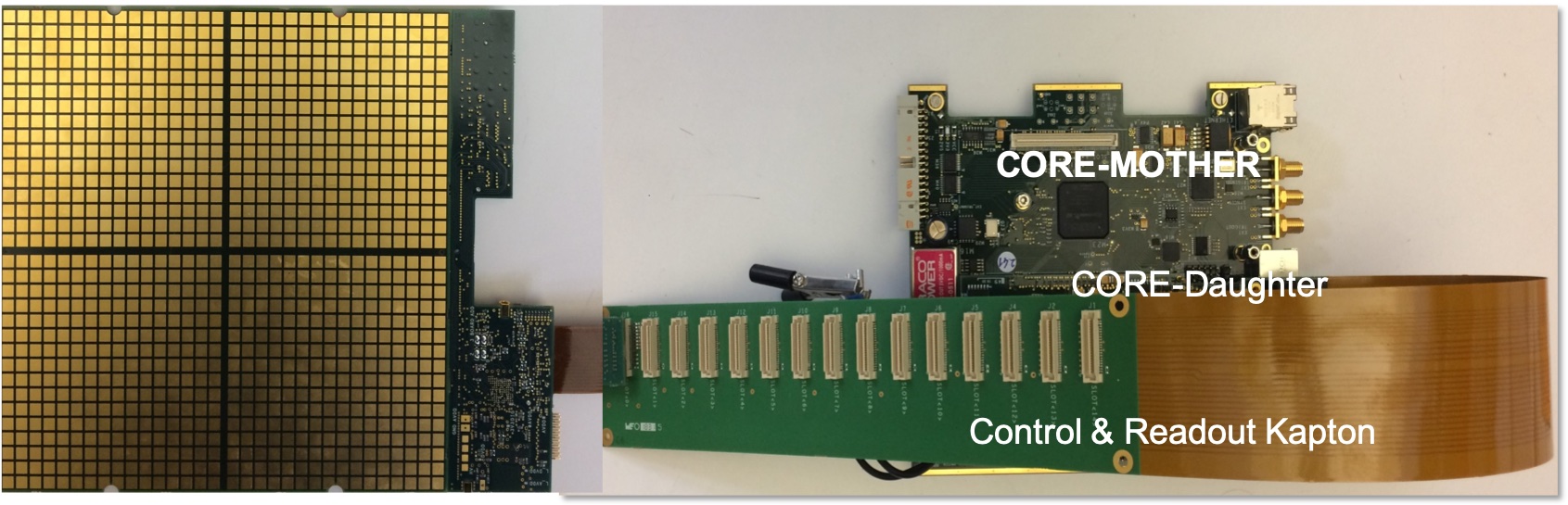}
\caption{}
\end{subfigure}
\caption{(a) Slab interface board for the digital readout and the power supply of the SiECAL layers. (b) Ensemble of one ASU with its slab interface board, control and readout kapton, and concentrator unit.}
\label{fig:det:SiWECAL_slcards}
\end{figure}

 A new slab interface board for the digital readout of the ASUs, located at the end of the slabs, has been developed to meet the ILD integration constraints (Figure~\ref{fig:det:SiWECAL_slcards} left). The board integrates the digital readout of the SiECAL layers and hosts the regulators for the power supply of the ASUs,
 With a lateral size of about $4\times 18\,{\rm cm^2}$, this board already meets the tight space requirements of ILD. The board has been successfully used in a beam test in summer 2019 at DESY~\cite{bib:talk-twepp-jj}.
 The slab interface boards are connected via a flat control and readout kapton cable to a concentrator unit as illustrated in the right part of Figure~\ref{fig:det:SiWECAL_slcards}. 
 The distinguished feature of this solution is that the kapton cable can be guided along the sides of an alveolar structure hence avoiding bulky cabling. 
 
\begin{figure}[t!]
\centering
\includegraphics[width=0.35\hsize]{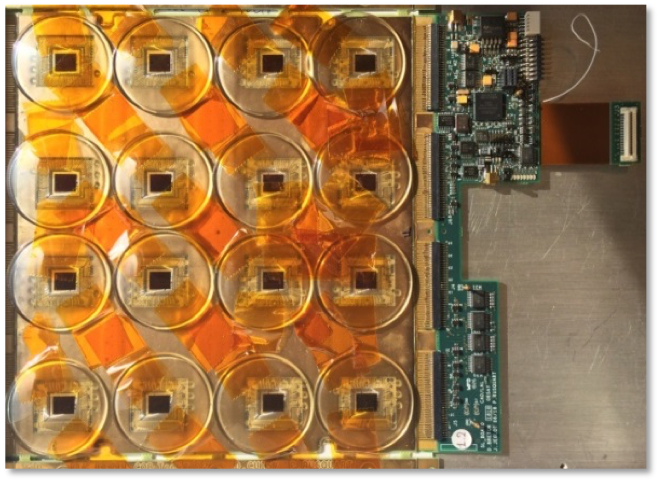}
\includegraphics[width=0.38\hsize]{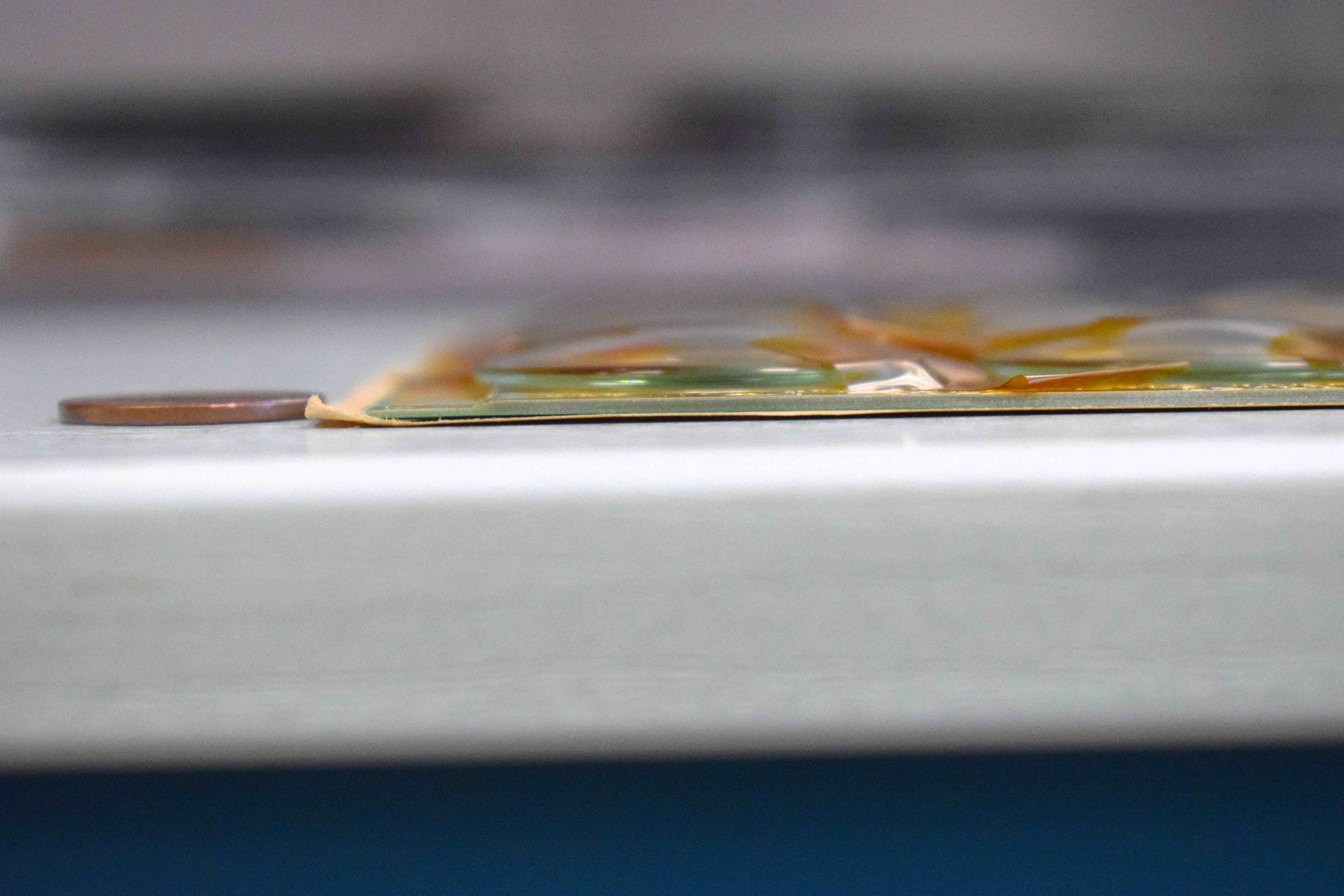}
\caption{(Left) COB version of an ASU connected to a slab interface board. (Right) Zoom onto the edge of a COB ASU indicating its thickness.}
\label{fig:det:SiWECAL_cob}
\end{figure}

An ultra-thin ASU with an overall thickness of 1.2\,mm has also been developed. In this version the ASICs are mounted in recessed cavities of the PCB in a so-called chip on board (COB) packaging (Figure~\ref{fig:det:SiWECAL_cob}). 
 First tests of the COB ASUs have been successfully performed at the 2019 beam test at DESY~\cite{bib:talk-twepp-jj}. Clean MIP spectra have been recorded and the noise level is competitive with that of the BGA based ASUs built up to now. This is remarkable since the design of the board leaves relatively little room for decoupling capacitors. The flatness of this card  meets the specifications, but the yield has to be studied with industry. Further tests will be performed and industrial aspects will  have to be investigated further.


The R\&D in the coming years will further emphasise system and integration aspects based on the current achievements. The planned R\&D includes a new version of the  ASIC with full zero suppressed readout, which implies an utmost stable pedestal level.


 The silicon technology developed for ILD has been adopted as the baseline for the electromagnetic end-cap section of the CMS High Granularity Calorimeter (HGCAL) upgrade~\cite{Collaboration:2293646}.
 The HGCAL layout is based on hexagonal readout modules with a technology similar to the ILD one. It incorporates a new version of the SKIROC ASIC with a sub-ns timing functionality which may also be of interest for the ILD detector.
 A HGCAL prototype of 27 layers has been successfully tested by CMS in a combined beam test at CERN with the AHCAL ILD prototype (next section). The full HGCAL represents a 10\% prototype of the ILD SiECAL w.r.t.\,the silicon diodes surface.
 Despite many differences between the ILD and CMS designs, the HGCAL construction will be a strong asset to validate the large scale assembly and fabrication processes for ILD.


\subsubsection{Scintillator option (ScECAL)}

Similar to the silicon option, the scintillator option of the electromagnetic calorimeter,
after the validation of the concept using the physics prototype,
has focused its R\&D towards a technological prototype 
with fully integrated detection layers. 
The design of the detection layer is based on an integrated readout board, 
called ECAL base unit (EBU), of $18\times18\,\mathrm{cm}^2$ 
with four SPIROC ASICs developed by OMEGA group\cite{Bouchel:2011zz} on which 144 scintillator strips 
($5\times45\times2\,\mathrm{mm^3}$ each) coupled to SiPMs are mounted.

Notable progress has recently been made on the SiPMs for the ScECAL. 
MPPCs with a smaller pixel pitch of 10 or $15\,\mu\mathrm{m}$ have been developed, which can provide a large dynamic range as required 
for the ScECAL\cite{ild:bib:hdmppc}. 
Further improvements have been made for the most recent small-pixel MPPCs, 
including reduced optical cross-talk by a trench structure between pixels, 
lower dark noise and higher photon detection efficiency. These have been confirmed by the prototype tests.

\begin{figure}[htb]
\centering
\includegraphics[width=0.7\hsize]{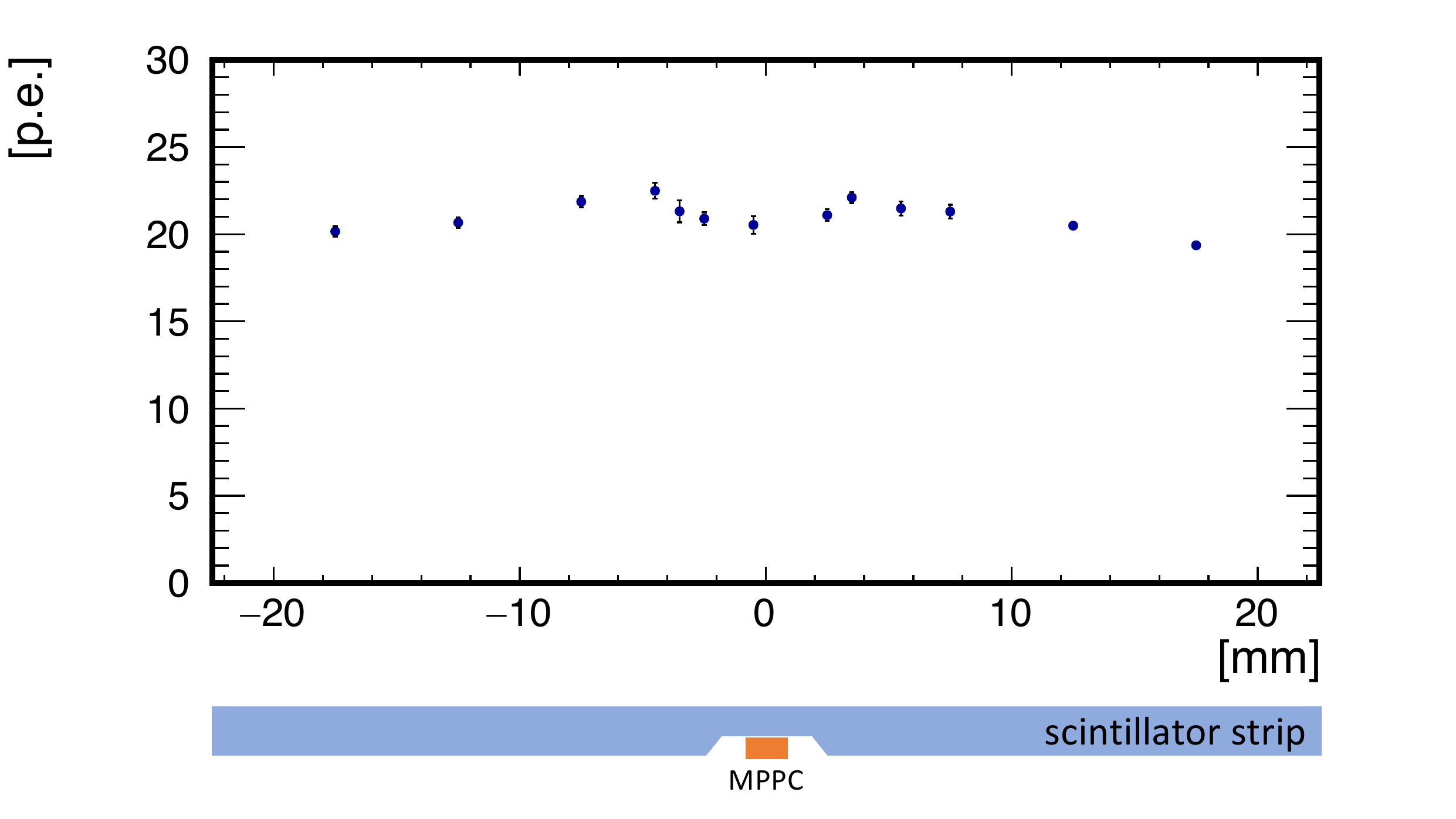}
\caption{Position dependence of the light yield for the scintillator 
strip with a dimple directly coupled to a SiPM on PCB}
\label{fig:det:ScWECAL_strip}
\end{figure}

In the previous prototype studies, the SiPM was attached to the side edge 
of the strip. 
New designs of the SiPM readout at the bottom side of the strip 
are being developed for more uniform response and a better compatibility 
with future large-scale production.
Especially a recently proposed design based on a strip 
with a dimple directly coupled to a surface-mounted SiPM on the PCB, similarly to the SiPM-on-tile technology of the AHCAL (see next section),
shows a promising performance with a reasonably high light yield and a uniform response 
over the strip length (Figure~\ref{fig:det:ScWECAL_strip}).
Another design of the SiPM readout is also being studied, where the strip 
is readout by two SiPMs set in coincidence, which reduces random noise and hence improves the signal to noise ratio. For this design a twice longer strip ($L=90\,\mathrm{mm}$) is expected to be used in order
not to increase the number of SiPMs.

Low cost and high light yield plastic scintillator materials are also being developed 
for the ScECAL.
The development focuses on the polystyrene-based scintillator 
produced by the injection moulding method, which is suitable 
for large-scale production. 
A reasonably high light yield of 65--70\% of that of 
the commercial PVT-based scintillator, has been achieved 
by optimizing the production parameters. 

\begin{figure}[htb]
\centering
\includegraphics[width=1.0\hsize]{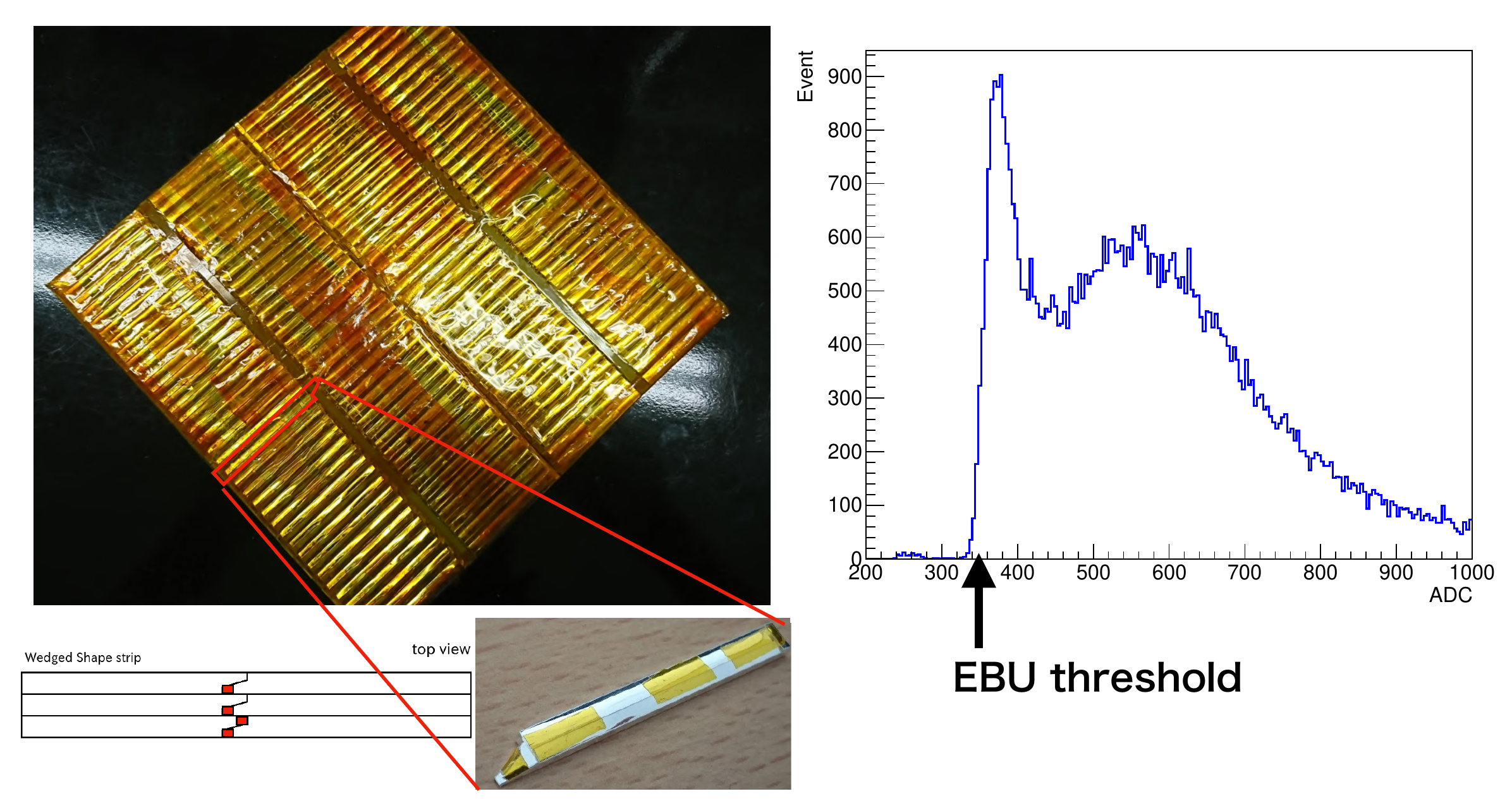}
\caption{(Left) Prototype detection layer with small pixel MPPCs  
   and (right) typical charge distribution obtained for positron beam 
where the MIP peak is well separated from the pedestal.}
\label{fig:det:ScWECAL_prototype}
\end{figure}

Detection layer prototypes have been developed 
with the small pixel MPPCs ($15\,\mu\mathrm{m}$ pixel pitch) 
as shown in Figure~\ref{fig:det:ScWECAL_prototype} left.
A prototype layer was tested in positron beams of 50--800\,MeV 
at the ELPH facility of the Tohoku University.
Figure~\ref{fig:det:ScWECAL_prototype} right shows the typical charge distribution 
obtained for the positron beam, with the MIP peak well separated from the pedestal.

A fully integrated technological prototype with 30 alternating absorber and detection 
layers is planned to be constructed 
as a joint effort with the ScECAL R\&D for CEPC 
to demonstrate the performance of the ScECAL technology 
and its scalability to the full-size detector.

\subsection{Hadronic Calorimeter}
\label{ild:sec:HCAL}

The technological status of both options considered for the highly granular multi-layer sampling hadronic calorimeter is summarised below. 

\subsubsection{Scintillator option (AHCAL)}

\begin{figure}[hbt]
\centering
\includegraphics[width=0.39\textwidth]{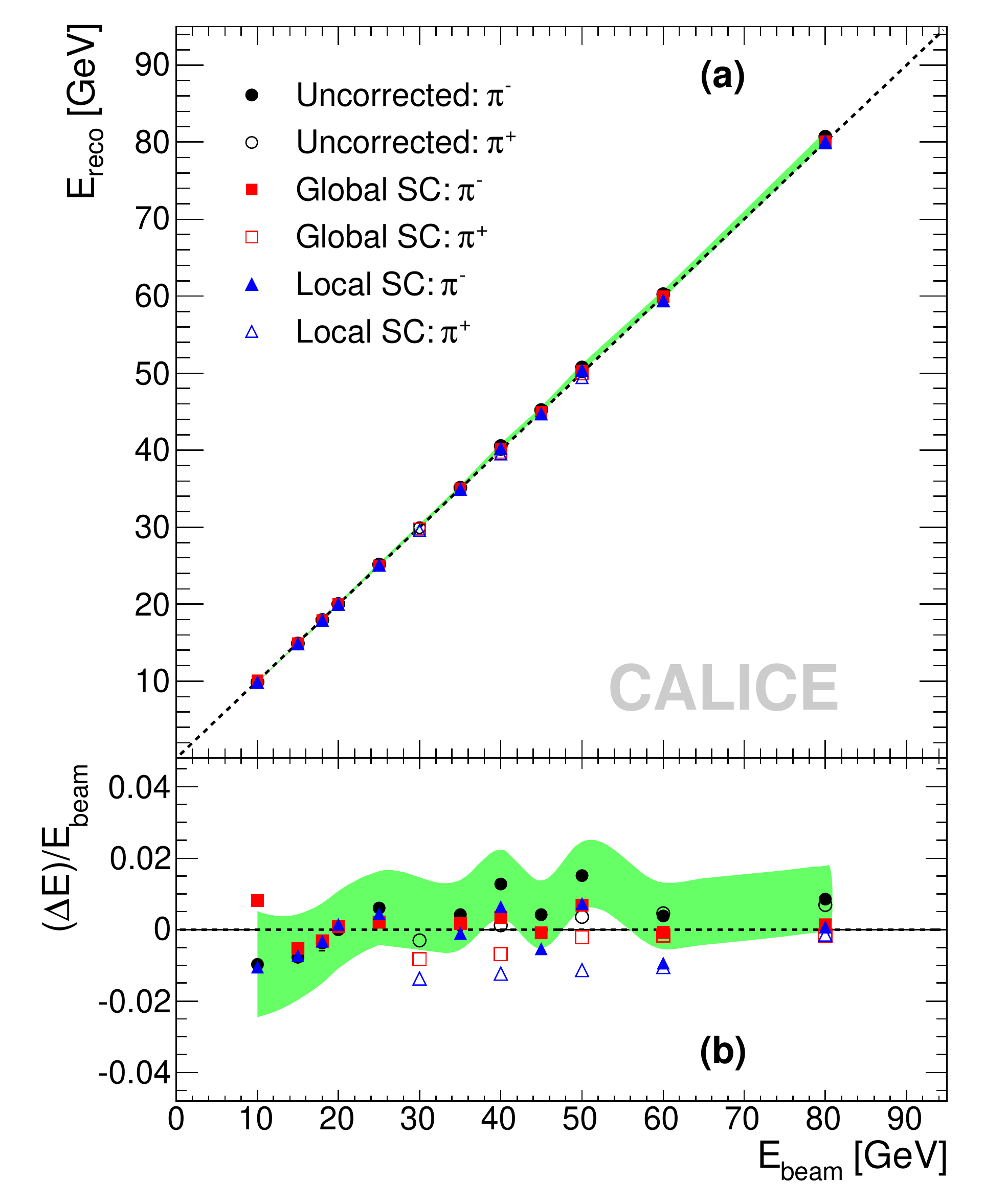}
\includegraphics[width=0.59\textwidth]{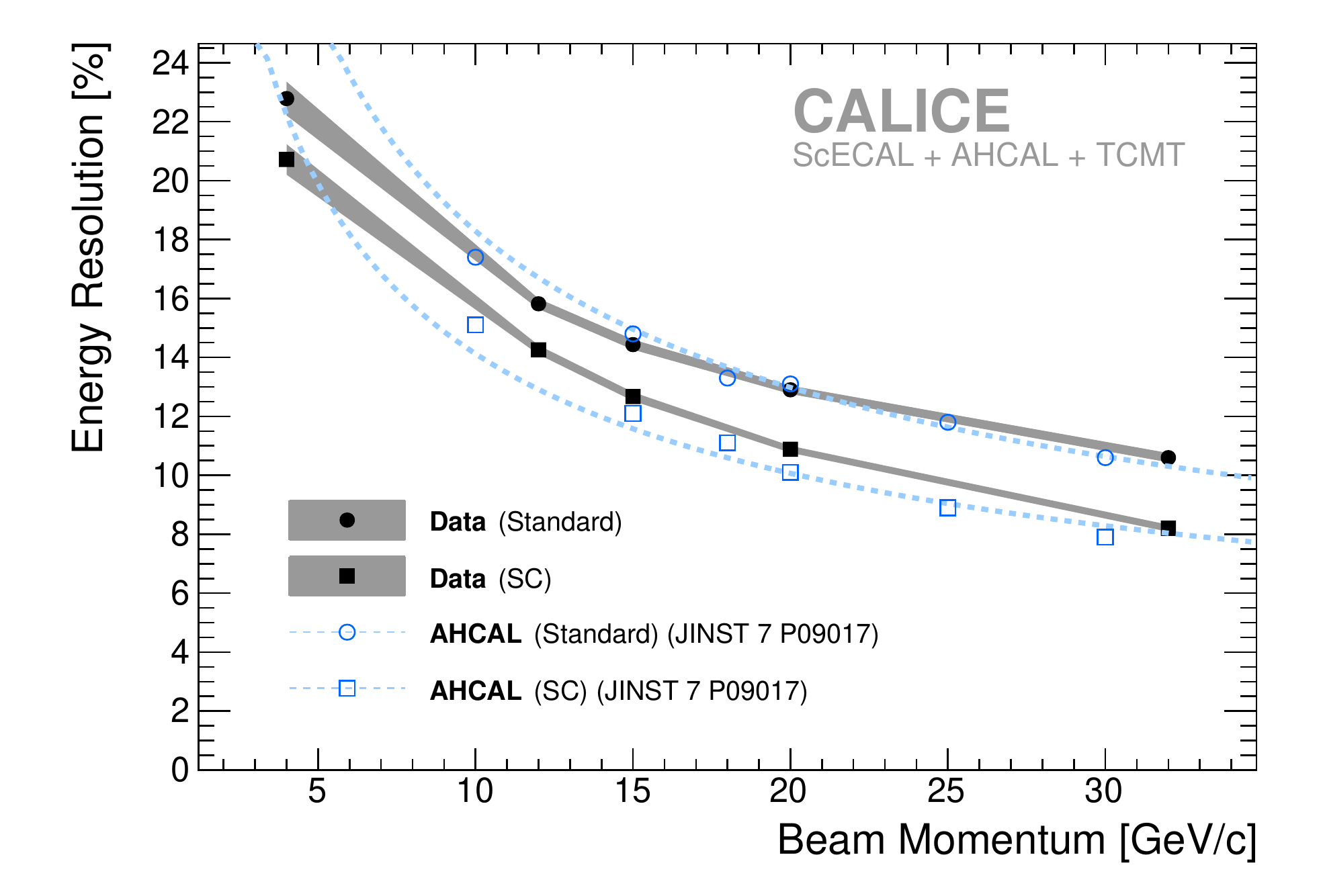}
\caption{Performance of the CALICE AHCAL physics prototype for pions. Left: Response without and with global or local software compensation for the AHCAL alone (with tail-catcher, from~\cite{Adloff:2012gv}). Right: resolution of the AHCAL with a scintillator ECAL in front ("Data") and in standalone mode ("AHCAL"), without ("Standard") and with ("SC") software compensation (from~\cite{Repond:2018flg})}
\label{fig:ahcal-linres}
\end{figure}

A CALICE AHCAL physics prototype was built and operated in 2006-2011, allowing to demonstrate the particle flow~\cite{Adloff:2011ha} and energy resolution performance~\cite{Adloff:2012gv} of the scintillator SiPM ("SiPM-on-Tile") technology for the DBD. In subsequently published papers~\cite{Adloff:2013vra,Adloff:2013kio,Adloff:2013jqa,Adloff:2014rya,Bilki:2014bga,Lucaci-Timoce:2013tkf,Price:2016sce} the adequate modelling of instrumental effects and shower evolution has been further established, and the hadronic energy resolution and linearity of a combined system with a scintillator tungsten ECAL in front was shown~\cite{Repond:2018flg} to be as good as that of the AHCAL alone, see Figure~\ref{fig:ahcal-linres}. The software used to model the detector effects such as photo-electron statistics in the SiPM has been validated by the test beam data and is also used in ILD, after adjusting for a small difference in sampling fraction only~\cite{Hartbrich:2016bbz}.

With the establishment of the principal viability of the AHCAL technology, the focus has shifted from the study of the physical performance characteristics of such a detector to the demonstration of the feasibility of the concept while satisfying the spatial constraints and scalability requirements of collider experiments such as ILD. For this purpose a new AHCAL technological prototype has been built. It is based on a scintillator tile design well-suited for mass production and automatic assembly, originally proposed in~\cite{Blazey:2009zz} and subsequently varied and optimised in further studies~\cite{Simon:2010hf, Liu:2015cpe}. 

The new AHCAL technological prototype~\cite{Sefkow:2018rhp} consists of a non-magnetic stainless steel absorber structure with 38 active layers and has 21888 channels. 
The structure has been produced from standard rolled plates, which had undergone a cost-effective roller-levelling procedure, ensuring a flatness of $\pm 1$~mm, demonstrated over an area of 2$\times$2~m$^2$. The plates were assembled using screws as foreseen for the full ILD structure in the TESLA mechanical layout. 
The basic unit of the active elements is the HCAL Base Unit HBU~\cite{Reinecke:2013zua}, with a size of 36 $\times$ 36 cm$^2$, holding 144 SiPMs controlled by four SPIROC2E ASICs \cite{Bouchel:2011zz}.  A key element of the electronics is the capability for power-pulsed operation.
In addition to dual-gain energy measurement, the electronics also provides a cell-by-cell auto trigger and time stamping at the few ns level in test beam operations. In operating conditions with shorter data-taking windows closer to the bunch train structure of linear colliders, sub-ns time resolution is achieved. 

\begin{figure}[hbt]
\centering
\includegraphics[width=0.39\textwidth]{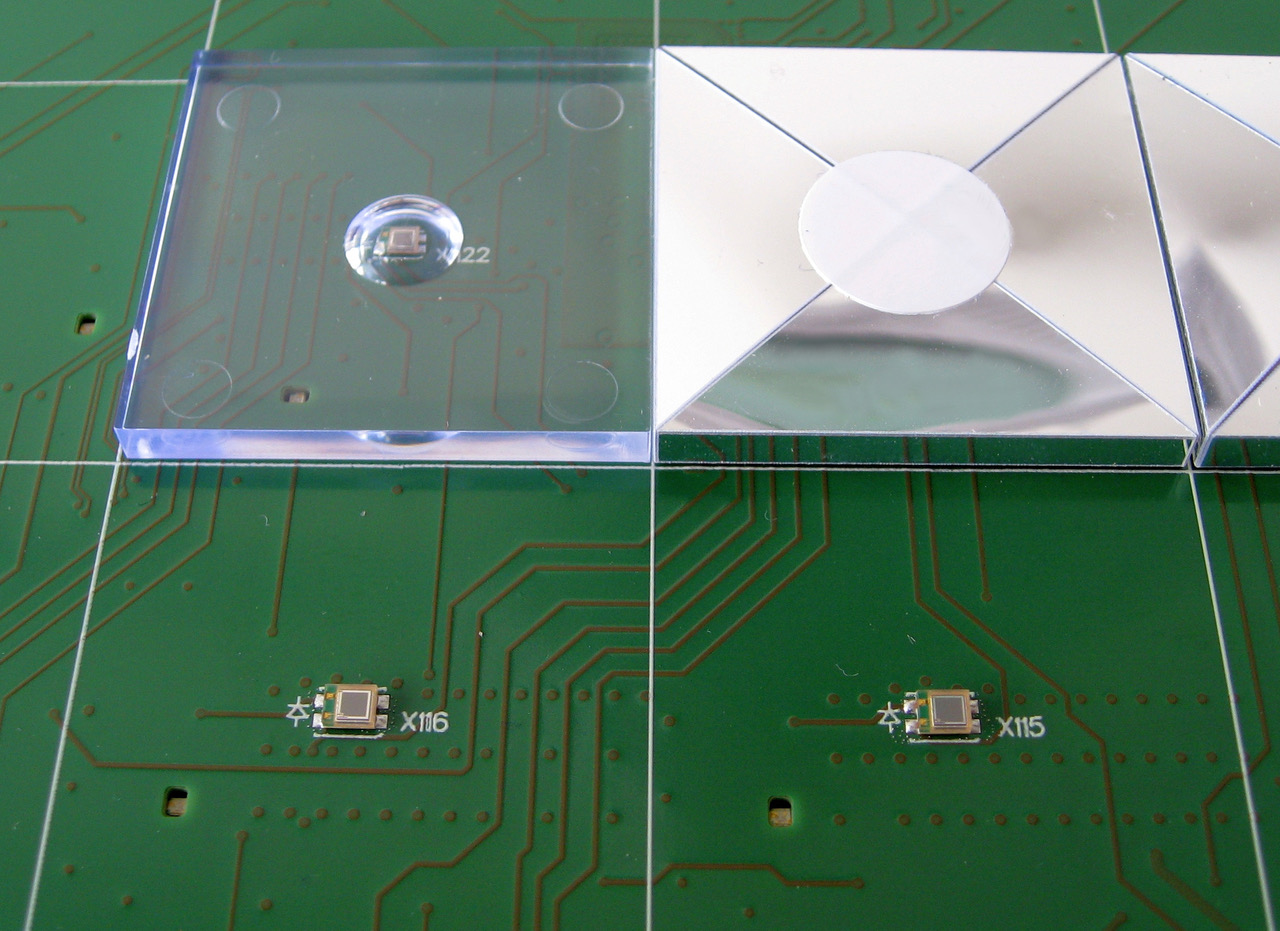}
\includegraphics[width=0.59\textwidth]{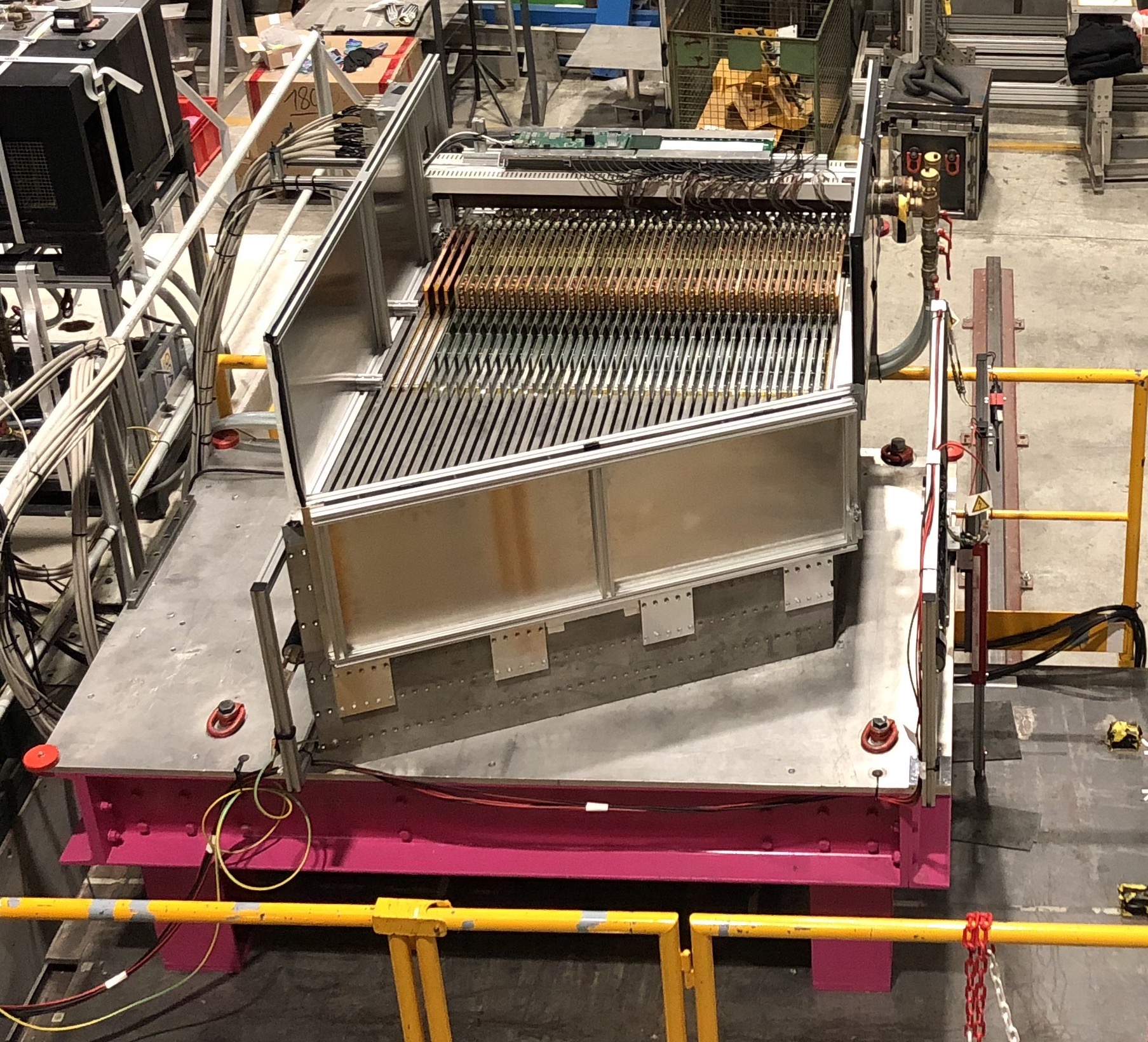}
\caption{The AHCAL technological prototype. Left: Read-out board (HBU) with SiPMs and (un-)wrapped tiles. Right: Prototype installed in the H2 beam line at the CERN SPS.} 
\label{fig:AHCAL-TileProto}
\end{figure}

The prototype uses Hamamatsu MPPC S13360-1325PE photon sensors and injection-moulded polystyrene scintillator tiles with a central dimple \cite{Liu:2015cpe} for optimal light collection, as shown in Figure~\ref{fig:AHCAL-TileProto}. 
%
Spot-samples of all SiPM lots, and each individual ASIC, had undergone semi-automatic testing procedures before soldering the HBUs \cite{Munwes:2634923}. The gain of the SiPMs was found to be uniform within~2.4\% when operated at a common over-voltage.
Without any further surface treatment, the scintillator tiles are wrapped in laser-cut reflective foil by a robotic procedure and mounted on the HBUs using a pick-and-place machine, after glue dispensing with a screen printer.   
The HBUs have been integrated into cassettes with interfaces for DAQ \cite{Kvasnicka:2017bpx}, LED pulsing and power distribution. The latter provides active compensation of temperature variations by automatic adjustments of the common bias voltage of the photon sensors in each layer. This was routinely used in test beam operation and stabilises the gain within $\pm 1\%$.
%
Data concentration, power distribution and cooling service systems of the prototype are also scalable to the full ILD detector. 

\begin{figure}[hbt]
\centering
\includegraphics[width=0.64\textwidth]{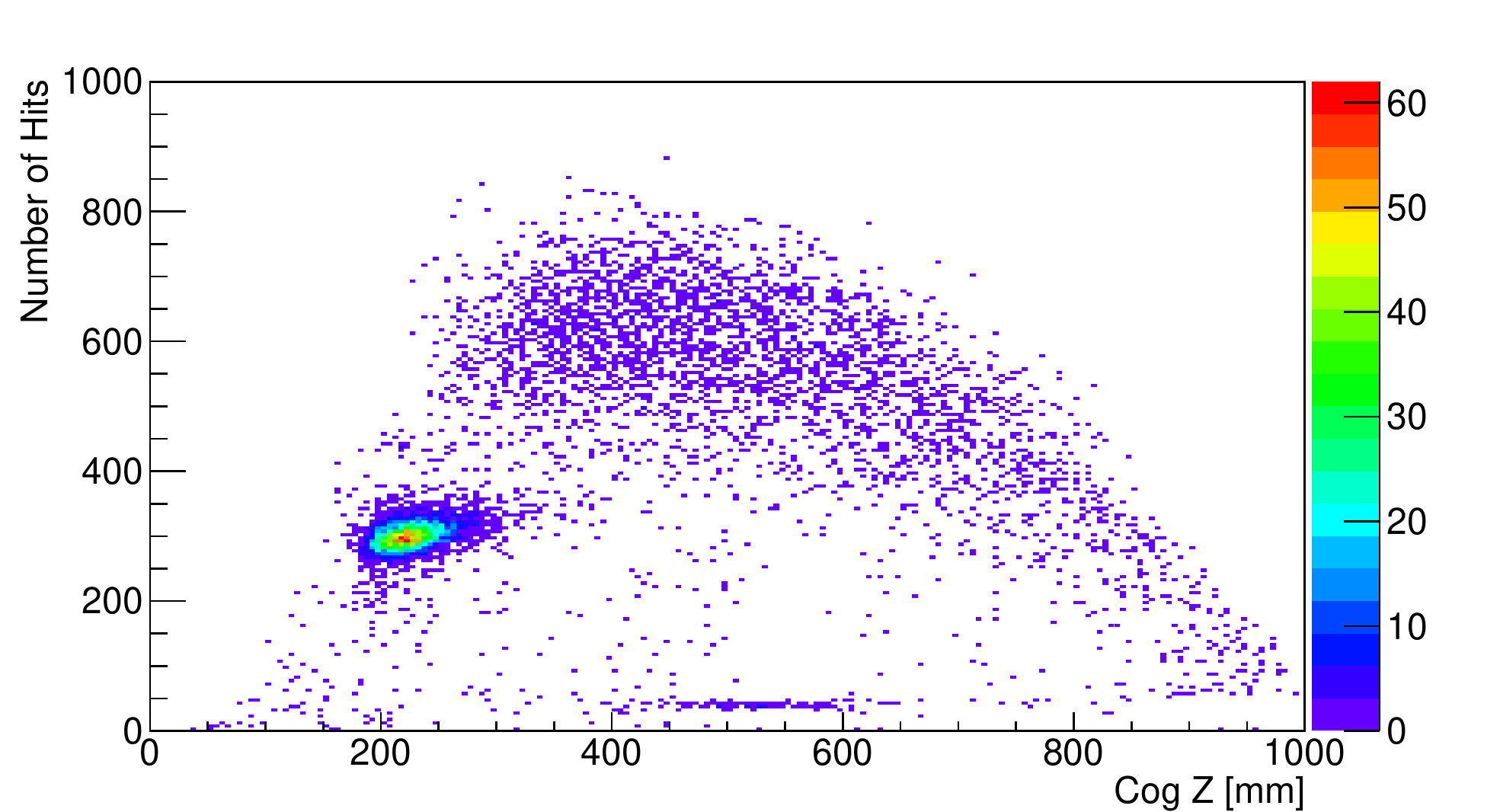}
\includegraphics[width=0.34\textwidth]{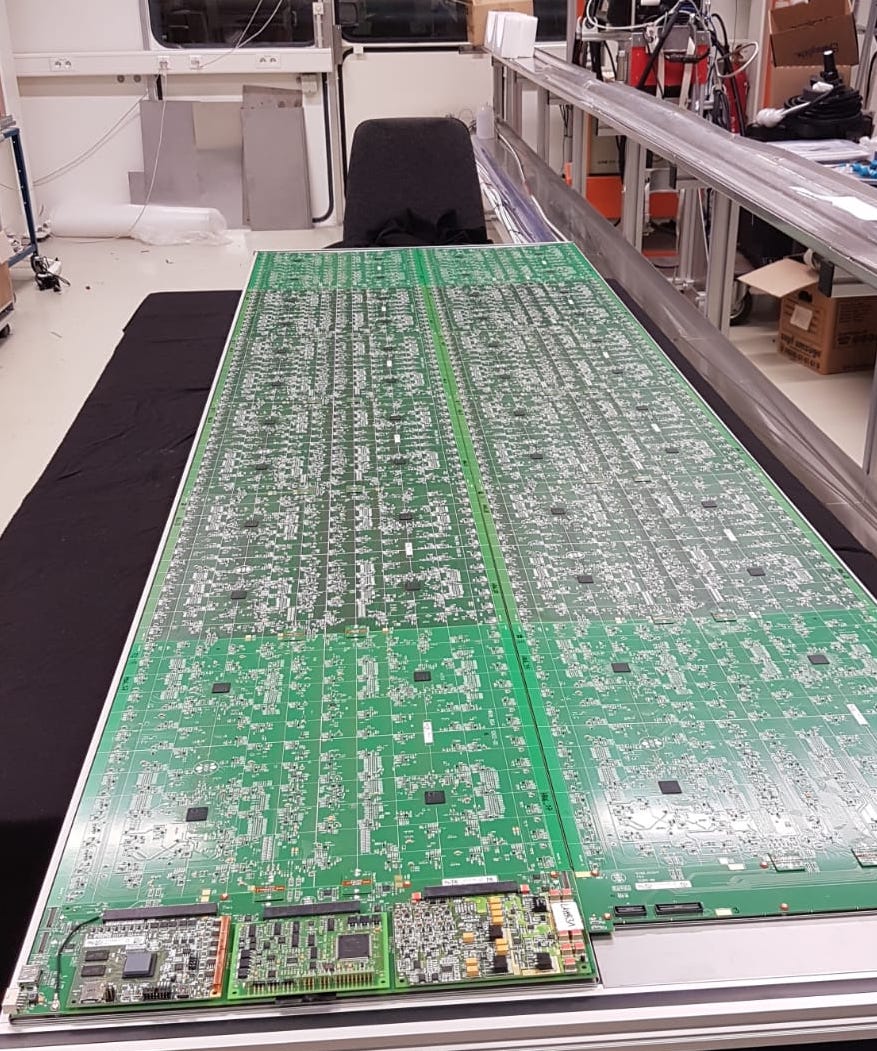}
\caption{Left: Distribution of the number of hits vs.\ the hit-energy weighted centre-of-gravity (cog) for a mixed beam in the AHCAL technological prototype. Right: Two slabs of HBUs with interfaces, corresponding to full ILD length in the TESLA configuration.} 
\label{fig:AHCAL-nhit-longslab}
\end{figure}

The AHCAL technological prototype was installed in the test beam for data taking at the CERN SPS, see Figure~\ref{fig:AHCAL-TileProto}.
During two periods in May and in June 2018, several $10^7$ events with muon tracks, as well as electron and pion showers in the energy ranges  10 -- 100~GeV and 10 -- 200~GeV, respectively, have been recorded. 
Figure~\ref{fig:AHCAL-nhit-longslab} left, from the quasi-instantaneous data quality monitoring, shows the distribution of the number of hits vs.\ the hit-energy weighted centre-of-gravity (cog) along the beam axis z for an electron run with a beam momentum of 100~GeV/c and admixtures of muons and hadrons. The different particle types populate different regions of the plot.
While electron showers are characterised by a relatively narrow distribution of number of hits and a cog near the front face of the detector, hadrons exhibit a wider distribution of the cog, and a larger number of hits, decreasing as the cog moves towards the rear of the detector, and leakage increases.
Muons appear as a narrow band with $\sim 38$ hits and a cog on z at about half the depth of the detector. 

%
%
The rich data sample collected in the two test beam periods in 2018  is being used for shower separation studies based on 5-dimensional reconstruction algorithms exploiting the high spatial, energy and time resolution of the prototype. 
While this is in progress, it can already be noted that in several aspects the  performance exceeds that of the physics prototype: the noise is a factor 100 lower, the dynamic range of the SiPMs is 3 times larger, and 99.96\% of the total 21888 channels are working.

The HBUs of the new prototype have also been used to build a large layer with two slabs of 6 HBUs each, corresponding to the full $\approx$2m length of an ILD barrel sector in the TESLA layout, see Figure~\ref{fig:AHCAL-nhit-longslab} right. The signal quality was unaffected, and the rise time of the power pulsing was within specifications. 

The AHCAL developments have also inspired the design of the scintillator section of the CMS end-cap calorimeter upgrade for the high luminosity phase of the LHC~\cite{Collaboration:2293646}, and the new prototype has been used together with silicon-instrumented sections in front in a common beam test, further illustrating the maturity of the technology. 
The new CMS endcap calorimeter will establish the SiPM-on-Tile technology in a collider environment at an intermediate scale between the AHCAL prototype and the full ILD detector. 

\subsubsection{RPC option (SDHCAL)}
\label{ild:sec:hcal:sdhcal}
\begin{figure}[t!]
\centering
\includegraphics[width=1.0\hsize]{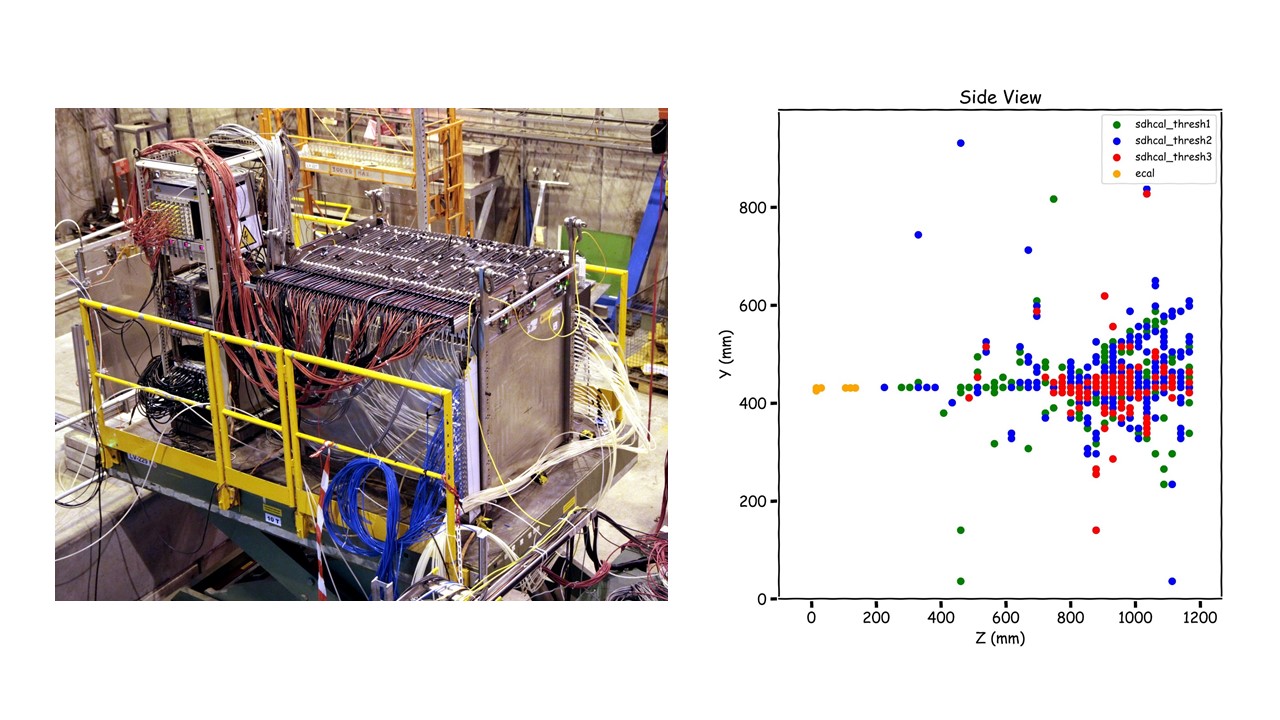}
\caption{Left: technological prototype of the Semi-Digital Hadronic Calorimeter in its beam line at CERN. Right: event display of a 70 GeV pion measured in a recent beam test combined with a few layers of SiECAL calorimeter.}
\label{fig:det:SDHCAL_proto}
\end{figure}

The SDHCAL technological prototype built in 2011 has been regularly tested in beams in the past years with various configurations, including a combined test with the SiECAL prototype in 2018 (section 5.2.4.1). The SDHCAL prototype consists of 48 single-gap RPC layers of 1 m$^2$ (Figure~\ref{fig:det:SDHCAL_proto} left). Each detection gap is instrumented with 6 Active Sensitive Units (ASU) made of a  50 x 33 cm$^2$ PCB with 24 "HARDROC" ASICs from OMEGA~\cite{Callier:2014uqa}. The RPC pad size is 1 cm$^2$ and the pad signals can be read out either in digital (1 bit and 1 threshold) or semi-digital (2 bits and 3 thresholds) modes (Figure~\ref{fig:det:SDHCAL_proto} right).

\begin{figure}[t!]
\centering
\includegraphics[width=1.0\hsize]{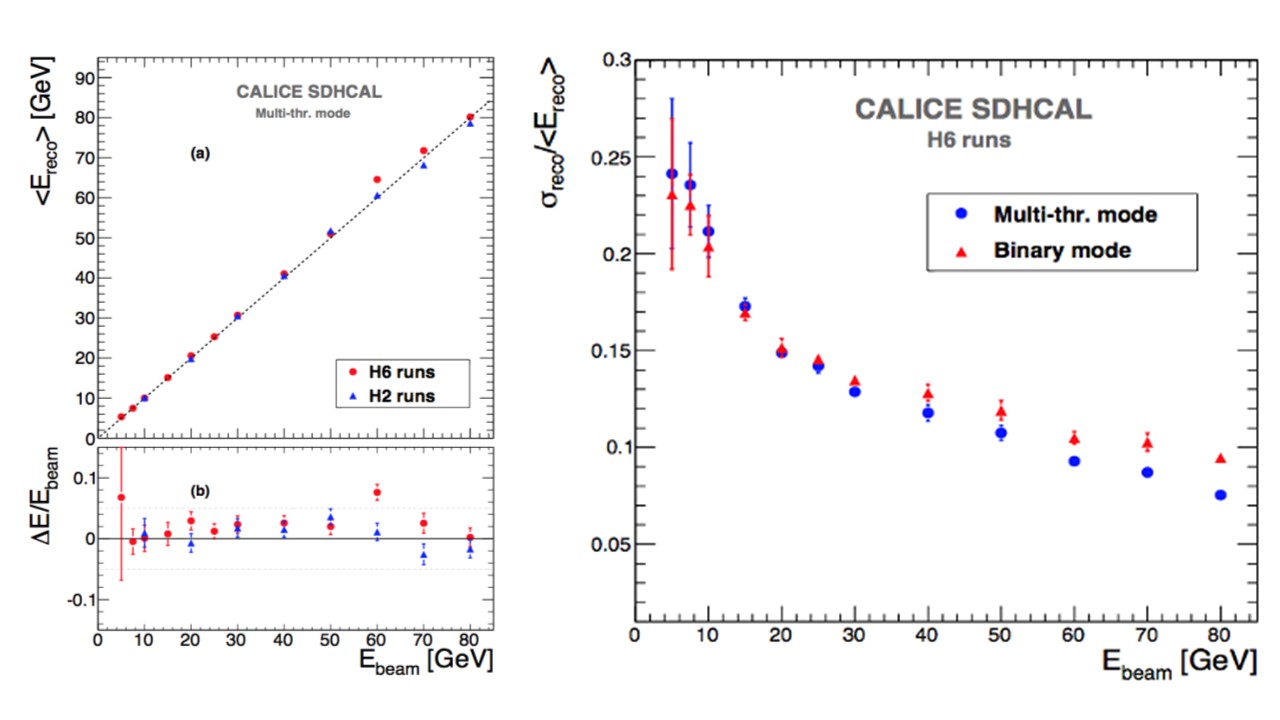}
\caption{Performance of the SDHCAL technological prototype as measured in beam tests. Left: linearity with the semi-digital readout option. Right: energy resolutions with the digital and semi-digital readout options in a hadron beam dominated by pions.}
\label{fig:det:SDHCAL_perf}
\end{figure}

The numerous data sets collected at CERN with high-energy hadron beams have been used to validate the performance of the technology. Special reconstruction methods adapted to the high granularity semi-digital structure of the calorimeter have been developed~\cite{Buridon:2016ill} to relate the energy estimation to the hit number and density. The current state of the performance is summarized in Figure~\ref{fig:det:SDHCAL_perf}. A good linearity is observed and the multi-threshold mode is found to mitigate saturation and improve the resolution at high energy. The description of the measured resolution by the simulation is however found to be sensitive to the description of the core of the hadronic showers, with a tendency for the Monte-Carlo to underestimate the performance due to harder cores in the showers~\cite{Deng:2016obt}. This point will require further tuning of the simulation.

\begin{figure}[t!]
\centering
\includegraphics[width=1.0\hsize]{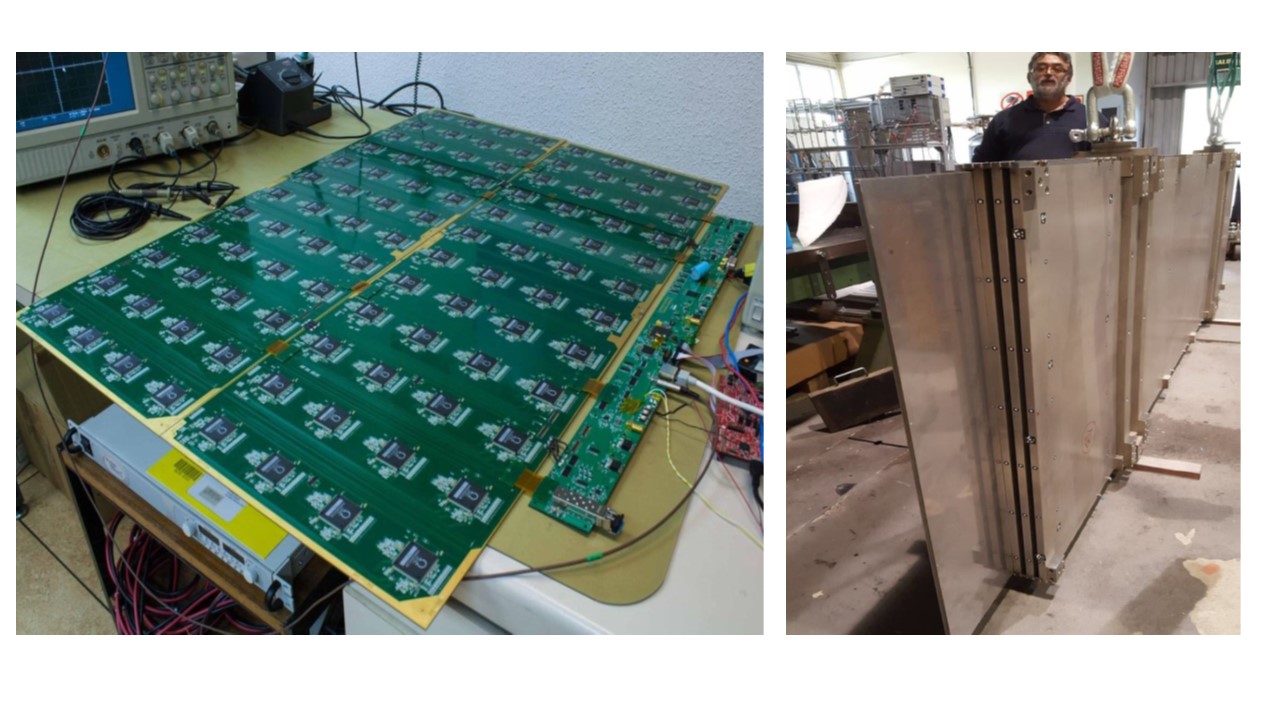}
\caption{SDHCAL ongoing developments for the final ILD dimensions: two active sensor units of 100 x 33 $cm^2$ chained to each other and connected to the outside with a new compact DAQ interface (left), and self-sustained welded structure of 4 calorimeter plates of 1x3 $m^2$   with the required mechanical tolerances (right).}
\label{fig:det:SDHCAL_dev}
\end{figure}

In recent years the SDHCAL teams have focused on adapting the technology to the full-size ILD requirements, in order to cover detection surfaces of up to 1 x 3 m$^2$ required by the ILD Hadronic Calorimeter in its "Videau" configuration (section 5.1.2). An improved RPC gas circulation system with better uniformity has been designed and validated with the construction of two large RPC's. Larger ASUs of 100 x 33 $cm^2$ have been designed with a new version of the HARDROC ASIC including zero-suppression (Figure~\ref{fig:det:SDHCAL_dev} left), and their interconnection improved to allow chaining of up to 9 ASUs. Efforts have also been invested in the manufacturing process of self-sustained hadron calorimeter structures with high precision mechanical tolerance as required by the RPC insertion. A method of "roller levelling" has been used to machine steel absorber plates with a high flatness. A high precision electro-welding has been used to build a first large size prototype of 4 calorimeter layers (Figure~\ref{fig:det:SDHCAL_dev} right) which has proven that gap size variations well below 1 mm can be reached on such large structures. 

For the longer term the option of multi-gap RPC's with a high timing resolution of $\approx$20 ps is prototyped based on the "PETIROC" ASIC~\cite{Fleury:2014hfa}. 

\subsection{Very Forward Detectors}
\label{chap:technologies:fcal}

In the past years the development of the ILD forward detectors has mostly been pursued by the FCAL R\&D Collaboration~\cite{ild:bib:FCAL}. Progress concerns mainly the LumiCAL calorimeter and, more recently, the BeamCAL sensors.

\begin{figure}[t!]
\centering
\includegraphics[width=1.0\hsize]{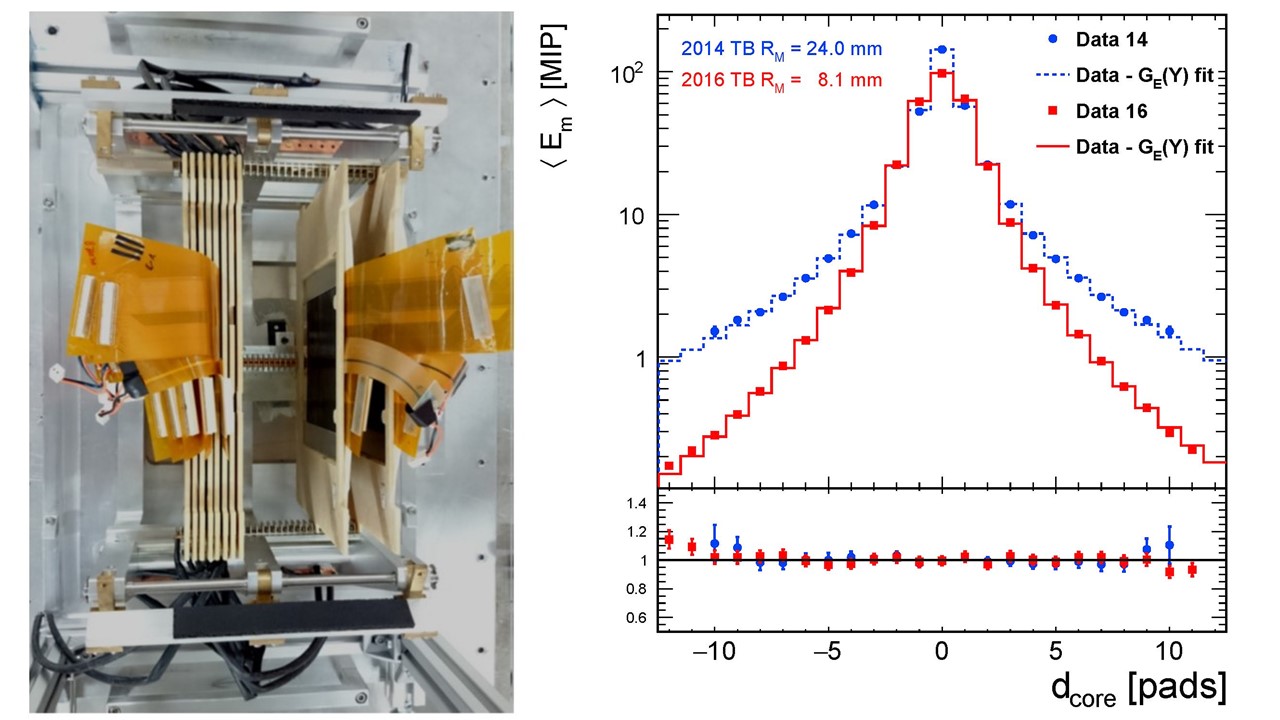}
\caption{LumiCAL compact prototype tested in the DESY electron beam in 2016 (left) and corresponding improvement in shower compactness achieved versus the 2014 prototype (right).}
\label{fig:det:LUMICAL_perf}
\end{figure}

Based on a specific ASIC developed after the DBD, calorimeter silicon sensitive layers have been built to assemble a first LumiCAL 4-layer tungsten calorimeter prototype and, two years later, a more compact 8-layer calorimeter prototype (Figure~\ref{fig:det:LUMICAL_perf} left). The two prototypes were tested with beam in 2014 and 2016, respectively. The test data~\cite{Abramowicz:2018vwb} confirm the expected significant improvement of the transverse compactness of the electromagnetic showers in the compact prototype compared to the earlier one (Figure~\ref{fig:det:LUMICAL_perf} right). 

A new ASIC "FLAME"~\cite{ild:bib:FLAME} based on 130 nm CMOS technology is currently under final validation. FLAME features the low power, in-situ digitisation and fast readout required by the final detector. A new $\approx$20- layer SiW calorimeter prototype based on FLAME, with specifications and configuration close to the final LumiCAL detector, is under construction and planned to be tested with beam. 

\begin{figure}[t!]
\centering
\includegraphics[width=0.8\hsize]{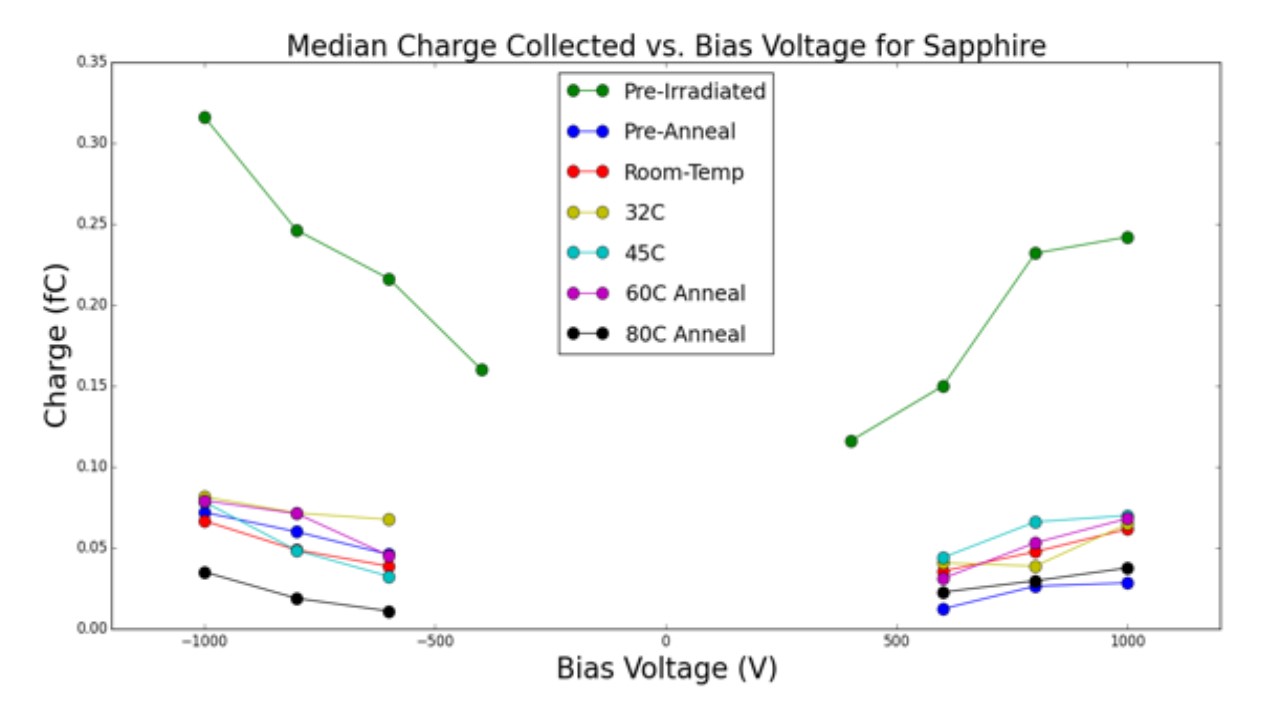}
\caption{results of irradiation tests performed at SLAC for sapphire sensors considered for the BeamCAL.}
\label{fig:det:BEAMCAL_rad}
\end{figure}

The LHCAL and BeamCAL calorimeters can be based on similar technologies as the LumiCAL, with radiation hardness requirements increasing as function of the sensor proximity to the beam. For the BeamCAL, new sensors such as sapphire are being considered. Irradiation campaigns are under way to characterize them (Figure~\ref{fig:det:BEAMCAL_rad}) and provide input for the final choice. 

\subsection{Iron Yoke Instrumentation}

\begin{figure}[t!]
\centering
\includegraphics[width=1.0\hsize]{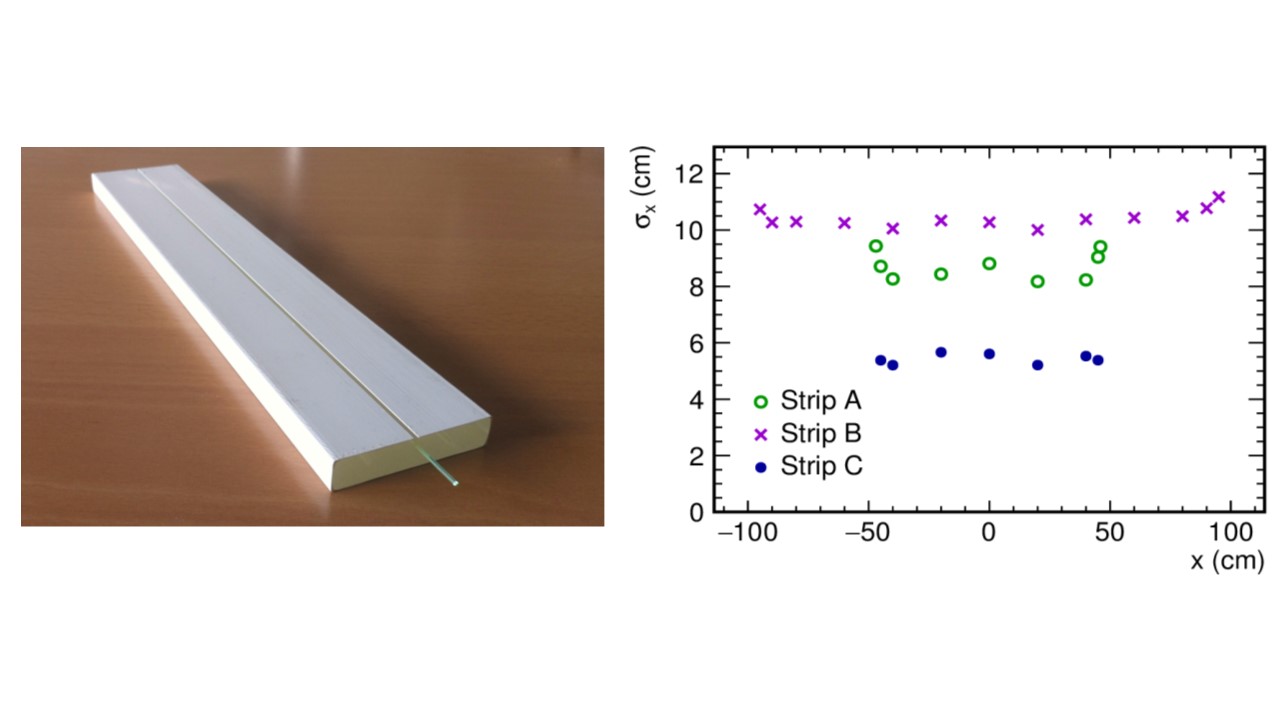}
\caption{Left: example of prototype scintillator bar built for the muon detector; right: longitudinal resolution on reconstructed muons as function of longitudinal coordinate: strip A and B as shown on the left of the figure with lengths of 1 and 2\,m respectively, strip C of 1\,m length with WLS fibers positioned on the small edge of the strip.}
\label{fig:det:Iron_proto}
\end{figure}

Dedicated studies have been conducted at FNAL in the past years to optimize the layout of scintillator bars adapted to muon detection. Prototypes have been built and tested with muon beams~\cite{Denisov:2015jjl}. They are all based on long scintillator bars with signal collected by WLS fibers and readout by SiPMs at both extremities. The transverse resolution of $\approx$1cm required for the muon momentum measurement is defined by the bar widths of a few cm. The longitudinal position is measured from the time difference of the signals at both extremities and depends on the WLS configuration. The two options under consideration described in section 5.1.2 (Figure~\ref{fig:det:yoke}) have been tested: longitudinal resolutions of 5 to 10 cm are measured and found roughly independent of the longitudinal position of the muon within the bar (Figure~\ref{fig:det:Iron_proto}). 

More studies are ongoing to develop low cost SiPMs also adapted to the measurement of the tails of high energy jets (tail catcher function). 

The RPC option for the iron yoke instrumentation was not specifically studied but would directly benefit from the RPC developments of the SDHCAL hadronic calorimeter option (section~\ref{ild:sec:hcal:sdhcal}).

\hfill

\chapter{ILD Global Integration}
\label{chap:integration}

This chapter is devoted to the diverse matters of global interest for the ILD integration, assembly and operation. These include the integration of the detector and its services within the experimental hall, the integration of subdetectors as a coherent detector, the mechanical behaviour of its components and mitigation of seismic movements, the design of the main magnet with its coil and iron yoke, mitigation of beam-related background, calibration strategies and data acquisition. All these topics still require a large amount of studies to finalise the ILD engineering design in its operational environment. There was however significant progress over the past years and the current status is summarised in the following sections.   

\section{External ILD integration}
\label{ild:sec:external_integration}

The proposed site for the ILC is located in the Kitakami mountains in the north of the Japanese main island Honshu. Dedicated studies are under way to adapt the generic ILC design, as described in the Technical Design Report, to the realities in this environment. For ILD, the arrangements of the surface and underground installations around the interaction point (IP) are of natural importance. The current conceptual design of the civil facilities and the plans for the detector related infrastructure and services have been coordinated between the relevant detector and ILC machine groups as well as with local experts.

\subsection{Site-related Infrastructure}

\begin{figure}[htbp]
\includegraphics[width=0.9\hsize]{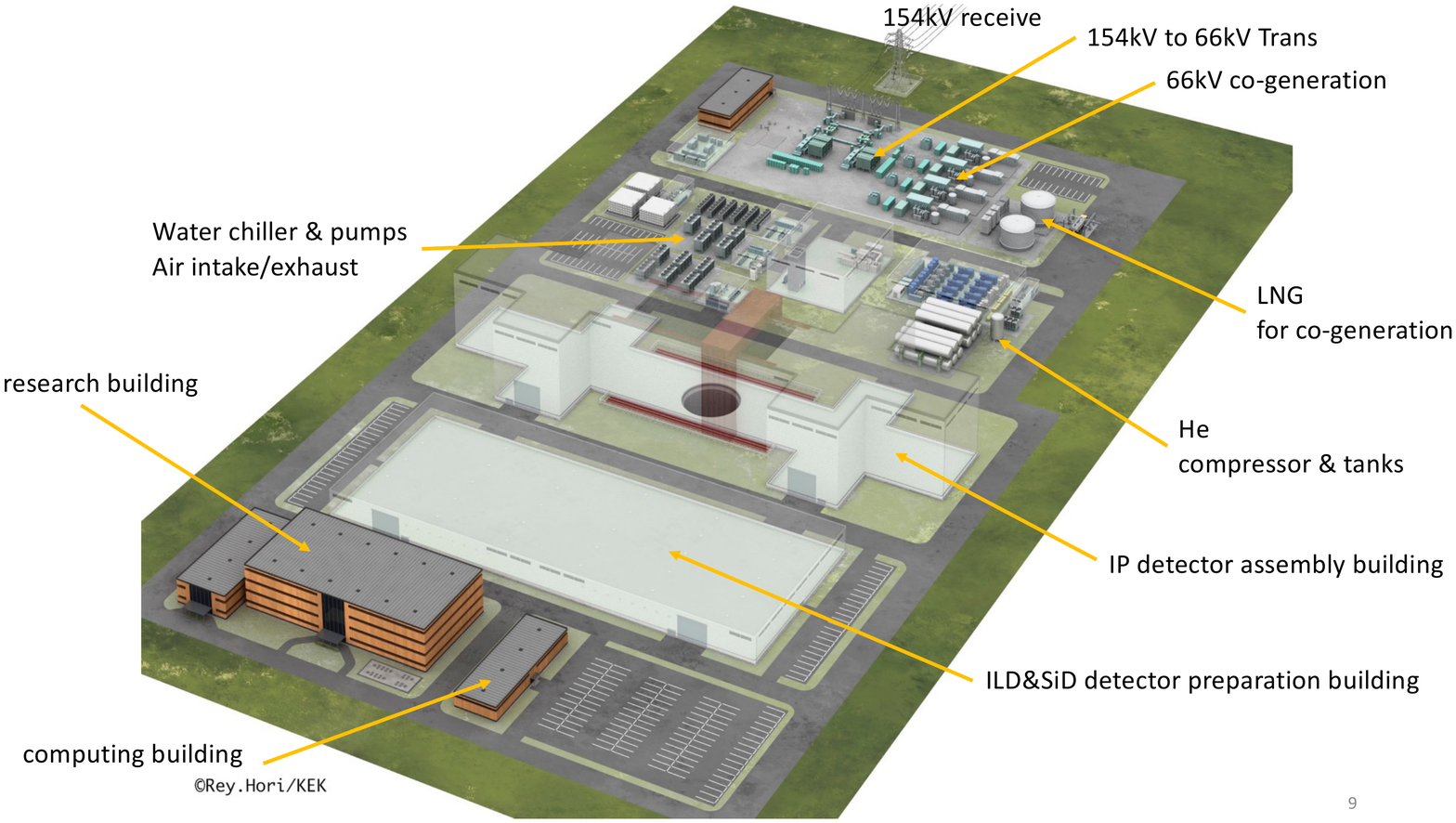}
\caption{\label{fig:integration:surface}Conceptual design (artist's view) of the surface facilities ("IP Campus") above the ILC interaction point~\cite{ild:bib:surface_facilities}. }
\end{figure}

ILD is foreseen to be assembled on the surface, similar to CMS at LHC. Figure~\ref{fig:integration:surface} shows a rendering of the surface installations above the ILC interaction point. At the heart is the detector assembly building that is located directly over the central shaft that gives access to the underground collider hall. A large gantry crane above the shaft allows for the lowering of the pre-instrumented detector parts into the underground area. A preparation building is foreseen where sub-detector elements can be assembled and tested. A research building and a computing building provide the infrastructure for the operation of the detectors at the IP Campus. 

\begin{figure}[h!]
\includegraphics[width=0.75\hsize]{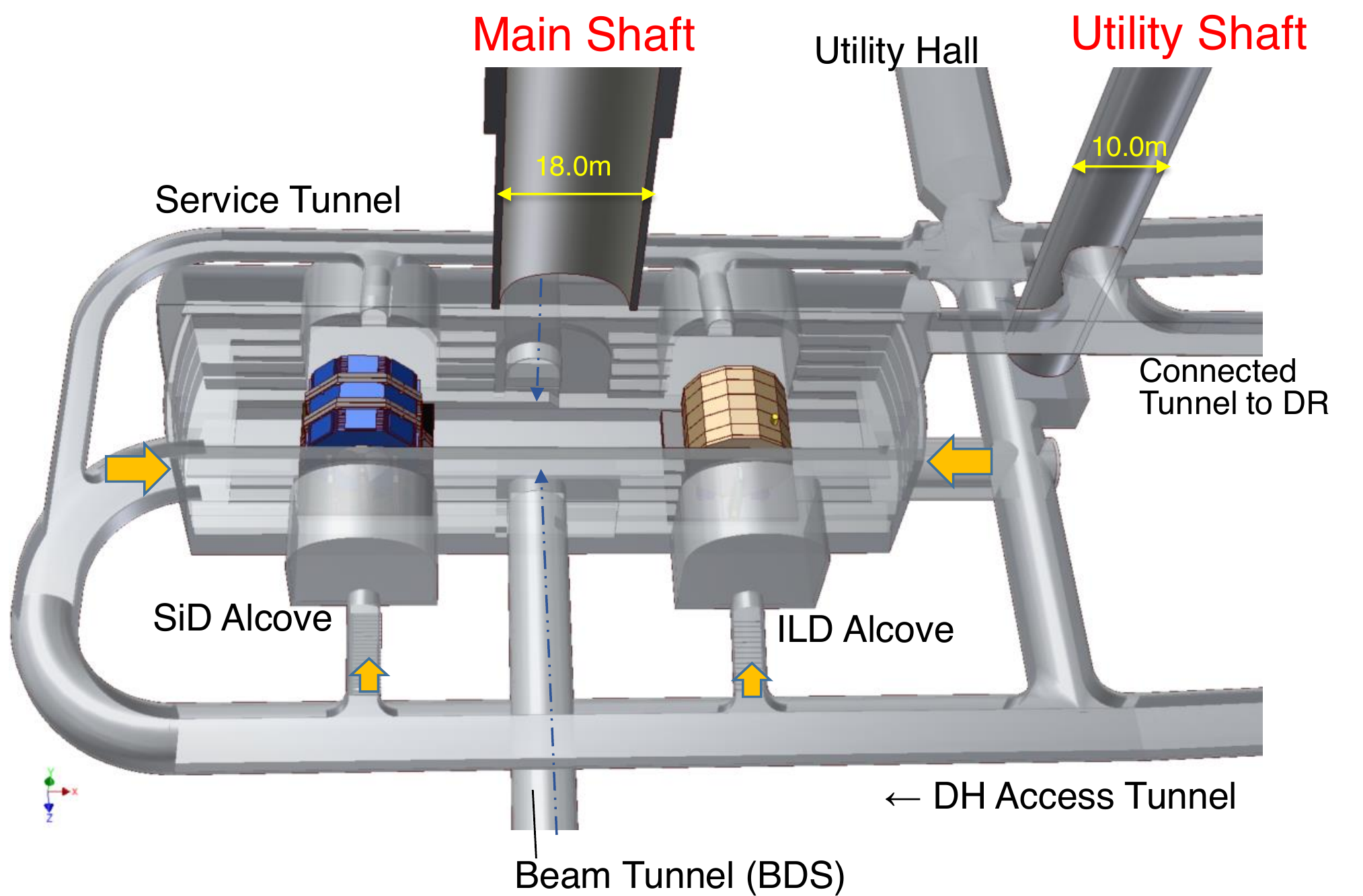}
\caption{\label{fig:integration:underground}Underground facilities with the detector hall, ILD and SiD in push-pull configuration, access tunnels and shafts~\cite{ild:bib:underground_facilities}. }
\end{figure}

\begin{figure}[h!]
\includegraphics[width=0.9\hsize]{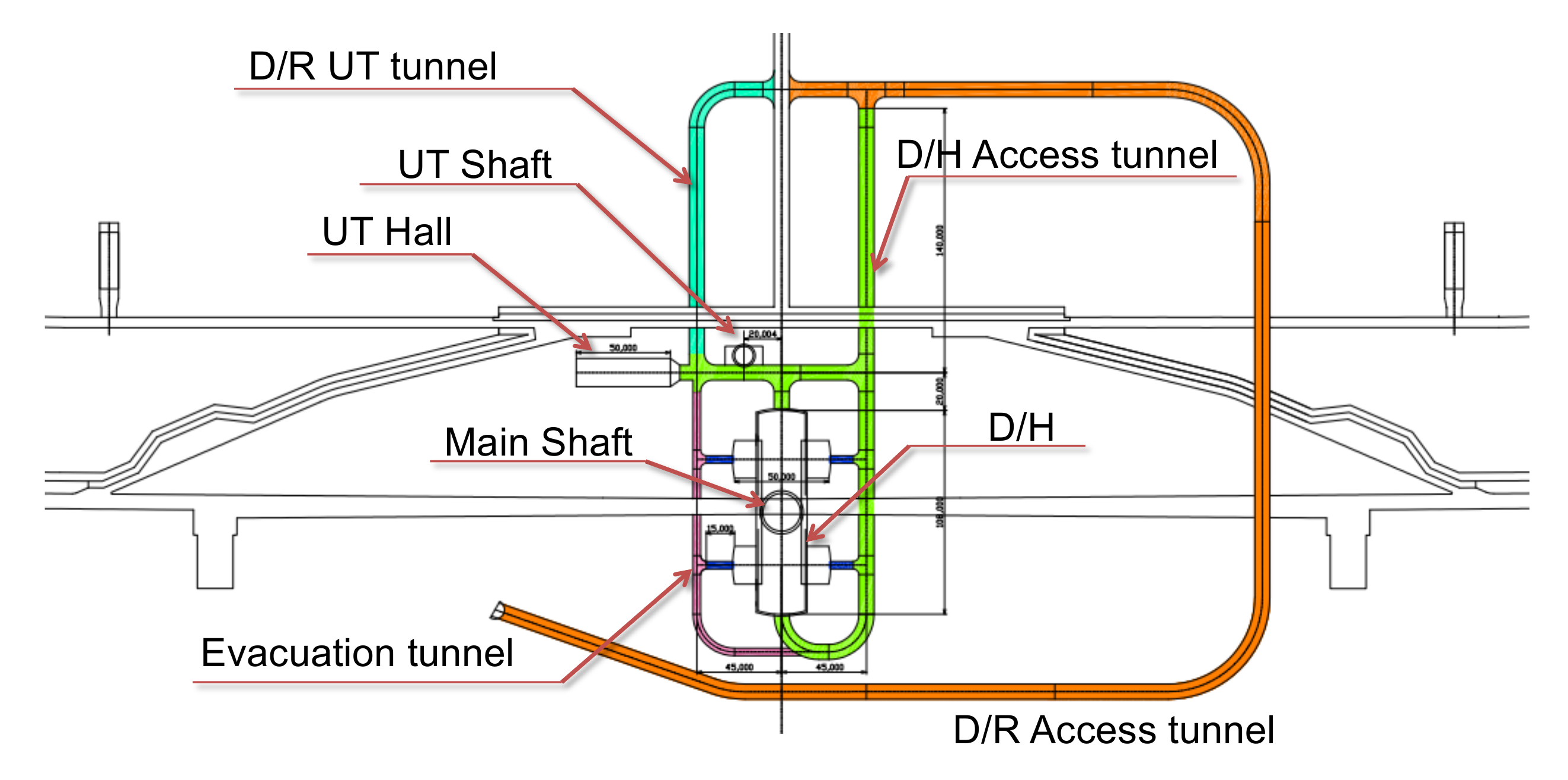}
\caption{\label{fig:integration:access}Access to the underground infrastructure is provided by two shafts, main shaft and utility (UT) shaft, and a system of access tunnels~\cite{ild:bib:Access} that serve the detector hall (D/H) as well as the damping rings (D/R). }
\end{figure}

The underground experimental hall is about 100~m below the surface and hosts two experiments, ILD and SiD, in a "push-pull" arrangement where both detectors share the same interaction region (Figure~\ref{fig:integration:underground}). The detectors are installed on movable platforms and can be rolled into or out of the beam line within a few hours. They can be opened and maintained in their parking positions. Access to the underground hall is provided by two vertical shafts and an access tunnel that allows for vehicles to drive directly into the underground area (Figure~\ref{fig:integration:access}).

\subsection{Detector Utilities and Cavern Ancillary Services}


In order to operate the ILD detector in the underground detector cavern, electricity, cooling water and other services have to be supplied from surface facilities. Though conceptual designs existed at the time of the DBD, more detailed knowledge about the requirements from the detector exists by now and should lead to an optimisation of the facilities layout and locations.

\subsubsection{Service Locations}
\label{ild:sec:service_locations}

\begin{figure}[h!]
\includegraphics[width=1.0\hsize]{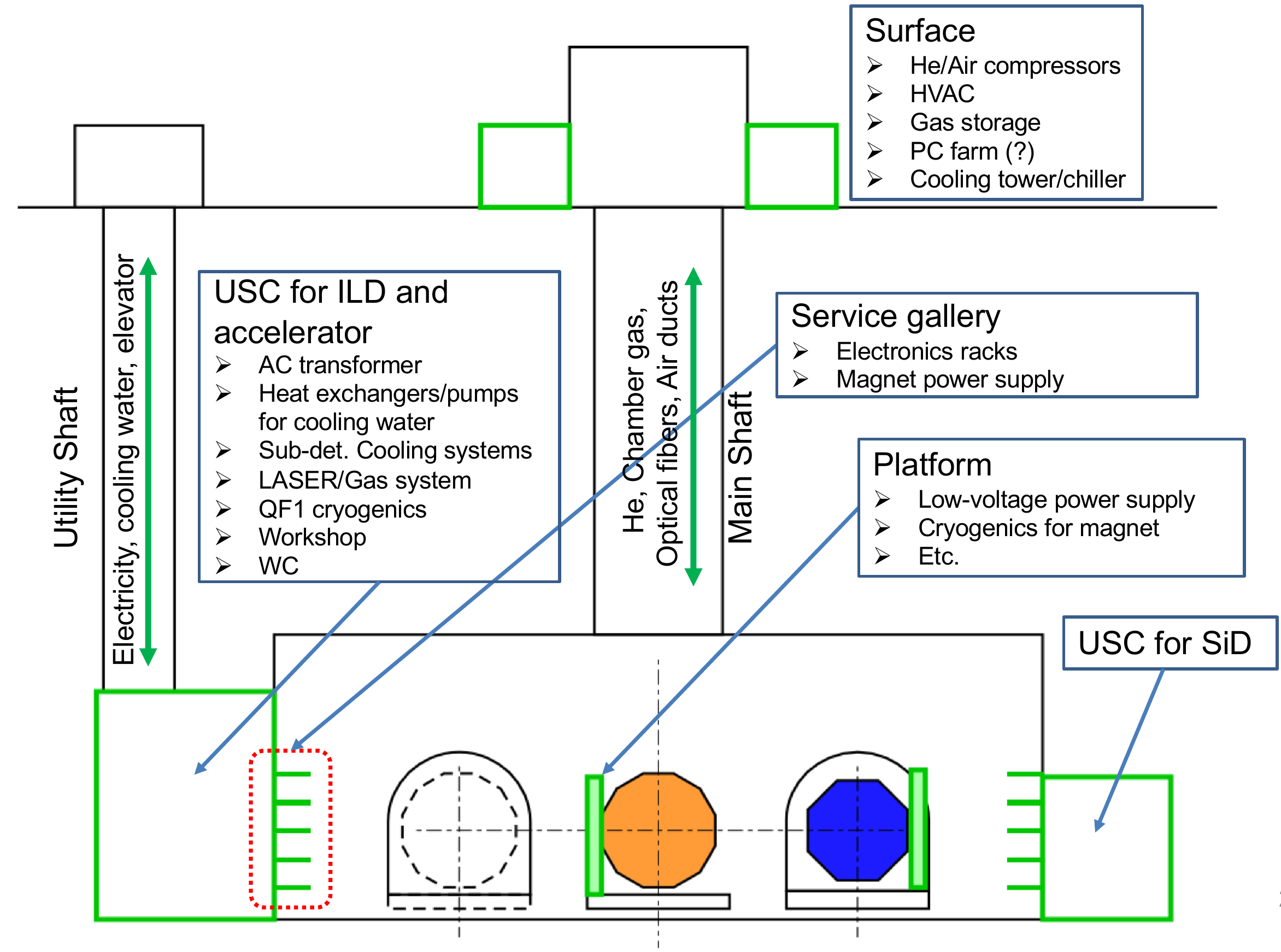}
\caption{\label{fig:integration:services}Schematic drawing of the possible locations for detector services. The Utility/Service cavern (USC) for the detector is proposed but not yet implemented into the ILC baseline design~\cite{ild:bib:services_figure}. }
\end{figure}

There are several possible locations for detector services: on the detector platform, on service galleries on the wall of the detector hall, in dedicated utility/service caverns (shown as "Utility Hall" / "UT Hall" in Figures~\ref{fig:integration:underground}, \ref{fig:integration:access}), and on surface. A possible configuration is shown in Figure~\ref{fig:integration:services}.
It is assumed that large or noisy apparatus such as transformers (6.6~kV$\rightarrow$400/200/100~V), heat exchangers and pumps for cooling water, sub-detector cooling plants, etc. should be located in the utility/service cavern. Cryogenic plant for the QF1 magnet is also supposed to be located in the utility/service cavern.

The utility/service cavern should be relatively close to the detector, but well isolated from the detector hall  regarding the noise, vibration, and radiation. Design of the facilities including caverns (CFS) for detector utilities/services has to be made based on requirements from the detector side. In order to clarify the requirements, a rough estimation on the ILD needs in electricity, cooling water, and space has been made (c.f. section~\ref{ild:sec:power}). 

\begin{figure}[h!]
\centering
\includegraphics[width=1.0\hsize]{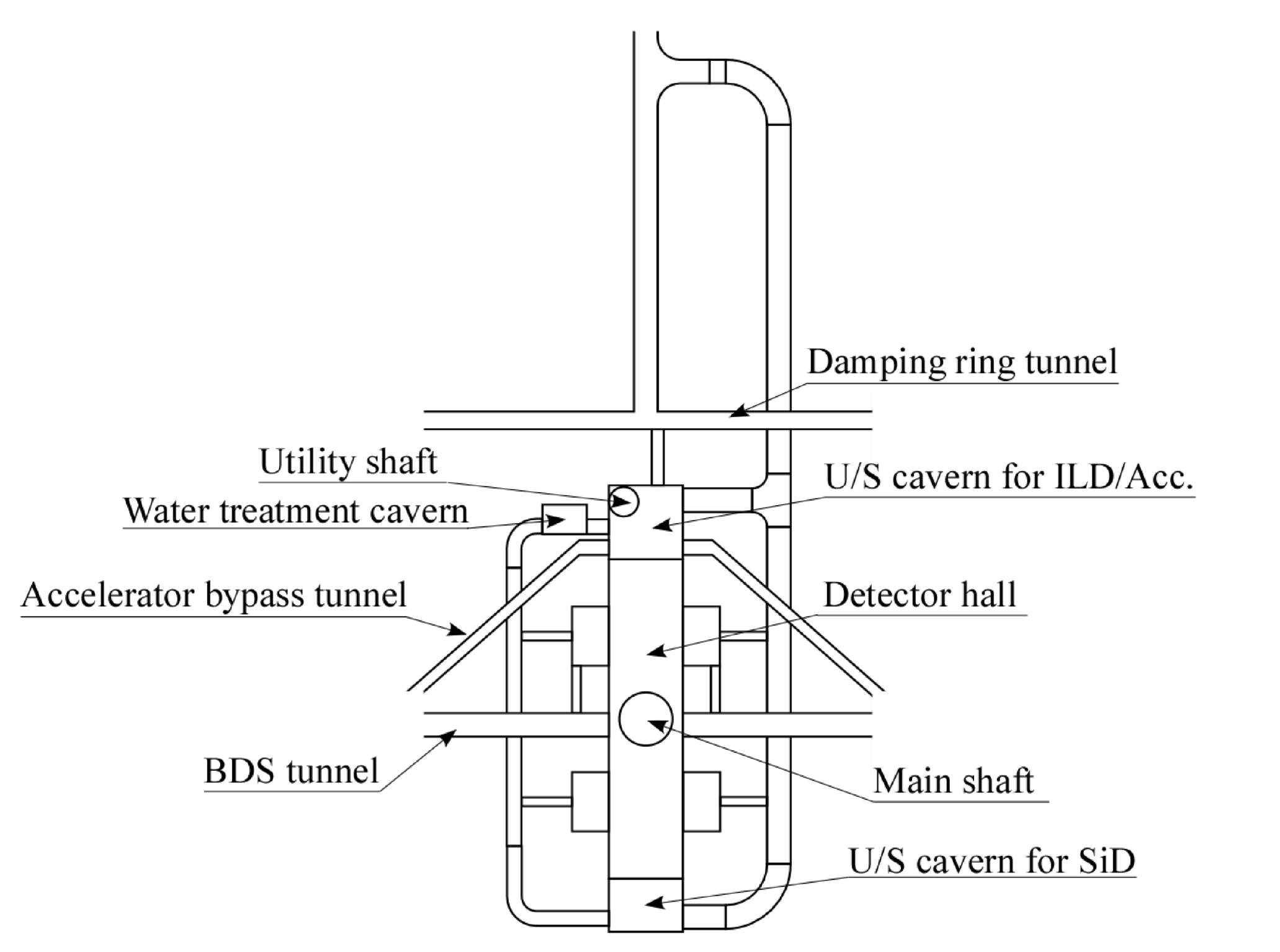}
\caption{\label{fig:integration:USC}A possible design of Utility/Service caverns. }
\end{figure}

The Design of the utility/service cavern is not fixed yet. In the baseline design (TDR-modified), the utility/service cavern had a dead-end as shown in Figure~\ref{fig:integration:access}. In addition, this cavern was supposed to be used for both detectors and accelerators, and did not have enough space for the estimated ILD needs. 
ILD proposes another design as shown in Figure~\ref{fig:integration:USC}. In this design, there are two utility/service caverns, one for the accelerator and ILD, and another one for SiD. The size of the accelerator/ILD utility cavern of 25m x 25m with five floors would be large enough for accelerator and ILD, and that of 25m x 18m with three floors would be large enough for SiD. Figure~\ref{fig:integration:services} is drawn assuming this utility/service cavern design.

\subsubsection{Power Consumption and Cooling Requirements}
\label{ild:sec:power}
The requirements for power and cooling at ILD have been studied in a bottom-up approach where the current information from all detector components has been collected and cross-checked.

A tentative estimation of the power consumption in the detector hall and surface facilities for ILD is listed in Table~\ref{tab:integration:power}. The total power consumption in the underground caverns is about 1~MW, dominated by the ILD solenoid and the machine components associated to the detector. 

\begin{figure}[h!]
\includegraphics[width=0.7\hsize]{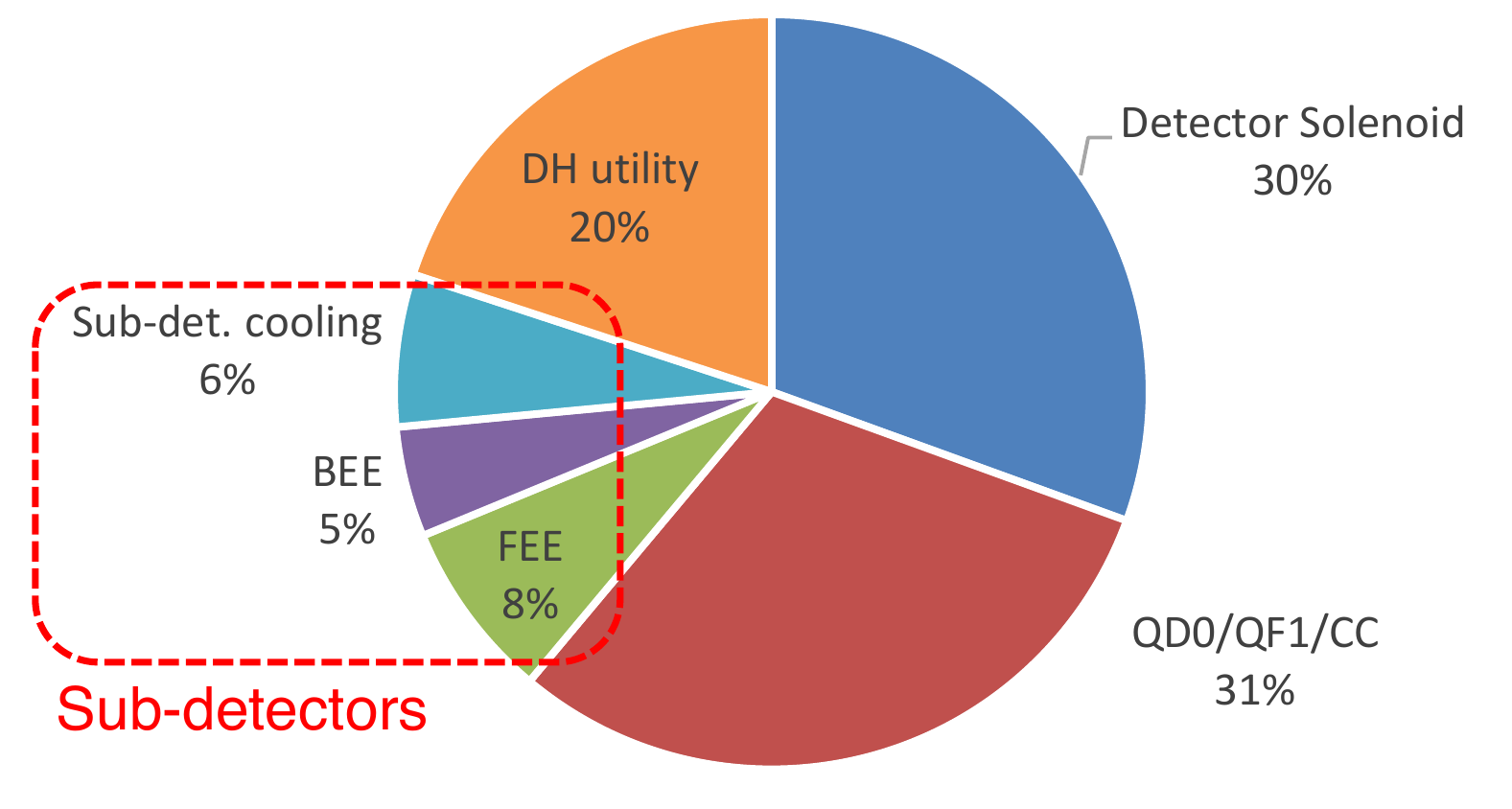}
\caption{\label{fig:integration:power}Distribution of the underground power consumption for ILD~\cite{ild:bib:services}. While the detector solenoid, the machine elements (magnets QD0/QF1 and the crab cavity system CC) use the major part of the power, the detector hall (DH) utilities and the sub-detectors cooling, back- and front-end electronics (BEE, FEE) contribute less. The total power consumption sums up to about 1~MW.}
\end{figure}

The cooling water which is necessary for extracting the power consumption is shown in Table~\ref{tab:integration:cooling}.
The sharing of the underground power is visualized in Figure~\ref{fig:integration:power}. It can be seen that the fraction of the power consumption that is coming from the sub-detectors is small.

\begin{table}[htb]
    \centering
    \begin{tabular}{l|l|c|c|c|c}
       \multicolumn{2}{l|}{{\bf Item}} & {\bf Power} &\multicolumn{3}{c}{ } \\ \hline
        \multirow{3}{*}{QD0/QF1/Crab Cavity} & Power Supply  & 150 & \multicolumn{3}{l}{ } \\
                            & Cold Box      & 150 & \multicolumn{3}{l}{ } \\
                            & He Compressor & 300 & \multicolumn{3}{l}{on surface}\\ \hline
        \multirow{3}{*}{Detector Solenoid} & Power Supply  & 250 & \multicolumn{3}{l}{ } \\
                            & Cold Box      & 50 & \multicolumn{3}{l}{ } \\
                            & He Compressor & 500 & \multicolumn{3}{l}{on surface}\\ \hline                      
        \multirow{10}{*}{Subdetectors} & {\bf Total}  & {\bf 161} & {\bf Front-end Electr.} & {\bf Back-end Electr.} & {\bf Cooling}\\ \cline{2-6}
        & Muon  & 12 & 5 & 5  & 2 \\
        & HCAL  & 45.5 & 27.5 & 8 & 10 \\
        & ECAL  & 40 & 20 & 12 & 8 \\
        & VFS   & 9 & 2 & 5 & 2 \\
        & SET   & 9 & 2 & 5 & 2 \\
        & TPC   & 16.2 & 15 & - & 1.2 \\
        & SIT   & 8 & 1 & 5 & 2 \\
        & FTD   & 8 & 1 & 5 & 2 \\
        & VTX   & 13.5 & 2 & 1.5 & 10 \\\hline
        \multicolumn{2}{l|}{Computer Farm}& 1000 & \multicolumn{3}{l}{on surface}\\ \hline
        \multicolumn{2}{l|}{Water Pump}& 25 & \multicolumn{3}{l}{}\\ \hline
        \multicolumn{2}{l|}{HVAC}& 600 & \multicolumn{3}{l}{on surface}\\ \hline
        \multicolumn{2}{l|}{Lighting}& 25 & \multicolumn{3}{l}{}\\ \hline
        \multicolumn{2}{l|}{Air Compressor}& 50 & \multicolumn{3}{l}{on surface}\\ \hline
        \multicolumn{2}{l|}{Platform Mover}& 100 & \multicolumn{3}{l}{}\\ \hline
        \multirow{2}{*}{Cranes} & 3x5t & 21 & \multicolumn{3}{l}{}\\
        & 40t & 50 & \multicolumn{3}{l}{}\\ \hline
        \multicolumn{2}{l|}{{\bf TOTAL}}& {\bf 3432} & \multicolumn{3}{l}{}\\ \hline
        \multicolumn{2}{l|}{Underground}& 982 & \multicolumn{3}{l}{}\\ \hline
    \end{tabular}
    \caption{Breakdown of the power consumption estimates for the ILD detector and appended ILC components (in kW)~\cite{ild:bib:services}. Power-pulsing operations are assumed for the ILD subdetectors.}
    \label{tab:integration:power}
\end{table}

\begin{table}[]
    \centering
    \begin{tabular}{m{1.4cm}|m{1.7cm}|m{0.6cm}|m{0.6cm}|m{0.7cm}|m{0.6cm}|m{0.6cm}|m{0.7cm}|m{0.6cm}|m{0.6cm}|m{0.7cm}}
    \multicolumn{2}{m{1.4cm}|}{}& \multicolumn{3}{c|}{Chilled Water} & \multicolumn{3}{c|}{Low-conductive Water} & \multicolumn{3}{c}{Normal Water} \\ \hline
    \multicolumn{2}{m{1.4cm}|}{Item} & Heat (kW) & $\Delta$T (K) & Flow (l/min) & Heat (kW) & $\Delta$T (K) & Flow (l/min) & Heat (kW) & $\Delta$T (K) &Flow (l/min) \\ \hline
    \multirow{2}{1.4cm}{QD0/QF1/ Crab Cav.} & Power Supply & & & & 150 & 10 & 214 & & & \\
    & Cold Box & & & & 150 & 10 & 214 & & & \\\hline
    \multirow{2}{1.4cm}{Detector Solenoid} & Power Supply & & & & 250 & 10 & 357 & & & \\
    & Cold Box & & & & 50 & 10 & 71 & & & \\\hline
    \multirow{9}{1.4cm}{Subdetectors} & Muon & 12 & 5 & 34 & & & & & \\
    & HCAL & 45.5 & 5 & 130 & & & & & & \\
    & ECAL & 40 & 5 & 114 & & & & & & \\
    & VFS & 9 & 5 & 26 & & & & & & \\
    & SET & 9 & 5 & 26 & & & & & & \\
    & TPC & 3 & 5 & 5 & & & & 13 & 5 & 38\\
    & SIT & 8 & 5 & 23 & & & & & & \\
    & FTD & 8 & 5 & 23 & & & & & & \\
    & VTX & 13.5 & 5 & 39 & & & & & & \\\hline
    \multicolumn{2}{m{2cm}|}{Pump} & 11 & 5 & 31 & 11 & 10 & 16 & 3.7 & 5 & 11\\ \hline
    \multicolumn{2}{m{2cm}|}{AC Transformer} & 49 & 5 & 140 & & & & & & \\ \hline
    \multicolumn{2}{m{2cm}|}{{\bf Total}} & 208 & & 595 & 611 & & 873 & 17 & & 48 \\ \hline
    \multicolumn{2}{l|}{{\bf Total Chilled Water}} & \multicolumn{3}{r|}{{\bf 595}} & \multicolumn{6}{r}{}\\ \hline
    \multicolumn{2}{l|}{{\bf Total Normal Temp. Water}} & \multicolumn{3}{r|}{} & \multicolumn{6}{r}{{\bf 921}}\\ \hline

    \end{tabular}
    \caption{Breakdown of the cooling water requirement estimates for ILD and appended ILC components~\cite{ild:bib:services}.}
    \label{tab:integration:cooling}
\end{table}

\vspace{2cm}
\FloatBarrier

\subsection{Access and Assembly}
\label{ild:sec:access}
The Kitakami site of the ILC is located in a mountainous and rural area, only accessible by road~(c.f.~Figure~\ref{ild:fig:ilc_site}). The nearest port, Kesennuma, is about 40~km away on the Pacific coast. All ILD parts need to be shipped via the central access roads into the mountains. Typically, street transport in Japan is limited to trucks with up to 25~t of mass (including the weight of the truck itself). The total mass of ILD is about 15,500~t, posing a specific challenge for the logistics infrastructure. Heavy-load transports of up to 70~t loads are possible in exceptional cases. For the transport of the ILD solenoid modules, this option is under study~(c.f.~Figure~\ref{ILD:fig:magnet_transport}).

\begin{figure}[h!]
\centering
\includegraphics[width=0.8\hsize]{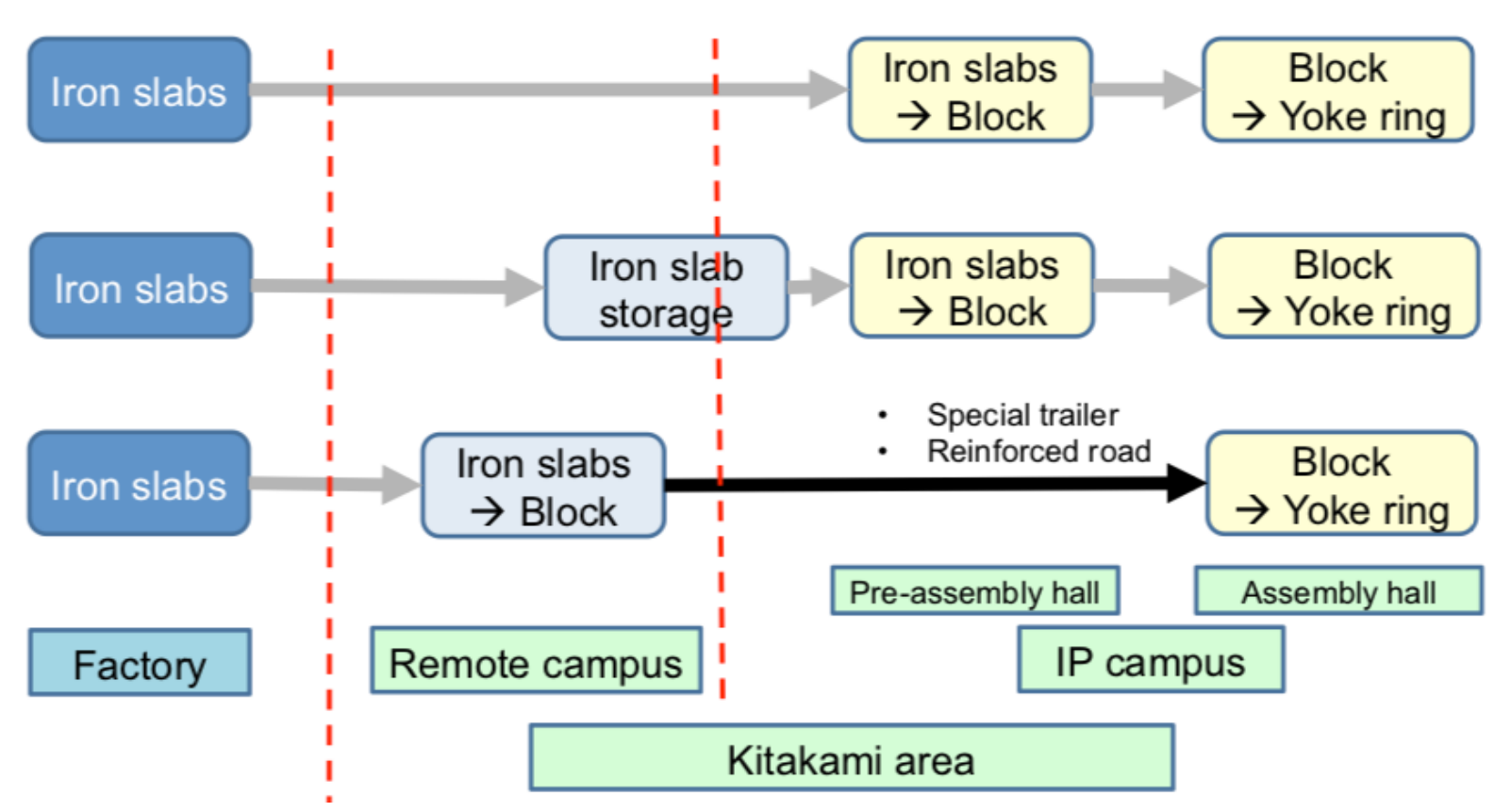}
\caption{\label{fig:integration:yoke_assembly}Options under study for the (pre)assembly of the ILD iron return yoke~\cite{ild:bib:ejade_mdi}.}
\end{figure}

The heaviest parts of ILD are the yoke rings. They need to be assembled on-site from pre-fabricated iron blocks. Figure~\ref{fig:integration:yoke_assembly} shows three options that are under study for the procedures. If a remote campus close to the interaction region could be made available, with a dedicated reinforced street in-between, then the pre-fabricated iron slabs from the factory could be assembled into blocks of up to 200~t there. The blocks could then be transported via the dedicated road to the interaction region for assembly of the yoke rings. Another option would be the construction of a pre-assembly hall at the interaction region. The construction of the iron block could then take place there, in the vicinity of the detector assembly hall, resulting in a reduced transportation problem. A remote area would still be useful as storage space. 


Figure~\ref{fig:integration:surface} shows a conceptual layout of the surface installations above the interaction region. The central detector assembly hall serves both detectors, SiD and ILD, and is located above the central access shaft to the underground areas (compare Figure~\ref{fig:integration:underground}). A pre-assembly building for SiD and ILD is located right next to it. Heavy and large parts, like the ILD yoke blocks, could be assembled in the pre-assembly area before they are moved to the central assembly hall for the installation of the yoke rings.

\begin{figure}[!ht]
\centering
\includegraphics[width=0.8\hsize]{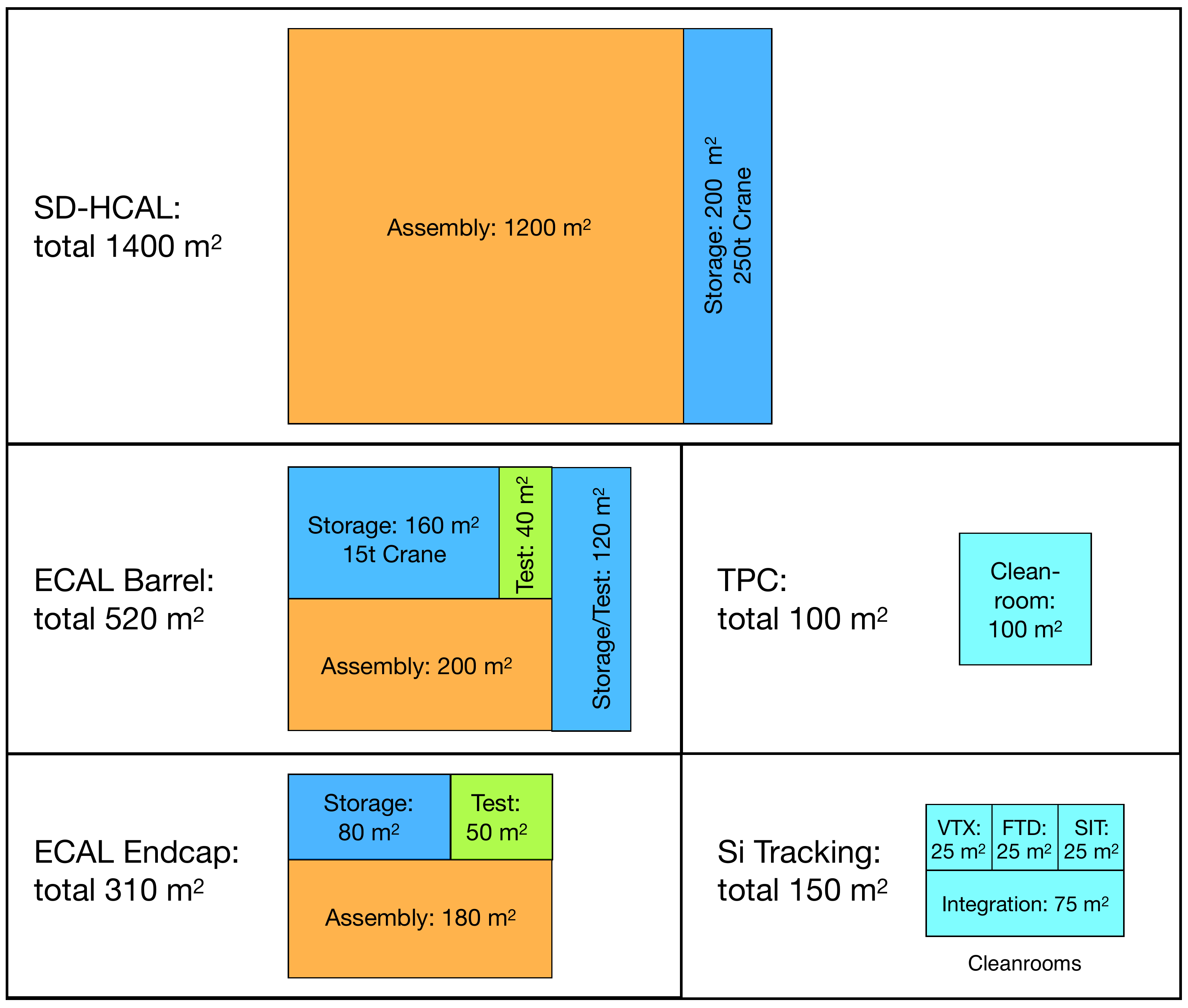}
\caption{\label{fig:integration:assembly_space}Result of a subdetector survey on assembly space requirements~\cite{ild:bib:ejade_mdi,ild:bib:assembly}.}
\end{figure}

A survey has been done within ILD to collect the requirements of the sub-detectors on assembly, storage and on-site testing space. Figure~\ref{fig:integration:assembly_space} shows the results from the trackers and the SiECAL/SDHCAL calorimeters (barrel and endcap). Though the information is at a very conceptual level, it serves as valuable input for the planners of the interaction region surface installations that need to be accommodated in the Kitakami mountains. 

The distribution of work between the main campus, a possible remote campus, and the IP campus depends on the exact availability and possible locations and needs to be worked out in detail, once the boundary conditions are better known.

It should be noted that the assembly of the detector takes only a limited amount of time, and that some of the foreseen buildings, such as the detector preparation building, could be made from temporary structures so that part of the occupied area could be re-cultivated again.

\begin{figure}[!ht]
\centering
\includegraphics[width=0.8\hsize]{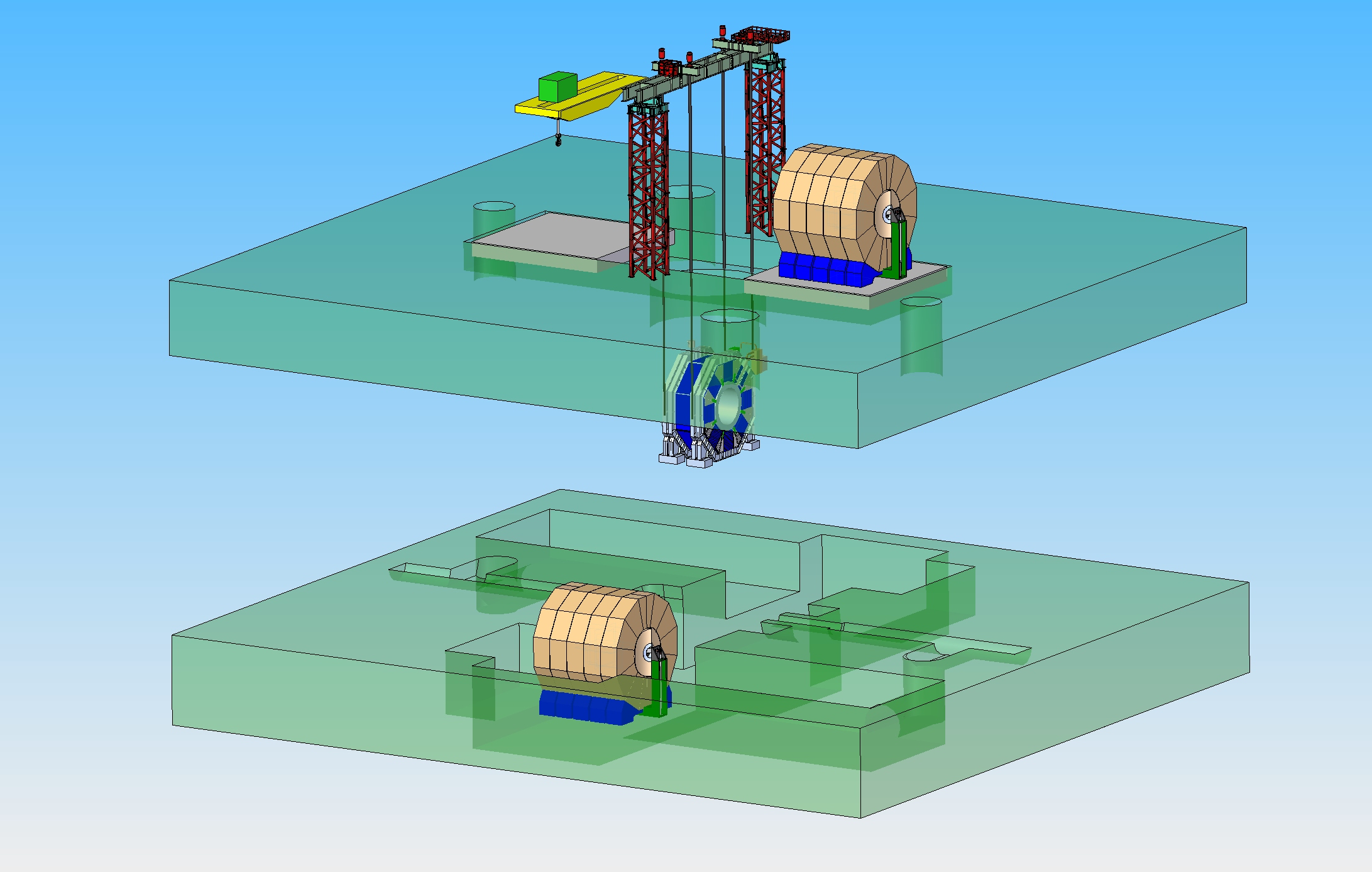}
\caption{\label{fig:integration:gantry_crane}Lowering of SiD and ILD main parts via the central shaft into the underground detector hall with the help of a heavy load gantry crane~\cite{ild:bib:gantry_crane}.}
\end{figure}

The final assembly of ILD and SiD should happen above ground in the central detector assembly hall. The major parts, the SiD barrel detector and the ILD yoke rings, have masses of up to 4500~t and will be lowered through the central access shaft into the underground areas with the help of a gantry crane. A similar procedure has been followed for the CMS detector at CERN and was proven to be very efficient. Figure~\ref{fig:integration:gantry_crane} shows an artists view of the SiD barrel detector hanging from a gantry crane.

\begin{figure}[h!]
\centering
\includegraphics[width=0.8\hsize]{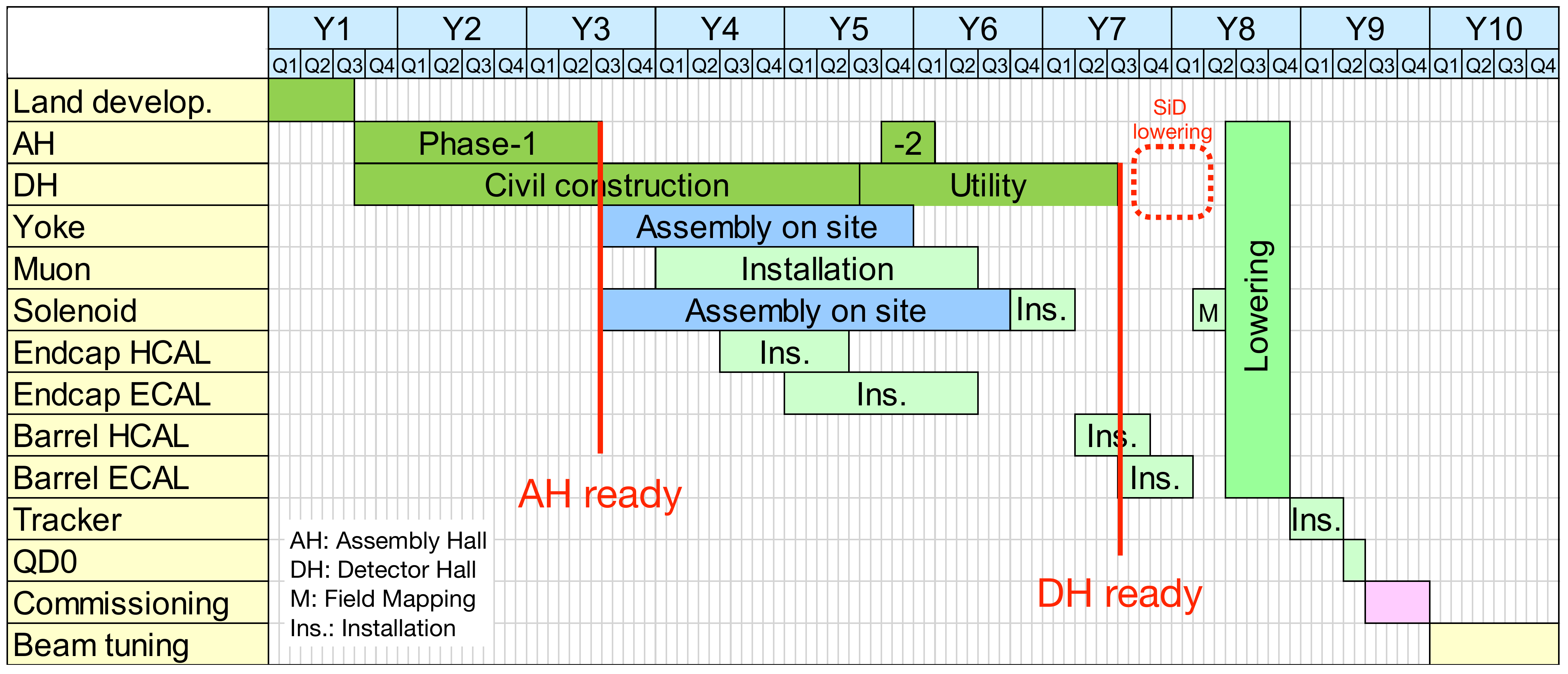}
\caption{\label{fig:integration:assembly_timeline}Tentative timeline for the detector construction and assembly~\cite{ild:bib:ejade_mdi, ild:bib:assembly}}
\end{figure}

The main advantage of the CMS-style surface assembly of the detectors lies in the decoupled time lines for the construction of the machine and the detectors. In this case, the on-site assembly can start as soon as the surface buildings are ready while the underground excavations and constructions can carry on. Figure~\ref{fig:integration:assembly_timeline} shows a rough schedule for the detector construction.
On-site assembly can start 2.5~years after the ground breaking when the surface buildings are ready. The lowering of the large detector parts only takes place about one year before the commissioning of the machine would start. The total technical construction and assembly time is about 9 years.
\section{Internal ILD integration}

The ILD detector concept underwent a dedicated effort to come to an integrated design that from subdetector level up to the major structures gives a conceptual engineering view of the full system. This includes realistic mechanical designs with tolerances, deformations and dead material zones as well as an integrated design of all detector services: electrical, cooling, gases, etc.

\subsection{ILD Mechanical Structure}

\begin{figure}[h!]
    \centering
    \includegraphics[width=0.8\hsize]{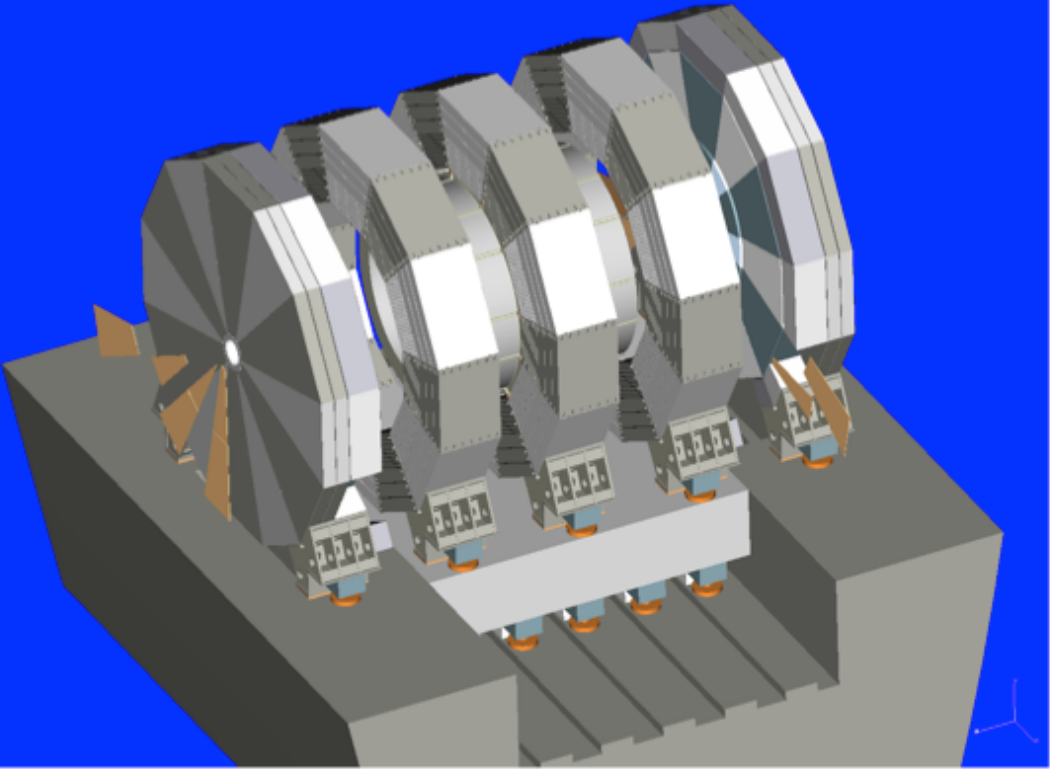}
    \caption{Mechanical structure of the ILD detector~\cite{ild:bib:ilddbd}.}
    \label{ILD:fig:mechanical_model}
\end{figure}

The mechanical structure of ILD has not changed since the DBD~\cite{ild:bib:ilddbd}. The main parts are the three barrel yoke rings and the two endcaps. The central barrel ring carries the solenoid coil in its cryostat; all barrel detectors are suspended from the cryostat. The two other central yoke rings can move over the cryostat. The endcaps can be opened as well. All rings can be moved using airpad systems. Figure~\ref{ILD:fig:mechanical_model} shows the mechanical model of the ILD yoke structure.

\subsection{ILD Services and Utilities}
\label{ild:sec:services}
Each component of ILD has requirements on services and utilities that are needed for operations and maintenance. This typically includes power and data lines, gas and cooling systems, guidances for laser beams, etc. All major support systems for those services, e.g. power supplies, cooling plants, lasers, DAQ computers, or gas systems are located outside of the detector, sometimes even far away (c.f.~section~\ref{ild:sec:service_locations}). General paths have been defined in the global detector structure where space is allocated for those services. The routing of those paths has to be designed to minimise the amount of gaps and dead material in the active detector areas, while at the same time provide enough space for the foreseen utilities. Three main pathways have been defined within ILD:
\begin{enumerate}
    \item The services of all barrel detectors are collected at both front-faces of the barrel, go around the solenoid cryostat and leave the detector through the gap between the central yoke ring and the neighbouring rings.
    \item The services of the endcap detectors (ECAL, HCAL, Muon) leave the detector along the endcap yoke ring.
    \item The services for the forward calorimeter systems (FCAL, ECAL ring) pass parallel to the beamline, outside of the QD0 magnet.
\end{enumerate}

\begin{figure}[h!]
    \centering
    \includegraphics[width=1.0\hsize]{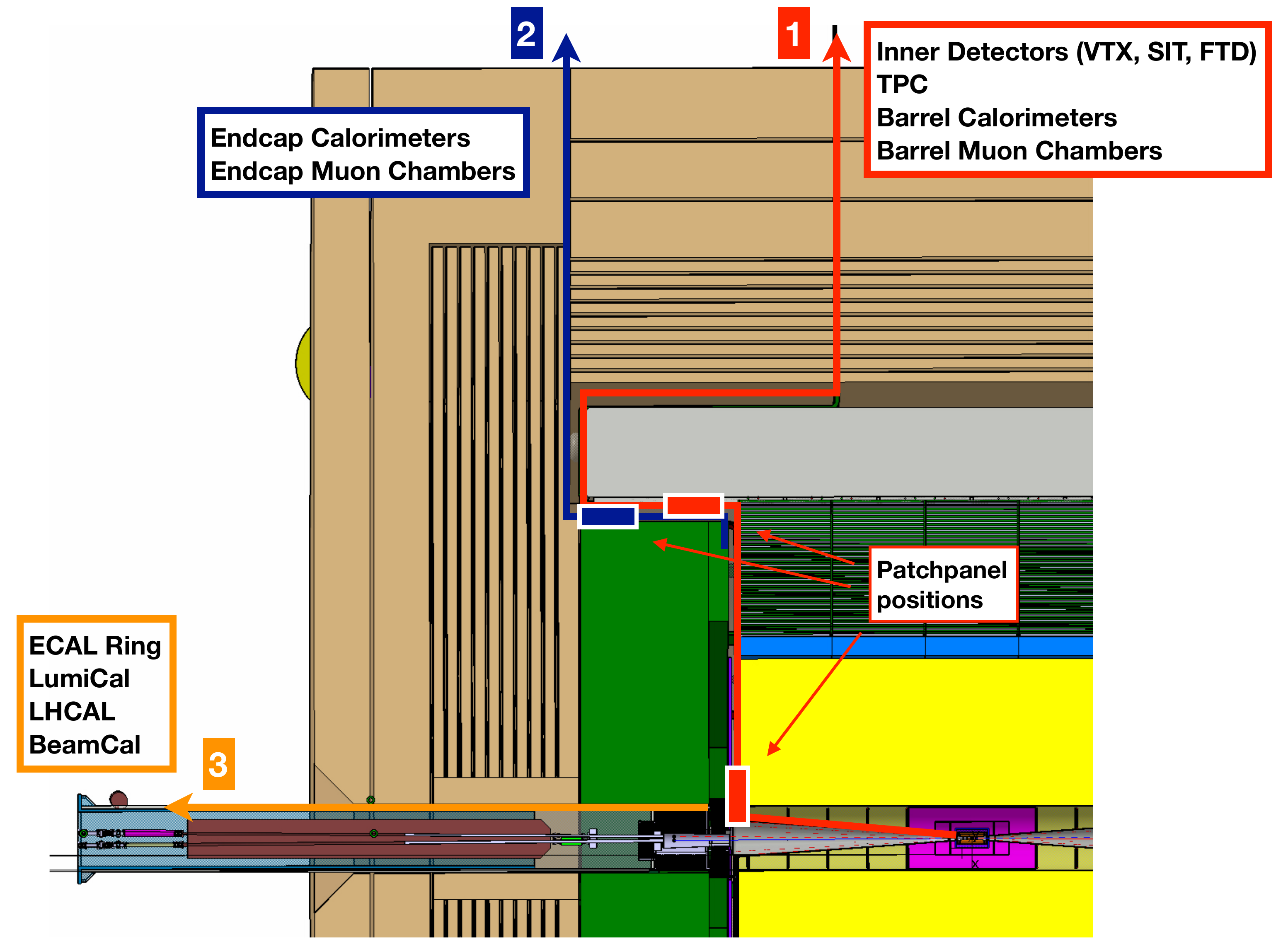}
    \caption{Service paths in the ILD detector and suggested positions for patch panels.}
    \label{ILD:fig:cable_paths}
\end{figure}

This scheme allows for the opening of the yoke endcaps as well as for moving the barrel yoke rings independently from each other. The front-end electronic systems of the subdetectors can often drive only a limited cable length. Therefore, space for additional patch panels, drivers, data concentrators needs to be provided inside the ILD detector. While the exact requirements for those are not known in each case, conceptual locations have been defined. Figure~\ref{ILD:fig:cable_paths} shows the general service paths and proposed locations for the patch panels in ILD.

\subsection{Inner Detector Integration}

\begin{figure}[h!]
    \centering
        \includegraphics[width=0.8\hsize]{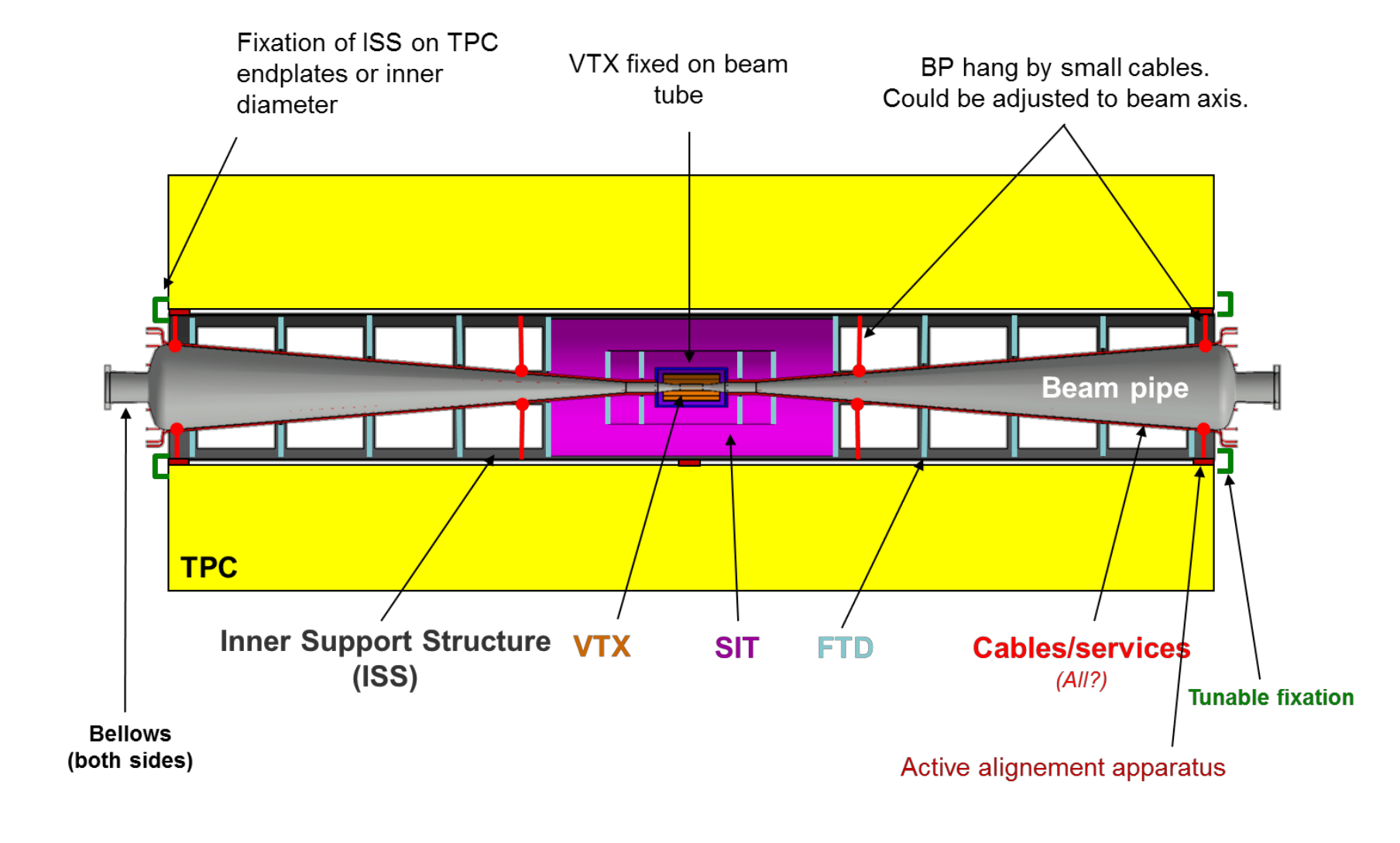}
    \caption{Schematic of the inner tracking detector system~\cite{ild:bib:inner_detector_integration}.}
    \label{ILD:fig:inner_detector_schematic}
\end{figure}

At the heart of ILD, directly at the interaction point, is the inner detector that comprises the beam pipe as well as the vertex detector and the inner silicon tracking devices, SIT and FTD~(c.f.~Figure~\ref{ILD:fig:inner_detector_schematic}).

\subsubsection{Mechanical Integration}
\begin{figure}[h!]
    \centering
        \includegraphics[width=0.8\hsize]{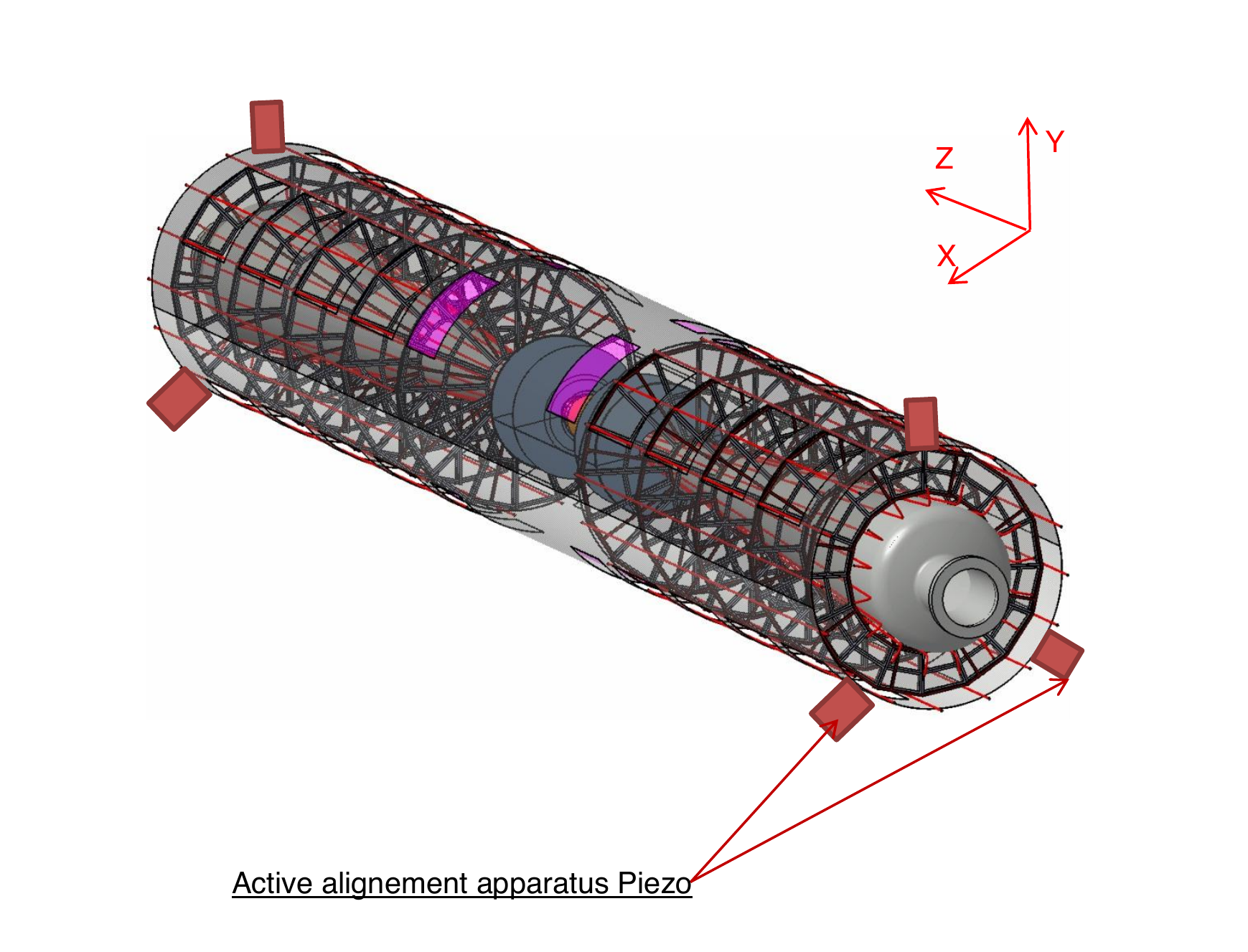}
    \caption{Engineering design of the inner detector~\cite{ild:bib:inner_detector_integration}.}
    \label{ILD:fig:inner_detector_integration}
\end{figure}

The vertex detector is suspended from the beam pipe that itself is carried together with the Forward Tracking Disks and the Si Intermediate Tracker from the Inner Detector Support Structure (ISS). The ISS is a support tube made out of carbon-fibre reinforced plastic and is suspended from the end flanges of the TPC. A piezo-based active alignment system (see Figure~\ref{ILD:fig:inner_detector_integration}) allows for the positioning of the ISS with a precision better than 0.01~mm~\cite{ild:bib:inner_detector_integration}, independently of the main ILD detector structure. This is required to adjust the beam pipe and the inner tracking devices with respect to the beam axis, to better precision than what can be achieved with the complete ILD detector, e.g.~after push-pull operations.

\subsubsection{Electrical Services and Cooling}
\begin{figure}[h!]
    \centering
        \includegraphics[width=0.8\hsize]{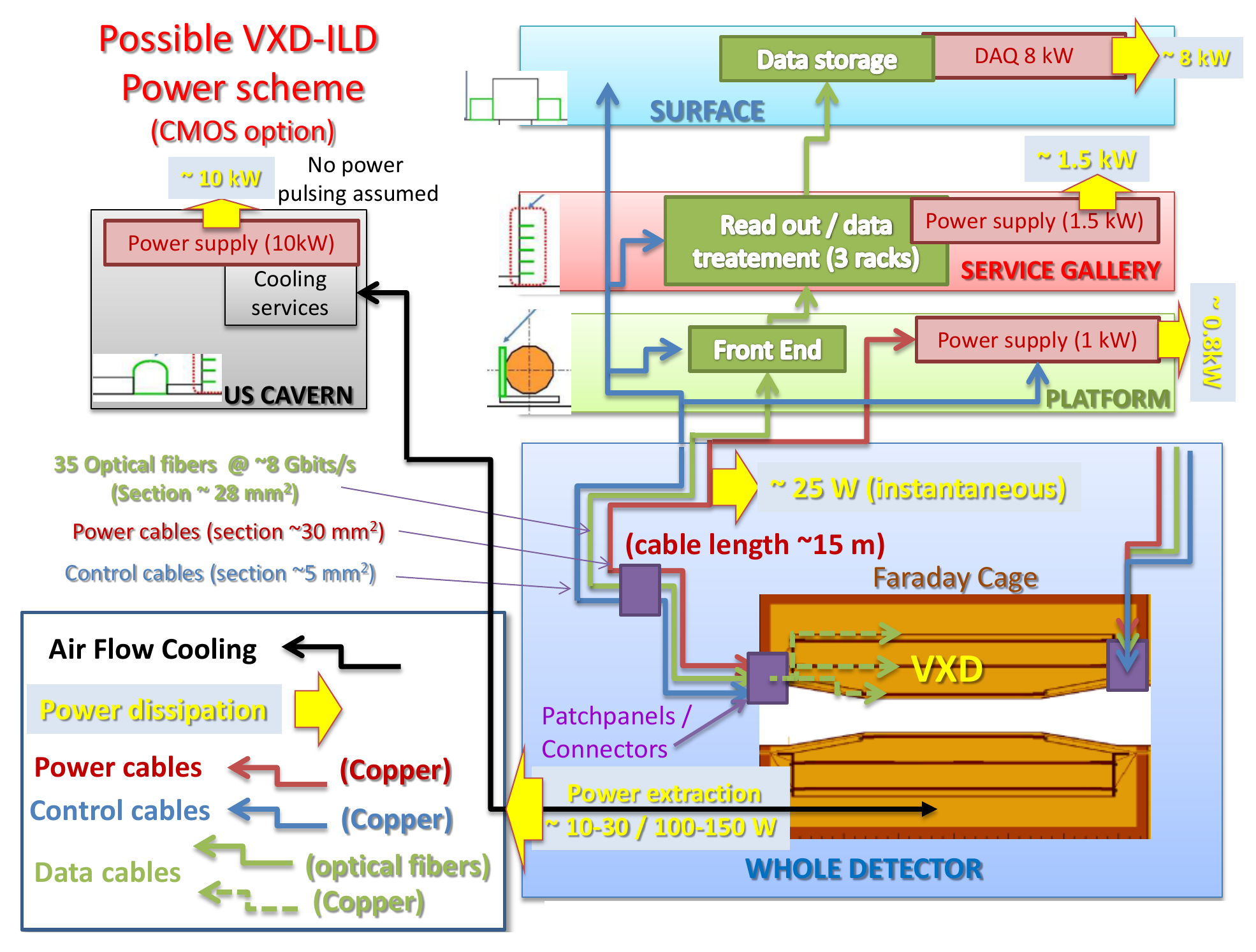}
    \caption{Diagram of a power scheme for the vertex detector (CMOS option)~\cite{ild:bib:VTX_integration}.}
    \label{ILD:fig:vtx_services}
\end{figure}

\begin{figure}[h!]
    \centering
        \includegraphics[width=0.8\hsize]{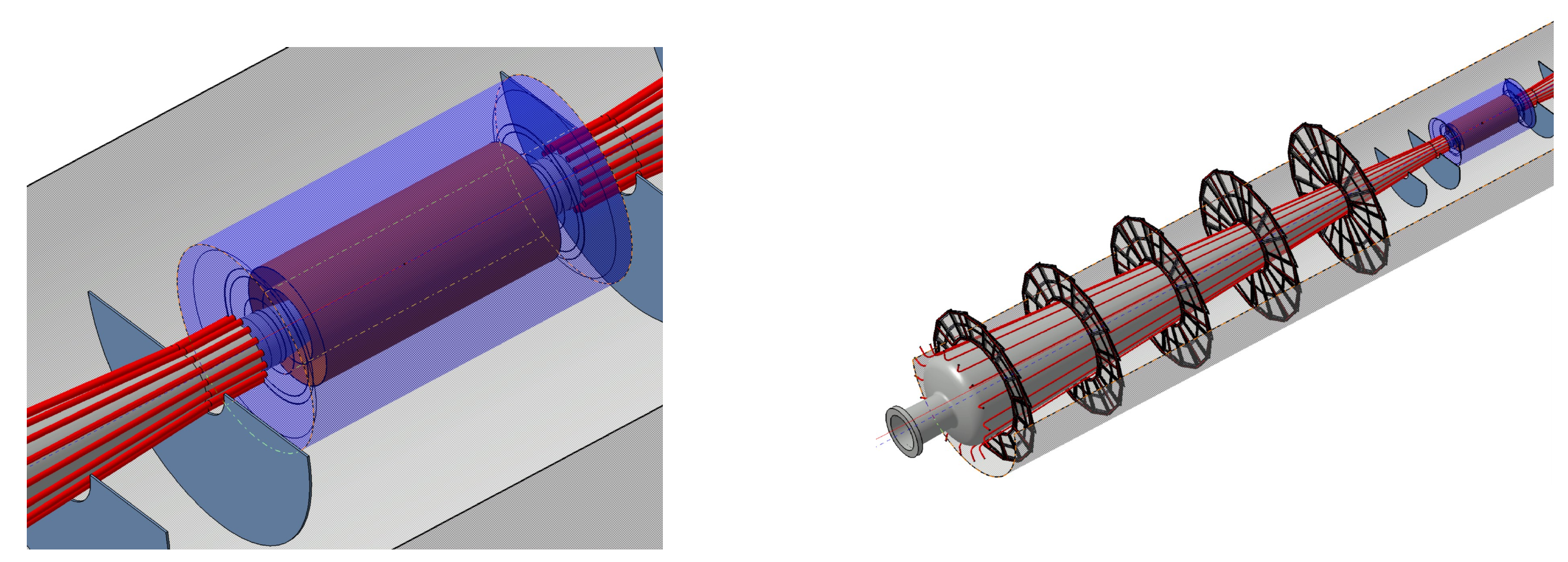}
    \caption{Cable placeholders for the inner SI detectors (VTX, SIT, FTD)~\cite{ild:bib:SI_integration}.}
    \label{ILD:fig:si_services}
\end{figure}

A concept has been developed for the power scheme of the vertex detector (CMOS version), see Figure~\ref{ILD:fig:vtx_services}. Copper based power and control cables as well as optical fibres for the data readout connect the vertex detector with patch panels at either ends of the ISS. From here, the cables are routed as described in section~\ref{ild:sec:services} to the outside of the detector. An engineering design for the details of the cabling and patch panels inside the ISS is still pending. Figure~\ref{ILD:fig:si_services} shows the place holders for the cables in the current model. The vertex detector is planned to be cooled using air flow, where the cooling pipes also need to follow the general services paths.

\subsection{TPC Integration}

\subsubsection{Mechanical integration}

The mechanical integration of the TPC is still under study, with two possible concepts being followed up: the TPC will be suspended either directly from the solenoid cryostat with the help of carbon ribbons or support struts, or from the absorber structure of the hadronic calorimeter. In the first case, the TPC would be decoupled from the mechanical properties of the calorimeters, at the price of having larger lever arms that might amplify vibrations. A longitudinal damping system would probably be required. In the second case, the lever arms would be much shorter, but the dynamic behaviour of the full system of the cryostat, hadronic and electomagnetic calorimeter as well as the TPC itself needs to be understood. 

\subsubsection{Electrical Services and Cooling}

\begin{figure}[h!]
    \centering
    \includegraphics[width=0.6\hsize]{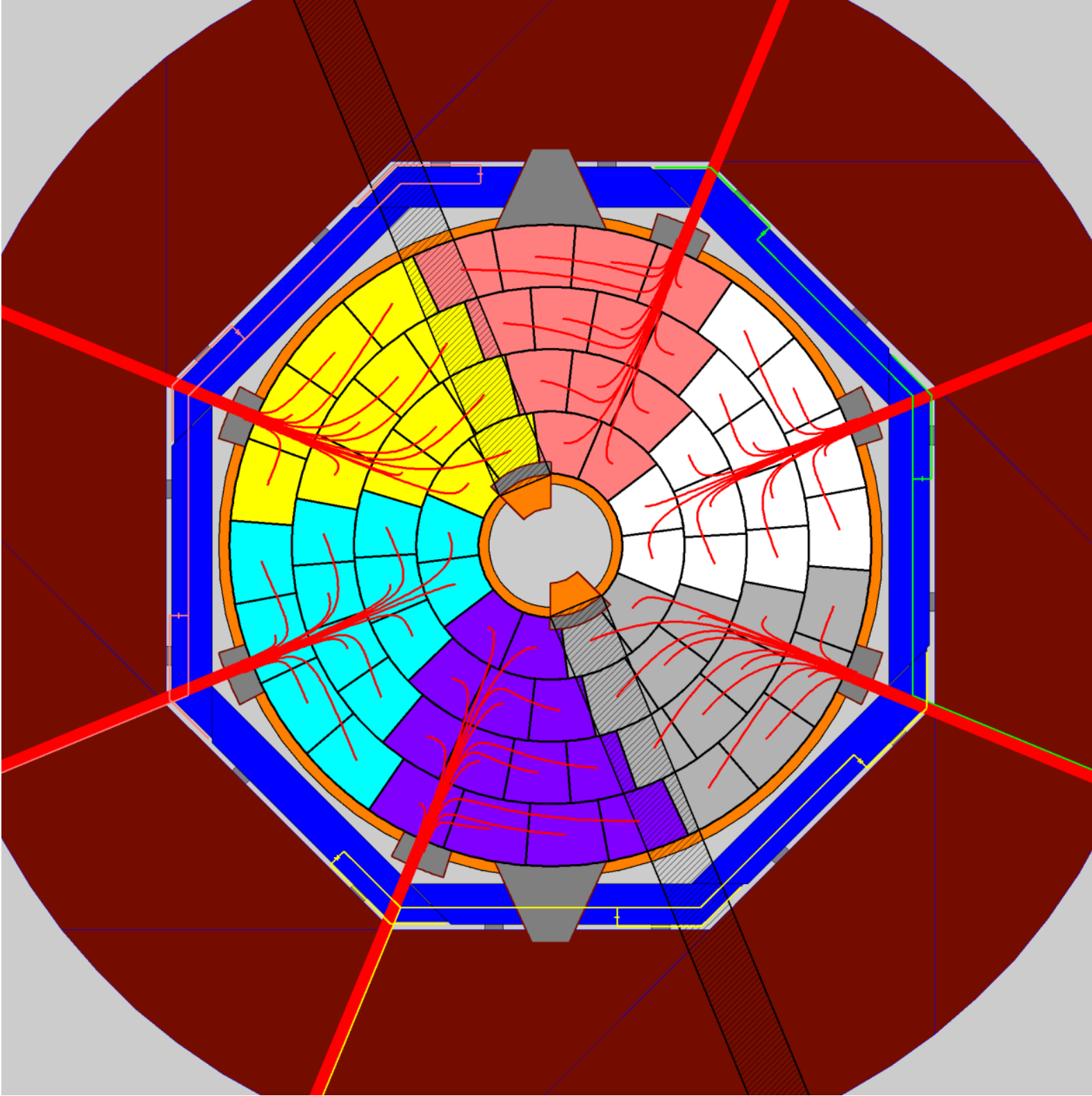}
    \caption{Sketch of the cable paths on the front-end of the TPC~\cite{ild:bib:TPC_ICD}.}
    \label{ILD:fig:tpc_cables}
\end{figure}

\begin{figure}[h!]
    \centering
    \includegraphics[width=1.0\hsize]{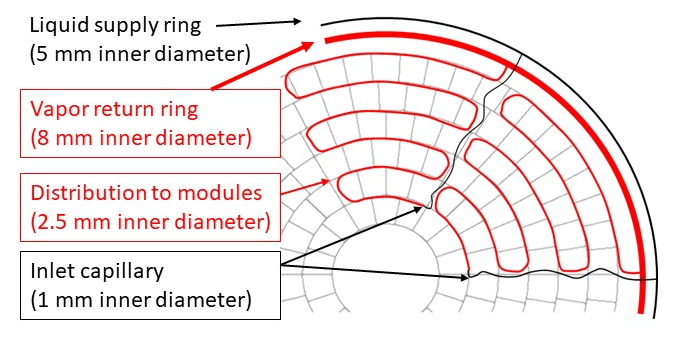}
    \caption{Sketch of a TPC cooling system with tube routing on the TPC end plate. Figure courtesy of Bart Verlaat, Nikhef.}
    \label{ILD:fig:tpc_cooling}
\end{figure}

\begin{figure}[h!]
    \centering
    \includegraphics[width=1.0\hsize]{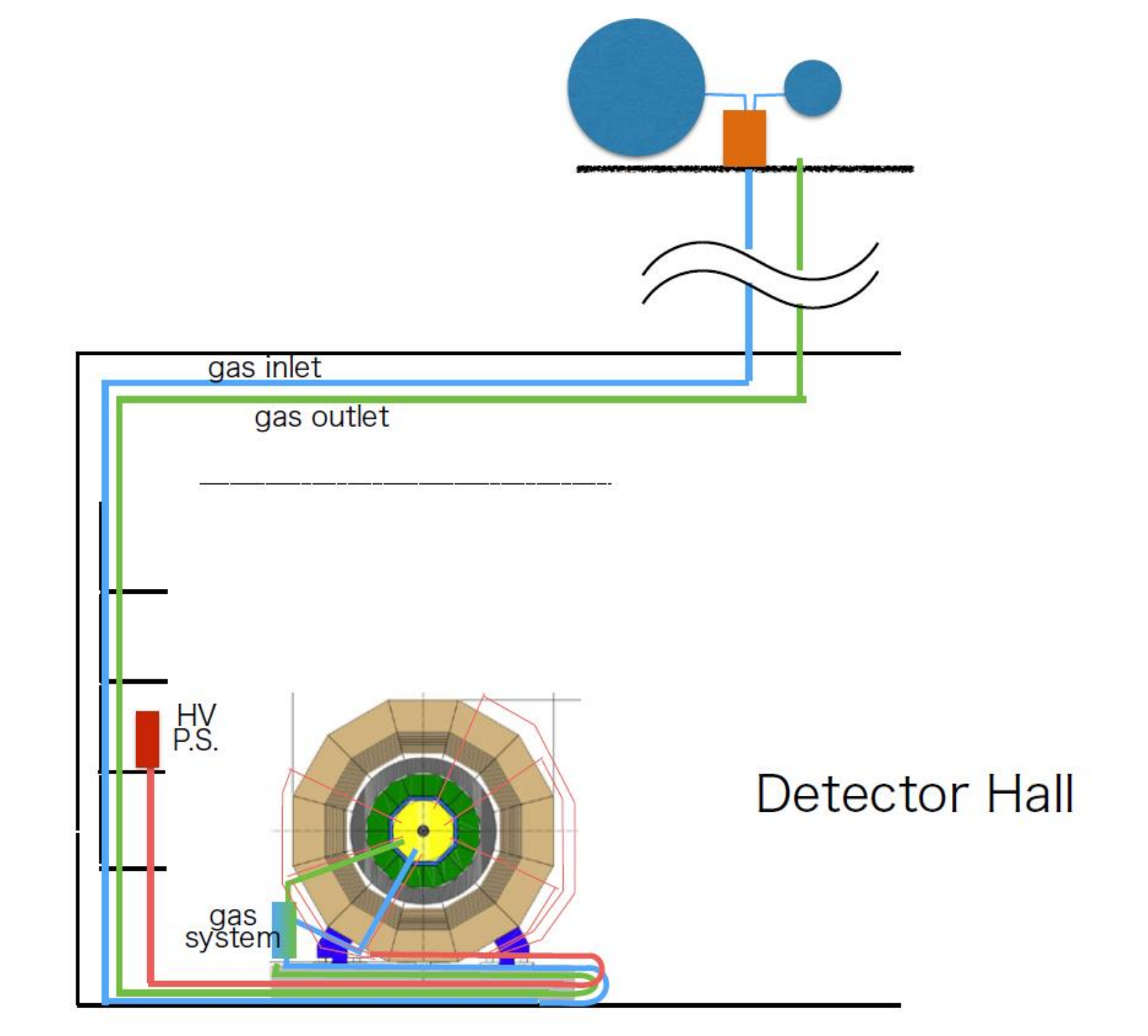}
    \caption{Gas and HV interfaces of the TPC~\cite{ild:bib:TPC_ICD}.}
    \label{ILD:fig:tpc_interfaces}
\end{figure}

The electrical services and the cooling pipes of the TPC start on both end plates and will be routed through gaps in the front-faces of the calorimeters, between the end-cap and barrel detectors (c.f.~Figure~\ref{ILD:fig:tpc_cables}). A cooling system based on 2-phase CO2 has been tested in 2014 and 2018 on a system of
7 Micromegas modules. Figure~\ref{ILD:fig:tpc_cooling} shows a solution with a 6-loop geometry. The external supplies of the TPC need to be accommodated in the detector environment: while a gas mixing and supply system will most probably be placed on the surface area, distribution sub-systems need to be closer to the detector, e.g.~on the detector platform. The high-voltage power supplies will be placed in the detector hall at reasonable cable-length distances. Figure~\ref{ILD:fig:tpc_interfaces} shows a schematic drawing of the TPC connections to the outer world.


\subsection{Electromagnetic Calorimeters Integration}

\subsubsection{Mechanical Integration}

\begin{figure}[h]
    \centering
        \includegraphics[width=0.5\hsize]{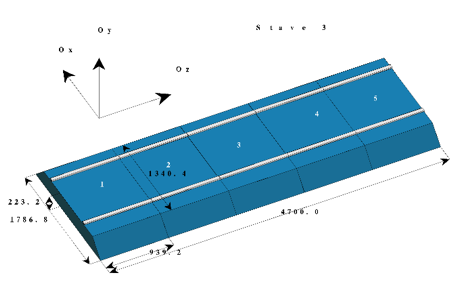}
        \includegraphics[width=0.3\hsize]{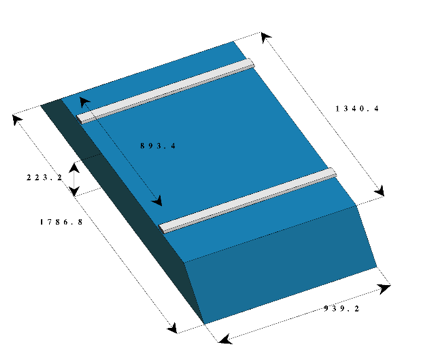}
    \caption{ECAL stave and module~\cite{ild:bib:SiECAL_ICD}}
    \label{ILD:fig:ECAL_Mechanics}
\end{figure}

The two options under study for the ILD electromagnetic calorimeters, SiECAL and ScECAL, share the same mechanical design as shown in Figure~\ref{fig:det:ECAL}. The ECAL barrel consists of eight staves that are built from five modules each~(c.f.~figure~\ref{ILD:fig:ECAL_Mechanics}). The staves are supported from the HCAL barrel sections. The ECAL endcaps are supported from the HCAL endcap detector.

\subsubsection{SiECAL Electrical Services and Cooling}

\begin{figure}[h!]
    \centering
        \includegraphics[width=0.6\hsize]{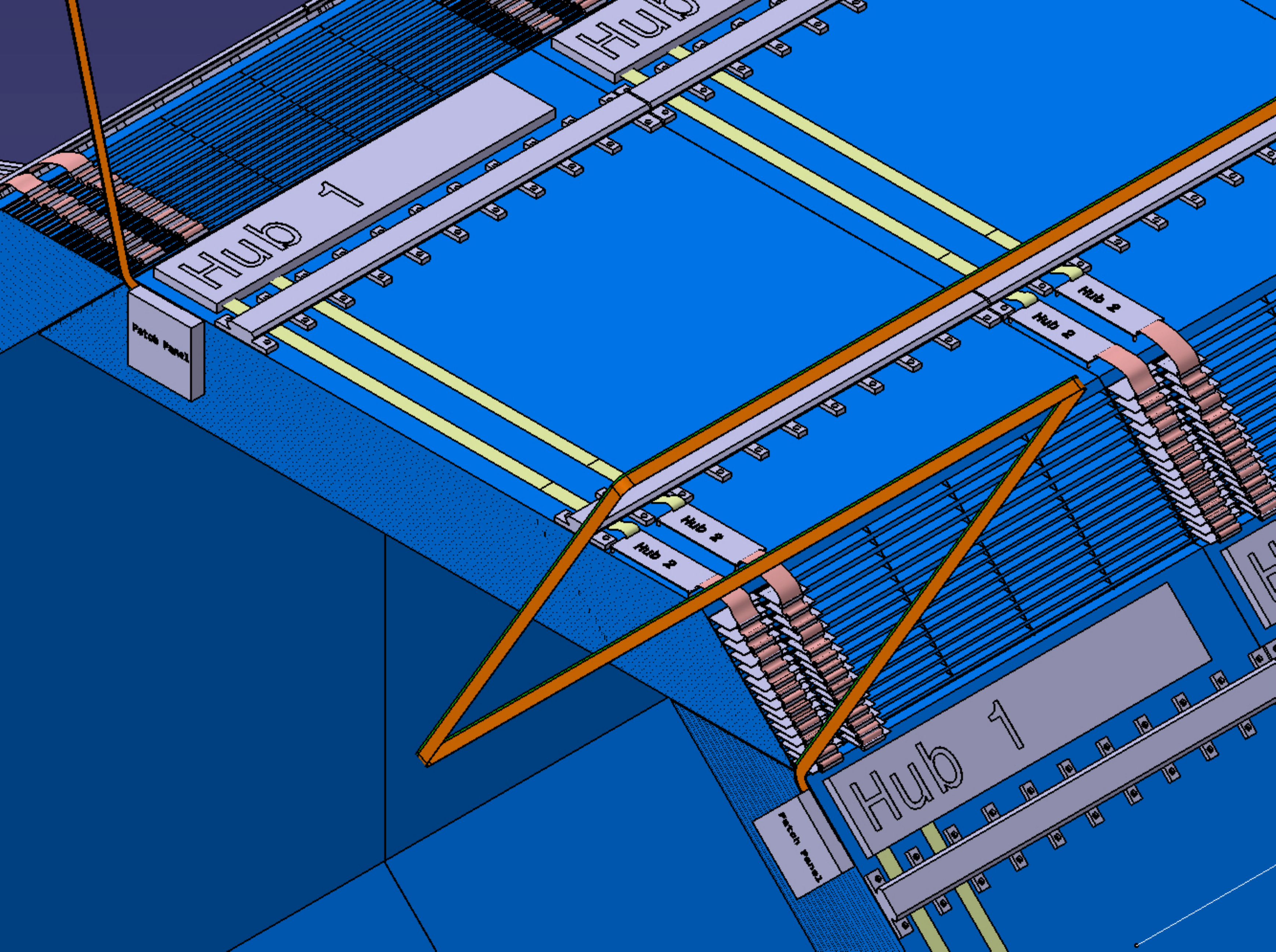}
    \caption{Schematic view onto SiECal Barrel Modules indicating services such as the hubs ("Hub 1", "Hub 2") for power distribution, data concentration and distribution of readout commands. The figure indicates the cables (orange) that arrive at the patch panels and the two type of hubs including their interconnection. The principle shown for one barrel module of a stave repeats for all barrel modules. For simplicity the connections between the patch panels and the Hub 1 is not shown~\cite{ild:bib:SiECAL_ICD}.}
    \label{ILD:fig:siecal_services}
\end{figure}

\begin{figure}[h!]
    \centering
        \includegraphics[width=0.8\hsize]{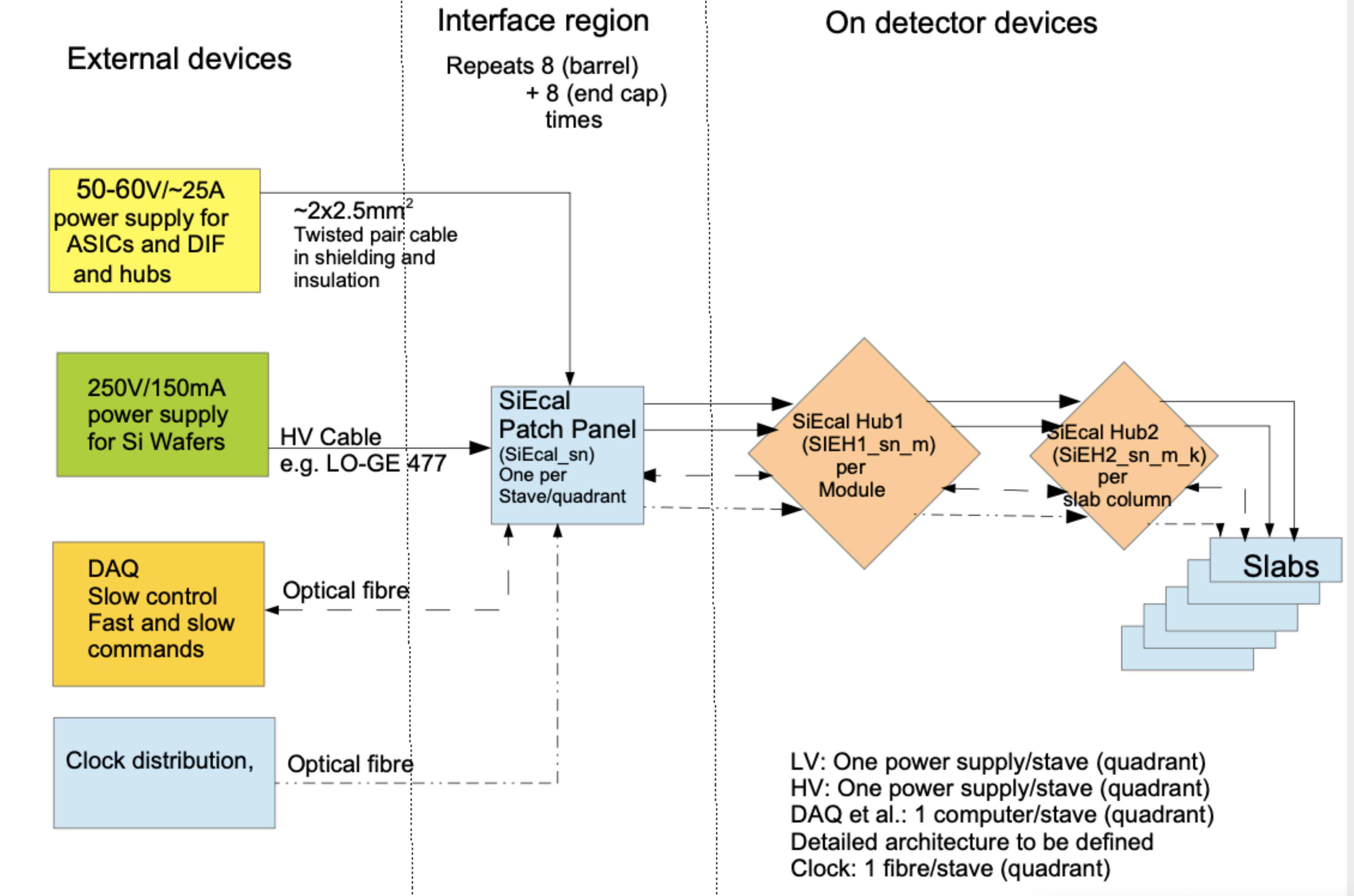}
    \caption{Block diagram of the electrical services for the SiECAL~\cite{ild:bib:SiECAL_ICD}.}
    \label{ILD:fig:siecal_block_diagram}
\end{figure}

A detailed integrated design of the SiECAL electrical services and of the cooling system has been developed. Figure~\ref{ILD:fig:siecal_services} shows a close-up view of the upper side of a SiECAL barrel stave. The front-end electronics is located at the end of the readout slabs. 
Service cables for power supply and readout arrive at a patch panel placed on the front-side of a stave from where they are distributed to big hubs on top of each module ("Hub 1" in Fig.~\ref{ILD:fig:siecal_services}). These hubs distribute the power and readout cables further to smaller hubs from where they are finally distributed to the readout slabs. Figure~\ref{ILD:fig:siecal_block_diagram} shows a block diagram of the electrical services for the SiECAL and the connections from "Hub 1" via a small number of patch panels to the outside power supplies, to the common clock and the DAQ. 

\begin{figure}[h!]
    \centering
        \includegraphics[width=0.8\hsize]{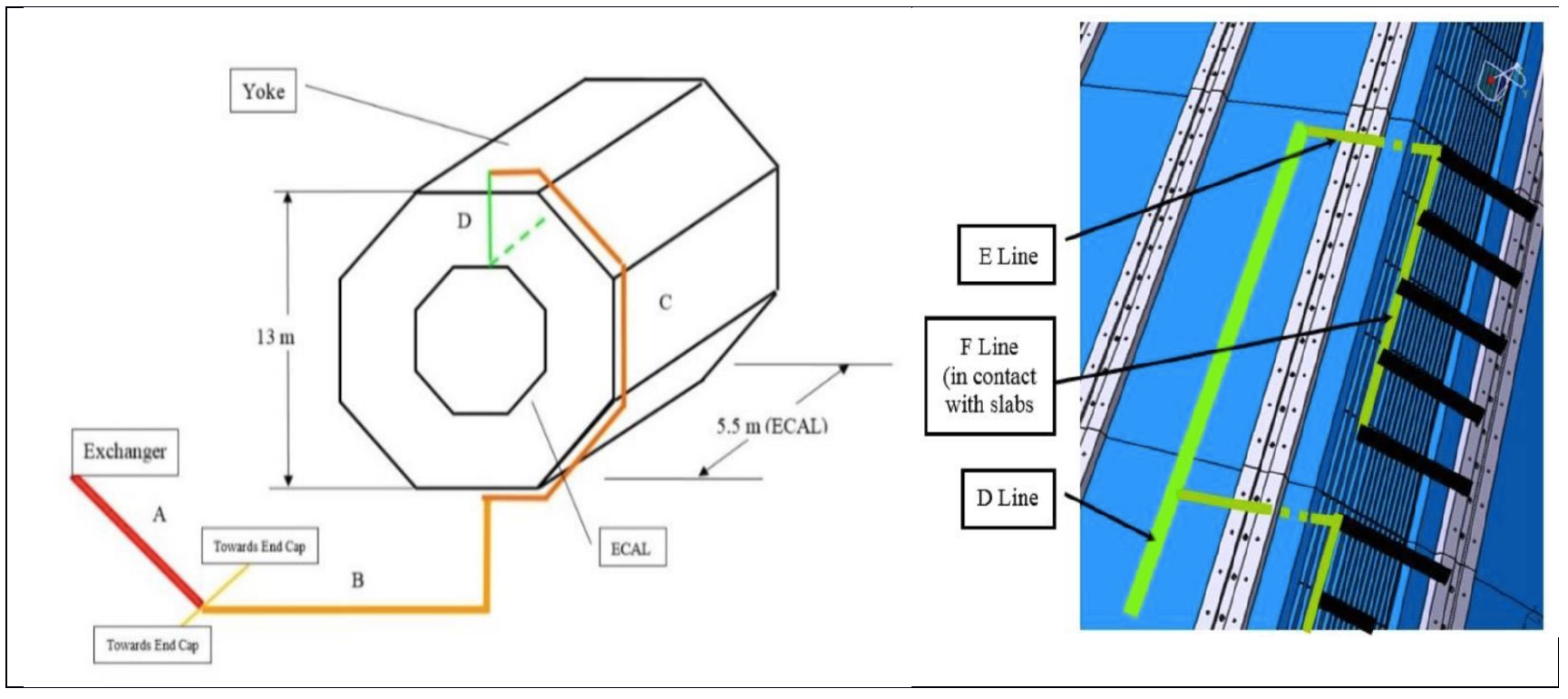}
    \caption{Conceptual design of the SiECAL cooling system~\cite{ild:bib:SiECAL_ICD}.}
    \label{ILD:fig:siecal_cooling}
\end{figure}

A conceptual design for the SiECAL cooling system is shown in Figure~\ref{ILD:fig:siecal_cooling}. The system foresees leakless water cooling, where a cooling station would be located outside of the ILD detector in the underground area. The water will be distributed via a hierarchical system of cooling lines ("A" to "F") to the barrel and endcap detectors.

\subsubsection{ScECAL Electrical Services}
\begin{figure}[h!]
    \centering
        \includegraphics[width=0.8\hsize]{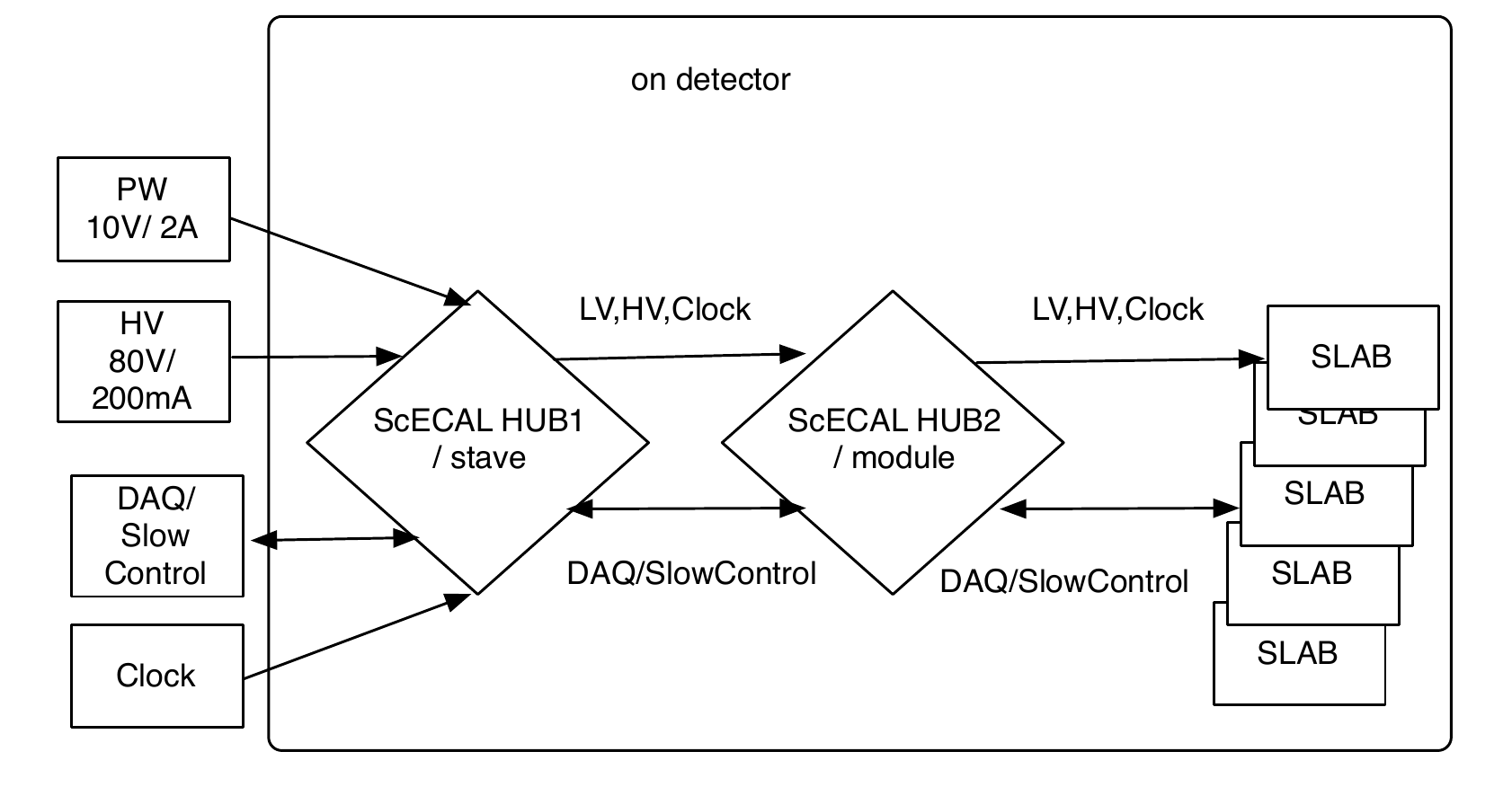}
    \caption{Block diagram of the electrical services for the ScECAL~\cite{ild:bib:ScECAL_ICD}.}
    \label{ILD:fig:scecal_block_diagram}
\end{figure}

The distribution of the electrical services for the ScECAL is very similar to the SiECAL case. Figure~\ref{ILD:fig:scecal_block_diagram} show a block diagram of the service distributions. Also in this case, two series of hubs are planned, one for each stave and one for each column of slabs, to distribute HV and signals accordingly.

\subsection{Hadronic Calorimeters Integration}
\subsubsection{Mechanical Integration}

Two absorber structures are under study for ILD, the so-called "TESLA" and "Videau" structures~(c.f.~Figure~\ref{fig:det:HCAL}). The main differences from the viewpoint of the detector integration are the mechanical behaviour, further discussed in section~\ref{ild:sec:mechanical_structures}, and the layout of the detector electrical and cooling services. The analogue AHCAL and the semi-digital SDHCAL can be adapted to both mechanical structures. In practice, the AHCAL has been designed with the "TESLA" structure in mind, while the SDHCAL is optimised for the "Videau" case.

\subsubsection{AHCAL Electrical Services and Cooling}
\begin{figure}[h!]
    \centering
        \includegraphics[width=0.8\hsize]{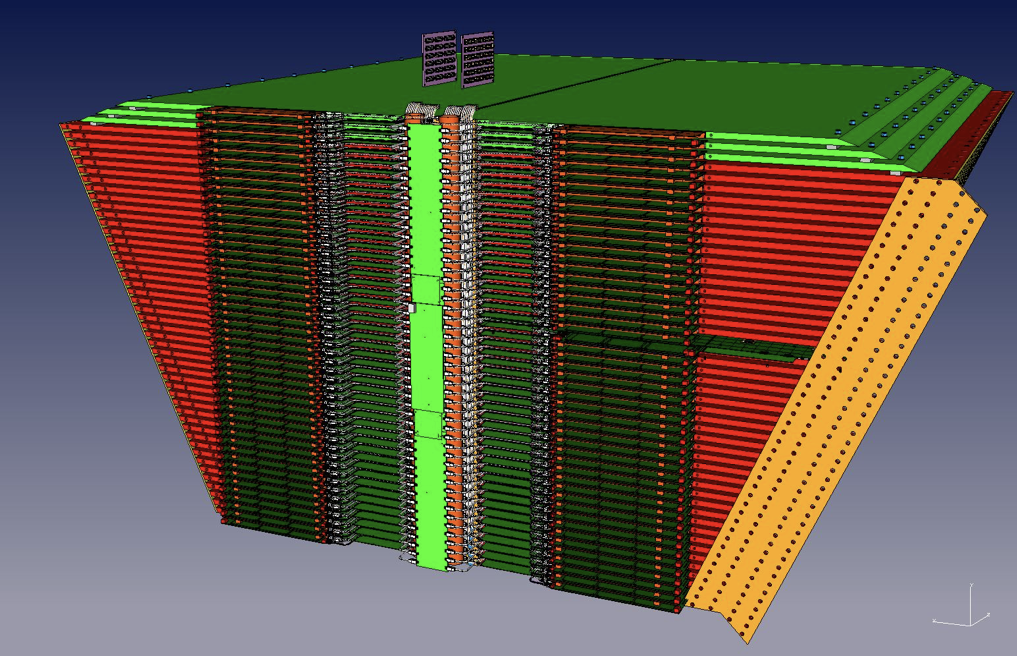}
    \caption{Engineering design of an AHCAL module with electrical and cooling services. Only one instrumented layer is shown for the wedge shape outer parts of the module.}
    \label{ILD:fig:ahcal_module_services}
\end{figure}

\begin{figure}[h!]
    \centering
        \includegraphics[width=0.8\hsize]{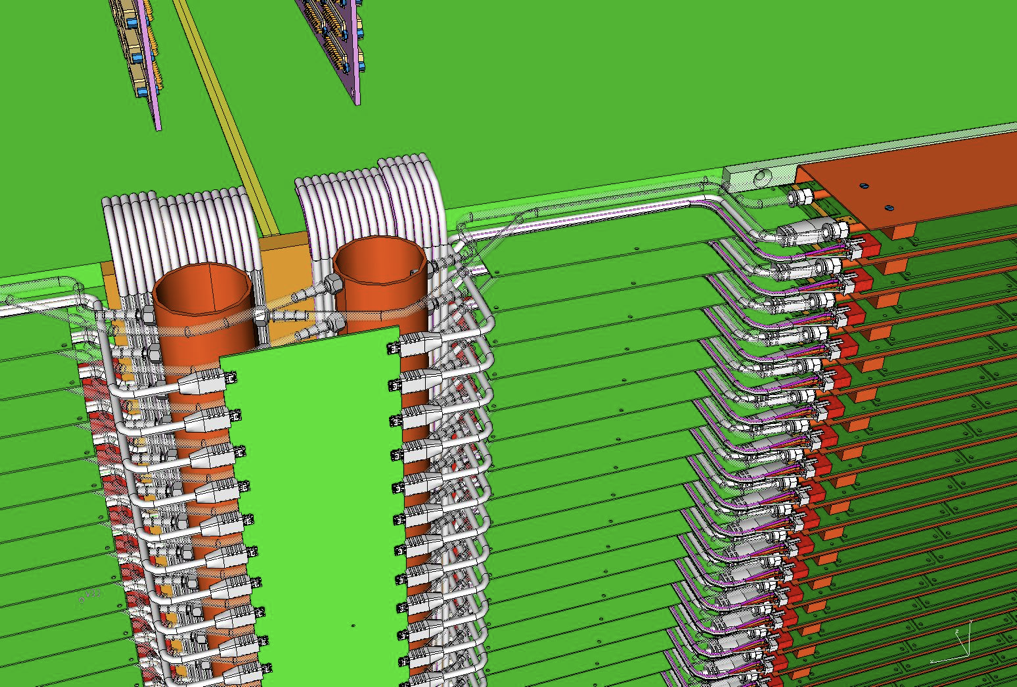}
    \caption{AHCAL module front face with readout boards, cables and cooling lines.}
    \label{ILD:fig:ahcal_services_closeup}
\end{figure}

\begin{figure}[h!]
    \centering
        \includegraphics[width=0.7\hsize]{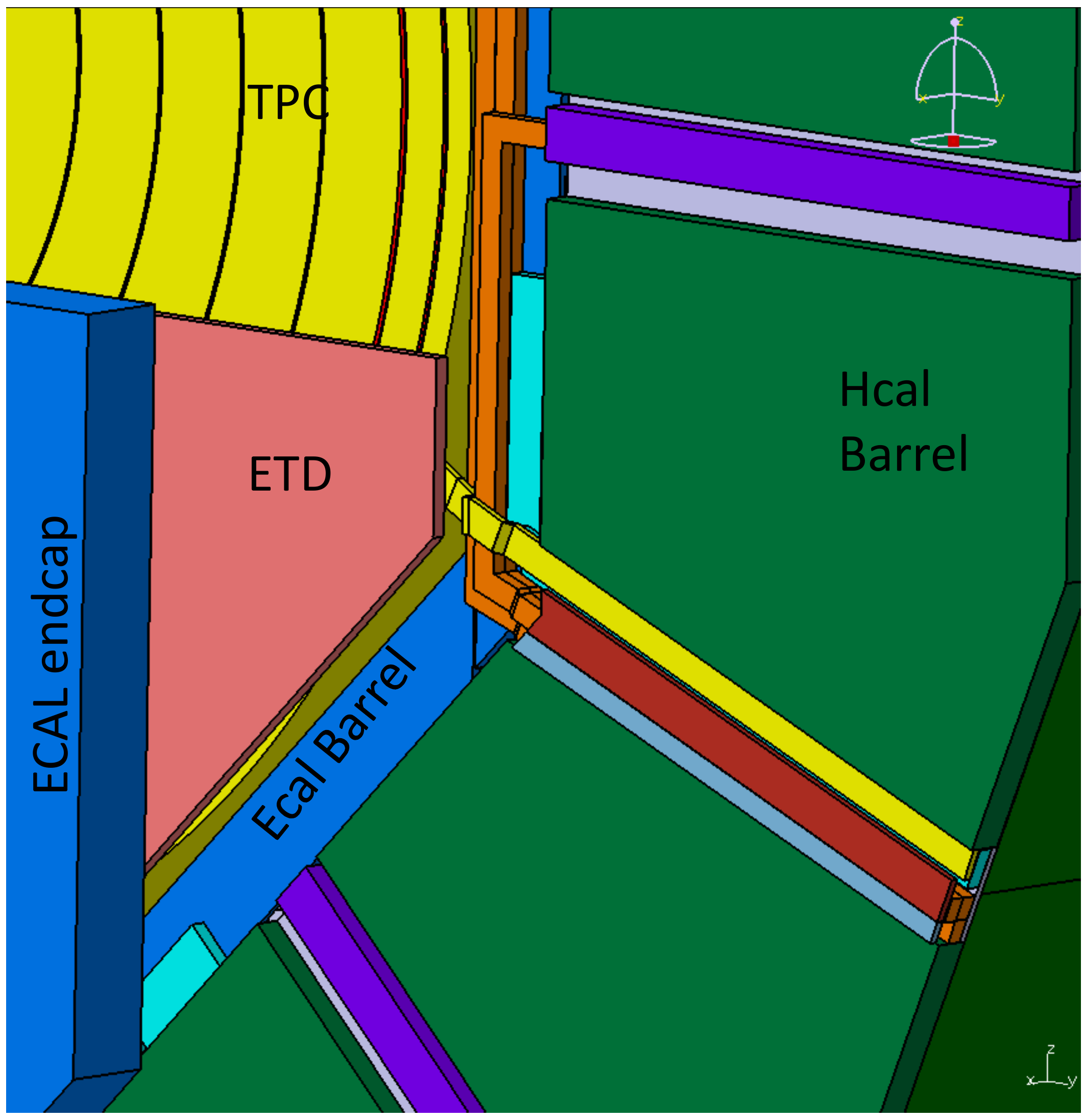}
    \caption{Service paths on the AHCAL barrel front. The domain tagged "ETD" corresponds to an empty space following removal of the Endcap Tracking Detector from the ILD baseline design (section 5.1.2).}
    \label{ILD:fig:barrel_services}
\end{figure}

An engineering model of the AHCAL barrel section in the "TESLA" structure has been designed and is shown in Figure~\ref{ILD:fig:ahcal_module_services}. Eight of these modules form a ring, two rings form the barrel detector. The active layers are read out via electronic boards that are located on the outside of the absorber structure. The data concentrators and power interfaces are accessible when the endcaps of the ILD detector are open.  
A close-up of the data concentator boards, the electrical services and the cooling system is shown in Figure~\ref{ILD:fig:ahcal_services_closeup}. This service concept has been implemented and successfully used with the AHCAL prototype (Figure~\ref{fig:AHCAL-TileProto}). The electrical and cooling lines of the ECAL, AHCAL, the TPC and the inner detector are routed via the gaps in the AHCAL front between the barrel and the endcap detector, as shown in Figure~\ref{ILD:fig:barrel_services}. 
\subsubsection{SDHCAL Electrical Services and Cooling}
\begin{figure}[h!]
    \centering
        \includegraphics[width=0.8\hsize]{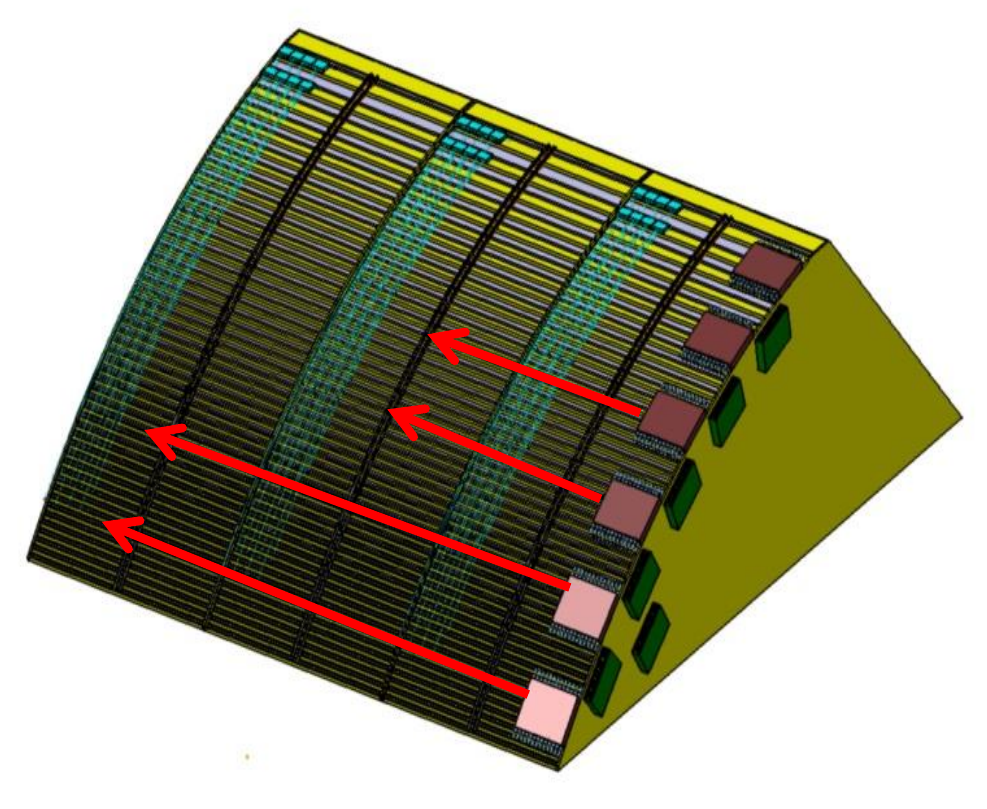}
    \caption{SDHCAL module with readout boards, cables and cooling lines.}
    \label{ILD:fig:sdcal_module_services}
\end{figure}

\begin{figure}[h!]
    \centering
        \includegraphics[width=0.6\hsize]{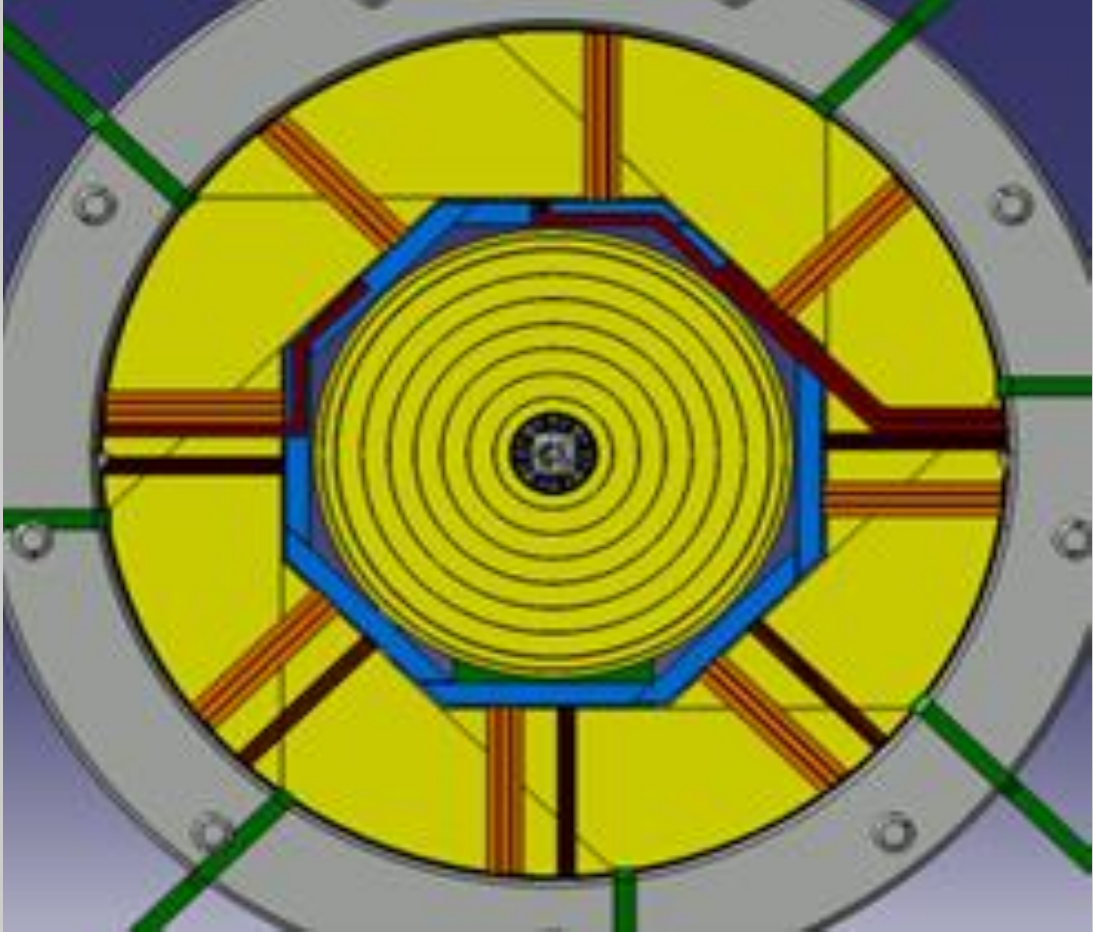}
    \caption{Service paths on the SDHCAL barrel front, in the "Videau" geometry.}
    \label{ILD:fig:sdhcal_barrel_services}
\end{figure}

The SDHCAL barrel detector consists of three or five rings of eight wedge-shaped modules in the Videau configuration. As the active layers are installed from the outside of the barrel structures, access requires the removal of the respective barrel rings from the detector solenoid. The cables and cooling services run on the outside of each module to the barrel front face, as shown in Figure~\ref{ILD:fig:sdcal_module_services}. The SDHCAL services are routed around the detector solenoid to the outside. The services of the inner detector, the ECAL and the TPC can be routed on the front side of the SDHCAL barrel detector, as shown in Figure~\ref{ILD:fig:sdhcal_barrel_services}.


\subsection{Very Forward System Integration}
\begin{figure}[h!]
    \centering
    \includegraphics[width=0.45\hsize]{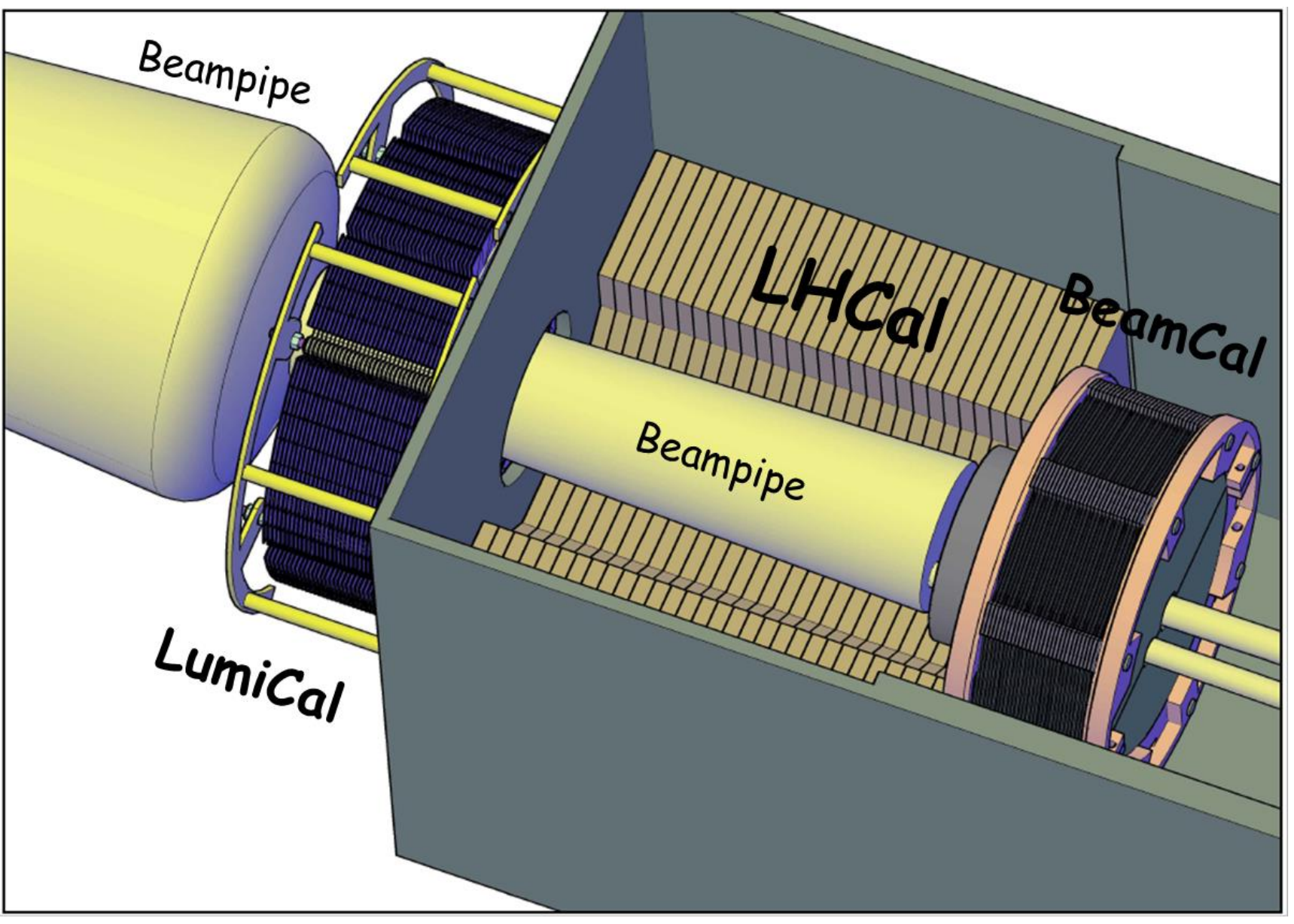}
        \includegraphics[width=0.45\hsize]{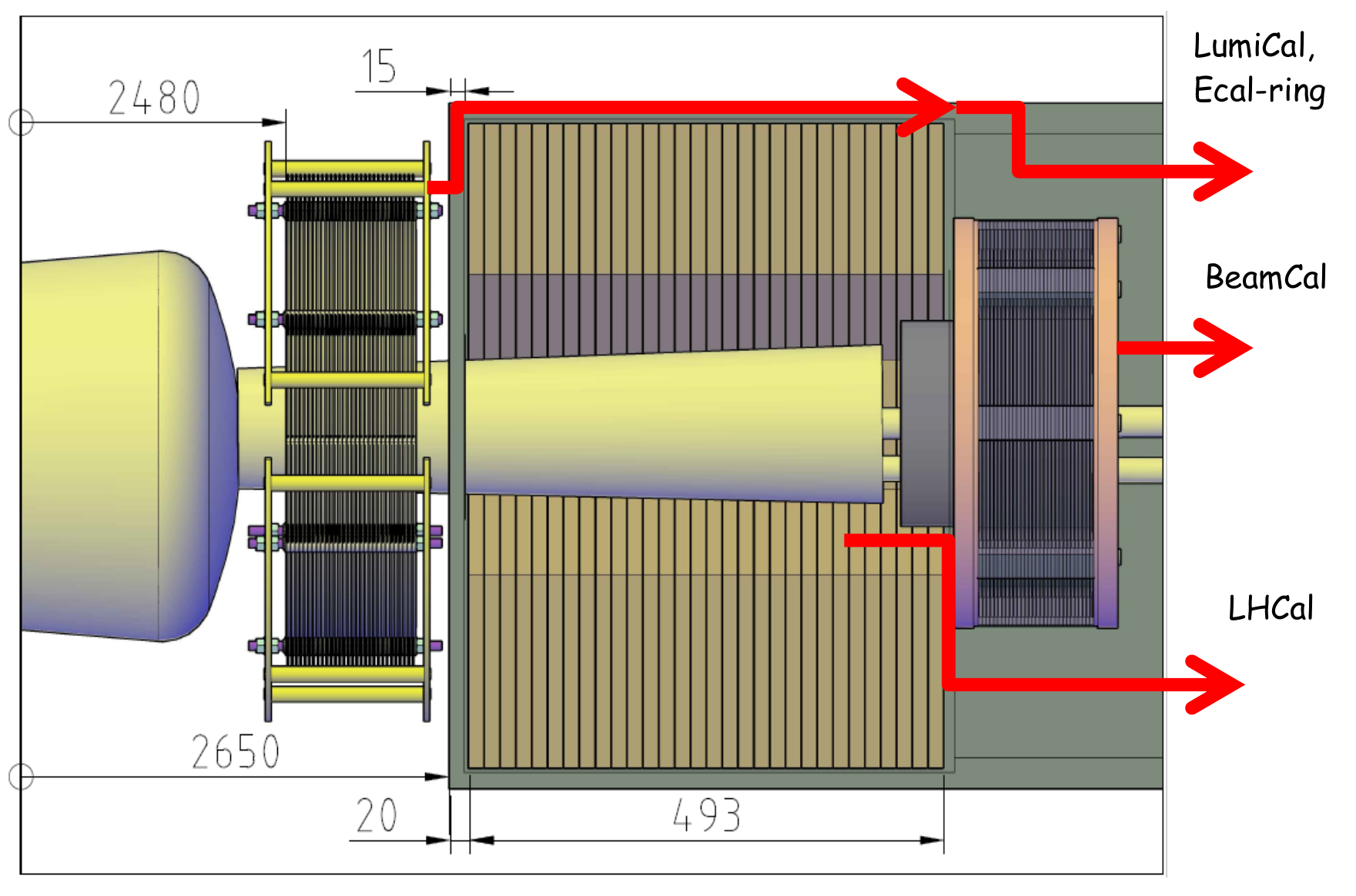}
    \caption{Design of the very forward systems with indication of service and cable paths~\cite{ild:bib:VFS_ICD}.}
    \label{ILD:fig:vfs_integration}
\end{figure}

The very forward systems, BeamCAL, LumiCAL and LHCAL, are carried by the support structure for the final focus quadrupole QD0. Figure~\ref{ILD:fig:vfs_integration} shows the mechanical layout and the conceptual service paths of the forward calorimeters.
\section{Mechanical structure studies}
\label{ild:sec:mechanical_structures}

The mechanical behaviour of the ILD components is crucial in two respects. On the one hand, the high-precision and hermeticity of the detector requires a precise relative adjustment of subdetector components within each other, with tight tolerances at the interfaces and boundaries. These aspects were studied by simulating static deformations of the components under gravity and other constraints. On the other hand large devices in Japan must obey strict rules as regards their behaviour in case of seismic events (see section~\ref{ild:sec:earthquake}). This was investigated by modelling the dynamic behaviour of components, including the computation of their "eigen modes" and their reaction to reference earthquake parameters from the foreseen ILC Kitakami site. Most of the attention has up to now been given to the calorimeter mechanical structure which governs the global stiffness of the ILD detector inside the coil. 


As mentioned in section~\ref{ild:sec:hcal} two options of the hadronic calorimeter, so-called "Videau" and "TESLA", are under consideration. In both cases, the electromagnetic modules are fixed to the inner plates of the hadronic wheels with two rails parallel to the z direction. Two critical aspects are of particular importance for the calorimeters: the respect of the tolerances of the thin azimutal clearance (2.5mm) between the electromagnetic modules, to avoid mutual contact and possible damage of the modules, and the flatness of the hadronic absorber plates which define the gaps in which the sensitive layers are introduced. The latter is particularly important for the SDHCAL instrumentation option since RPC's require a high level of flatness. Both Videau and TESLA mechanical options have been simulated in detail and the results provide input for further optimization of the layouts.

\begin{figure}[t!]
\centering
\includegraphics[width=1.0\hsize]{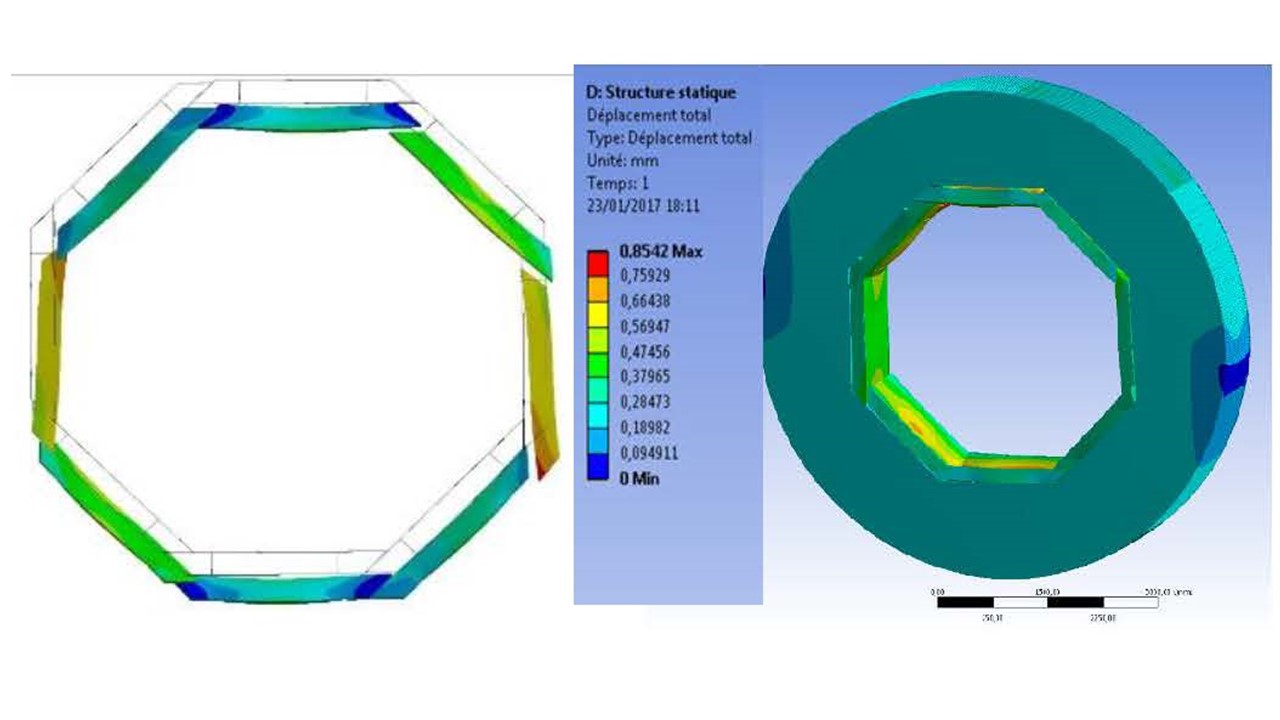}
\caption{\label{fig:integration:Videau_deformations}Static deformations of the calorimeter "Videau" structure (right, in mm) and zoom on the electromagnetic modules displacements with a magnification factor of 750 (left).}
\end{figure}

\subparagraph{\textbf{Videau simulations:}} The static behaviour of a full Videau calorimeter wheel was simulated with a shell model. In order to save computing time, the electromagnetic modules were approximated by a 3D solid model. This simplification was validated separately by a comparison to a single module shell model simulation. The results are shown in Figure~\ref{fig:integration:Videau_deformations}: the hadronic structure turns out to be very stiff with largest deformations of a fraction of a mm. This is due to the vertical flanges of the Videau modules which strongly rigidify the overall structure. The electromagnetic modules are also only slightly distorted: the largest deformation stays below 1mm and the azimutal clearance between modules is reduced to 2.3 mm in the worst case, well within the required tolerances. 

One advantage of the Videau layout is to avoid a projective dead zone at a polar angle of $90^0$, but the number of dead zones in z is 4 for the baseline number of 5 wheels. One question is whether one could reduce this number of dead zones by reducing the number of wheels to 3. Mechanical simulations show that for 3 wheels it is possible to keep deformations at the boundary between the electromagnetic modules within specifications, provided the inner hadronic absorber plate thickness is slightly increased.

First dynamic simulations of the Videau structure were also performed with the computation of eigen modes. They show that the overall calorimeter barrel behaves as a rigid structure under oscillating accelerations, with little variation of the z distance between the wheels.

\subparagraph{\textbf{TESLA simulations:}} 

In the ILD DBD the TESLA mechanical layout was initiall designed with corners at $0^{o}$, $45^{o}$, $90^{o}$ etc. in azimuth, which corresponded to the optimal configuration as regards mechanical stiffness. In order to facilitate the silicon sensor tiling of the ECAL endcap, the structure has since then been rotated by $22.5^{o}$. Since there are no vertical disks, the behaviour of the structure now resembles that of a Roman arc. 

\begin{figure}[t!]
\centering
\includegraphics[width=1.0\hsize]{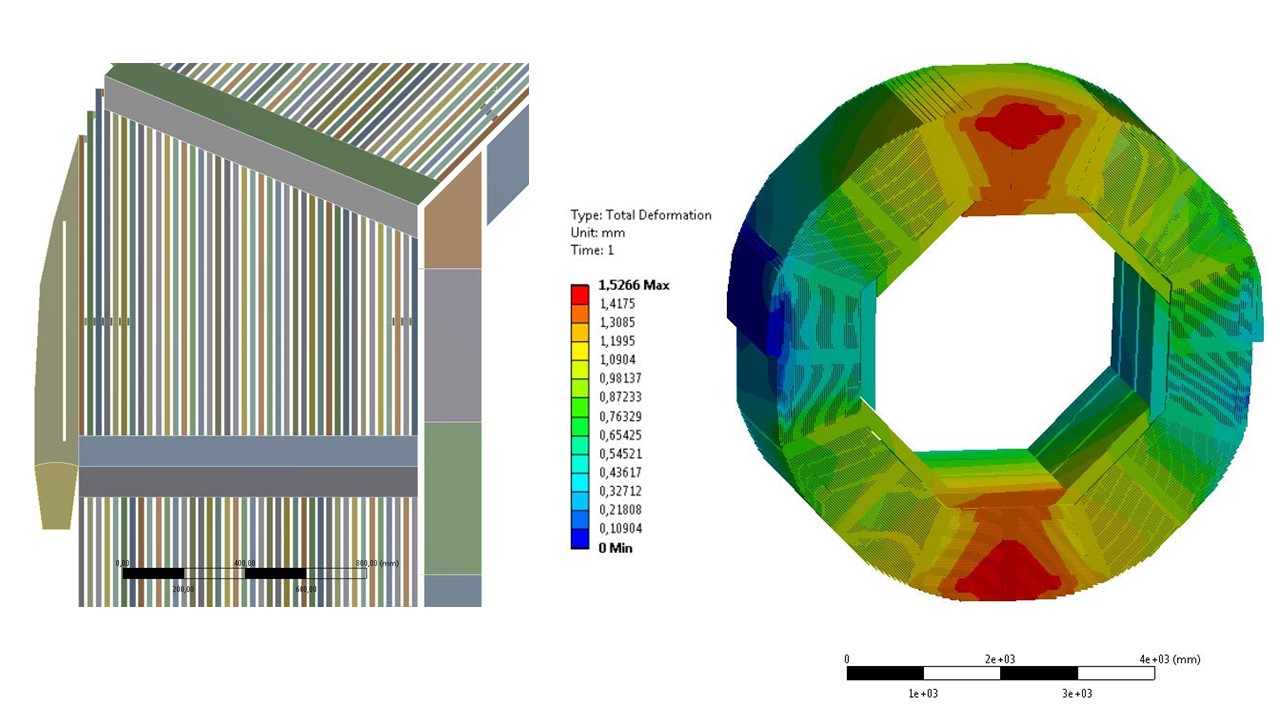}
\caption{\label{fig:integration:Tesla_deformations}Zoom on the consolidation of the TESLA structure with spacers (left) and resulting static deformations of the structure (right).}
\end{figure}

The mechanical behaviour of the barrel structure with the new orientation has been simulated with shell and 3D models. The simulations showed deformations of up to 5~mm, requiring actions to reduce them. A minimised number of 10~mm wide additional spacers have been introduced between absorber plates in critical points -- near the fixtures of the ECAL and near the supports from the cryostat -- (Figure~\ref{fig:integration:Tesla_deformations} left), resulting in static deformations of less than 2~mm everywhere (Figure~\ref{fig:integration:Tesla_deformations} right). 

The 3D model has also been used to obtain the first 200 vibration eigen modes. However, calculating the response of the system to an external excitation exceeds computational limitations and requires more advanced methods. 
The so-called component mode synthesis (CMS) method has been used, in which the detailed sub-structure of calorimeter modules is replaced by blocks which however preserve the overall mechanical behaviour of the module at specially configured boundaries. 
A simplified AHCAL model has been used to validate this model against a full 3D simulation up to the response to frequency sweeps and to real Japanese Earth quake data. 
The application of this model to the re-optimised AHCAL structure is in progress. 





\section{Coil and yoke studies}

A conceptual design of the ILD solenoid, the yoke and the cryostat had been developed for the ILD DBD~\cite{ild:bib:Magnet_Note, ild:bib:Cryostat_Note}. Recently, studies have been done to optimise the engineering design of the magnet, to study the magnetic properties of the large and small ILD detector models, and to explore possible cost reductions.

\subsection{Magnet Engineering Studies}
A common study by KEK, Toshiba and Hitachi has been started to better understand the engineering challenges in the design and manufacturing of the ILD solenoid coil and the Anti-DID, which adds a small additional dipole field to the solenoid for better suppression of background particles in the detector~(c.f. section~\ref{ild:sec:beam_backgrounds} and \cite{ild:bib:anti-did}).

\begin{figure}[h!]
    \centering
    \includegraphics[width=0.8\hsize]{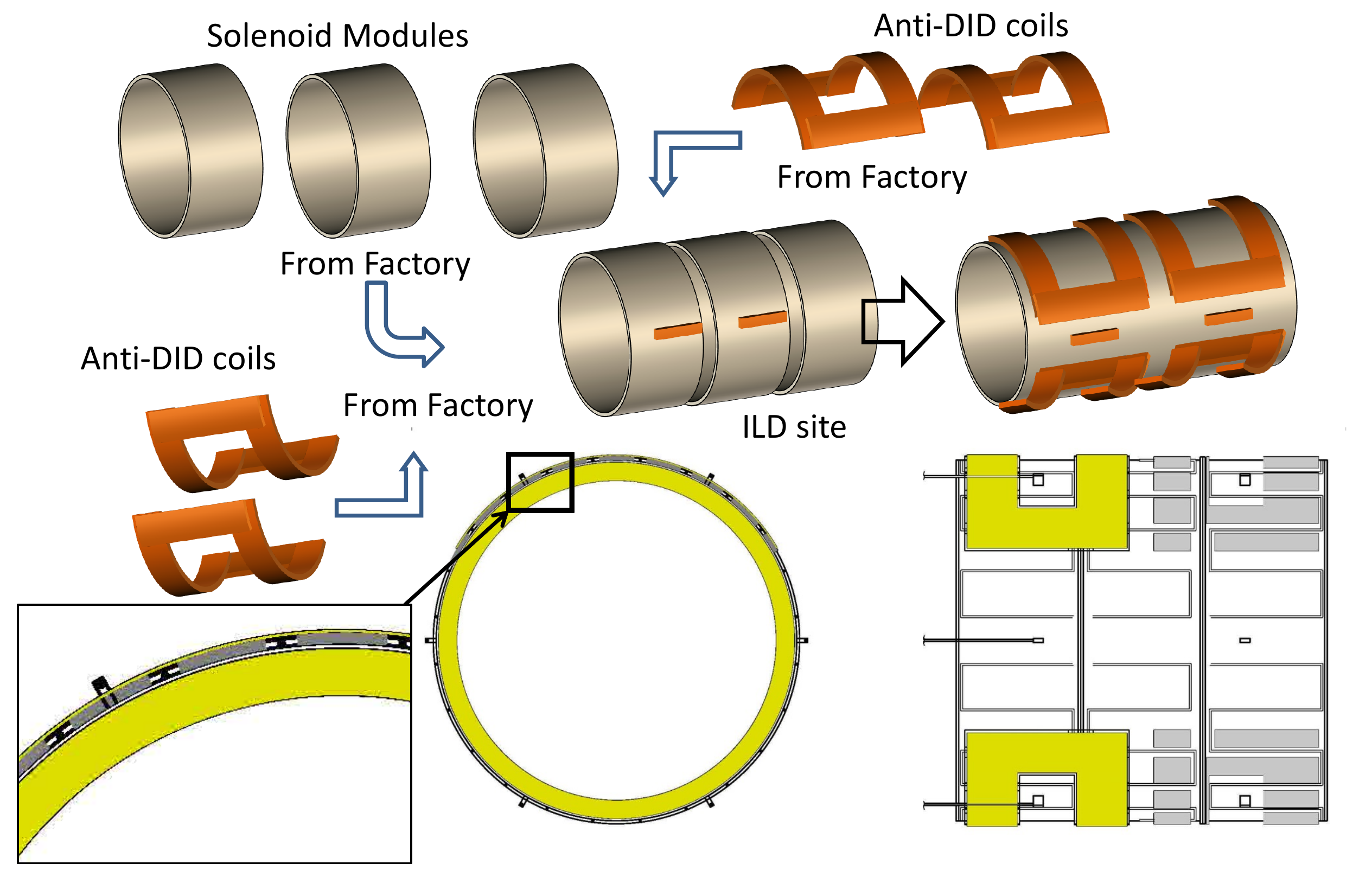}
    \caption{Illustration of the manufacturing process of the ILD solenoid and the Anti-DID coils~\cite{ild:bib:Solenoid_Manufacturing}.}
    \label{ILD:fig:solenoid_manufacturing}
\end{figure}

\begin{figure}[h!]
    \centering
    \includegraphics[width=0.8\hsize]{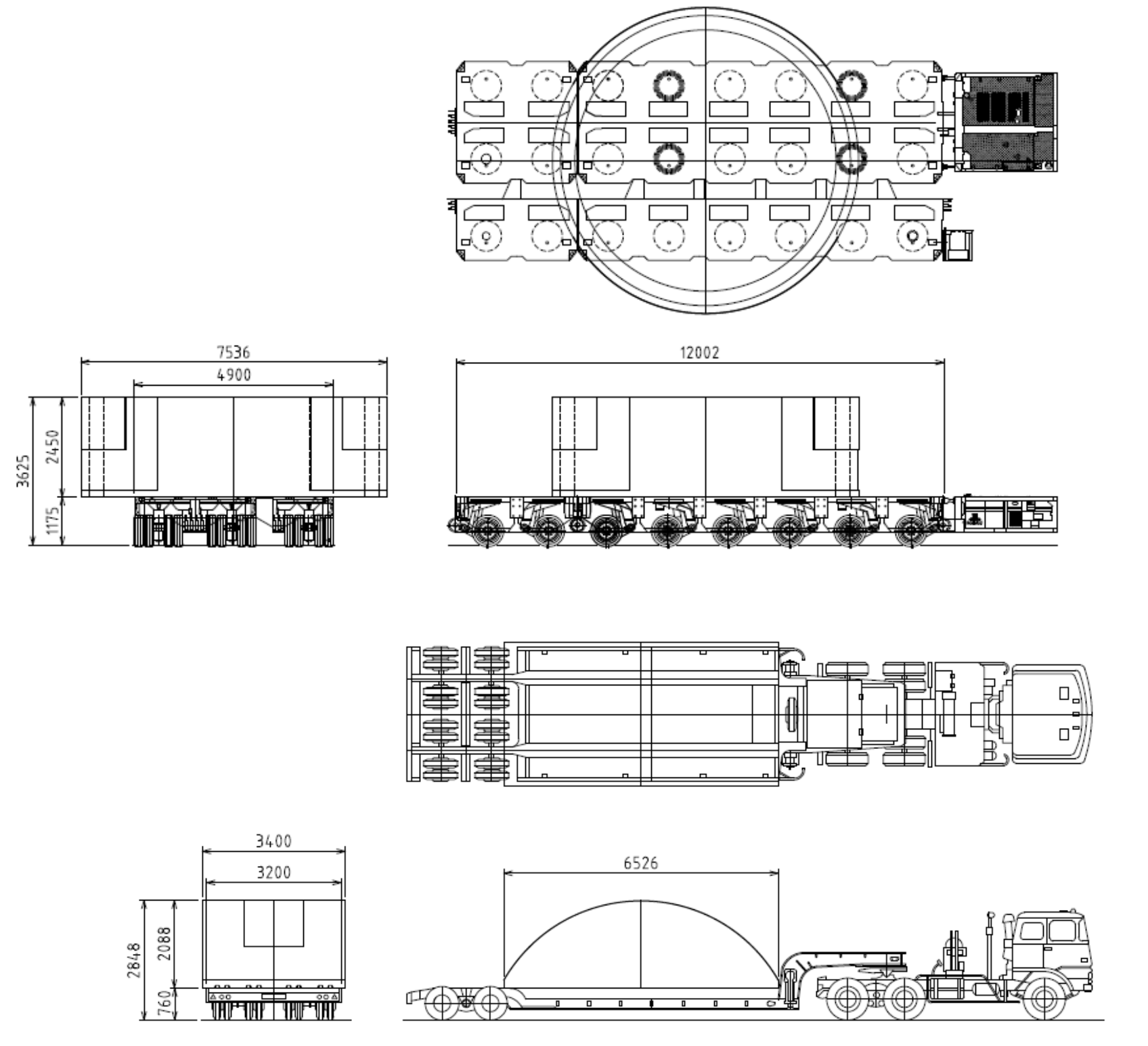}
    \caption{Transport of coil components: while a solenoid module requires transportation by a specialised Jumbo Carrier (top), the Anti-DID modules would fit on a low flatbed trailer (bottom)~\cite{ild:bib:Solenoid_Manufacturing}.}
    \label{ILD:fig:magnet_transport}
\end{figure}

Figure~\ref{ILD:fig:solenoid_manufacturing} shows a conceptual design of the manufacturing process. While the three solenoid modules and the Anti-DID coils would be fabricated in industry, transportation boundary conditions require the assembly of the complete magnet on or close to the final detector site.
A case study by Toshiba concluded that the main solenoid modules can be transported on the road with the use of a specialised Jumbo Carrier while the modules of the Anti-DID would fit on a flatbed trailer~(Figure~\ref{ILD:fig:magnet_transport}).

\begin{figure}[h!]
\begin{subfigure}{0.49\hsize} \includegraphics[width=\textwidth]{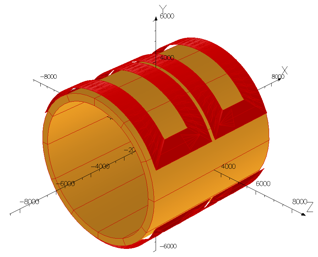}
\caption{ \label{ild:fig:anti_did_mechanics}}
 \end{subfigure}
\begin{subfigure}{0.49\hsize} \includegraphics[width=\textwidth]{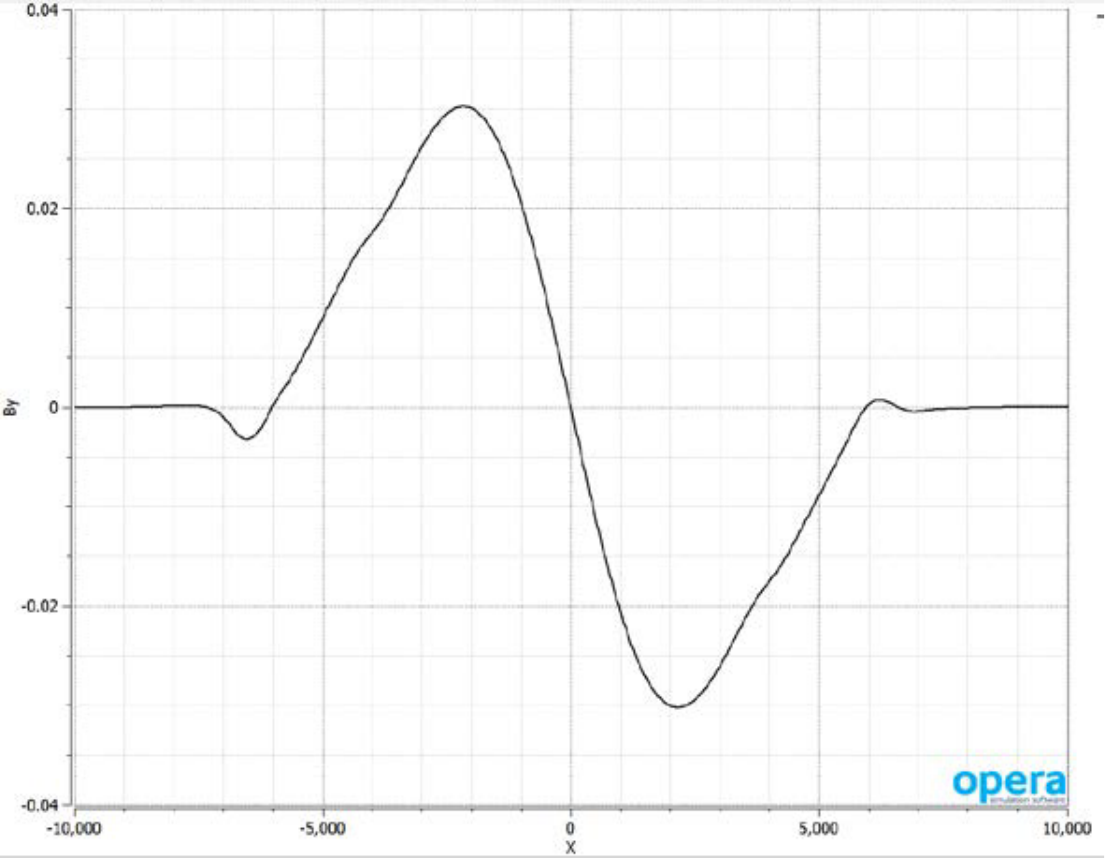}
\caption{  \label{ild:fig:anti-did-field}}
 \end{subfigure}
\caption{(a) Conceptual design study of the Anti-DID coils. (b) Simulated Anti-DID field (in T)~\cite{ild:bib:anti-did-design}}
\label{ILD:fig:anti_did_design}
\end{figure}

 Figure~\ref{ILD:fig:anti_did_design} shows a conceptual design of the Anti-DID coils and the resulting simulated dipole field with a maximum value of 0.036~T. Splitting the Anti-DID coils in four makes construction at the manufacturer and transport to the ILC site easier (c.f.~Figure~\ref{ILD:fig:magnet_transport}).

\subsection{Field Optimisation Studies}

\begin{figure}[t]
\begin{center}
\begin{subfigure}{0.75\hsize} \includegraphics[width=\textwidth]{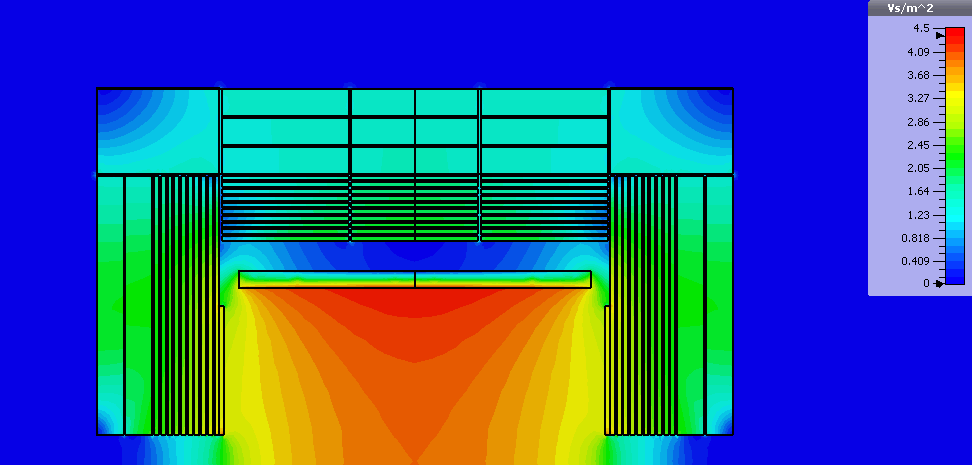}
\caption{ \label{ild:fig:magnet_nominal_map}}
 \end{subfigure}
\hspace{0.03\textwidth}
\begin{subfigure}{0.75\hsize} \includegraphics[width=\textwidth]{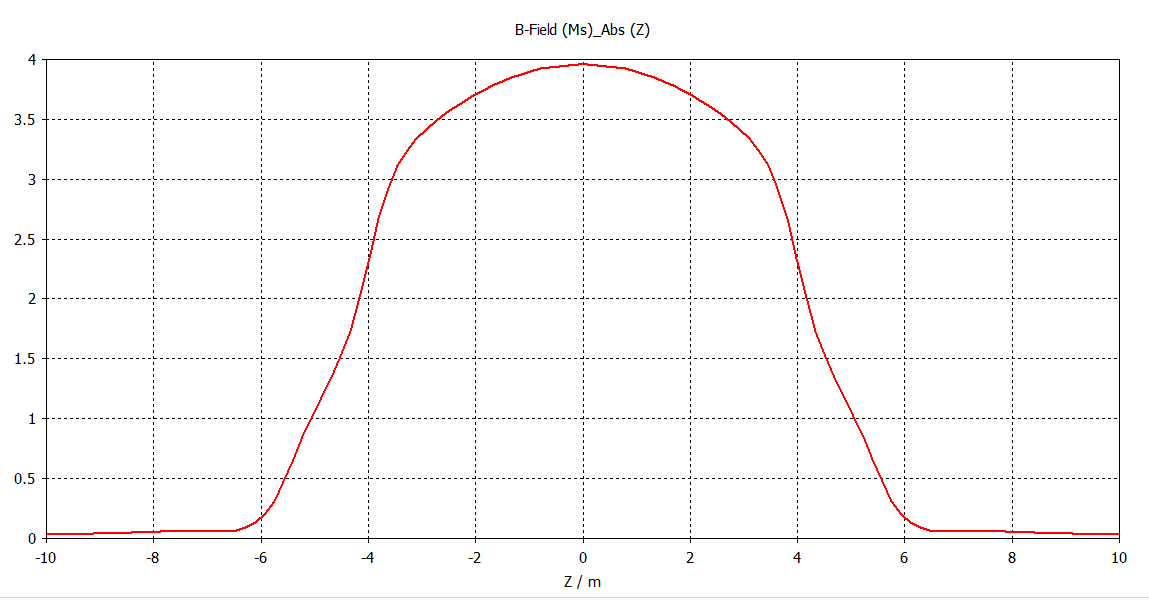}
\caption{  \label{ild:fig:magnet_nominal_field}}
 \end{subfigure}
\end{center}
\caption{(a) Field map of the ILD solenoid for the large ILD model with a maximum field of 4~T. (b) Field distribution along the detector axis (in T)~\cite{ild:bib:Magnet_Simulations}}
\label{ILD:fig:magnet_nominal}
\end{figure}

\begin{figure}[t]
\begin{center}
\begin{subfigure}{0.75\hsize} \includegraphics[width=\textwidth]{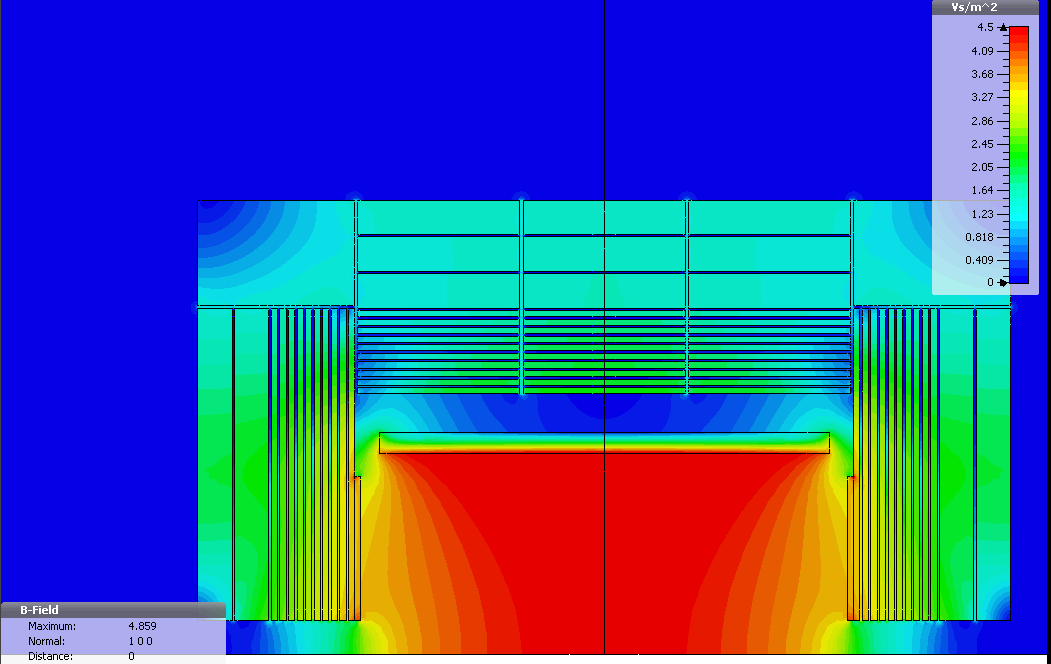}
\caption{ \label{ild:fig:magnet_small_map}}
 \end{subfigure}
\hspace{0.03\textwidth}
\begin{subfigure}{0.75\hsize} \includegraphics[width=\textwidth, height = 6cm]{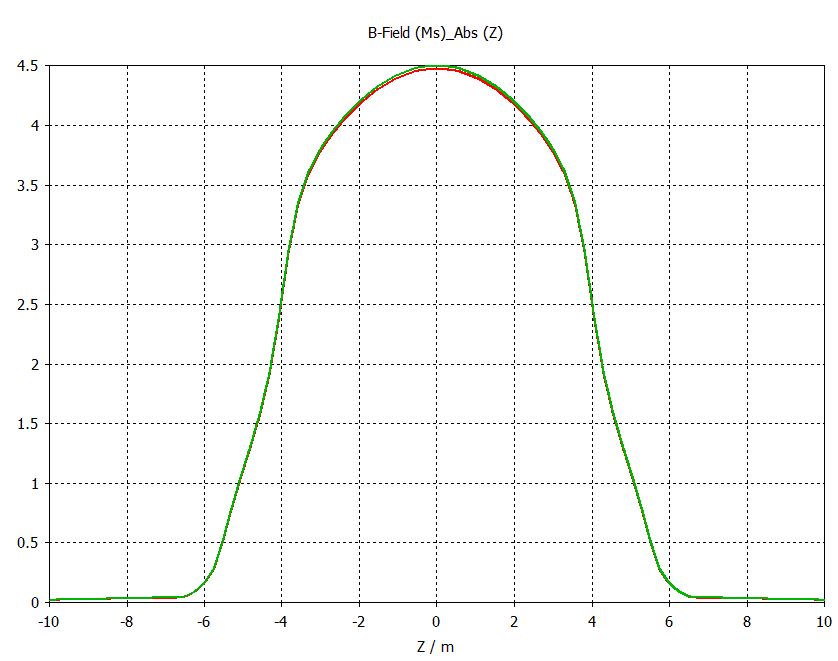}
\caption{  \label{ild:fig:magnet_small_field}}
 \end{subfigure}
\end{center}
\caption{(a) Field map of the ILD solenoid for the small ILD model with a maximum field of 4.5~T. (b) Field distribution along the detector axis (in T)~\cite{ild:bib:Magnet_Simulations}}
\label{ILD:fig:magnet_small}
\end{figure}

The coil and yoke system should provide at the same time a high magnetic field in the central region and low stray field on the outside of the detector. The nominal fields for the large (small) ILD detector model are 3.5T (4T). To allow for safety margins and possible extensions to higher field values, the coil design has been developed to reach 0.5T more in each case. Figure~\ref{ILD:fig:magnet_nominal} therefore shows the field map and the field distribution for the large ILD detector model and a maximum field of 4T. The corresponding distributions for the small ILD detector model are shown in Figure~\ref{ILD:fig:magnet_small}. The field maps are the result of 3D magnetic field simulations. They are implemented in the simulation tools of the detector (chapter 7) to allow for reconstruction with maximum realism when this is required by specific studies such as beam background effects (section 6.5). 

\begin{figure}[t]
\begin{center}
\begin{subfigure}{0.75\hsize} \includegraphics[width=\textwidth]{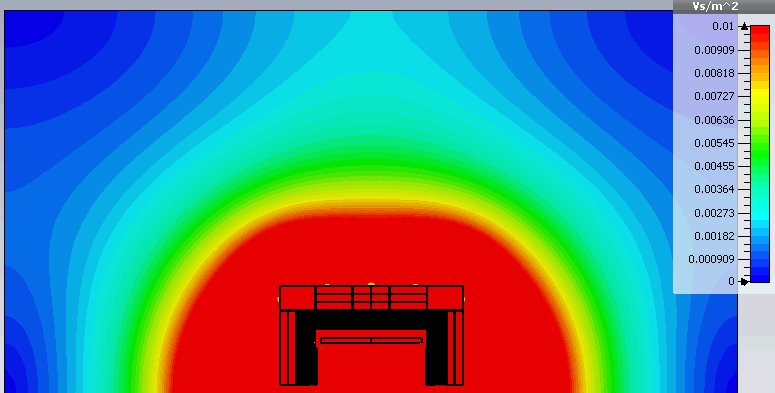}
\caption{ \label{ild:fig:magnet_nominal_stray_map}}
 \end{subfigure}
\hspace{0.03\textwidth}
\begin{subfigure}{0.75\hsize} \includegraphics[width=\textwidth]{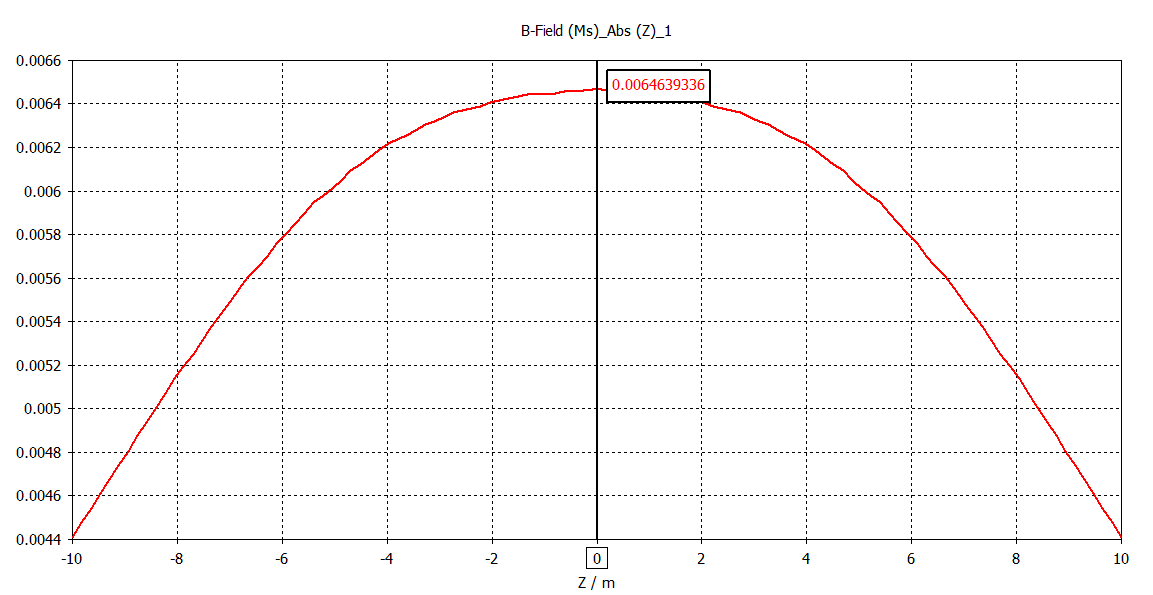}
\caption{  \label{ild:fig:magnet_nominal_stray_field}}
 \end{subfigure}
\end{center}
\caption{(a) Map of the ILD solenoid strayfield for the large ILD model with a maximum central field of 4~T. (b) Field distribution at a distance of 15~m parallel to the detector axis (in T)~\cite{ild:bib:Magnet_Simulations}.}
\label{ILD:fig:magnet_nominal_stray}
\end{figure}

\begin{figure}[h!]
\begin{center}
\begin{subfigure}{0.75\hsize} \includegraphics[width=\textwidth]{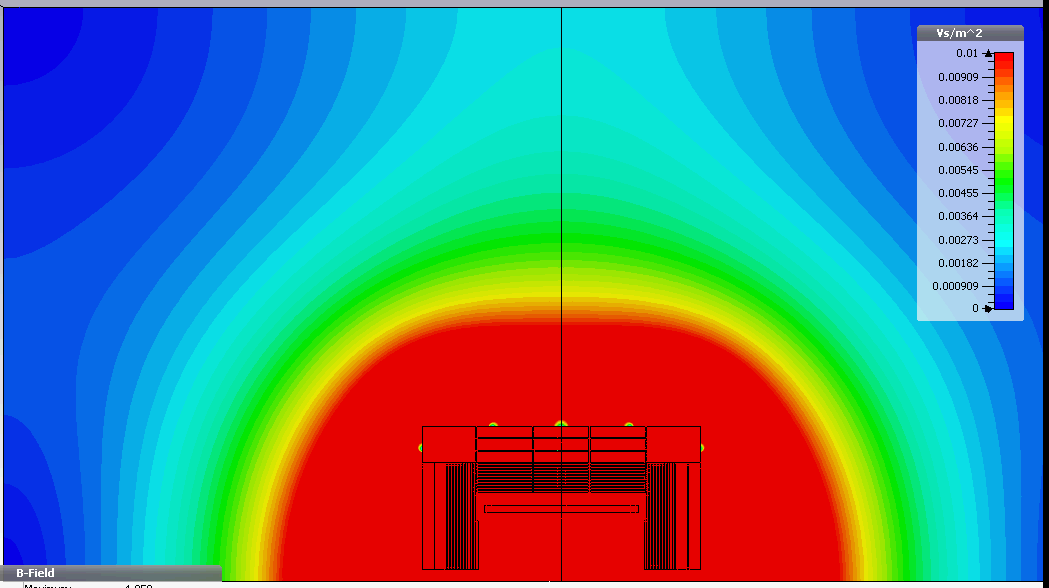}
\caption{ \label{ild:fig:magnet_small_stray_map}}
 \end{subfigure}
\hspace{0.03\textwidth}
\begin{subfigure}{0.75\hsize} \includegraphics[width=\textwidth, height =6cm]{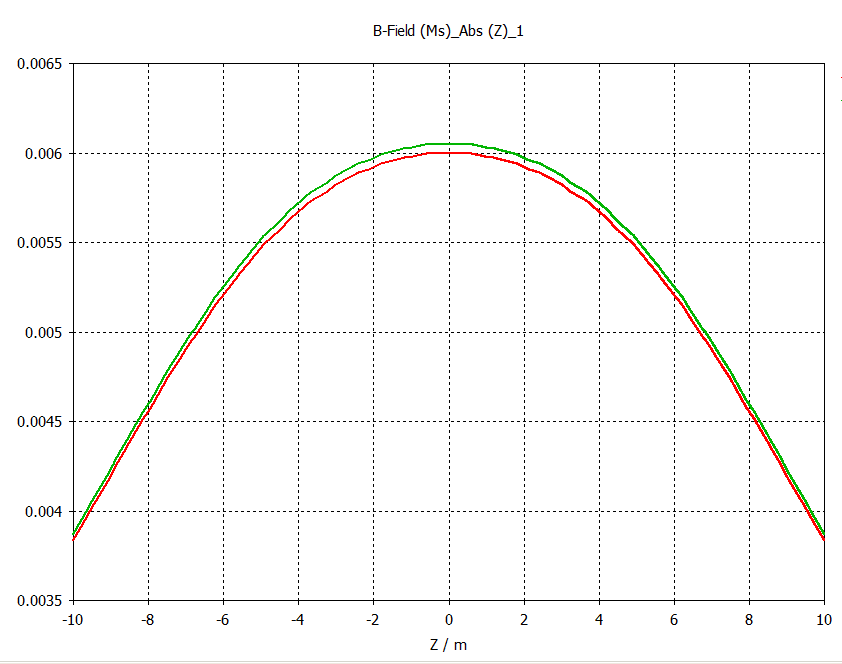}
\caption{  \label{ild:fig:magnet_small_stray_field}}
 \end{subfigure}
\end{center}
\caption{(a) Map of the ILD solenoid strayfield for the small ILD model with a maximum field of 4.5~T. (b) Field distribution (red curve) at a distance of 15~m parallel to the detector axis (in T)~\cite{ild:bib:Magnet_Simulations}.}
\label{ILD:fig:magnet_small_stray}
\end{figure}

The ILD detector is expected to share its experimental environment with another detector, e.g. SiD, in a push-pull scenario. This model of operations assumes that each detector can run its magnetic field while still allowing the other detector crew to work on their detector with the use of standard iron-based tools. Therefore a limit on the maximum magnetic stray fields has been agreed upon at a maximum of 5~mT at a distance of 15~m from the detector axis~\cite{Parker:2009zz}. This requirement is the main driver for the thickness of the iron yoke of ILD. Figures~\ref{ILD:fig:magnet_nominal_stray} and \ref{ILD:fig:magnet_small_stray} show the simulated stray fields for the large and the small detector model at 4~T and 4.5~T maximum central field.

Simulations have been done with different codes and the results are very consistent with an average value for the large detector of 5.6$\pm$0.6~mT for the maximum stray field at 15~m~\cite{ild:bib:Magnet_Simulations}. The stray fields for the small detector model are very similar. 

The stringent limits on the stray fields are a cost issue for ILD as they drive the thickness of the iron return yoke. If the thickness would be reduced by 60~cm, the maximum stray fields at 15~m would increase to 9.3$\pm$ 0.8~mT but at the same time reduce the cost for the yoke by about 20\% for the large ILD detector model~\cite{ild:bib:Magnet_Simulations}.

\begin{figure}[htb]
    \centering
    \includegraphics[width=0.5\hsize]{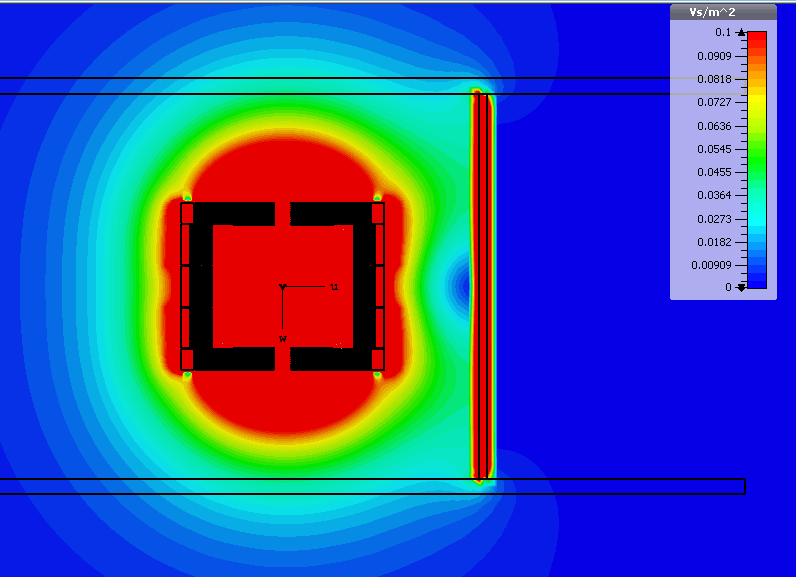}
    \caption{Field map of the ILD solenoid for the large ILD model with an iron shielding wall in the direction of the other detector in the experimental hall~\cite{ild:bib:Magnet_Simulations}.}
    \label{ILD:fig:magnet_wall}
\end{figure}

A cost reduction of in total $\approx$50\% can be reached if the iron yoke is reduced further to a thickness of 2.04~m (including gaps). The stray fields at 1~m distance from the outside of the yoke would be at the level of 100~mT which is acceptable for the operation of magnetically sensitive equipment and which is well below limits given by human safety regulations. The far stray fields in the direction of the other detector need to be shielded in that case, e.g. by using a magnetic shielding wall. Figure~\ref{ILD:fig:magnet_wall} shows the simulation of the large ILD detector model magnetic fields with a reduced yoke and a shielding wall.

Though this concept looks attractive, further studies are required. The shielding wall needs to be mobile and has to follow the push-pull movements. An engineering solution needs to be developed. Another problem is the radiation shielding. In the present design, the thick iron yoke serves as a radiation shield and allows for access to the outer side of the detector while the ILC delivers beams to the interaction region~\cite{ild:bib:Radiation_Hall}. A reduced yoke might be too thin to serve for this purpose alone and additional radiation shielding, e.g. made from concrete, might need to be added. 

\section{\label{sec:beam:background}Beam background studies}
\label{ild:sec:beam_backgrounds}
The ILD detector response is affected by three main sources of beam-related background: beamstrahlung emitted at the crossing point of the electron and positron bunches, halo muons produced upstream of the detector along the beamline, and low energy neutrons back-scattered from the beam dumps. Their impact on the detector occupancies has been quantified and possible mitigation investigated\cite{ild:bib:Machine_Backgrounds,ild:bib:schuetz_thesis}. Results are similar for both options of the ILD detector and are shown here for the large version.

\subsection{Beamstrahlung}

Beamstrahlung photons are radiated from the ILC beam particles in the interaction with the electromagnetic fields of the strongly focused oncoming bunches. These photons can convert to $e^+e^-$ pairs with a broad energy spectrum. Most of the low-energy pairs are confined close to the z axis by the solenoid magnetic field but, due to the crossing angle between the colliding beams, a significant fraction hit the very forward detectors near the incoming and outgoing beam pipes. The central detectors are also affected, both directly by the $e^+e^-$ pairs, and indirectly through back-scattering of low-energy particles from the forward detectors. It was proposed to mitigate these effects by the addition of a small dipole field to the main solenoidal field. This so-called "anti-DID" field can be tuned to guide a large fraction of the  $e^+e^-$ pairs into the outgoing beampipe. 

\begin{figure}[t!]
\begin{subfigure}{0.48\textwidth}
\includegraphics[width=1.0\hsize]{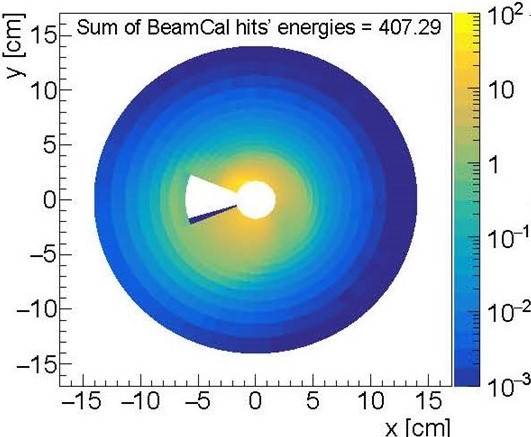}
\caption{}
\end{subfigure}
\begin{subfigure}{0.48\textwidth}
\includegraphics[width=1.0\hsize]{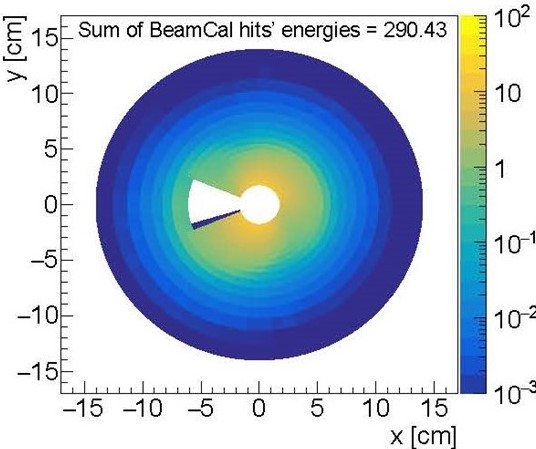}
\caption{}
\end{subfigure}
\caption{\label{fig:integration:beamcal}Beamstrahlung energy deposition in the BeamCAL without (a) and with (b) anti-DID field for the large ILD version at ILC-250 (arbitrary units). }
\end{figure}

These effects have been quantified for the updated 250 GeV ILC conditions (ILC250, see Section~\ref{ild:sec:ilc}), using a full simulation of the beamstrahlung particle production and of their tracking within ILD~\cite{ild:bib:Machine_Backgrounds}. Special care has been taken to ensure stable results by adapting the GEANT4 tracking parameters to low energy particles. A realistic field map of both solenoid and anti-DID fields has been used. The response of the most affected detector, the BeamCAL, is shown in Figure~\ref{fig:integration:beamcal} with and without anti-DID. A positive effect of the anti-DID is clearly visible: when the anti-DID field is applied, the particle distribution is more symmetric around the central beampipe, and the overall energy deposition in BeamCAL is reduced by about 30\%.

\begin{table}
\begin{center}
\begin{tabular}{ l c c c }
\toprule
LAYER: & hits/BX & hits/BX & hits/BX/$cm^2$ \\
& No anti-DID & anti-DID & anti-DID \\
& mean $\pm$ RMS & mean $\pm$ RMS & mean $\pm$ RMS \\
\midrule
VXD 1: & 1400 $\pm$ 780 & 910 $\pm$ 360 & 6.6 $\pm$ 2.6 \\
VXD 2: & 970 $\pm$ 560  & 540 $\pm$ 210 & 4.0 $\pm$ 1.5 \\
VXD 3: & 150 $\pm$ 80   & 130 $\pm$ 60  & 0.21 $\pm$ 0.10 \\
VXD 4: & 110 $\pm$ 60   & 110 $\pm$ 50  & 0.18 $\pm$ 0.09 \\
VXD 5: & 44 $\pm$ 30    & 40 $\pm$ 26   & 0.04 $\pm$ 0.03 \\
VXD 6: & 39 $\pm$ 27    & 34 $\pm$ 24   & 0.04 $\pm$ 0.03 \\
\midrule
FTD 1: & 42 $\pm$ 30 & 38 $\pm$ 26 & 0.043 $\pm$ 0.030 \\
FTD 2: & 27 $\pm$ 19 & 24 $\pm$ 15 & 0.029 $\pm$ 0.019 \\
FTD 3: & 62 $\pm$ 45 & 40 $\pm$ 27 & 0.014 $\pm$ 0.010 \\
FTD 4: & 42 $\pm$ 33 & 25 $\pm$ 17 & 0.009 $\pm$ 0.007 \\
FTD 5: & 29 $\pm$ 23 & 18 $\pm$ 13 & 0.007 $\pm$ 0.005 \\
FTD 6: & 16 $\pm$ 13 & 9 $\pm$ 7   & 0.004 $\pm$ 0.003 \\
FTD 7: & 10 $\pm$ 8  & 6 $\pm$ 5   & 0.003 $\pm$ 0.003 \\
\midrule
SIT 1: & 51 $\pm$ 37 & 24 $\pm$ 16 & 0.0032 $\pm$ 0.0023 \\
SIT 2: & 49 $\pm$ 36 & 21 $\pm$ 12 & 0.0029 $\pm$ 0.0017 \\
SIT 3: & 77 $\pm$ 56 & 34 $\pm$ 24 & 0.0014 $\pm$ 0.0010 \\
SIT 4: & 71 $\pm$ 54 & 31 $\pm$ 21 & 0.0013 $\pm$ 0.0009 \\
\midrule
SET 1: & 39 $\pm$ 28 & 15 $\pm$ 10 & 0.00003 $\pm$ 0.00002 \\
SET 2: & 46 $\pm$ 36 & 18 $\pm$ 12 & 0.00003 $\pm$ 0.00002 \\
\bottomrule
\end{tabular}
\end{center}
\caption{\label{ild:tab:BGhits}Number of beamstrahlung hits per bunch crossing in the silicon trackers without (left) and with (middle and right) anti-DID field for the large ILD version at ILC250.}
\end{table}

A similar improvement is seen for the central detectors in table~\ref{ild:tab:BGhits}. The most affected are the inner layers of the vertex detector, where the background is equally shared between direct hits rather uniformly distributed, and back-scattered particles showing hot spots in azimuth. Background hits further away from the beampipe are dominated by direct $e^+e^-$ pairs. The anti-DID has little effect on direct $e^+e^-$ hits but helps to suppress back-scattered particles and associated hot spots in the central detectors.

For the small ILD version, the overall beamstrahlung background hit rates are reduced by $\approx$10\%  thanks to better confinement within the beampipe by the higher solenoid field of 4T.



\subsection{Halo muons}

The electrons and positrons of the beam halo produce high-energy penetrating muons parallel to the beam by interacting with beamline components such as collimators. Beam simulations~\cite{Keller:2019aak} predict that a fraction of $10^{-3}$ of the incident beam electrons hit a machine component of the beam delivery system and may therefore produce muons. Five magnetized muon spoilers and one optional larger magnetized muon wall are foreseen to deviate these muons outside the ILC experimental hall. The resulting rate of muons in ILD has been simulated for ILC250 and ILC500. 

\begin{figure}[t!]
\includegraphics[width=1.0\hsize]{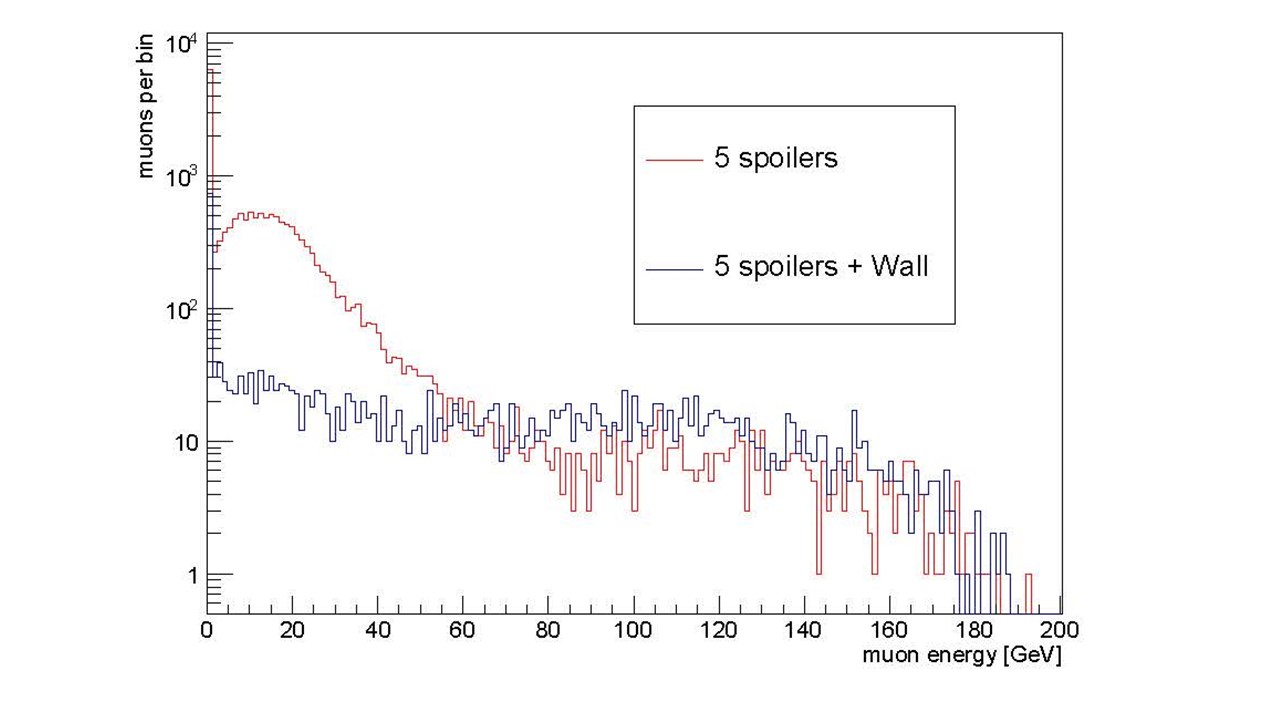}
\caption{\label{fig:integration:muons}Energy spectrum of muons entering the ILC experimental hall for ILC500, normalised to a full single train of 2625 bunches assuming a beam halo fraction of $10^{-3}$. Red curve: muon filtering with 5 spoilers only; Blue curve: muon filtering with an additional muon wall~\cite{ild:bib:schuetz_thesis}.}
\end{figure}

Figure~\ref{fig:integration:muons} shows the energy spectra of the muons crossing the ILD detector for the two filtering options. The muon wall significantly reduces the flux of the lower energy muons below a few 10 GeV. At ILC500 the rate of halo muons crossing ILD is of the order of 4 (resp. 0.6) muons/BX without (resp. with) the optional muon wall. At ILC250 the corresponding numbers are of the order 1 and 0.03 muons/BX for the two filtering options, respectively.

These halo muons will affect the occupancy of the detector but should be easily identified and subtracted from the physics events using topological and timing information.

\subsection{Backscattered neutrons from beam dumps}

A potential additional source of beam-related background consists in particles back-scattered from the electron and positron beam dumps located 300m from the interaction point in both beam directions. This background is expected to be dominated by low energy neutrons which can propagate over a long distance. Their contribution has been fully simulated with FLUKA in the context of SiD, including a detailed description of the beamline up to the beam dump as well as of the dump system itself~\cite{ild:bib:schuetz_thesis}. The incoming neutron fluxes close to SiD are expected to be the same as for ILD and were interfaced to the ILD detector simulation to estimate their impact in the detector. The simulation was normalised to the dump of one full electron bunch on the +z side of the detector.

\begin{figure}[t!]
\centering
\begin{subfigure}{0.40\textwidth}
\includegraphics[width=1.0\hsize]{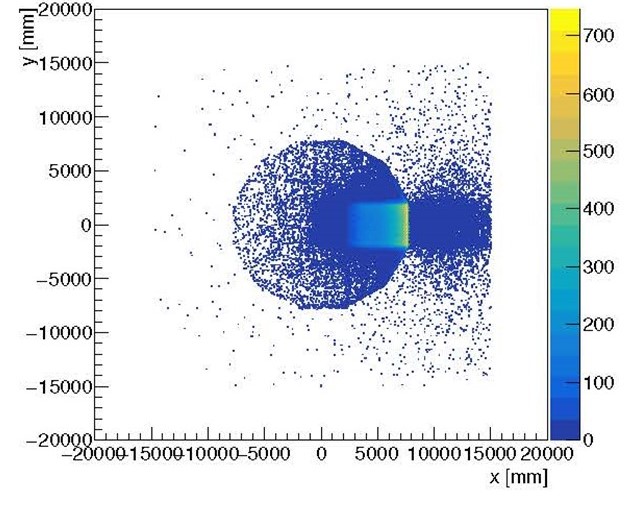}
\caption{}
\end{subfigure}
\begin{subfigure}{0.48\textwidth}
\includegraphics[width=1.0\hsize]{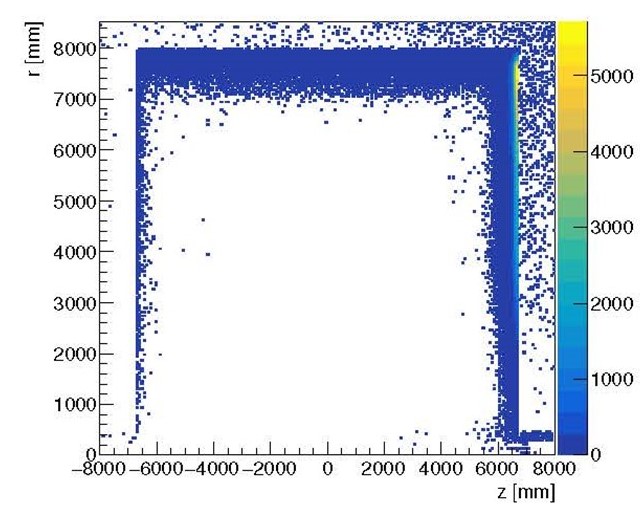}
\caption{}
\end{subfigure}
\caption{\label{fig:integration:neutrons}Map of the stopping points in ILD of neutrons back-scattered from the beamdump: (a) transverse coordinates and (b) longitudinal coordinates. The maps are normalised to the dump of one electron bunch on the +z side of ILD~\cite{ild:bib:schuetz_thesis}.}
\end{figure}

The total number of neutrons reaching the ILD detector region is found to be O($10^6$) per electron bunch dump. These neutrons have very low momenta of a few 10 MeV and most of them do not deposit a detectable signal in the detector. Their propagation in ILD was simulated and the map of their stopping points in the detector is shown in Figure~\ref{fig:integration:neutrons}. The results are in line of what is found for SiD~\cite{ild:bib:schuetz_thesis}: most neutrons are absorbed in the external layers of the ILD iron yoke, with a very small fraction reaching the external layers of the very forward calorimeters. Their energy depositions concentrate close to their stopping points: they are very small ($\approx$0.1 MIP on average) and asynchronous with respect to the bunch crossings. No visible signal is seen in the ILD vertex detector and central tracker. This indicates that the neutron background should not be a critical issue within its current state of understanding.

\section{Data acquisition}

The ILD data acquisition takes advantage of the relatively low rate of relevant physics $e^+e^-$ interactions (O(0.1 hadronic evt)/BX) and of the long idle periods between bunch trains (chapter 3) to provide a triggerless readout of the detector. The subdetector data of all BXs of a given bunch train are collected before the next bunch train and processed offline as a single data set for identification, bunch tagging and reconstruction of the individual interactions.

\subsection{DAQ architecture}

\begin{figure}[t!]
\centering
\includegraphics[width=0.75\hsize]{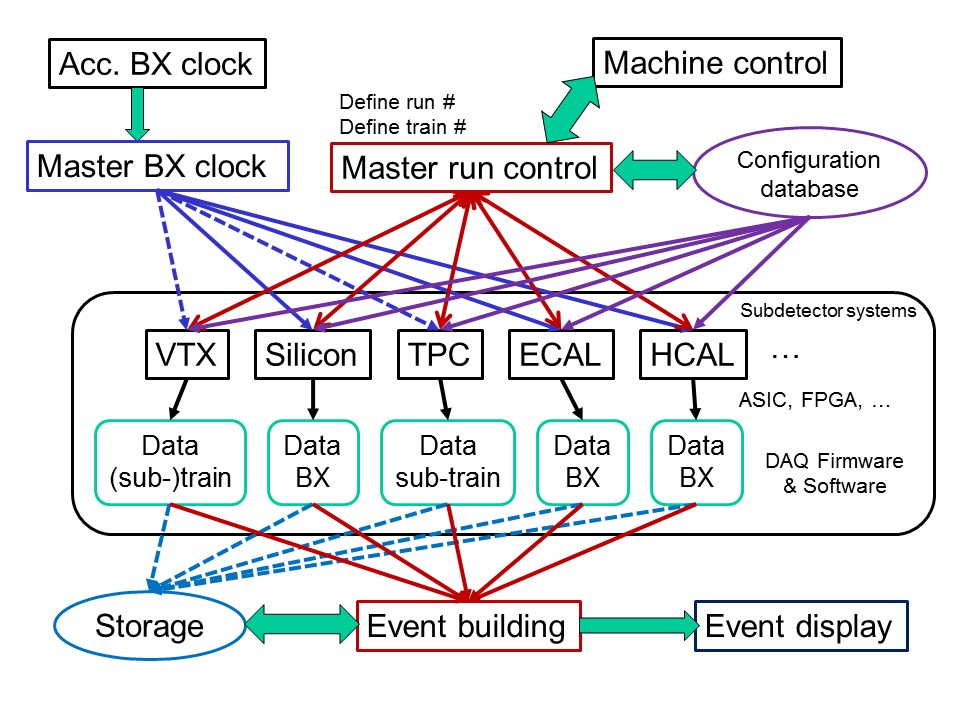}
\caption{\label{fig:integration:DAQ_architecture}Global architecture of the ILD data acquisition system.}
\end{figure}

The overall organisation of the DAQ system is summarized in Figure~\ref{fig:integration:DAQ_architecture}. 

The processing of subdetector data is first performed locally with a high parallelism within FPGA and ASIC boards which each typically treat O(100) subdetector channels. The individual data are zero-suppressed, amplified, time stamped and digitised, depending on the subdetector, and stored in local pipelines dimensioned adequately for a full bunch train. The "event building" is performed between bunch trains by gathering all subdetector data of a given bunch train into a single bunch train data set. The bunch train data sets are stored locally on the IP Campus (section 6.1) and transferred to an offline computing farm which may be located on the main campus. In the farm each bunch train data set is processed individually by one processor in order to identify individual events, tag their bunch crossings and perform calibration and reconstruction. This scheme caters for the different time resolutions of the subdetectors, which range from individual bunch tagging as done e.g. in the calorimeters, to a full train integration time as in the TPC.

One critical component of the system is the clock distribution to the subdetector front-ends: this has to be done on the overall ILD detector with a precision better than 10 ps to enable time-of-flight measurements. On the other hand the overall data flow requirements are moderate. They have not changed significantly compared to the estimates reported in~\cite{ild:bib:ilddbd}, and correspond to a raw data rate of O(100MB)/train for a storage data size of O(10PB)/year. The data flow is dominated by high granularity detectors exposed to high beam background close to the beampipe (VTX and BeamCAL). If needed it could be further reduced by local partial event processing ahead of the event building.

\subsection{DAQ R\&D}

The front-end processing has benefited from the development of subdetector technological prototypes including final readout front-end components as reported in section 5.2. ASIC and FPGA boards with the required ILD specifications are now available for most subdetectors.

A number of common DAQ aspects relevant for the final ILD central DAQ have been developed \cite{ild:bib:AIDADAQ} within the AIDA-2020 programme. They include both hardware and software components which have been used for different detectors or combinations of detectors in beam tests involving not only ILD prototypes but also detectors developed for the HL-LHC. 

\begin{figure}[t!]
\includegraphics[width=1.0\hsize]{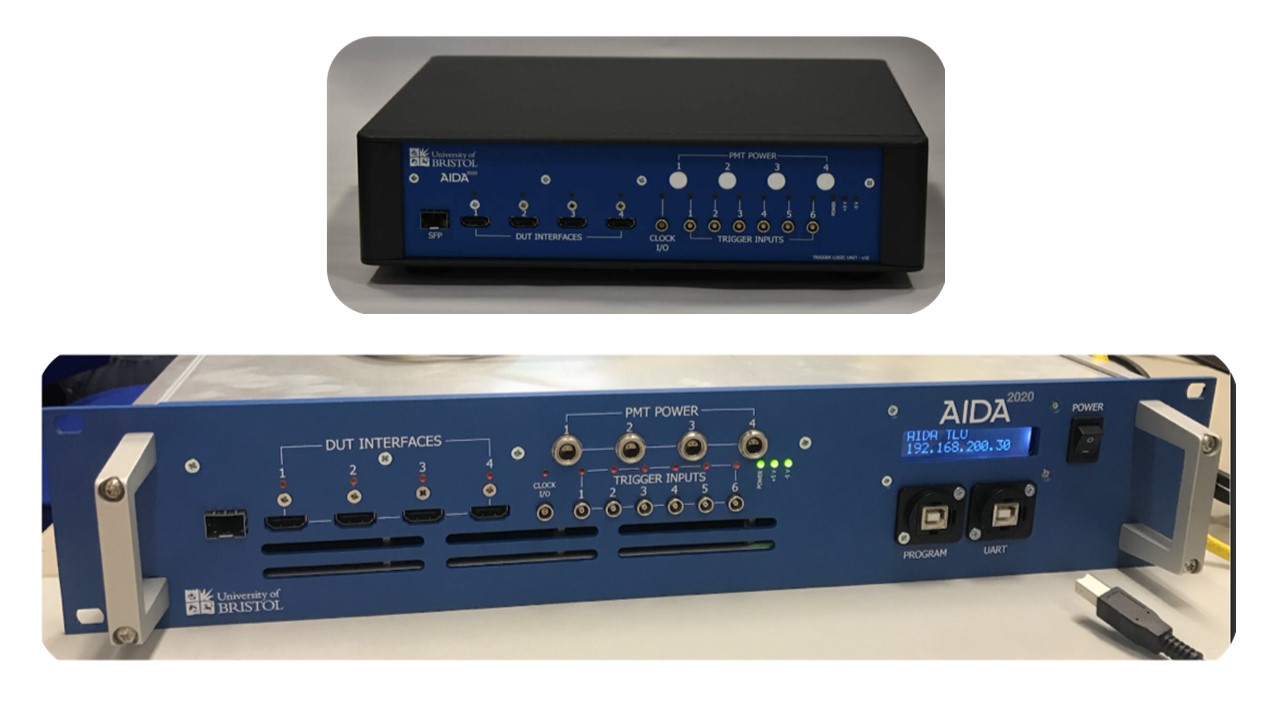}
\caption{\label{fig:integration:DAQ_TLU}The AIDA-2020 newly developed Trigger Logic Unit. Top: table top enclosure. Bottom: 19-inch rack-mounted enclosure. The two enclosures have identical inner components and provide the same functionalities.}
\end{figure}

The core hardware component of the central DAQ system, the Trigger Logic Unit (TLU), provides a common interface for synchronisation and control of subdetectors. A new TLU has been developed to distribute signals to multiple detectors in beam tests. The AIDA-2020 TLU (Figure~\ref{fig:integration:DAQ_TLU}) has been extended over the previous version to be able to synchronise detectors with differing trigger and readout schemes, such as the CALICE calorimeters, pixel detectors and tracking devices, as well as to be able to operate at a higher particle flux. Therefore, data from different detectors corresponding to the same particle in a test beam can be combined. The TLU is used extensively in test-beam facilities at CERN and DESY and will continue to do so thereby serving the needs of many detectors also beyond ILC R\&D. 

Within AIDA-2020 DAQ software packages have also been developed for both data collection and monitoring. The EUDAQ2 software is an extension of the EUDAQ software developed for the EUDET pixel beam telescope. It supports detectors with different trigger schemes and different readout speeds, enabling combined beam tests of very different detectors to be performed. The software has been tested and used in several beam tests so far, notably for the AHCAL (also in conjunction with the CMS HGCAL), and for tracker modules of the ATLAS upgrade.  The TLU is also fully integrated into the EUDAQ2 framework. The monitoring software, DQM4HEP, originally developed for the SDHCAL, has been used for data quality monitoring and slow control monitoring. This software is a generic development which has extendibility built in and so can be used to monitor data from any detector in any format. It has already been successfully used by the AHCAL+beam telescope and SDHCAL+SiECAL combined beam tests. 

These DAQ developments have allowed multiple detectors to be more easily integrated, sparing time for understanding technical aspects of the detector technologies and the physics performances.  Further detector developments will also benefit from this. The work also informs our understanding of integration and DAQ issues for the final ILD.

\section{Calibration/ Alignment procedures}

The ILD detector is a high-precision instrument. It can reach its designed level of performance only if the sub-systems can be fully calibrated, if the different parts can be aligned relative to each other, and if the calibration and the alignment of the system can be maintained over a long time. 

For the purpose of this document all tasks which concern the internal description of a sub-detector response are defined as "calibration". Relative positioning of internal parts within a sub-detector, and positioning of sub-detectors relative to each other, are defined as "alignement". 

A detailed calibration and alignment concept of the ILD detector has been presented in the letter-of-intent~\cite{ild:bib:ILDloi} and remains valid. Here the main arguments are repeated and their implementation updated wherever appropriate. In particular significant progress has been made since the publication of the LOI in the experimental demonstration of calibration and alignment using prototypes of the different sub-detectors. 

Data play a central role in the overall calibration and alignment strategy of ILD. Particles recorded during physics running are a powerful tool for the relative alignment of different sub-detectors. They are supplemented by detailed and high precision measurements taken during the construction phase of ILD, and  augmented in some cases by special data taken either from dedicated calibration or alignment systems, or in special calibration runs. 

A particular challenge for ILD is the proposed push-pull scheme. Push-pull implies that the detector is frequently moved from the interaction position into a parking position, and back. Calibration and alignment need to be re-established after each move within a short time, ideally of less than a day, to minimise the loss of integrated luminosity due to re-calibration. 
Data also play a central role in the re-establishment of the calibration and alignment after a push-pull move. A fast re-establishment of the full calibration is only possible if external means, e.g., metrology or built in sensors, first allow a quick re-establishment of the base calibration of ILD. Based on data recorded during a normal physics run, the full calibration and alignment is then re-established and re-calculated during the physics running, without loosing additional luminosity. Only the very first part, needed to find the base-calibration would potentially result in lost data. 

It is important to understand the required tolerances which are different for different sub-systems. Using a simple track model the dependence of track parameters on alignment tolerances have been studied: 
\begin{itemize}
    \item coherent displacement of the VTX: $2.8 \mu$m;
    \item coherent displacement of the SIT: $3.5 \mu$m;
    \item coherent displacement of the SET: $6 \mu$m;
    \item coherent displacement of the TPC: $3.6 \mu$m;
    \item coherent displacement of the ECAL: $100 \mu$m;
    \item coherent displacement of the HCAL: $1000 \mu$m.
\end{itemize}
These numbers, which need to be confirmed by further more detailed studies, nevertheless issue important guidelines when optimising the different sub-detectors. 

Another central ingredient into the overall precision reachable for the reconstruction is the degree of knowledge of the magnetic field in ILD. Uncertainties on the size and the direction of the field will directly translate into errors on track parameters. The field knowledge will come from two sources. First, high precision measurements of the magnetic field and its direction will be performed during the building of the detector. These measurements should be able to reach a relative precision of $dB/B < 10^{-4}$.
Second, once data are collected stiff tracks will provide a continuous source of tracks for the further relative calibration of the magnetic field, and the correction of local distortions. 

The calibration of the calorimeter systems will rely strongly on test-beam based calibration runs before installation. Independent of which technology will be selected in the end, it is envisioned that each module of both the ECAL and HCAL will be exposed to test-beam previous to installation, in order to provide a base calibration. Significant experience with prototypes over many years has shown that the calibration of the ILD calorimeter is robust over time, and can be tracked based on samples of real data. 

The needs for calibration and alignment have to be taken into account from the start of the development of the overall ILD concept. The design of the mechanical structure, of the supports, and of the integration strategy all will have an impact on the calibration and alignment, and will need to be considered. 

Overall it is thought that the current design and integration of ILD will allow to build a detector which can be calibrated to the required precision, aligned mostly based on data, and maintained for long times in a well calibrated and well aligned state. Nevertheless it should be noted that the detailed technical validation and demonstration of the feasibility of the different concepts is still needed before a final design of ILD can be frozen. 

\section{Earthquake Safety}
\label{ild:sec:earthquake}
\begin{figure}[h!]
\centering
\includegraphics[width=0.8\hsize]{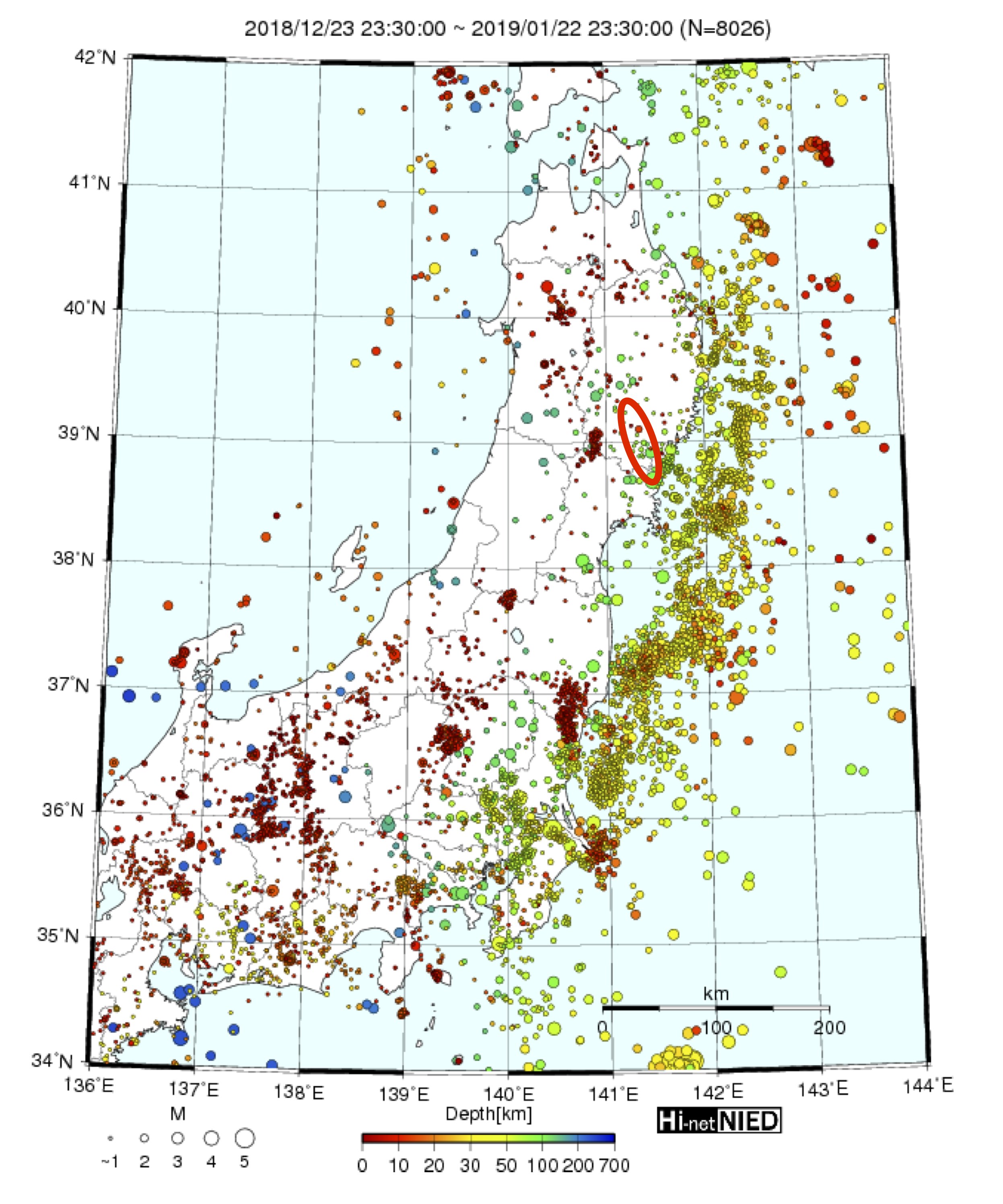}

\caption{\label{ild:fig:integration:earthquake_map}Map of northern Honshu with all detected earthquakes between 23rd December 2018 and 23rd January 2019 (30 days)~\cite{ild:bib:hi-net}. The ILC site is indicated by the red ellipse.}
\end{figure}

Japan is one of most seismically active regions in the world. The proposed site for the ILC in the Kitakami mountains of northern Honshu has been especially selected putting emphasis on benign seismic conditions. Earthquake history has been taken into account as well as recent surveys on active tectonic faults. Nevertheless, earthquakes can occur and need to be taken into account. Figure~\ref{ild:fig:integration:earthquake_map} shows all earthquakes that were detected during 30 days in winter 2018/2019 in northern Honshu. The ILD interaction region is in the region around (39N, 141,5E).

\subsection{Structural Design}

Emphasis has been put on the design of the ILD detector with respect to earthquake safety in view of operability of the detector as well as disaster prevention. General rules are provided by the ISO3010 standard: "Bases for design of structures -- seismic actions on structures". The ISO standard states two basic principles:
\begin{itemize}
\item The structure should not collapse nor experience other similar forms of structural failure due to severe earthquake ground motions that could occur at the site (ultimate limit state: ULS).
\item The structure should withstand moderate earthquake ground motions which may be expected to occur at the site during the service life of the structure with damage within accepted limits (serviceability limit state: SLS).
\end{itemize}

The design seismic forces on mechanical structures can be determined by taking into account normalised design response spectra, weighted with load factors that take into account the respective state, ULS or SLS, local conditions and structural factors. In addition, required degrees of reliability of the structures can be taken into account.

\begin{figure}[t!]
\begin{center}
\includegraphics[width=0.8\hsize]{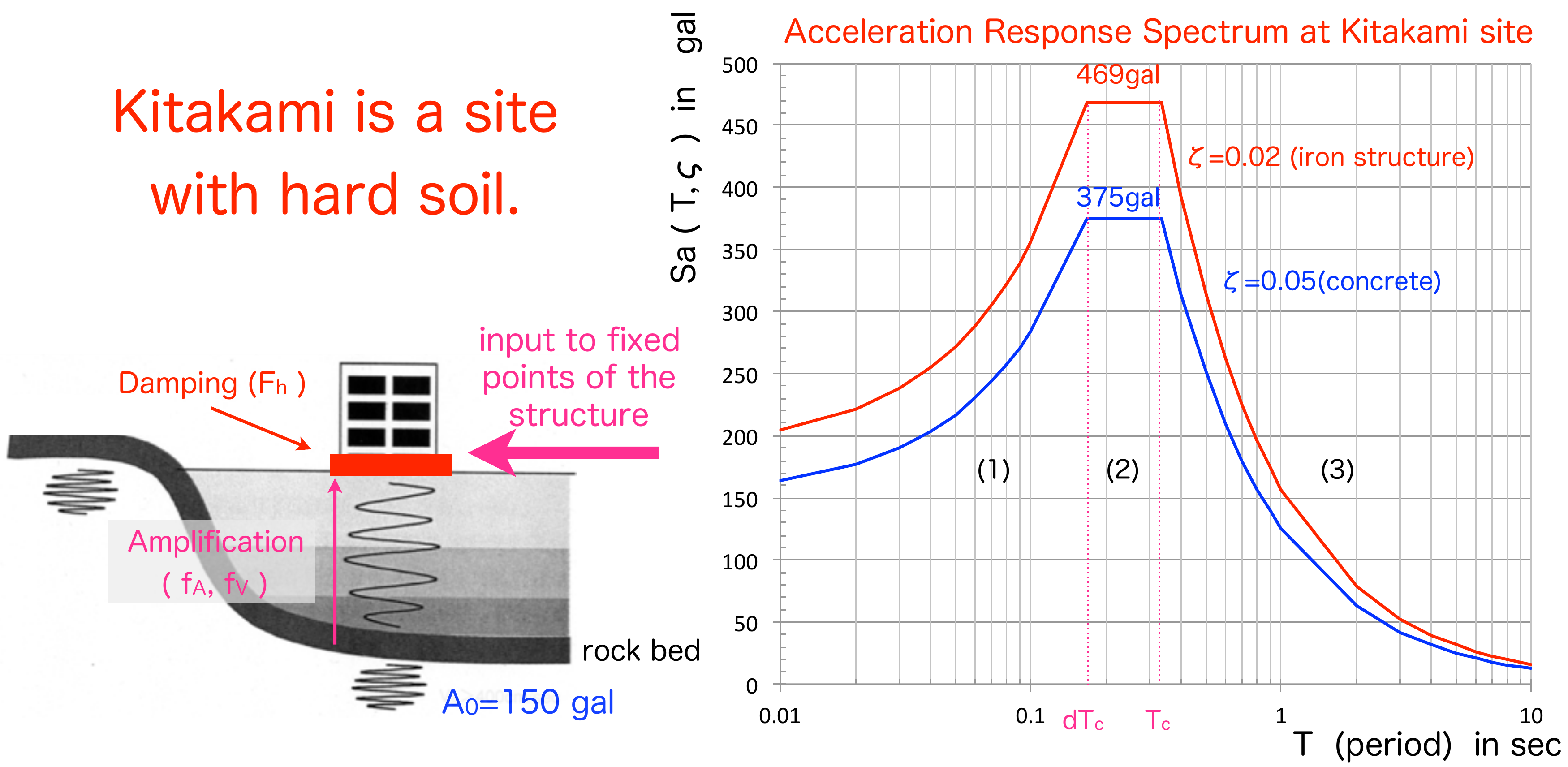}
\end{center}
\caption{\label{ild:fig:integration:earthquake_spectra}Standard response spectra for earthquakes in Kitakami (hard soil) for structures with different damping behaviour~\cite{ild:bib:earthquake}. A maximum acceleration of 150~gal (1 gal = 1 cm/s$^2$) for an earthquake with a recurrence frequency of 100~years has been assumed.}
\end{figure}

An acceleration response spectrum for the SLS state at the Kitakami site is shown in Figure~ \ref{ild:fig:integration:earthquake_spectra} for structures with different damping behaviours. In the ULS state, the accelerations are assumed to be larger by about a factor of two. These type of standard spectra serve as input for the structural studies described in section~\ref{ild:sec:mechanical_structures}.

\subsection{Seismic Isolation}

\begin{figure}[h!]
\begin{center}
\begin{subfigure}{0.9\hsize} \includegraphics[width=\textwidth]{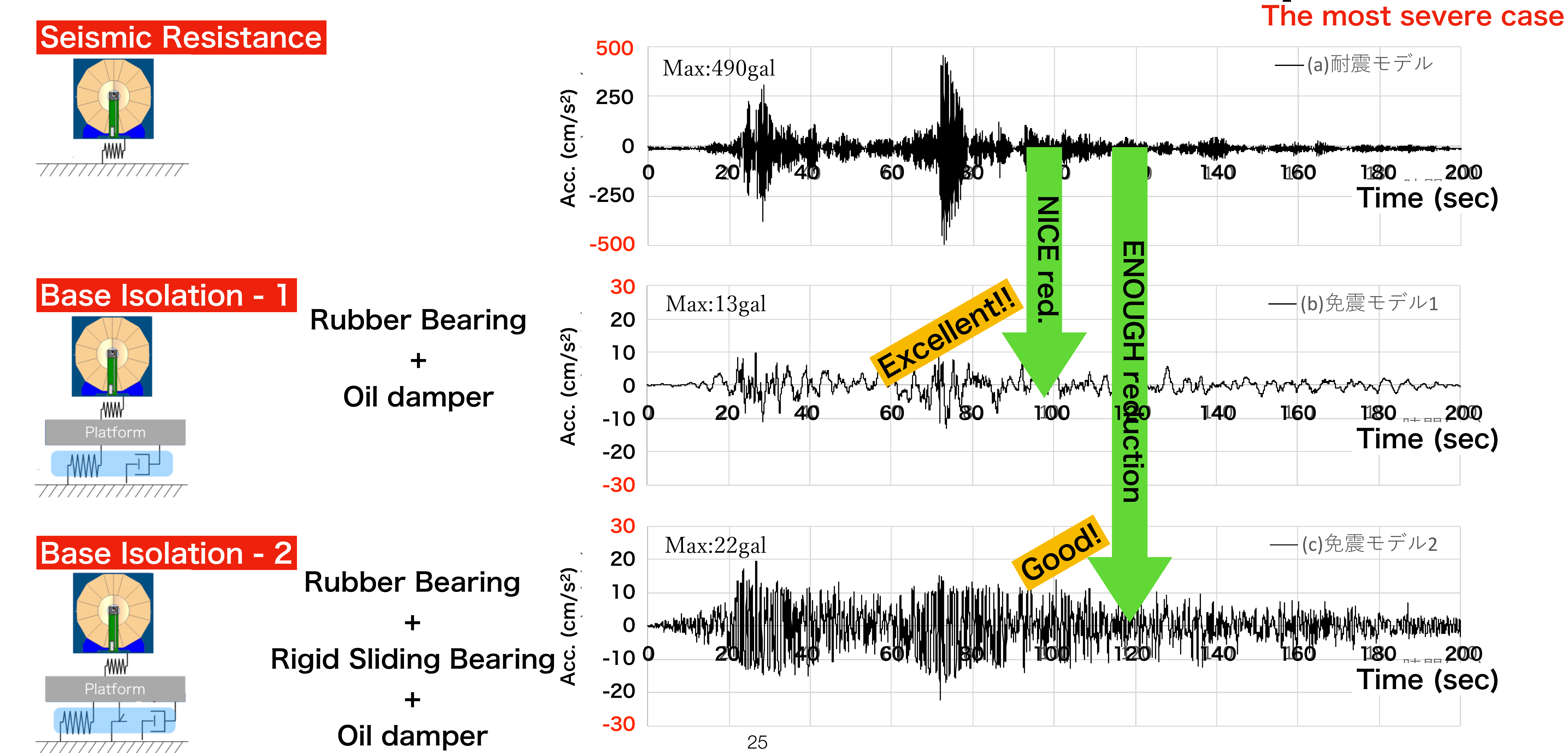}
 \caption{ \label{ild:fig:integration:damping_earthquake}}
 \end{subfigure}
\hspace{0.03\textwidth}
\begin{subfigure}{0.9\hsize} \includegraphics[width=\textwidth]{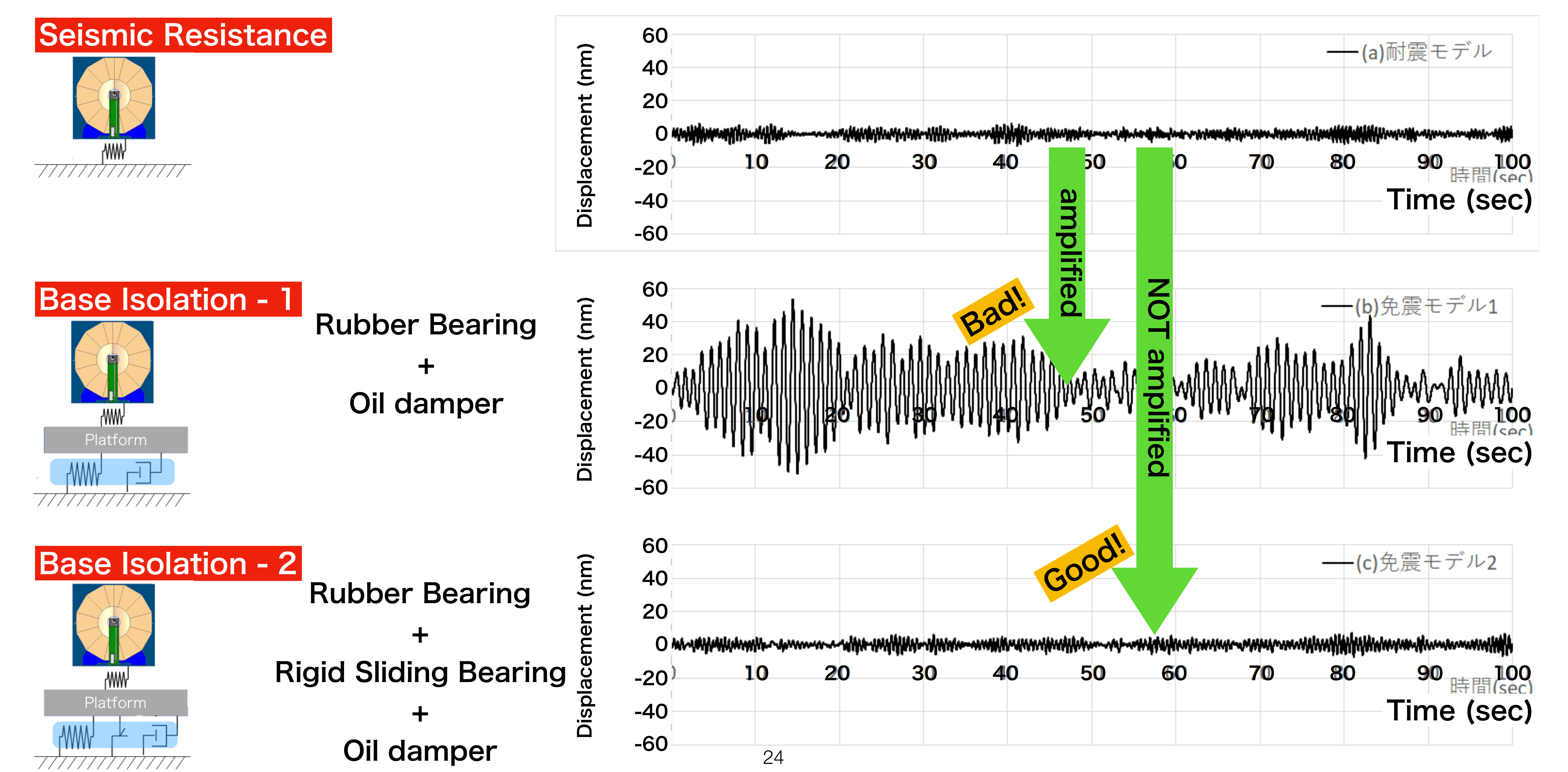}
 \caption{  \label{ild:fig:integration:damping_microtremor}}
 \end{subfigure}
\end{center}
\caption{
(a) Simulated accelerations on the ILD detector in case of a catastrophic earthquake without base isolation (top) with an rubber bearing/oil damper system (middle) and with an additional rigid sliding bearing (bottom). Acceleration spectra from the 2011 Tohoku earthquake have been used as a case study for a most severe case~\cite{ild:bib:Seismic_Damping}.
(b) Simulated displacements of the ILD detector in case of microtremors without base isolation (top) with an rubber bearing/oil damper system (middle) and with an additional rigid sliding bearing (bottom) \cite{ild:bib:Seismic_Damping}.
}
\end{figure}

Seismic isolation could be an attractive solution to mitigate the risk of earthquake induced accidents. Base isolation systems are in use in many cases in seismic active countries like Japan, under buildings, bridges and other critical infrastructure. A recent study~\cite{ild:bib:Seismic_Damping} has explored the possibility to install seismic damping systems underneath the ILD detector, e.g. under the detector platform. Standard isolation systems consist of rubber bearings in combination with oil dampers. Figure~\ref{ild:fig:integration:damping_earthquake} shows that accelerations acting on the ILD detector could be damped by a factor of $\approx$35 in case of a catastrophic earthquake. At the same time, such a system unfortunately increases the displacements of the detector in case of seismic mircrotremors that occur frequently in Japan, also at the Kitakami site~(c.f.~Figure~\ref{ild:fig:integration:earthquake_map}). An extended system that adds a rigid sliding bearing could mitigate the problem. Here, friction keeps the detector coupled to the ground at lower acceleration forces. Figure~\ref{ild:fig:integration:damping_microtremor} shows that no amplification of microtremors happens in that case. At the same time, the damping behaviour in case of major earthquakes is still in the order of a factor of $\approx$22 better than without base isolation~(Figure~\ref{ild:fig:integration:damping_earthquake}). 

A seismic base isolation system underneath the detector platform could therefore mitigate seismic risks for the ILD detector. However, earthquakes could also occur when the detector is still under construction or has been opened and moved partially away from the platform. Systematic risk evaluations and the development of mitigation strategies for the full lifetime of ILD still need to be done.

\section{Technical Documentation}

\begin{figure}[t!]
\centering
\includegraphics[width=0.8\hsize]{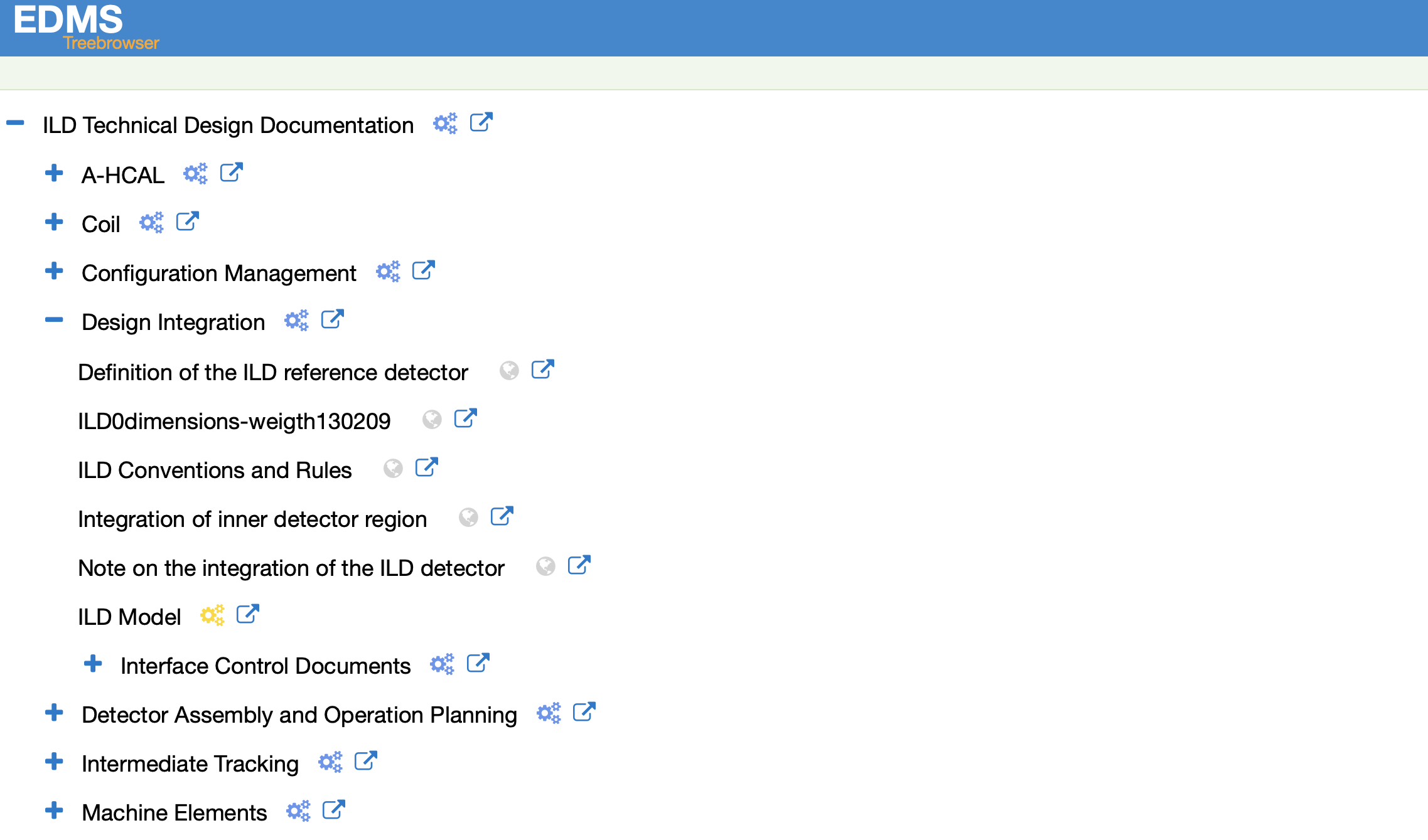}

\caption{\label{ild:fig:integration:edmsdirect}Part of the treebrowser of the ILD technical documentation Work Breakdown Structure in the ILC Engineering Data Management System. The top level node "Design Integration" is shown expanded.}
\end{figure}

The technical description of the ILD detector concept comprises specification and design documents as well as 3D engineering models, interface descriptions and drawings. All the documents that form the backbone of the ILD technical documentation are stored in the ILC Engineering Data Management System ILC-EDMS~\cite{ild:bib:edms}. As this powerful system is made for experts and requires appropriate attention, an easy accessible frontend ("EDMSdirect") has been made available. The ILD technical documentation is organised in a Work Breakdown Structure (WBS) that is mapped on a tree browser that can be accessed on the web~\cite{ild:bib:edmsdirect}. Figure~\ref{ild:fig:integration:edmsdirect} shows the tree browser for the ILD technical documentation. All WBS top nodes can be expanded by clicking on them. In the Figure, this has been done for the node "Design Integration".

\begin{figure}[h!]
\centering
\includegraphics[width=0.8\hsize]{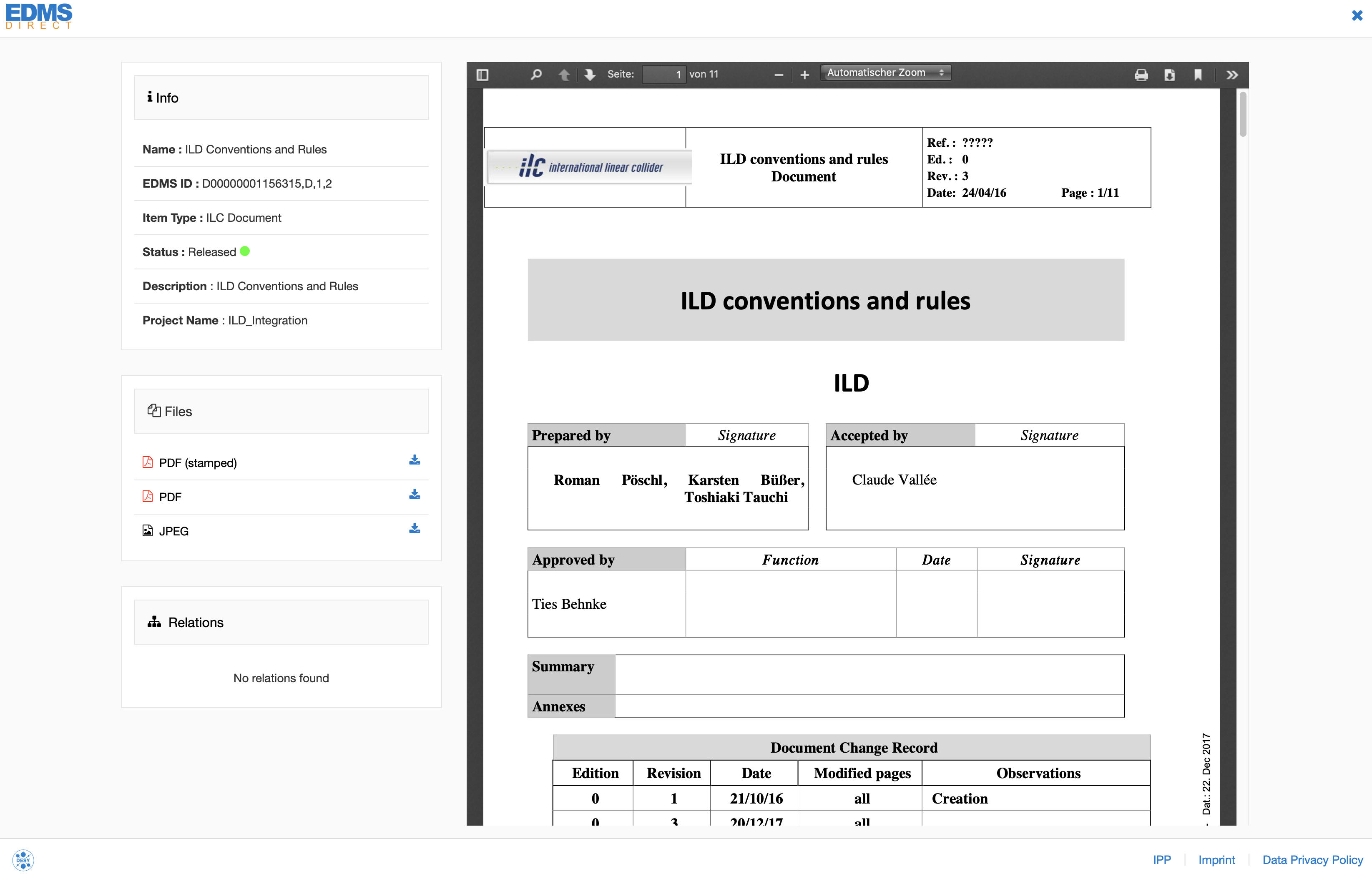}

\caption{\label{ild:fig:integration:edmsdirect_document}Example document (here:"ILD Conventions and Rules") in the document browser of EDMSdirect.}
\end{figure}

Figure~\ref{ild:fig:integration:edmsdirect_document} shows the document browser that opens for the documents stored in the EDMSdirect system. Shown here is the "Conventions and Rules" document that belongs to the previous mentioned WBS node "Design Integration". A preview of the document is shown in the document browser. The document browser allows for preview of the respective documents as well as downloads of pdf or source files, depending on the authorisation of the users. Documents can be made worldwide visible as well as protected for selected users.

\subsection{Interface Control Documents}
The technical integration of the ILD detector is based on a set of Interface Control Documents which describe the interfaces of each subdetector to its environment:
\begin{itemize}
    \item Mechanical Interface
    \begin{itemize}
        \item Local coordinate system
        \item Mechanical concept
        \item Critical dimensions
        \item Weights
        \item Positioning and alignment constraints
    \end{itemize}
    \item Electrical Interface
    \begin{itemize}
        \item Block and connection diagrams
        \item List of connectors
        \item Cabling and connection sheets
        \item Grounding circuits
        \item Power consumption
    \end{itemize}
    \item Gas and Liquid Interfaces
    \item Thermal Interfaces
    \item Test Interfaces
\end{itemize}
The Interface Control Documents are living documents that are updated as the design and understanding of the subdetector technologies advance. All documents are stored in EDMS~\cite{ild:bib:edms} and are accessible via the treebrowser of the ILD technical documentation~\cite{ild:bib:edmsdirect}.

A dedicated document describes the ILD Conventions and Rules (c.f.~Figure~\ref{ild:fig:integration:edmsdirect_document}). It contains a definition of the global ILD coordinate system, unit conventions, naming and numbering conventions as well as mechanical, electrical and cooling constraints and rules. This document is also available on~\cite{ild:bib:edmsdirect}. 
The interface descriptions together with the conventions and rules contain the information that has been used in creating the integrated model of the ILD detector.

%
%

\newcommand{\CPP}{C\nolinebreak\hspace{-.05em}\raisebox{.4ex}{\tiny\bf +}\nolinebreak\hspace{-.10em}\raisebox{.4ex}{\tiny\bf +}}

\chapter{ \label{chap:modelling}Physics and Detector Modelling}

Accurate and detailed modelling of the physics interactions as well as the detector
response are crucial for making realistic predictions about the expected physics and detector
performance. The ILD software for detector simulation, reconstruction and analysis is entirely
based on the common linear collider software ecosystem called \emph{iLCSoft}~\cite{bib:ilcsoft}.
It will be described in more detail in the next sections.

\section{\label{sec:generator}Modelling of ILC Conditions and Physics Processes}

Large, realistic Monte Carlo samples with the full Standard Model physics ($E_{cms}=500~\GeV$) as well as various
BSM scenarios have been created for the purpose of detector optimization and performance evaluation, to be presented in detail in
chapter~\ref{chap:performance}. The input samples for the full detector simulation
are created with the Whizard~\cite{Kilian:2007gr} event generator.
Whizard uses tree-level matrix elements and loop corrections to generate events with the final state partons and leptons
based on a realistic beam energy spectrum, the so called \emph{hard sub-process}. The hadronization into the visible final state
is performed with Pythia~\cite{Sjostrand:2006za} tuned to describe the LEP data.
The beam energy input spectrum is created with Guinea-Pig~\cite{Schulte:1998au}, a dedicated simulation program for computing
beam-beam interactions at linear colliders - see also the discussion in section~\ref{sec:beam:conditions}.
%
\thisfloatsetup{floatwidth=\textwidth,capposition=below}
\begin{figure}[h!]
  \begin{subfigure}{0.49\hsize}
    \includegraphics[width=\textwidth]{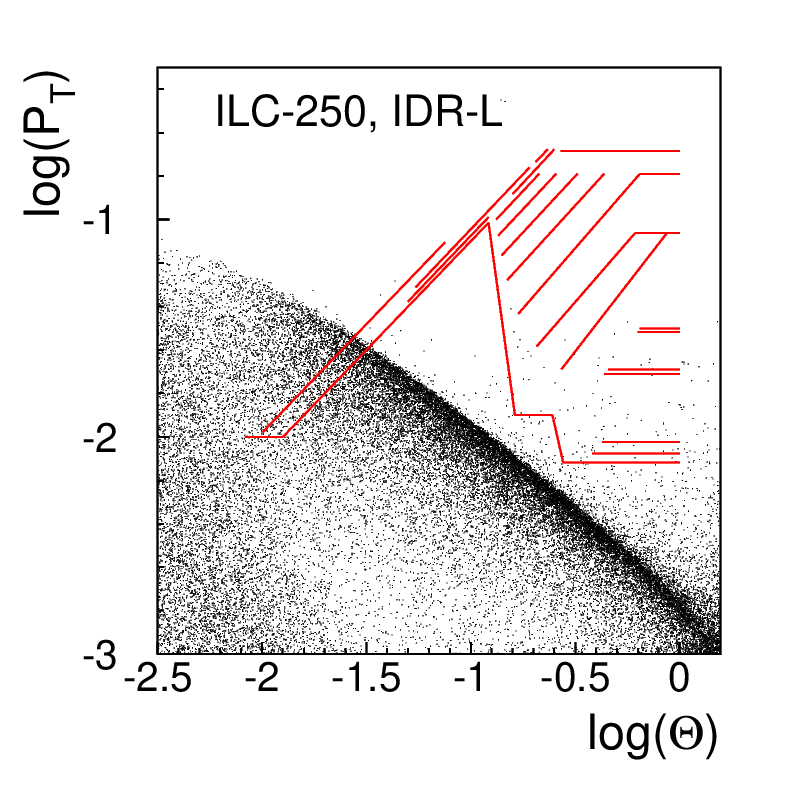}
    \caption{ \label{fig:pair_bg_cone_250}}
  \end{subfigure}
  \begin{subfigure}{0.49\hsize}
    \includegraphics[width=\textwidth]{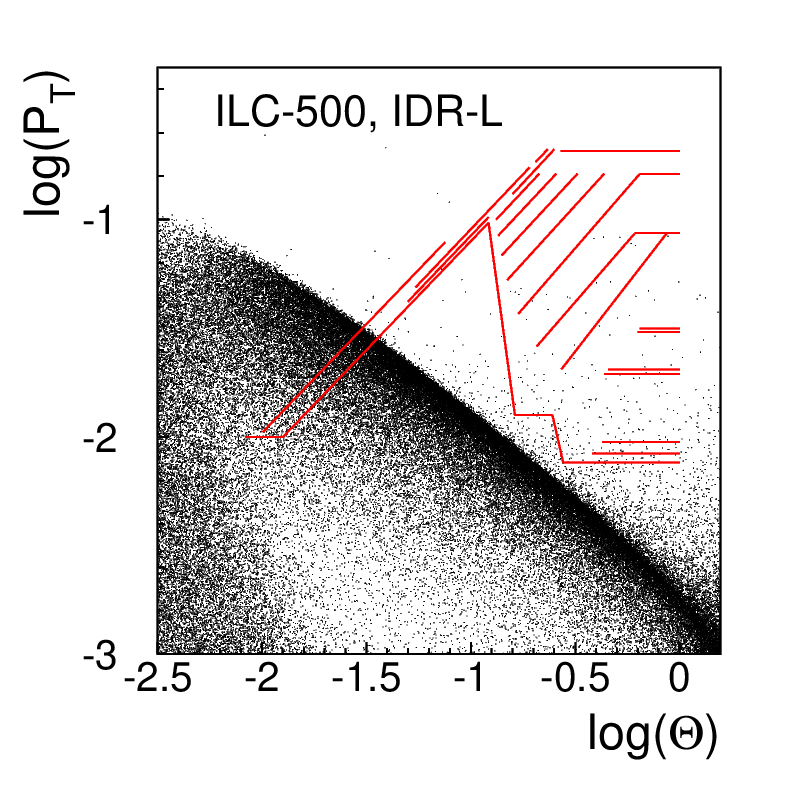}
    \caption{ \label{fig:pair_bg_cone_500}}
  \end{subfigure}
\caption{\label{fig:pair_bg_cone} Cones of incoherent $\Pep\Pem$-pairs in the ILD detector for $E_{cms}=250~\GeV$ (a) and $E_{cms}=500~\GeV$ (b)
  as created with GuineaPig . Shown is $\log{p_t}$ of the particles,
  corresponding to the largest radial extend of the helical trajectory as a function of $\log{\theta}$.
  Also shown are the inner detector elements of the ILD detector (horizontal lines represent
  barrel elements and diagonal lines represent end-cap elements). All detector layers are well outside of the background cone,
  except for the face of the BeamCal and the beam pipe endcap in front of it (the two leftmost diagonal lines in the plot).}
\end{figure}
The two main effects of the strong beam-beam interactions are the energy loss due to
beamstrahlung leading to the available luminosity spectrum (see Fig~\ref{fig:ilc:ecmspect}) and the creation of
incoherent $\Pep\Pem$-pairs that are the source of the dominating background at the ILC.
These electrons and positrons are predominantly created in a forward cone as shown in Fig~\ref{fig:pair_bg_cone}.
It is this cone that restricts the minimal allowed radius of the innermost layer of the vertex detector.

Another source of background at the ILC are $\Pgamma \Pgamma \to hadrons$ events, due to beamstrahlung photons.
These type of events are generated for $\Pgamma \Pgamma$ cms-energies from \unit{300}{\MeV} to \unit{2}{\GeV} with a dedicated generator based
on ~\cite{Chen:1993dba}, whereas for higher energies Pythia is used.

For the large Monte Carlo data sets both backgrounds are overlaid to the actual physics event that is simulated. For the incoherent pairs, only
\emph{reconstructable tracks}, i.e. those that leave at least three hits in the tracking detectors, are overlaid as these constitute an irreducible
background. The majority of the pair particles do not leave enough hits to be reconstructed as particle tracks and therefore give rise to a
large number of seemingly uncorrelated background hits. As shown in section~\ref{sec:performance:tracking}, these additional hits have little
effect on the resulting track finding efficiency and therefore can safely be omitted for the large scale production.
The situation is different for the BeamCal, which is hit by a very large number of pair particles at every bunch crossing as discussed in
section~\ref{sec:beam:background}. Here we us the detailed and complete background map for every bunch crossing, as simulated for the
case of an anti-DID.
$\Pgamma \Pgamma \to hadrons$ events are randomly overlaid according to Poisson-distributions describing the expected number of events per bunch
crossing (BX) for the four different combinations of the photon virtuality as shown in table~\ref{tab:ild_aalowpt}. Also shown in the table are
the spread of the z-position of the vertex and its mean value. It is interesting to note that the $\Pgamma \Pgamma \to hadrons$ events with exactly
one virtual photon have a non-zero offset of the vertex z-position. On average $1.1$~events per BX are overlaid.

\begin{table}[htbp]
\renewcommand{\arraystretch}{1.25}

\centering\small
\begin{tabular}{llll}
\hline
process type & Vertex z offset (\micron) & Vertex z sigma (\micron) & expected events per BX \\
\hline \hline
WW &	$0$ 	    &    $196.8$          &     $0.211$  \\
WB &	$- 42.22$    &   $186.0$          &     $0.246$  \\
BW &	$+ 42.22$    &   $186.0$          &     $0.244$  \\
BB &	$0$ 	    &    $169.8$          &     $0.351$  \\
\hline
\end{tabular}
\caption{Key parameters used in the overlay of $\Pgamma \Pgamma \to hadron$ background at $\sqrt{s}=500~\GeV$ for the four different combinations of photon
  virtualities. W denotes a virtual photon and B a real photon.\label{tab:ild_aalowpt} }
\end{table}

\section{\label{sec:det-sim}Detector Simulation}

The main core software tools in iLCSoft used by ILD are the common event data model and persistency tool LCIO~\cite{Gaede:2003ip},
the \CPP\ application framework Marlin~\cite{Gaede:2006pj} and the recently added generic detector description toolkit
DD4hep~\cite{Frank:2014zya,Frank:2015ivo}. DD4hep provides a single source of information for describing the detector geometry, its
materials and the readout properties of individual sub detectors. Various components of DD4hep provide different functionalities.
Here we use DDG4, the interface to full simulations with Geant4~\cite{Agostinelli:2002hh} and DDRec the specialized view into the
geometry needed for reconstruction.
%
%
\thisfloatsetup{floatwidth=\textwidth,capposition=below}
\begin{figure}[t!]
  \begin{subfigure}{0.40\hsize}
    \includegraphics[width=0.95\textwidth]{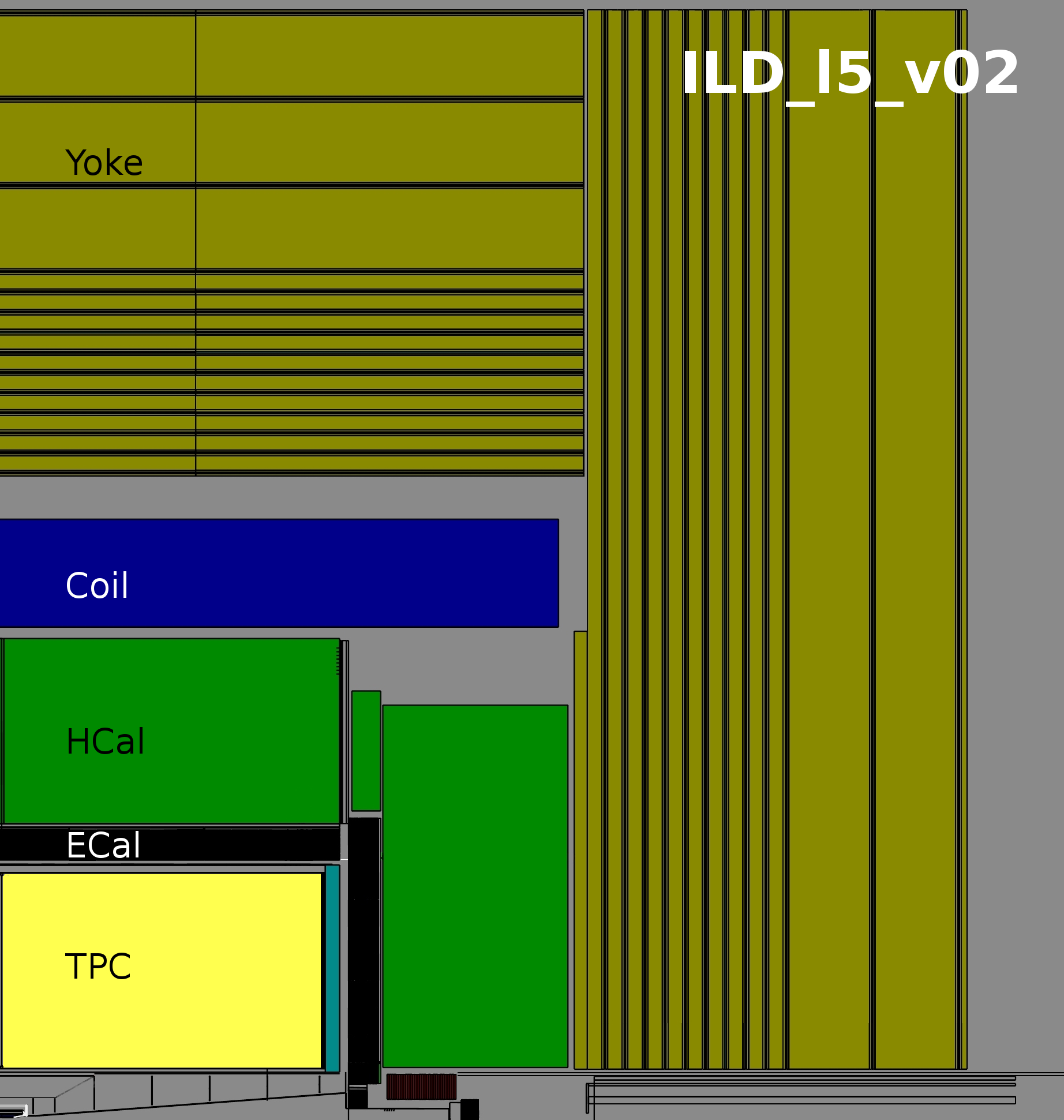}
    \caption{ \label{fig:sim_model_quad}}
  \end{subfigure}
  \begin{subfigure}{0.60\hsize}
    \includegraphics[width=0.95\textwidth]{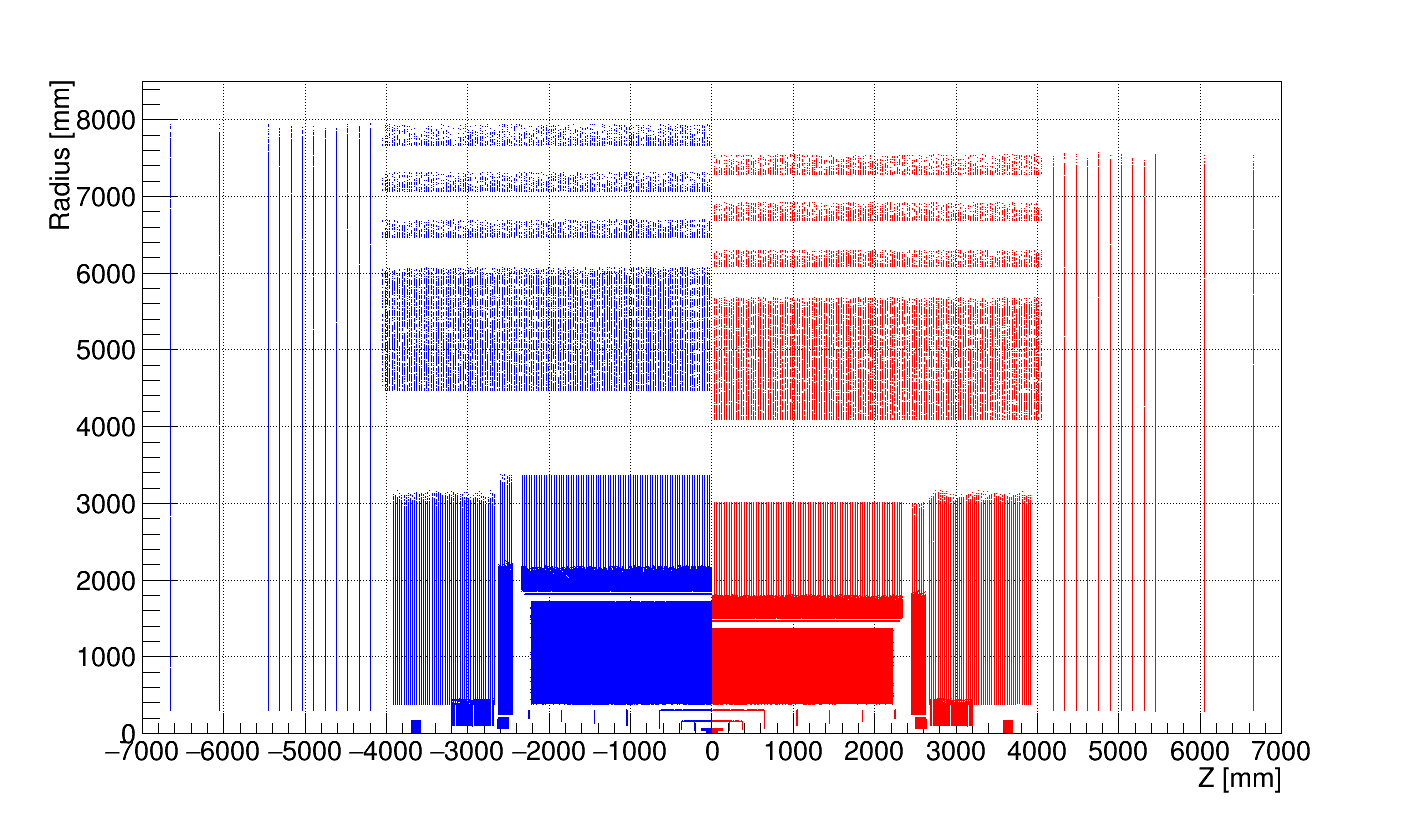}
    \caption{ \label{fig:sim_hits_LS_rz}}
  \end{subfigure}
  \caption{\label{fig:sim_models}(a): Quadrant view of the (large) ILD simulation model. Not labelled in the figure are the inner
  silicon tracking detectors: VTX, SIT and FTD, the outer silicon tracker SET as well as the forward calorimeters LumiCAL, LHCAL and BeamCAL.
  (b): Simulated hits in the large and small ILD simulation models in the $rz$-view.
  }
\end{figure}


\subsection{\label{sec:det-models}ILD Simulation Models}

\begin{figure}[b!]
  \begin{subfigure}{0.49\hsize}
    \includegraphics[width=\textwidth]{Modelling/fig/ILD_l5_v02_matbudget_tracker_85deg.pdf}
    \caption{ \label{fig:sim_matbudget_x0}}
  \end{subfigure}
  \begin{subfigure}{0.49\hsize}
    \includegraphics[width=\textwidth]{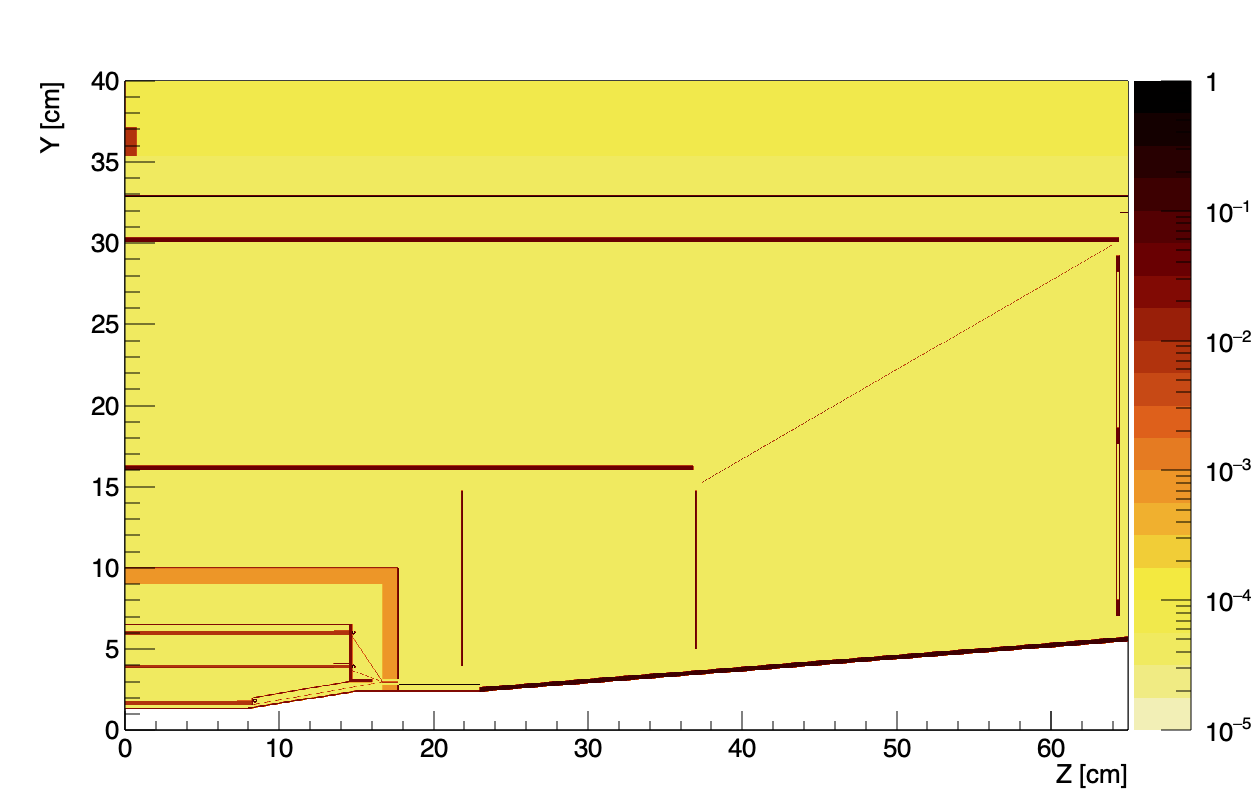}
    \caption{ \label{fig:sim_materialscan_vxd}}
  \end{subfigure}
  \caption{(a): Integrated radiation lengths of the tracking detectors in the ILD simulation models.
    (b): Material scan in inner tracking region of the simulation model showing detector components of the VTX, SIT and FTD as well
    as dead material from the beam pipe, support structures, cables and services. Plotted is the local material budget per bin in units of X0
    with an arbitrary scaling factor applied.) }
\end{figure}

For detector optimization studies two ILD simulation models, \emph{ ILD\_l5\_v02} (large)
and \emph{ ILD\_s5\_v02} (small) with the dimensions described in tables~\ref{ild:tab:barrelpara} and~\ref{ild:tab:endcappara},
have been implemented using DD4hep and released in a dedicated software package, lcgeo~\cite{bib:lcgeo}.
Note that for simplicity in the remainder of this document these two simulation models are referred to as \textbf{IDR-L} and \textbf{IDR-S}.
A quadrant view of the large simulation model is shown in Fig.~\ref{fig:sim_model_quad} together with a plot of simulated
hits in the large and small detector model for comparison in Fig.~\ref{fig:sim_hits_LS_rz}. The main differences between the two models (see chapter 4) are the reduced radii of
the TPC, the barrels of Ecal, Hcal, Yoke and the Coil and the increased B-field of \unit{4.0}{\tesla} for the small model,
compared to \unit{3.5}{\tesla} for the large model.
All other detector dimensions, in particular the longitudinal extents, and all material thicknesses have not been modified.
Thus, the two models have a different aspect ratio.
Detailed B-field maps with and without anti-DID have been created for both models and were used in the background studies presented in
section~\ref{sec:beam:background}. For the benchmark performance studies shown in chapter~\ref{chap:performance}
simplified solenoidal B-fields are used instead.
Considerable effort has been invested into making the ILD simulation models as realistic as possible, in particular by
\begin{itemize}
\item following the exact dimensions and layout of detector elements from engineering models
\item implementing correct material properties
\item implementing precise descriptions of the actual detector technology
\item adding realistic amounts of dead material from supports and services, such as cables and cooling pipes
\item introducing realistic gaps and imperfections into the subdetectors
\end{itemize}
Great care has been taken to include realistic material estimates, established by the detector R\&D groups,
in particular in the tracking region where the material budget has a direct impact on the detector performance.
As pointed out above, this includes dead material from supports and services.
Fig.~\ref{fig:sim_matbudget_x0} shows the  material budget in the ILD tracking volume resulting from the
individual tracking subdetectors including dead material and services. The spike at the edge of the tracking
fiducial volume at $\theta \approx 5^{\circ}$ is due to the shallow crossing of cables routed along the
conical part the beam pipe. Fig.~\ref{fig:sim_materialscan_vxd} shows the material distribution in the
inner tracking region close to the IP.


\thisfloatsetup{floatwidth=\textwidth,capposition=below}

\subsection{\label{sec:hybrid-sim}Hybrid Simulation}

\begin{figure}[b!]
\centering
  \begin{subfigure}{0.50\hsize}
    \includegraphics[width=\textwidth]{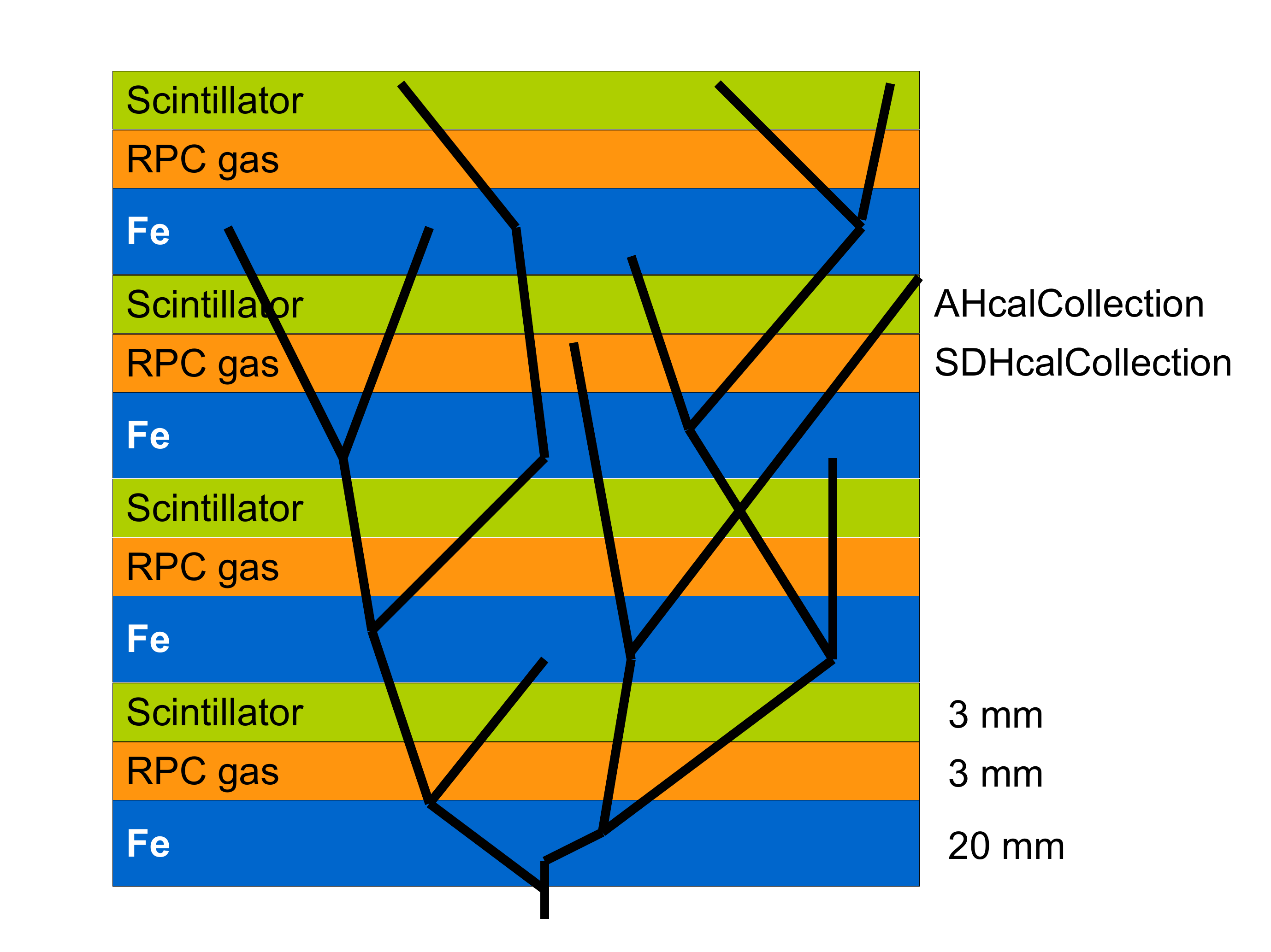}
    \caption{ \label{fig:sim_hybrid_schema}}
  \end{subfigure}
  \begin{subfigure}{0.49\hsize}
    \includegraphics[width=\textwidth]{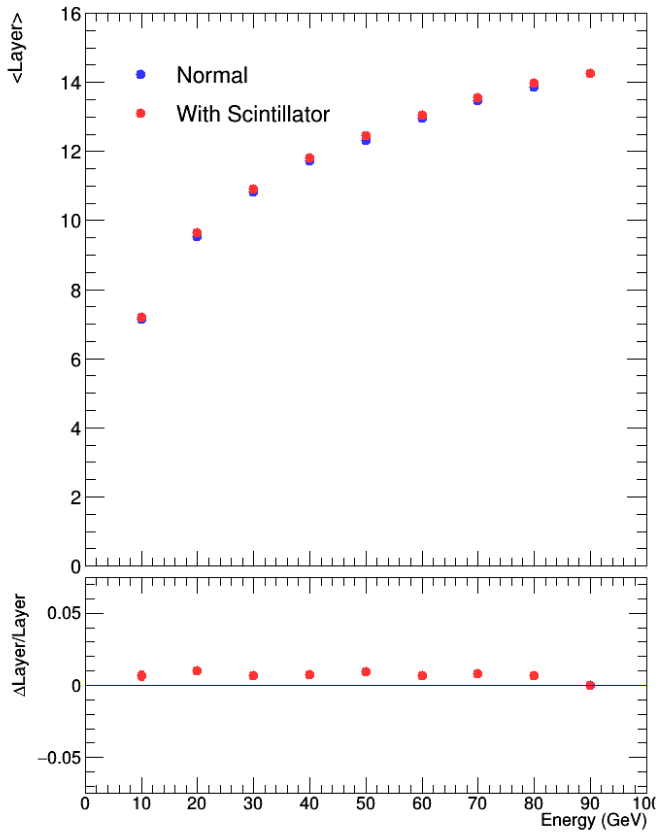}
    \caption{ \label{fig:sim_hybrid_longitudenal_profile}}
  \end{subfigure}
  \caption{(a): Schematic view of the \emph{hybrid simulation} model for two different
    calorimeter technologies using scintillator and RPCs respectively.
    (b): Longitudinal shower profile for the individual simulation of the RPCs (\emph{normal}) compared to
    the hybrid simulation (\emph{with scintillator}).}
\end{figure}

In order to be able to study and compare the different calorimeter technologies proposed for ILD
(see sections ~\ref{ild:sec:ECAL} and~\ref{ild:sec:HCAL}) \emph{hybrid simulation} models have been implemented for the
ILD simulation models, where two different readout technologies are implemented in the
gaps of the sandwich absorber structure as shown in Fig.~\ref{fig:sim_hybrid_schema}.
%
%
In this approach the respective other technology replaces the electronics and services present in the gaps in a
real calorimeter, resulting in a very similar material budget in the passive part of the layer.
The calorimeter shower development is additionally, almost entirely dominated by the absorber structure. Therefore,
this approach provides results that are equivalent to the stand alone simulation of each individual technology to
better than 1\%, as shown in Fig.~\ref{fig:sim_hybrid_longitudenal_profile} for the longitudinal shower profile. Equivalent results have
been obtained for other parameters, such as the total number of hits and the transverse shower profile.
With this approach one can simulate a large set of Monte Carlo events for
several calorimeter technologies using only little more CPU time than would be needed for simulating just one
technology choice and at the same time one can compare different technologies with identical physics events.
This hybrid simulation scheme has been implemented for the HCAL with the options of scintillator and RPC based readout as well as for
the ECAL with the options of silicon and scintillator based readout.

\section{\label{sec:reco}Event Reconstruction}
The detailed simulation with Geant4 provides hit objects with the exact amount and position of energy
depositions in individual sensitive detector elements, such as silicon pixels or calorimeter cells.
In the \emph{Digitization} step all relevant effects from the detector technology and the readout electronics
are applied to the simulated hits.

\subsection{Digitization}
For silicon strip-and pixel detectors as well as the ILD-TPC, the hit positions are smeared according to
resolutions that have been established from test beam campaigns for the different sensor technologies,
thereby including effects from charge sharing, clustering and position reconstruction.
Table~\ref{tab:ild_trk_res} shows the point resolution parameters used in the current ILD simulation models.
%
\begin{table}[htbp]
\renewcommand{\arraystretch}{1.25}

\centering\small
\begin{tabular}{llcl}
\hline
 Subdetector &  \multicolumn{3}{c}{ Point Resolution }  \\
\hline
        VTX    &  $ \sigma_{r\phi,z}  $  & $=$ &  \unit{3.0}{\micron} (layers 1-6) \\
        SIT    &  $ \sigma_{r\phi,z}  $  & $=$ &  \unit{5.0}{\micron} (layers 1-4) \\

        SET    &  $ \sigma_{r\phi}$      & $=$ &  \unit{7.0}{\micron} (layers 1-2, $\phi_{stereo} = \unit{7}{^\circ}$ )\\

       $\mathrm{FTD}_{Pixel}$     &  $\sigma_{r,r_\perp}$         & $=$ & \unit{3.0}{\micron} (layers 1-2) \\

       $\mathrm{FTD}_{Strip}$       &  $ \sigma_{r\phi}   $ & $=$ &  \unit{7.0}{\micron}  (layers 3-7, $\phi_{stereo} = \unit{7}{^\circ}$)   \\

       TPC    &  $ \sigma^2_{r\phi} $ & $=$ & $ \bigl( 50^2+900^2\sin^2\phi + \bigl( (25^2/22)\times$  \\
              &                      &     &   $(4~\rm{T}/B)^2\sin\theta\bigr) (z/\cm) \bigr)\,\micron^2$  \\
               &  $ \sigma^2_{z}    $ & $=$ & $ (400^2+80^2\times (z/\cm)) \,\micron^2 $ \\
               &   \multicolumn{3}{c}{ where $\phi$ and $\theta$ are the azimuthal and} \\
               &   \multicolumn{3}{c}{ polar angle of the track direction } \\
\hline
\end{tabular}
\caption[Simulated ILD tracking point resolutions.]{Effective point resolutions as used in the digitization of
  the ILD tracking detectors.
  The parameterization for the TPC takes into account geometric effects due to the direction of the track with
  respect to the pad row. All numbers and shown, have been established by the R\&D groups and have been demonstrated with
  test beam data.
        \label{tab:ild_trk_res} }
\end{table}
All point-resolutions have been demonstrated in test beams or reflect the current \emph{state of the art}.
In the TPC hit digitization, simulated hits that are closer than the established double-hit resolution
of 2~mm in $r\phi$ and 5~mm in $z$ are merged into one.
For the silicon detectors this treatment is not necessary, due to the expected low occupancies.
During the calorimeter digitization two calibration factors are applied to every simulated calorimeter hit:
\begin{itemize}
\item first, the deposited energy is normalized to the mean energy deposited by a minimum ionizing particle (MIP) in this particular calorimeter
  sub detector
\item  then the deposited, i.e. \emph{visible}, energy in units of MIPs is converted to a \emph{total cell energy} in \GeV, such that the
  sum of all hit energies corresponds to the incident particle's energy
\end{itemize}

The calibration of these factors is an iterative procedure. In a first step, the mean energy deposited by MIPs is computed from single \Pmuon-events.
In a second step, single particle events with \Pphoton and \PKzL at  \unit{10}{\GeV} are simulated and fully reconstructed with the \emph{particle flow algorithm}
and the resulting energy is used to compute updated calibration factors. This last step is repeated until the reconstructed energy agrees with the true particle
energy to within a given margin.
Different from the MC samples produced in the context of the DBD~\cite{ild:bib:ilddbd}, Birks' Law, describing the non-linear light yield at large loss rates in scintillator-based calorimeters, has not been applied during the production of the IDR-L and IDR-S MC samples due to a technical issue. Previous studies have shown that the effect on the energy resolution is very small and therefore there is no effect on the comparison of the two detector models expected.
Dedicated digitizers take into account
effects of non-uniformity of the light yield for scintillators as well as cross-talk between
neighboring channels.

\subsection{Track reconstruction}

\begin{figure}[h!]
\centering
\begin{subfigure}{0.48\textwidth}
\includegraphics[width=1\hsize]{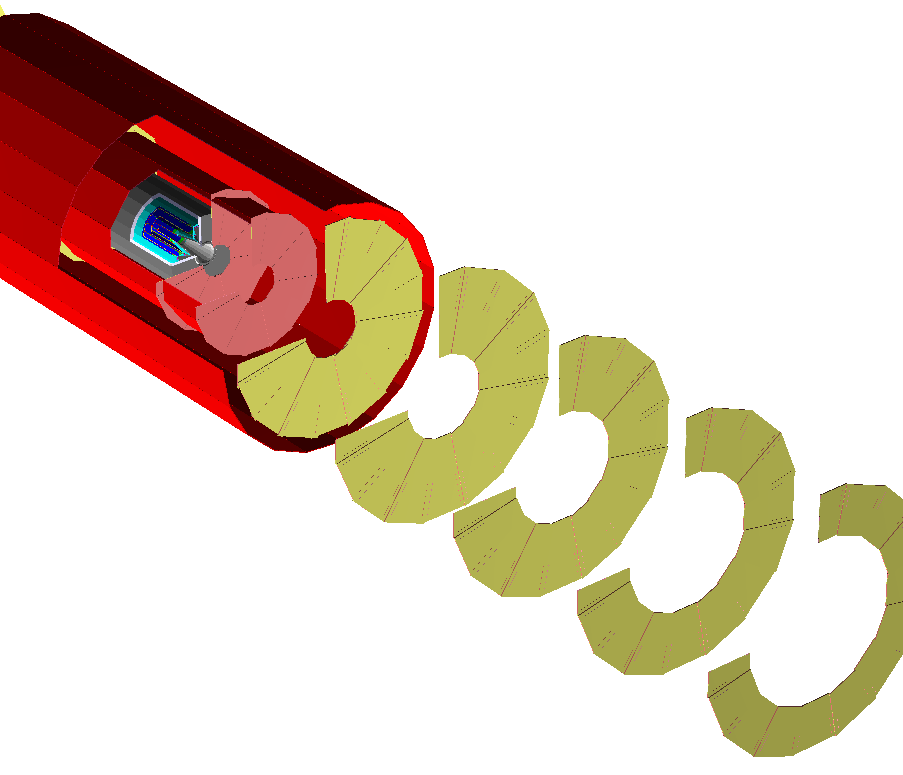} 
\caption{}
\end{subfigure}
\begin{subfigure}{0.48\textwidth}
\includegraphics[width=1\hsize]{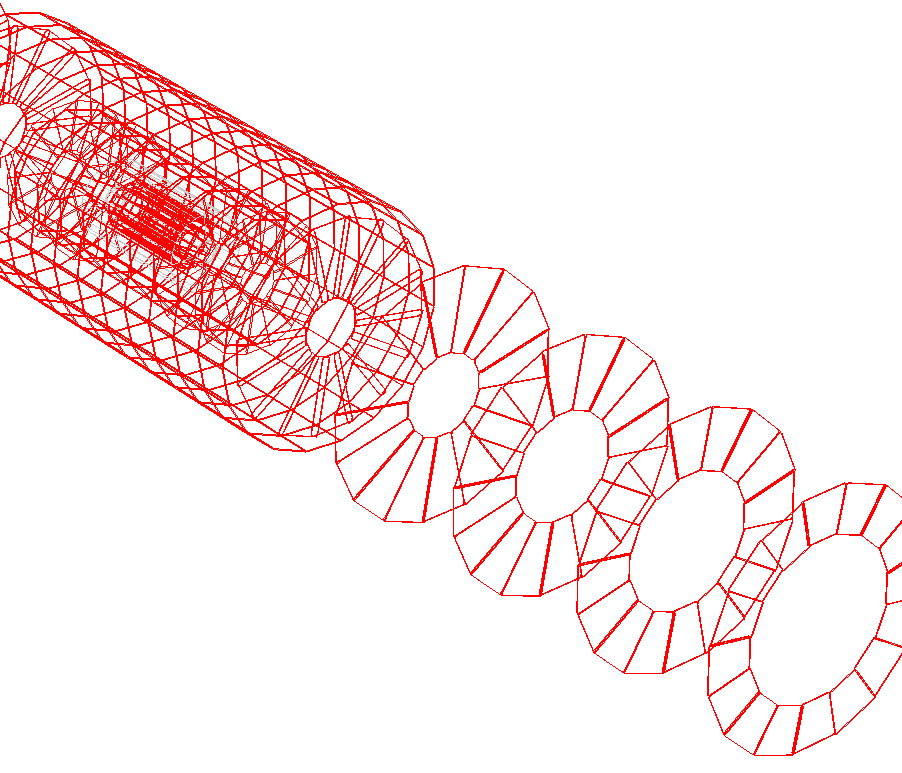}
\caption{}
\end{subfigure}
\caption{\label{fig:inner_trk_surfaces} (a) Inner tracking detectors in the ILD simulation model.
  (b) Surfaces used for track reconstruction, attached to the volumes of the simulation model.}
\end{figure}

The ILD track reconstruction is described in more detail in~\cite{Gaede:2014aza}. The \emph{pattern recognition}
step is carried out independently in three regions, the:
\begin{itemize}
\item inner Si-trackers  VTX and SIT ( and partly FTD)\\
  brute-force triplet seeding followed by a road search using extrapolations to the next layer
\item forward Si-tracker: FTD \\
  a Cellular-Automaton finding many track candidates; reduced to a unique and consistent set through the use of a Hopfield Network.
\item TPC \\
  topological clustering in the outer TPC pad row layers for seeding, followed by a Kalman-Filter based road search inwards
\end{itemize}
In a final step the track candidates and segments are combined into a unique set and then, after assignment of left-over hits, a final
refit is performed with a Kalman filter.
The correct reconstruction of the kinematics of charged particles requires a sufficiently detailed description of the material
the particles have traversed in order to correctly account for effects of energy-loss and multiple-scattering in the fit.
The DD4hep component DDRec provides dedicated surface classes for track reconstruction and fitting. These surface classes provide the
geometric information of the corresponding measurement surfaces as well as material properties, averaged automatically from
the detailed model. Surfaces are also used to account for effects from dead material layers, such as support structures or cables and services.
Fig.~\ref{fig:inner_trk_surfaces} shows the tracking surfaces used for the inner tracking detectors of ILD.
%

\subsection{Particle Flow}

After the reconstruction of charged particle tracks the  \emph{particle flow algorithm} (PFA) is applied
to reconstruct particle showers in the calorimeters.  PFA aims at reconstructing every individual particle
created in the event taking the best available measurement for each given type, i.e.:
\begin{itemize}
\item charged particles\\
  using the momentum measured in the tracking detectors with the excellent resolution described in sec.~\ref{sec:system-performance}
\item photons\\
  measured in the ECAL with an energy resolution of $\sigma(E)/E \sim  17\% / \sqrt{(E/\rm{GeV})}$
\item neutral hadrons\\
  measured predominantly in the HCAL\footnote{hadronic showers often start in the ECAL and might extend into the Muon system -
    this is taken into account in PandorPFA} with an energy resolution of $\sigma(E)/E \sim  50\% / \sqrt{(E/\rm{GeV})}$ 
\end{itemize}

 The best jet energy measurement in hadronic events would be achieved if the above algorithm would work perfectly. However in reality
 there is always confusion in the assignment of individual \emph{CalorimeterHits} to Clusters and showers as well as in the assignment
 of tracks to clusters. The best PFA implementation to date is PandoraPFA~\cite{Marshall:2015rfa},  interfaced to Marlin in a dedicated package
 DDMarlinPandora. The ILD reconstruction with Pandora also utilizes the instrumented return yoke and the forward calorimeters.
 The input to PandoraPFA are the reconstructed tracks, candidates for  kinks and  $V0$s as well as all digitized calorimeter hits.
 A number of sophisticated clustering algorithms are then applied in an iterative way, thereby optimizing the track-cluster matching
 based on the momentum-energy consistency. The output of PandoraPFA is a list of reconstructed particles, typically referred to as
 \emph{particle flow objects (PFO)}.

\subsection{\label{sec:model:hlr}High Level Reconstruction}

After having reconstructed all the individual particles in the event, the next step in the processing is the reconstruction of
primary and secondary vertices. This is carried out in iLCSoft with the LCFIPlus~\cite{Suehara:2015ura} package.
The primary vertex of the event is found in a \emph{tear-down} procedure, starting with all tracks and gradually removing tracks with
the largest $\chi^2$~-contribution, up to a given  $\chi^2$~-threshold. Thereby, the constraints from the expected beam spot
$(\sigma_x=516~\rm{nm}, \sigma_y=7.7~\rm{nm},\sigma_z \sim 200~\mu\rm{m}~~\rm{at}~~~E_{cms}=250~\rm{GeV})$ are taken into account .
In a second step LCFIPlus tries to identify secondary vertices, applying suitable requirements for invariant masses, momentum directions
and $\chi^2$s. Secondary vertices and optionally isolated leptons can be used by LCFIPlus for jet clustering, aiming at high efficiency for correctly
identifying heavy flavor jets. The actual jet clustering is then performed by using a cone-based clustering with a Durham-like algorithm~\cite{Catani:1991hj}.
Alternatively users can use $k_T$ jet clustering algorithms from the Fastjet~\cite{Cacciari:2006sm} library that is interfaced to Marlin in a
dedicated package MarlinFastJet. LCFIPlus also provides algorithms for jet flavor tagging using boosted decision trees (BDTs) based on suitable
variables from tracks and vertices. A palette of additional high level reconstruction algorithms is used for physics analyses:
\begin{itemize}
\item particle identification using dE/dx, shower shapes and multi-variate methods
\item $\gamma\gamma$-finders for the identification of $\pi^0$ and $\eta$-mesons
\item reconstructed particle to Monte-Carlo truth linker for cross checking analysis and reconstruction efficiencies
\item tools for jet clustering using Monte-Carlo truth information
\item processors for the computation of various event shapes
\end{itemize}

\section{\label{sec:monte-carlo}Monte Carlo Production on the Grid}

\begin{figure}[t!]
\includegraphics[width=1.0\hsize]{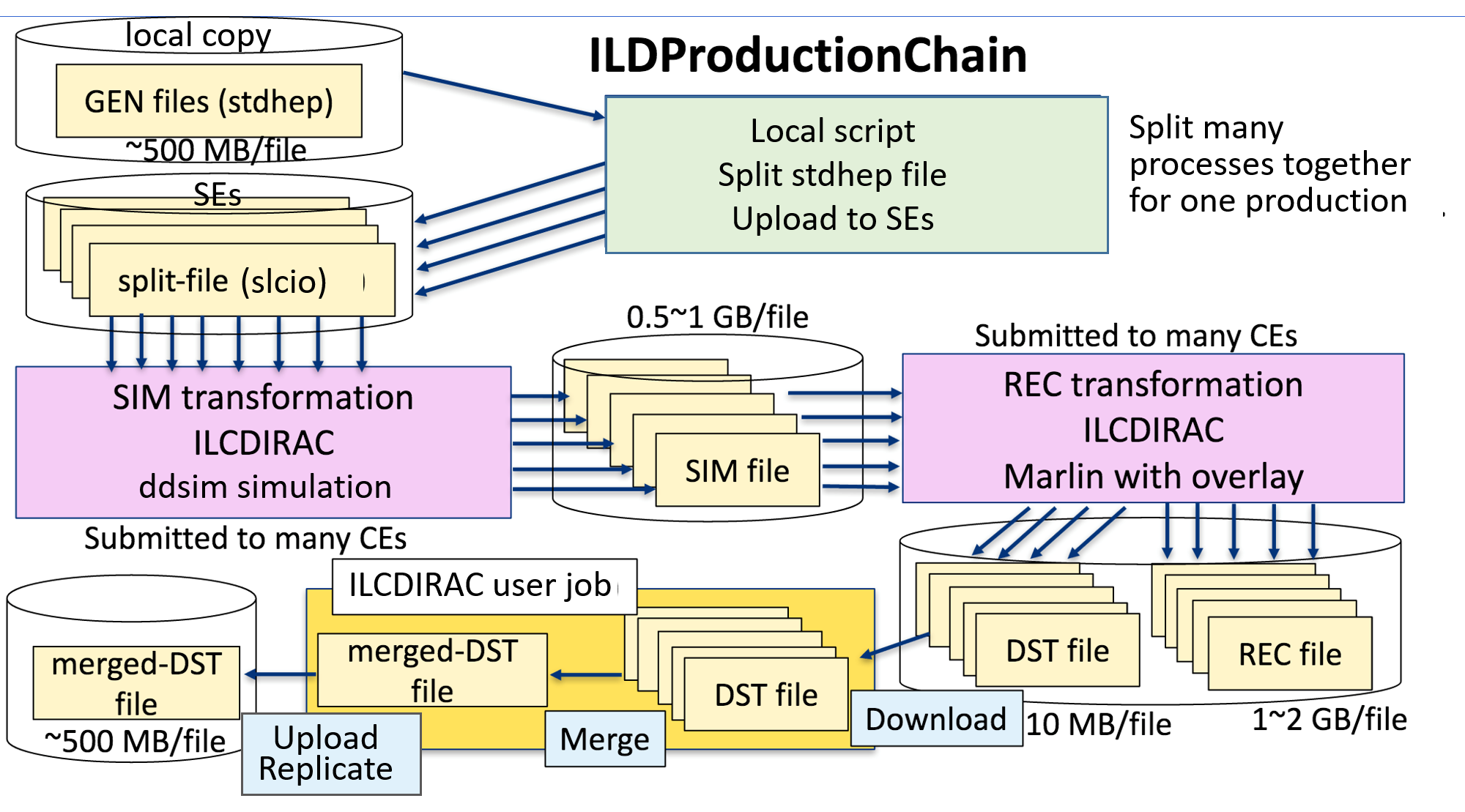}
\caption{\label{fig:sim_ild_mcprod} Schematic view of the Monte Carlo production system used for ILD.}
\end{figure}

The linear collider community uses the ILCDirac~\cite{Grefe:2014sca} toolkit for large scale Monte Carlo production on the Grid.
ILCDirac is highly configurable and ILD uses a dedicated production chain~\cite{Miyamoto:2019xve}, a schematic view of which is shown in Fig.~\ref{fig:sim_ild_mcprod}.
The Monte Carlo production is split into four main steps:
\begin{itemize}
\item GenSplit\\
  Split generator file to many files with small number of event files so that simulation and reconstruction jobs complete in adequate CPU time and 
produce a managable size of output files.  This has the advantage that the same input files can also be used for fast simulation programs such as SGV~\cite{Berggren:2012ar}.
\item Simulation\\
  Simulation of the detector response to the particles generated in the events using ddsim. ddsim is a python application that uses the Geant4 gateway
  DDG4 together with the detailed detector simulation model.
\item Reconstruction\\
  Full event reconstruction with the algorithms described in sec.~\ref{sec:reco}, writing out detailed \emph{REC}-files with all available information,
  including digitized hits and a much reduced \emph{DST}-file format.
\item{Merging of DST files}\\
  The rather small DST-files are merged into larger files for easier handling.
\end{itemize}
In order to investigate the effect of reducing the detector dimensions on the physics performance, two large Monte Carlo data sets for the large and small
ILD simulation models have been produced. The produced data sets correspond roughly to $500~\invfb$ at $E_{cms}=500~\GeV$ with the exact numbers of
events processed for the different classes shown in table~\ref{tab:mcprod_evtnum}.

\begin{table}[htbp]
\renewcommand{\arraystretch}{1.25}

\centering\small
\begin{tabular}{lcl}
\hline
 event class  &  description & events processed \\ 
\hline 
2f &   two fermion  final states &  $60.0 \times 10^6$ \\
4f &  four fermion final states & $22.6 \times 10^6$ \\
5f &  five fermion final states & $4.01 \times 10^6$ \\
6f &   six fermion  final states &  $13.8 \times 10^6$ \\
aa\_4f & two fermion by $\gamma\gamma$ interaction & $1.63 \times 10^6$ \\
higgs & higgs process & $3.97 \times 10^6$ \\
np & new physics process & $3.25 \times 10^6$ \\
\hline
aa\_lowpt &  $\Pgamma \Pgamma \to hadrons$ background  &  $2.50 \times 10^6$ \\
seeablepairs &   $\Pep\Pem$-pair background    &  $1.00\times 10^5$ BXs \\
calibration & single particle, $q\bar{q}$ events & $27.71\times 10^6$ \\
\hline
6f(WW) &  dedicated 6f sample at $E_{cms}=1~\TeV$ &  $1.75 \times 10^6$ \\

\hline
\end{tabular}
\caption{\label{tab:mcprod_evtnum} Number of Monte Carlo events produced for the different event classes. 
Approximately the same number of events were produced for the large and the small ILD simulation model.
The sum of events produced for the two models are shown in the table.} 
\end{table}


\thisfloatsetup{floatwidth=\textwidth,capposition=below}

\renewcommand{\fix}[1]{\textcolor{red}{\texttt{#1}}} 


\chapter{Detector and Physics Performance}
\label{chap:performance} 
\label{ild:sec:performance}
The overall performance of the ILD detector is a combination of the excellent
resolutions and efficiencies of the individual sub-detectors and the
sophisticated reconstruction and analysis algorithms described in the previous chapter.
This chapter reviews the ILD performance, beginning with the pure system performance that
is achieved from reconstructing individual long lived particles, followed by the high-level
reconstruction of jets and physics objects, ending with selected detector physics benchmarks.
Where applicable this performance is presented for the large (IDR-L) and the small (IDR-S)
detector models. Unless stated otherwise, the silicon based readout option for the ECAL and the
scintillator based readout option for the HCAL of the hybrid simulation described in section~\ref{sec:hybrid-sim}
have been used in the event reconstruction.

\section{\label{sec:system-performance} System performance}
\newcommand{\rmsn}{\mathrm{rms}_{90}}

\subsection{Tracking}\label{sec:performance:tracking}

The efficient identification and precise reconstruction of charged particle
tracks is crucial for the overall physics performance of the ILD detector.
The goal for the asymptotic momentum resolution~\cite{ild:bib:perfgoal::barklow} is
\begin{equation*}
\sigma_{1/p_T} \approx 2\times 10^{-5}~\text{GeV}^{-1}.
\end{equation*}
ensuring that the Higgs mass measurement from $\Pep\Pem \rightarrow H, Z\rightarrow\Pmuon\APmuon$ events
is dominated by the beam energy spread rather than the detector resolution.

\begin{figure}[htbp]
\begin{subfigure}{0.49\hsize} 
 \includegraphics[width=\hsize]{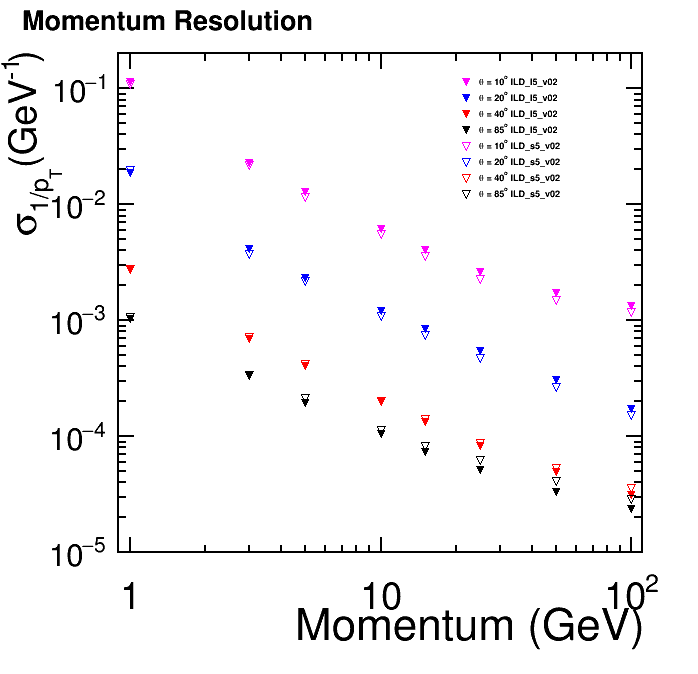}
 \caption{ \label{fig:perf:trk_pt}}
 \end{subfigure}
\begin{subfigure}{0.49\hsize} 
 \includegraphics[width=\hsize]{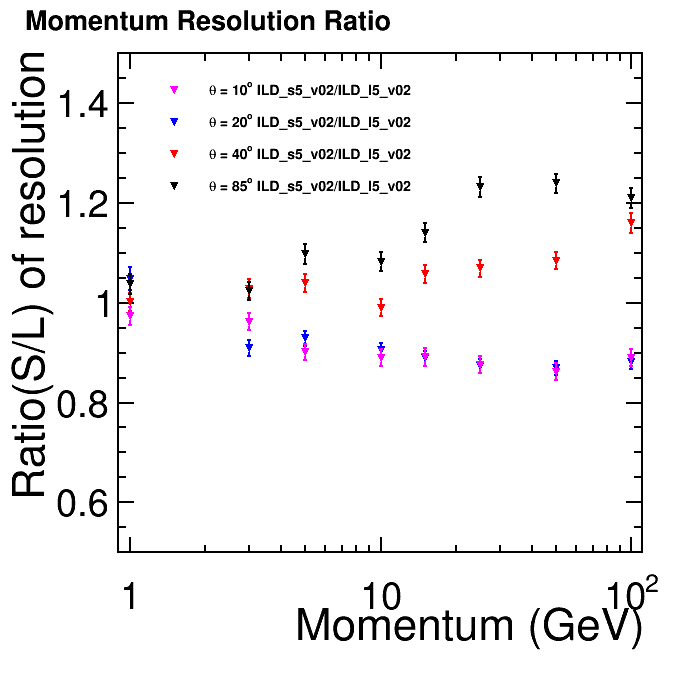}
 \caption{  \label{fig:perf:trk_ptcmp}}
 \end{subfigure}
\begin{subfigure}{0.49\hsize} 
 \includegraphics[width=\hsize]{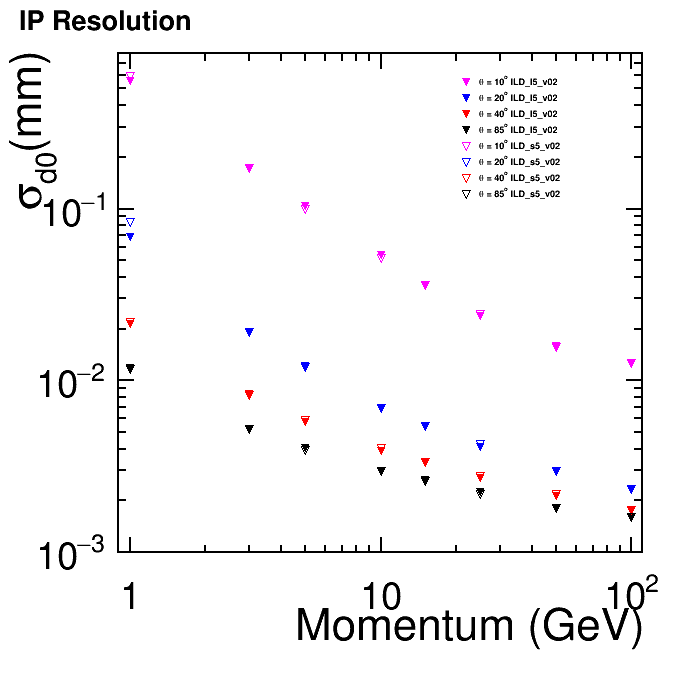}
 \caption{ \label{fig:perf:trk_d0}}
 \end{subfigure}
\begin{subfigure}{0.49\hsize} 
 \includegraphics[width=\hsize]{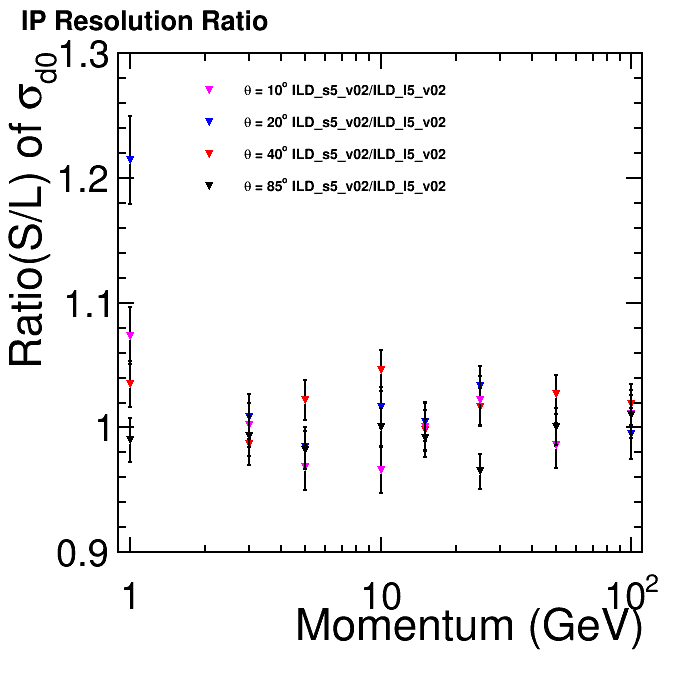}
 \caption{  \label{fig:perf:trk_d0cmp}}
 \end{subfigure}
\begin{subfigure}{0.49\hsize} 
 \includegraphics[width=\hsize]{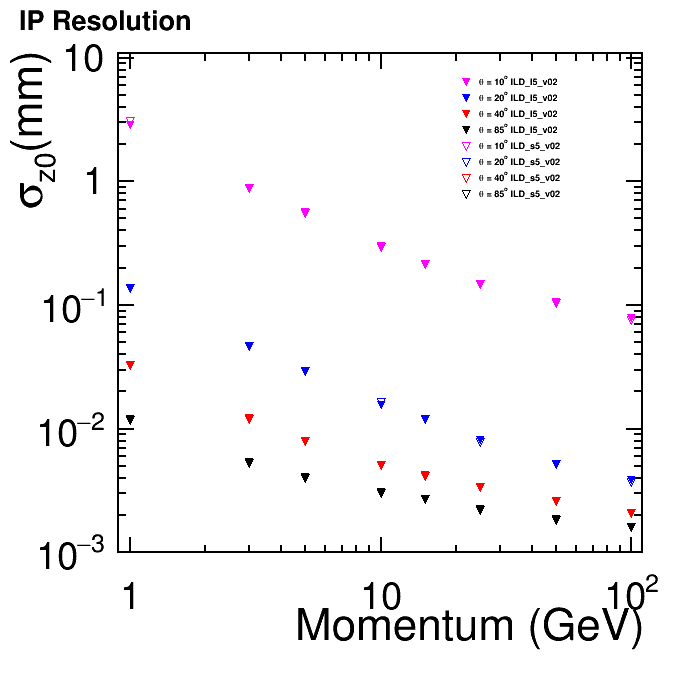}
 \caption{ \label{fig:perf:trk_z0}}
 \end{subfigure}
\begin{subfigure}{0.49\hsize} 
 \includegraphics[width=\hsize]{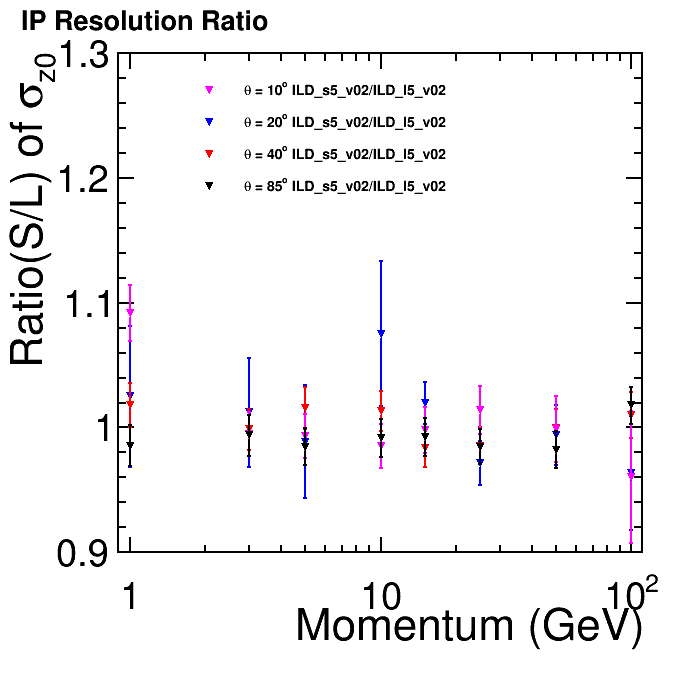}
 \caption{  \label{fig:perf:trk_z0cmp}}
 \end{subfigure}
\caption{
  Tracking resolutions for single muons for the large and small ILD detector models.
  (a) Inverse transverse momentum resolution $\sigma_{1/p_T}$ as a function of momentum and the ratio $small/large$ in (b).
  (c) Impact parameter in the $r\phi$-plane $\sigma_{d0}$ as a function of momentum and the ratio $small/large$ in (d).
  (e) Impact parameter along the $z$-axis $\sigma_{z0}$  as a function of momentum and the ratio $small/large$  in (f).
}
\label{fig:perf:trkres}
\end{figure}

To evaluate the tracking resolution of the ILD detector models, samples of single $\Pmu$-events at fixed
momenta (p=\unit{1,3,5,10,15,30,100}{\GeV} ) and polar angles ($\theta=10,20,40,85^\circ$) have been
run through the full simulation and reconstruction as described in
sections~\ref{sec:det-sim} and~\ref{sec:reco} with the single point resolutions given in table~\ref{tab:ild_trk_res}.
Fig.~\ref{fig:perf:trkres} shows the results for the inverse transverse momentum resolution
$\sigma_{1/p_T}$ and the impact parameter resolutions $\sigma_{d0}$ and $\sigma_{Z0}$ in the 
$r\phi$-plane and along the $z$-axis respectively for the large and small detector models ILD-L and ILD-S.
The performance goal for the asymptotic momentum resolution is met by both detector modes as can be seen in
Fig.~\ref{fig:perf:trk_pt}. Fig.~\ref{fig:perf:trk_ptcmp} shows the ratio of the resolution for the small
and large detector. In the barrel region the larger detector is slightly better, due to the larger number of hits and
the corresponding larger lever arm, despite the lower B-field. In the forward region this behaviour
is reversed as here the same number of hits are available due to the identical detector geometry in this region
and it is the higher B-field that causes a better measurement of the curvature and thus the momentum.
The impact parameter resolutions for the large and small detector models are equal to within a few percent,
as can be seen from Fig.~\ref{fig:perf:trk_d0cmp} and~\ref{fig:perf:trk_z0cmp}, except at low momenta in the
forward region, where the large detector is better, as here the higher B-field and increased curvature
lead to a reduced lever arm for the small detector.
In the barrel region, one observes very similar resolutions for the impact parameter in the $r\phi$-plane
and that along the $z$-axis, whereas in the forward region, where less VTX-hits contribute, $\sigma_{Z0}$
is significantly worse than  $\sigma_{d0}$ (see Fig.~\ref{fig:perf:trk_d0} and~\ref{fig:perf:trk_z0})
%
%

%
%
\begin{figure}[htbp]
\begin{subfigure}{0.49\hsize} 
 \includegraphics[width=\hsize]{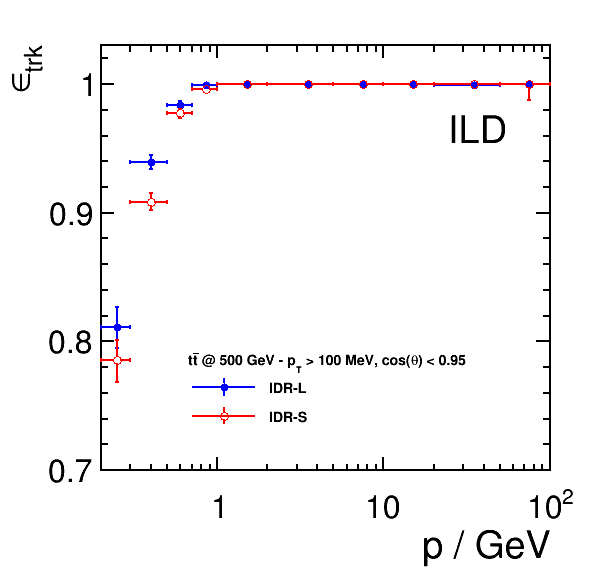}
 \caption{ \label{fig:perf:trkeff_p}}
 \end{subfigure}
\begin{subfigure}{0.49\hsize} 
 \includegraphics[width=\hsize]{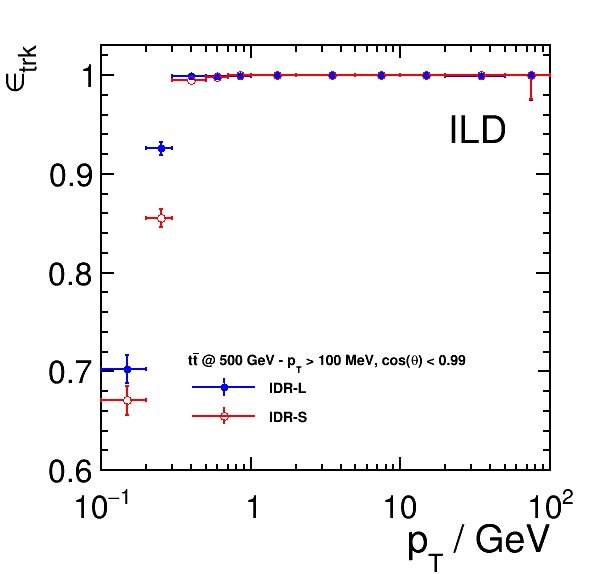}
 \caption{  \label{fig:perf:trkeff_pt}}
 \end{subfigure}
\begin{subfigure}{0.49\hsize} 
 \includegraphics[width=\hsize]{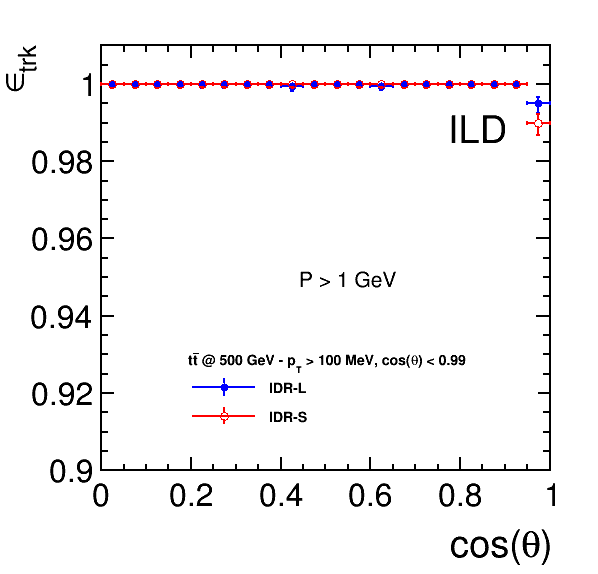}
 \caption{ \label{fig:perf:trkeff_th}}
 \end{subfigure}
\begin{subfigure}{0.49\hsize} 
 \includegraphics[width=\hsize]{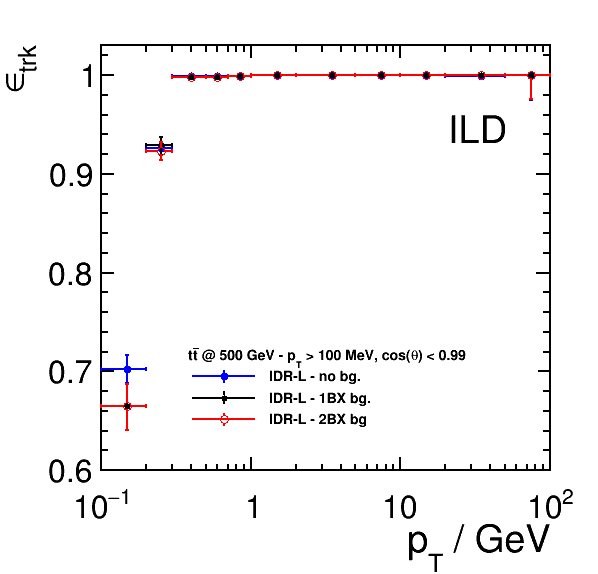}
 \caption{  \label{fig:perf:trkeff_bg}}
 \end{subfigure}
\begin{subfigure}{0.49\hsize} 
 \includegraphics[width=\hsize]{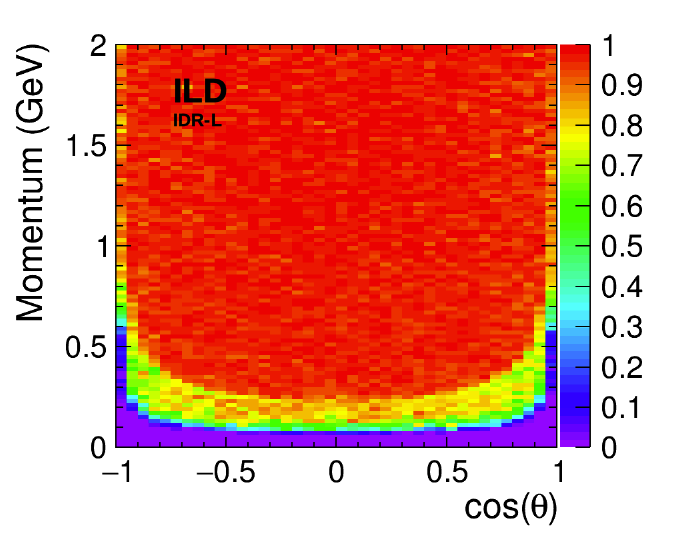}
 \caption{ \label{fig:perf:trkeff_2D_p}}
 \end{subfigure}
\begin{subfigure}{0.49\hsize} 
 \includegraphics[width=\hsize]{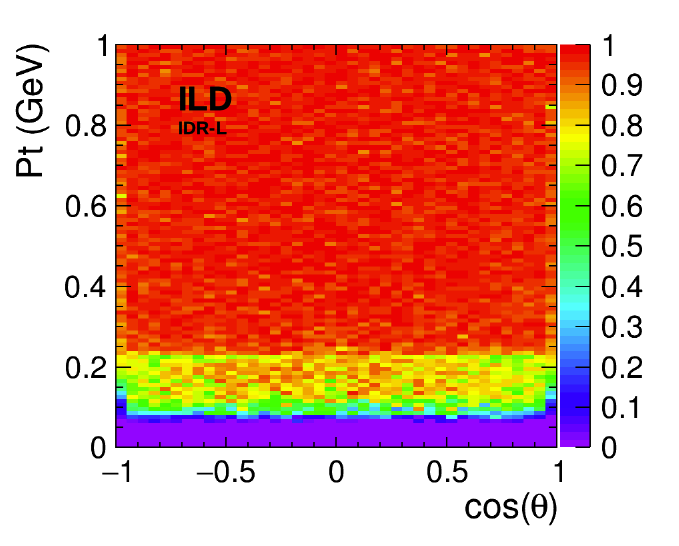}
 \caption{  \label{fig:perf:trkeff_2D_pt}}
 \end{subfigure}
\caption{
  Track finding efficiency for prompt ($r_{vertex}<\unit{10}{\cm}$) tracks in $t \bar t$-events at 500 GeV as a function of kinematic variables for the large and small detector:
  (a) as a function of momentum $p$  (b) as a function of transverse momentum $p_T$ (c) as a function of $\cos(\theta)$).
  The effect on the efficiency of overlaying hits from 1BX and 2BX of pair background is shown in (d). The tracking efficiency as a
  function of $\cos(\theta)$ and either momentum or transverse momentum is shown for the large model in (e) and (f) respectively. 
}
\label{fig:perf:trkeff}
\end{figure}

The track finding efficiency of the algorithms described in section~\ref{sec:reco} is evaluated with $\Ptop\APtop$-events
at $E_{cms}=\unit{500}{\GeV}$ fully simulated and reconstructed for both detector models.
The tracking efficiency is defined as the ratio of the number of correctly reconstructed tracks with a hit purity higher than 75~\% to the
number of Monte Carlo tracks that are created within \unit{10}{\cm} of the {\em IP}, left at least 4 hits in the detector and
did not decay in flight. The resulting efficiency is shown in Fig.~\ref{fig:perf:trkeff} as a function of momentum $p$ in (a),
transverse momentum $p_T$ in (b) and $\cos(\theta$) in (c). As can be seen in (c), the tracking is almost perfect for particles
with $p>\unit{1}{\GeV}$ and $p_T>\unit{100}{\MeV}$ down to $\cos(\theta) \approx 0.95$ and better than 99~\% in the very forward
direction. At low momenta and in the forward direction, the small detector performs slightly worse than the larger variant, which
is due to the larger B-field that causes larger curvatures of the tracks. In the barrel region some of this effect could potentially
be alleviated by moving the innermost layer of the vertex closer to the beam, which would be possible as the background cone from
pair particles is also forced to lower radii by the higher B-field.

The qualitative dependency of the efficiency on momentum and polar angle together is shown in the 2D plots (e) and (f) for the large
detector model. The efficiency is flat as a function of transverse momentum in all but the very forward ($\cos(\theta)>0.95$)
part and almost perfect above $p_T>\unit{250}{\MeV}$, also in that region. This is expected as the track finding in this region has to be done
across several sub-detectors (VTX, SIT, FTD) with mixed barrel and disk layouts, naturally leading to slightly larger inefficiencies.
An optimization of the geometrical layout of the inner silicon tracking detectors for ILD with respect to an improved pattern recognition
capability would be the subject of future studies.
A small inefficiency persists in the very forward region also for higher momenta.
The above studies have been done with the same background overlaid as for the large Monte Carlo samples.
In (d) the tracking efficiency is shown also with complete and detailed simulations of one and two bunch crossings of pair
background overlaid. Only a very small degradation of the tracking efficiency at the lowest momenta is observed.

\subsection{Particle Flow performance and JER}
The performance of the Particle Flow Algorithm (PFA) is evaluated with dedicated  $\PZ\rightarrow \Pquark\APquark$-events, $\Pquark \in [\Pqu,\Pqd,\Pqs]$,
where the \PZ is chosen to have the desired mass of twice the jet energy and decays at rest. This allows to distinguish the measurement of the
actual jet energy resolution (JER) from other effects like ISR-photons or confusion in jet clustering that occur in more complex,
realistic events.
%
%
\begin{figure}[htbp]
\begin{subfigure}{0.49\hsize}
 \includegraphics[width=\hsize]{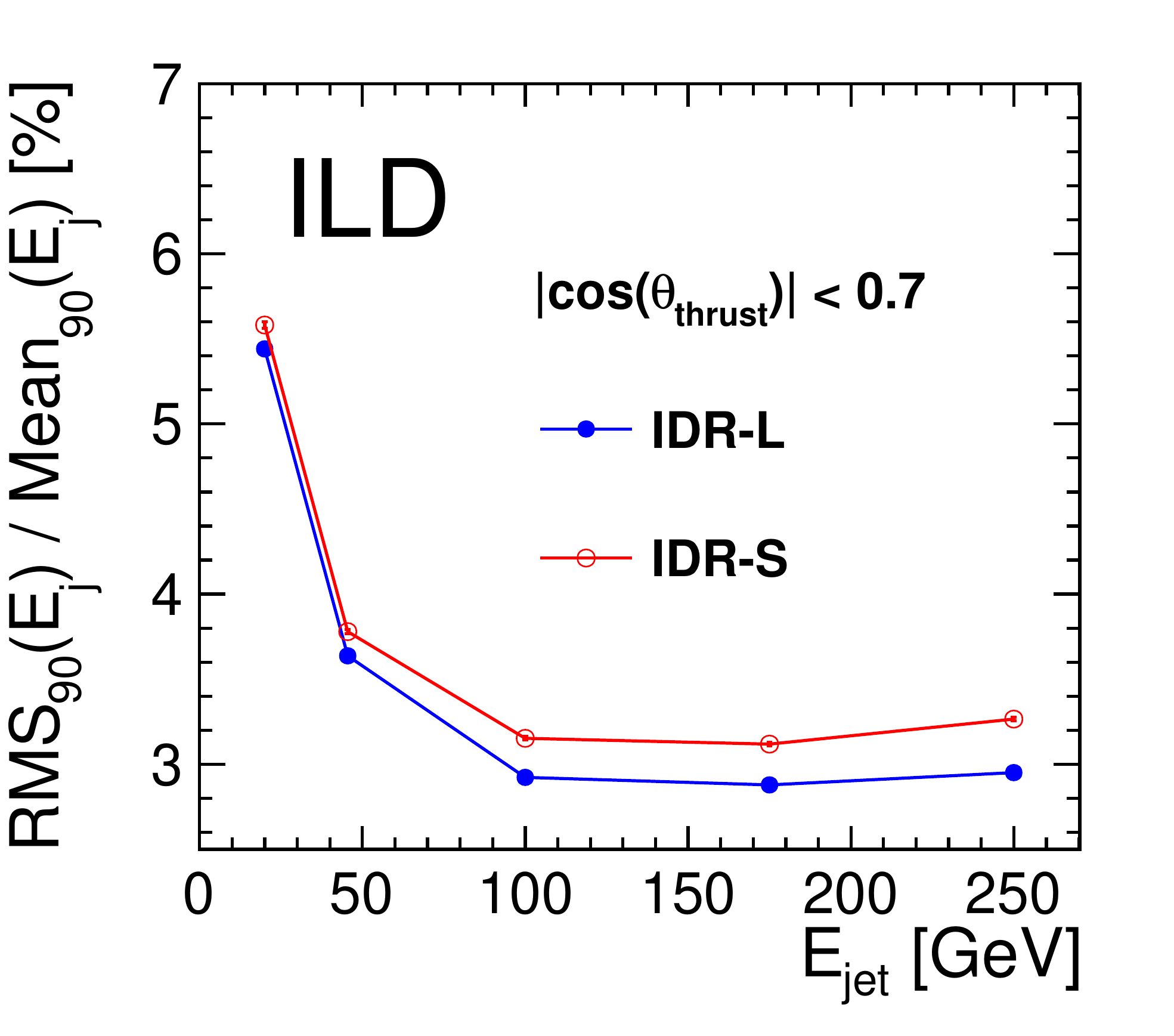}
 \caption{ \label{fig:perf:pfa_jer}}
 \end{subfigure}
\begin{subfigure}{0.49\hsize}
 \includegraphics[width=\hsize]{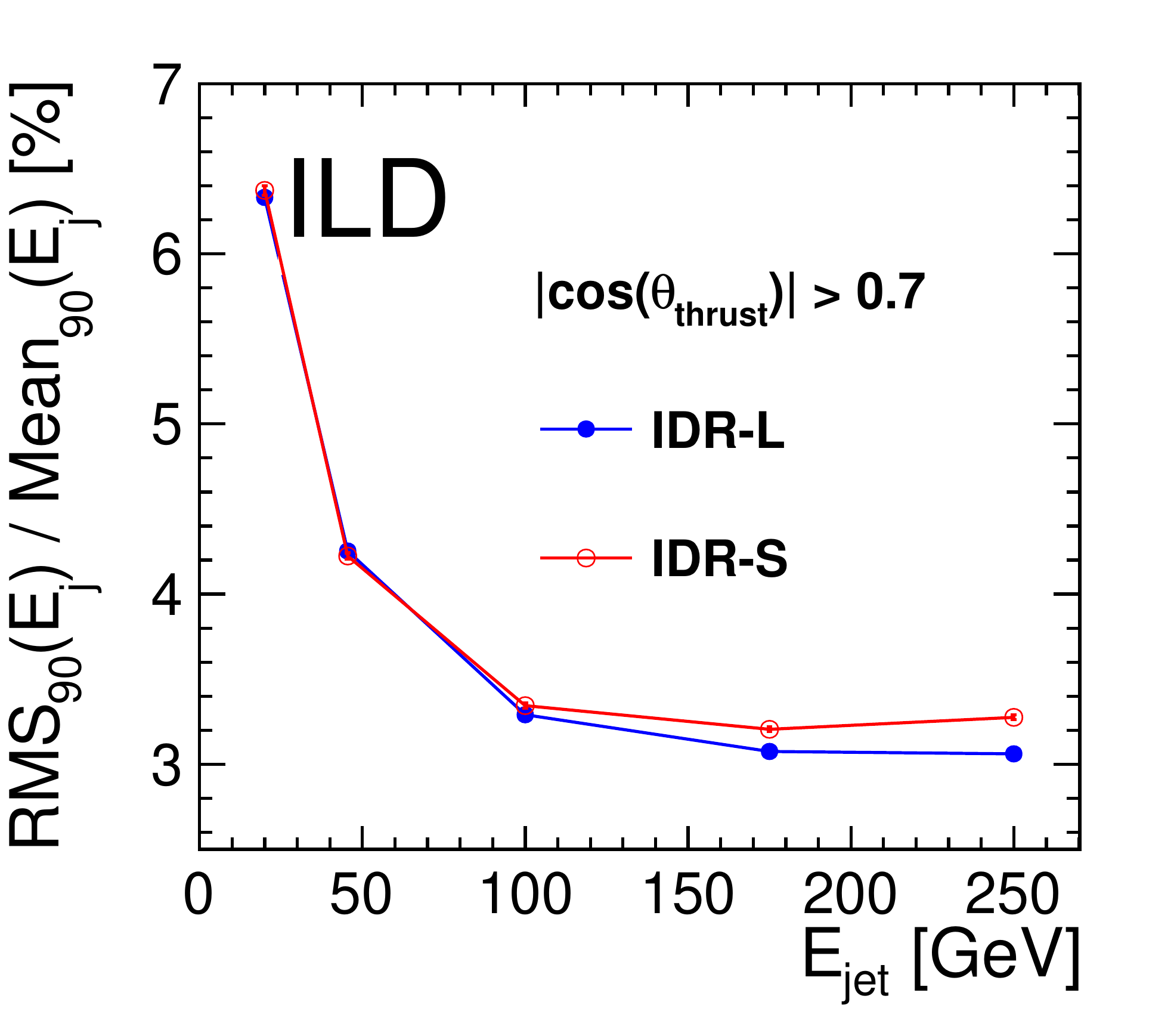}
 \caption{  \label{fig:perf:pfa_jer_endcap}}
 \end{subfigure}
\begin{subfigure}{0.49\hsize}
 \includegraphics[width=\hsize]{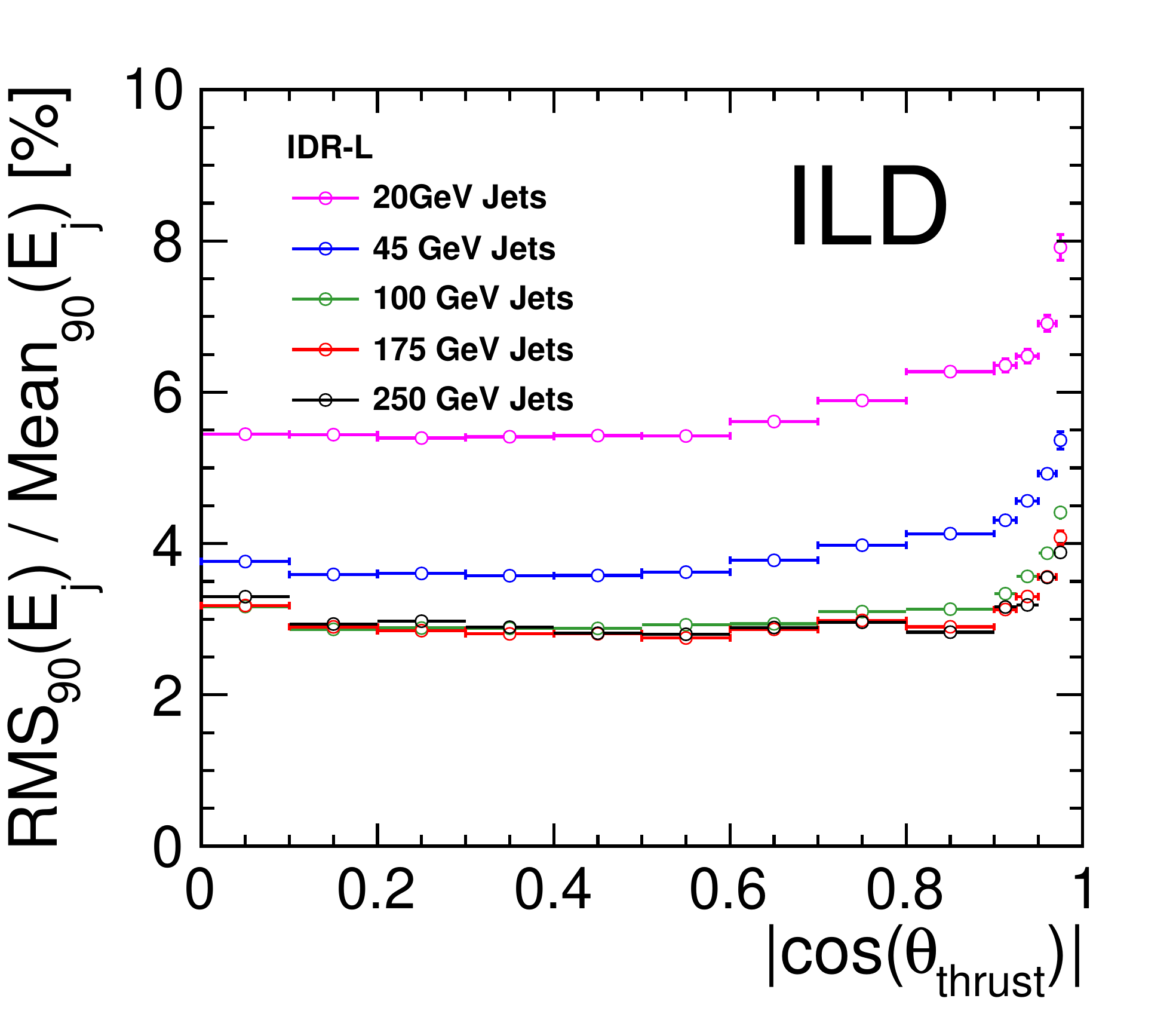}
 \caption{ \label{fig:perf:pfa_costh}}
 \end{subfigure}
\begin{subfigure}{0.49\hsize}
 \includegraphics[width=\hsize]{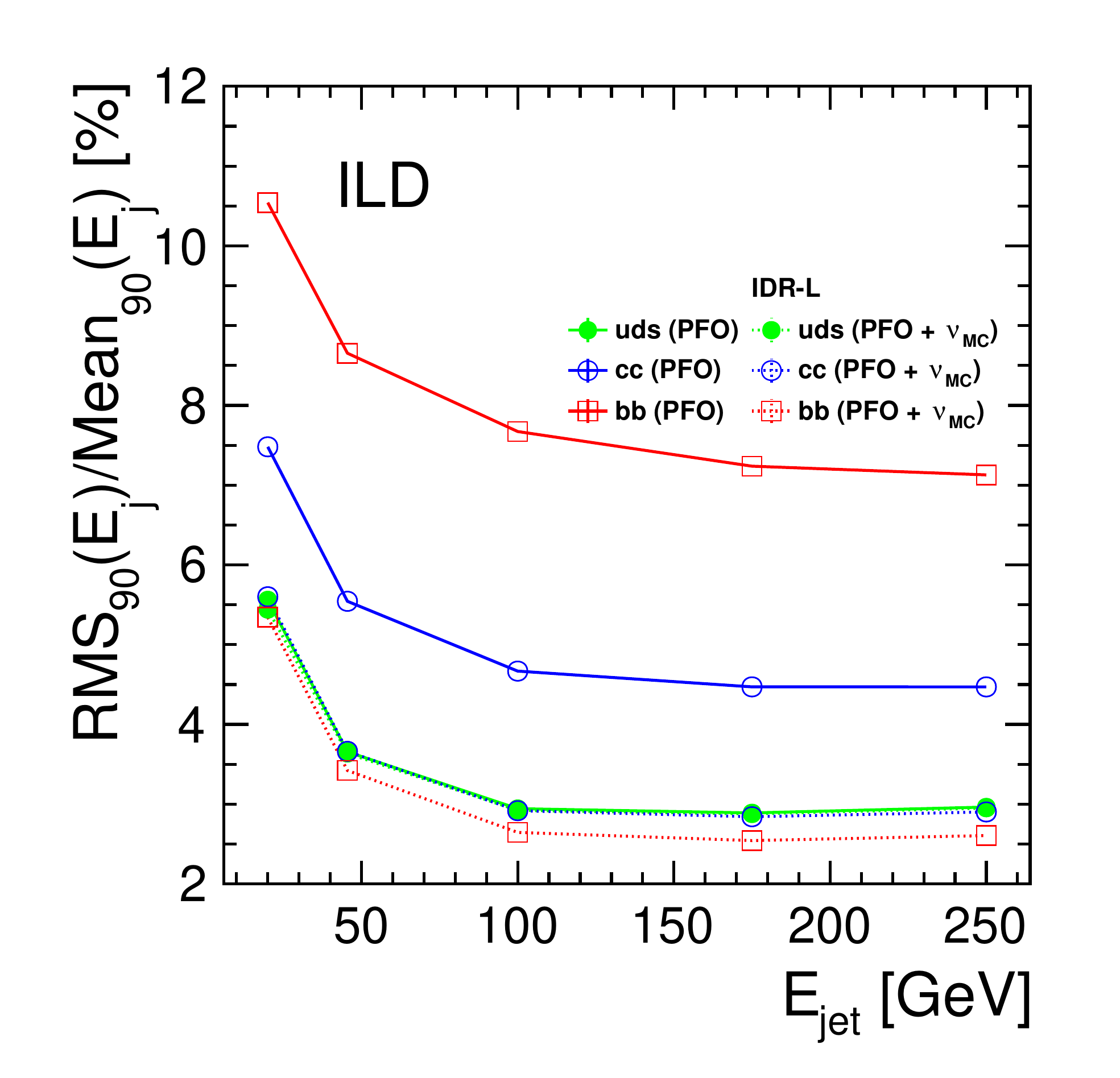}
 \caption{  \label{fig:perf:pfa_udscb}}
 \end{subfigure}
\begin{subfigure}{0.49\hsize}
 \includegraphics[width=\hsize]{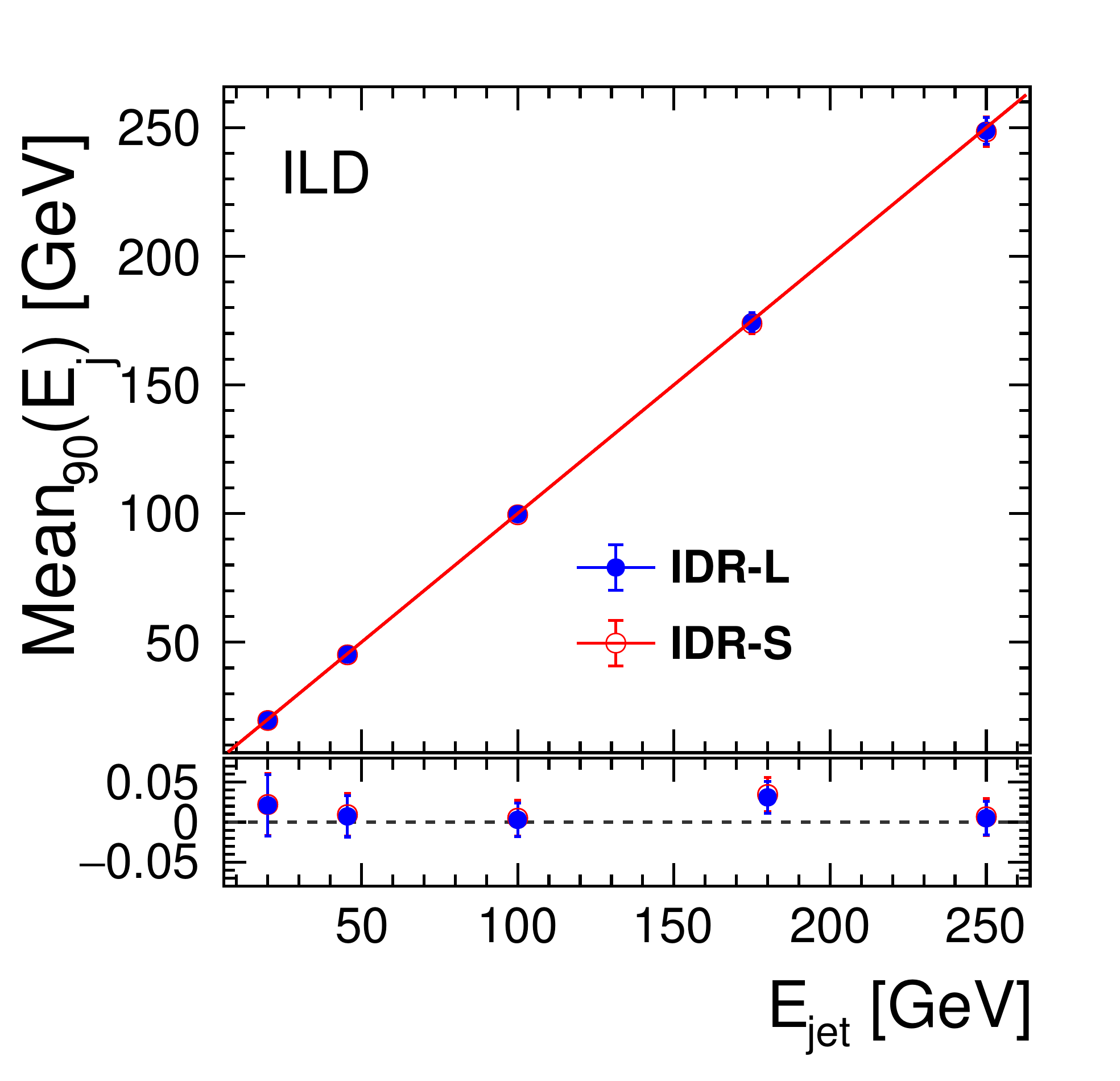}
 \caption{  \label{fig:perf:pfa_jes}}
 \end{subfigure}
\caption{
  Jet energy resolution (JER), evaluated as defined in eq.~\ref{ild:eq:jer} for $\PZ\rightarrow \Pquark\APquark$-events and $\Pquark \in [\Pqu,\Pqd,\Pqs]$.
  (a) comparison of the JER for the large and small ILD detector in the barrel region with
  $|\cos(\theta_{thrust})|<0.7$  (b) the same for the endcap region with  $0.7<|\cos(\theta_{thrust})|<0.98$
  (c) JER as function of the polar angle ({\em thrust}-axis) of the event. (d) JER for $\Pqu,\Pqd,\Pqs$ di-jet events together with $\Pqc\APqc$ and $\Pqb\APqb$ events,
  with (dashed lines) and without (solid lines) correcting the $\Pneutrino$ energies using Monte Carlo truth information. (e) Jet energy scale for the large and small model
  for barrel events.
  }
\label{fig:perf:pfa}
\end{figure}
The jet energy resolution is then evaluated as 
\begin{equation}\label{ild:eq:jer}
\frac{ \sigma_{E_{jet}} } { E_{jet}}  :=  \frac{\rmsn(E_{jet})}{ \mathrm{mean}_{90}(E_{jet})}
\end{equation}
The $\rmsn$ is defined to be the $rms$ of the central $90\%$ of the distribution and the $\mathrm{mean}_{90}$ is its mean value.
This measure is robust against large tails and
should be multiplied by a factor of $\sim1.1$ to obtain an equivalent Gaussian analyzing power\cite{ild:bib:PandoraPFA}.
Fig.~\ref{fig:perf:pfa_jer} shows the JER as a function of the jet energy for selected energies $E_{jet}=\unit{(20, 45, 100, 175, 250)}{\GeV}$
for the large and small detector model in the barrel region ($|\cos(\theta_{thrust})|<0.7$) . For $E_{jet}\geq \unit{45}{\GeV}$ the resolution is better than 4\% and
approaches 3\% (3.2\%) for the large (small) model at higher energies. In the forward region ($|\cos(\theta_{thrust})|>0.7$) the JER is
slightly worse and the difference between large and small detector is less pronounced as can bee seen in Fig.~\ref{fig:perf:pfa_jer_endcap}. In order to
ensure that the jets are fully contained in the acceptance of the detector, an additional cut of  $|\cos(\theta_{thrust})|<0.98$ is applied.
The JER as a function of the polar angle of the jets (the {\em thrust}-axis of the di-quark events) is shown in Fig.~\ref{fig:perf:pfa_costh} for the
large detector model. The observed dependency is rather flat throughout the barrel region, increasing somewhat after the barrel-end-cap transition
with a visible rise in the very forward region, similar for all energies. The effect of heavy flavor quarks on the jet energy resolution is shown in
Fig.~\ref{fig:perf:pfa_udscb}, where the resolutions is plotted for $\Pqu,\Pqd,\Pqs$ di-quark events together with $\Pqc\APqc$ and $\Pqb\APqb$ events for the large model.
The observed degradation in the JER for the heavy quark jets can be fully attributed to the missing energy carried by neutrinos,
as can be seen from the dashed lines, were the energy is corrected for the $\Pneutrino$~-energies using Monte Carlo truth information.
The linearity of the jet energy measurement is shown in Fig.~\ref{fig:perf:pfa_jes} to be better than 5\% in the barrel region.

\subsection{Vertexing}

\begin{figure}[htbp]
\begin{subfigure}{0.49\hsize}
 \includegraphics[width=\hsize]{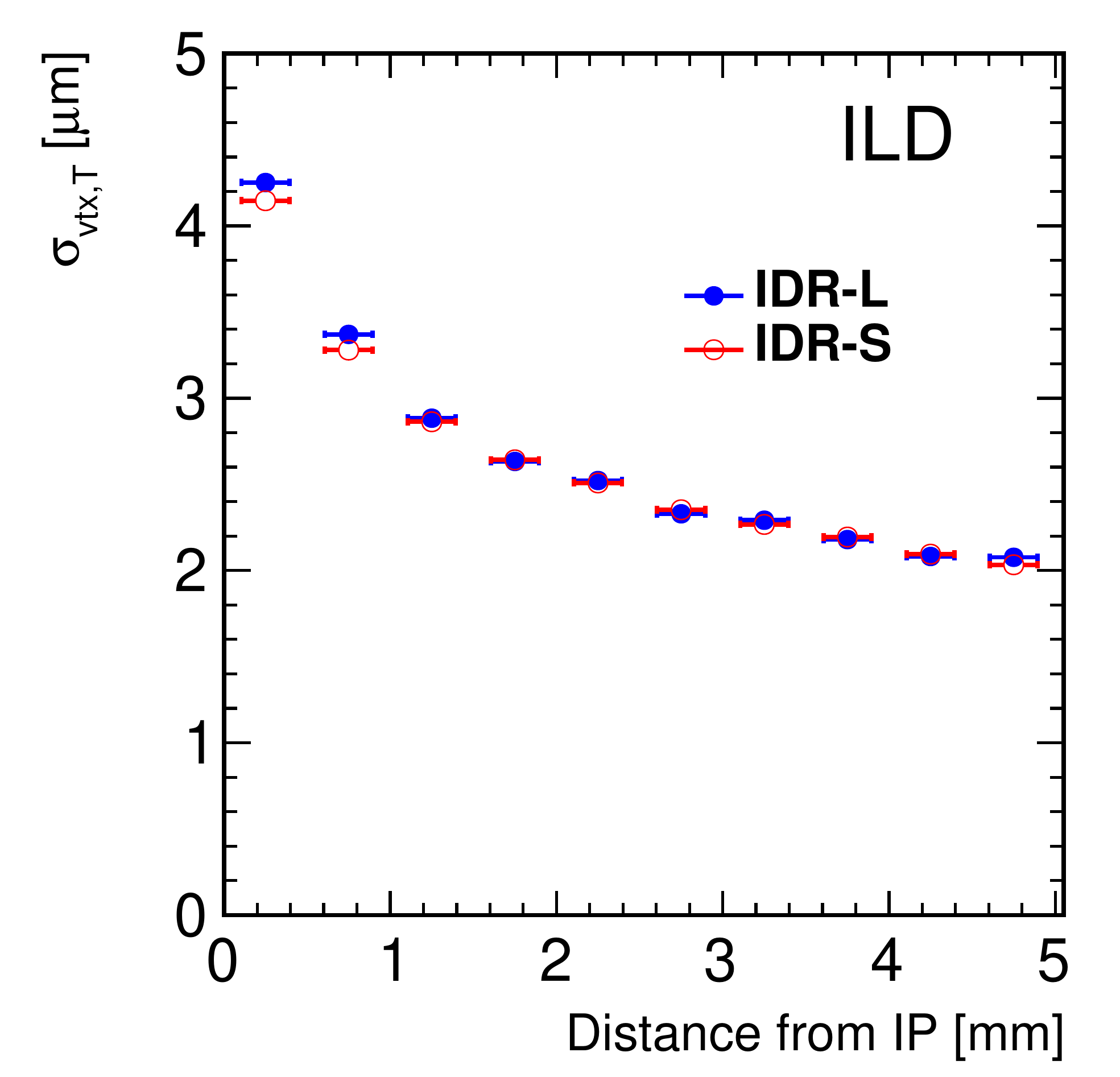}
 \caption{ \label{fig:perf:svtx_r}}
 \end{subfigure}
\begin{subfigure}{0.49\hsize}
 \includegraphics[width=\hsize]{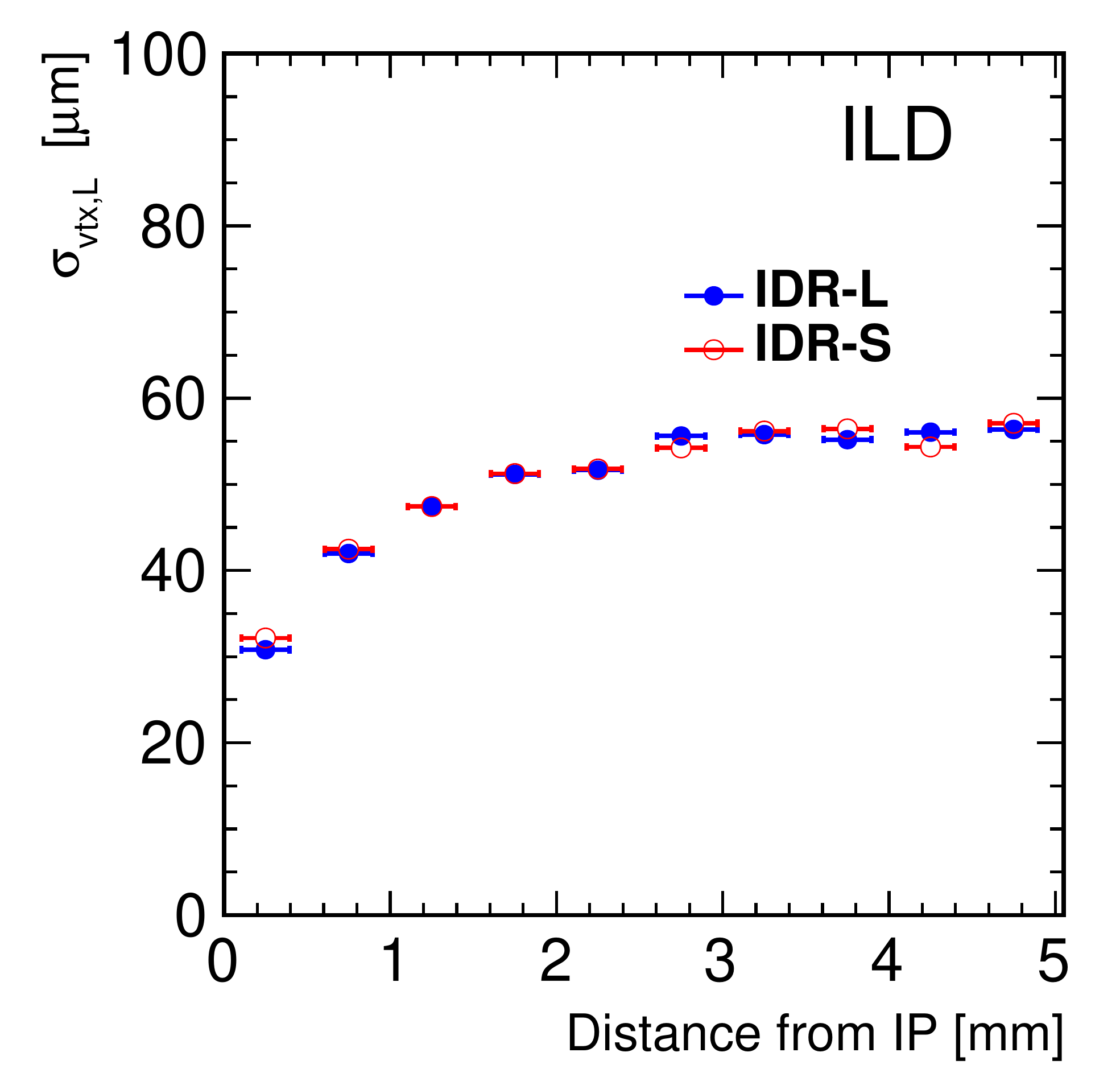}
 \caption{  \label{fig:perf:svtx_z}}
 \end{subfigure}
\begin{subfigure}{0.49\hsize}
 \includegraphics[width=\hsize]{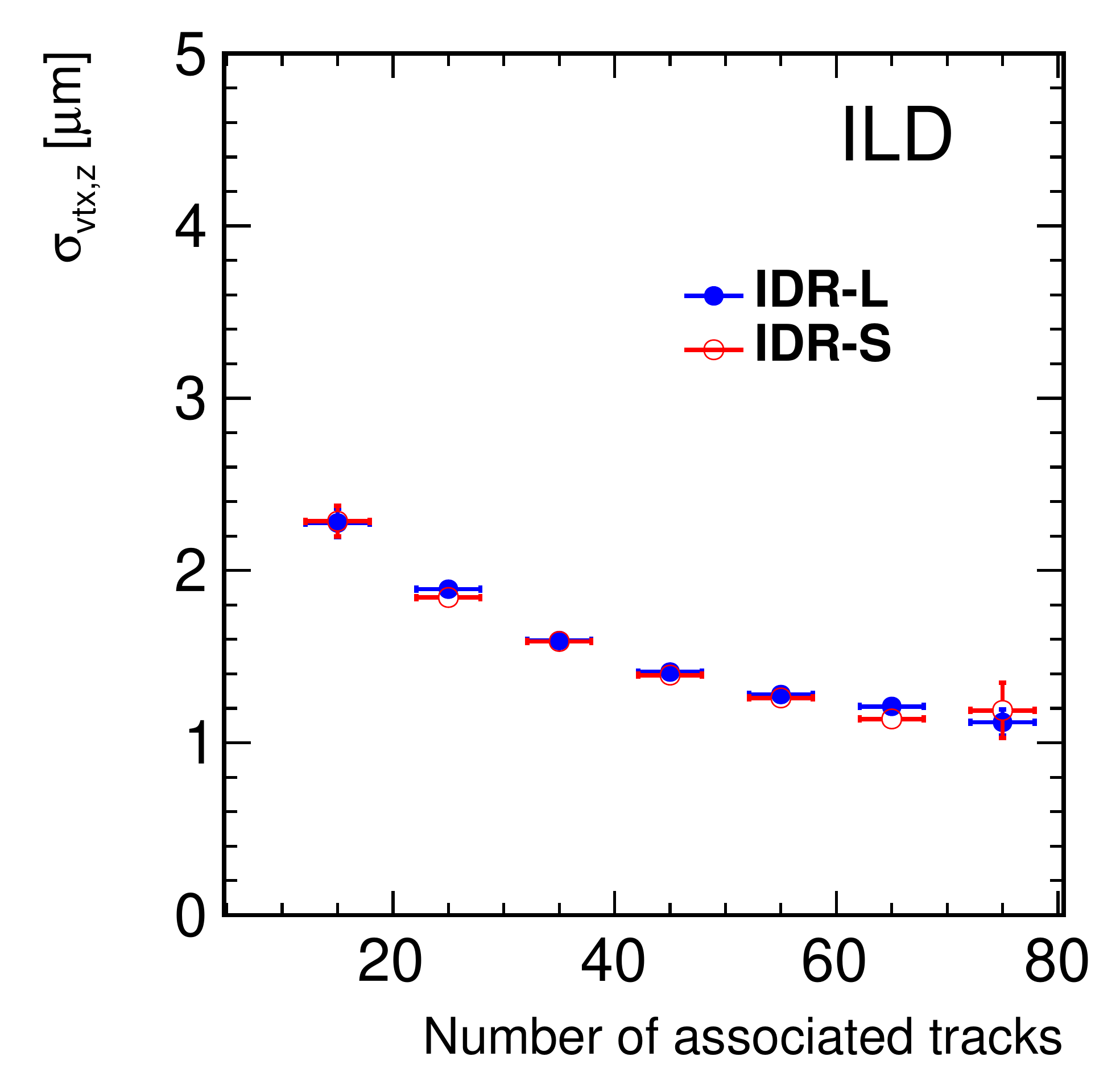}
 \caption{  \label{fig:perf:pvtx_z}}
 \end{subfigure}
\caption{ Vertex resolutions for the large and small ILD detector models for  $\Pep\Pem \rightarrow 6~\Pqc$ events at $\sqrt{s}=\unit{500}{\GeV}$.
  (a) Resolution of the transverse component of secondary vertices as a function of the distance from the $IP$ and the same for the longitudenal component in (b).
  (c) Resolution of the z-coordinate of the primary vertex as a function of the number of tracks used in the fit. 
  Note that the radial component of the primary vertex is known extremely well to \unit{O(10)}{\nm} from the beam-spot position.
}
\label{fig:perf:vtxres}
\end{figure}

The correct identification of heavy flavour decays in jets requires a precise measurement of the coordinates of secondary vertices.
The vertex resolution is studied with the dedicated $\Pep\Pem \rightarrow 6~\Pqc$ events that are also used for the flavour tagging
training and performance studies, described in section~\ref{sec:perf:hlr:lcfi}.
The resolution is measured along the longitudinal direction of the jet $\vec{e}_L$
and two transverse directions $\vec{e}_{T1}$ and $\vec{e}_{T2}$, defined as:
\begin{equation}
\vec{e}_L = \frac{ \vec{p}_c }{|\vec{p}_c| }~; ~~~~~~~  \vec{e}_{T1} = \vec{e}_L \times \vec{z} ~; ~~~~~~~  \vec{e}_{T2} = \vec{e}_L \times \vec{e}_{T1}
\end{equation}
whith the 3-momentum of the $\Pqc$-quark $\vec{p}_c$ taken from Monte Carlo truth.
The result is shown for the transverse (a) and longitudinal (b) components of the secondary vertices as a function of the distance from the $IP$
in Fig.~\ref{fig:perf:vtxres}.
Fig.~\ref{fig:perf:pvtx_z} shows the resolution of the primary vertex' z-position as a function of the number of associated tracks.
The resolution is better than \unit{3}{\micron} for low multiplicity events and
approaches \unit{1}{\micron} for high multiplicity events.  The radial component of the primary vertex is already known
extremely well from the beam-spot position to \unit{O(10)}{\nm}. The overall vertexing performance for the large and small detector models
is very similar, as expected already from the single track impact parameter resolutions shown in Fig.~\ref{fig:perf:trkres} (c)-(f).

\subsection{Charged Particle identification}
\label{sec:perf:sys:pid}
\begin{figure}[htbp]
\begin{subfigure}{0.49\hsize}
 \includegraphics[width=\hsize]{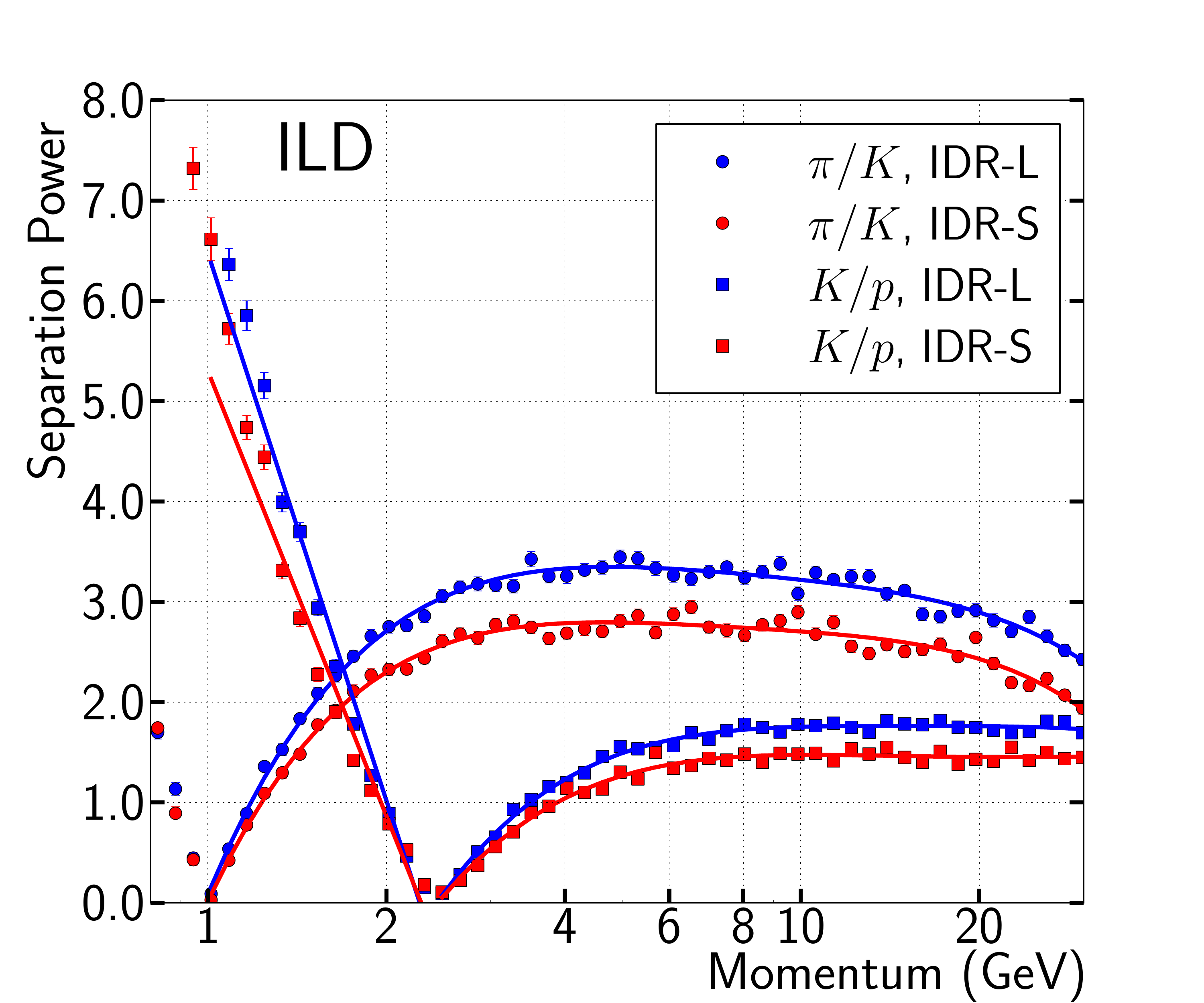}
 \caption{ \label{fig:perf:dedx_sep}}
 \end{subfigure}
\begin{subfigure}{0.49\hsize}
 \includegraphics[width=\hsize]{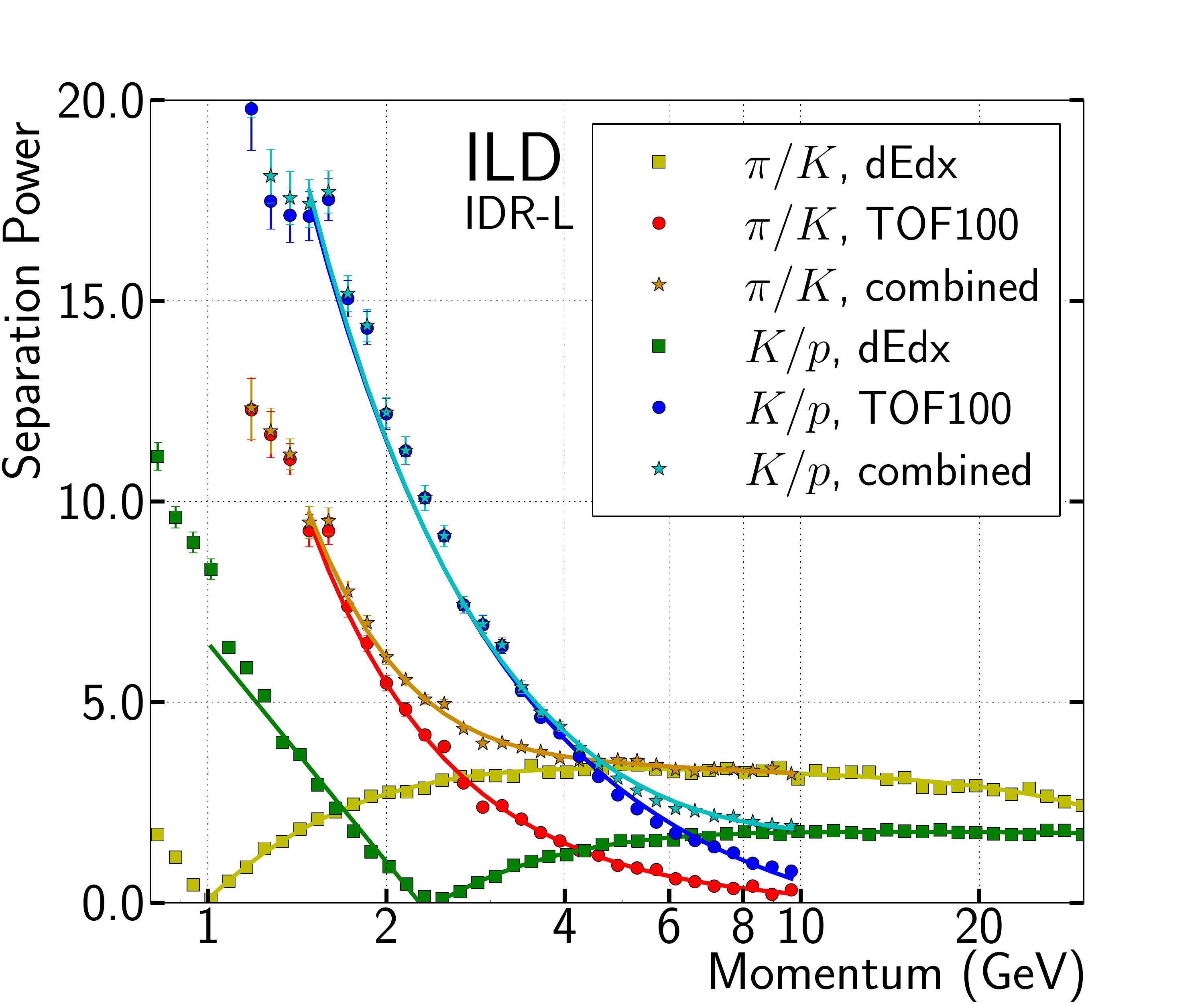}
 \caption{  \label{fig:perf:dedxtof_sep}}
 \end{subfigure}
\caption{ (a) particle separation power (eq.~\ref{ild:eq:seppow}) for $\Ppi/\PK$ and $\PK/\Pp$ based on the $dE/dx$ measurement in the TPC.
  (b) improvement of the same separation power if combined with a {\em time-of-flight (TOF)} estimator from the first ten Ecal layers,
  where $\eta_{dE/dx,TOF}=\eta_{dE/dx} \oplus \eta_{TOF}$. The curves are shown to guide the eye.
}
\label{fig:perf:dedxtof}
\end{figure}

%
%
Measuring the energy loss of charged particles in the ILD-TPC provides a powerful tool for identifying the type of the particle.
\begin{figure}[htbp]
\begin{center}
\includegraphics[width=0.8\hsize]{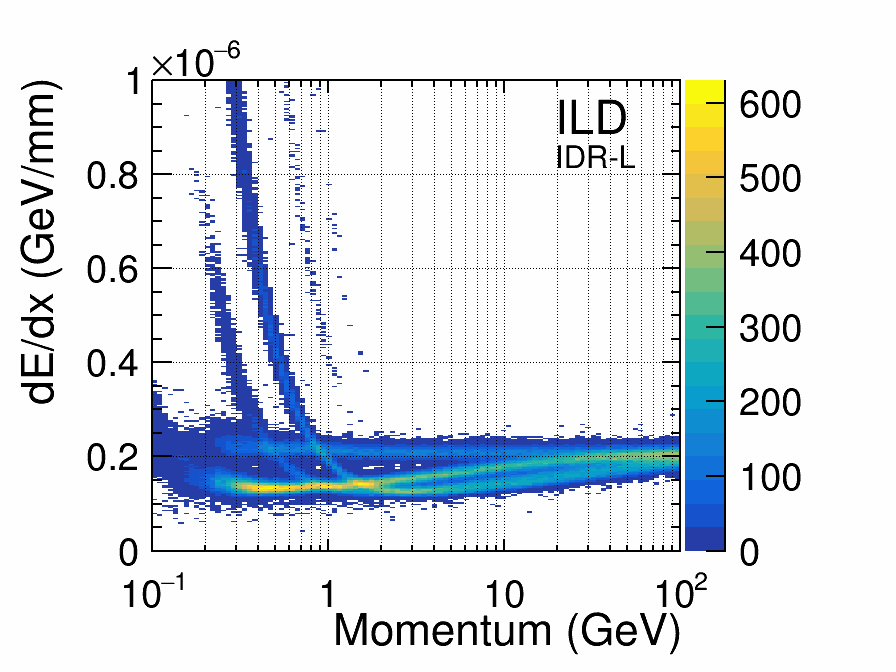}
\end{center}
\caption{\label{fig:perf:dedx_tpc}
  $dE/dx$ as a function of particle momentum as reconstructed from a full simulation of single particle events ($\Pe, \Pmu, \Ppi, \PK$ and $\Pproton$)
  in the TPC of the large ILD detector model. The particles were simulated with a logarithmic momentum distribution and isotropic direction. Spurious
  entries in the bands for more massive particles, such as the deuteron, as well as entries from low momentum particles, below the TPC acceptance,
  are due to secondaries created in the events.
}
\end{figure}
Fig.~\ref{fig:perf:dedx_tpc} shows the $dE/dx$ reconstructed from a truncated mean for charged particle tracks in the TPC as a function of
the particle momentum, clearly revealing the bands of the most abundant particle types $\Pepm, \Pmupm, \Ppipm, \PKpm$ and $p^{\pm}$.
By fitting Gaussian distributions, with mean $\mu(p)$ and standard deviation $\sigma(p)$, to individual bands in momentum bins one can
define a separation power $\eta_{A,B}$ for distinguishing the two particle types A and B:
\begin{equation}
\eta_{A,B}(p) = \frac{ |\mu_A(p) - \mu_B(p)| } { \sqrt{ \frac{1}{2} ( \sigma^2_A(p) + \sigma^2_B(p) )  }  }
\label{ild:eq:seppow}
\end{equation}
Figure~\ref{fig:perf:dedxtof} shows the separation power
for $\Ppi,\PK$ and $\PK,\Pp$  based on the $dE/dx$ measurement in the TPC of the large and small detector model (a)
and the possible improvement that could be achieved by combining it with a  {\em time-of-flight (TOF)} measurement in the large detector (b).
As a proof of concept a possible TOF estimator is computed here. It uses the first ten calorimeter hits in the Ecal that are closest to the straight line,
resulting from extrapolation of the particle's momentum into the calorimeter, assuming an individual time resolution of $100$~ps
per hit\footnote{While this time resolution seems realisticly possible, it has to be noted, that so far it has not yet been demonstrated
 in a test beam prototype.}. The results optained clearly motivate further studies on timing measurements and optimised TOF estimators
in the ILD calorimeters.
%
%

\begin{figure}[b!]
  \includegraphics[width=0.6\hsize]{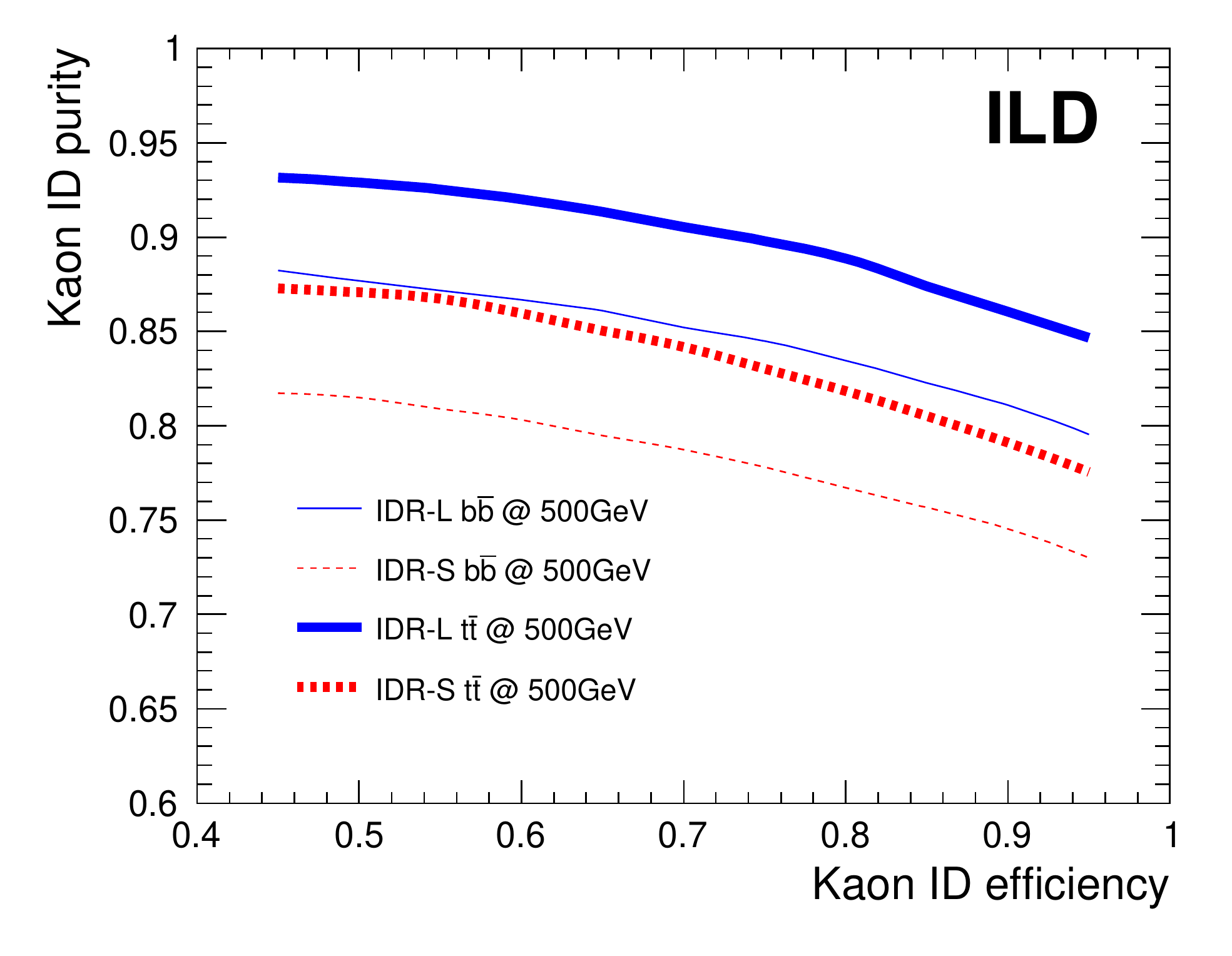}
  \caption{\label{fig:perf:KaonID}
    Efficiency-purity curves for charged Kaon identification in the context of the $t\bar{t}$ and $b\bar{b}$ analyses. For the same
    efficiency, the large detector reaches a $5\%$ higher purity due to its larger number of TPC measurements, which results in a better $dE/dx$
    resolution. Details on the analyses can be found in Sec.~\ref{subsec:bench:bbbar} and~\ref{subsec:bench:ttbar}.
  }
\end{figure}

An example in the context of physics analyses can be seen in Fig.~\ref{fig:perf:KaonID}. It compares the performance of the charged Kaon identification based on $dE/dx$ for the large and small detector models, as obtained from the $b\bar{b}$ and $t\bar{t}$ benchmarks described in Sec.~\ref{subsec:bench:bbbar} and~\ref{subsec:bench:ttbar}, respectively. For the same efficiency, the large detector reaches a $5\%$ higher purity due to its larger TPC radius, which results in a better $dE/dx$ resolution.

\thisfloatsetup{floatwidth=\SfigwFull,capposition=beside}
Additional methods for charged particle identification using calorimeter shower shapes have also been developed and used in specific physics analyses.

\subsection{BeamCal reconstruction}

\begin{figure}[htbp]
\begin{subfigure}{0.49\hsize}
 \includegraphics[width=\hsize]{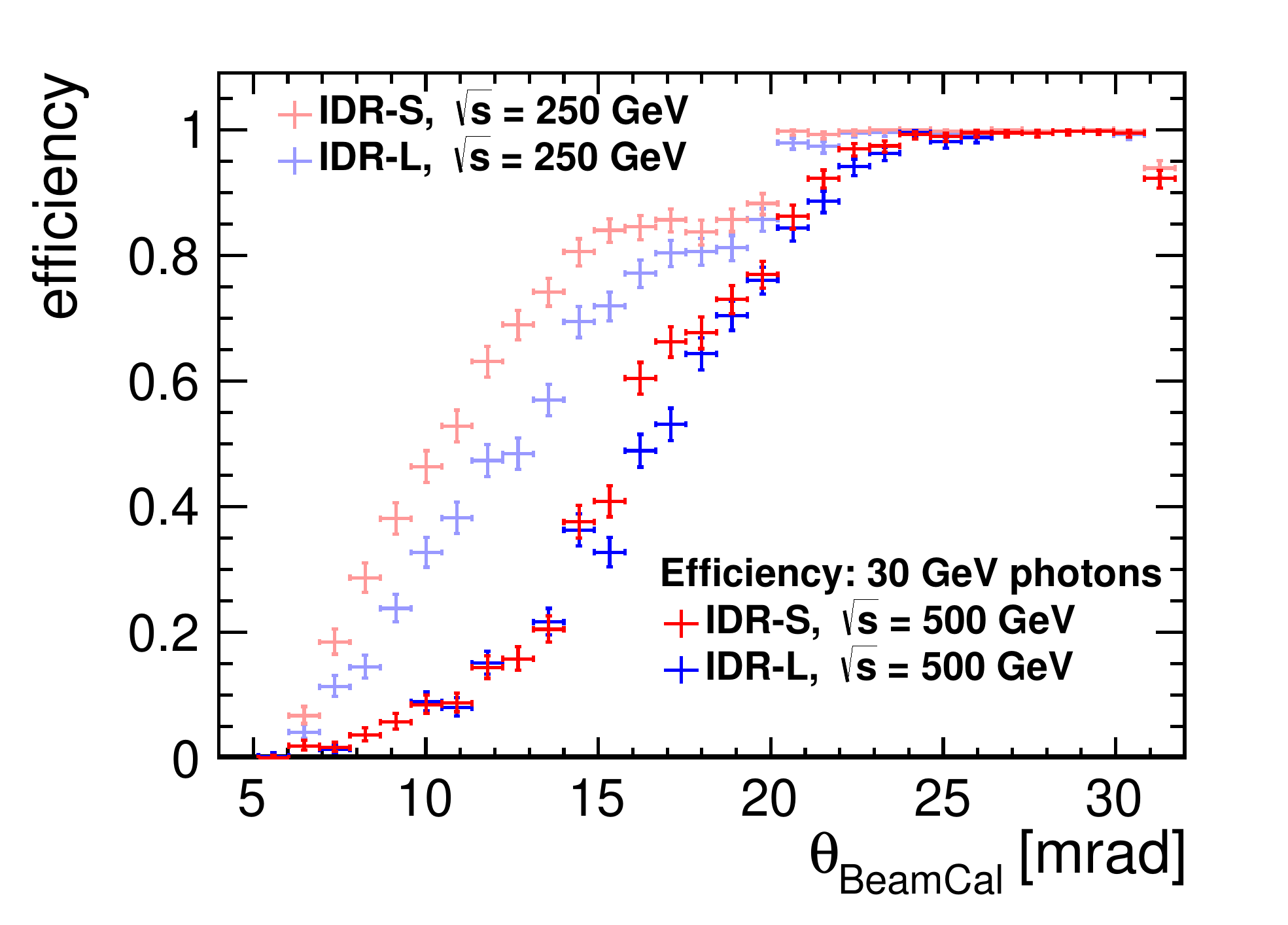}
 \caption{ \label{fig:perf:beamcal_eff}}
 \end{subfigure}
\begin{subfigure}{0.49\hsize}
 \includegraphics[width=\hsize]{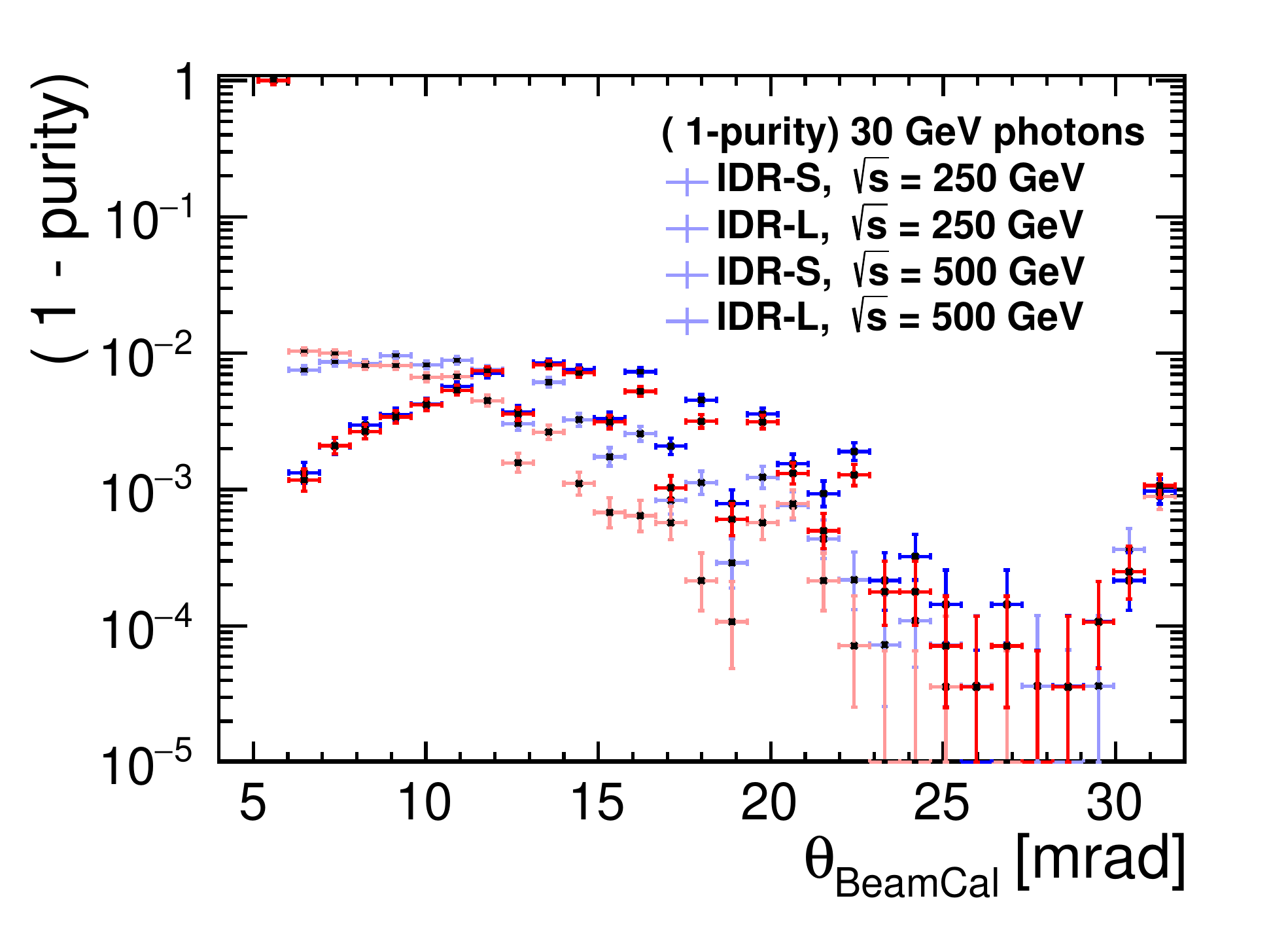}
 \caption{  \label{fig:perf:beamcal_fake}}
 \end{subfigure}
\caption{ (a) Reconstruction efficiency for single \unit{30}{\GeV} photons in the large and small detector models for $E_{cms}=\unit{500}{\GeV}$  and
  $E_{cms}=\unit{250}{\GeV}$ as a function of the polar angle.
  (b) Fraction of clusters that are falsely reconstructed from background hits (1-purity) as a function of the polar angle. 
}
\label{fig:perf:beamcal}
\end{figure}

Many analyses require the efficient measurement of electromagnetic showers at small polar angles in the BeamCal.
Due to the large amounts of background from pair particles (c.f.~\ref{sec:beam:background}), the BeamCal requires a special reconstruction algorithm
for identifying showers in the presence of background.
Fig.~\ref{fig:perf:beamcal_eff} shows the achieved reconstruction efficiency for single \unit{30}{\GeV} photons for the large and small detector models
for $E_{cms}=\unit{500}{\GeV}$  and $E_{cms}=\unit{250}{\GeV}$ respectively. Due to the significantly larger backgrounds at \unit{500}{\GeV} the efficiency
rises slower with polar angle than at \unit{250}{\GeV} and there is almost no difference observed between the two models.
In contrast at lower center of mass energy the small detector shows better performance, as expected due to the reduced background as an effect of the
higher B-field. Here a plateau of the efficiency at $\approx 80\%$~is reached between \unit{15}{\mrad} and \unit{20}{\mrad} due to the keyhole opening
of the BeamCal front face.
Fig.~\ref{fig:perf:beamcal_fake} shows the fraction of clusters that are falsely reconstructed from background hits (1-purity) as a function of the polar angle.
It is at most 1\% in the region of low efficiency and drops to below one permille in the region where the full efficiency is reached.
%
%


\section{\label{sec:HLR-performance} High-level Reconstruction Performance}
\subsection{Flavour-Tag Performance}
\label{sec:perf:hlr:lcfi}
\begin{figure}[htbp]
\begin{subfigure}{0.49\hsize}
\includegraphics[width=\textwidth]{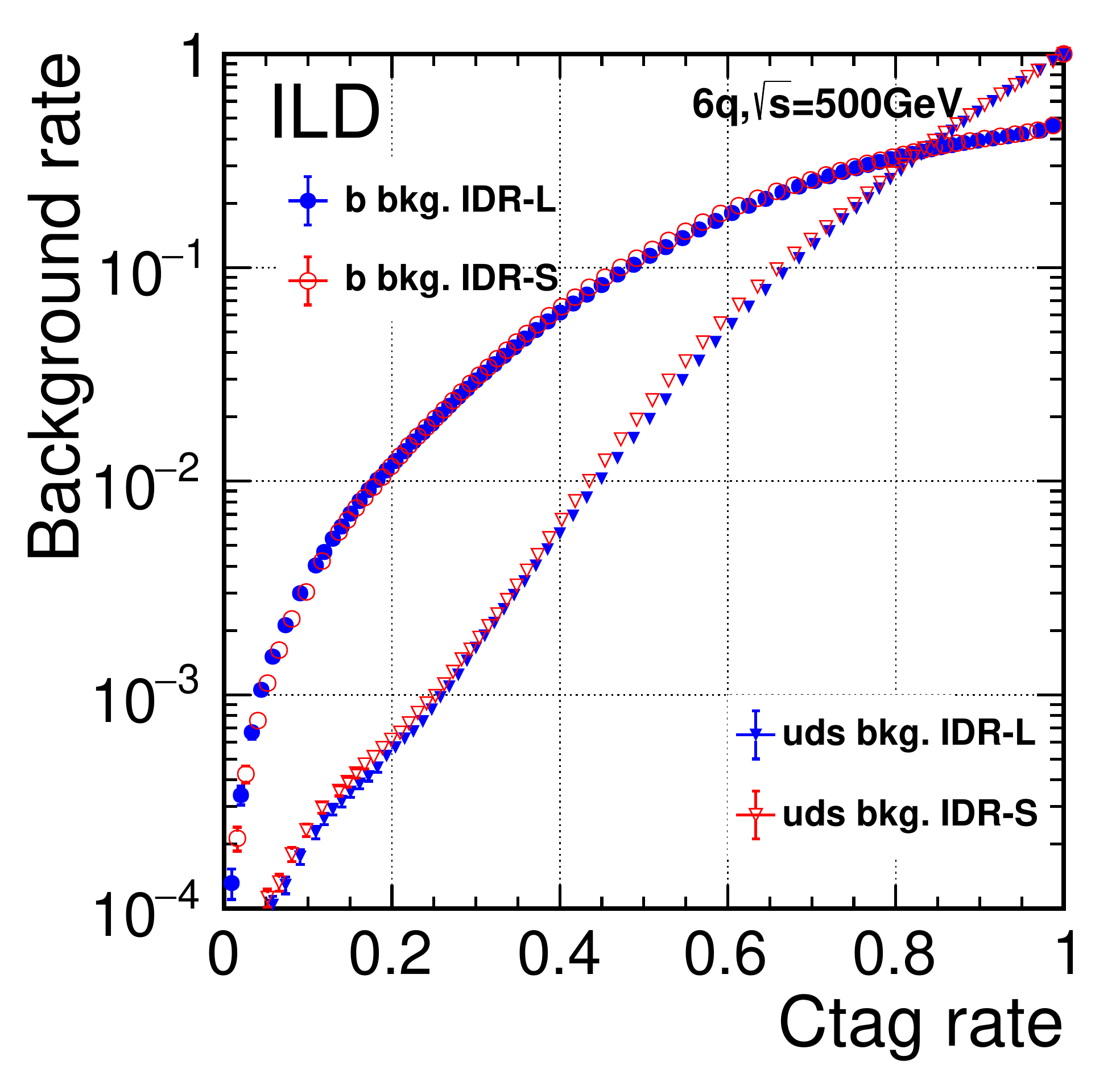}
 \caption{ \label{fig:HLR-ctag_perf}}
 \end{subfigure}
\begin{subfigure}{0.49\hsize}
\includegraphics[width=\textwidth]{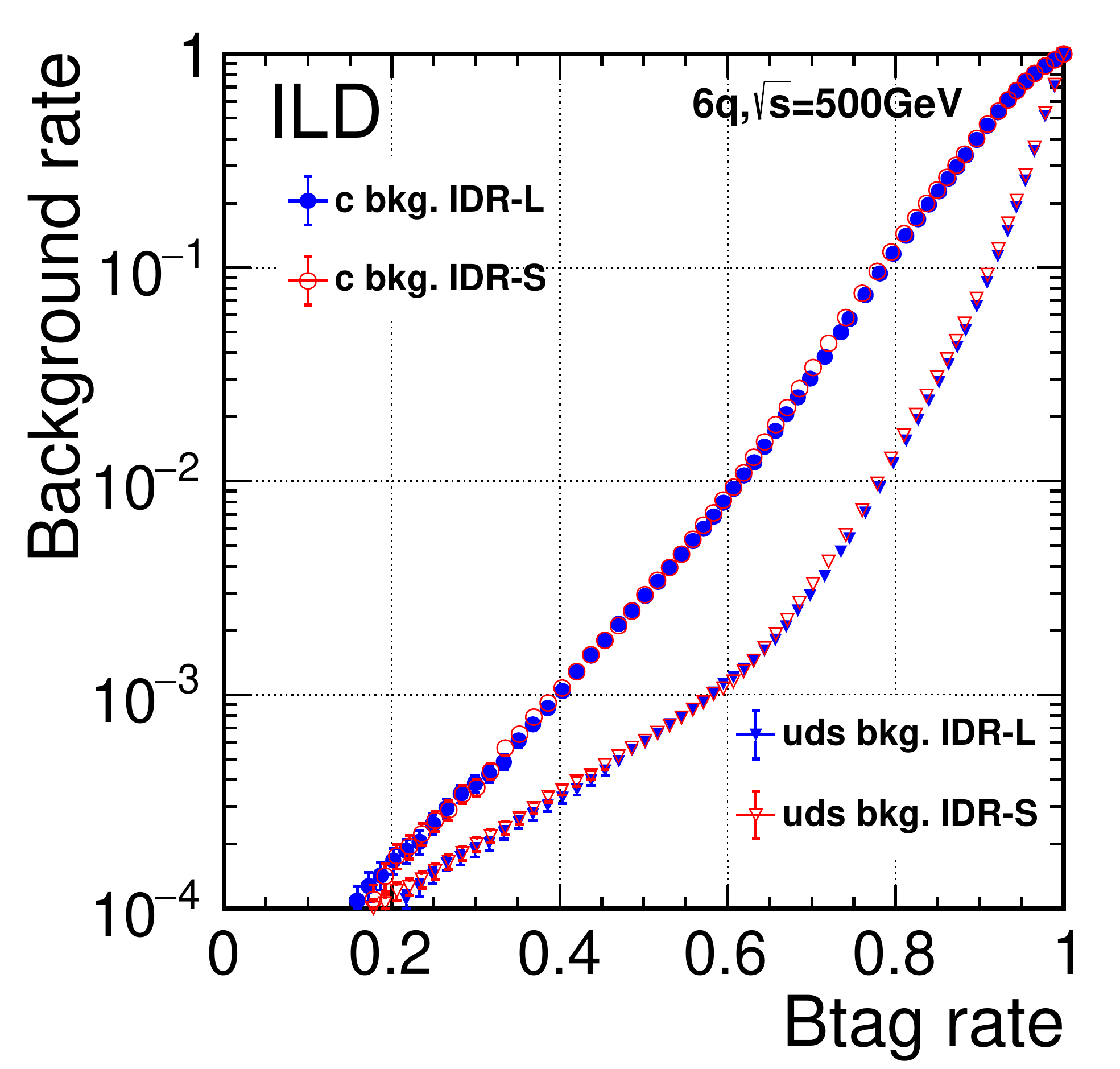}
 \caption{  \label{fig:HLR-btag_perf}}
 \end{subfigure}
\caption{Flavour tag performance for the large and small ILD detector models. (a) background rate as a function of the c-tagging efficiency
  for b-quark and light flavour quark jets. (b) background rate as a function of the b-tagging efficiency for c-quark and light flavour quark jets. }
\label{fig:HLR-flavtag}
\end{figure}

The efficient identification of heavy flavour jets in hadronic events is an indispensable ingredient to many important physics analyses, such as
the $\PH\rightarrow\Pqc\APqc$ and $\PH\rightarrow\Pqb\APqb$ branching ratio  measurements.
The LCFIPlus tool, described in section~\ref{sec:model:hlr}, is used for flavour tagging with BDTs. The training of the BDTs is done with
$\Pep \Pem \rightarrow 6~\Pquark$ events at $\sqrt{s}=\unit{500}{\GeV}$, where all quarks are chosen to have the same flavour, mostly
from $\Pep \Pem \rightarrow \PZ\PZ\PZ$ events. The jets are predominantly produced in the central region of the detector in these events at the given center of mass energy.
The performance is evaluated with a sub-sample of the same type of events that has not been used for training the BDT.
The resulting performance is shown in Fig.~\ref{fig:HLR-flavtag} for the large and small ILD detector model.
In (a) the background rate as a function of the c-tagging efficiency for b-quark and light flavour quark jets is plotted and (b) shows the
 background rate for c-quark and light flavour quark jets as a function of the b-tagging efficiency.
As expected from the impact parameter and vertex resolutions there are no significant differences observed between the large and small ILD variants
for the flavour tagging performance.
The flavour tagging performance has been seen to vary with the jet energy and jet multiplicity~\cite{Suehara:2015ura}.
The optimal performance in a physics study can be achieved by retraining the BDTs for the corresponding signal event topology.

\subsection[Hadronically decaying tau ID]{Hadronically decaying {\protect$\tau$ ID}}
\label{sec:perf:hlr:tau}

The correct identification of $\tau$ lepton decay modes is of particular importance in the extraction of observables sensitive to the $\tau$ lepton spin direction: examples are measurements of the $\tau$ polarisation and Higgs CP based on spin correlations in $H \to \tau \tau$ decays.
Hadronic $\tau$ decays typically offer the highest sensitivity to the spin, due the presence of a single neutrino. 
The identification of these hadronic decay modes can be factorised into the charged and neutral components. 
We focus here in particular on understanding the identification of the neutral part, which consists largely of photons from neutral pion (and to a lesser extent other neutral meson) decays. 
In the decays of highly boosted taus, these photons are typically rather close to both a charged particle and one or more additional photons produced in the same $\tau$ decay. 
The separation between these photons in the calorimeter depends on its inner radius, while the distance to charged particles additionally depends on the magnetic field strength. 
We can therefore expect some differences in performance for the large and small ILD detector models.

The performance of $\tau$ decay mode identification was studied in $\tau$-pair production events at a centre-of-mass energy of $500$\,GeV. 
These very highly boosted $\tau$ decays are the most challenging to reconstruct due to the small distance between particles in the highly collimated $\tau$ decay jets.
The standard ILD reconstruction algorithms were applied to these events.

\begin{figure}[htbp]
\begin{subfigure}{0.49\hsize} 
\includegraphics[width=\textwidth]{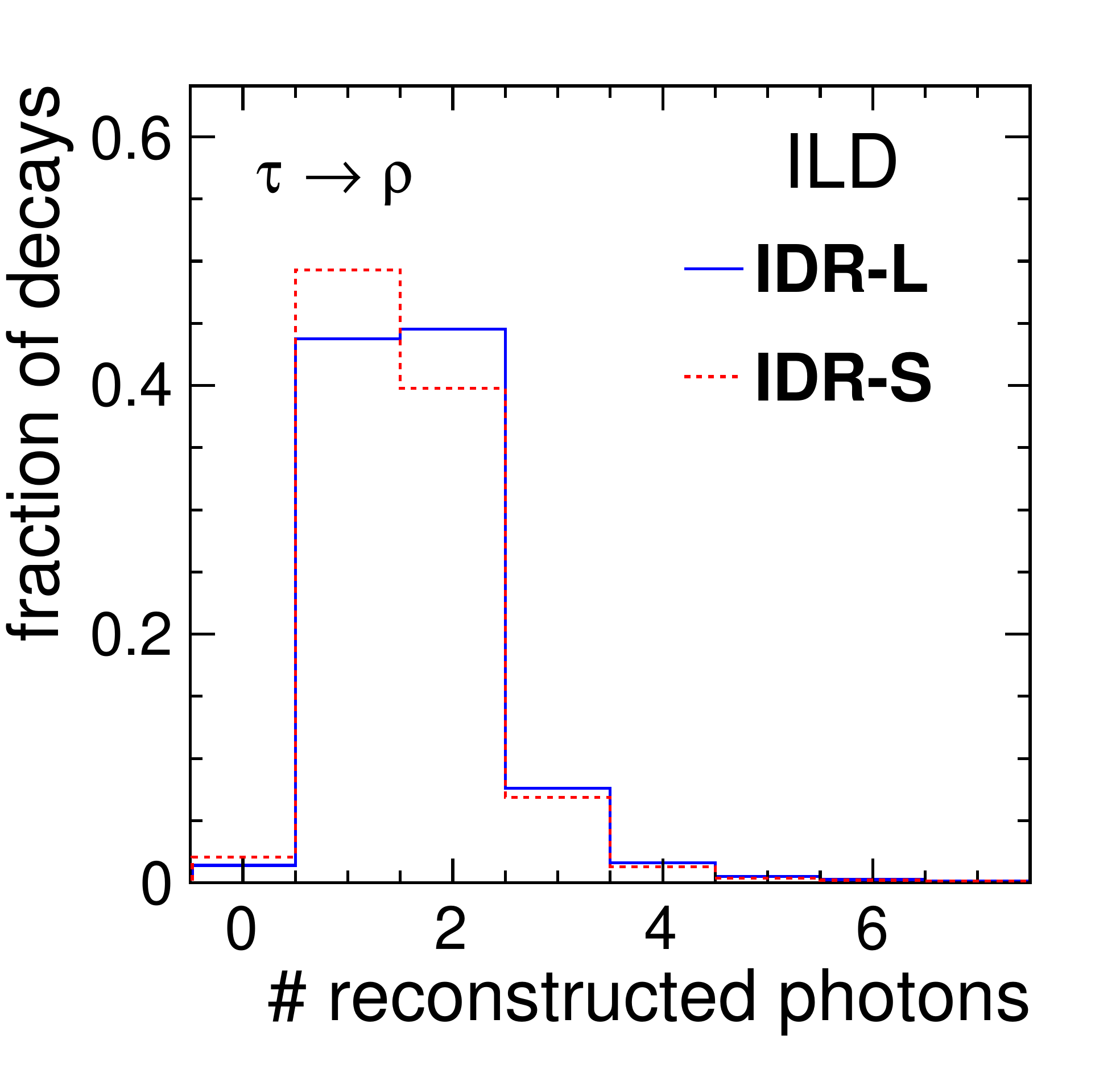}
 \caption{ \label{fig:HLR-tauID:ngamma}}
 \end{subfigure}
\begin{subfigure}{0.49\hsize} 
\includegraphics[width=\textwidth]{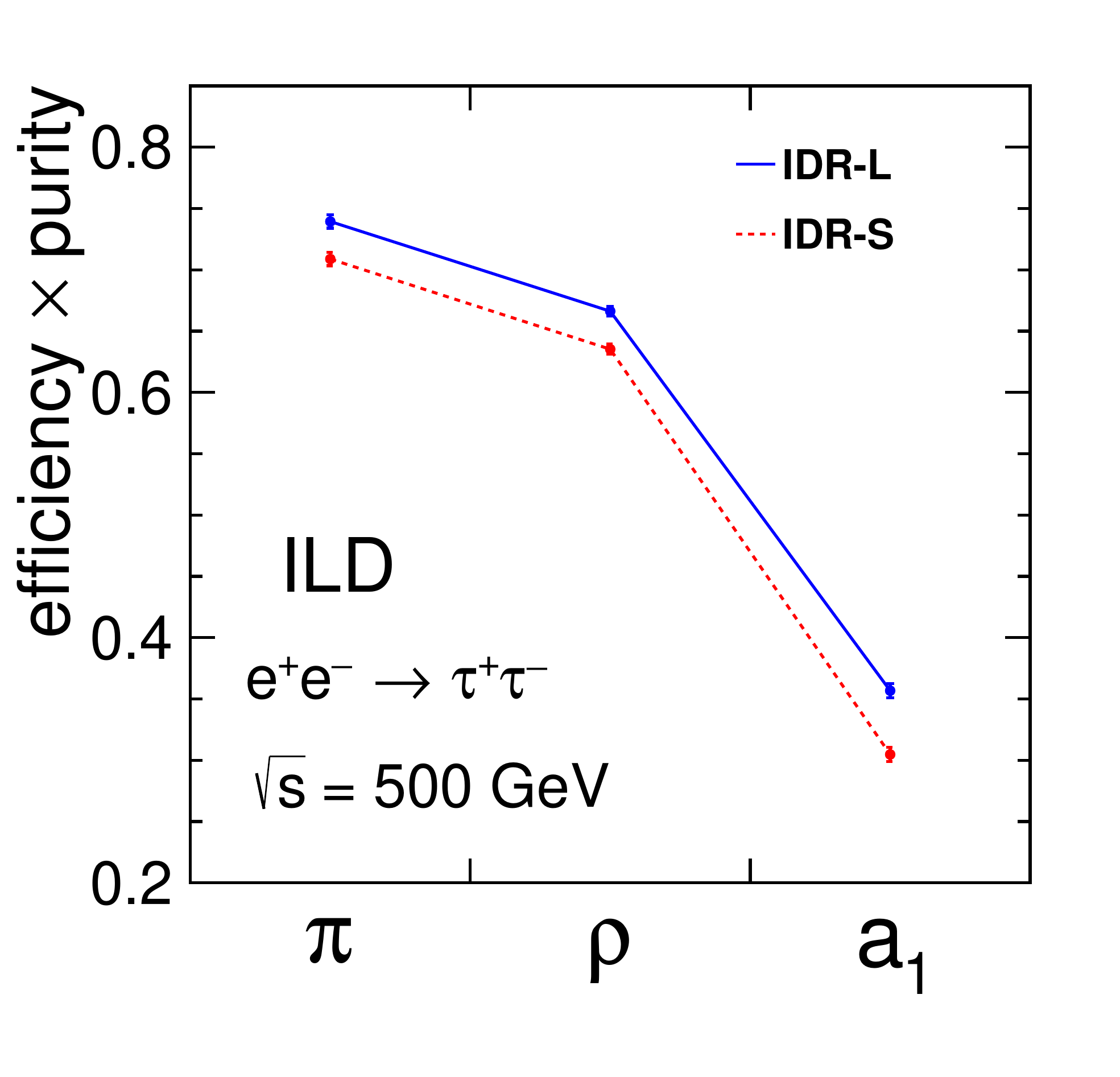}
 \caption{  \label{fig:HLR-tauID:effpur}}
 \end{subfigure}
\caption{Hadronic decay mode identification for isolated $\tau$-leptons with momenta near $250$\,GeV.
(a) The number of reconstructed photon PFOs in  $\tau \to \rho \nu \to \pi^\pm \pi^0 \nu$.
(b) The performance of a simple $\tau$ decay mode identification algorithm.
}
\label{fig:HLR-tauID}
\end{figure}

In each event, two high momentum, back-to-back, charged \emph{particle flow objects (PFOs)} were identified as $\tau$ jet "seeds". Figure~\ref{fig:HLR-tauID} shows the number of photon PFOs identified in a cone around these $\tau$ seed tracks, in the case when the $\tau$ lepton decayed to $\pi^\pm \pi^0 \nu$.
In these events, exactly two photons are expected in the vast majority of cases. 
The distribution shows that it is challenging to reconstruct both photons: in around half of the cases only a single photon cluster was reconstructed, which is due to the merging of the two photons into a single reconstructed particle. 
A difference is seen between the two detector models, with the large version somewhat more often correctly resolving the two photons, which can be understood as being due to the larger ECAL radius increasing the distance between the photons' electromagnetic showers.

A simple cut-based approach to the identification of single-prong hadronic $\tau$ decays was developed, based on the number of reconstructed photon PFOs, and the invariant mass of this set of photon PFOs both alone and together with the charged PFO around which the $\tau$ candidate jet was built.
The ability of this algorithm to distinguish $\tau \rightarrow \pi^\pm \nu$ ("$\pi$"), $\tau \rightarrow \pi^\pm \pi^0 \nu$ ("$\rho$"), and $\tau \rightarrow \pi^\pm \pi^0 \pi^0 \nu$ ("$\mathrm{a_1}$") decays is shown in Fig.~\ref{fig:HLR-tauID}. 
The product of selection efficiency and purity varies between around 30\% and 75\% among these three decay modes.
The large detector model performs slightly better, as expected thanks to the larger inner ECAL radius.

\subsection[J/Psi reconstruction]{$J/\psi$ reconstruction}
With the excellent particle flow, particle identification and vertexing capabilities
of ILD discussed in the previous sections, there is a great potential to reconstruct
various exclusive decay chains of short-lived baryons and mesons. This is of special
interest in case the ILC will be operated at the $Z$ pole, but also at higher energies we expect further improvements to flavour tagging and jet energy resolution
once this potential is fully exploited in reconstruction and analyses. This is an area where new algorithmic developments could still lead to a significant improvement
of the ILD performance.

At the current stage, we only provide a very simple example as a proof of principle, namely the reconstruction of $J/\psi \to \mu^+\mu^-$ decays. Due to its extremely well-known mass (known to 3.6\,ppm~\cite{Tanabashi:2018oca}), the $J/\psi$ is an important standard candle, e.g.\ for calibration of the tracker momentum scale.

\begin{figure}[htbp]
\begin{center}
 \includegraphics[width=0.75\textwidth]{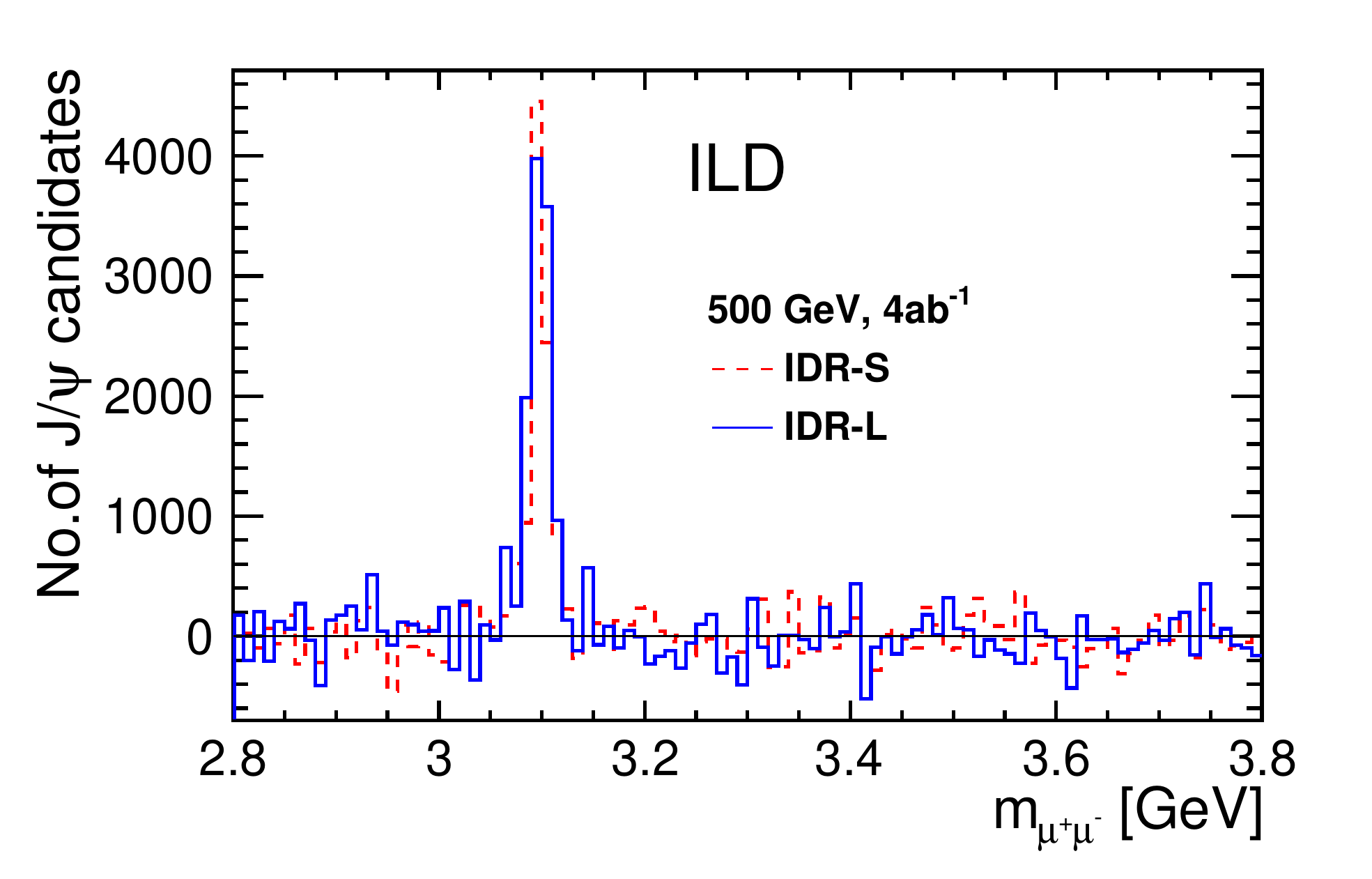}
\end{center}
\caption{$J/\psi \to \mu^+\mu^-$ candidates as reconstructed with IDR-L and IDR-S for ILC500 as defined in Sec.~\ref{sec:benchmarks:lep}.}
\label{fig:hlr:jpsi}
\end{figure}

Figure~\ref{fig:hlr:jpsi} shows the invariant mass spectrum of reconstructed muon pairs in the $J/\psi$ region, comparing the large and small detector. All available
SM MC events from the optimisation production at $\sqrt{s}=500$\,GeV (c.f.\ Sec.~\ref{chap:modelling}) have been used and weighted to the ILC500 conditions as defined in Sec.~\ref{sec:benchmarks:lep}. The combinatorial background has been determined from like-sign muon pairs and has been subtracted from the opposite-sign pairs, leading to the fluctuations around zero in the off-peak regions.
At $500$\,GeV, most $J/\psi$ candidates are produced in the forward direction, with $|\cos{\theta}|>0.8$ and at rather low transverse momenta, typically below $30$\,GeV.
Due to the larger $B$-field, the small detector has a smaller acceptance at low $p_t$ than the large detector, with about $10$k vs $12$k recontructed $J/\psi$'s compared to about $20$k available at generator level. On the other hand, the better momentum resolution of the small detector in the forward region leads to a more narrow peak.



\section{\label{sec:benchmarks} Physics Benchmarks}


The performance of the two ILD detector models IDR-L and IDR-S has been
evaluated on a number of physics benchmarks discussed in this section. 
Despite the fact that a 250-GeV version of the ILC is currently under political consideration in Japan, the detector benchmarking has been performed at higher center-of-mass energies, mostly at 500\,GeV, and in one case even at 1\,TeV. This choice has been made in order to make sure that both detector models perform adequately also under the more challenging experimental conditions of the higher energy stages, and in order to cover e.g.\ a wider range of jet, lepton and photon energies.

The benchmarks presented in this section have been chosen to illustrate many performance aspects with a minimum number of physics channels, and are not meant to cover the {\em complete} physics case. The main focus of the analysis work was not always the pure optimisation for utmost physics performance, but rather to better understand and highlight the role of individual performance aspects and their interplay. For all benchmarks, the MC samples described in Sec.~\ref{chap:modelling} have been used. In particular, the $\gamma\gamma \to $ low $p_t$ hadron as well as $e^+e^-$ pair background in the detector acceptance has been overlaid as described in Sec.~\ref{sec:generator}.

Unless stated otherwise, the analyses performed for and since the time of the DBD remain valid in their physics message. For an up-to-date review of the ILC physics case, based on ILD detector simulations, see e.g.~\cite{Bambade:2019fyw}.

\subsection{Luminosity, Energy and Polarisation for the Physics Benchmarks}
\label{sec:benchmarks:lep}
The benchmark studies are based on the Monte-Carlo samples created with the conditions and software described in Sec.~\ref{chap:modelling}. About 150 million events have been produced in total as detailed in Tab.~\ref{tab:mcprod_evtnum}. 

All results have been scaled to the integrated luminosity and beam polarisation of the full H20 running scenario originally defined in~\cite{Bambade:2019fyw}, which is --- for the higher center-of-mass energies --- unchanged since~\cite{Barklow:2015tja}. In particular, the H20 scenario comprises 4\,ab$^{-1}$ at 500\,GeV, with beam polarisation absolute values of 80\% for the electron and 30\% for the positron beam. The total luminosity is shared between the different polarisation sign combinations according to $f(-+,+-,++,--) = (40\%,40\%, 10\%, 10\%)$. This configuration is referred to as ILC500 in the following.
Analoguously, 8\,ab$^{-1}$ are considered at 1\,TeV, with absolute polarisation values of 80\% for the electron and 20\% for the positron beam, with the same $f(-+,+-,++,--) = (40\%,40\%, 10\%, 10\%)$ sharing.  This configuration is referred to as ILC1000 in the following.

\subsection{Hadronic Branching Ratios of the Higgs Boson}

The determination of the branching ratios of the Higgs boson into $b\bar{b}$, $gg$ and in particular to $c\bar{c}$ is one of the unique items
on the menu of future $e^+e^-$ colliders. These measurements crucially depend on 
an excellent flavour tag, c.f.\ Sec.~\ref{sec:perf:hlr:lcfi}, enabled
by vertex detectors with micrometer point resolution and a first layer placed as close as 1.6\,cm to the beam line, as well as the tiny, nanometer size beam spot of the ILC. The full performance of the ILC on Higgs branching ratio measurements, combining all final states,  can be found in~\cite{Bambade:2019fyw}.

As a benchmark, the $\nu \bar{\nu} H \to \nu \bar{\nu} jj$ final state, including both the dominant contribution from $WW$ fusion as well as a small contribution from $ZH\to\nu\bar{\nu} H$, was chosen in order to minimize the impact of other performance aspects like e.g.\ jet clustering. Thus the target physics observable here is $\sigma(\nu\bar{\nu} H)\times BR(H\to b\bar{b} / c\bar{c} / gg)$. With the full 500\,GeV data set, about 200000 $H \to b\bar{b}$ events would be produced in this final state alone, while about 30000 and 10000 $H \to gg$ and $H \to c\bar{c}$ would be available, respectively. In the 
limit of 100\% signal efficiency and zero background, this would correspond to statistical precisions on $\sigma \times BR$ of 0.2\%, 0.6\% and 1\% for $H \to b\bar{b}$, $H \to gg$ and $H \to c\bar{c}$, respectively.

\begin{figure}[htbp]
\begin{center}
 \includegraphics[width=0.5\textwidth]{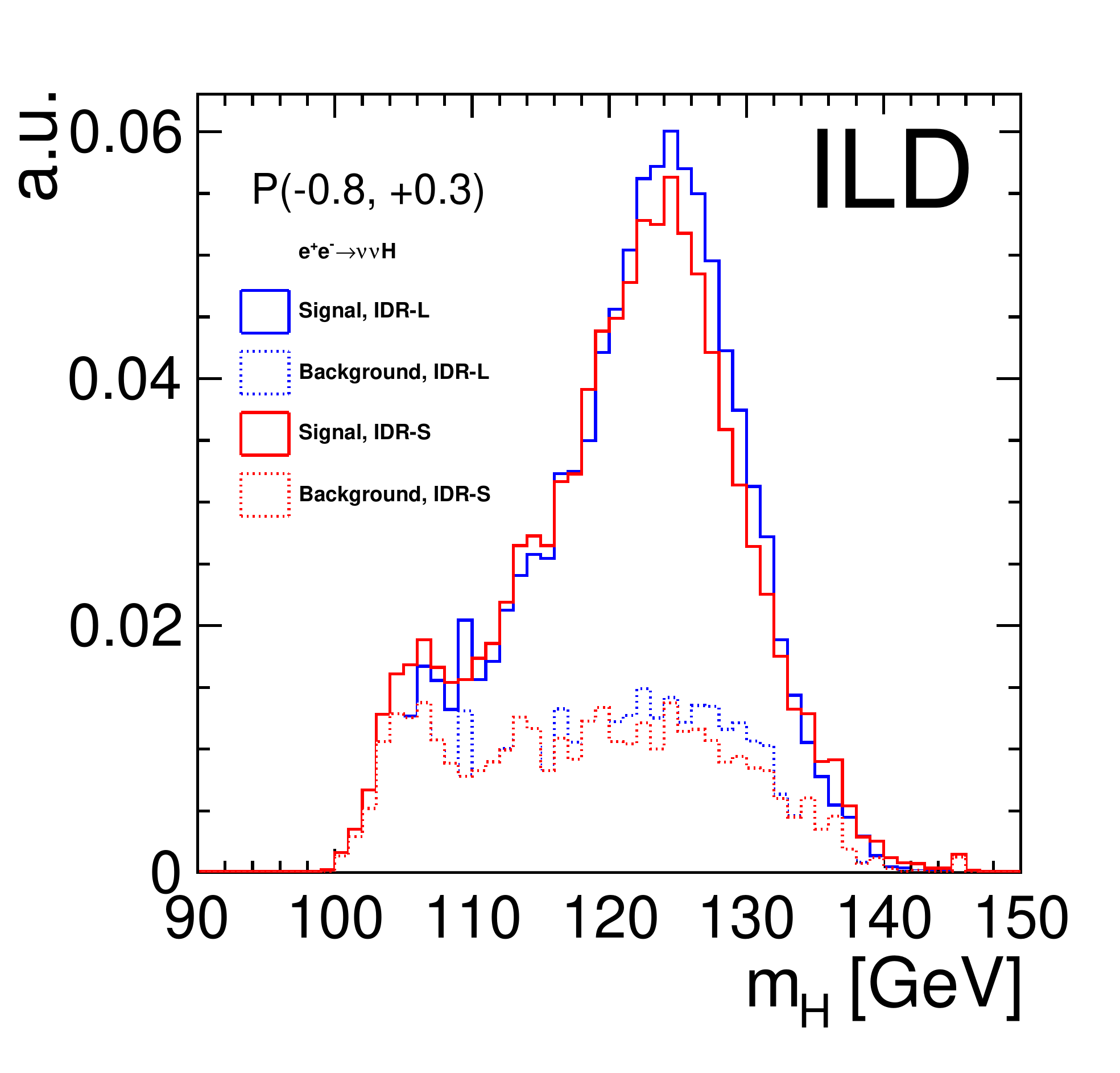}
\end{center}
\caption{Reconstructed Higgs mass distribution after the kinematic selection. The signal is shown on top of the remaining SM background for the $P(e^-,e^+)=(-80\%,+30\%)$ data set of ILC500 as defined in Sec.~\ref{sec:benchmarks:lep}. The histograms have been normalized such that $S+B$ has an integral of $1$ in order to allow a better shape comparison between the two detector models.
}
\label{fig:Hbbccgg:mh}
\end{figure}

The benchmark analysis is documented in detail in~\cite{ILDNote:Hbbccgg}, and follows earlier analyses~\cite{Mueller:2016exq,Ono:2013voc,Ono:2013sea}.  After a cut-based preselection, the kinematic selection of $\nu \bar{\nu} H \to \nu \bar{\nu} jj$ events is refined by a multi-variate approach. Up to this point, no flavour-tag information is used.
Figure~\ref{fig:Hbbccgg:mh} shows the distributions of the reconstructed Higgs mass at this stage of the analysis for the signal on top of the remaining SM backgrounds, comparing both detector models. In this plot, the relative normalisation of signal and background is given by the respective cross sections, but the total histograms for $S+B$ have been normalised to an integral of $1$ in order to allow a shape comparison between IDR-L and IDR-S.

\begin{figure}[htbp]
\begin{center}
 \includegraphics[width=\textwidth]{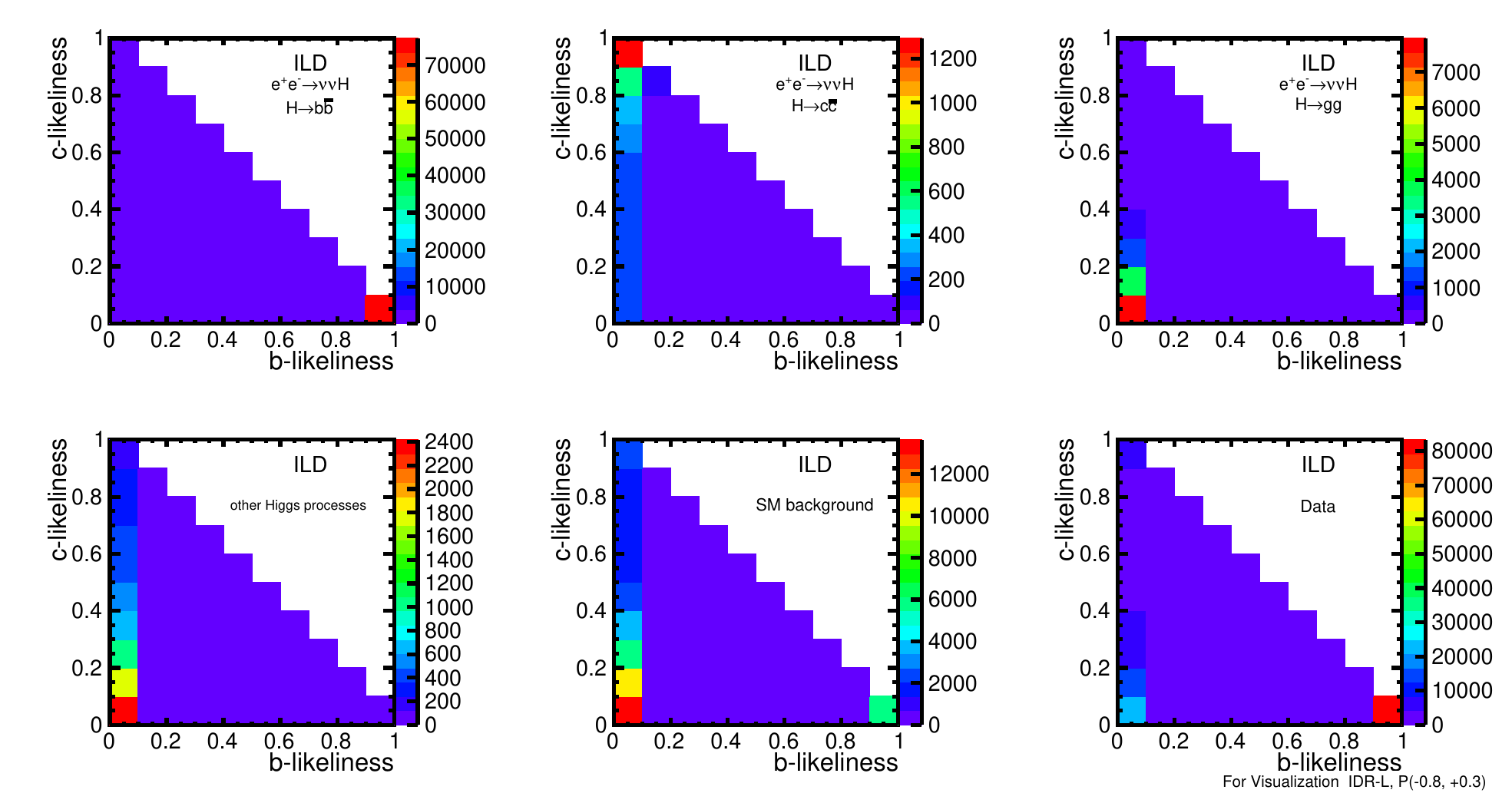}
\end{center}
\caption{Visualisation of the flavour tag performance in $\nu\bar{\nu} H$. The panels show the 2D distributions of $c$- vs $b$-likeliness separately for $H \to b\bar{b}$, $H \to c\bar{c}$, $H \to gg$, $H \to$other, the SM background and their mix expected in data for the $P(e^-,e^+)=(-80\%,+30\%)$ data set of ILC500 as defined in Sec.~\ref{sec:benchmarks:lep}.
}
\label{fig:Hbbccgg:likeli}
\end{figure}

\begin{figure}[htbp]
\begin{subfigure}{0.49\hsize} 
 \includegraphics[width=\textwidth]{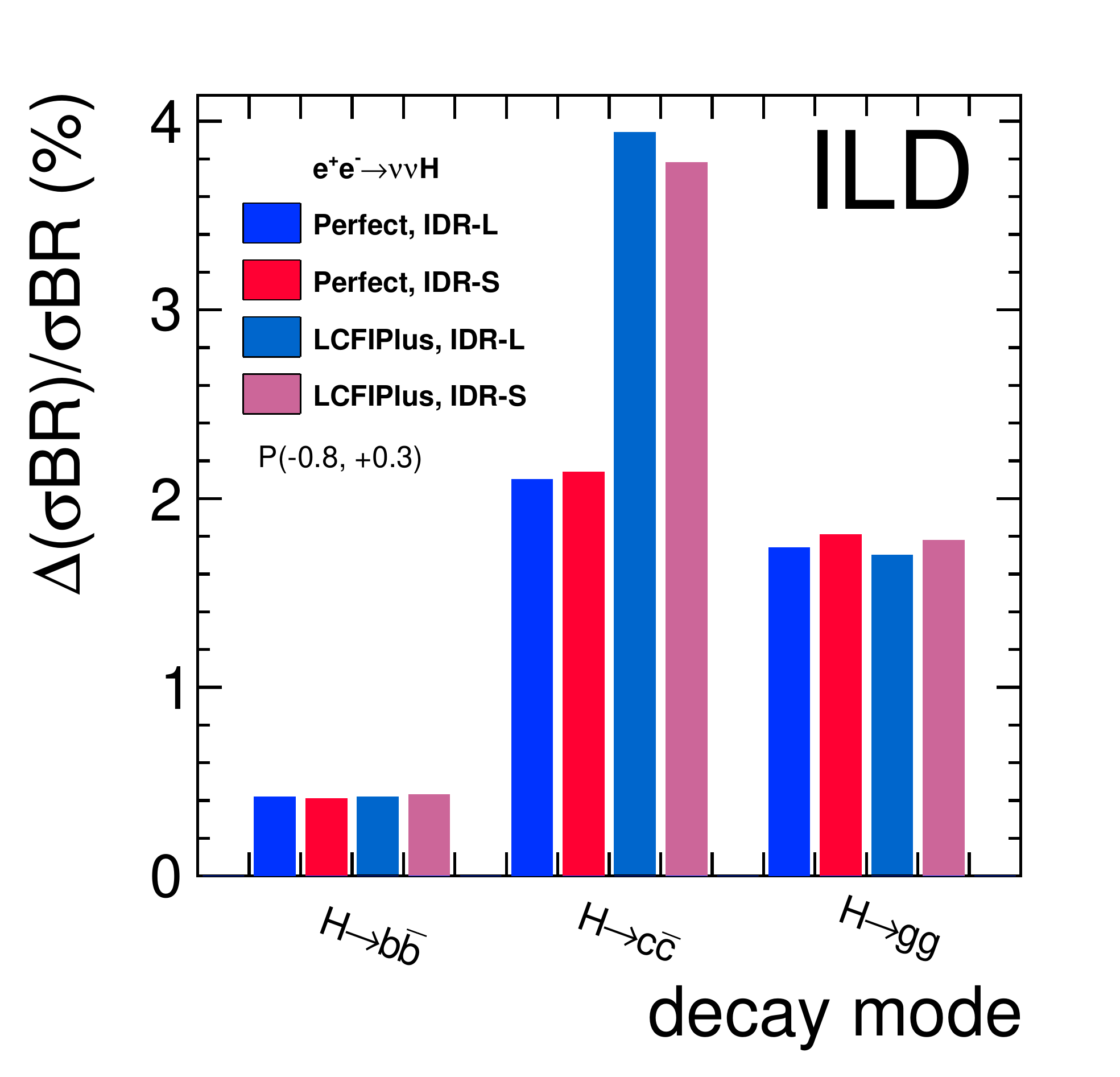}
 \caption{ \label{fig:Hbbccgg:BR:cheat}}
 \end{subfigure}
\begin{subfigure}{0.49\hsize} 
 \includegraphics[width=\textwidth]{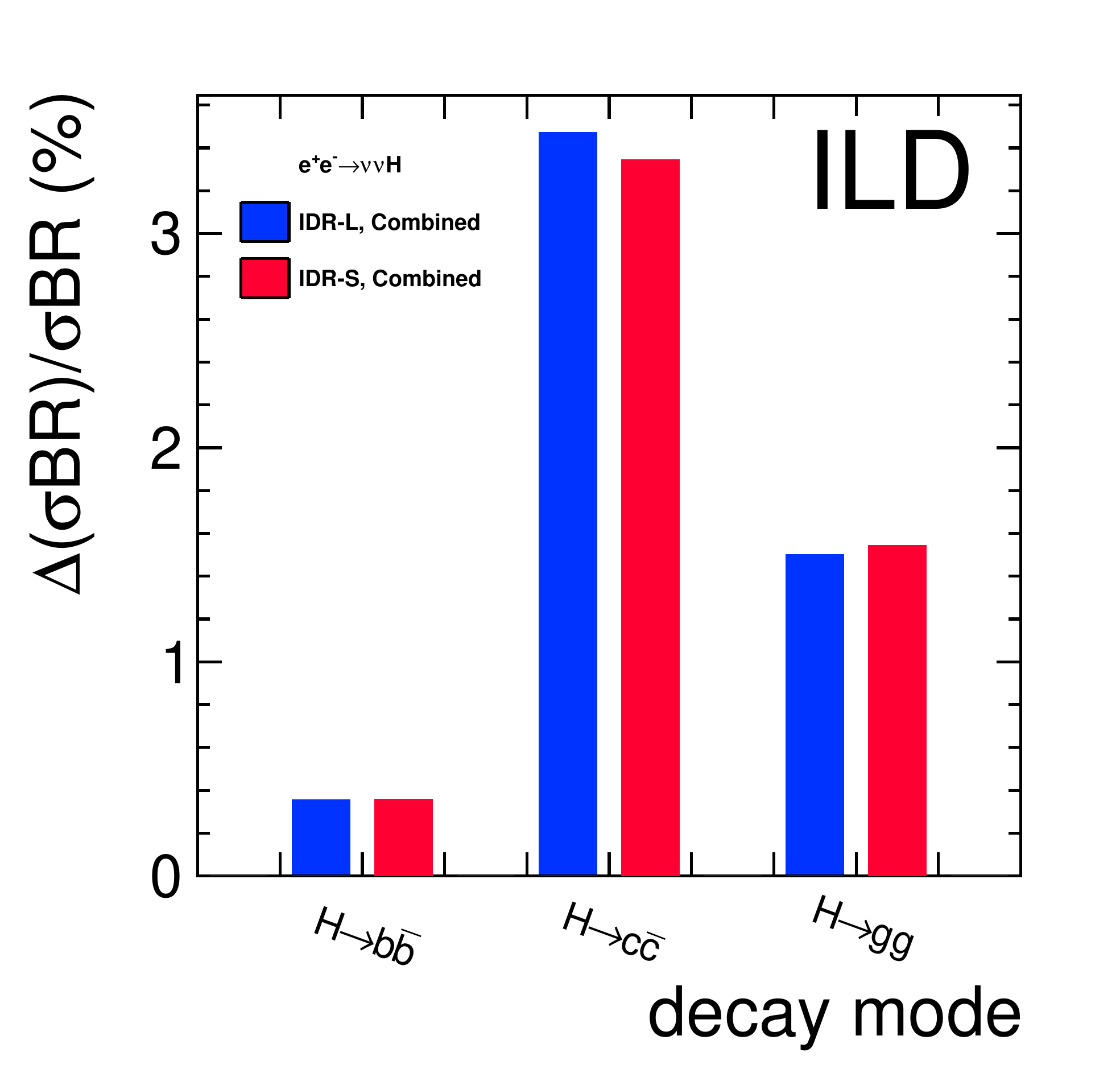}
 \caption{  \label{fig:Hbbccgg:BR:LS}}
 \end{subfigure}
\caption{
(a) Comparison of current and ideal flavour tag performance for the $P(e^-,e^+)=(-80\%,+30\%)$ data set of ILC500 as defined in Sec.~\ref{sec:benchmarks:lep}. The precision on $H \to c\bar{c}$ is most sensitive to the actual flavour tag performance.
(b) Comparison of IRD-L and IDR-S for all polarisation configurations of ILC500 as defined in Sec.~\ref{sec:benchmarks:lep} combined.
}
\label{fig:Hbbccgg:BR}
\end{figure}

Figure~\ref{fig:Hbbccgg:likeli} shows the 2D distributions of $c$- vs $b$-likeliness (c.f.\ Sec.~\ref{sec:perf:hlr:lcfi}) for the different Higgs decay modes. The $x$-likeliness (for $x=c$, $b$ or $bc=c/(b+c)$) is obtained by combining the flavour tag output for the two individual jets, $x_1$ and $x_2$, according to
\begin{equation}
x = \frac{x_1 \cdot x_2}{(x_1\cdot x_2+(1-x_1)(1-x_2))}
\label{eqn:flavourtag}
\end{equation}
The ``data'' distribution is then fitted in 3D template approach in order to determine the contained fractions of the various hadronic Higgs decay modes, where the 3rd dimension is the $bc$-likeliness. Due to the limited available statistics of background MC, a much smaller number of bins than displayed in Fig.~\ref{fig:Hbbccgg:likeli} was used~\cite{ILDNote:Hbbccgg}. The resulting precisions from this template fit are displayed in Fig.~\ref{fig:Hbbccgg:BR}.  Fig.~\ref{fig:Hbbccgg:BR:cheat} shows the actual results for IDR-L and IDR-S with the $P(e^-,e^+)=(-80\%,+30\%)$ data only. For comparison,  the corresponding result obtained for a perfect flavour tag is also displayed. The comparison shows that for $H \to b\bar{b}$ and $H \to gg$, the current flavour tag performance yields a close to perfect identification of these final states. For $H \to c\bar{c}$, however, the
real flavour tag performs worse by a factor of two. On the other hand, for a worse flavour separation, especially the expected precision for $H \to c\bar{c}$ degrades rapidly~\cite{ILDNote:Hbbccgg},
thus the performance of the ILD detector and reconstruction is crucial for the ability to measure $H \to c\bar{c}$.

Figure~\ref{fig:Hbbccgg:BR:LS} finally compares for IDR-L and IDR-S the precisions combined over all data sets of ILC500 as defined in Sec.~\ref{sec:benchmarks:lep}. It shows a rather equivalent performance of both detector models. In the case of $H \to c\bar{c}$, which is most sensitive to the detector performance, the smaller detector model actually performs a little better: Due to the stronger magnetic field of IDR-S, less tracks from $\gamma\gamma \to$ low-$p_t$ hadron overlay are reconstructed, which helps the charm tag to perform a bit better.

\subsection{Higgs Mass from \texorpdfstring{$ZH \to ll b\bar{b}$}{ZH -> llbb}}

The single most precise method to measure the Higgs mass is the recoil analysis at $\sqrt{s}=250$\,GeV~\cite{Yan:2016xyx}, for which a precision of $14$\,MeV has been projected for the full ILC250 data set~\cite{Fujii:2017vwa}. At $\sqrt{s}$ much higher than the $ZH$ production threshold, the recoil
technique suffers substantially from the increasing amount of ISR and beamstrahlung. In addition, the momenta of the muons from $Z\to \mu\mu$ increase in magnitude and thus are measured less precisely. Still, the data collected at higher centre-of-mass energies can be exploited effectively when using the fully-reconstructable decay modes of the Higgs boson in combination with kinematic constraints\footnote{In $x$ direction the initial momentum is not zero due to the crossing angle of the beams, but given by $\sqrt{s} \cdot \sin{14\,\mathrm{mrad}}$.} from the known initial state. E.g.\ the constraints $\sum_i{p_{i_y}}=0$ and $\sum_i{p_{i_x}}=3.5$\,GeV can replace the measured energies of the Higgs decay products, so that only the {\it angles} of the Higgs decay products and the momenta of the muons from $Z\to\mu^+\mu^-$ enter in the mass reconstruction. Thus the technique is independent of systematic uncertainties as e.g.\ associated with the $b$-jet energy scale. The detailed description of the technique can be found in~\cite{ILDNote:MH}.

\begin{figure}[htbp]
\begin{subfigure}{0.49\hsize} \includegraphics[width=\textwidth]{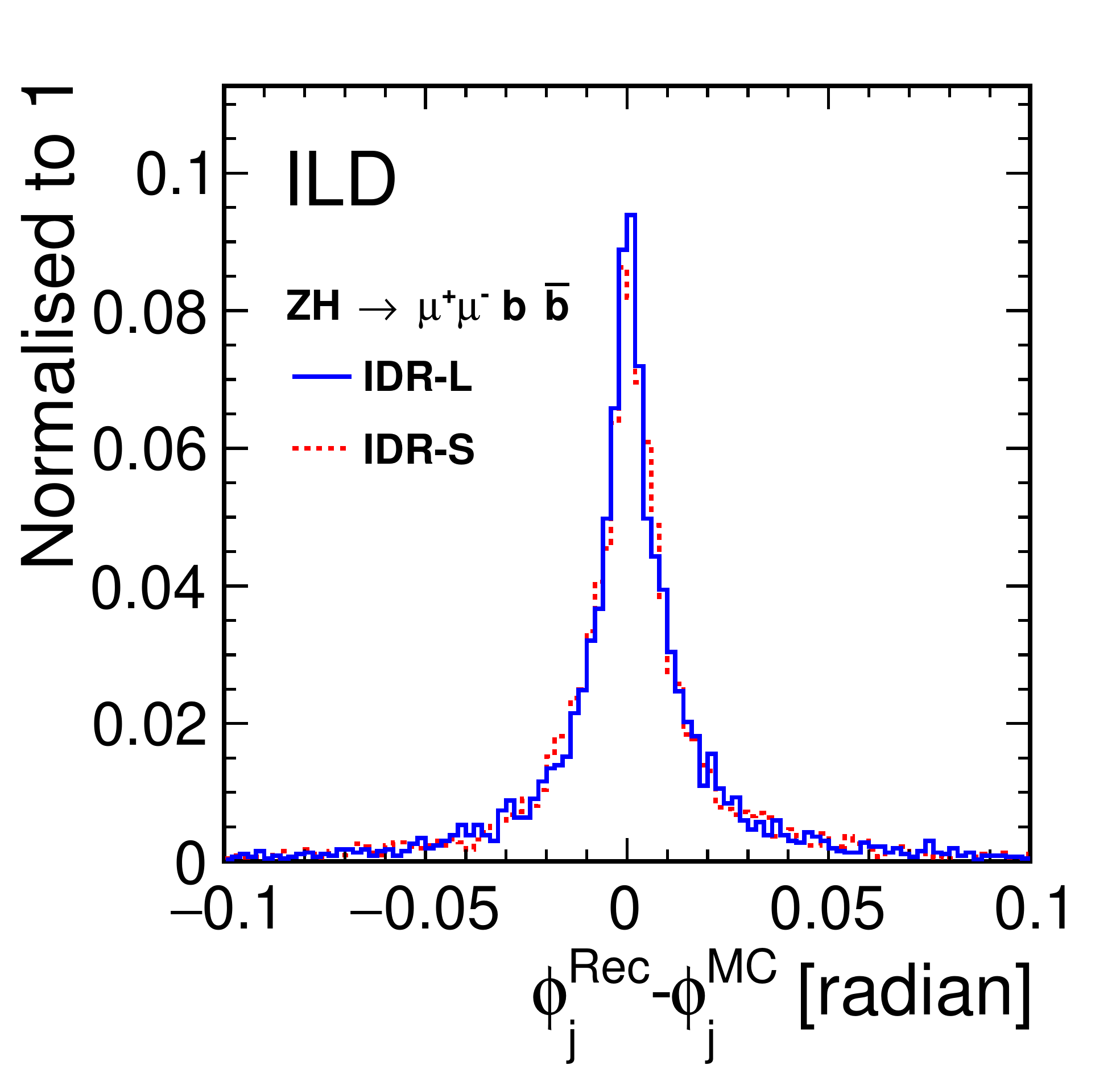}
 \caption{ \label{fig:mh:res:phi_LS}}
 \end{subfigure}
\begin{subfigure}{0.49\hsize} \includegraphics[width=\textwidth]{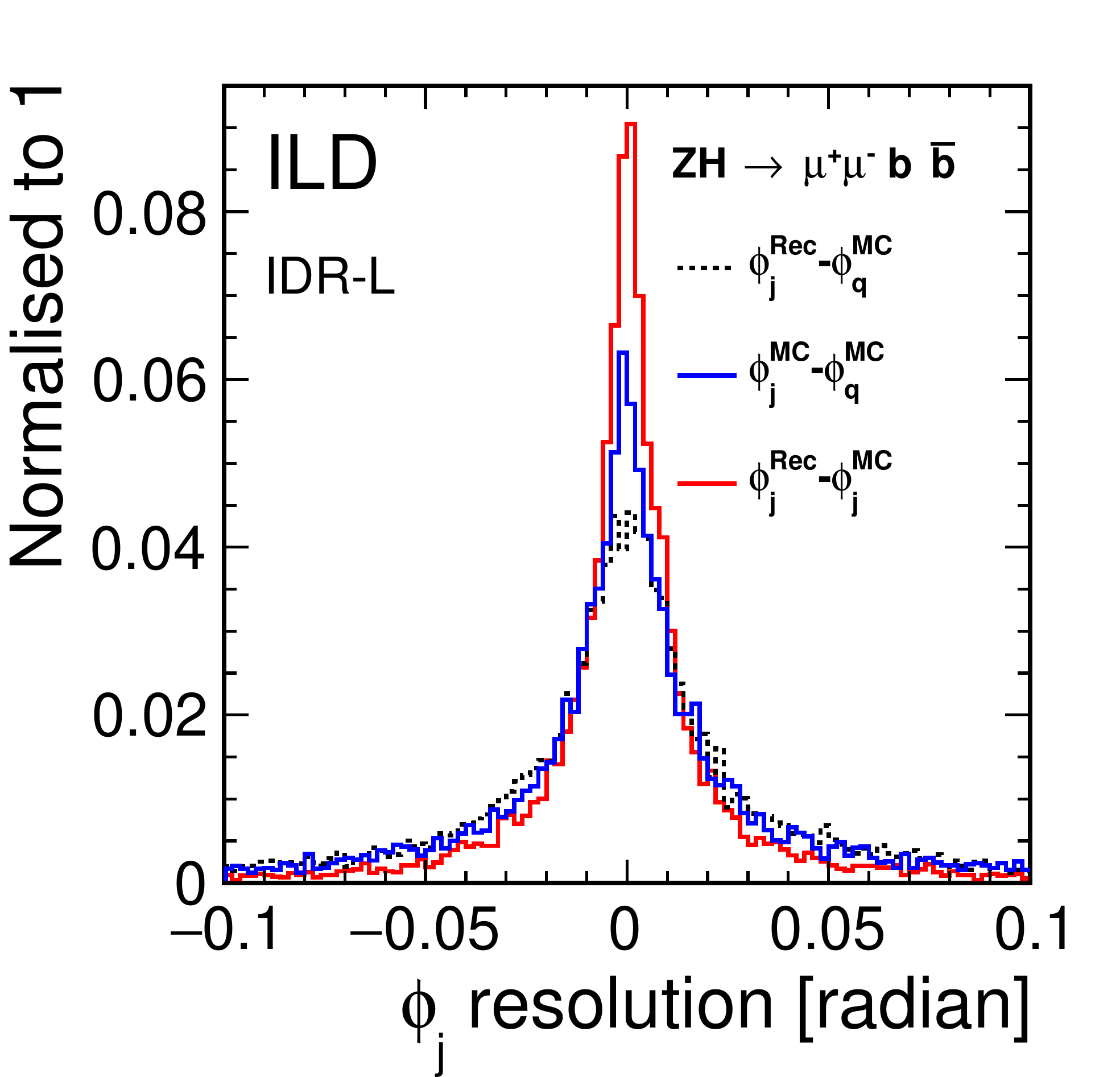}
 \caption{  \label{fig:mh:res:phi_hadronisation}}
 \end{subfigure}
\caption{Reconstruction of the jet azimuthal angle in $ZH \to \mu^+ \mu^- b\bar{b}$: 
(a) Resolution obtained with IDR-L and IDR-S for the jet azimuthal angle
(b) Influence of hadronisation on the jet azimuthal angle
}
\label{fig:mh:res:phi}
\end{figure}

\begin{figure}[htbp]
\begin{subfigure}{0.49\hsize} \includegraphics[width=\textwidth]{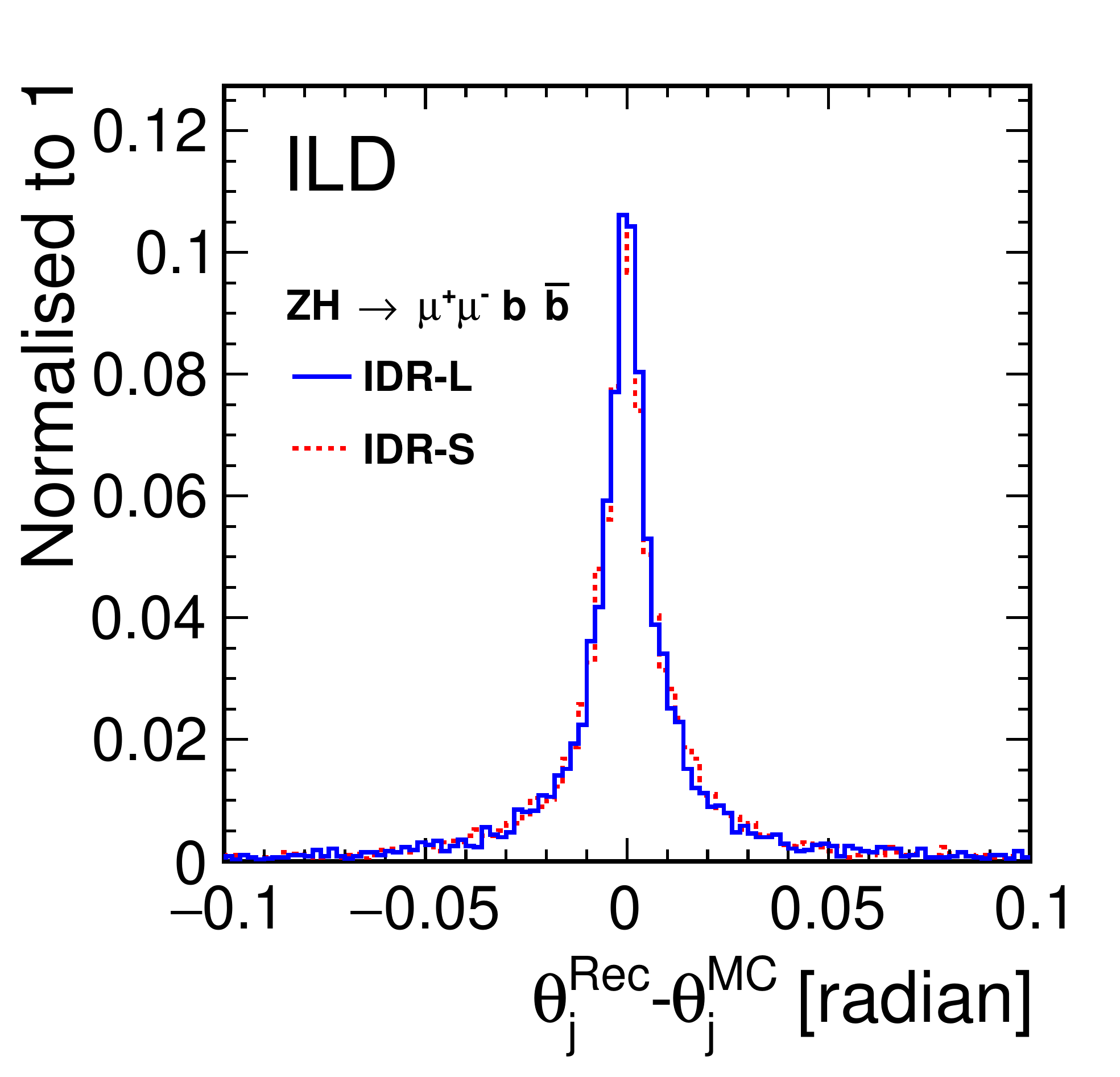}
 \caption{  \label{fig:mh:res:theta}}
 \end{subfigure}
\begin{subfigure}{0.49\hsize} \includegraphics[width=\textwidth]{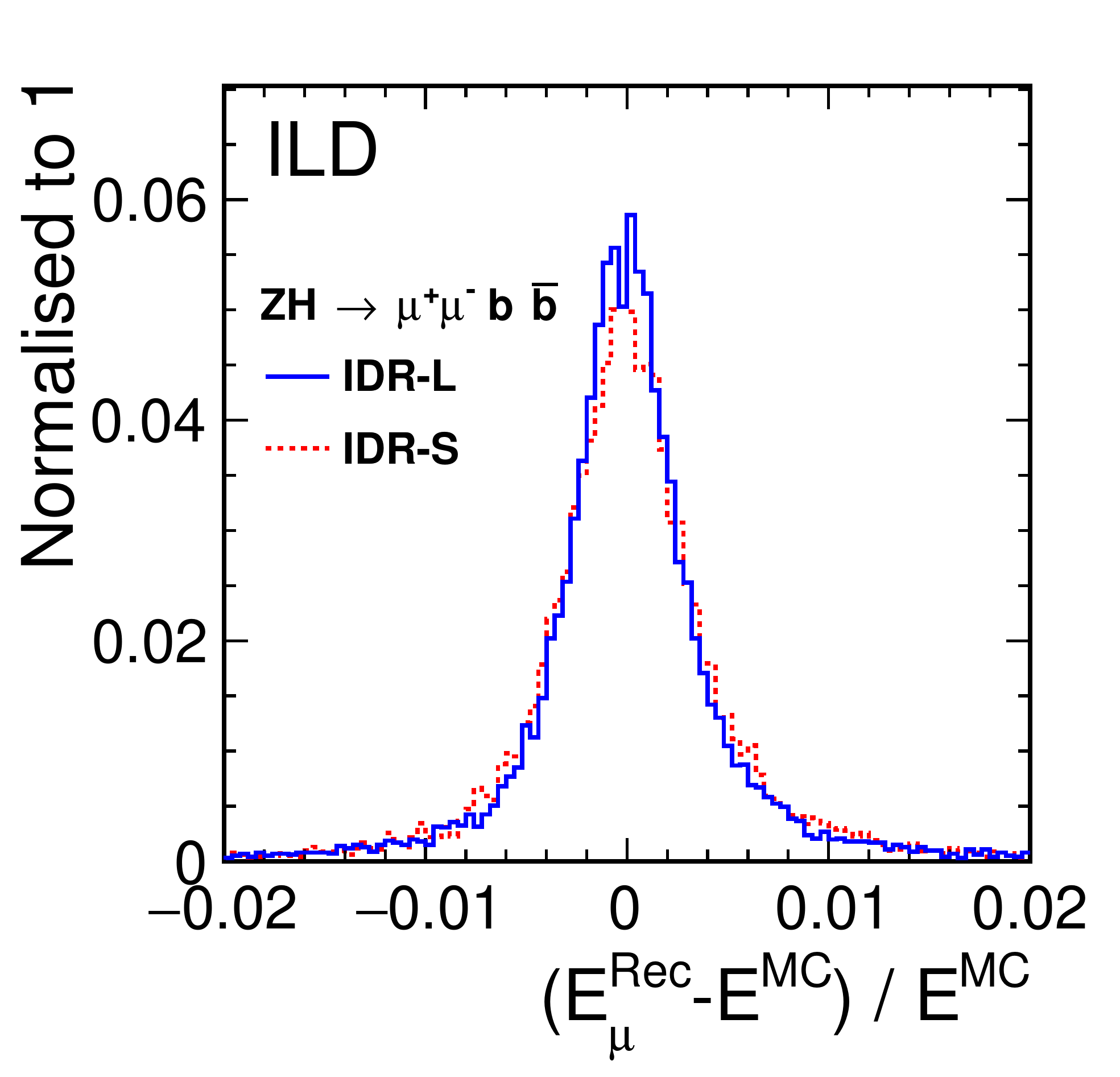}
 \caption{  \label{fig:mh:res:Elep}}
 \end{subfigure}
\caption{Resolutions obtained  in $ZH \to \mu^+ \mu^- b\bar{b}$ with IDR-L and IDR-S for 
(a) the jet polar angle
(b) the muon energy, after recovery of FSR photons. 
}
\label{fig:mh:res:etheta}
\end{figure}

The resolutions on the kinematic quantities which enter the Higgs mass reconstruction, namely azimuthal and polar angles of the jets and the muon energy, are compared in Figs.~\ref{fig:mh:res:phi_LS} and~\ref{fig:mh:res:etheta} for IDR-L and IDR-S. For this, 
the jet angles obtained from clustering the MC particles after hadronisation serve as truth reference, so that the detector performance is singled out from other influences.  Figure~\ref{fig:mh:res:phi_hadronisation} illustrates the effect of hadronisation by comparing to the quark-level direction taking IDR-L as example. It shows that the detector resolution (blue histogram, same as in Fig.~\ref{fig:mh:res:phi_LS}) is comparable, but
smaller than the hadronisation effect (red histogram). While the angular resolutions are very similar for both detector models, the muon energy resolution is worse for the small detector, reflecting the somewhat worse $p_t$ resolution for high-momentum tracks in the central region of the detector, c.f.\ Fig.~\ref{fig:perf:trkres}. 
 
\begin{figure}[htbp]
\begin{subfigure}{0.49\hsize} \includegraphics[width=\textwidth]{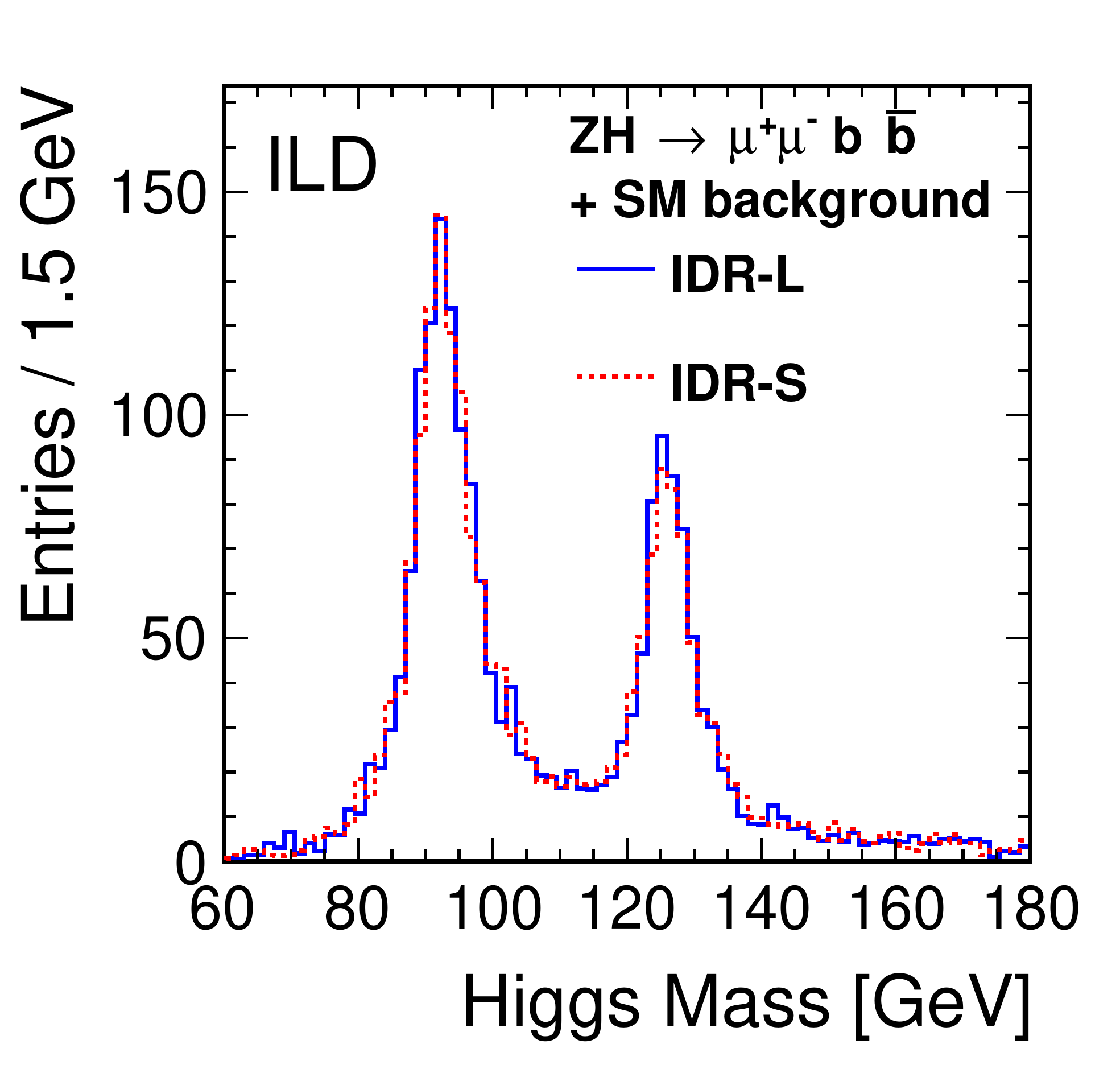}
 \caption{ \label{fig:mh:mass:mumu}}
 \end{subfigure}
\begin{subfigure}{0.49\hsize} \includegraphics[width=\textwidth]{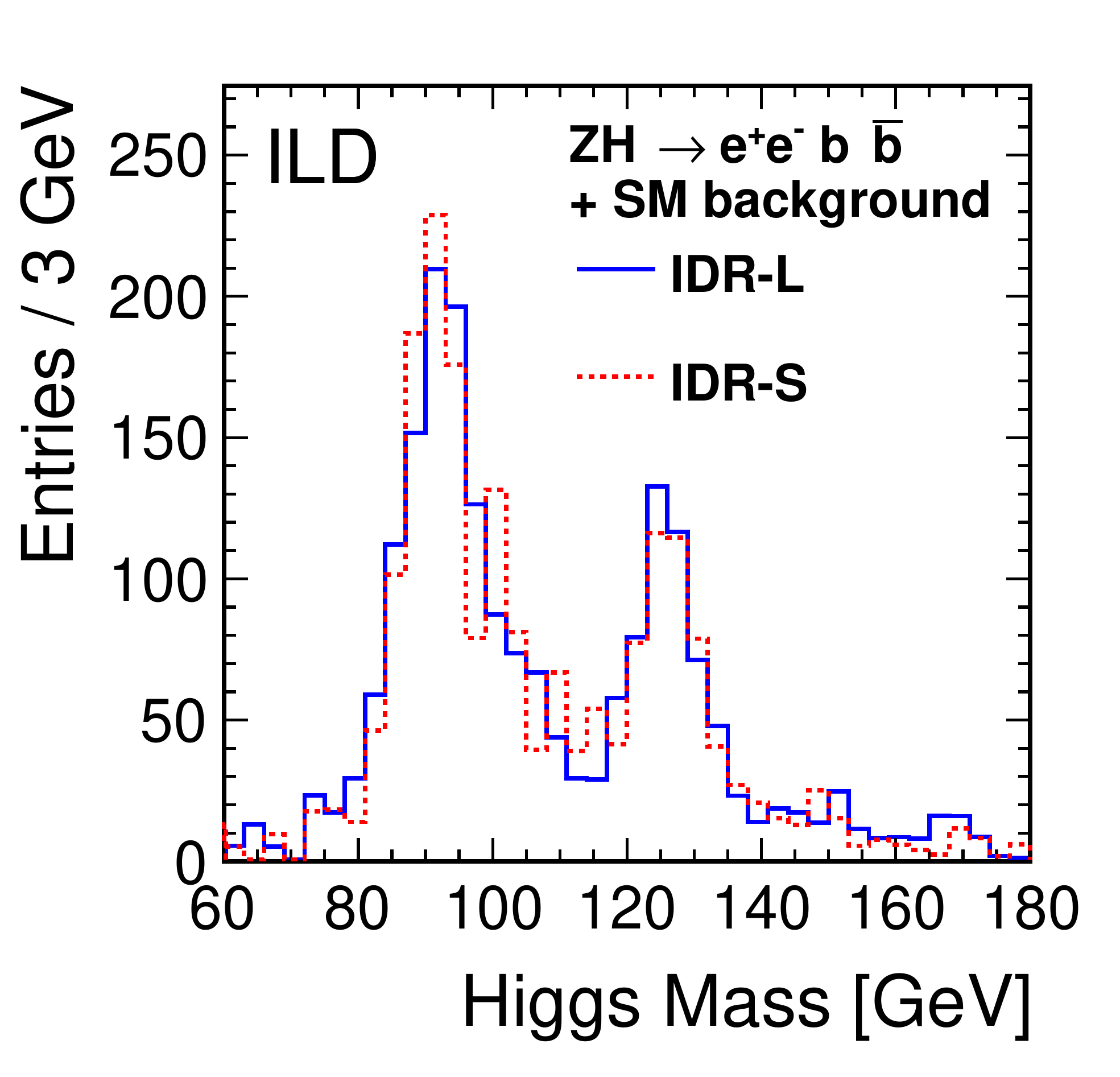}
 \caption{  \label{fig:mh:mass:ee}}
 \end{subfigure}
\caption{Reconstructed Higgs mass distribution in $ZH \to l^+ l^- b\bar{b}$ for signal and background for IDR-L and IDR-S at ILC500 as defined in Sec.~\ref{sec:benchmarks:lep} for
(a) the muon channel
(b) the electron channel (here with limited MC statistics, thus a reduced number of bins)
}
\label{fig:mh:mass}
\end{figure}

Figure~\ref{fig:mh:mass} shows the propagation of this effect to the actual physics observable, i.e.\ the reconstructed mass of the Higgs candidates. IDR-L and IDR-S are compared for the muon channel in Fig.~\ref{fig:mh:mass:mumu} and for the electron channel in Fig.~\ref{fig:mh:mass:ee}. In both cases signal and all backgrounds from the SM and other $ZH$ modes are combined, with clear peaks visible around the nominal Higgs and $Z$ boson masses. In particular in the muon channel, the worse momentum resolution of the small detector leads to the peaks being visibly wider. The uncertainty on the Higgs mass is extracted by fitting these distributions following the approach used for the recoil analysis~\cite{Yan:2016xyx}. For ILC500 as defined in Sec.~\ref{sec:benchmarks:lep}, this results in an uncertainty on the measured Higgs mass of $66$\,MeV in the case of IDR-L and $81$\,MeV for IDR-S, an increase of about $22\%$.

\subsection{Branching Ratio of \texorpdfstring{$H \to \mu^+\mu^-$}{H -> mm}}

In the SM, the decay of the Higgs boson into a pair of muons is a very rare decay, with a branching ratio of $2.2 \times 10^{-4}$. In order to identify this small signal, the achievable mass resolution, thus the precision to which the momenta of the two muons are measured, plays a crucial role. For the purpose of detector benchmarking, we consider only 
$\sigma(\nu\bar{\nu} H)\times BR(H\to \mu^+\mu^-)$ as observable, which isolates best the effect of the muon momentum resolution. At center-of-mass energies above $350$\,GeV, the $\nu\bar{\nu} H$ channel gives the leading contribution to the combined result, while at $250$\,GeV $q\bar{q} H$ dominates. The ILD result for combining all channels at $250$\,GeV and $500$\,GeV, based on DBD MC samples, can be found in~\cite{Kawada:2019isz}.

\begin{figure}[htbp]
\begin{subfigure}{0.49\hsize}  
\includegraphics[width=\textwidth]{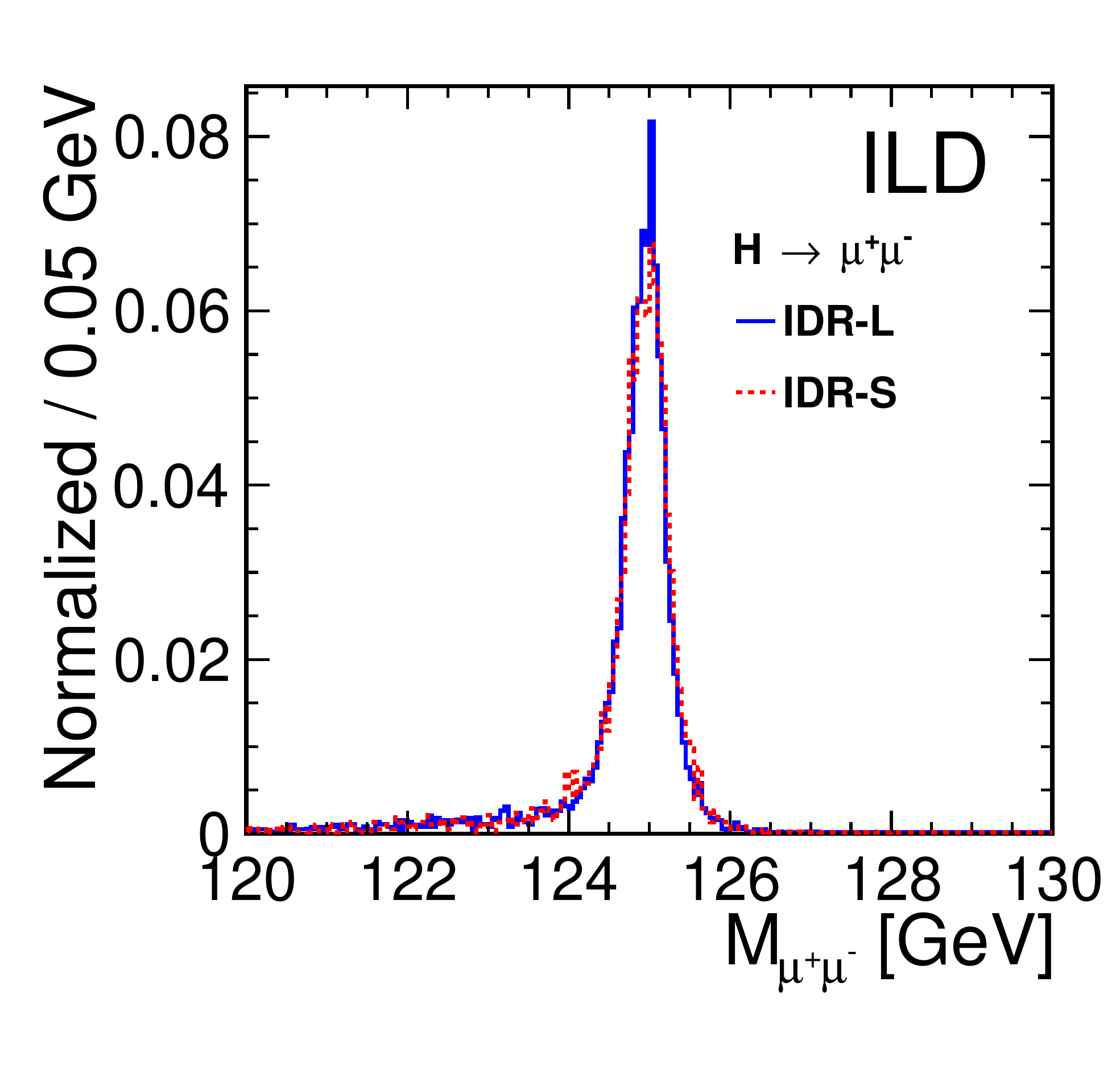}
\caption{ \label{fig:Hmumu:mass}}
 \end{subfigure}
\begin{subfigure}{0.49\hsize}  \includegraphics[width=\textwidth]{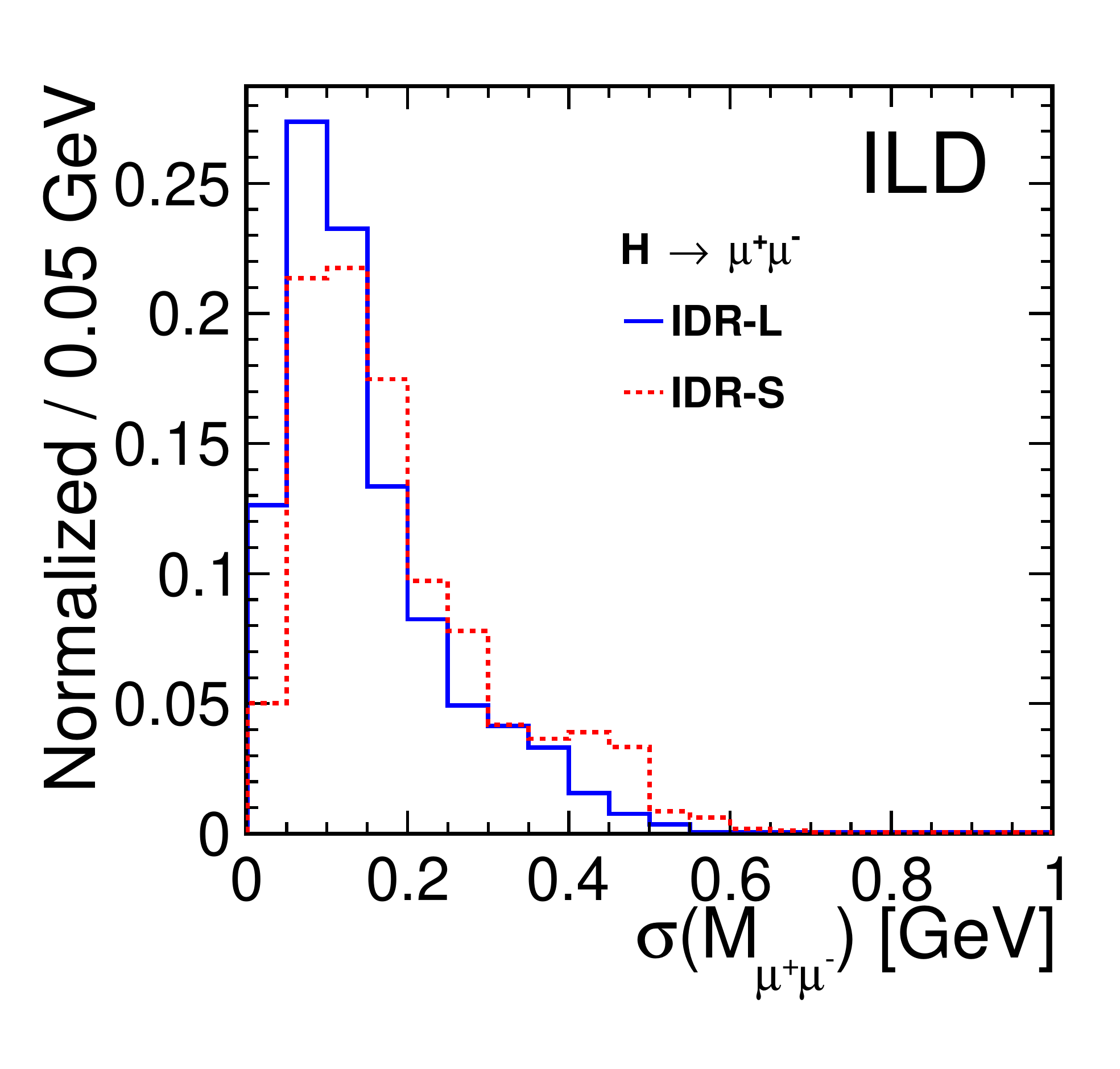}
 \caption{  \label{fig:Hmumu:sigma}}
 \end{subfigure}
\caption{$\nu\bar{\nu} H \to \nu\bar{\nu} \mu^+\mu^-$ benchmark for ILC500 as defined in Sec.~\ref{sec:benchmarks:lep}:
(a) di-muon invariant mass for all selected signal events
(b) event-by-event mass uncertainty as calculated from the track covariance matrices for the case that both muons are in the barrel region of the detector ($|\cos{\theta}| < 0.7$)
}
\label{fig:Hmumu}
\end{figure}

The selection~\cite{ILDNote:Hmumu} targets events with substantial missing four-momentum plus 
two well-reconstructed, oppositely charged muons. Kinematic and angular variables are exploited in an MVA analysis. Figure~\ref{fig:Hmumu:mass} shows the di-muon invariant mass distribution for all selected signal events. In case of the small detector model, the mass peak is about $10\%$ wider than for the large detector. This originates from the combination of the muons from the Higgs decay being highly energetic and rather central, plus the better momentum resolution of the large detector for high-momentum tracks in the barrel. 
This effect is seen more clearly in Fig.~\ref{fig:Hmumu:sigma}, which compares the event-by-event uncertainty of the di-muon invariant mass, as calculated from the track covariance matrices, for the selected signal events with both muons in the barrel region of the detector ($|\cos{\theta}| < 0.7$).

For the $500$\,GeV data set of the H20 scenario, the asymptotic precision on cross section times branching ratio for the case of $100\%$ efficiency and no backgrounds would be 13\%.  

After the event selection, about $33$ signal events, corresponding to a selection efficiency of $58\%$, remain over a evenly distributed background of about $1100$ events (counted in a mass window between $120$ and $130$\,GeV) for ILC500 as defined in Sec.~\ref{sec:benchmarks:lep}. Finally, the expectation values for the number of signal events observable above the backgrounds as well as for its uncertainty are obtained from many toy MC fits to the di-muon invariant mass spectrum~\cite{ILDNote:Hmumu}. The obtained precisions on $\sigma(\nu\bar{\nu} H)\times BR(H\to \mu^+\mu^-)$ are $40.2$\% for IDR-L and $41.3$\% for IDR-S. The relative difference of $2.8\%$ is consistent with the expectation of a $\sim 10\%$ difference in the signal peak width over a flat background. Either number, however, is about a factor $3$ worse than the asymptotic precision for the case of $100\%$ efficiency and no backgrounds, which would be 13\%. The difference is vastly dominanted by the remaining ``irreducible'' backgrounds with two muons and two neutrinos from $W$ pairs decaying either directly to muons or via tau-leptons. While there is certainly room for improvement in rejecting these backgrounds, this will factorize from the impact of the signal peak width, as long as the background remains flat in the discriminating variable.

\subsection{Sensitivity to \texorpdfstring{$H \to $ invisible}{H -> invisible}}

The decay of the Higgs boson into invisible particles is of particular interest because
it could give important clues about the nature of Dark Matter. As a detector benchmark for testing the impact of the jet energy resolution, in particular the hadronic decay mode of the $Z$ boson is considered here. Thus the physics observable will be the upper limit on $\sigma(q\bar{q} H)\times BR(H \to \mbox{inv.})$.

The event selection~\cite{ILDNote:Hinv} targets events which are consistent with  a di-jet plus missing four-momentum topology, where the di-jet invariant mass should be compatible with the $Z$ boson mass. The jet finding step also serves to reject PFOs from overlay of $\gamma\gamma \to $ low-$p_t$ hadron events. The final discriminating variable is the
invariant mass of the ``invisible'' four-momentum recoiling against the $Z$ boson. 

\begin{figure}[htbp]
\begin{subfigure}{0.49\hsize} 
 \includegraphics[width=\textwidth]{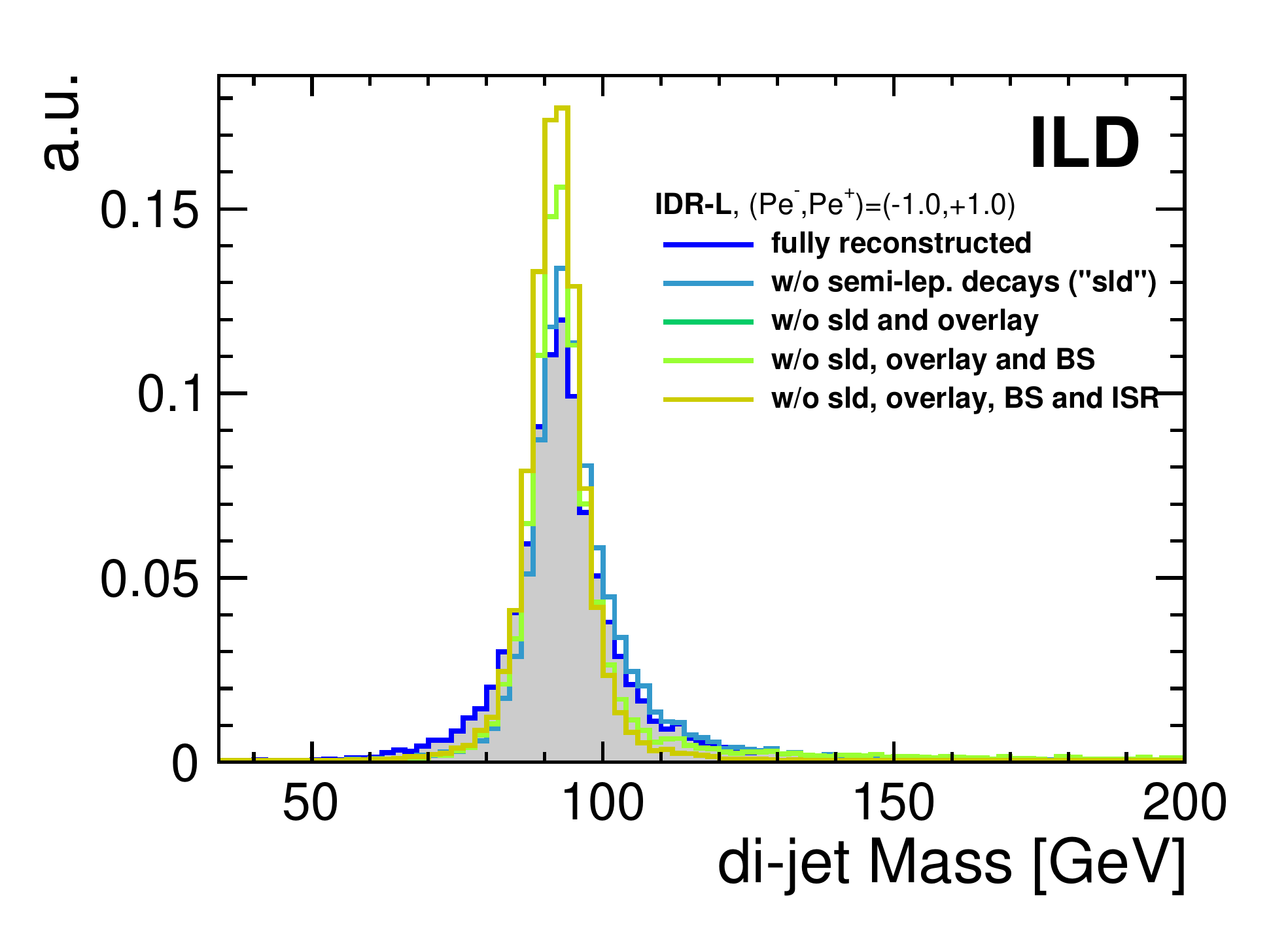}
 \caption{ \label{fig:Hinv:cheat:mjj}}
 \end{subfigure}
\begin{subfigure}{0.49\hsize} 
\includegraphics[width=\textwidth]{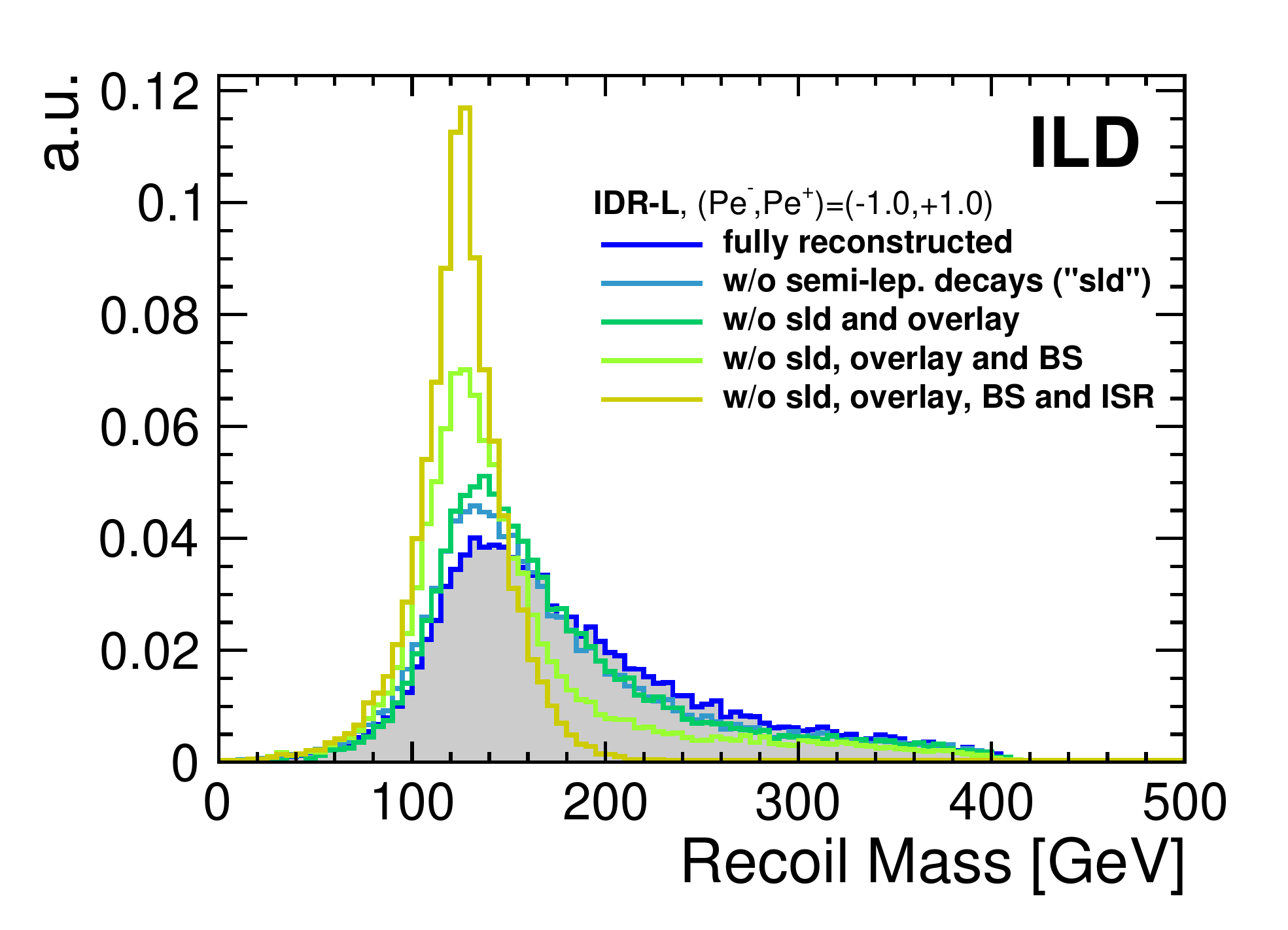}
 \caption{  \label{fig:Hinv:cheat:mrec}}
 \end{subfigure}
\caption{Impact of various effects on
(a) the invariant di-jet mass and
(b) the recoil mass,
shown for the example of the large detector model.
}
\label{fig:Hinv:cheat}
\end{figure}

\begin{figure}[htbp]
\begin{subfigure}{0.49\hsize} 
 \includegraphics[width=\textwidth]{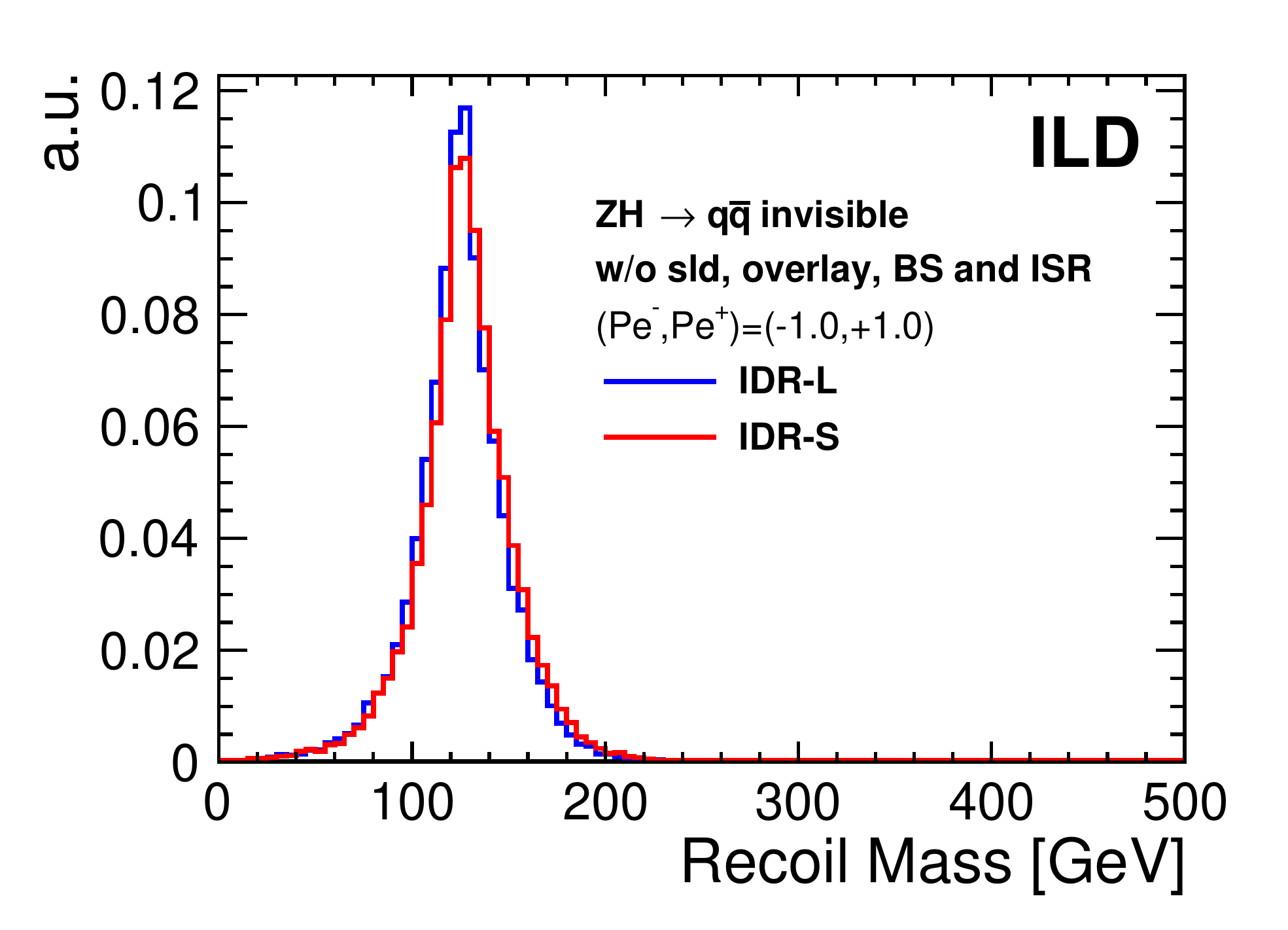}
 \caption{ \label{fig:Hinv:comp:cheatmrec}}
 \end{subfigure}
\begin{subfigure}{0.49\hsize} 
 \includegraphics[width=\textwidth]{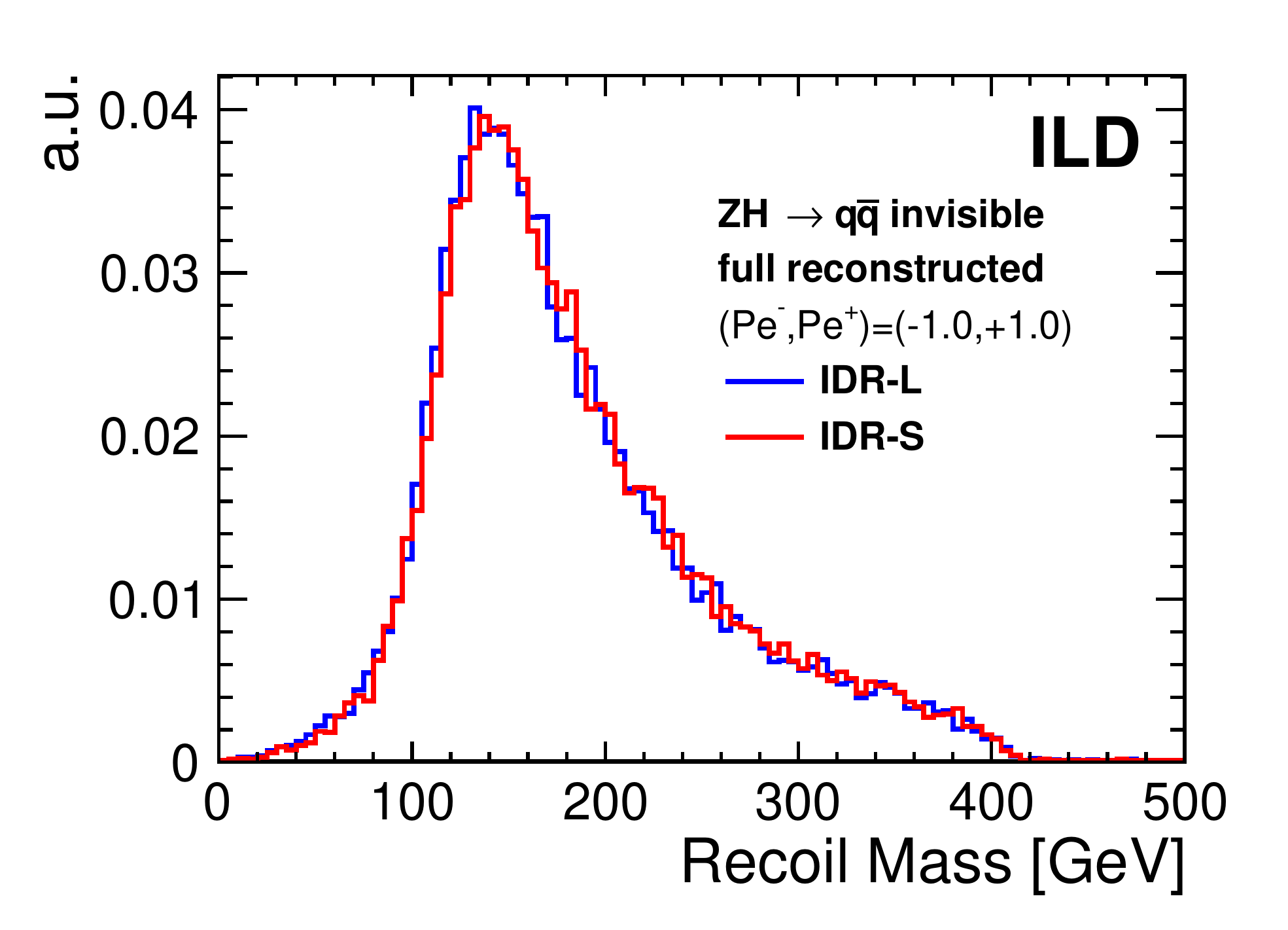}
 \caption{  \label{fig:Hinv:comp:mrec}}
 \end{subfigure}
\caption{Comparison of the recoil mass distributions for IDR-L and IDR-S
(a) when cheating semi-leptonic decays, overlay removal, beam spectrum and ISR
(b) at full reconstruction level.
}
\label{fig:Hinv:comp}
\end{figure}

Figure~\ref{fig:Hinv:cheat:mjj} shows the di-jet invariant mass for the selected signal events at various levels of realism from the full reconstruction to cheating all effects apart from the detector and particle flow performance. The first step of partial cheating removes jets with semileptonic heavy flavour decays. In a detector like ILD shower shapes in the highly-granular calorimeter and specific energy loss information from the TPC should allow an
excellent identification of leptons in jets, which, combined with secondary vertex information
should allow to significantly improve scale and resolution for heavy flavour jets with semileptonic decays. However, the corresponding reconstruction algorithms are still under development and thus could not be applied here. Similarly, work is ongoing to improve the
removal of overlay backgrounds, see e.g.\ Sec.~\ref{subsec:bench:higgsino} and Ref.~\cite{Boronat:2014hva}. Thus, we expect that with future reconstruction improvements, a performance similar to the case ``w/o sld and overlay'' could be reached. The beam spectrum (BS) by construction does not affect the invariant di-jet mass. ISR, on the contrary, can lead to photons in the detector. In this analysis, no attempt has been made to identify the corresponding particle flow objects. Therefore, also a large part of the effect of ISR on the di-jet mass should be recoverable with a more sophisticated analysis. 

The corresponding situation for the recoil mass is shown in Fig.~\ref{fig:Hinv:cheat:mrec}.
Here, ISR and BS have a large impact since they lead to a deviation of the actual initial state of the hard interaction from the naive assumption. Since this effect is dominated by photons from ISR and BS which escape undetected along the beam pipe, no attempt has been made to correct the kinematics of those events in which a photon is detected. See e.g.\ Sec.~\ref{subsec:bench:extraH} for an analysis where such a correction is applied. 

The recoil mass distributions obtained with the large and small detector model are compared in Fig.~\ref{fig:Hinv:comp}. Fig.~\ref{fig:Hinv:comp:mrec} shows the situation at the current full reconstruction level, while Fig.~\ref{fig:Hinv:comp:cheatmrec} cheats the effect of semi-leptonic decays (``w/o sld''), overlay removal (``overlay''), beam spectrum (``BS'') and ISR. In both cases, the recoil  mass is slightly shifted to higher values in case of the small detector, due to differences 
in the calibration of the particle flow for the two models. In addition, the cheated recoil mass distribution is a bit wider for the small detector, by about 15\% when considering the gaussian core of the peak, as expected from its slightly worse
JER, c.f.\ Fig.~\ref{fig:perf:pfa_jer} for the barrel and Fig.~\ref{fig:perf:pfa_jer_endcap} for the endcaps.

\begin{figure}[htbp]
\begin{center}
 \includegraphics[width=0.75\textwidth]{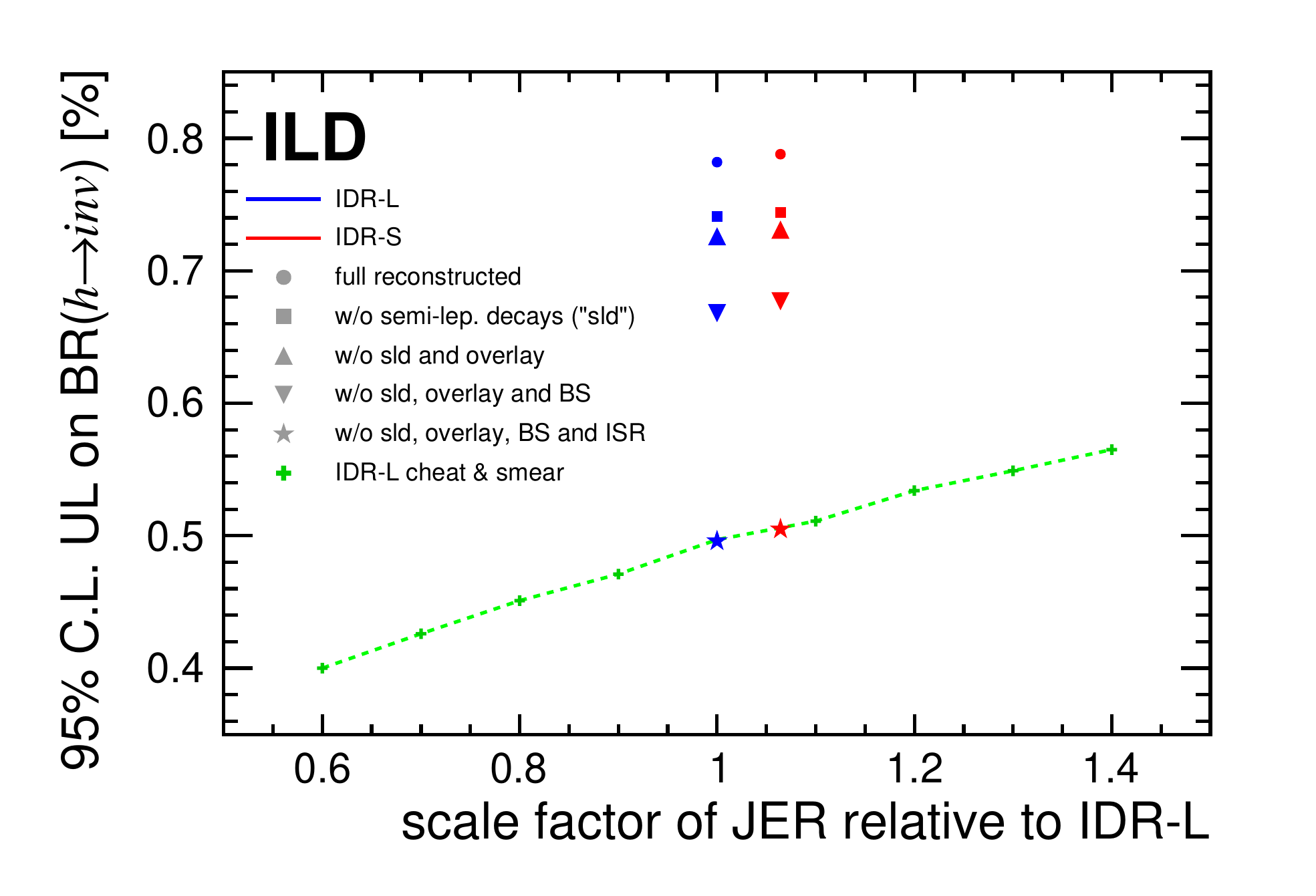}
\end{center}
\caption{Upper limit on $BR( \to $ invisible) at $95\%$ C.L. for ILC500 as defined in Sec.~\ref{sec:benchmarks:lep} as a function of the jet energy resolution. The blue and red symbols show the results obtained from simulation of the IDR-L and IDR-S detector models, respectively, in full reconstruction and at various levels of cheating. The green crosses are obtained by varying the JER up and down w.r.t.\ IDR-L.
}
\label{fig:Hinv:BRlimit}
\end{figure}

The results in terms of the physics observable, namely the $95\%$ C.L.\ upper limit on $\sigma(q\bar{q} H)\times BR(H \to \mbox{inv.})$, are summarized in Fig.~\ref{fig:Hinv:BRlimit} for both detector models at the various cheating levels. In the case of full reconstruction, the upper limit is at 0.78\%  for IDR-L  and at  0.79\%  for IDR-S, corresponding to a relative change of about 1\%. When isolating the effect of the particle flow performance by cheating all other aspects, the limit would be 0.50\% (0.51\%) for IDR-L (IDR-S), i.e.\ a relative change of about 2\%. Also displayed is an estimate of how
the cheated results would change when scaling the JER up and down. This is achieved by fitting the fully reconstructed recoil mass distribution, scaling its width and then generating toy distributions according to the scaled functions. The scale factor with respect to the IDR-L mass resolution is given on the horizontal axis. This clearly shows that
larger variations of the JER, by 20\% or so, have a clear impact on this physics analysis.
In the case of $\sqrt{s}=250$\,GeV, the impact of ISR and BS is much smaller, increasing the relative contribution from the JER.

\subsection{\texorpdfstring{$\tau$}{Tau} polarisation \texorpdfstring{in $e^+e^- \to \tau^+\tau^-$}{in e+e- -> tau tau}}

As shown in Sec.~\ref{sec:perf:hlr:tau}, the smaller ECAL inner radius of the small detector model slightly reduces the ability to identify the correct number of photons in
highly-boosted $\tau$ decays. Using the product of efficiency times purity as a figure of merit, this leads to a $5\%$ worse identification of $\tau \to \pi \nu$ and $\tau \to \rho \nu$ decays, while the identification of $\tau \to a_1 \nu$ decays deteriorates by about $15\%$ relative, c.f.\ Fig.~\ref{fig:HLR-tauID}. 

In order to evaluate the impact of this difference in a physics example, the measurement of the $\tau$ polarisation in $e^+e^- \to \tau^+\tau^-$ has been studied, looking specifically at events with no significant ISR, so those {\em not} returning to the $Z$ pole. For the $\tau \to \pi \nu$ channels, the magnitude of the $\pi^{\pm}$ momentum can directly be used to extract the polarisation. In case of the $\tau \to \rho \nu$ decay, a polarimeter vector is constructed from the momenta of the $\pi^{\pm}$ and the $\pi^0$. A detailed description of the analysis and the polarisation extraction can be found in~\cite{ILDNote:tautau}.

\begin{figure}[htbp]
\begin{center}
 \includegraphics[width=0.55\textwidth]{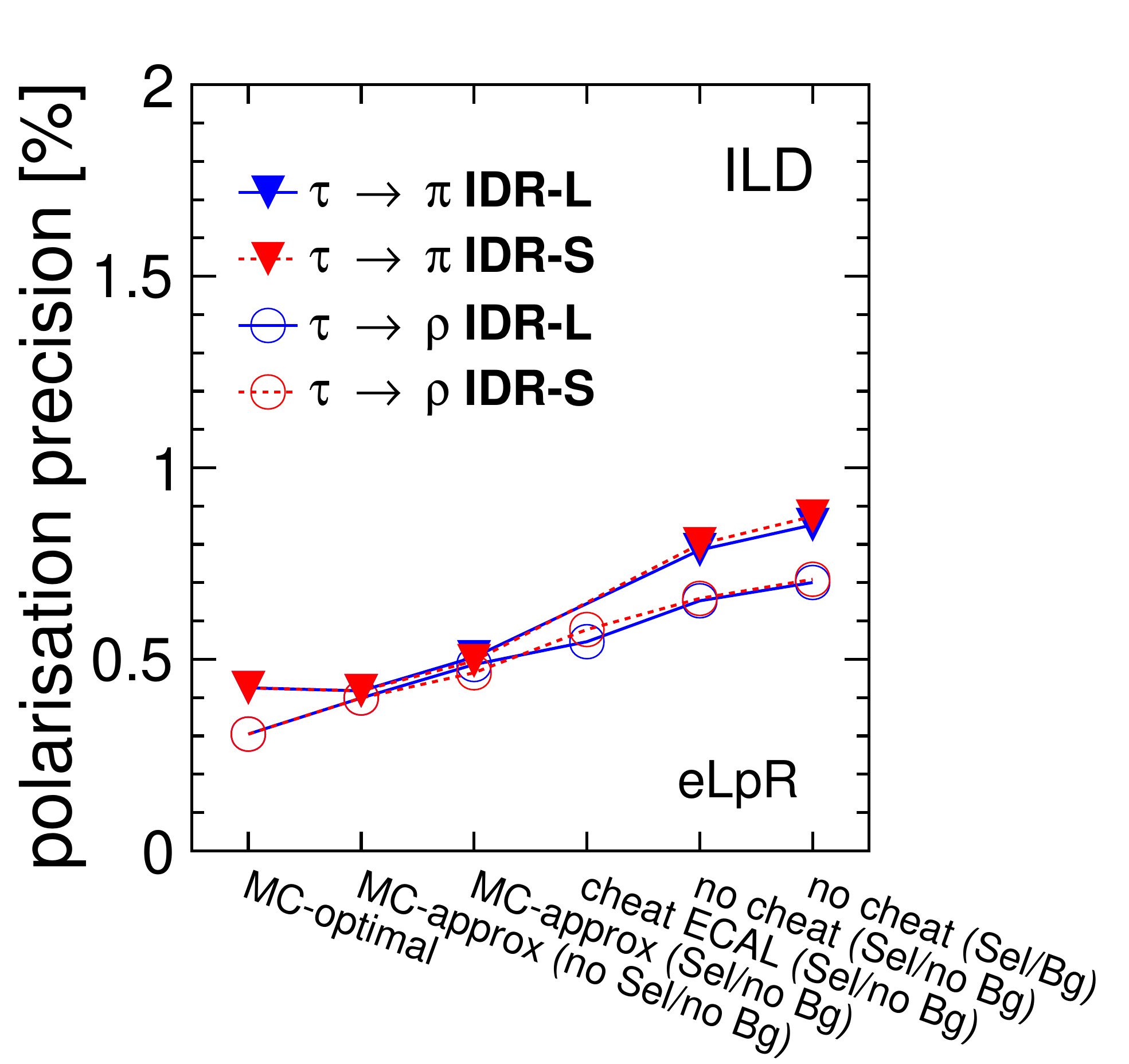}
\end{center}
\caption{Precision on the $\tau$ polarisation achieved with IDR-L and IDR-S at various levels of cheating (see text) based on the $\pi$ and $\rho$ channels in the $P(e^-,e^+)=(-80\%,+30\%)$ data set of ILC500 as defined in Sec.~\ref{sec:benchmarks:lep}.}
\label{fig:tautau:taupol}
\end{figure}

Figure~\ref{fig:tautau:taupol} illustrates the precision on the $\tau$ polarisation achieved with IDR-L and IDR-S  based on the $\pi$ and $\rho$ channels in the $P(e^-,e^+)=(-80\%,+30\%)$ data set, which dominates the combined precision. Various
levels of cheating are shown, starting from the optimal result when using all MC information, including the neutrino momentum. The next entry shows by how much the performance in the $\rho$ channel degrades by the approximate definition of the polarimeters used here\footnote{In principle, improved methods can be used, which however need further investigation~\cite{ILDNote:tautau}.}, followed by the application of the selection efficiency. The last three steps use the fully simulated and reconstructed events, apart from the entry ``cheat ECAL'', which uses MC information for the $\pi^0$ in the $\rho$ channel. In the last step, the full SM background is added. For the $\pi$-channel, the most significant effect occurs when reducing the number of signal events according to the selection efficiency of about $55\%$ observed in the full analysis. In case of the $\rho$-channel, all steps contribute at a similar level to the final result. Overall, the differences between IDR-L and IDR-S are very small. 


\subsection{Hadronic \texorpdfstring{$WW$ and $ZZ$}{WW and ZZ} separation in Vector Boson Scattering} 

Vector boson scattering is an important process for testing the unitarisation of $WW$ scattering by the Higgs boson, as well as for measuring quartic gauge couplings, and thereby probing for anomalous contributions. Among all relevant final states, the reaction $e^+e^- \to \nu\nu VV \to \nu\nu qqqq$, where $VV$ can be $WW$ or $ZZ$, poses a particular challenge to the detectors and reconstruction algorithms, since it requires the separation of the hadronic $W$ and $Z$ decays without the ability to exploit kinematic constraints, e.g.\ on the
total event energy, due to the two invisible neutrinos.

\begin{figure}[htbp]
\begin{subfigure}{0.49\textwidth} \includegraphics[width=\textwidth]{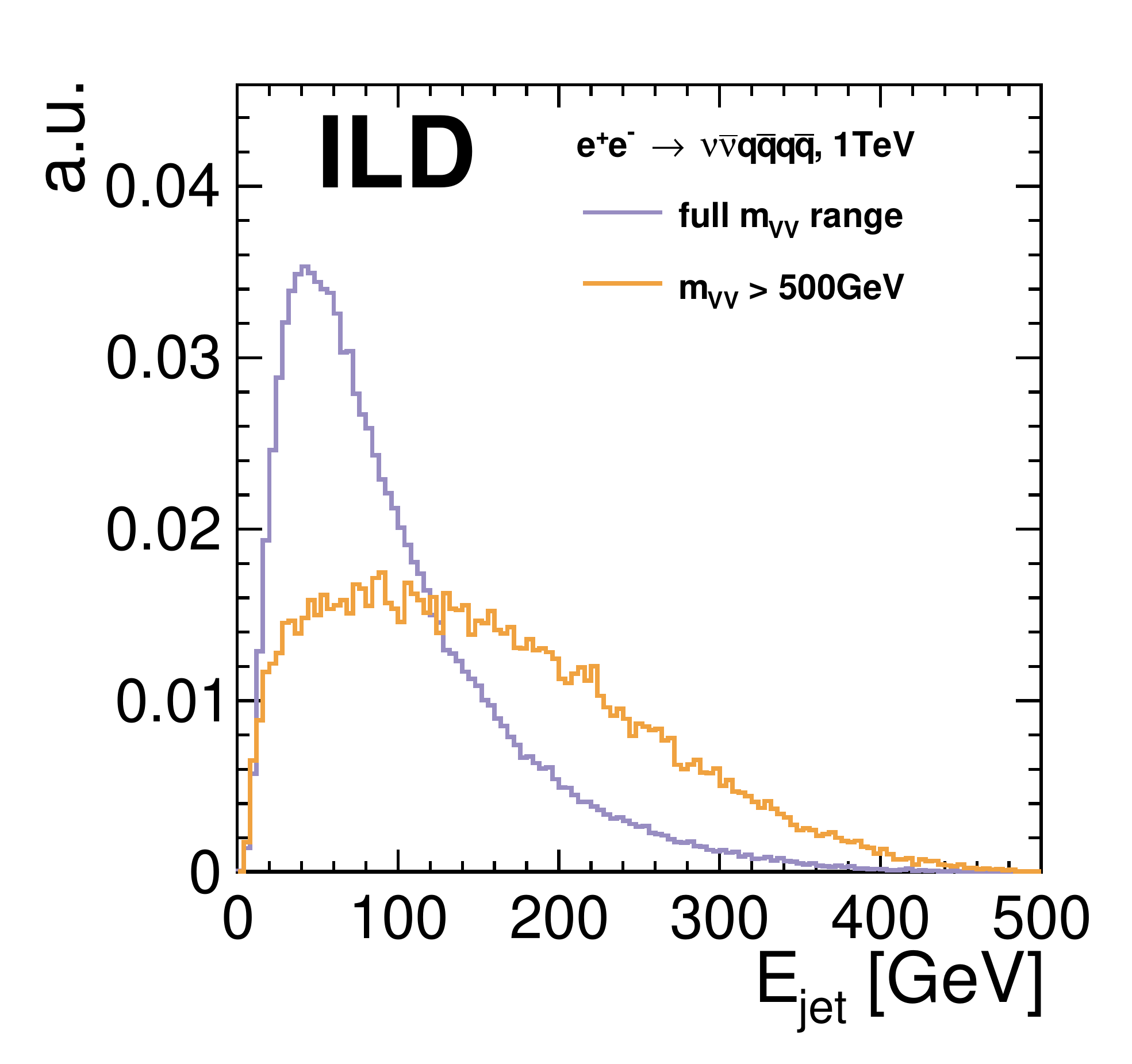}
 \caption{ \label{fig:qgc:truejet:E}}
 \end{subfigure}
\begin{subfigure}{0.49\textwidth} \includegraphics[width=\textwidth]{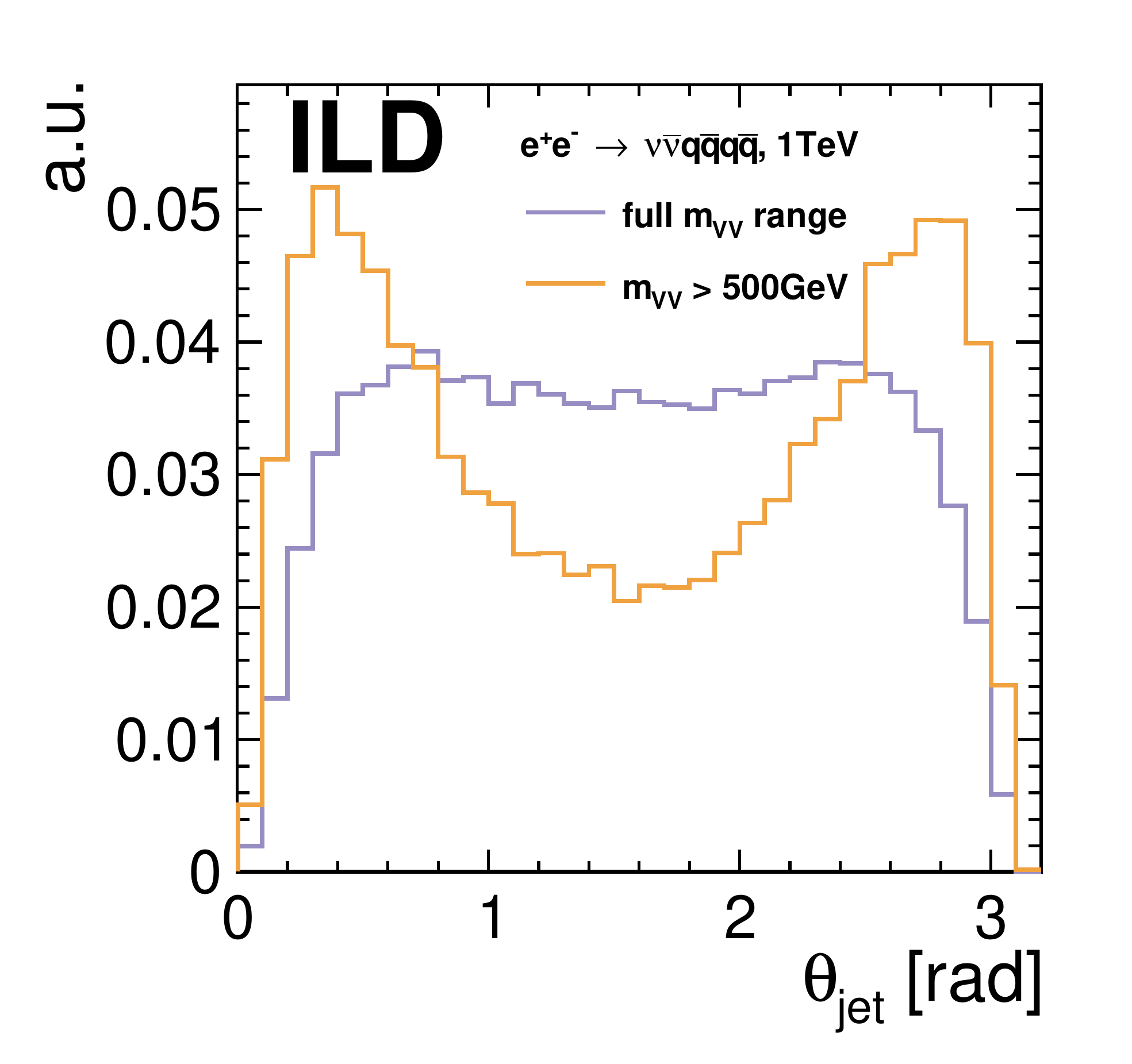}
 \caption{  \label{fig:qgc:truejet:theta}}
 \end{subfigure}
\caption{(a) Energy and (b) polar angle distributions of the jets from $e^+e^- \to \nu\nu qqqq$ at $\sqrt{s}=1$\,TeV on truth level.
}
\label{fig:qgc:truejet}
\end{figure}

This benchmark, at a center-of-mass energy of 1\,TeV, has already been studied in full detector simulation for the ILD LoI~\cite{ild:bib:ILDloi}. All relevant aspects, in particular the full overlay from $\gamma\gamma \to$ low-$p_t$ hadrons and from $e^+e^-$ pair background, are now included for the first time, and all quark flavours are considered in the final state~\cite{ILDNote:QGCs}. Figure~\ref{fig:qgc:truejet} illustrates the ranges of jet energies and polar angles probed by this benchmark, both inclusively and for the case of high di-boson invariant
masses, which is the phase space most sensitive to new physics effects. 

\begin{figure}[htbp]
\begin{subfigure}{0.48\hsize} \includegraphics[width=\textwidth]{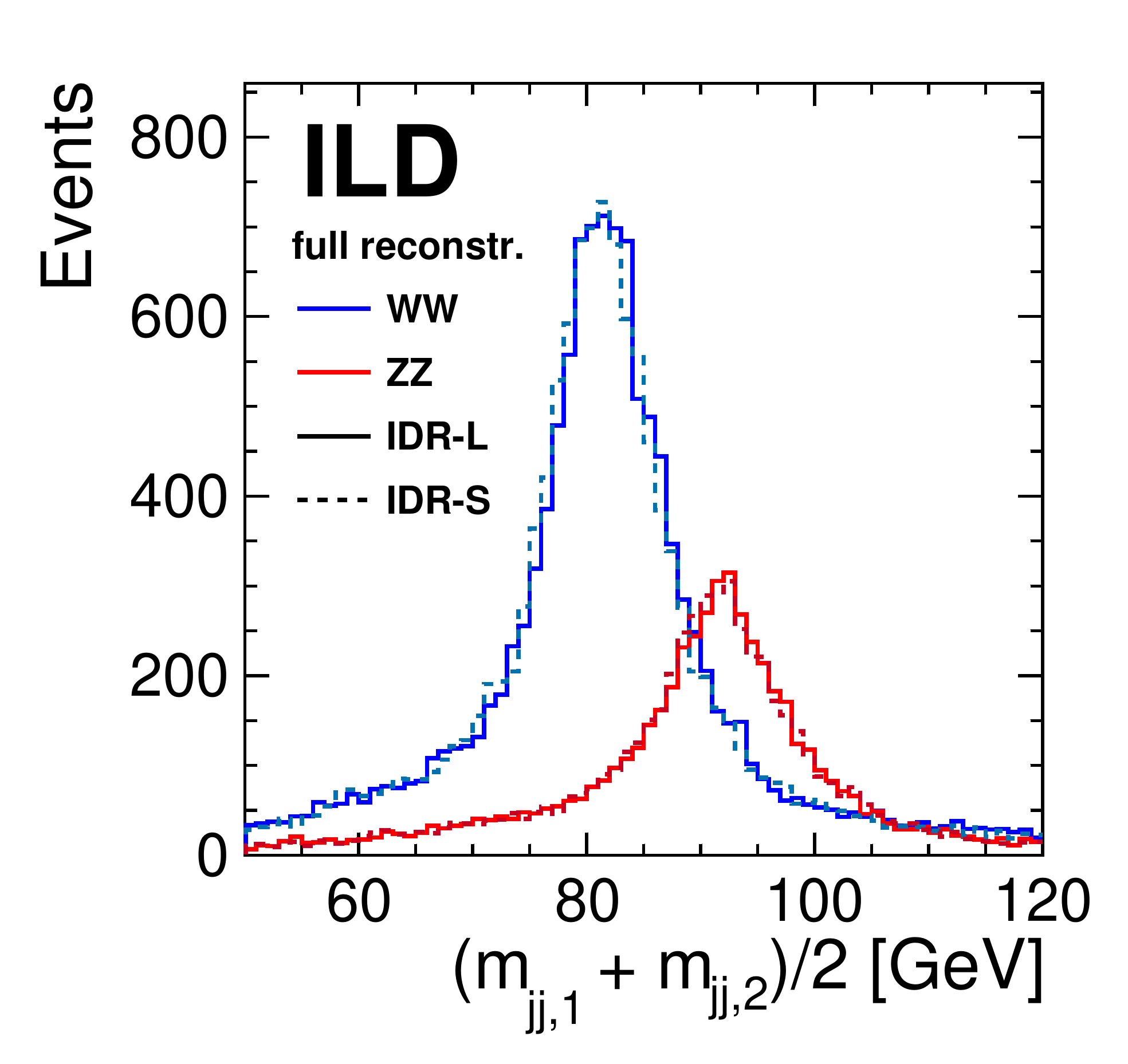}
 \caption{ \label{fig:qgc:rec:1d}}
 \end{subfigure}
\begin{subfigure}{0.48\hsize} \includegraphics[width=\textwidth]{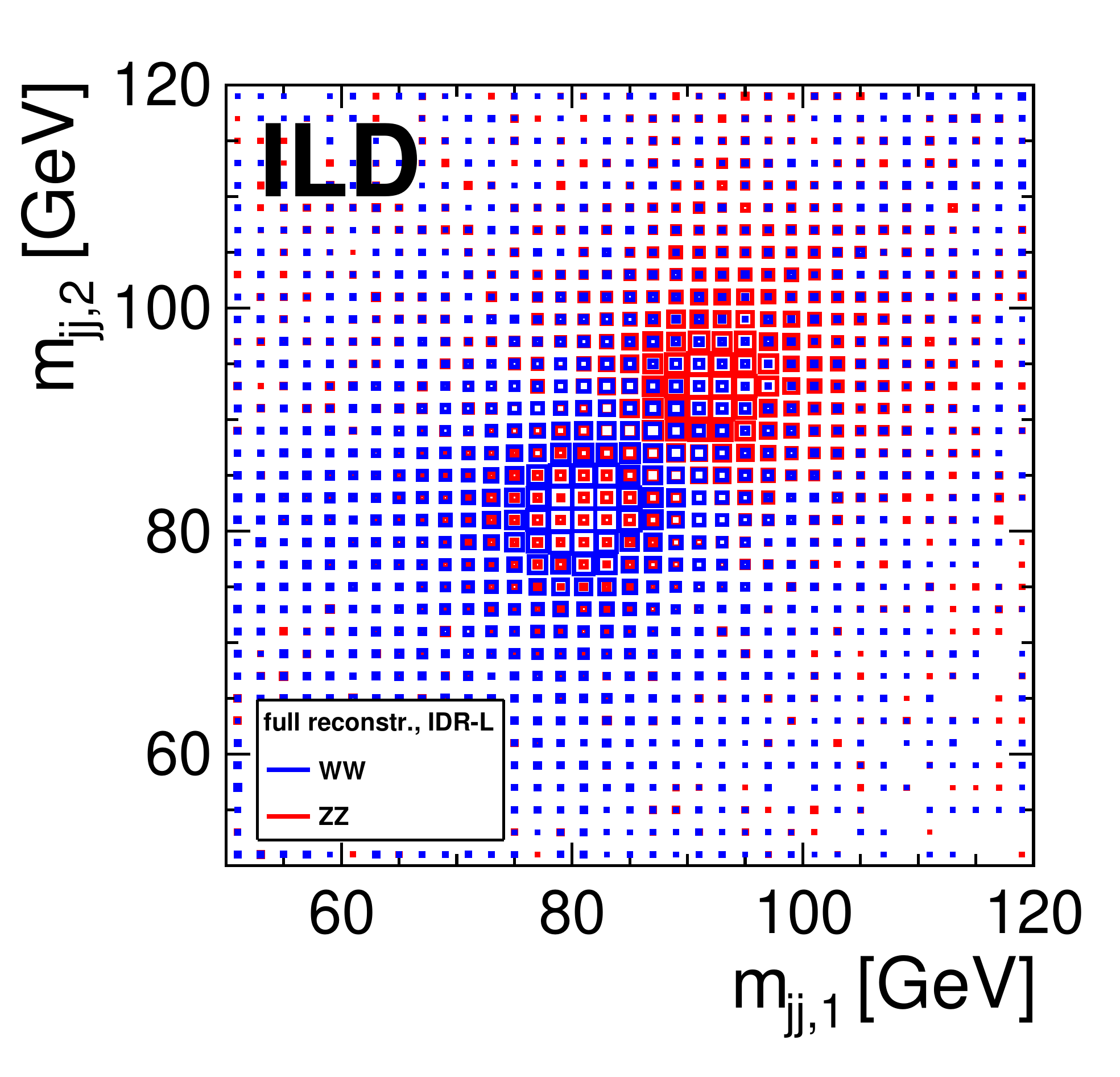}
 \caption{  \label{fig:qgc:rec:2d}}
 \end{subfigure}
\caption{Dijet masses in $e^+e^- \to \nu\nu WW$ (blue) and $e^+e^- \to \nu\nu ZZ$ (red) events as obtained from the current full reconstruction for ILC1000 as defined in Sec.~\ref{sec:benchmarks:lep}.
(a) Average of the two di-jet masses per event.
(b) 2D illustration of the two di-jet masses per event. 
}
\label{fig:qgc:rec}
\end{figure}

Jet clustering is performed in a two-step procedure, first by requesting $4$ jets with a cone radius of $1.3$ from an exclusive $k_t$ algorithm in order to remove PFOs from the overlay backgrounds. In the second step, all PFOs in the four jets are reclustered into $4$ jets with the (inclusive) $ee$ $k_t$ algorithm. The jets are paired into two boson candidates by minimizing the mass difference between the two bosons. The resulting $W$ and $Z$ mass distributions
are shown in Fig.~\ref{fig:qgc:rec}. Fig.~\ref{fig:qgc:rec:1d} shows the average di-jet mass per event, comparing ILD-L and ILD-S, while Fig.~\ref{fig:qgc:rec:2d} shows the 2-dimensional distribution in the mass plane of the two invariant di-jet masses for ILD-L. At this level, no significant difference between the detector models can be observed.

\begin{figure}[htbp]
\begin{subfigure}{0.49\hsize} \includegraphics[width=\textwidth]{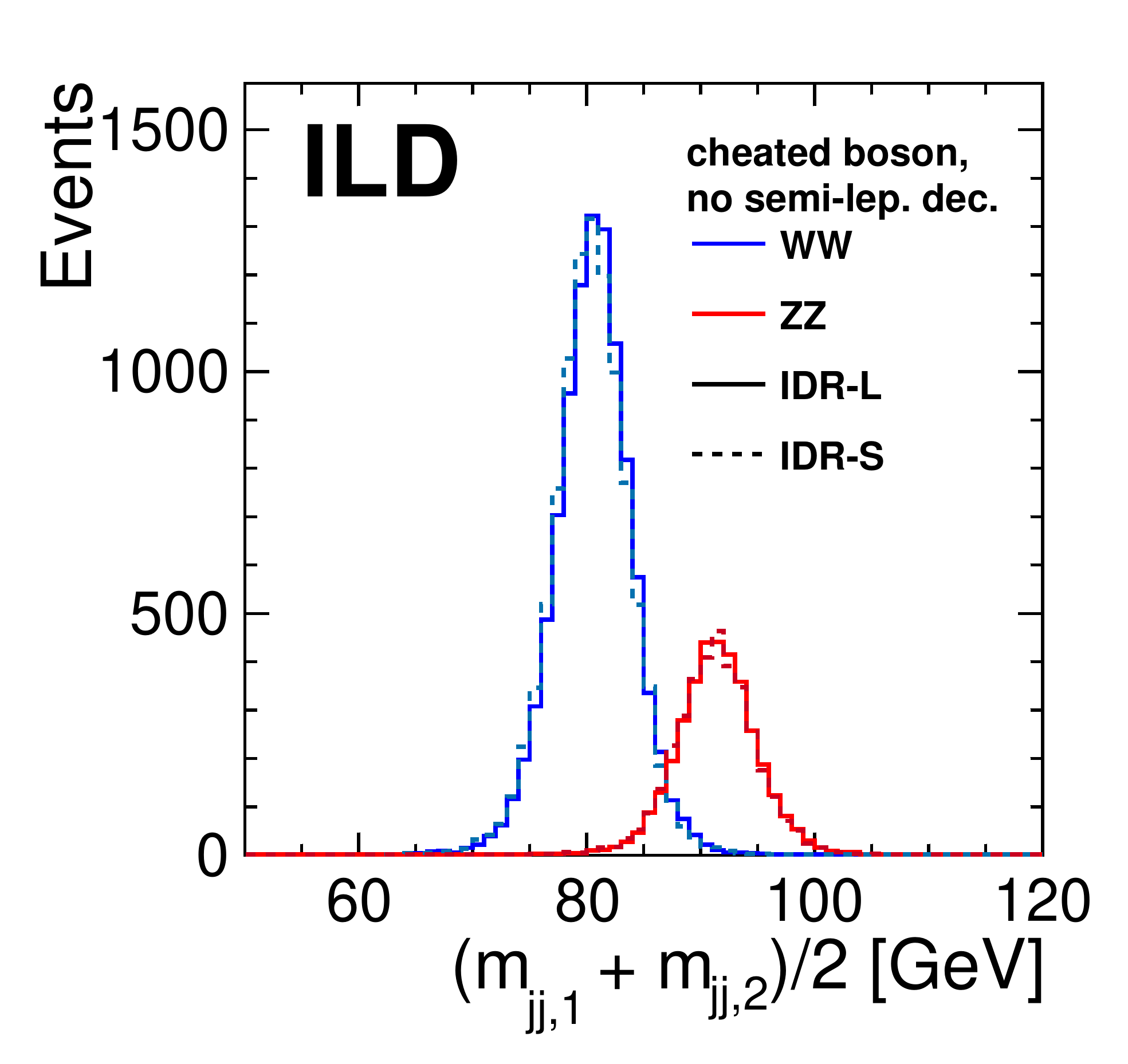}
 \caption{ \label{fig:qgc:cheat:1d}}
 \end{subfigure}
\begin{subfigure}{0.47\hsize} \includegraphics[width=\textwidth]{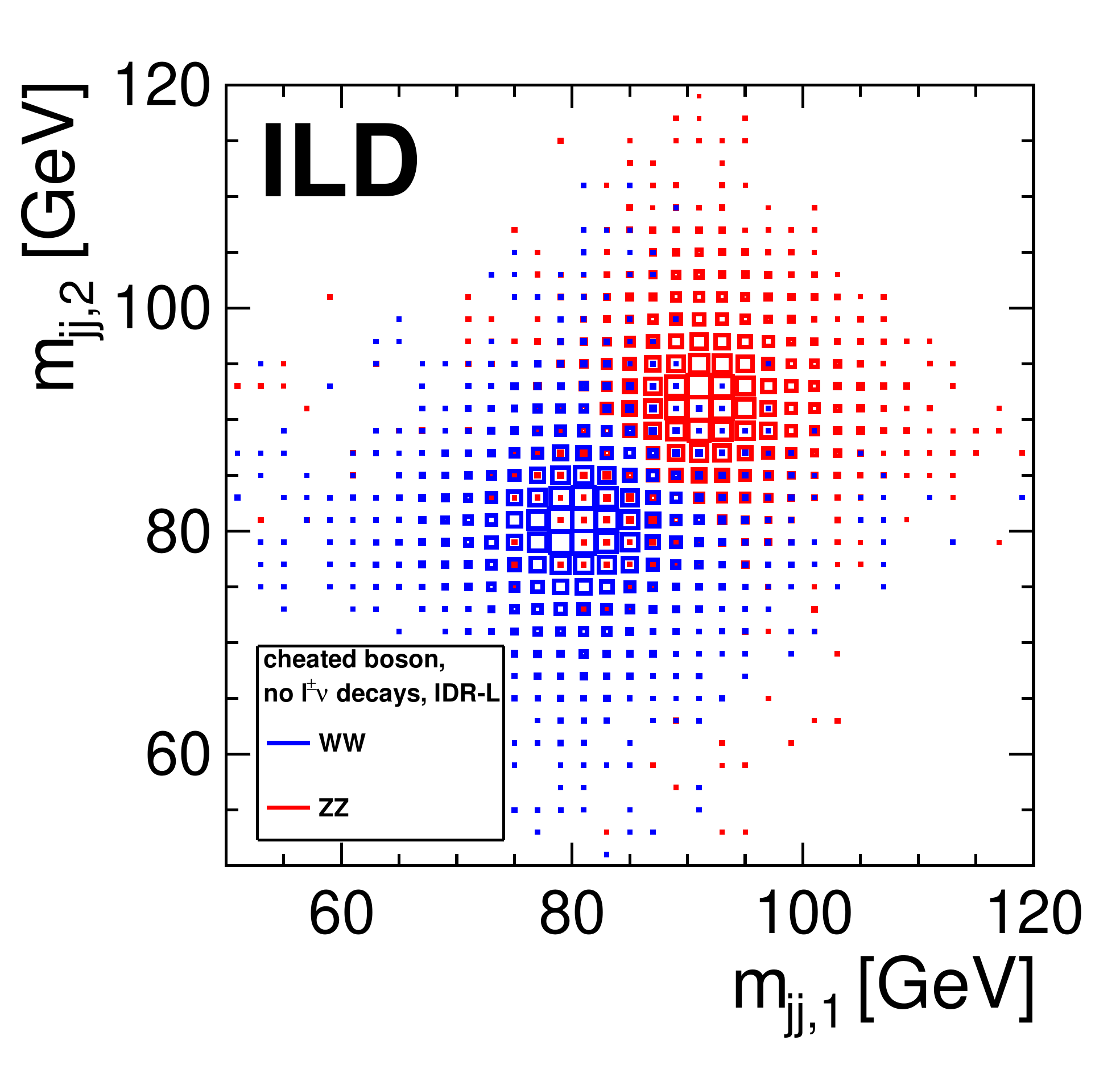}
 \caption{  \label{fig:qgc:cheat:2d}}
 \end{subfigure}
\caption{Dijet masses in $e^+e^- \to \nu\nu WW$ (blue) and $e^+e^- \to \nu\nu ZZ$ (red) events as obtained for ILC1000 as defined in Sec.~\ref{sec:benchmarks:lep} when cheating the jet clustering and excluding events where one (or more) jets contain semi-leptonic charm or beauty decays.
(a) Average of the two di-jet masses per event. 
(b) 2D illustration of the two di-jet masses per event.
}
\label{fig:qgc:cheat}
\end{figure}

\begin{figure}[htbp]
\begin{center}
\begin{subfigure}{0.5\hsize} \includegraphics[width=\textwidth]{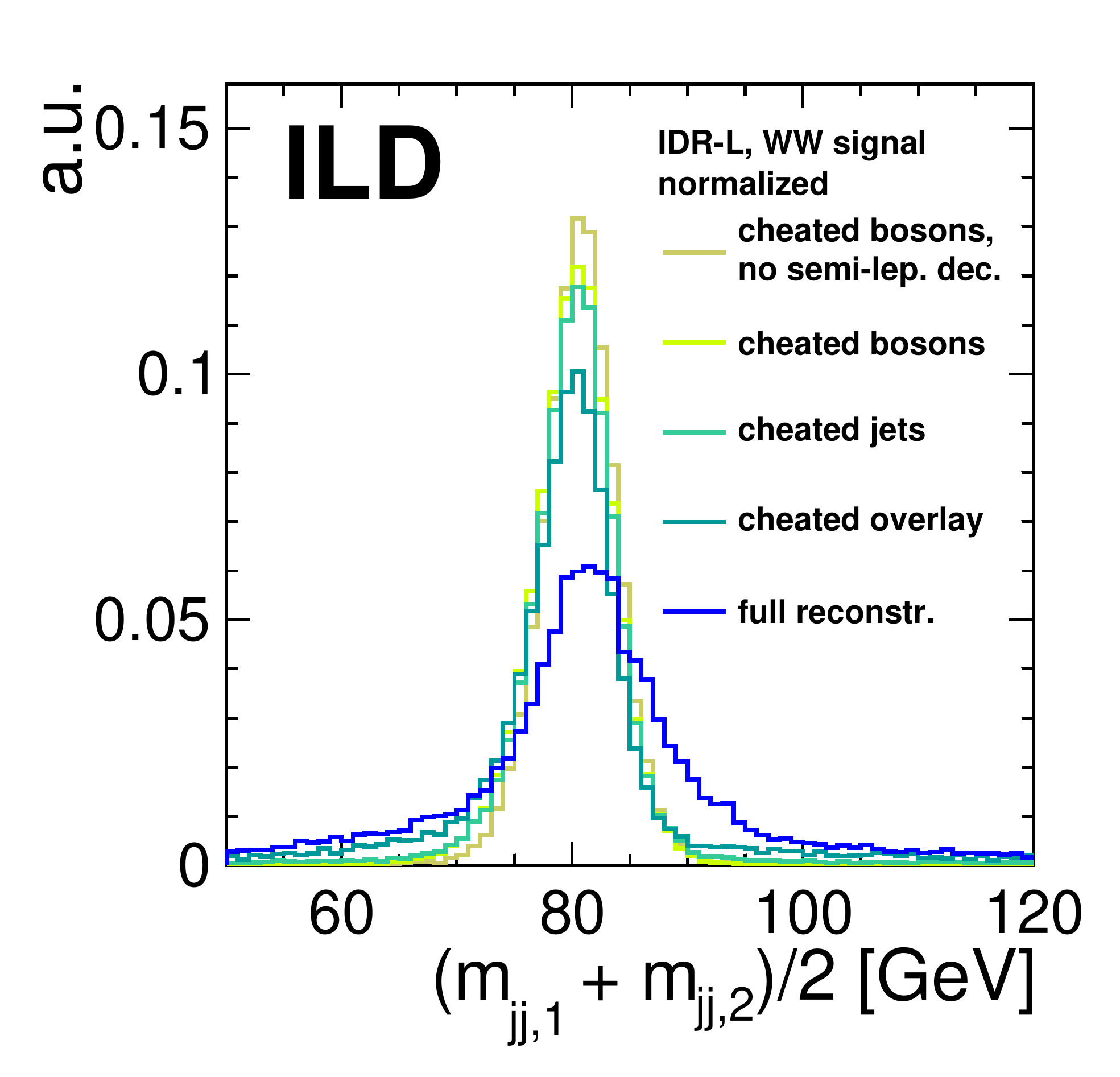}
 \caption{ \label{fig:qgc:cheat:WW}}
 \end{subfigure}
\begin{subfigure}{0.49\hsize} \includegraphics[width=\textwidth]{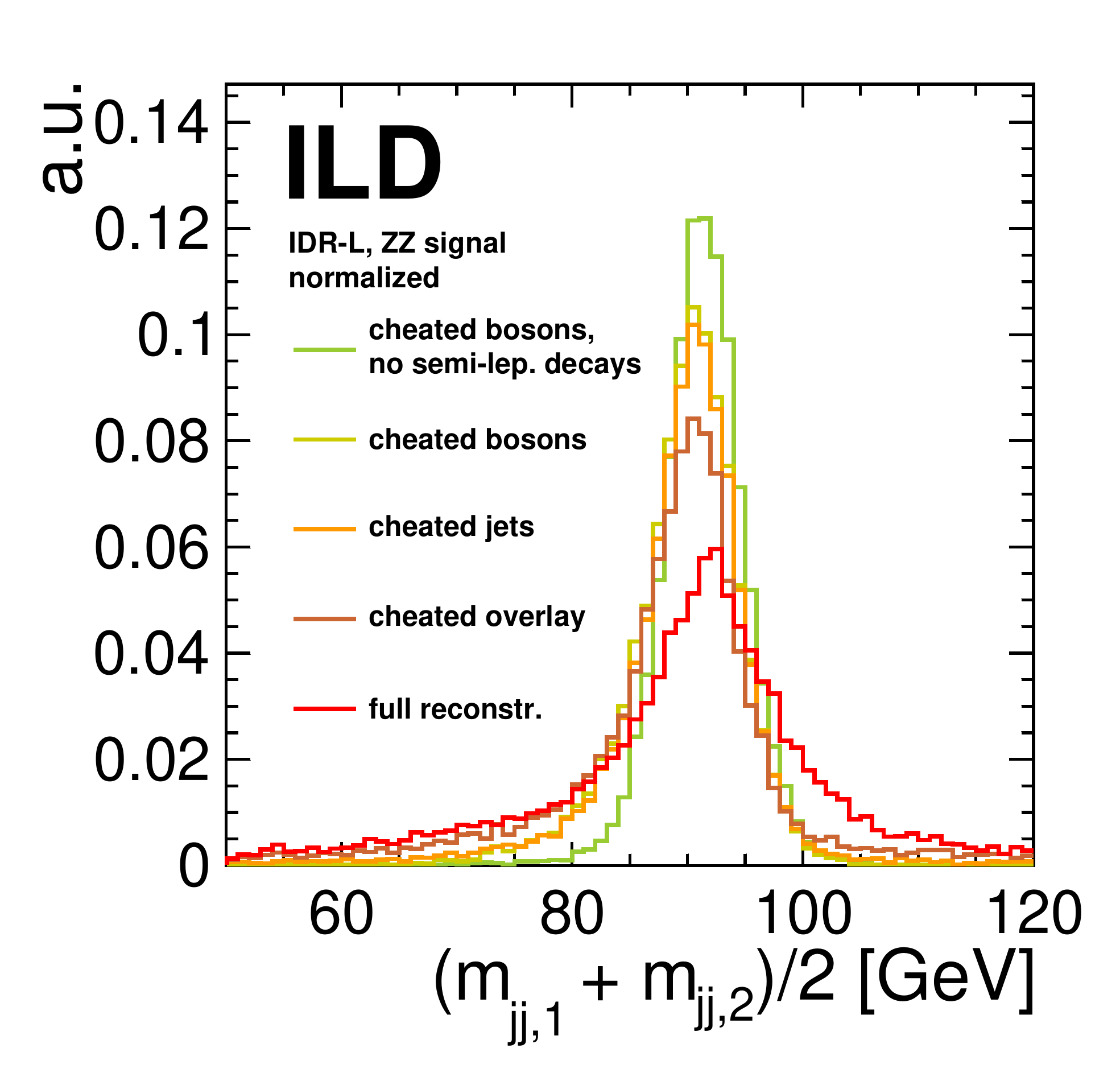}
 \caption{  \label{fig:qgc:cheat:ZZ}}
 \end{subfigure}
\begin{subfigure}{0.495\hsize} \includegraphics[width=\textwidth]{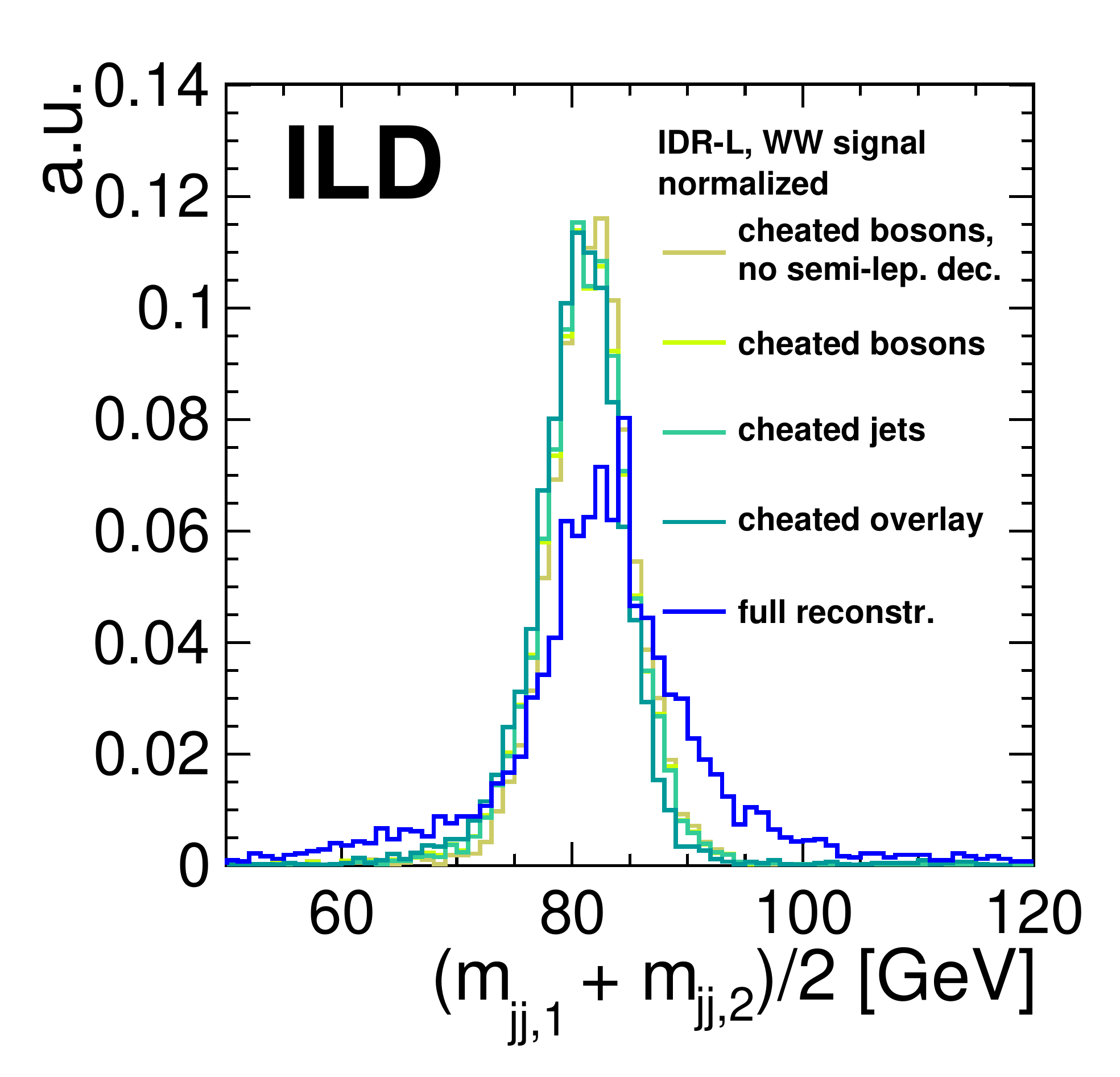}
 \caption{ \label{fig:qgc:cheat:WWQ2}}
 \end{subfigure}
\begin{subfigure}{0.495\hsize} \includegraphics[width=\textwidth]{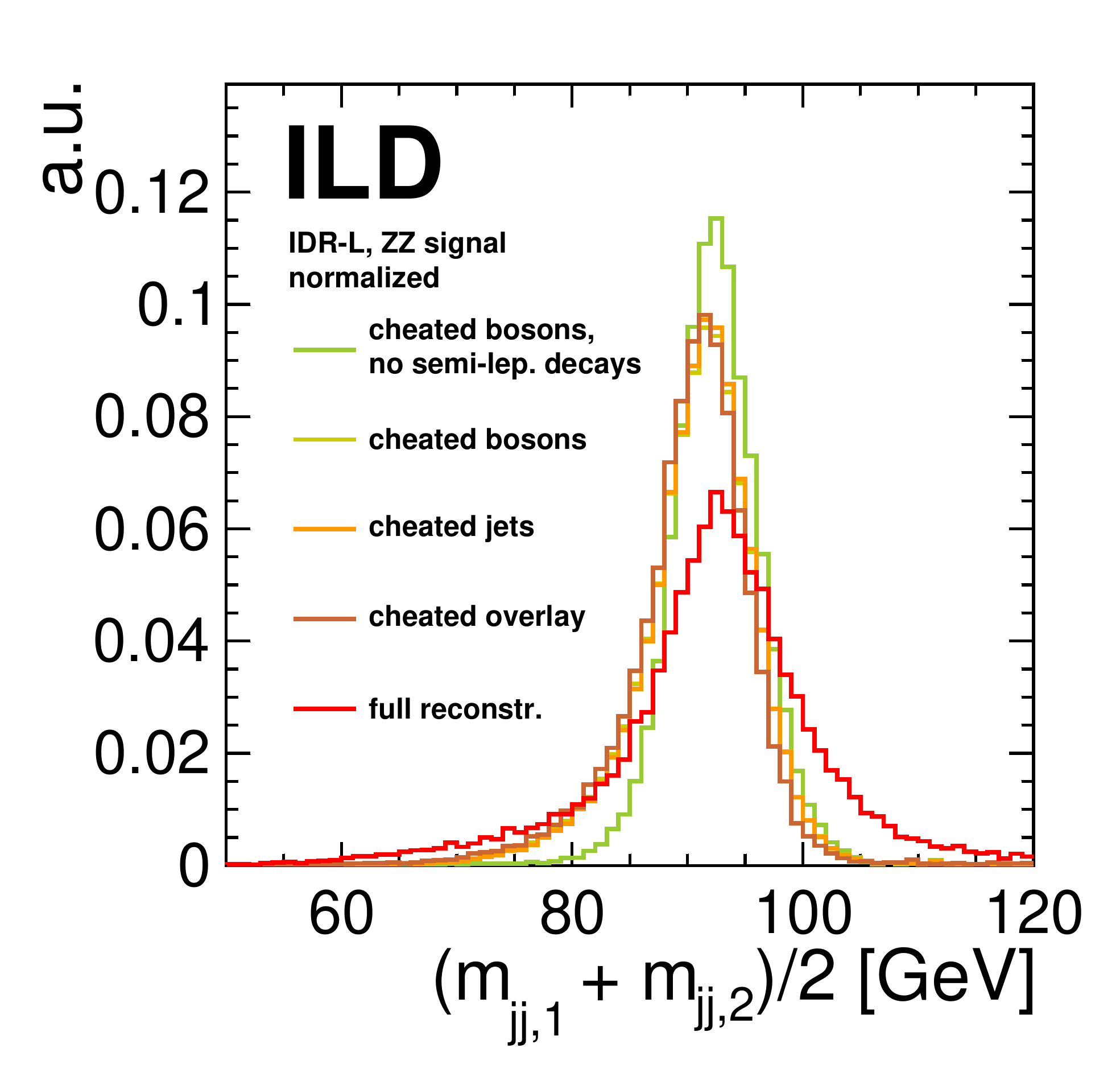}
 \caption{  \label{fig:qgc:cheat:ZZQ2}}
 \end{subfigure}
\end{center}
\caption{Average di-jet masses as obtained in full reconstruction at various levels of cheating.\\
(a) inclusive $e^+e^- \to \nu\nu WW$ events. 
(b) inclusive $e^+e^- \to \nu\nu ZZ$ events.
(c)  $e^+e^- \to \nu\nu WW$ events with $M(WW)>500$\,GeV. 
(d)  $e^+e^- \to \nu\nu ZZ$ events with $M(ZZ)>500$\,GeV.\\
}
\label{fig:qgc:cheatWWZZ}
\end{figure}

Figure~\ref{fig:qgc:cheat} shows the analogous distributions obtained when cheating the jet clustering (incl.\ the overlay removal), the jet pairing and when excluding events with semi-leptonic decays of heavy quarks, so that only the effects of the natural widths of the bosons, of fragmentation and hadronisation as well as the JER itself remain. Also here, no striking difference between the models can be seen, which leads to the conclusion that on this event sample, which is dominated by events with rather low invariant masses of the di-boson system, the effect of the slightly worse JER of ILD-S is hidden beneath the width and fragmentation/hadronisation effects, since these are only remaining influences -- besides the JER -- in Fig.~\ref{fig:qgc:cheat}.

Nevertheless, the differences between Fig.~\ref{fig:qgc:rec} and Fig.~\ref{fig:qgc:cheat} are striking. Therefore, we investigated the impact of the size of various contributions individually as shown in Fig.~\ref{fig:qgc:cheatWWZZ}, for ILD-L only. For the inclusive $WW$ and $ZZ$ samples, shown in Figs.~\ref{fig:qgc:cheat:WW} and~\ref{fig:qgc:cheat:ZZ}, the dominant effect is the residual of the non-perfect overlay removal, followed by the jet clustering itself and the semi-leptonic decays. Non-perfect jet pairing only plays a minor role. This can be compared to the situation found when only considering events with high $WW$/$ZZ$ invariant masses, shown in Figs.~\ref{fig:qgc:cheat:WWQ2} and~\ref{fig:qgc:cheat:ZZQ2}. In this case, the impact of non-perfect jet clustering is reduced to a negligible level. Instead, the effect of the residual overlay from low-$p_t$ $\gamma \gamma \to $hadron events dominates. For the $ZZ$ events, also the missing neutrinos from semi-leptonic heavy quark decays play a visible role. 
These results demonstrate the need for development of more sophisticated high-level reconstruction algorithms, in particular for the overlay removal, the jet clustering and the identification and correction of semi-leptonic heavy flavour decays. For all these cases promising tools are under development, see e.g.\ Sec.~\ref{subsec:bench:higgsino} and Ref.~\cite{Boronat:2014hva}.

\subsection{Photon Energy Scale Calibration from \texorpdfstring{$e^+e^- \to \gamma Z \to \gamma \mu^+\mu^-$}{e+e- -> aZ -> mumu}}
\label{subsec:bench:gammaZ}

Di-fermion production, with or without radiative return to the $Z$ pole, is an integral part of the ILC physics case. In addition, the radiative return events offer an important
opportunity to cross calibrate the energy scales of various subdetectors. As a detector benchmark, we chose here the example of calibrating the photon energy scale against the momentum scale of the tracker. Thereby, the momenta and angles of the muons as well as the 
polar and azimuthal angle of the photon serve as input from which the energy of the photon and the amount of energy lost in beamstrahlung and collinear ISR are determined by
requiring conservation of energy and $p_y$ between initial and final state. It should be stressed that it is not necessary to apply a $Z$ mass constraint, which would introduce an additional uncertainty due to the large natural width of the $Z$ resonance. A full description of this and alternative methods can be found in~\cite{ILDNote:gammaZ}.

\begin{figure}[htbp]
\begin{subfigure}{0.49\hsize} 
 \includegraphics[width=\textwidth]{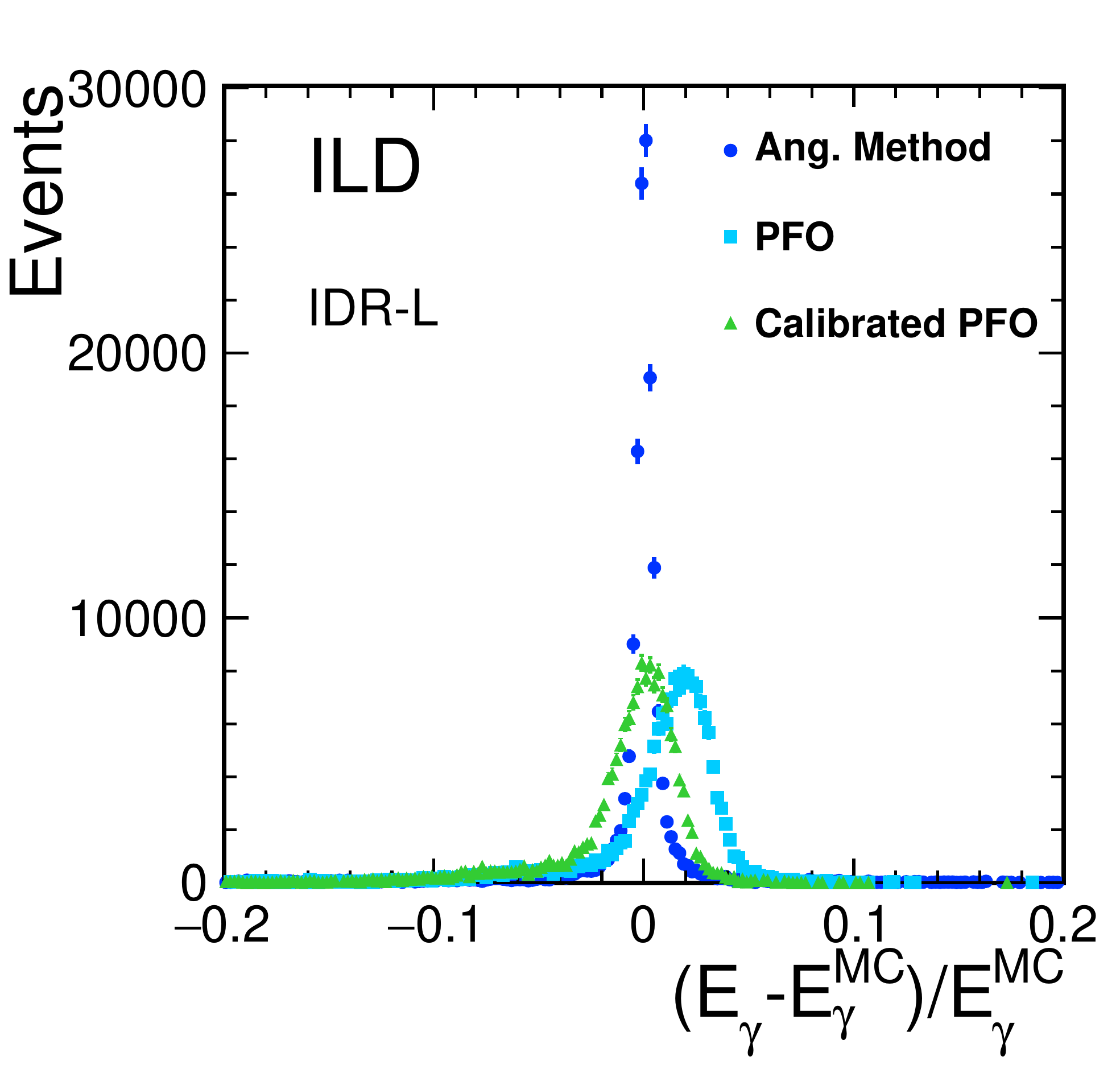}
 \caption{ \label{fig:gammaZ:meanE:allE}}
 \end{subfigure}
\begin{subfigure}{0.49\hsize} 
 \includegraphics[width=\textwidth]{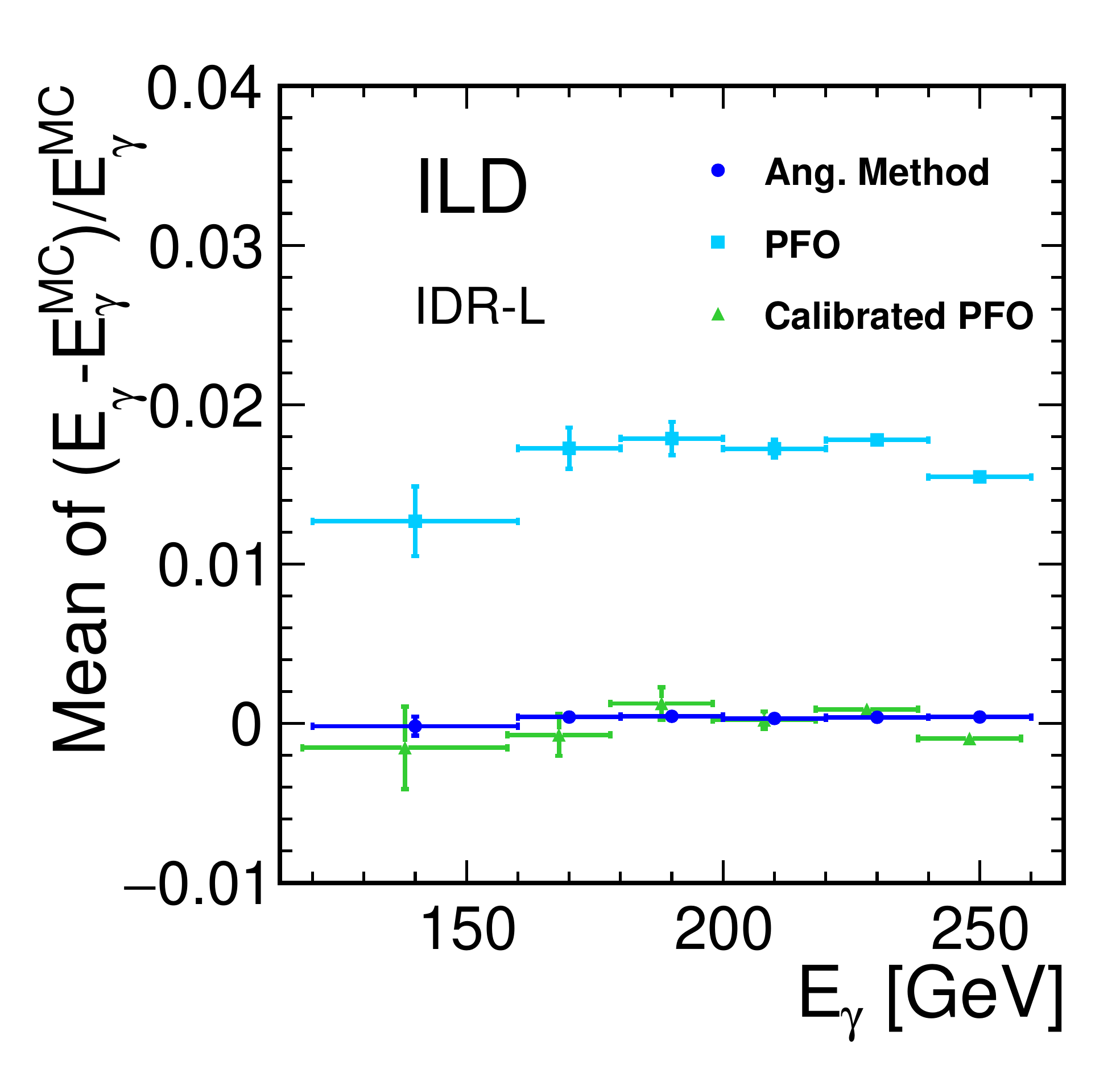}
 \caption{  \label{fig:gammaZ:meanE:vsE}}
 \end{subfigure}
\caption{
Deviation of the photon energy from its true value when using non-perfectly calibrated PFO-level energies (cyan), when calculating the photon energy from the $\mu$ momenta and kinematic constraints (blue, ``angular method'') and after calibrating the mean PFO-level w.r.t.\ the mean obtained from the $\mu$ momenta (green).  
(a) Distribution of deviation for all photons in the sample.
(b) Mean deviation in bins of the photon energy.
}
\label{fig:gammaZ:meanE}
\end{figure}

Figure~\ref{fig:gammaZ:meanE} illustrates the power of this method by application to a non-perfectly calibrated photon reconstruction in the large detector model, both inclusively for all photons (Fig.~\ref{fig:gammaZ:meanE:allE}) and in bins of the photon energy (Fig.~\ref{fig:gammaZ:meanE:vsE}). Since the $e^+e^- \to \mu^+\mu^-\gamma$ sample
is dominated by radiative returns to the $Z$ pole, the majority of photons has high energies close to $241$\,GeV.

\begin{figure}[htbp]
\begin{subfigure}{0.49\hsize} 
 \includegraphics[width=\textwidth]{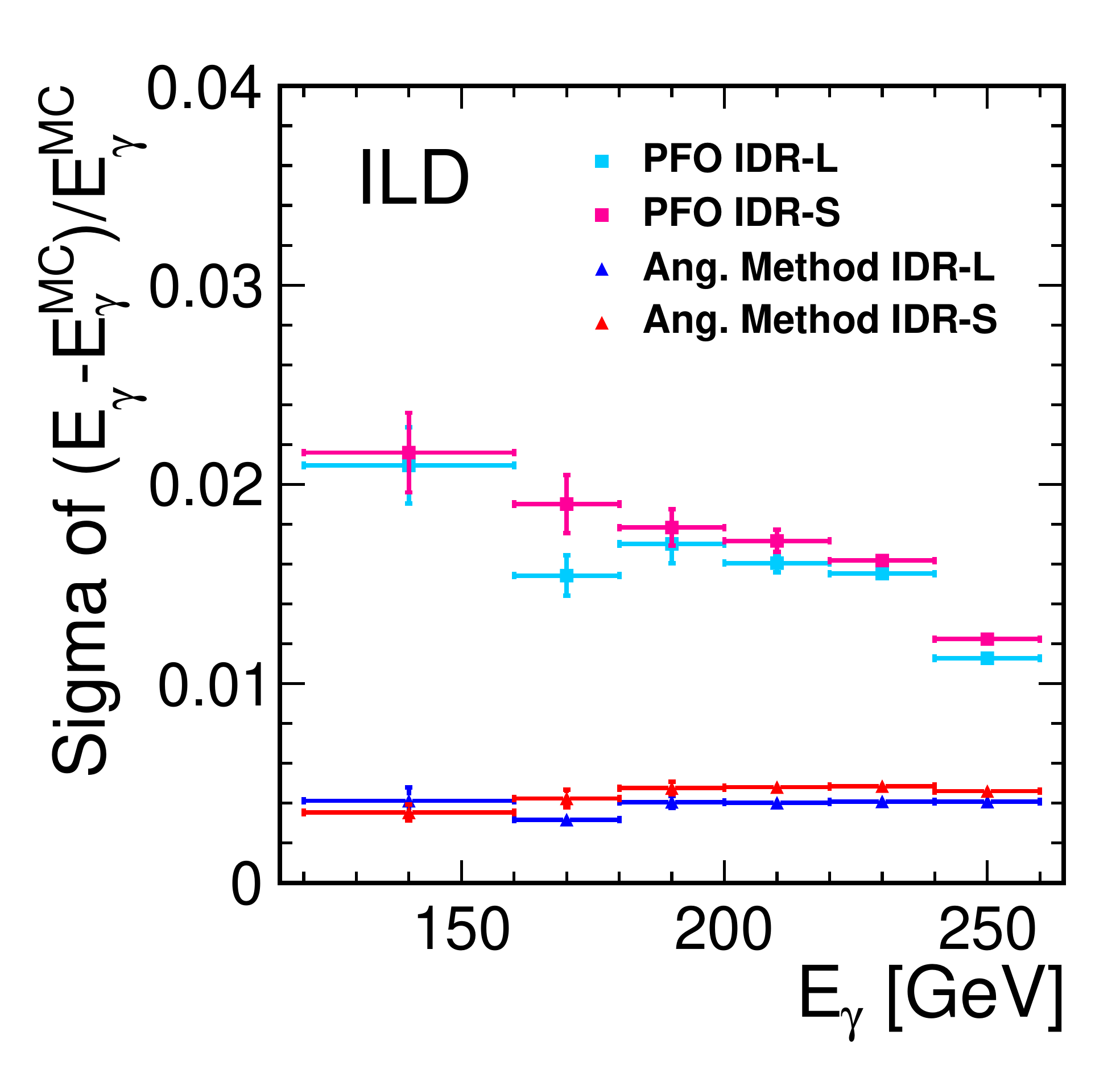}
 \caption{ \label{fig:gammaZ:sigmaE:rel}}
 \end{subfigure}
\begin{subfigure}{0.49\hsize} 
 \includegraphics[width=\textwidth]{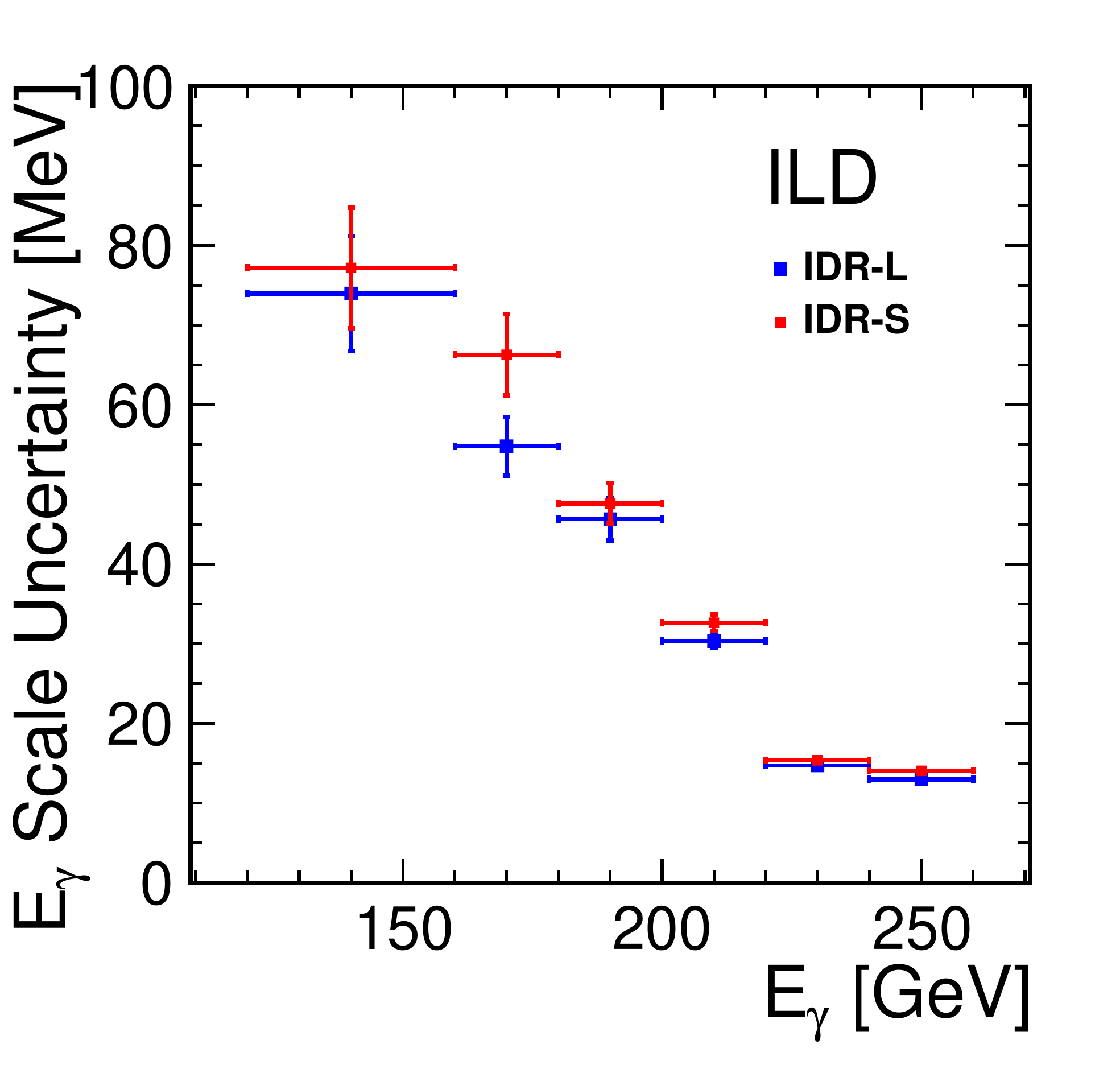}
 \caption{  \label{fig:gammaZ:sigmaE:abs}}
 \end{subfigure}
\caption{Uncertainty on the photon energy scale calibration as a function of the photon energy for IDR-L and IDR-S.
(a) Relative uncertainty when using non-perfectly calibrated PFO-level energies (cyan/magenta) and when calculating the photon energy from the $\mu$ momenta and kinematic constraints (blue/red, ``angular method'').
(b) Absolute uncertainty from the angular method in MeV.
}
\label{fig:gammaZ:sigmaE}
\end{figure}

\begin{figure}[htbp]
\begin{subfigure}{0.49\hsize} 
 \includegraphics[width=\textwidth]{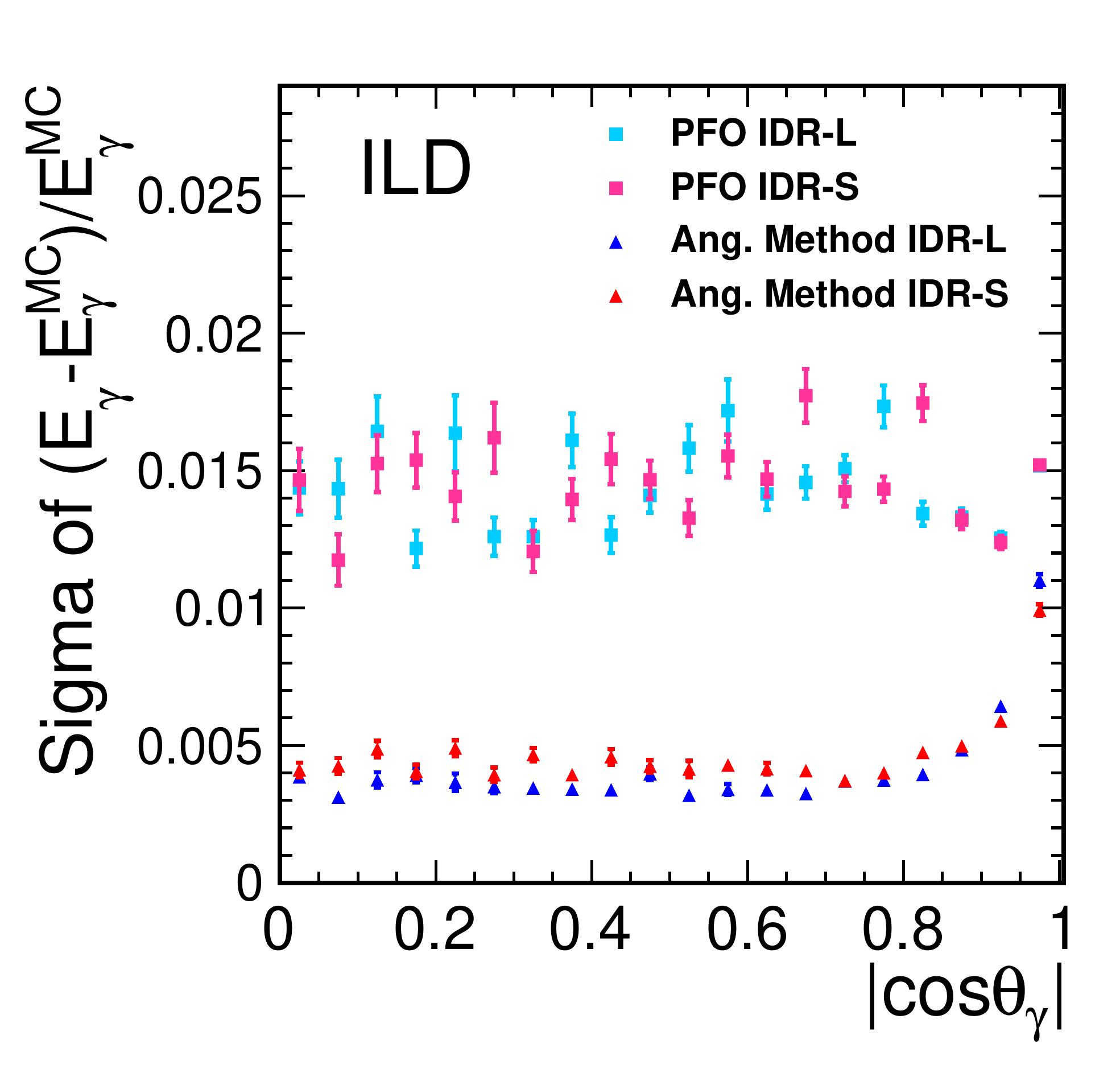}
 \caption{ \label{fig:gammaZ:angles:theta}}
 \end{subfigure}
\begin{subfigure}{0.49\hsize} 
 \includegraphics[width=\textwidth]{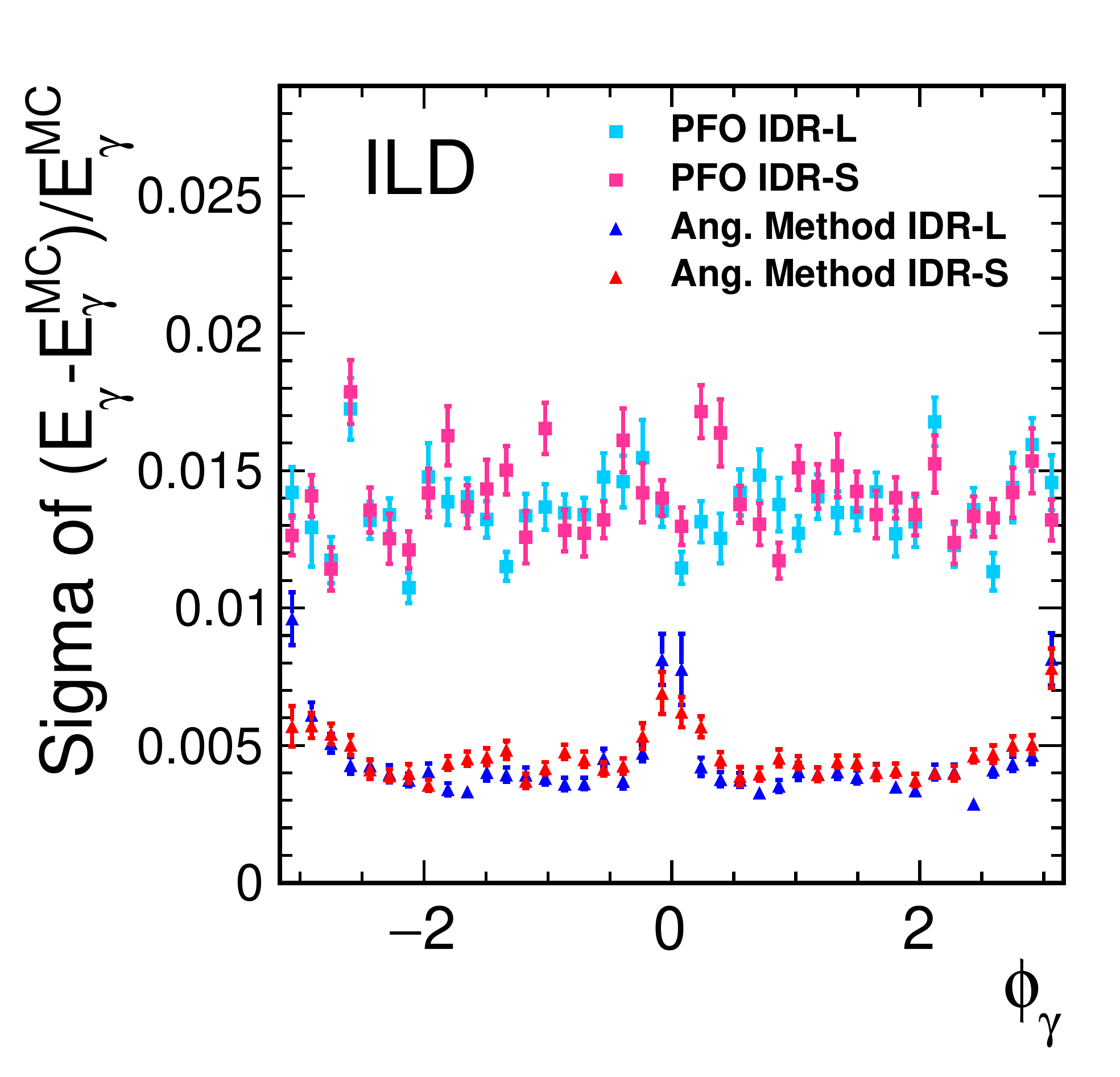}
 \caption{  \label{fig:gammaZ:angles:phi}}
 \end{subfigure}
\caption{
Resolution of the photon energy when using non-perfectly calibrated PFO-level energies (cyan/magenta) and when calculating the photon energy from the $\mu$ momenta and kinematic constraints (blue/red, ``angular method'').  
(a) as a function of the polar angle
(b) as a function of the azimuthal angle
}
\label{fig:gammaZ:angles}
\end{figure}

The resolution of the angular method, i.e.\ the width of the blue distribution in Fig.~\ref{fig:gammaZ:meanE:allE} is shown for IDR-L and IDR-S in Fig.~\ref{fig:gammaZ:sigmaE:rel} as a function of the photon energy. This translates into
an absolute uncertainty on the photon energy scale calibration of about $10$\,MeV for high-energy photons, as shown in Fig.~\ref{fig:gammaZ:sigmaE:abs}, for a perfectly calibrated and aligned tracking system and when integrating over the whole calorimeter. 
The angular dependencies of the resolution of this method are shown in Fig.~\ref{fig:gammaZ:angles}. As a function of the polar angle, Fig.~\ref{fig:gammaZ:angles:theta} clearly shows the effect of the better momentum resolution of IDR-L for central high-momentum tracks, while in the two most forward bins, the small detector performs better due to its higher magnetic field. As a function of the azimuthal angle, there is no particular region where the performance of the two detector models differs, as can be seen in Fig.~\ref{fig:gammaZ:angles:phi}. The modulation of the resolution with $\phi$ is an effect of using the $p_y$ constraint only, which is the easiest method as the $p_y$ conservation is not influenced by the beam crossing angle. The modulation will be reduced when also $p_x$ balance is exploited e.g.\ in a kinematic fit.

\subsection{\texorpdfstring{$A_{FB}$ and polarised cross sections from $e^+e^- \to b\bar{b}$}{AFB and ALR from ee -> bb}}
\label{subsec:bench:bbbar}

The measurement of the polar angle spectrum and hence of the forward backward asymmetries of $b$-quarks requires to distinguish the $b$-jet from the $\bar{b}$-jet. The two most important techniques for this are the reconstruction of the charge at the secondary vertex, and the identification of charged Kaons in the $b$-decay chain. While the first method requires a complete reconstruction of all tracks from the secondary vertices, the Kaon ID hinges upon a special
feature of ILD, namely the measurement of the specific energy loss $dE/dx$ in the TPC. In order to arrive at a reliable $b$-charge measurement, a consistent double-tag is required for each event, allowing for all four possible combinations of the two techniques: Either each of the two $b$-tagged jets contributes one tag (``K+K'', ``Vtx+Vtx'', ``Vtx+K (diff. jet)''), or both charge tags are found in the same $b$-tagged jet (``Vtx+K (same jet)''). The efficiency of the Kaon identification is compared for both detector models in Fig.~\ref{fig:perf:KaonID} in Sec.~\ref{sec:perf:sys:pid}. All details of the event reconstruction and selection can be found in~\cite{ILDNote:bbtt}.

\begin{figure}[htbp]
\begin{subfigure}{0.49\hsize} 
\includegraphics[width=\textwidth]{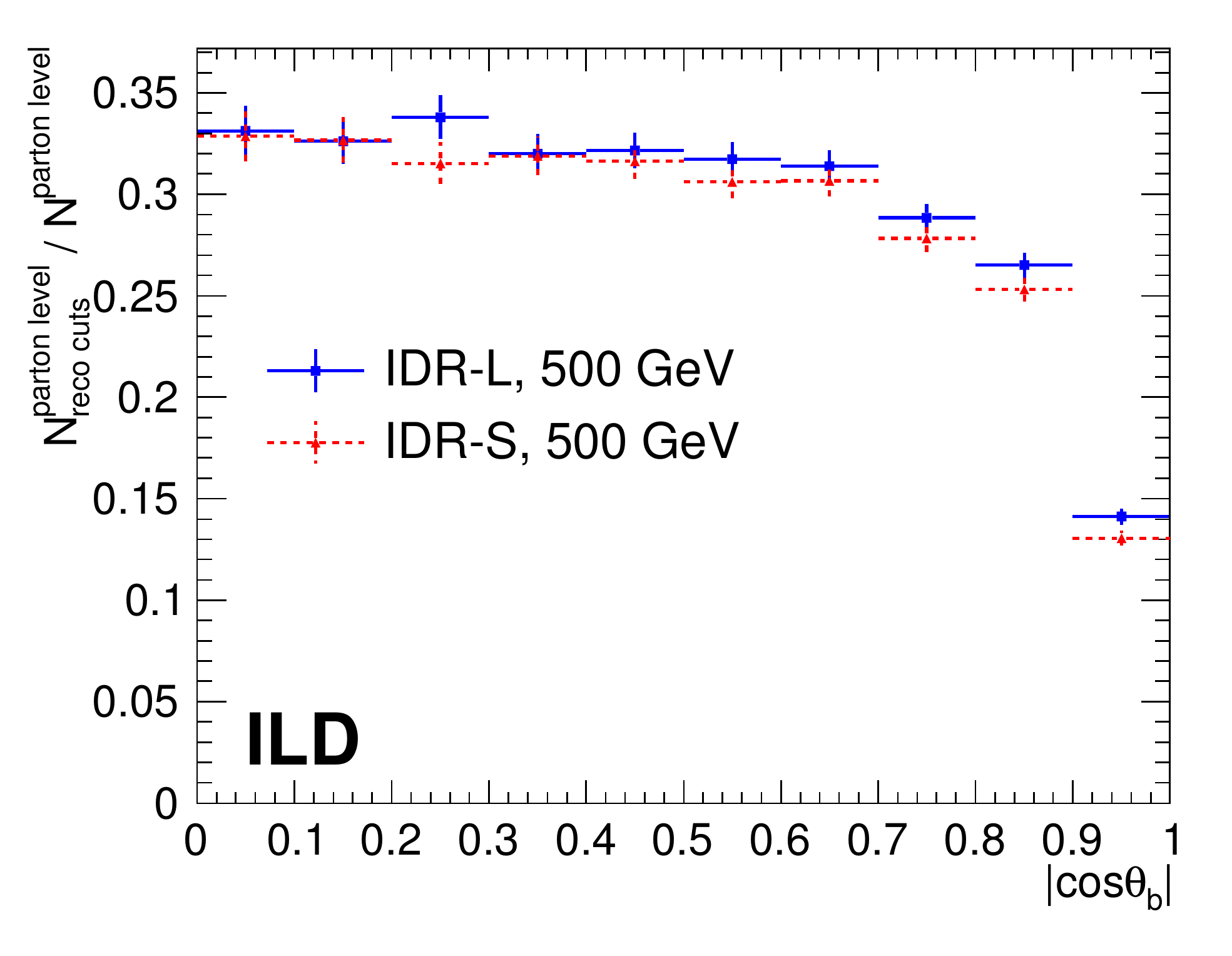}
 \caption{ \label{fig:bbbar:effipur:effi}}
 \end{subfigure}
\begin{subfigure}{0.49\hsize} 
\includegraphics[width=\textwidth]{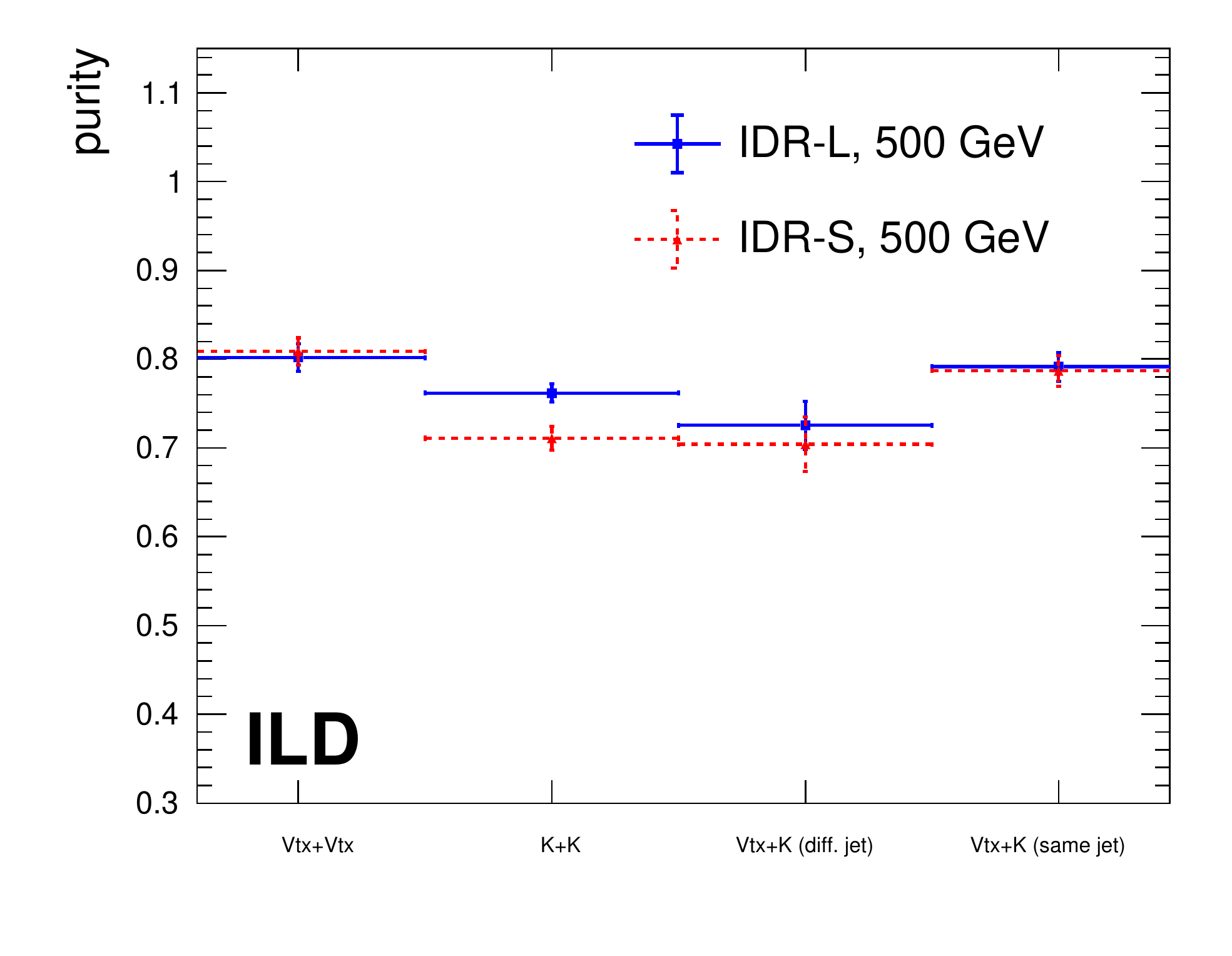}
 \caption{  \label{fig:bbbar:effipur:pur}}
 \end{subfigure}
\caption{
(a) Acceptance of the $e^+e^- \to b\bar{b}$ analysis as a function of $\cos{\theta}$ of the $b$-quark for IDR-L and IDR-S.
(b) Purity of the four different categories for charge tagging for IDR-L and IDR-S. 
}
\label{fig:bbbar:effipur}
\end{figure}

Figure~\ref{fig:bbbar:effipur:effi} compares the acceptance of the $b$-jet reconstruction for the large and the small version of ILD.
For $|\cos{\theta_b}|>0.5$, corresponding to a large part of the endcap 
region, the acceptance of IDR-L is about 1\% larger than for IDR-S. The purity of the four combinations of charge-ID is shown in Fig.~\ref{fig:bbbar:effipur:pur}. While the charge-ID via vertex charge performs identically for both detectors, the Kaon-charge ID yields a higher purity for IDR-L due to the larger radius of the TPC, which improves the $dE/dx$ resolution. 

\begin{figure}[htbp]
\begin{center} \includegraphics[width=0.55\textwidth]{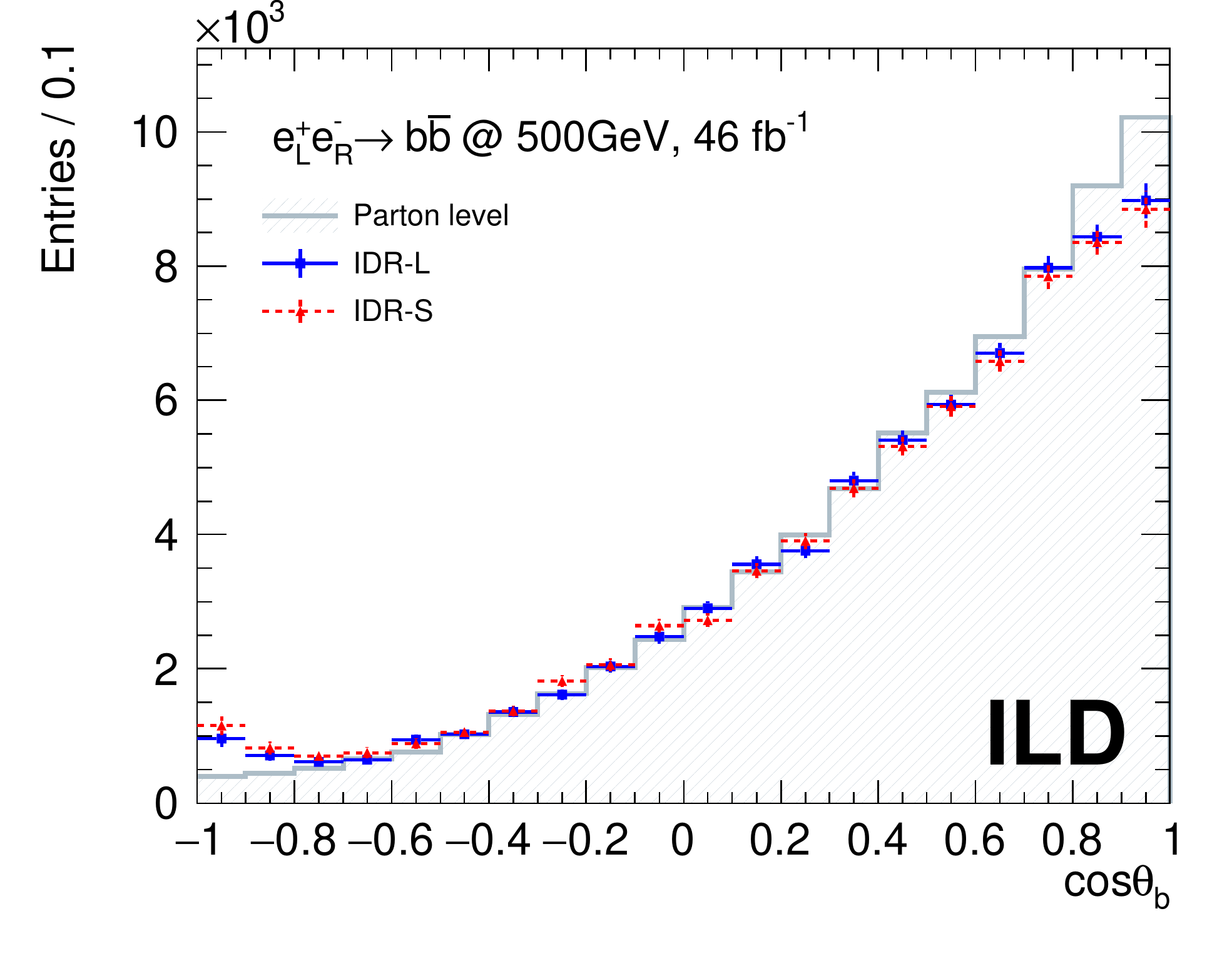}
\end{center}
\caption{Generator-level distribution of $\cos{\theta_b}$ and the corresponding reconstructed distributions for IDR-L and IDR-S. The distribution is shown for a simulated sample of $46$\,fb$^{-1}$ for pure $e^-_L e^+_R$ beams.}
\label{fig:bbbar:result}
\end{figure}

The reconstructed  $|\cos{\theta_b}|$ distribution is shown in Fig.~\ref{fig:bbbar:result} for an integrated luminosity of $46$\,fb$^{-1}$ with purely left-handed electrons and right-handed positrons. The distributions obtained for both detectors are compared with the parton level. The results for both detector models agree within the statistical uncertainty given by the size of the simulated sample. However the migration from larger to smaller $\cos{\theta_b}$ seems to be larger in case of the small detector model.

\subsection{\texorpdfstring{$A_{FB}$ and polarised cross sections from $tt \to bb qql\nu$}{AFB and ALR from tt -> bbqqlv}}
\label{subsec:bench:ttbar}

The forward-backward asymmetry and the polarised cross sections of $e^+e^- \to t\bar{t}$ are crucial ingredients to the determination of the couplings of top quark to the photon and the $Z$ boson. In the semi-leptonic channel, which represents 43.5\% of all $t \bar t$ decays, the lepton charge is available to distiguish between the $t$ and the $\bar{t}$ quark. In the fully hadronic case, representing 46.5\% of all $t\bar t$ decays, $t$ and the $\bar{t}$ can only be distinguished by identifying the $b$ and $\bar{b}$-jets. For this purpose, the same techniques as presented in the previous subsection can be used, namely the charge at the secondary vertex and the charge of a Kaon in the jet (if present). While these techniques are common between the $e^+e^- \to b\bar{b}$ and $e^+e^- \to t\bar{t}$ cases, the two benchmarks probe very different momentum ranges for the Kaon as well as for the tracks forming the secondary vertex: The $b\bar{b}$ case offers a sample of back-to-back jets with typically $200-250$\,GeV each, while the $t\bar{t}$ case, the majority of the $b(\bar{b})$ jets have momenta of less than $50$\,GeV. Furthermore, the 6-fermion final state presents a much more busy environment than the 2-fermion final state.
Therefore, the two channels are complementary in terms of evaluating the detector performance. This is for instance reflected by the $5\%$ higher purity of the Kaon ID in case of the $t\bar{t}$ benchmark, c.f.\ Fig.~\ref{fig:perf:KaonID} in Sec.~\ref{sec:perf:sys:pid}.

As a $t\bar{t}$ benchmark for this document, the semi-leptonic channel was chosen, because it also tests the lepton (charge) reconstruction. Nevertheless, the Kaon and vertex charge tags can be applied on the $b(\bar{b})$-jets in addition. The inclusion of the The full description of the analysis can be found in~\cite{ILDNote:bbtt}. 

\begin{figure}[htbp]
\begin{center} \includegraphics[width=0.55\textwidth]{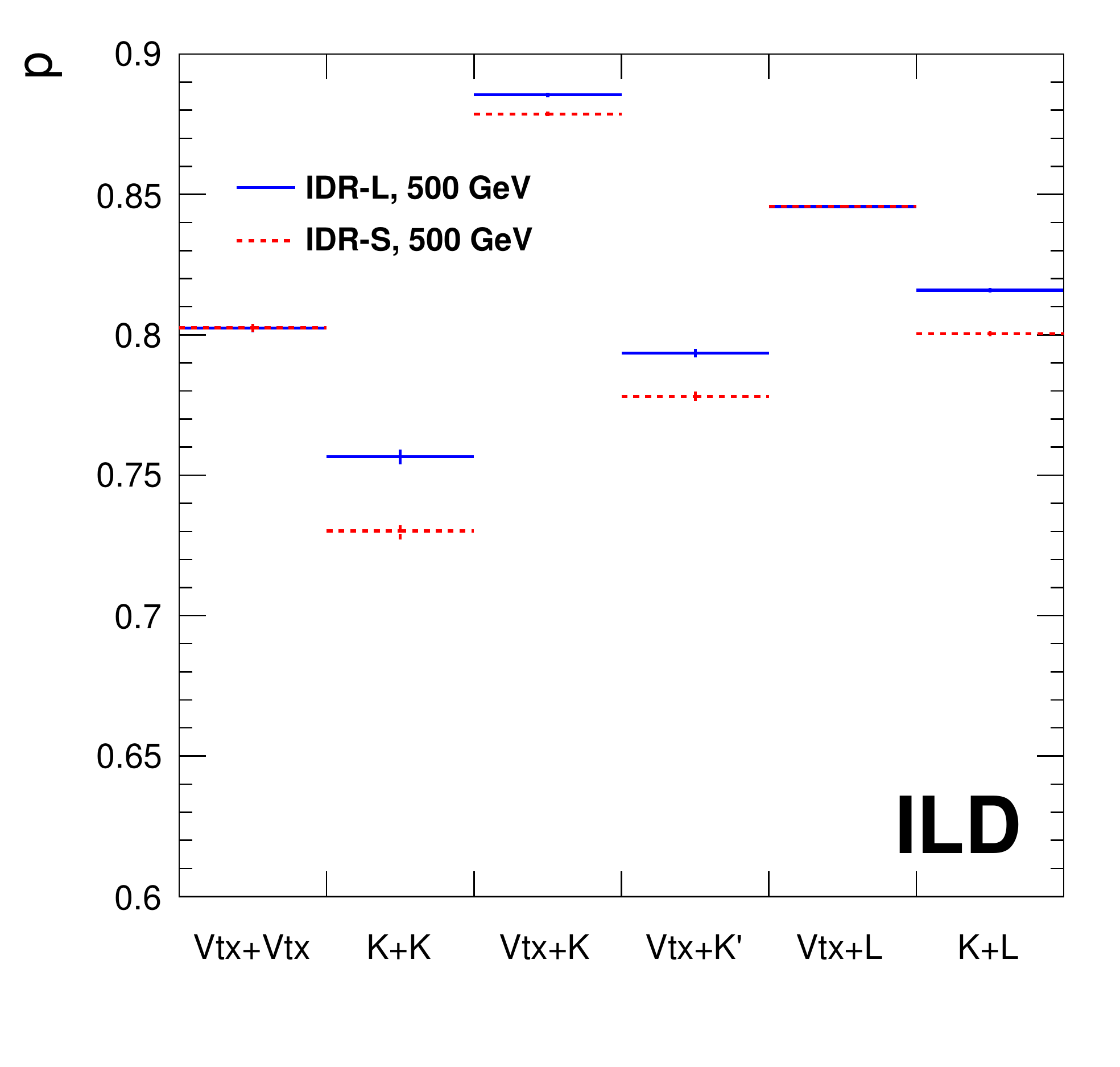}
\end{center}
\caption{Purity of the different double-tag methods to identify the charge of the $t$/$\bar{t}$ quarks. ``Vtx+K'' and `Vtx+K$'$'' refer to the vertex and Kaon charge tags in different jets and in the same jet, respectively. As expected, the double Kaon tag shows the largest difference between the two detector models due to the different TPC radius.}
\label{fig:ttbar:effi}
\end{figure}

Like in the $e^+e^- \to b\bar{b}$ case, an event is considered only if two independent charge tags give consistent information. They can be from the lepton and one of the $b$-tagged jets (``K+L'', ``Vtx+L'') from both $b$-tagged jets (``K+K'', ``Vtx+Vtx'', ``Vtx+K'') or even from the same $b$-tagged jet (``Vtx+K$'$''). The purities of several double-tag combinations are compared for the large and small detector models in Fig.~\ref{fig:ttbar:effi}. While the size of the detector has no direct influence on the semi-leptonic analysis, especially the Kaon-ID based methods achieve a higher purity in case of the large detector. Overall, the combination of all the double-tag methods
increases the available statistics by about $40\%$ compared to using the lepton charge alone, from $22\%$ to $30\%$. 

If the lepton tag is (artificially) ignored, the channel is a perfect proxy for estimating the performance achievable in the fully hadronic channel. The jet-based double-tag methods alone yield an efficiency of about $15\%$. This means that from a future inclusion of the fully hadronic channel in the analysis a $50\%$ increase in the statistics available for the asymmetry and cross-section measurements.

\begin{figure}[htbp]
\begin{subfigure}{0.475\hsize} 
\includegraphics[width=\textwidth]{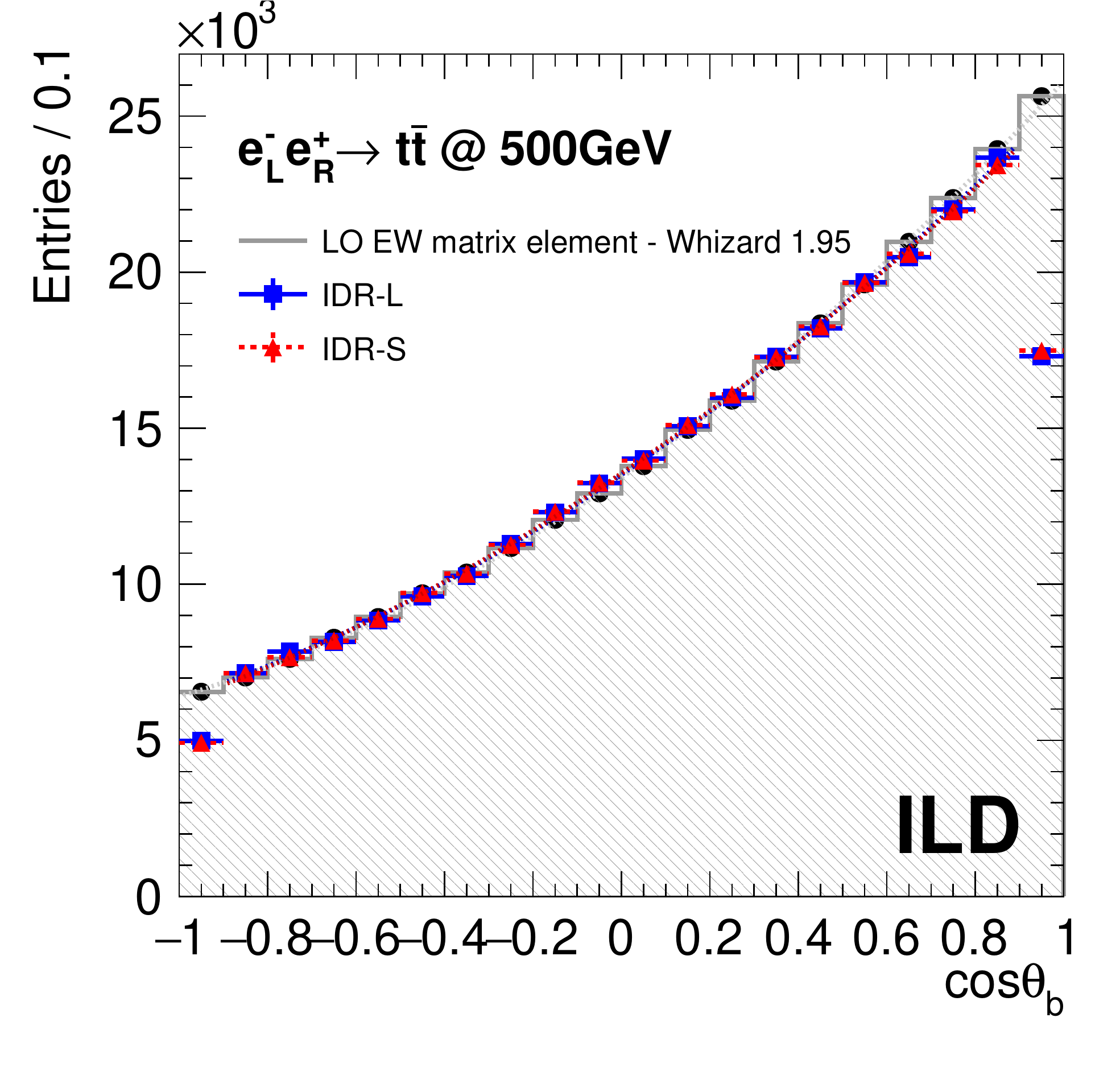}
 \caption{ \label{fig:ttbar:costhetab}}
 \end{subfigure}
\begin{subfigure}{0.475\hsize} 
\includegraphics[width=\textwidth]{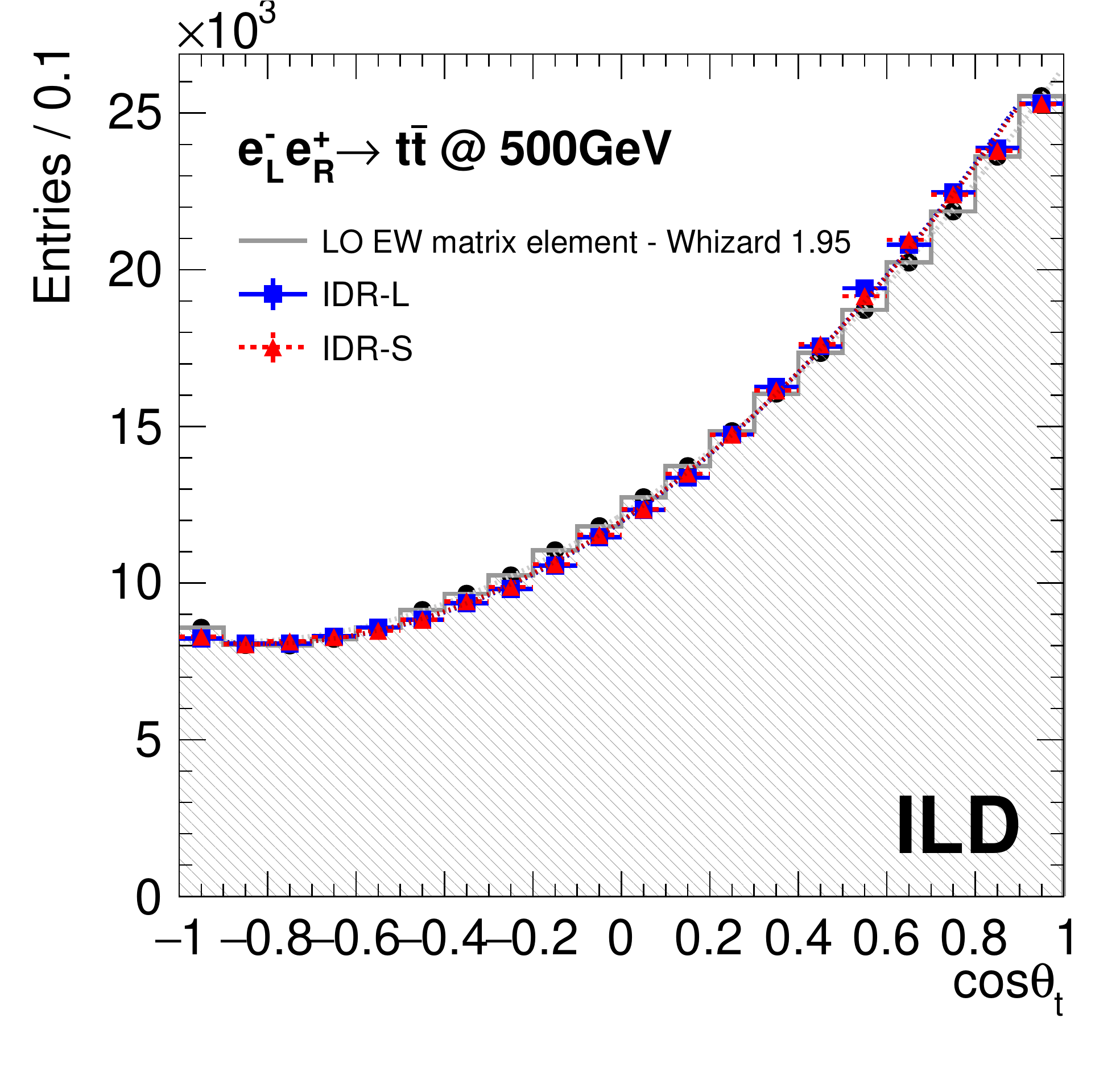}
 \caption{  \label{fig:ttbar:costhetat}}
 \end{subfigure}
\caption{Polar angle distributions at leading order matrix element level and at reconstruction level for IDR-L and IDR-S. The distributions are shown for pure $e^-_L e^+_R$ data, in arbitrary absolute normalisation. Results corresponding to ILC500 as defined in  Sec.~\ref{sec:benchmarks:lep} are given in Tab.~\ref{tab:AFBtt}.
(a) Polar angle of the $b$-quark $\cos{\theta_b}$. 
(b) Polar angle of the $t$-quark $\cos{\theta_t}$.
}
\label{fig:ttbar:result}
\end{figure}

The reconstructed  polar angle ($\cos{\theta_t}$) distribution is shown for the purely left-handed electron and purely right-handed positron case in Fig.~\ref{fig:ttbar:costhetat}. Thereby, $\cos{\theta_t}$ is the polar angle of the three-momentum calculated from the two reconstructed tops as $\vec{p}_{t\bar{t}} = \vec{p}_t-\vec{p}_{\bar t}$. Figure~\ref{fig:ttbar:costhetab} shows the -- analoguously calculated -- polar angle spectrum of the $b$ quark that is emitted from the $t$ decay. This observable has been studied for the first time for the IDR and adds to the physics potential of ILD.  Both, the polar angle spectrum of the $t$ quark and of the $b$ quark,  as reconstructed with the two detector models under study are compared to the leading-order electroweak matrix element prediction. No difference is found here between the two detector models.
 
\begin{figure}[htbp]
\begin{subfigure}{0.475\hsize} 
\includegraphics[width=\textwidth]{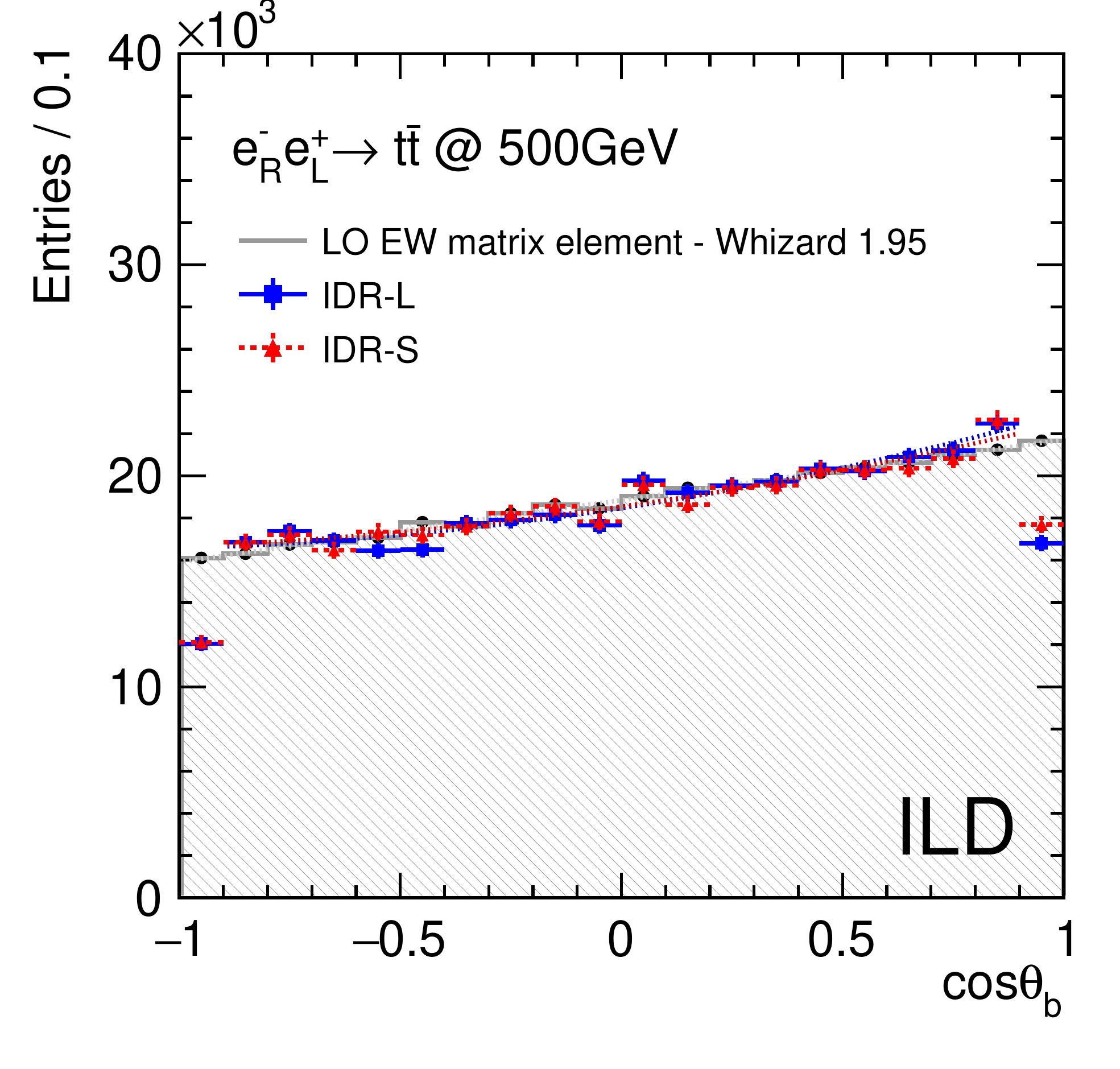}
 \caption{ \label{fig:ttbar:costhetabRL}}
 \end{subfigure}
\begin{subfigure}{0.475\hsize} 
\includegraphics[width=\textwidth]{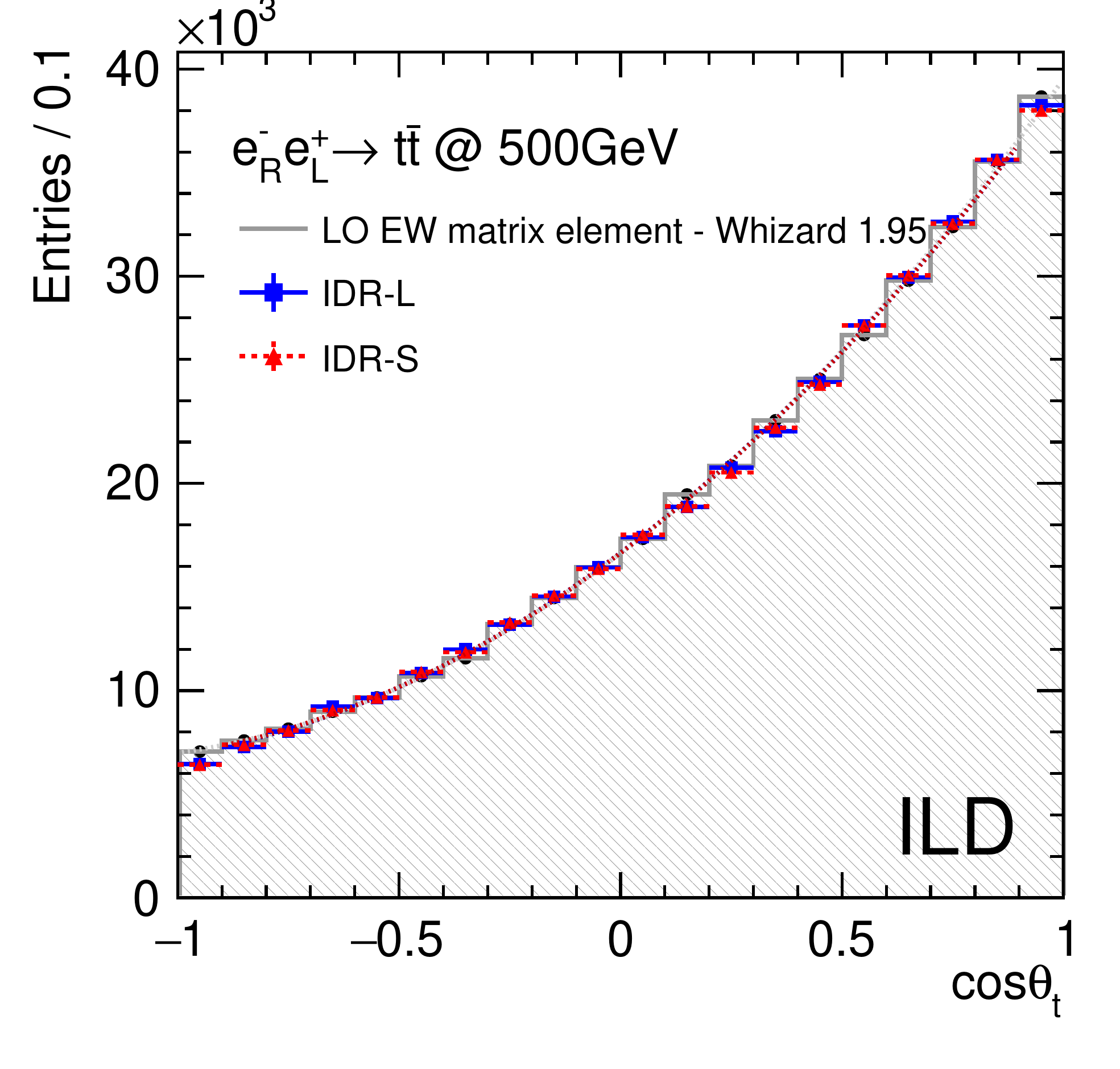}
 \caption{  \label{fig:ttbar:costhetatRL}}
 \end{subfigure}
\caption{Polar angle distributions at leading order matrix element level and at reconstruction level for IDR-L and IDR-S. The distributions are shown for pure $e^-_R e^+_L$ data, in arbitrary absolute normalisation. Results corresponding to ILC500 as defined in  Sec.~\ref{sec:benchmarks:lep} are given in Tab.~\ref{tab:AFBtt}.
(a) Polar angle of the $b$-quark $\cos{\theta_b}$. 
(b) Polar angle of the $t$-quark $\cos{\theta_t}$.
}
\label{fig:ttbar:resultRL}
\end{figure}

The analogous distributions are shown in Fig.~\ref{fig:ttbar:resultRL} for the opposite beam helicity configuration, again compared to the leading-order electroweak matrix element level. Also here, no difference is found here between the two detector models.  Note the difference in the polar angle spectra of the $b$ quark between Figs.~\ref{fig:ttbar:costhetab}, and~\ref{fig:ttbar:costhetabRL} which is due to the $V-A$ interaction at the $tbW$ vertex.

The resulting precisions on the polarised cross sections as well as on $A_{FB}$ as expected for the $P(e^-,e^+)=(\pm 80\%, \mp 30\%)$ data sets of ILC500 as defined in Sec.~\ref{sec:benchmarks:lep} are equal for the two detector models as given in Tab.~\ref{tab:AFBtt}.

\begin{table}[htb]
\begin{center}
\begin{tabular}{|c|c|c|c|}
\hline
 & $P(e^-,e^+)$ & $(\delta\sigma / \sigma)_{stat} [\%]$ & $(\delta A_{FB}^t / A_FB^t)_{stat} [\%]$ \\
\hline
\multirow{2}{*}{IDR-L/S} &  $(-80\%,+30\%)$ & 0.17 & 0.7 \\
                         &  $(+80\%,-30\%)$ & 0.25 & 0.53 \\
\hline
\end{tabular}
\end{center}
\caption{Statistical uncertainties on the polarised cross section and the top forward-backward asymmetry obtained in this analysis for the ILC500 as defined in Sec.~\ref{sec:benchmarks:lep}.}
\label{tab:AFBtt}   
\end{table}    

\begin{figure}[htbp]
\begin{center} \includegraphics[width=0.6\textwidth]{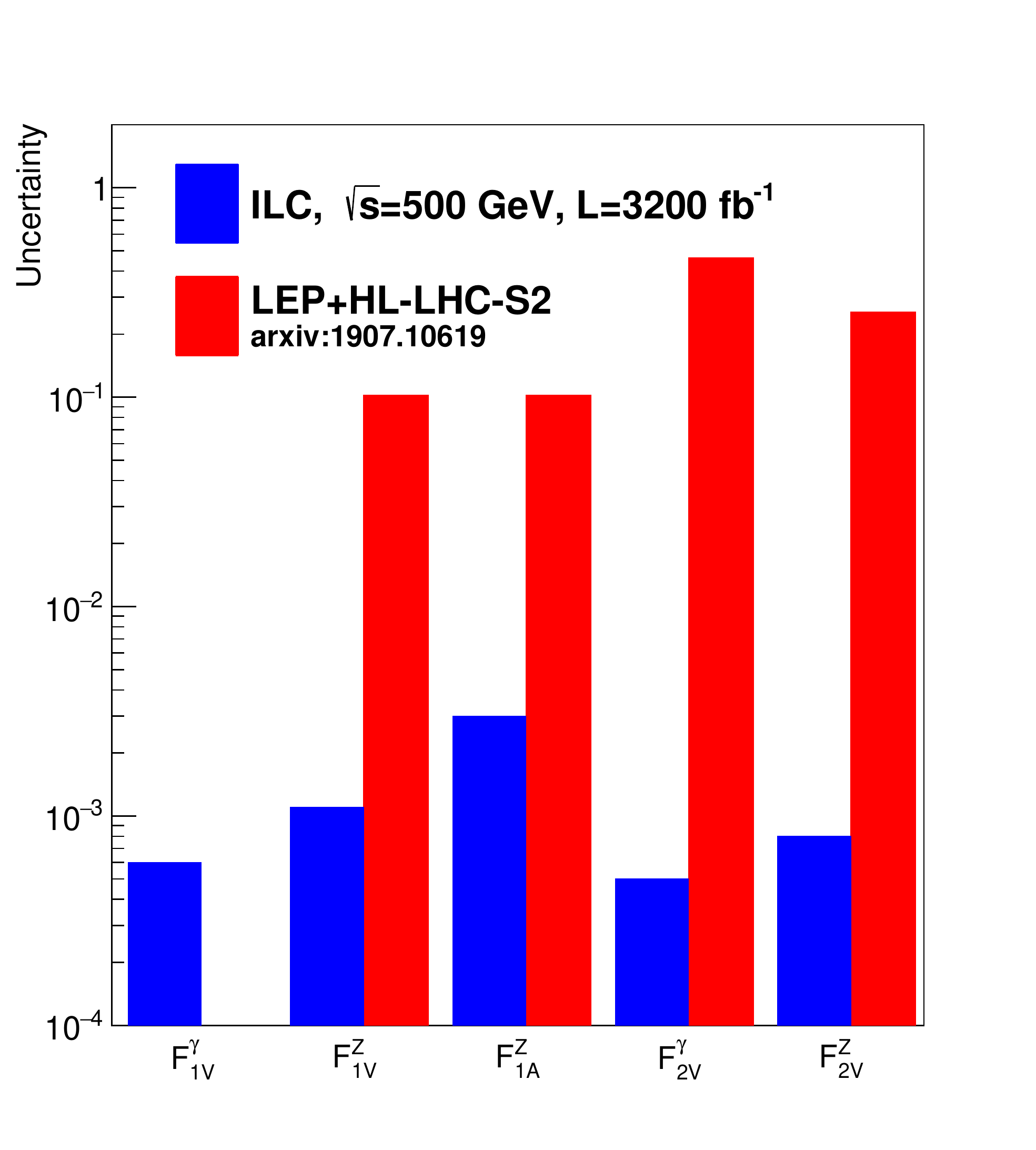}
\end{center}
\caption{Expected precision on electromagnetic form factors of the top quark from this study, based on ILC500 as defined in Sec.~\ref{sec:benchmarks:lep}, compared to recent projections for the HL-LHC~\cite{Durieux:2019rbz}.}
\label{fig:ttbar:formfac}
\end{figure}

In order to put these results into perspective, they have been translated into $1\sigma$ precisions on the electromagnetic form factors of the top quark. These are displayed in Fig.~\ref{fig:ttbar:formfac} in comparison to the most recent projections for HL-LHC. 
The projections for HL-LHC are derived from the {\em individual} constraints of EFT Wilson coefficients presented in Tab.\,C2.3 of Ref.~\cite{Durieux:2019rbz} (the most favorable scenario for HL-LHC). This figure clearly demonstrates the superiority of a linear $e^+e^-$ collider with polarised beams operated at an adequate center-of-mass energy.             

\subsection{Search for extra Scalars in \texorpdfstring{$e^+e^- \to ZS^0$}{e+e- -> ZS^0}}
\label{subsec:bench:extraH}

Searches for additional Higgs bosons or other new scalar particles, denoted here generically with $S^0$, are theoretically well-motivated, and benefit from an $e^+e^-$ collider sensitivity highly
complementary to that of hadron colliders. Due to the SM-likeness of the $125$-GeV Higgs boson, additional Higgs-like scalars are expected to have a suppressed coupling to the $Z$ boson. Nevertheless, they can 
be searched for using the same recoil technique which is the basis of the decay-mode independent measurement
of the total Higgsstrahlungs cross section. While in principle all visible decay modes of the $Z$ boson can
be exploited in this search, $Z\to\mu^+\mu^-$ as the cleanest mode has been chosen as a benchmark here~\cite{ILDNote:extraH}.
The two most important detector performance aspects for this analysis are the momentum resolution for the two
muons and the ability to detect and identify ISR photons in the detector. The latter applies in particular for the lower Higgs masses, where $M_{S^0} + M_Z << \sqrt{s}$ and thus significant photon radiation occurs frequently.
If the photon is detected, then the event kinematics can be corrected accordingly, which improves the separation of signal and background significantly.

\begin{figure}[htbp]
\begin{subfigure}{0.475\hsize} 
\includegraphics[width=\textwidth]{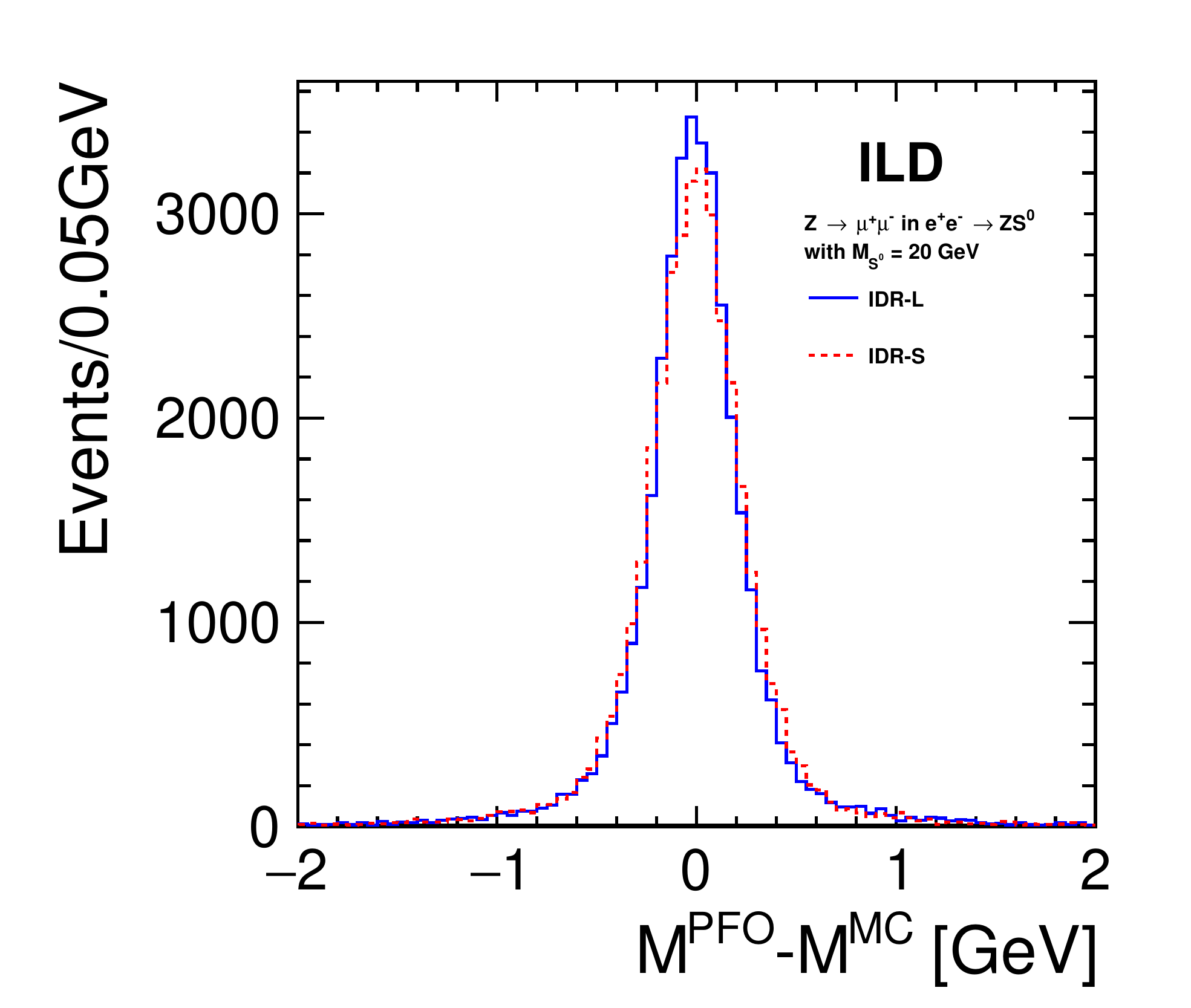}
 \caption{ \label{fig:extraH:Mdiff:mh20}}
 \end{subfigure}
\begin{subfigure}{0.475\hsize} 
\includegraphics[width=\textwidth]{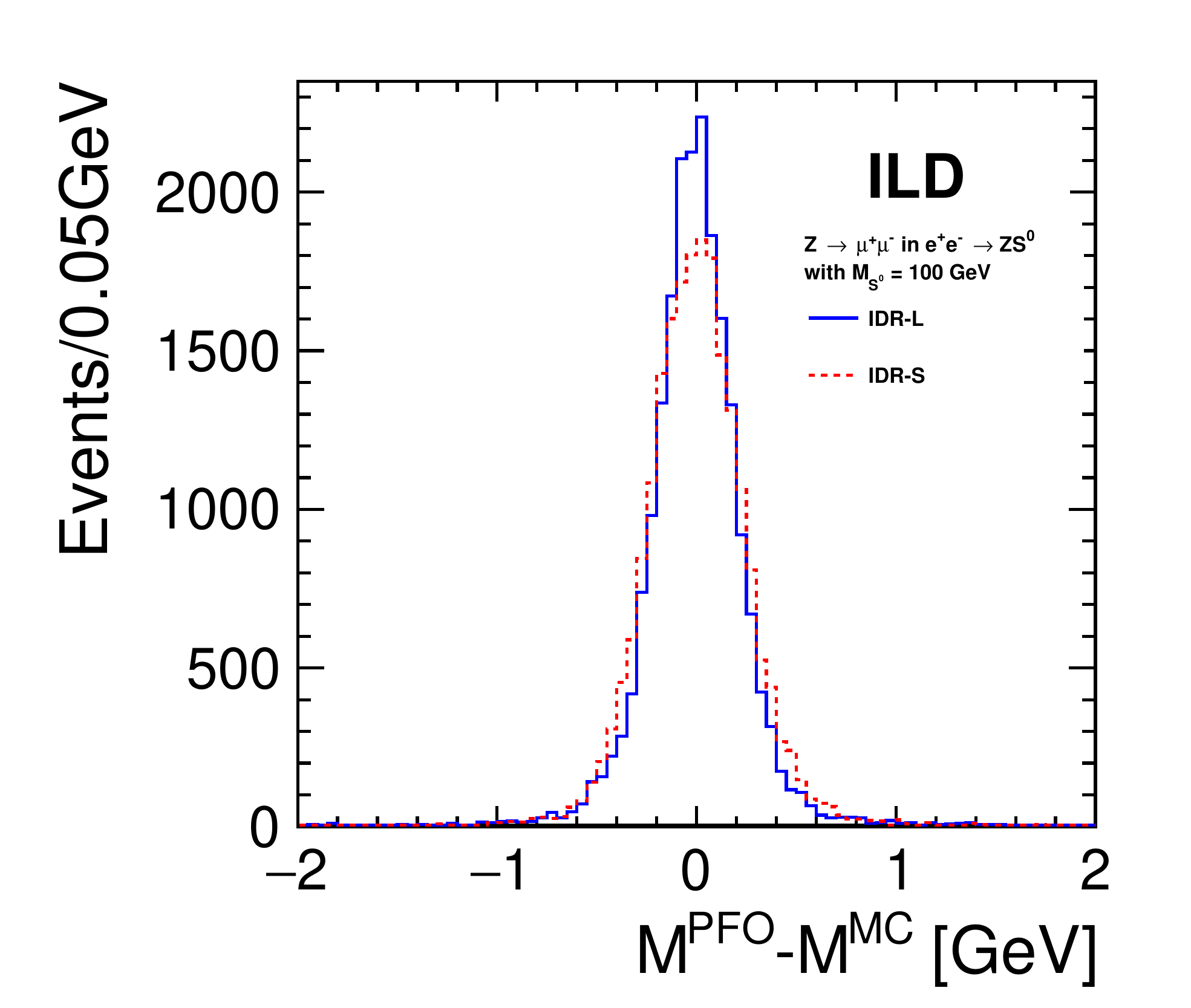}
 \caption{  \label{fig:extraH:Mdiff:mh100}}
 \end{subfigure}
\begin{subfigure}{0.475\hsize} 
\includegraphics[width=\textwidth]{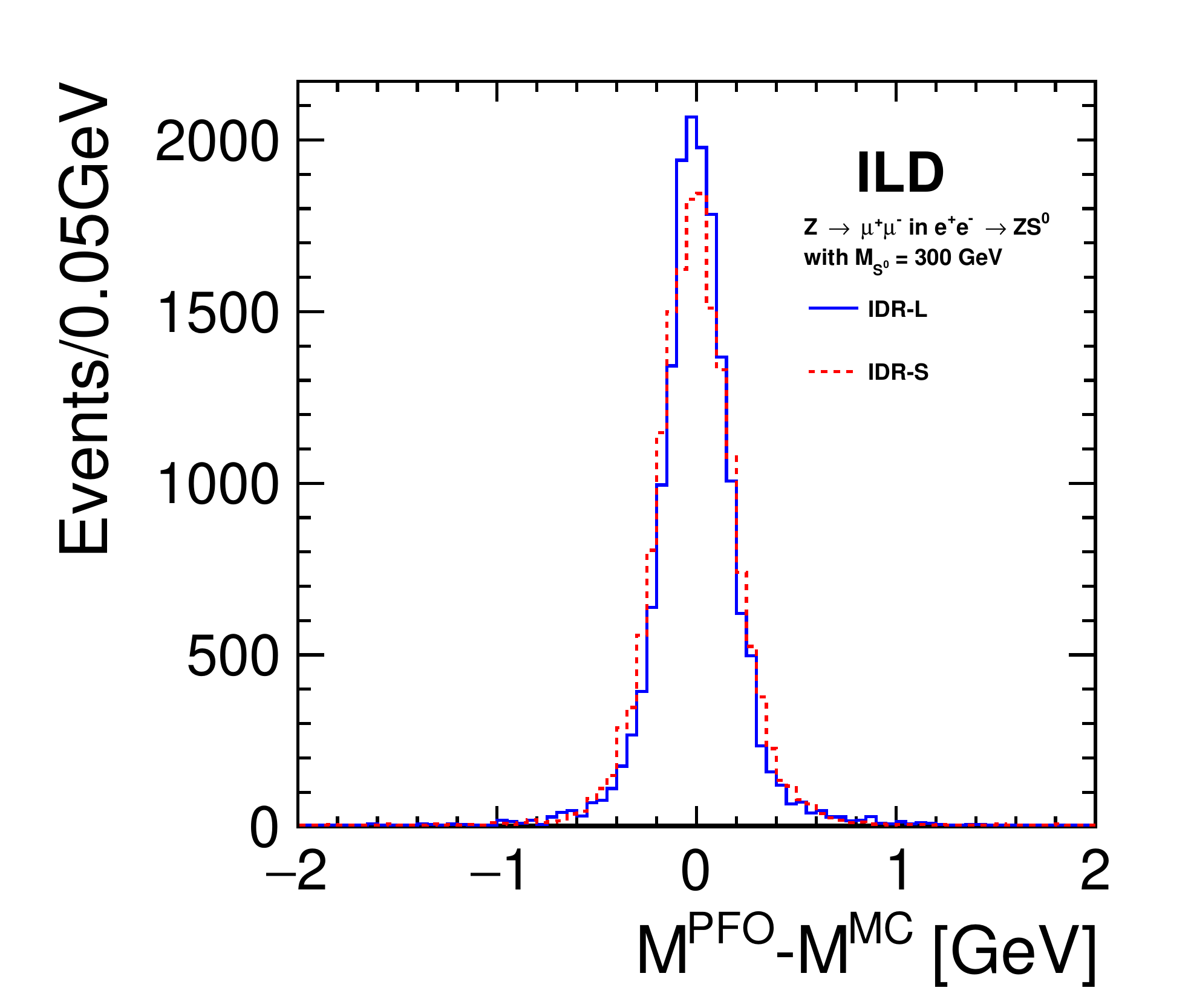}
 \caption{ \label{fig:extraH:Mdiff:mh300}}
 \end{subfigure}
\begin{subfigure}{0.475\hsize} 
\includegraphics[width=\textwidth]{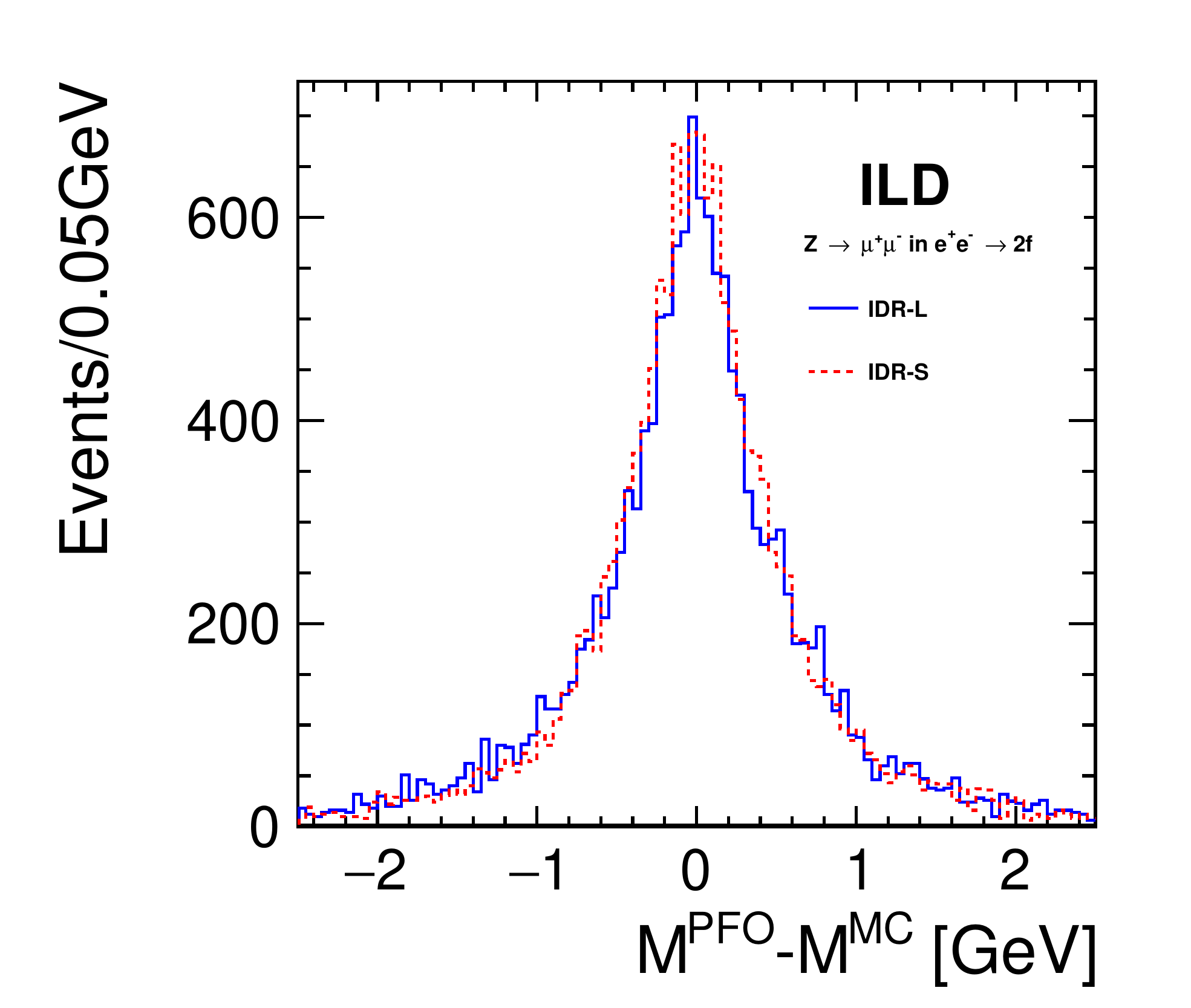}
 \caption{  \label{fig:extraH:Mdiff:2f}}
 \end{subfigure}
\caption{Event-by-event difference between reconstructed and generated di-muon mass for different types of events sampling different ranges in polar angle and momentum of the muons  in IDR-L and IDR-S:
(a) $ZS^0$ signal with $M_{S^0} = 20$\,GeV
(b) $ZS^0$ signal with $M_{S^0} = 100$\,GeV
(c) $ZS^0$ signal with $M_{S^0} = 300$\,GeV
(d) $2f$ background.
}
\label{fig:extraH:Mdiff}
\end{figure}

Figure~\ref{fig:extraH:Mdiff} shows the event-by-event difference between the reconstructed and generated di-muon mass for signal events with different scalar masses as well as for the 2-fermion background. The underlying distributions of momentum and polar angle differ significantly between the various samples. Therefore also the impact of the choice of detector model varies from sample to sample: With higher scalar masses, the $Z$ is less and less boosted, thus the muons are more and more central and back-to-back, resulting in an overall better mass reconstruction and a better performance of the large detector. In case of the $2$-fermion background, about half of the events return to the $Z$ pole, resulting in forward boosted $Z$'s with a small opening angle between the also forward-going muons. In those events which do not return to the $Z$, the muons have higher momenta than in the
signal samples. In combination, the mass reconstruction in the $2$-fermion events is worse, but when comparing
the two detector models, the better momentum resolution of ILD-S in the forward region and its worse resolution
in the barrel roughly cancel.

\begin{figure}[htbp]
\begin{subfigure}{0.475\hsize} 
\includegraphics[width=\textwidth]{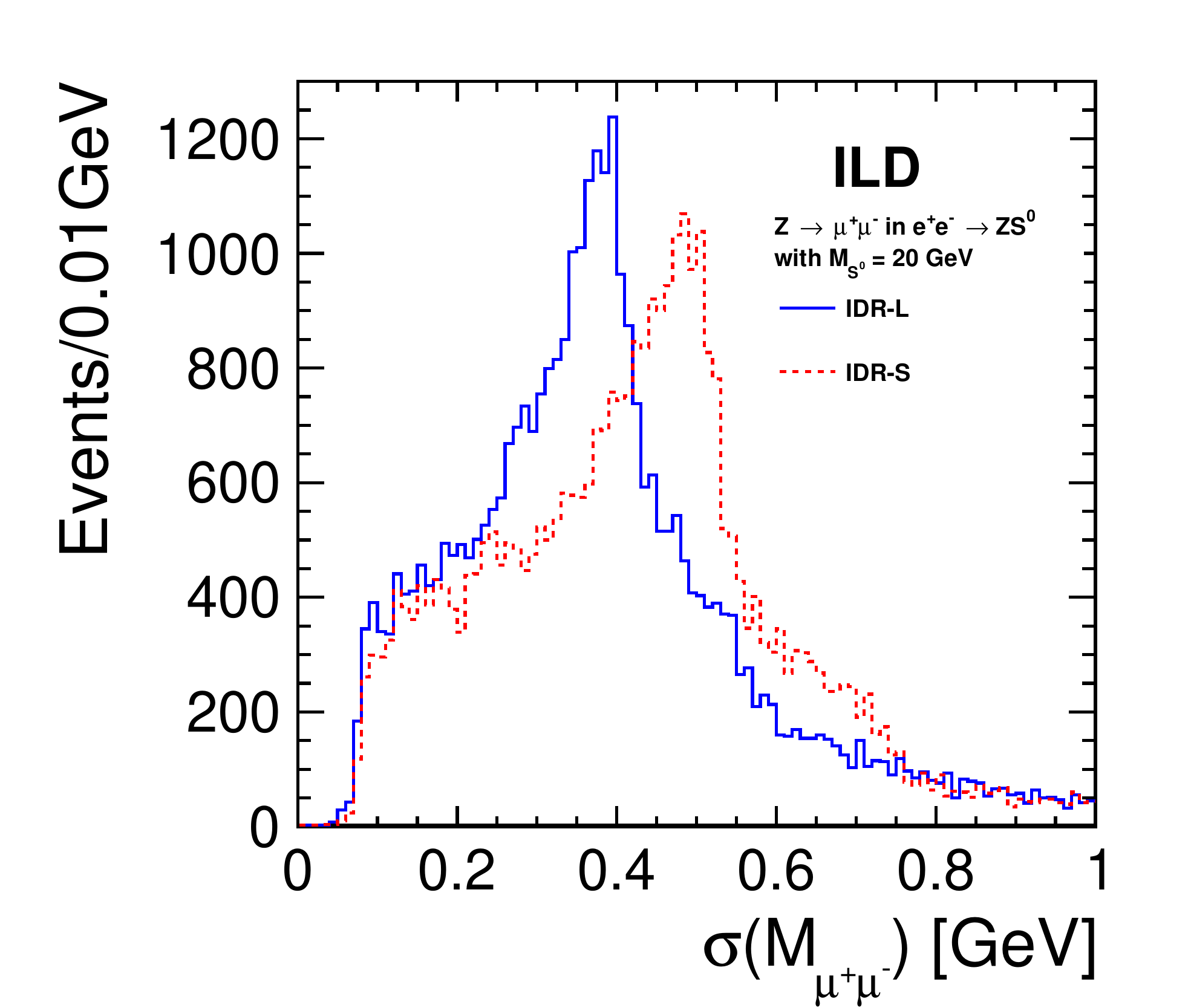}
 \caption{ \label{fig:extraH:Msigma:mh20}}
 \end{subfigure}
\begin{subfigure}{0.475\hsize} 
\includegraphics[width=\textwidth]{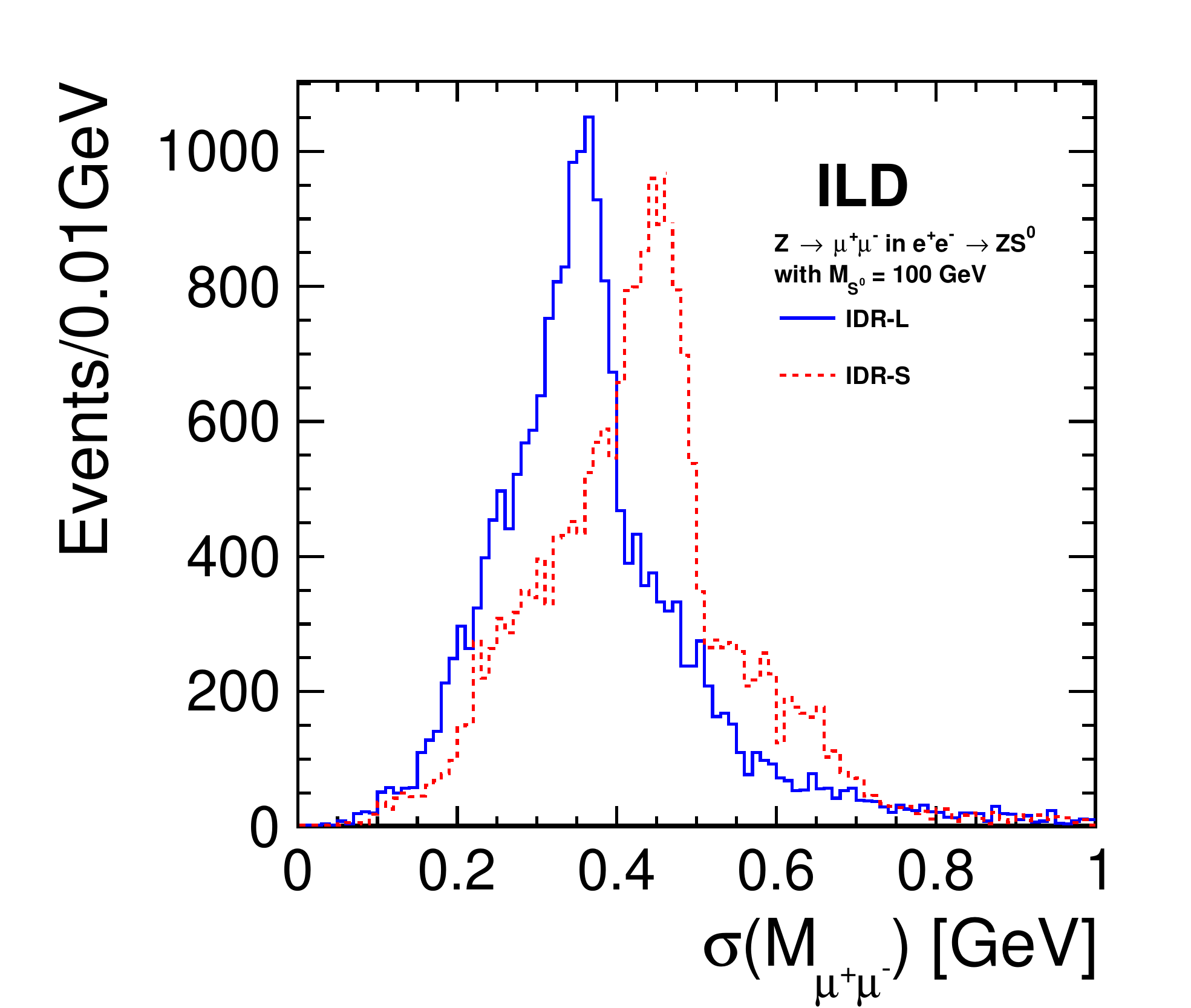}
 \caption{  \label{fig:extraH:Msigma:mh100}}
 \end{subfigure}
\begin{subfigure}{0.475\hsize} 
\includegraphics[width=\textwidth]{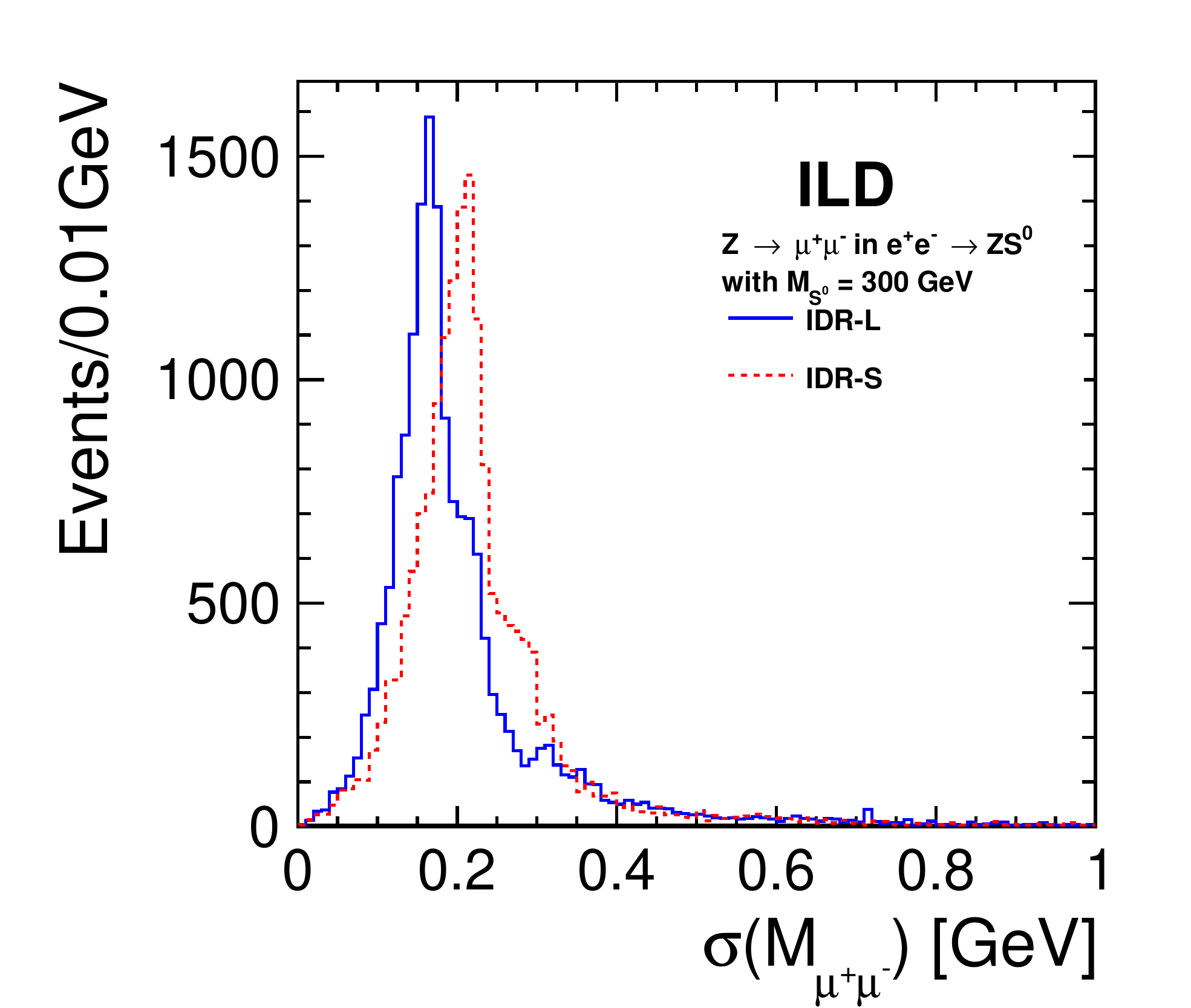}
 \caption{ \label{fig:extraH:Msigma:mh300}}
 \end{subfigure}
\begin{subfigure}{0.475\hsize} 
\includegraphics[width=\textwidth]{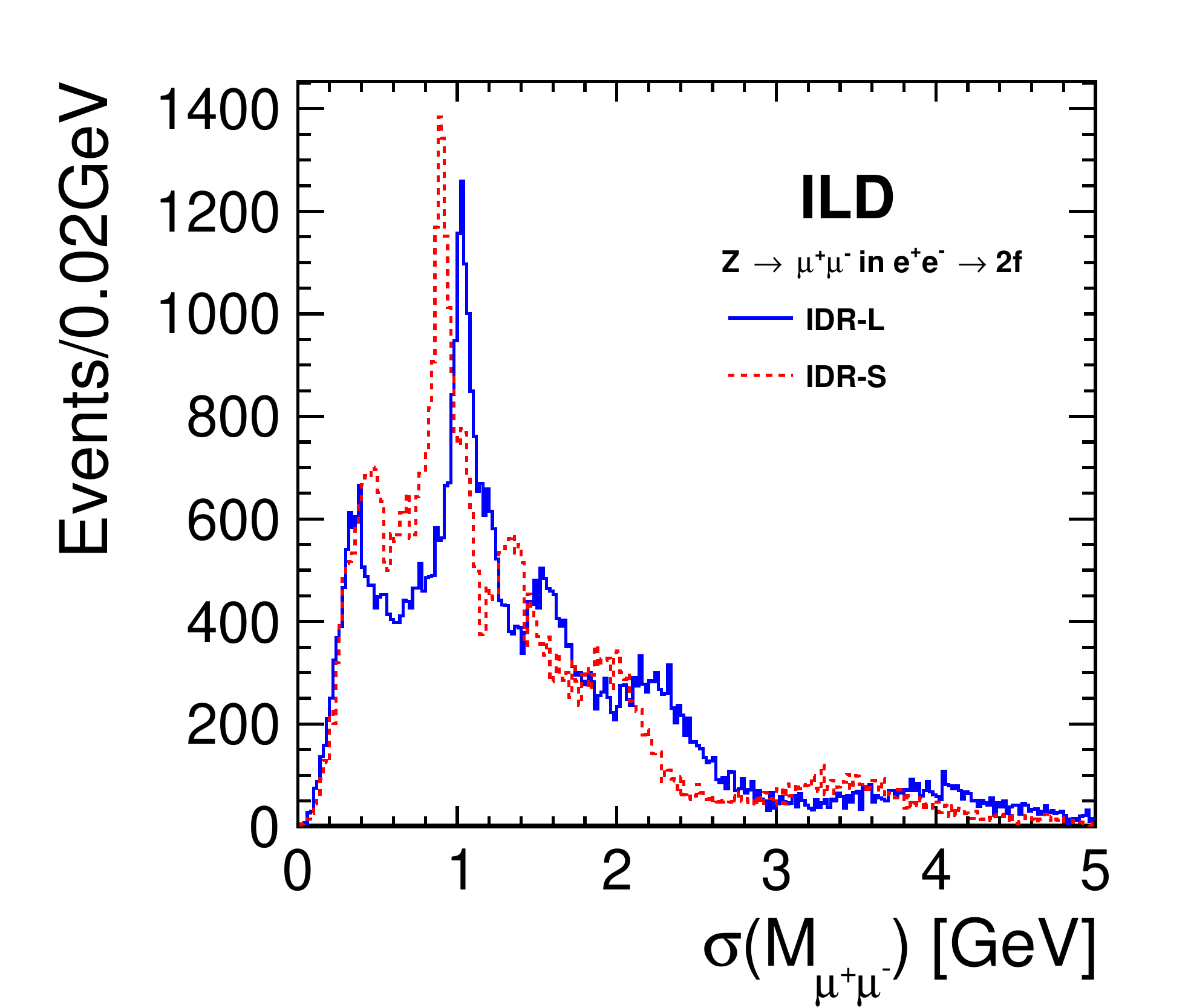}
 \caption{  \label{fig:extraH:Msigma:2f}}
 \end{subfigure}
\caption{Event-by-event uncertainty on the invariant di-muon mass calculated by error propagation from the track fit for different types of events sampling different ranges in polar angle and momentum of the muons in IDR-L and IDR-S:
(a) $ZS^0$ signal with $M_{S^0} = 20$\,GeV
(b) $ZS^0$ signal with $M_{S^0} = 100$\,GeV
(c) $ZS^0$ signal with $M_{S^0} = 300$\,GeV
(d) $2f$ background.
}
\label{fig:extraH:Msigma}
\end{figure}

This can be seen much more clearly from the event-by-event mass uncertainty as calculated from the errors on 
the track parameters, shown in Fig~\ref{fig:extraH:Msigma} for the same four event samples. In case of the scalar
signal, the large detector always gives a significantly better resolution. In case of the $2$-fermion background, there is a small population at the smallest uncertainties which behaves like the signal, while for cases of the larger uncertainty, the small detector performs better than the large detector.

%

\begin{figure}[htbp]
\begin{center} 
\begin{subfigure}{0.49\hsize} 
 \includegraphics[width=\textwidth]{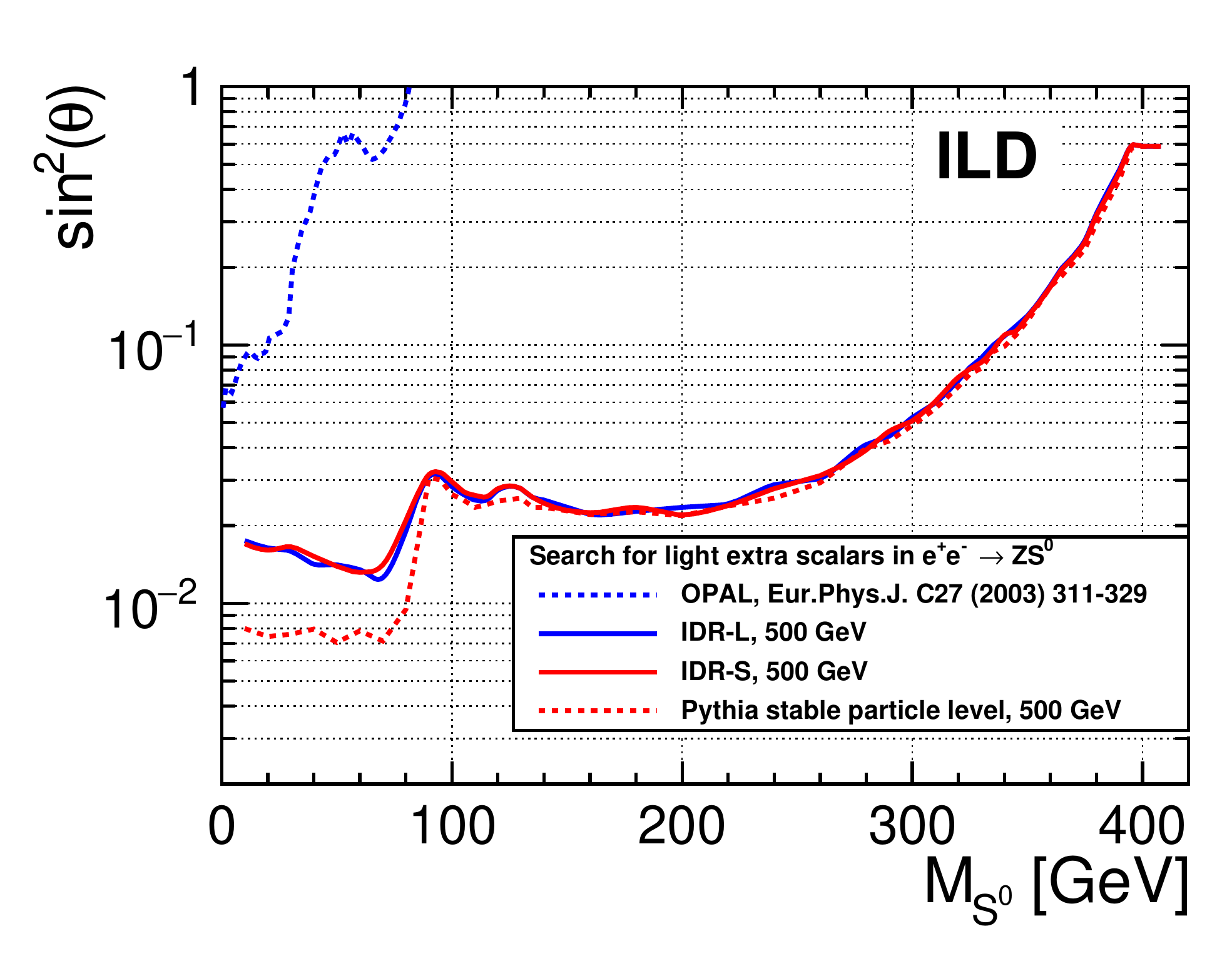}
 \caption{500\,GeV, IDR simulation\label{fig:extraH:limit:500}}
 \end{subfigure}
\begin{subfigure}{0.49\hsize} 
 \includegraphics[width=\textwidth]{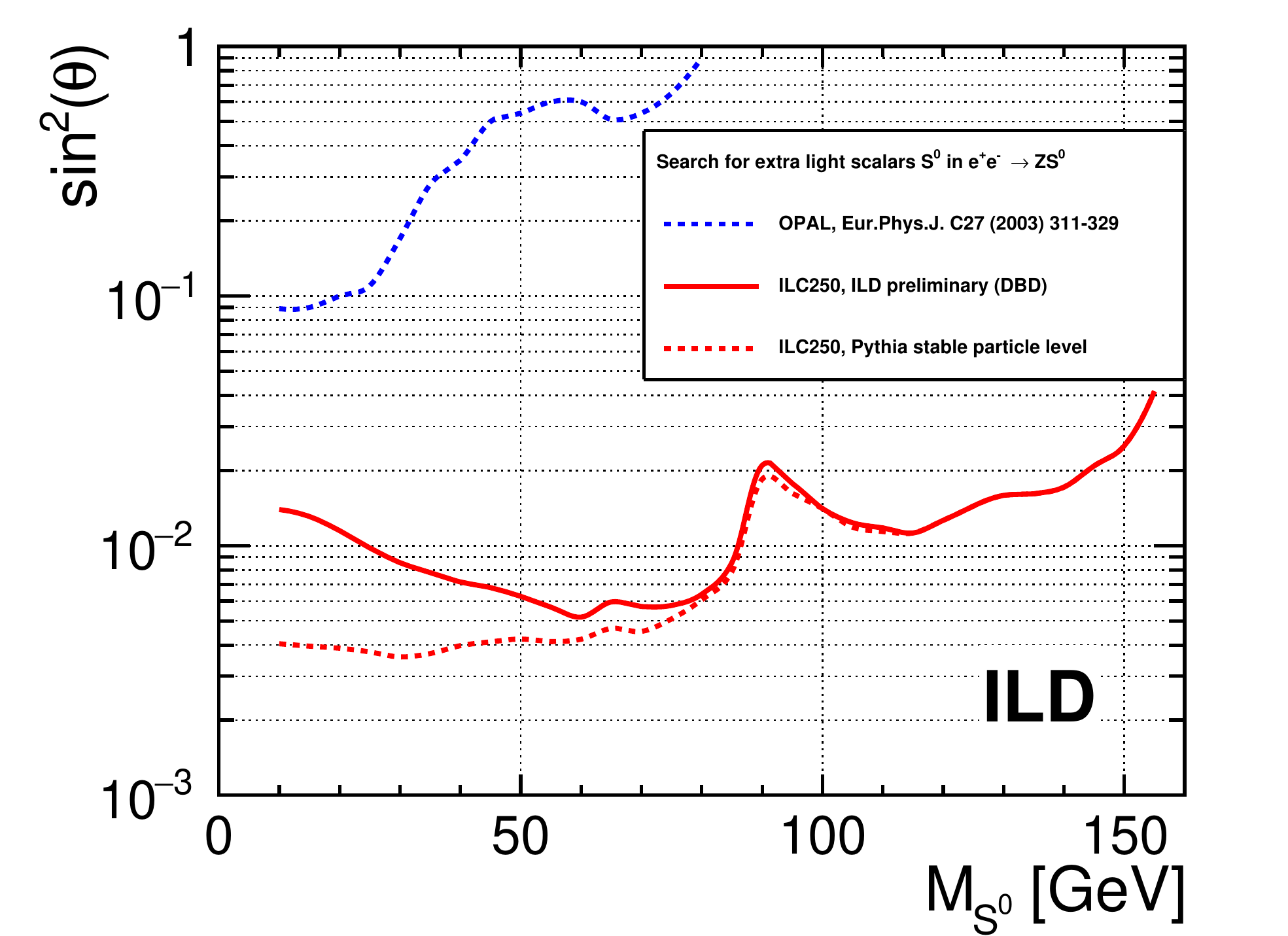}
 \caption{250\,GeV, DBD simulation\label{fig:extraH:limit:250}}
 \end{subfigure}
\end{center}
\caption{Expected sensitivity of (a) ILC500 as defined in Sec.~\ref{sec:benchmarks:lep} at the $95\%$ CL for IDR-L and IDR-S compared to the existing limit from LEP~\cite{Abbiendi:2002qp}. (b) Corresponding result for the ILC run at $\sqrt{s}=250$\,GeV (2\,ab$^{-1}$, $|P(e^+,e^-)|=(30\%,80\%)$, $f(-+,+-,++,--) = (45\%,45\%, 5\%, 5\%)$), which probes significantly smaller mixing angles $\sin^2{\theta}$ for scalar masses below 140\,GeV~\cite{FIPnote:ESU_BSM}.
In both cases, the difference between the ``Pythia stable particle level'' (i.e.\ MC truth after hadronisation) and the full reconstruction results at small masses is mainly due to imperfect identification of ISR photons.}
\label{fig:extraH:limit}
\end{figure}

Figure~\ref{fig:extraH:limit:500} shows the resulting $95\%$ CL sensitivity of ILC500 as defined in Sec.~\ref{sec:benchmarks:lep} on the mixing of the new scalar with the SM-like Higgs boson, $\sin^2{\theta}$, as a function of the scalar mass $M_{S^0}$. Also shown is the corresponding observed limit by the OPAL collaboration~\cite{Abbiendi:2002qp}. Since the OPAL result is based on a combination of the $Z\to e^+e^-$ and $Z\to \mu^+\mu^-$ channels, the ILD projections have been scaled by a factor $1/\sqrt{2}$ to allow a more direct comparison. This assumes that the electron channel will reach a similar sensitivity than obtained for the case of muons, which is supported by the roughly similar performance of the two channels in the measurement of the total $ZH(125)$ cross section via the recoil technique~\cite{Yan:2016xyx}. 

No significant difference between IDR-L and IDR-S is observed at this level, since the effect of the beam energy spectrum covers the differences in momentum resolution. At low $M_{S^0}$, the performance reached with either IDR-L or IDR-S is somewhat worse than expected from the MC truth after hadronisation (``Pythia stable particle level'') due to imperfect recognition of ISR photons. Note that for scalar masses below about $150$\,GeV, the ILC run at $\sqrt{s}=250$\,GeV probes significantly smaller mixing angles $\sin^2{\theta}$, as can be seen in Fig.~\ref{fig:extraH:limit:250}. Also at $\sqrt{s}=250$\,GeV it has been observed that the detector performance matters most for $M_{S^0}<80$\,GeV, dominated by the ability (of the current reconstruction) to reconstruct and identify ISR photons~\cite{FIPnote:ESU_BSM}.

\subsection{Search for low \texorpdfstring{$\Delta M$}{DeltaM} Higgsinos}
\label{subsec:bench:higgsino}

\begin{figure}[htbp]
\begin{center}
\begin{subfigure}{0.525\hsize} 
\includegraphics[width=\textwidth]{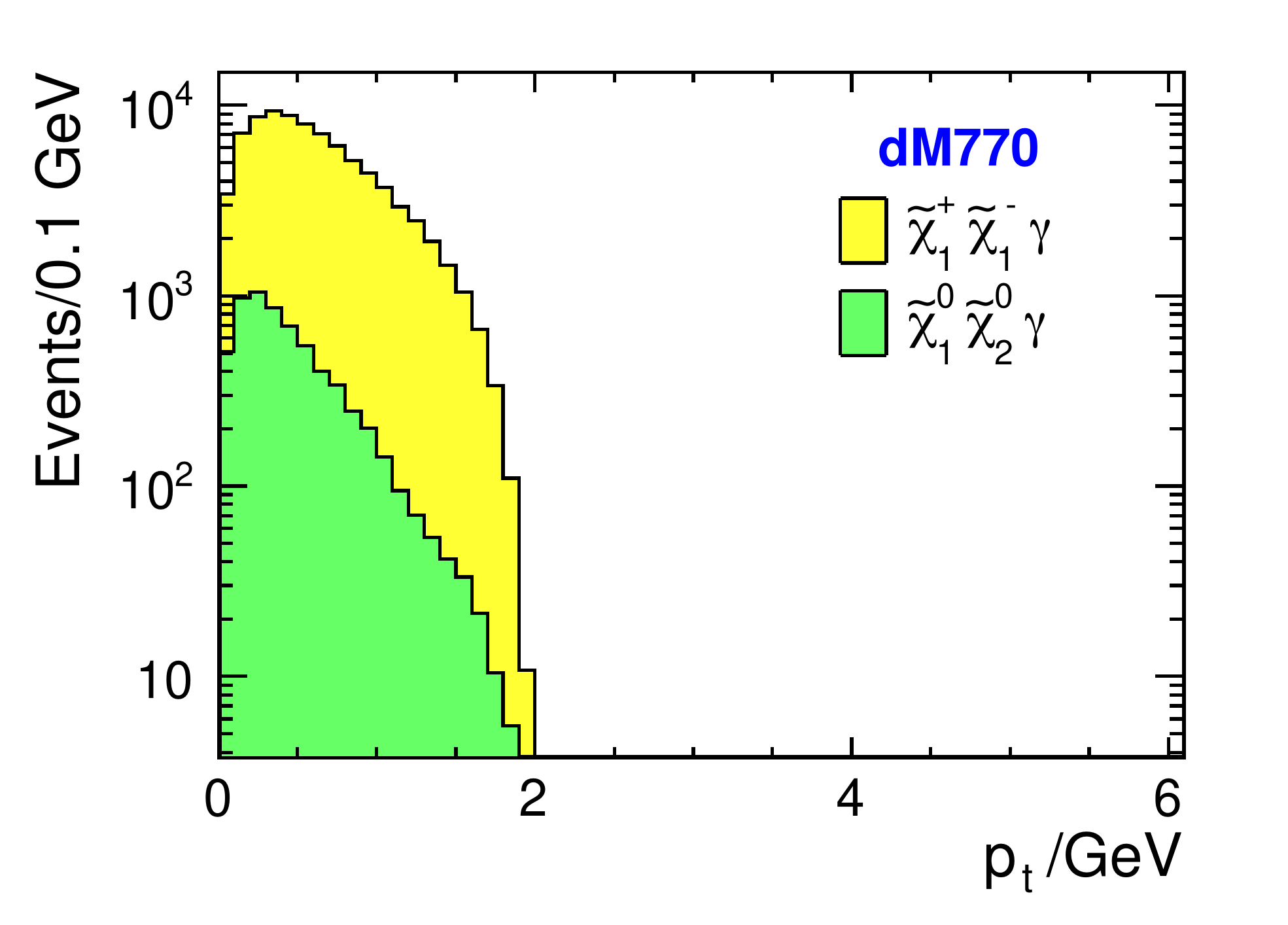}
 \caption{  \label{fig:higgsino:trkeffi:pt}}
\end{subfigure}
\begin{subfigure}{0.455\hsize} 
 \includegraphics[width=\textwidth]{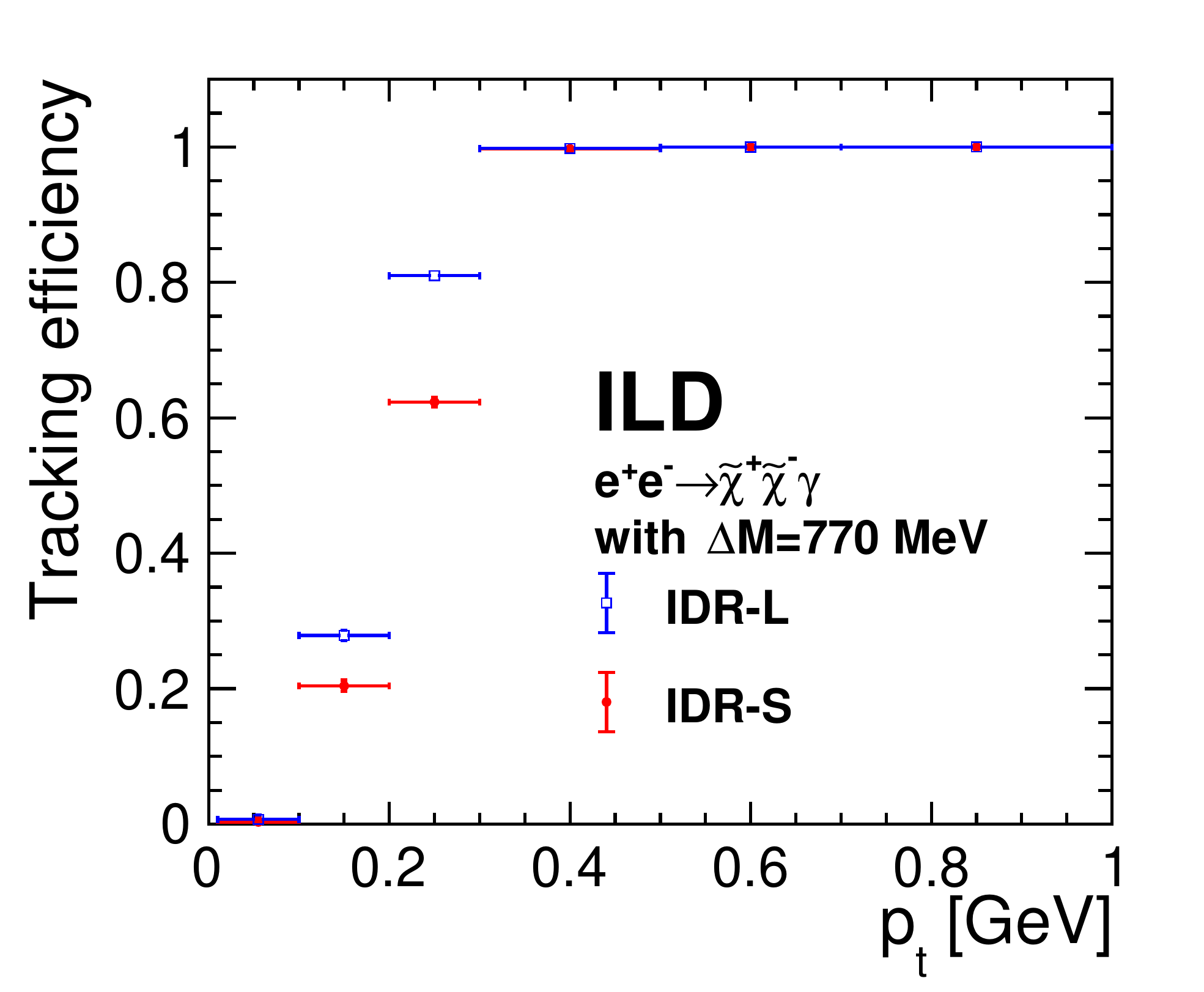}
 \caption{ \label{fig:higgsino:trkeffi:effi}}
 \end{subfigure}
\end{center}
\caption{(a) Transverse momentum distribution of the visible decay products in the $\Delta M =770$\,MeV chargino sample (yellow) and in the corresponding neutralino sample (green) of the same SUSY benchmark at $\sqrt{s}=500$\,GeV~\cite{Berggren:2013vfa}. (b) Tracking efficiency obtained in IDR-L and IDR-S for the chargino decay products as a function of their transverse momentum.
}
\label{fig:higgsino:trkeffi}
\end{figure}

New particles with at most electroweak interactions and small mass differences in the decay chain are among the prime examples of discovery opportunities at future $e^+e^-$ colliders. A prominent example of such kind of
new physics are higgsinos, which are expected to be light in natural SUSY models and tend to have small mass
splittings, in particular when the gaugino masses are much heavier. One of the model points studied previously in fast detector simulation~\cite{Berggren:2013vfa}, with a chargino mass of about 167\,GeV and a mass splitting of only $770$\,MeV between the chargino and the LSP, is used as a detector performance benchmark here. For this mass splitting, the chargino decays to more than $99\%$ into a single charged particle and the LSP. Figure~\ref{fig:higgsino:trkeffi:pt} shows the transverse momentum ($p_t$) distribution of the charged particles in $e^+e^- \to \tilde{\chi}^+_1 \tilde{\chi}^-_1$ events at a center-of-mass energy of $500$\,GeV.

The tracking efficiency  as obtained in these events is displayed in Fig.~\ref{fig:higgsino:trkeffi:effi} as a function of the charged particle transverse momentum. 
For transverse
momenta below $p_t = 300$\,MeV, a clear difference can be seen between the two detector models due to the higher magnetic field of IDR-S. For this specific model point, which lies -- with $2 M(\tilde{\chi}^{\pm}_1) \approx 334$\,GeV -- significantly below the kinematic limit, the efficiency to detect all charged decay products is around $60$\% and differs only by $2$\% between IDR-L and IDR-S, thanks to the sufficient boost of the charginos. For chargino masses closer to the kinematic limit, however, the typical $p_t$ will be even lower, 
and the difference visible in Fig~\ref{fig:higgsino:trkeffi:effi} will lead to a larger effect.
The low $p_t$ tracking efficiency could be recovered at least partially by reducing the radii of the vertex detector layers for IDR-S, which is possible since the higher magnetic field also confines the pair background at smaller radii.

\subsection{WIMP Search in the Mono-Photon Channel}

The search for WIMP production via the mono-photon signature is another example of a search highly complementary to the HL-LHC, which predominantly probes couplings of WIMPs to quarks. This channel has been chosen as a  benchmark specifically in order to test the photon reconstruction and the performance of the BeamCAL, which plays an essential role in vetoing background from Bhabha scattering. The benchmark analysis is described in more detail in~\cite{ILDNote:WIMPs} and closely follows~\cite{Habermehl:2018yul}.
\begin{figure}[htbp]
\begin{subfigure}{0.49\hsize} 
\includegraphics[width=\textwidth]{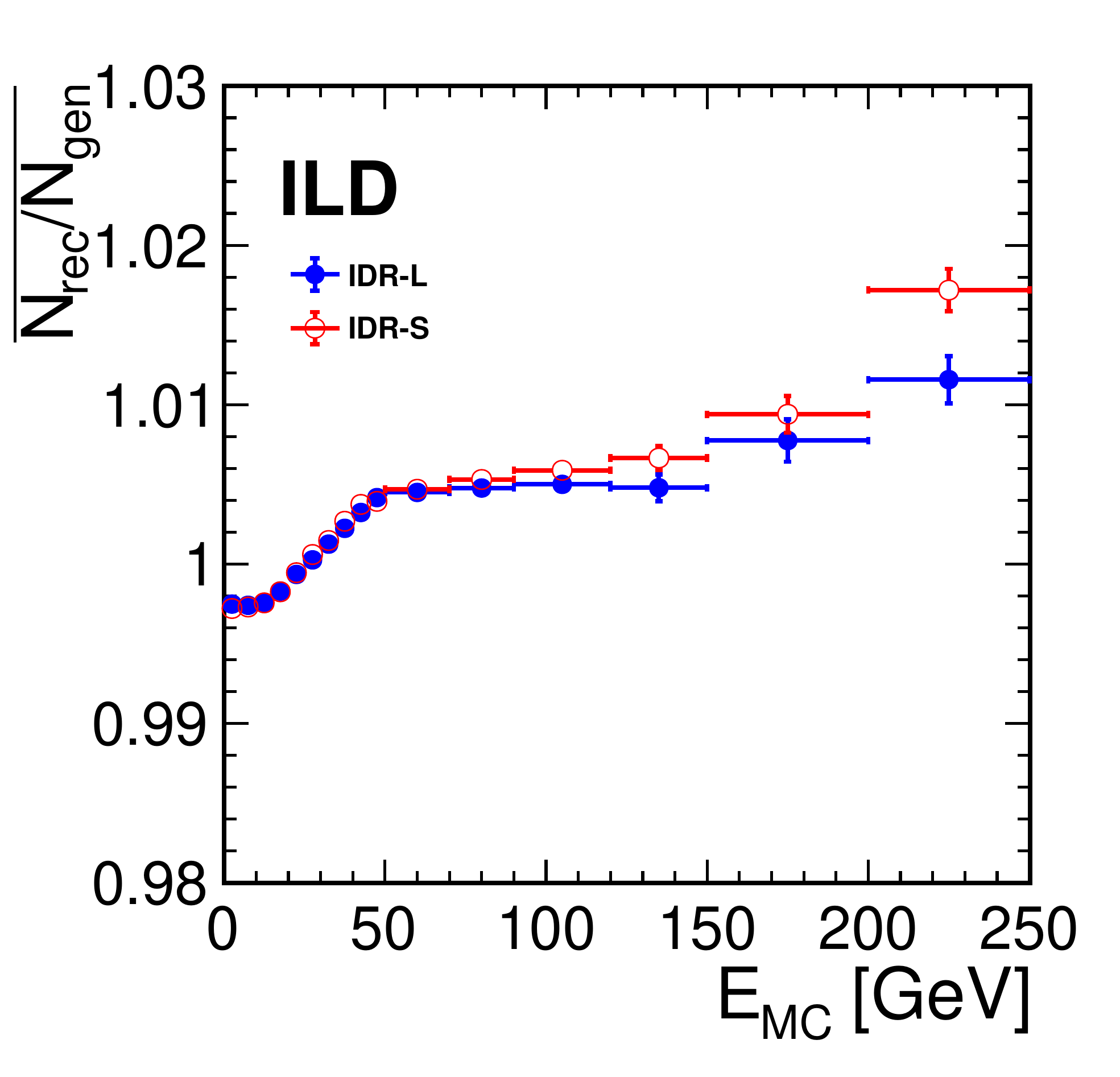}
 \caption{ \label{fig:WIMP:Ngamma:Egood}}
 \end{subfigure}
\begin{subfigure}{0.49\hsize} 
\includegraphics[width=\textwidth]{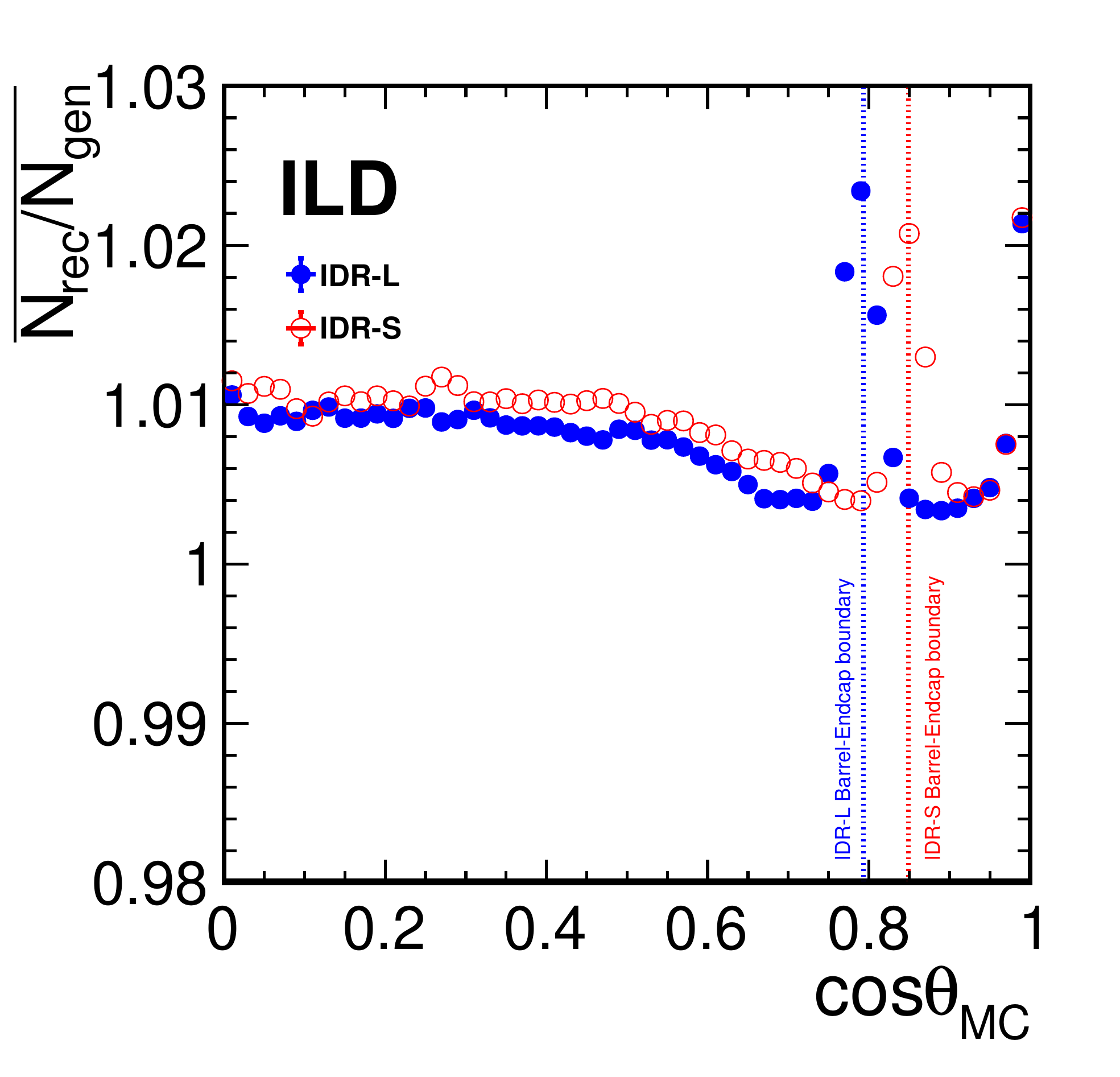}
 \caption{  \label{fig:WIMP:NPFO:theta}}
 \end{subfigure}
\caption{Ratio of recontructed and generated PFOs per $\nu\nu\gamma$ event for ILC500 as defined in Sec.~\ref{sec:benchmarks:lep}
(a) based on PFOs identified as photons as function of the photon energy, excluding the corner region between barrel and endcap ($|\cos{\theta}|\simeq 0.8$)
(b) based on all PFOs as function of the photon polar angle.}

\label{fig:WIMP:Ngamma}
\end{figure}

Figure~\ref{fig:WIMP:Ngamma} shows the ratio of reconstructed and generated photons per $\nu\nu\gamma$ event as a function of
the photon's energy and polar angle. While it has been shown in~\cite{Habermehl:2018yul} that the actual energy
resolution is not very critical in this analysis, it is important that the number of photons is correct, since the signature of a single photon and no other significant detector activity forms the basis of the selection.
As can be seen in Fig.~\ref{fig:WIMP:Ngamma:Egood}, the average number of reconstructed photons per generated photon tends to be a few permille below 1 at low photon energies. With increasing photon energy, the tendency of the particle flow reconstruction to split a photon into two increases, but stays below the level of 1\% for IDR-L and below 2\% for IDR-S. In both cases, the corner region between barrel and endcap has been excluded, because the current reconstruction software tends to split PFOs in this region. This can be clearly seen in 
Fig.~\ref{fig:WIMP:NPFO:theta} which shows the ratio of reconstructed PFOs vs generated particles as a function of $\cos{\theta}$. Also indicated are the locations of the barrel-endcap transition for IDR-L and IDR-S, which occurs at different polar angles due to the different aspect ratios of the two detector models.  All the features in these two plots are expected to be reducable with further development of the particle flow reconstruction.

\begin{figure}[htbp]
\begin{subfigure}{0.49\hsize} 
\includegraphics[width=\textwidth]{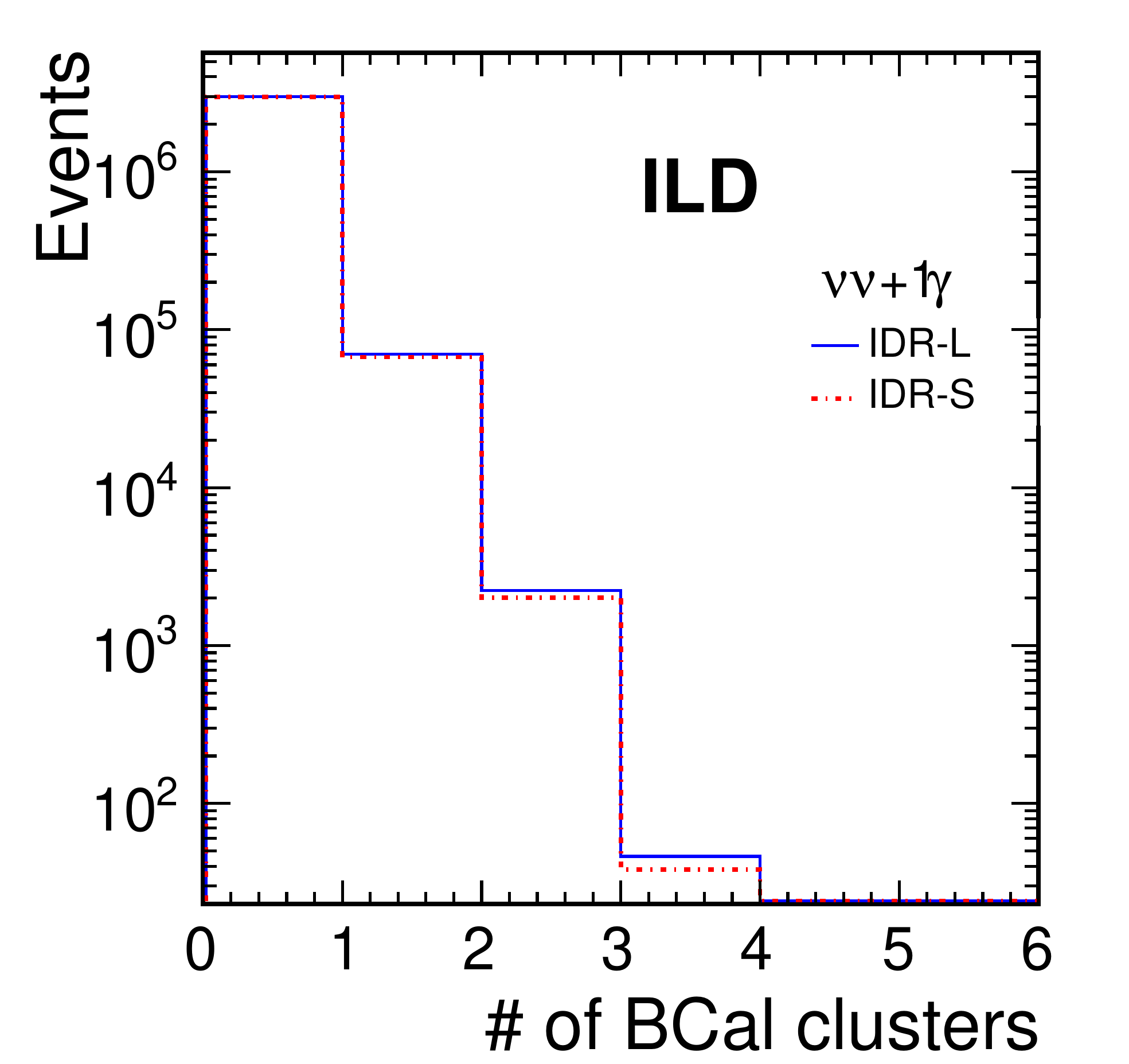}
 \caption{$\nu\bar{\nu}\gamma$ \label{fig:WIMP:BCal:IDR-L}}
 \end{subfigure}
\begin{subfigure}{0.49\hsize} 
\includegraphics[width=\textwidth]{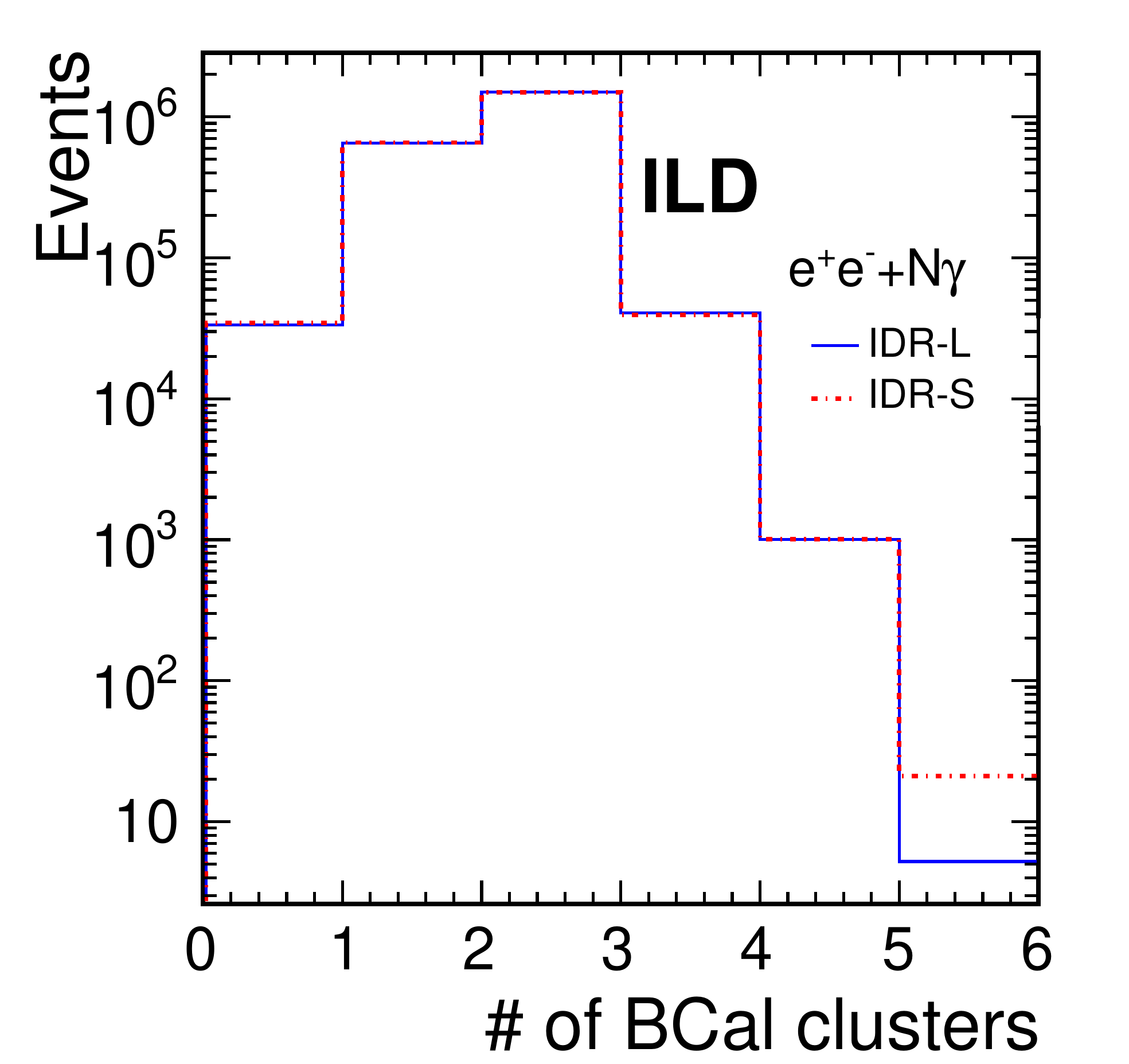}
\caption{$e^+e^- N\gamma$ \label{fig:WIMP:BCal:IDR-S}}
 \end{subfigure}
\caption{Number of BeamCAL clusters reconstructed in radiative neutrino and radiative Bhabha events for IDR-L and IDR-S, for ILC500 as defined in Sec.~\ref{sec:benchmarks:lep}. No significant difference between the detector models is observed.
}
\label{fig:WIMP:BCal}
\end{figure}

Figure~\ref{fig:WIMP:BCal} compares the number of reconstructed clusters in the BeamCal of the two detector models. Ideally, the $\nu\bar{\nu}\gamma$ events, which serve as proxy for the WIMP signal, should have no BeamCal cluster, while for the majority of the $e^+e^-\gamma$ events at least one BeamCal cluster should be found. Events with at least one cluster in the BeamCal will be rejected in the analysis. Due to its higher $B$ field, the background from $e^+e^-$ pairs is confined to somewhat smaller radii in case of IDR-S. This, however, does not lead to a significant effect on the number of reconstructed clusters in the BeamCal.

\begin{figure}[htbp]
\begin{center}
\begin{subfigure}{0.49\hsize} 
 \includegraphics[width=\textwidth]{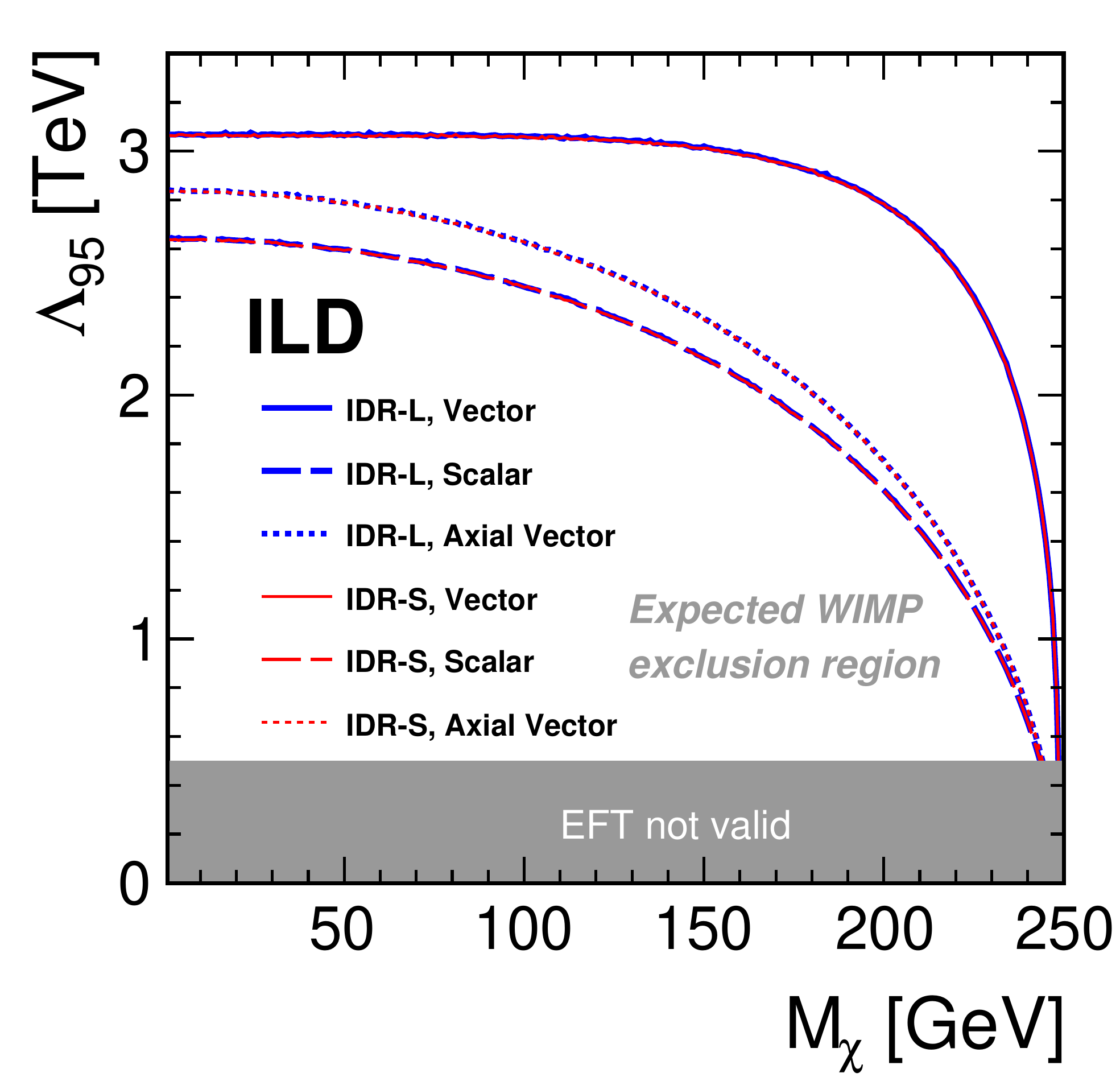}
 \caption{linear mass scale\label{fig:WIMP:limit:std}}
 \end{subfigure}
\begin{subfigure}{0.49\hsize} 
 \includegraphics[width=\textwidth]{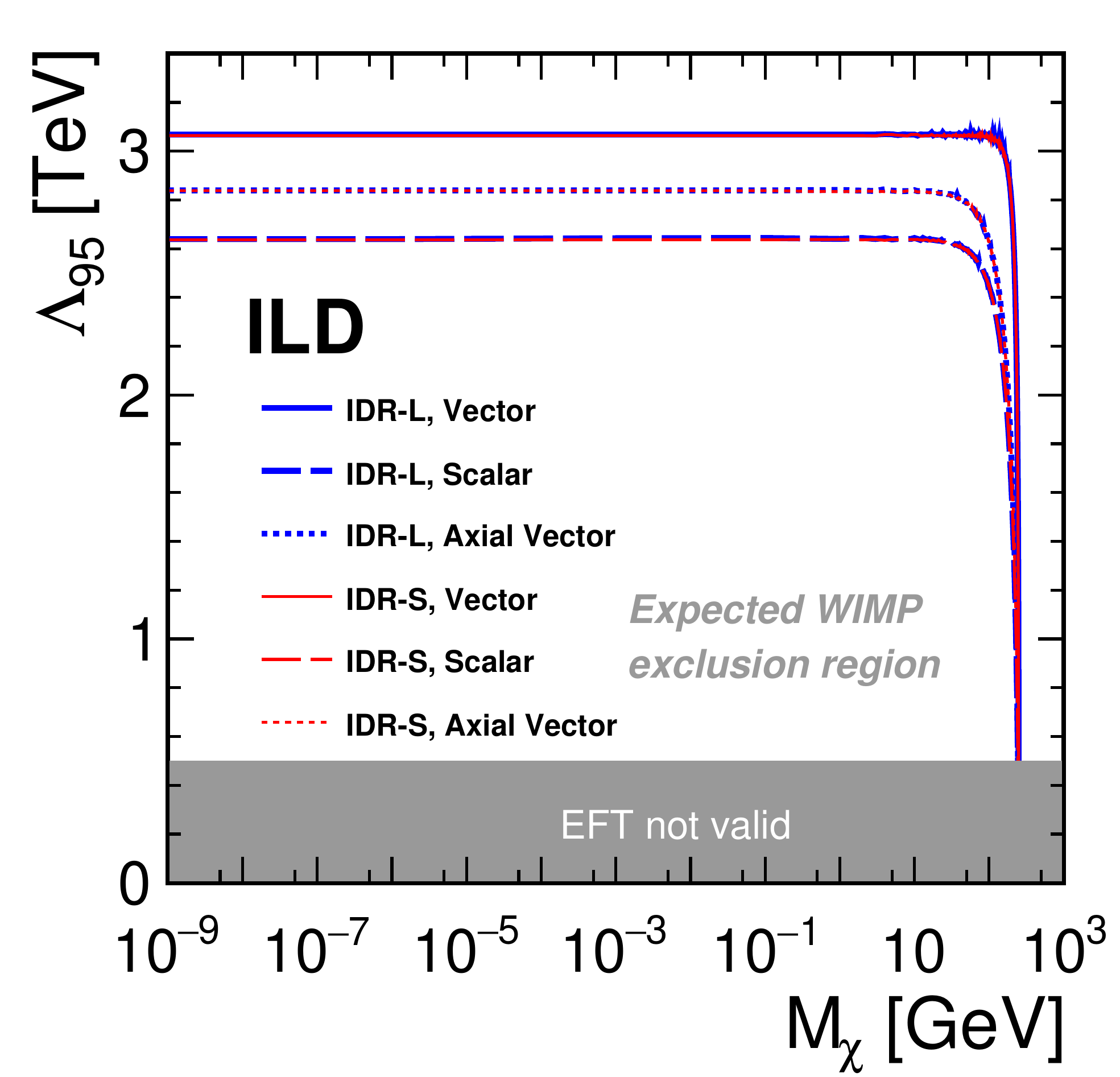}
 \caption{logarithmic mass scale\label{fig:WIMP:limit:lowM}}
 \end{subfigure}
\end{center}
\caption{Expected sensitivity of ILC500 as defined in Sec.~\ref{sec:benchmarks:lep} at $95$\% CL in the plane of the new physics scale $\Lambda$ vs the WIMP mass $M_{\chi}$ for the cases of a vector, axial-vector or scalar operator describing the WIMP-SM interaction.}
\label{fig:WIMP:limit}
\end{figure}

The final result of the analysis is compared in Fig.~\ref{fig:WIMP:limit:std} for the two detector models for scalar, vector, and axial-vector operators for WIMP masses down to $1$\,GeV. At this level, no difference between IDR-L and IDR-S is perceivable. Figure~\ref{fig:WIMP:limit:lowM} extends the projection down to masses of $1$\,eV in order to illustrate the sensitivity to light feably interacting particles.


\chapter{Costing}
\label{chap:costing}
This chapter presents an updated costing of the ILD detector, corresponding to its latest baseline design and dimensions as used in the simulations for performance evaluation. The method is very similar to that used in the costing exercise of the DBD, with two notable differences. Two size options are now costed: the large model IDR-L, very similar to the DBD baseline, and the small model IDR-S where the outer radius of the TPC has been reduced by about 30cm. In addition, the required manpower is now included in the costs, in an attempt to identify the in-kind laboratory manpower necessary to assemble the detector. 

The costing can now benefit from the construction of significant technological prototypes of the main subdetectors (chapter 5), as well as from spin-off detectors starting to be built for e.g. HL-LHC. The cost difference of the two models combined with the differences observed in their respective performances (chapter 8) will allow a better evaluation of the impact of the detector size than the simple scaling laws shown in the DBD. 

\section{The method}
\begin{figure}[h!]
\centering
\includegraphics[width=1.0\hsize]{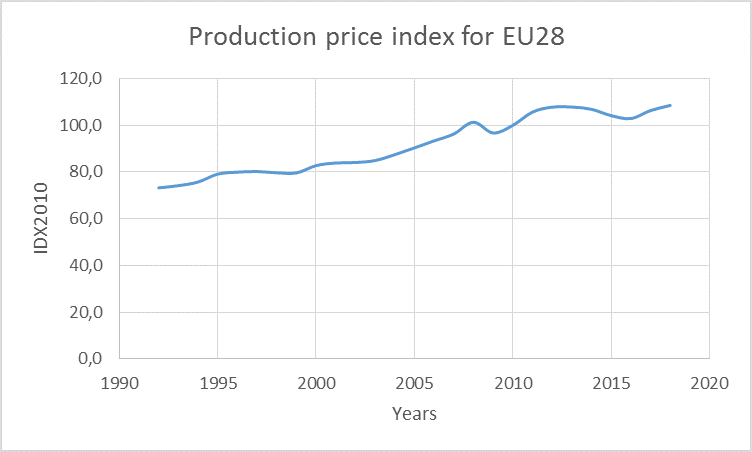}
\caption{Manufactured products price evolution over the 28 countries of UE.}
\label{price_index}
\end{figure}

The DBD costing had been made in an "ILC currency", the ILCU, in an effort to have a costing coherent between ILD, SiD and the accelerator. This implied making translations from different currencies using exchange rates and, in most cases, "Purchase Power Parities" (PPP). At that time an ILCU was 0.97 Euros using PPP's. Within the current exercise Euros(2018) are used as currency unit. When originating from Japan, e.g. for silicon diode matrices, prices in Euros were provided by the vendors. Extrapolation of the DBD estimates to the present is made by converting the DBD ILCU's to Euros(2013) using PPP's as was done at that time (0.97 Euros for 1 ILCU), and propagating the cost to 2018 using as inflation rate the evolution of manufactured products in Europe, amounting to 3\% between 2013 and 2018, as shown in figure~\ref{price_index}. The two factors compensate and 1 DBD ILCU turns out to be equivalent to 1 Euro(2018). 

Except for the currency, the method used to establish the costing is totally similar to that of the DBD. It rests on the detailed knowledge of the fabrication processes and on the prices provided by the numerous prototypes built over the past years.  The idea is to identify the cost drivers, often very sensitive to strong price evolution, and to have a precise Work Breakdown System (WBS) identifying the procurements, the tooling and the fabrication operations. The WBS allows to estimate the manpower, which is now included in the costing. The manpower is twofold: in house laboratory  manpower mostly linked to the follow-up of the operations, but also to some construction work when high quality is required for small quantity items; industrial manpower which has to be estimated on top of the material costs in case no industrial offer is available for a given component. The manpower costs are converted into Euros(2018) assuming 80k\texteuro~per FTExYear as a mean unit cost over the different types of competences. The industrial manpower is incorporated into the material costs, whereas the in house manpower is shown separately as an estimation of the in-kind contributions of the participating institutes. 
It should be noticed that spares are most of the time not included. More generally no contingency is applied.

\section{Subdetector costing}
The cost of each ILD subdetector is reviewed for the IDR-L and IDR-S models. The inner and forward detectors which have the same configuration in both options (VTX, SIT, FTD, ECAL ring, LumiCAL, LHCAL, BeamCAL) have only one quote, whereas others (TPC, ECAL, HCAL, Coil, Yoke and Iron Instrumentation) are costed for their two size versions. The subdetector baseline costing is made for the layouts described in section 5.1.2 and used in the performance evaluations of chapter 8. In some cases a costing is also estimated for modified or upgraded designs corresponding to the ongoing studies reported in section 5.2.   

The most expensive subdetector, the SiECAL, has received special attention in updating its costing of both material and manpower contributions, based on an updated detailed WBS. Some other subdetectors have fewer new pieces of information. For those which have only material costs available, the manpower costing has been estimated assuming the same manpower/material ratio as for the SiECAL. When no update is available w.r.t the DBD version, the costing is propagated from the DBD estimate and simple scaling laws are used for the small version.

In the following, when a cost estimate appears as two numbers inside brackets, the first number corresponds to the material cost and the second one to the in house manpower.

\subsection{VTX}
The Vertex Detector costing has been fully revisited 
for its CMOS option based on recent detectors built for various experimental projects (see section 5.2.1). The results are summarised in  table~\ref{vertex_cost}. The cost for this option is (2.96, 1.45) M\texteuro. A DEPFET option has also been reexamined leading to a similar cost of (3.44, 1.5) M\texteuro.



\begin{table}\hspace*{-0cm}\small
\begin{tabular}[h!]{ l p{0.1\hsize}p{0.1\hsize}p{0.1\hsize} p{0.1\hsize}p{0.1\hsize}p{0.1\hsize} }
\toprule
\multicolumn{7}{ l }{{\bf Vertex detector}}\\
\midrule
Cost   & Sensors & Mechanics & Electronics & Services & Installation & Total \\
\midrule
Material    & 1.15   &  0.45   &  0.49    & 0.77 & 0.10 & 2.960 \\
Manpower    & 0.10   & 0.50    & 0.40     & 0.25 & 0.20 & 1.450 \\
\midrule
Total      & 1.25   &  0.95   &  0.89    & 1.02 & 0.30 & 4.410 \\
 \bottomrule
\end{tabular}
\caption{\label{vertex_cost}Elements of cost of the vertex detector (CMOS option) in M\texteuro.}
\end{table}


\subsection{SIT}
In its baseline design used for estimating the physics potential of the detector, the SIT is made of strips as in the DBD. The cost of this strip version could be estimated similarly to the SET (see later), using an areal extrapolation from the CMS tracker tempered by a reasonable account of items which do not scale with surface. This extrapolation is well compatible with the costs of the smaller silicon strip beam telescope LYCORIS recently built for the DESY TPC test setup (section 5.2.3). The resulting estimated cost of the baseline strip version of the SIT is (5, 1.5)M\texteuro.

The SIT costing has also been revisited in detail in a version based on CMOS pixels. The inputs are the same as used for the VTX CMOS costing,  
and the results summarised in table~\ref{SIT_cost}. It can be observed that the total cost of (7.55, 2) M\texteuro~ of the pixel SIT does not scale with the area compared with the VTX: the cost factor is about 3 for an area factor of about 20.


\begin{table}\hspace*{-0cm}\small 
\begin{tabular}[h!]{ l p{0.1\hsize}p{0.1\hsize}p{0.1\hsize} p{0.1\hsize}p{0.1\hsize}p{0.1\hsize} }
\toprule
\multicolumn{7}{ l }{{\bf Silicon Inner Tracking}}\\
\midrule
Cost   & Sensors & Mechanics & Electronics & Services & Installation & Total \\
\midrule
Material    & 3.82   &  0.76   & 1.28    & 1.58 & 0.11 & 7.55 \\
Manpower    & 0.20   & 0.50    & 0.80    & 0.30 & 0.20 & 2.00 \\
\midrule
Total      & 4.02   &  1.26   &  2.08    & 1.88 & 0.31 & 9.55 \\
\bottomrule
\end{tabular}
\caption{\label{SIT_cost}Elements of cost of the SIT (CMOS pixel option) in M\texteuro.}
\end{table}

\subsection{FTD}
There is no full update since the DBD. The current FTD baseline configuration (chapter 5) includes 4 disks using pixel technology and 10 disks using strip technology. The cost of the sole 4 pixel disks in the CMOS version is estimated to (1.1, 0.2) M\texteuro, whereas an option using DEPFET has been estimated to (2.05, 1.5) M\texteuro.
The cost of the strip disks is inferred to be (3.7, 1.1) M\texteuro~ similarly as for the strip SIT. The global cost for the baseline option is then of the order of (4.8, 1.3) M\texteuro. 

A direct comparison to the DBD estimate is possible only at the level of the global inner silicon tracking (SIT + FTD). The DBD was quoting 2.3 M\texteuro~ without manpower when the new estimate is (9.8, 2.8) M\texteuro.

The cost of a fully pixelised FTD made of 14 disks with CMOS pixels is estimated similarly to the pixel SIT case to (9.1, 1.7) M\texteuro.


It has to be noted that the cost of the structure (called ISS) holding the beam tube, the vertex detector, the SIT and the FTD has not been estimated.

\subsection{Forward Calorimetry (FCAL)}

Three forward calorimeters, namely the BeamCAL, the LumiCAL and the LHCAL, have been reexamined in detail in their new configuration with the new beam optics (see chapter \ref{chap:technologies:fcal}). The updated costing includes consideration of the mechanical elements, sensors, ASICs, front-end electronics, power supplies, data acquisition, tooling and manpower, separately for each of the subdetectors. Ancillary systems such as specific fanouts (for LumiCAL and LHCAL) and a laser positioning system needed for LumiCAL are also included. The resulting costs are summarised in table~\ref{FCals_summary}. They correspond to a total of 8.44 M\texteuro~ and 6 FTE Years equivalent to 0.48M\texteuro.


The forward region also includes an ECAL ring making the transition from the LumiCAL to the ECAL endcap. Its costing has been updated based on a rough silicon area extrapolation from the SiECAL costing (see below). The ECAL rings silicon area is about  1\% of the full SiECAL, barrel plus end-caps. The electronics is more similar to that of the LumiCAL and the manpower is taken identical to that needed for the LumiCAL.  The estimated cost of the two rings is then (1.5, 0.16) M\texteuro.  

The whole forward calorimetry, including BeamCAL, LumiCAL, LHCAL and ECAL ring, and referred to as FCAL in Figures ~\ref{fig:det:DBD_cost_sharing} and ~\ref{Costing:Small_cost_sharing}, is then estimated to (10, 0.6) M\texteuro. The cantilevered beam structure holding the QD0 focussing quadrupoles but also the forward calorimetry has not been costed in the detector part.


 \begin{table}\hspace*{-0cm}\small 
\begin{tabular}[h!]{ l p{0.2\hsize}p{0.2\hsize}p{0.2\hsize} }
\toprule
& BeamCAL & LumiCAL & LHCAL \\
\midrule
Mechanics              & 0.65   & 0.64   & 0.99   \\
Connectivity           &        & 0.09   & 0.14   \\
Sensors                & 0.90   & 1.60   & 1.20  \\
Laser system           &        & 0.07   &       \\
Front-end ASICs        & 0.18   & 0.28   & 0.22   \\
Front-end electronics  & 0.10   & 0.15   & 0.05  \\
Power supplies         & 0.08   & 0.17   & 0.08    \\
Data acquisition       & 0.22   & 0.34   & 0.22  \\
Tooling                & 0.03   & 0.03   & 0.03   \\
\midrule
Total                  & 2.15  & 3.37   & 2.93  \\
\midrule
Manpower (FTE x Year)  &2       &2       & 2 \\
\bottomrule
\end{tabular}
\caption{\label{FCals_summary}Elements of cost of the forward calorimeters in M\texteuro~ and manpower in FTE x Year.}
\end{table}

\subsection{TPC}
The quoted TPC price of the DBD corresponds to 36 M\texteuro(2018) including $\approx$5 M\texteuro~ of manpower. The TPC dimensions  were very close to those of today's large IDR-L option of ILD.


Since then two TPC projects with similar technologies are progressing towards construction : ALICE with GEM readout, and T2K/ND280 High Angle with resistive anode Micromegas readout. The T2K/ND280 project has allowed to update the TPC costing in the Micromegas option with many details, including manpower and taking into account requirements imposed by Japanese regulations. Similarly the ALICE project has allowed an update for the GEM option. The construction of a new field cage for the DESY TPC test setup (section \ref{chap:technologies:tpc}) has also strongly modified the estimate of the field cage cost. The costs can be split into a technology-independent part and a technology-dependent part.

The technology-independent part comprises the electrostatic system for 1.8 M\texteuro~ (including test and shipping), the TPC hanging system for 0.2 M\texteuro~, the ancillaries for 2.7 M\texteuro~ (CO$_2$ compressor, TPC gas system, laser system, power supplies, their packing and shipping) and the overall management for 1.2 M\texteuro. The total technology-independent cost estimate is therefore (4.8, 1.1) M\texteuro.

The technology-dependent part consists of everything which is sensitive to the pad size and the module size. It includes the end plates, the readout modules, the electronics, the cables and pipes as well as the assembly. The manpower is estimated to be equal in the two solutions. The Micromegas option costing is currently estimated to (11.6, 3.9) M\texteuro, whereas the GEM option amounts to (20.4, 3.9) M\texteuro.

The updated estimate of the total TPC cost is therefore (16.4, 5) M\texteuro~ in the Micromegas option and (25.2, 5) M\texteuro~ in the GEM option. 
The largest estimate of (25.2, 5) M\texteuro~ is conservatively kept as the baseline cost. The IDR-L estimate is scaled down to (19, 3.6) M\texteuro~ for the IDR-S version.

\subsection{SET}
For the DBD the external silicon tracker, SET and ETD, was costed globally. The endcap part, the ETD, is no longer part of the ILD design. Considering that the cost should be roughly proportional to the detector area and that the number of silicon layers was three in the ETD and only two in the SET, the  cost of the SET can be inferred from the DBD to be around 13.4 M\texteuro~(2018) manpower excluded.

An estimate of the barrel part (SET) can now be more accurately derived from the cost of the CMS tracker which is also made of double-layer silicon strips. The CMS tracker has a total instrumented area of 220 $m^2$, with a cost of about 275kCHF per $m^2$ of detector, including silicon sensors, front-end electronics, cooling and cabling as well as mechanics. The ancillaries such as back-end electronics, cooling plant and power supplies account for 16MCHF globally. This can be considered  as 73kCHF per $m^2$ which add to the 275 to give 306 k\texteuro~ per $m^2$.
The SET area in the IDR-L ILD model is about 52.9 $m^2$. Using the CMS areal cost results in a value of about 16 M\texteuro~which, taking into account items which do not scale with surface, can be converted into a conservative SET material cost estimate of 20 M\texteuro. The in house manpower is estimated to 2 M\texteuro~ providing a total of 22 M\texteuro.
Using a simple scaling law the IDR-S SET is then estimated to (16, 1.5) M\texteuro~ for an area of 42.7 $m^2$. 

\subsection{ECAL}
\subsubsection{Silicon option of the ECAL (SiECAL)}
The DBD SiECAL costing corresponded to 157 M\texteuro(2018) excluding manpower, with about half of the cost due to the silicon diode matrices.

The updated costing of the SiECAL is based on a very detailed WBS informed from the fabrication processes of the SiECAL technological prototypes (chapter 5). This WBS follows a detailed and chronological fabrication description with the procurements for the different operations, the needed tooling and the fabrication steps, including manpower and duration, under the constraint that none should have a duration longer than two years. Most of the tooling has been kept at the same cost for the different ILD size options, which is slightly at the advantage of the large model for comparison of the costings between the different sizes.

Apart for a slight tungsten cost rise, the main difference with the DBD estimate comes from the evolution of the diode matrices estimation. In the updated costing the offer to CMS for the HGCAL matrices is used. The resulting total SiECAL cost reaches 119 M\texteuro~ excluding manpower. The in house manpower has also been estimated to 131 FTExYears, equivalent to 11 M\texteuro, which results in a total amount of 130 M\texteuro. It should be noted that possible future improvements such as a high timing resolution have not been taken into consideration for this costing.

Using similar inputs the small SiECAL version is costed to 92 M\texteuro~ for material. Including 9 M\texteuro~ for manpower, the total cost is 101 M\texteuro. 
A version of the small SiECAL with a slightly coarser sampling has also been estimated. For this version the number of active layers is reduced from 30 to 26 (section 5.2.4.1). The motivation for such a configuration is dictated by technical reasons rather than cost, but it is worth noting that the impact of the reduced sampling on the energy resolution is expected to be compensated by an increase of the silicon thickness to 725 micrometers. The reduced sampling provides a further cost reduction to (81, 8) M\texteuro.  

\begin{figure}[h!]
\begin{tabular}{ccc}
\includegraphics[width=0.3\hsize]{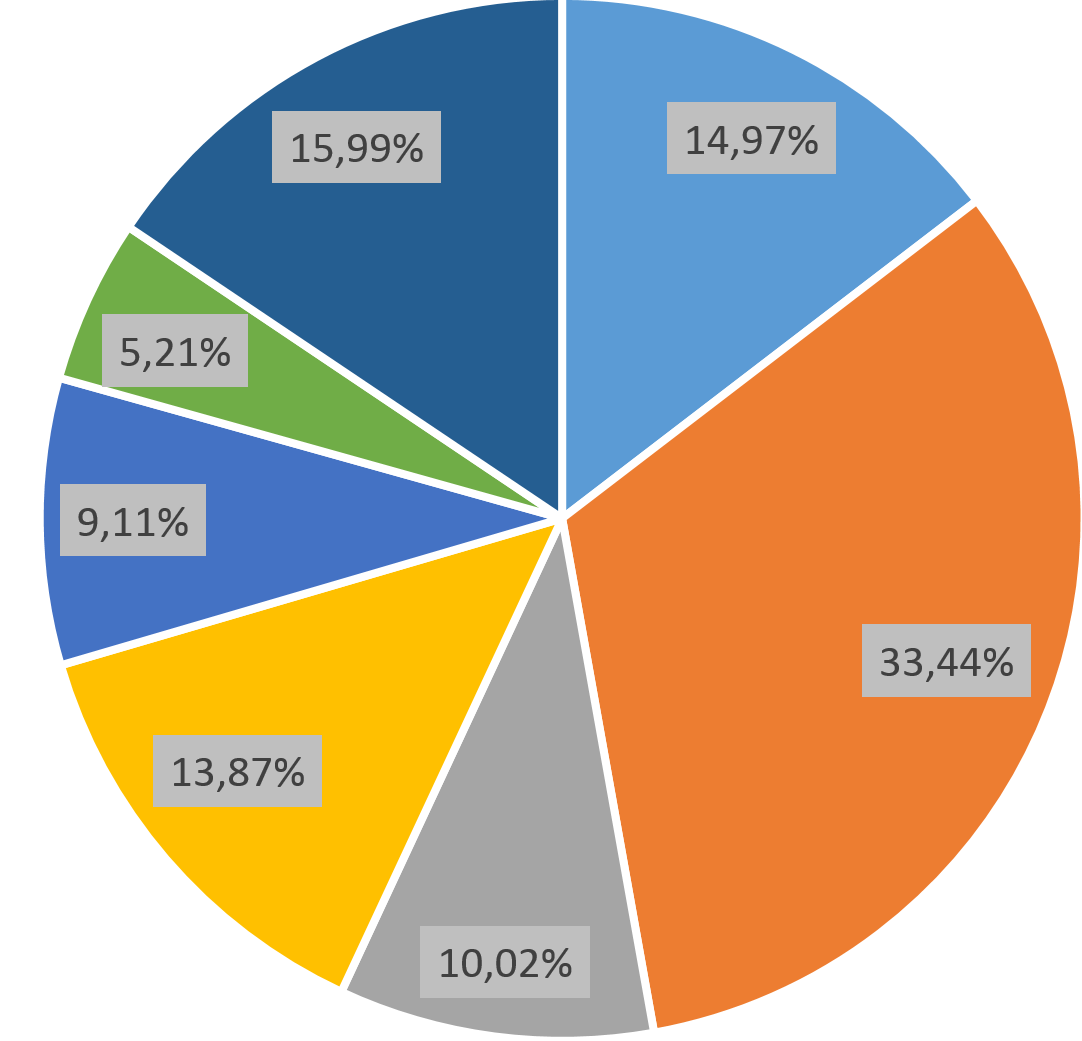}&
\includegraphics[width=0.3\hsize]{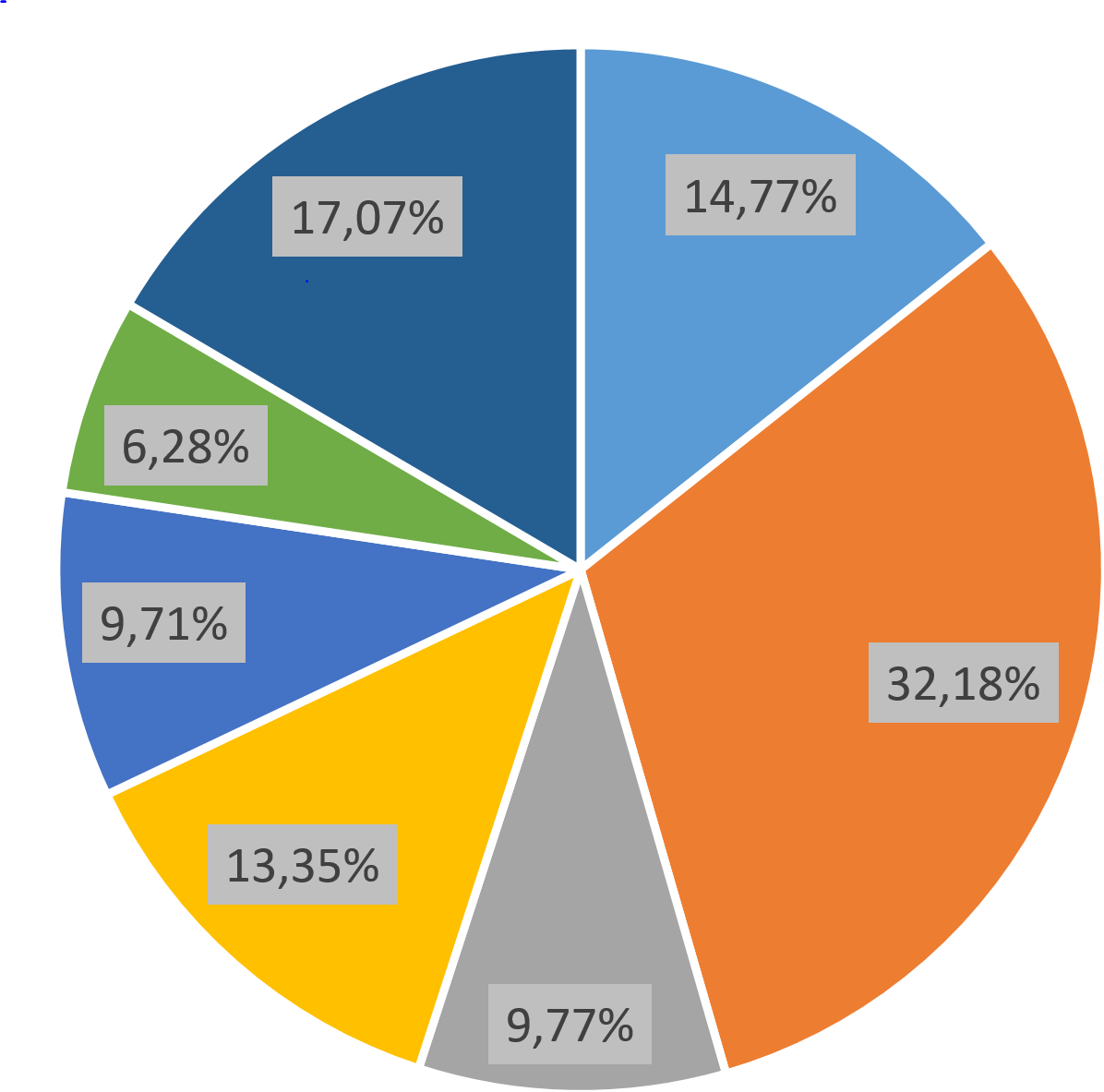}&
\includegraphics[width=0.3\hsize]{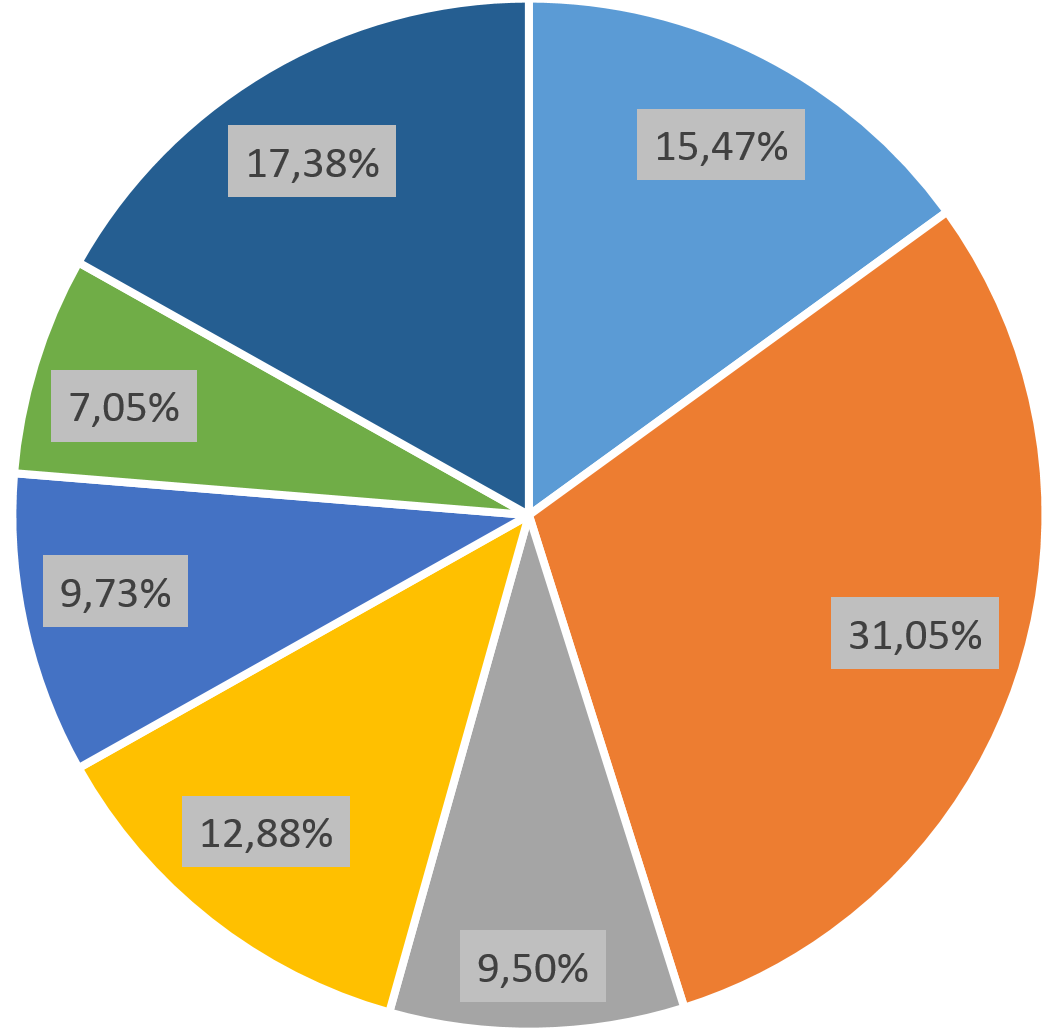}
\end{tabular}
\includegraphics[width=0.8\hsize]{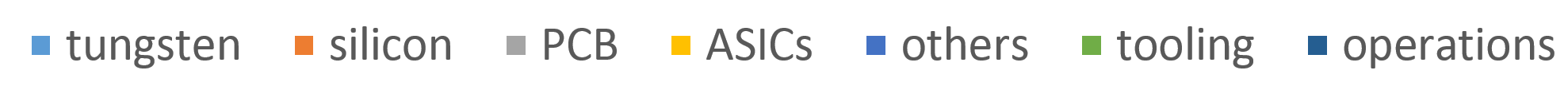}
\centering
\caption{Contributions of the different items to the SiECAL cost in the case of the IDR-L (left), IDR-S (middle), IDR-S with reduce sampling (right). The item "operations" includes the in-house manpower discussed in the main text as well as operations by the sub-contractors.}
\label{fig:det:ECal_Si_cost_sharings}
\end{figure}

The various global costings are summarised in table~\ref{ECal_summary}. A breakdown of the contributions of the main items to the cost is shown in Figure~\ref{fig:det:ECal_Si_cost_sharings} for the three models. The silicon matrices cost still dominates the procurement, but other items such as tungsten, printed boards and ASICs also represent significant parts. The in-house manpower is around 10\% of the cost.

\begin{table}\hspace*{-0cm}\small 
\begin{tabular}[h!]{ l p{0.2\hsize}p{0.2\hsize}p{0.2\hsize} }
\toprule
Version& Material & Manpower & Total \\
\midrule
IDR-L                       & 119   & 11    & 130   \\
IDR-S                       & 92    &  9    & 101   \\
IDR-S with reduced sampling & 81    &  8    & 89  \\
\bottomrule
\end{tabular}
\caption{\label{ECal_summary}Estimates for the three versions of the SiECAL (M\texteuro).}
\end{table}

\subsubsection{Scintillator version of the ECAL (ScECAL)}
The ScECAL material costing from the DBD is considered to be still valid and corresponds to 74 M\texteuro(2018) for the IDR-L version. The in-house assembly manpower has since then been estimated to 11.5 M\texteuro. Applying simple scaling laws the ScECAL IDR-S version is costed to (62.5, 9.5) M\texteuro.

\subsection{HCAL}

\subsubsection{Analog option (AHCAL)}
Since the DBD the application of the SiPM-on-tile technology to the CMS endcap calorimeter upgrade allows to consolidate the costing further. With about 400000 channels the CMS HGCAL represents an intermediate scale between the 22000 channel AHCAL prototype and the ILD full AHCAL. 
The cost envelopes for the CMS SiPMs are consistent with the scaling assumed for ILD at the DBD times and so far confirmed in informal contacts with vendors. The AHCAL material costing from the DBD is therefore considered to be still valid. It amounts to 44.9 M\texteuro(2018) for the IDR-L version. 
For the IDR-S model, a simple scaling corresponds to a factor 0.84 with a resulting material cost of 37.7 M\texteuro.

No detailed manpower cost estimate has been done. Assuming the same manpower/material ratio as for the SiECAL (9\%) the in-house assembly manpower is estimated to 4 M\texteuro~ for the IDR-L model and 3.4 M\texteuro~ for the IDR-S model.

\subsubsection{Semi-digital option (SDHCAL)}

The SDHCAL material costing from the DBD is considered to be still valid and corresponds to 44.8 M\texteuro(2018) for the IDR-L version. Applying simple scaling laws the SDHCAL IDR-S version is costed to 37.6 M\texteuro~ for material.

No detailed manpower cost estimate has been done. Assuming the same manpower/material ratio as for the SiECAL (9\%) the in-house assembly manpower is estimated to 4 M\texteuro~ for the IDR-L model and 3.4 M\texteuro~ for the IDR-S model.

Note: since the costs for the AHCAL and SDHCAL options are very close to each other they will no longer be distinguished in the HCAL contribution to the global ILD costing.

\subsection{Magnet}
The costing of the magnet (coil and yoke) has been fully revisited. As for the DBD the anti-DID option (section 6.4) is not taken into account since its configuration is not fully defined. The main source of the current evaluation consists of the documentation of the CMS magnet, which has a similar size as that of ILD. When extrapolating to ILD the CHF costs from CMS have been converted into Euros at the exchange parity of the CMS magnet construction time, and the manufactured products price evolution in Europe to 2018 taken into account. The resulting costs are summarised in table~\ref{magnet_cost}. One important change compared to the DBD estimation is the strong reduction in the yoke cost. At the time of the DBD, an iron price of 6ILCU/kg had been agreed upon with SiD and CLIC, whereas the price payed for CMS corresponds to 3.5 \texteuro(2018)/kg. 
In summary the magnet system cost is estimated to (88.5, 9.4) $\approx$ 98 M\texteuro.

No attempt to cost the IDR-S option has been made. A priori the cost of the small version is expected to be reduced. However the impact of the higher nominal field of 4T on the coil winding and on the flux return is not trivial and may counteract the reduced sizes. Keeping the small magnet cost similar to the large one may be a good approximation for the time being.

The magnet costing is likely to still evolve significantly in the future. For the coil, industrial offers are expected from the Japanese companies which are currently studying it, including the anti-DID option. For the yoke, updated designs may significantly reduce the cost, by up to 50\%, as discussed in section 6.4. The yoke final cost will strongly depend on the final stray field constraints retained for ILD, as well as on the evolution of the iron market prices until ILD construction.


\begin{table}\hspace*{-0cm}\small 
\begin{tabular}[h!]{ l p{0.2\hsize}  p{0.1\hsize} }
\toprule
\textbf{Magnet system} & \textbf{88.5}\\
\midrule
\textbf{Coil} & \textbf{28.5}\\
\midrule
Conductor and winding & 22\\
Internal cryogenics and suspension &  0.2\\
Suspension system & 0.3 \\
Tooling, assembling & 5.4 \\
Qualification and partial testing & 0.6\\
\midrule
\textbf{Ancillaries for coil} & \textbf{11.5}\\
\midrule
Cryogenics and vacuum & 6.5\\
Electrical power circuits & 0.9\\
Control and safety systems& 1.4 \\
Engineering (transport to cavern) & 2.1\\
Integration in cavern& 0.3 \\
Field mapping&0.3\\
\midrule
\textbf{Yoke and vacuum tank} & \textbf{48.5}\\
\midrule
Yoke steel including works and vacuum tank& 39.6\\
Support &1.2\\
Moving system& 2.6\\
Assembly& 4.8\\
Photogrammetry and survey& 0.3 \\
\bottomrule
\end{tabular}
\caption{\label{magnet_cost}Elements of cost of the magnet system in M\texteuro.}
\end{table}


\subsection{Iron yoke instrumentation}
There is no update since the DBD, where the quoted iron instrumentation material cost corresponded to 6.5 M\texteuro(2018) for instrumentation of the 14 layers of the large IDR-L version. Using simple scaling laws the Iron instrumentation of the IDR-S option is estimated to 6.0 M\texteuro.


No assembly tooling and manpower has been estimated. Based on the SiECAL detector these contributions can be estimated to 0.6 M\texteuro~ for both the large and small options. 

\section{Global ILD costing.}
The subdetector costs are summarised in table~\ref{cost_summary} for their baseline design. The two ILD models IDR-L and IDR-S are considered separately and global sums computed using the SiECAL option for the ECAL.
The inner system which offers no difference between the two sizes sums up to 23.2 + 4.9 = 28.1 M\texteuro. The outer system sums up to 304.1 + 32.0 = 336.1 M\texteuro~ for IDR-L and 259.2 + 27.1 = 286.3 M\texteuro~ for IDR-S.
The item labelled "global" includes the central DAQ, integration and transportation. These contributions have not been revisited since the DBD and have some overlap with items in the subdetector sections. Their total amount is directly reproduced from the DBD.

\begin{table}[h!]\hspace*{-0cm}\small
\begin{tabular}{ l p{0.15\hsize}p{0.12\hsize}p{0.12\hsize} p{0.12\hsize}p{0.12\hsize}}
\toprule
\bf {ILD COSTING (M\texteuro~ 2018)}& \bf {IDR-L} & \bf              &  \bf {IDR-S}&\bf \\
\bf {Item}      & \bf material & \bf manpower &  \bf material &\bf manpower \\
\midrule
Beam tube& \it0.5 &  \it0.& idem&idem \\
VTX        & 2.96  &1.45  &  idem &idem \\
SIT        & 5.0   &1.5 & idem&idem\\
FTD       & 4.8   &1.3  & idem &idem  \\
LumiCAL & 3.37  & 0.16& idem&idem\\
ECAL ring & 1.5 & 0.16 & idem&idem\\
LHCAL   & 2.93  & 0.16&idem& idem\\
BeamCAL & 2.15  & 0.16& idem&idem\\
\bf{Inner part total} &\bf{23.2}&\bf{4.9}&\bf{idem}&\bf{idem}\\
\midrule
TPC & 25.2 & 5 & 19 & 3.6\\
SET    & 20& 2&16&1.5\\
SiECAL & 119 & 11.0 & 92. & 8.7\\
ScECAL & \it74 & 11.5 & \it62.5 & 9.5\\
AHCAL  & \it44.9 & 4 & \it37.7 & 3.4\\
SDHCAL & \it44.8 & 4 & \it37.6 & 3.4\\
Coil and ancillaries &  40 & 4& idem & idem\\
Yoke and vacuum tank &  48.5 & 5.4& idem & idem \\
Iron instrumentation  &  \it6.5 & 0.6 & 6 & 0.54\\
\bf{Outer part total}&\bf{304.1}&\bf{32.0}&\bf{259.2}&\bf{27.1}\\
\midrule
\bf{Global (DAQ, integration, transport)}  &\bf{\it14.6} &\bf{\it0.16}& \bf{idem}& \bf{idem}\\
\midrule
\bf{Total} & \bf{342}   &  \bf{37.0}  & \bf{297.0} & \bf{32.2}  \\
\midrule
\bf{Grand Total including manpower}    & \bf{379}   &    & \bf{329} &   \\
 \bottomrule
\end{tabular}
\caption{\label{cost_summary}Global cost estimate of ILD in its baseline design (M\texteuro~ 2018). The numbers in italic are extrapolated from the DBD. The manpower has been translated from FTExYears to Euros. The total sums are computed with the SiECAL option.
}
\end{table}
Table ~\ref{cost_summary} presents only the costing of the IDR-L and IDR-S baseline design. Some variants such as fully pixelised inner detectors, reduced sampling in the SiECAL or reduced return yoke have been discussed and costed above. Their individual impacts on the cost of the small IDR-S ILD version are summarised in table ~\ref{cost_variants}, together with the result of taking all of them into account.
\begin{table}[h!]\hspace*{-0cm}\small
\begin{tabular}{ccccc}
\toprule
\bf {Baseline design} & \bf Fully pixelized inner trackers &  \bf Reduced ECAL sampling&\bf Reduced yoke&\bf Altogether\\
\midrule
329&337&318&320&316\\
\bottomrule
\end{tabular}
\caption{\label{cost_variants}IDR-S costing taking into account several variants discussed in the text, individually and altogether, in M\texteuro.}
\end{table}

\section{Comparison to the DBD cost estimate and discussion.}
 
\begin{figure}[h!]
\begin{tabular}{cc}
\includegraphics[width=0.47\hsize]{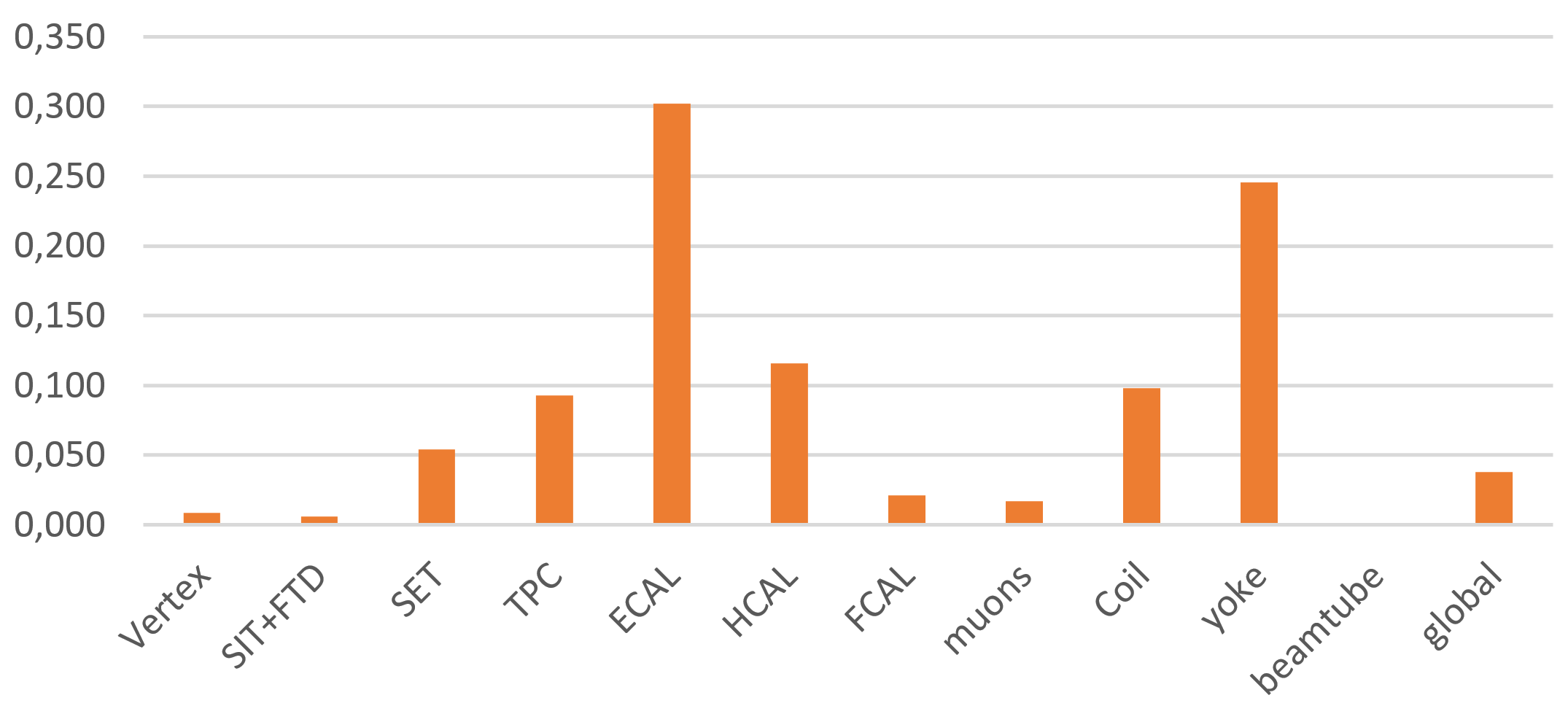}&
\includegraphics[width=0.47\hsize]{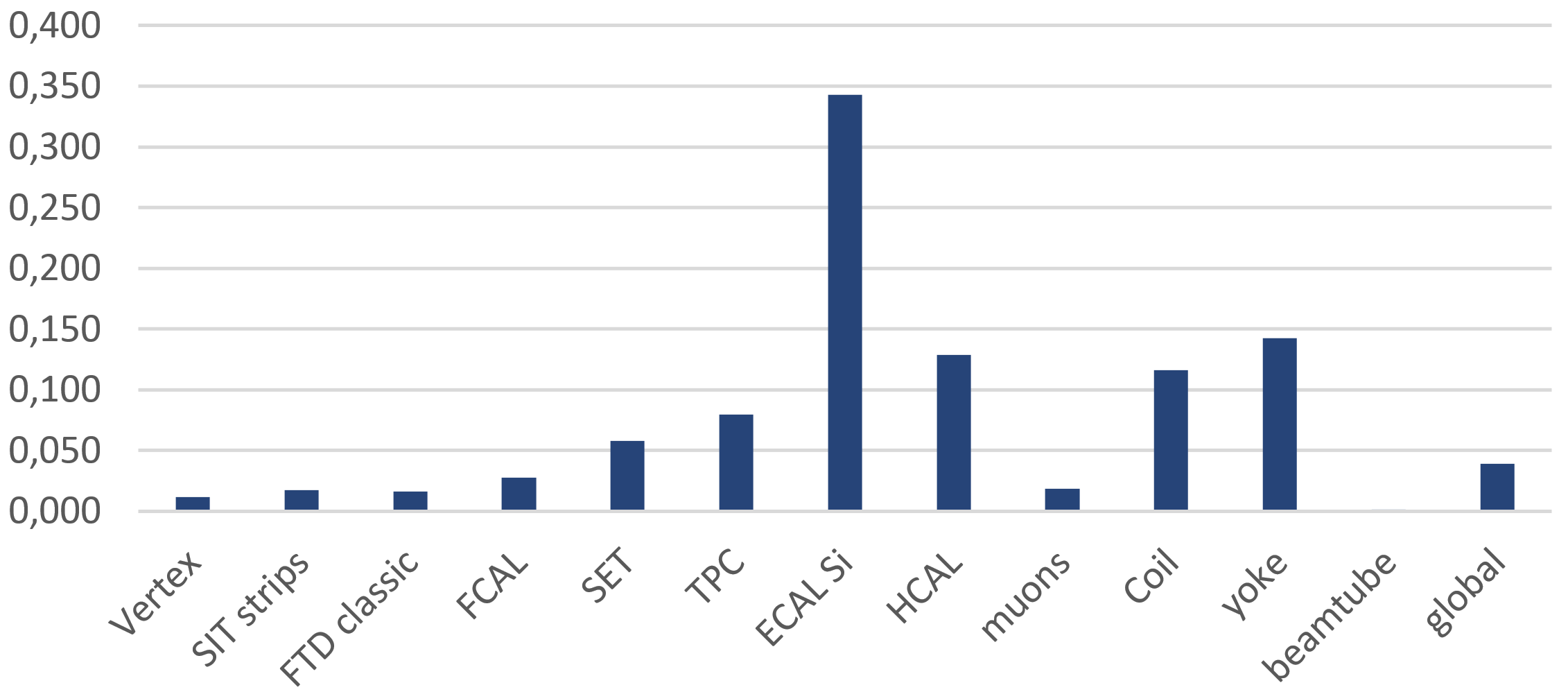}
\caption{ILD cost sharing as documented in the DBD (left), and for the large IDR-L version (right)}
\label{fig:det:DBD_cost_sharing}
\end{tabular}
\end{figure}

 The DBD ILD total cost corresponded to 392 M\texteuro(2018), with a relative sharing between subdetectors shown on Figure~\ref{fig:det:DBD_cost_sharing} left. The same figure shows on the right the sharing for IDR-L.
 
\begin{figure}[h!]
\begin{tabular}{cc}
\includegraphics[width=0.47\hsize]{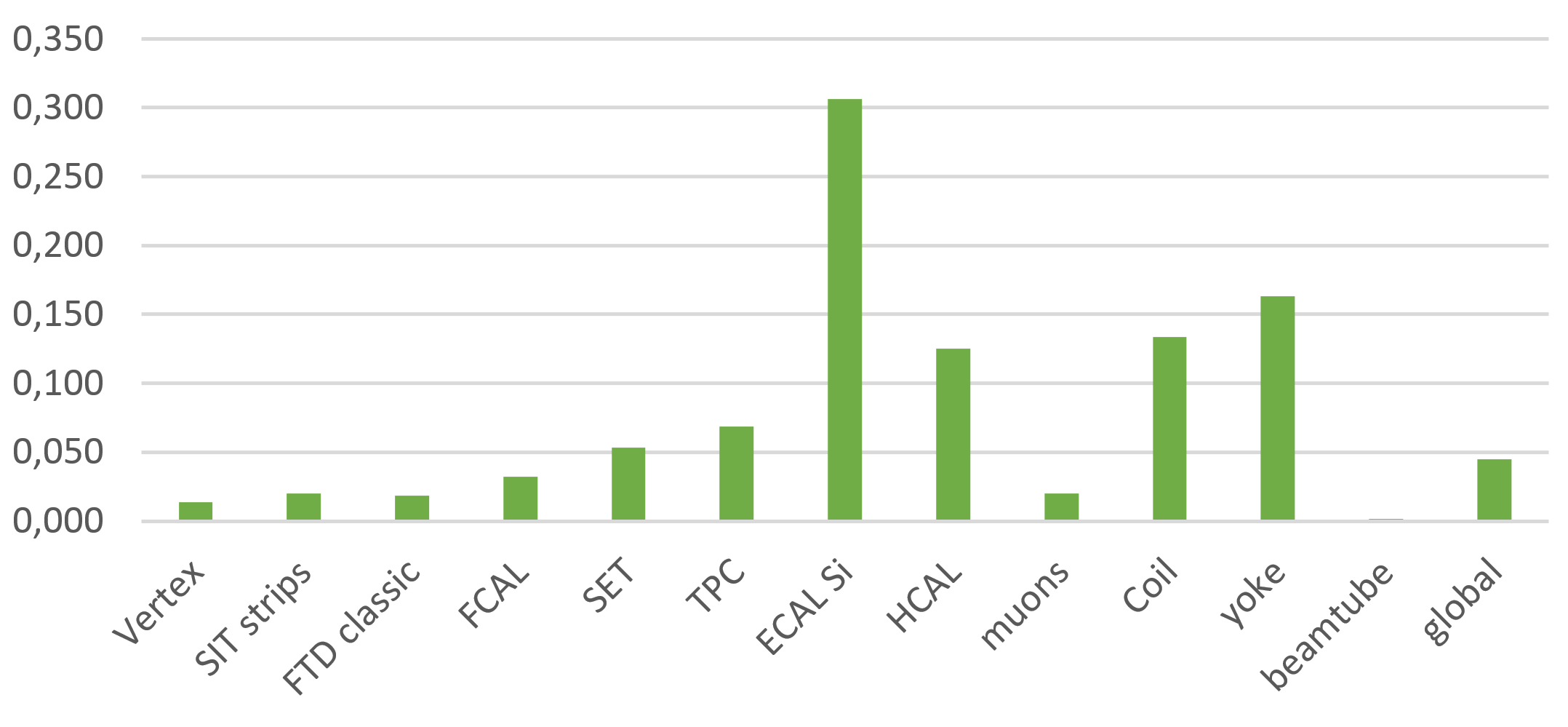}&
\includegraphics[width=0.47\hsize]{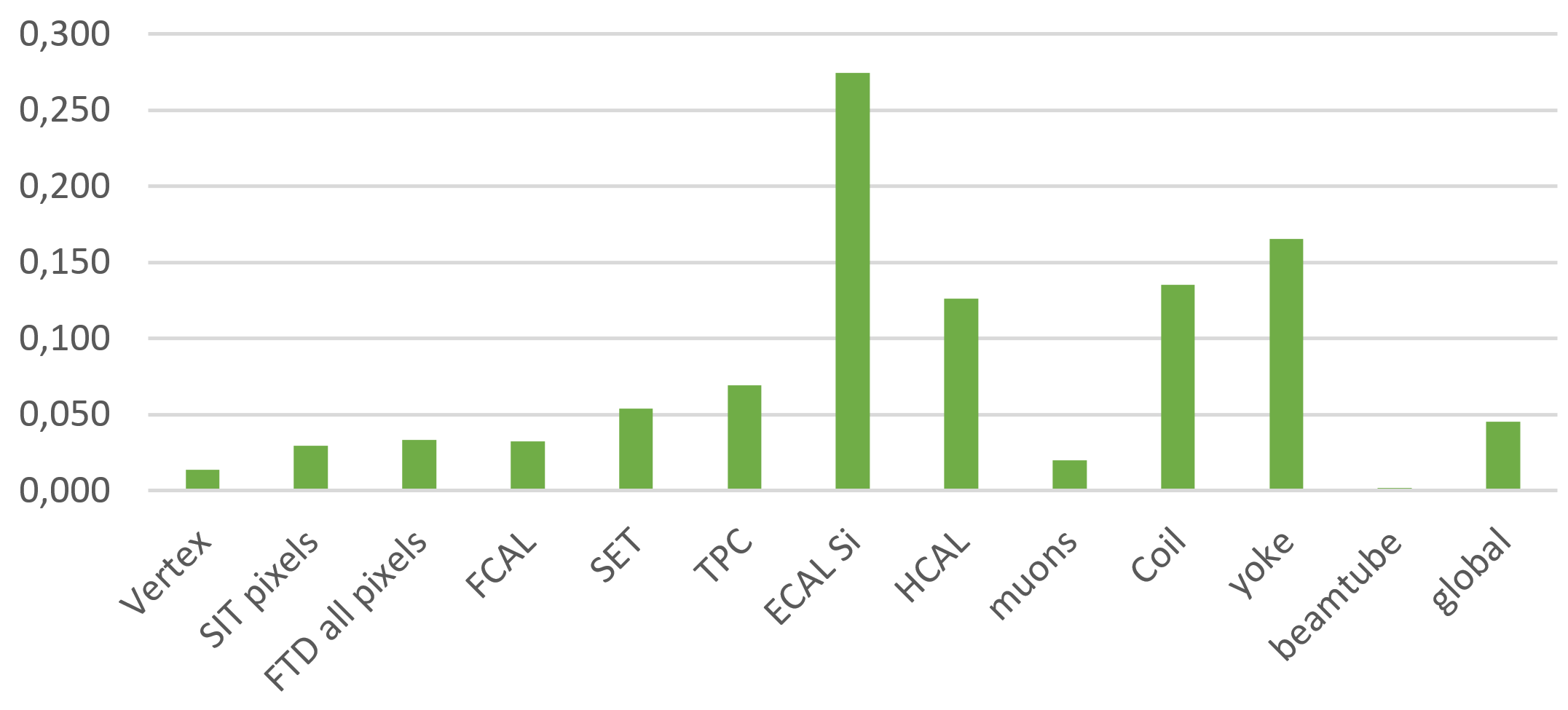}
\caption{ILD cost sharing for the IDR-S small version in its baseline design (left), and modified with full pixel inner tracking and SiECAL reduced sampling for a total cost of 326M\texteuro~ (right). }
\label{Costing:Small_cost_sharing}
\end{tabular}
\end{figure}

 The updated cost sharing for IDR-S is shown on Figure \ref{Costing:Small_cost_sharing} left. As an illustration of the potential effect of design changes the cost sharing is also shown on Figure \ref{Costing:Small_cost_sharing} right for ILD-S with SIT and FTD totally equipped with pixels, and sampling reduced to 26 layers in the SiECAL.
 
When comparing the updated estimations to the DBD one, it should be reminded that the latter did not include in house manpower in most cases, and that the DBD ECAL was a mix of the Silicon and Scintillator options. 
One important message of the new estimation is that the dominant SiECAL contribution has reduced significantly using latest information from the ongoing spin-off projects of the technology, and that further reduction could be achieved by reducing the sampling. In the same vein the yoke cost, which has been kept identical for the two options, could be reduced by a significant amount depending on its design and the stray field which can be accepted. 

The size reduction of IDR-S provides a saving of about 50M\texteuro~ compared to IDR-L. This $\approx$15\% cost reduction will have to be considered in regards with the performance comparison of the two size options.

In summary, the ILD technology maturation since the DBD times has comforted the global detector cost estimation in the direction of lower costs. New detection features w.r.t the current baseline design, such as enhanced pixel tracking and high resolution timing, may increase the costs in the future. However several options for further cost reduction have been identified and require further studies to assess their impact.

\chapter{Summary and outlook}
\label{chap:summary}

This paper presents the concept and the design status of the ILD detector. Its content provides a self-consistent update of the previous ILD documents, in particular the DBD~\cite{ild:bib:ilddbd}. 

The updated ILD concept is the result of an effort of many people over a significant time. The proposed detector is now developed to a point where all construction issues are under control, even though a significant amount of engineering remains to be done. 

ILD has always worked in very close cooperation with major R\&D collaborations such as CALICE, LCTPC or FCAL. Many solutions proposed for ILD result from this common effort. In most cases significant demonstrations and proof-of-concept experiments could be done, and are used to validate the soundness of the designs of ILD. 

The ILD concept relies on particle flow as the central method to reconstruct events at an ILC. This has far-reaching consequences for the design of the detector. An imaging calorimeter, with a large number of cells both in transverse and in the longitudinal direction, is essential. This very powerful calorimeter is associated to a tracking system which combines powerful and highly redundant tracking with excellent vertexing capabilities. A central and unique part of the ILD concept is a large volume time projection chamber in the tracking system, which delivers excellent efficiency for the tracking over a large solid angle. The complete tracking system has been optimised for low material and a solid angle acceptance as close as possible to $4 \pi$.

Together with the R\&D collaborations, the past few years since the publication of the DBD have been used to further develop the subdetector technologies and, more importantly, push the different technologies closer to a point of readiness for use in a large scale detector. This has been particularly successful for the very ambitious ILD calorimeters, where most proposed technologies have recently tested large scale prototypes, built in a way which would be scalable to the complete experiment. 

ILD has developed two versions of its detector concept, a large and a small version. In this document, many results are shown as a function of these two models, as a means to better understand the scaling of both performance and cost of ILD. No decision has been taken at this moment as to whether the ILD detector would follow more closely the large or the small design. This will also depend largely on the further development of the ILC project, its ultimate definition, and the boundary conditions set to realise the project. 

Both versions of ILD have been the subject of study for a number of benchmarking reactions with detailed simulations. These were selected in a way to probe the capabilities of the two designs as precisely as possible. The impact of the detector design on the results is clearly demonstrated in the benchmarking section of this document. 

In addition ILD has pursued a range of analyses to probe and demonstrate the physics potential of ILC (and ILD). These analyses are not all described in this document, but are available from the recent review on science at the ILC \cite{Bambade:2019fyw}.

In its current form the ILD detector concept is a well proven and tested concept for a detector at an electron-positron collider, and in particular at the ILC collider. 
In the coming years the ILD subdetector technologies will be further consolidated thanks to the ongoing construction of new detectors, out of which several spin-offs of ILC-oriented R\&D. Among current projects of particular interest for ILD are the ALICE upgrade, CBM and BELLE-II vertex and inner tracker detectors (Si-pixel technology), the ALICE upgrade and T2K/ND280 TPCs (TPC readout technologies) and the CMS upgrade HGCAL detector (calorimetry technologies). The ILD Collaboration is also currently performing a full simulation of the ILC collisions at a c.m.s energy of 250 GeV (as opposed to the 500 GeV collisions used in the present benchmarking), in order to revisit the detector performance in its latest design for the baseline ILC program. This will also allow to assess the potential impact of possible new features such as extended pixel tracking, a TOF functionality or a calorimetry high-resolution timing.  
ILD therefore remains committed to go forward to proposing a detector at an electron-positron collider in general, ILC in particular, once the facility is moving into the next project phase. 

At the time of writing this report, no final decision on the construction of the ILC accelerator or any other electron-positron collider has been reached. Japan has announced that it will study the ILC proposal in depth, and has started discussions with international partners to understand possible project schemes, including cost sharing issues. At the same time, in Europe, the ongoing discussion on the particle physics strategy for the next decade, which includes several options for electron-positron colliders at the energy frontier, is coming close to conclusion. 
This paper therefore provides a timely documentation of the current state of the ILD detector, and defines a possible configuration with which ILD could enter into a proposal at the ILC, should ILC move forward.

\backmatter
%

\bibliographystyle{utphys_mod}
\bibliography{ILD-master,ILD-software-refs,ILD-benchmark-refs}

\providecommand{\href}[2]{#2}\begingroup\raggedright\begin{thebibliography}{100}

\bibitem{ild:bib:ILDloi}
{ ILD Concept Group}, T.~Abe {\em et al.}, ``{The International Large Detector:
  Letter of Intent}'' \href{http://arxiv.org/abs/1006.3396}{{\tt
  arXiv:1006.3396 [hep-ex]}}.
FERMILAB-LOI-2010-03, FERMILAB-PUB-09-682-E, DESY-2009-87, KEK-REPORT-2009-6.

\bibitem{ild:bib:ilddbd}
H.~Abramowicz {\em et al.}, ``{The International Linear Collider Technical
  Design Report - Volume 4: Detectors}''
\href{http://arxiv.org/abs/1306.6329}{{\tt arXiv:1306.6329 [physics.ins-det]}}.

\bibitem{bib:sid:loi}
H.~Aihara, P.~Burrows, M.~Oreglia, E.~L. Berger, V.~Guarino, J.~Repond,
  H.~Weerts, L.~Xia, J.~Zhang, Q.~Zhang, {\em et al.}, ``{SiD Letter of
  Intent}''
\href{http://arxiv.org/abs/0911.0006}{{\tt arXiv:0911.0006 [physics.ins-det]}}.

\bibitem{bib:4th:loi}
N.~Akchurin {\em et al.}, ``{Another Detector for the International Linear
  Collider}''
\href{http://arxiv.org/abs/1307.5495}{{\tt arXiv:1307.5495 [physics.ins-det]}}.

\bibitem{Cepeda:2019klc}
{ HL/HE WG2 group}, M.~Cepeda {\em et al.}, ``{Higgs Physics at the HL-LHC and
  HE-LHC}''
\href{http://arxiv.org/abs/1902.00134}{{\tt arXiv:1902.00134 [hep-ph]}}.

\bibitem{Ref:Decoupling}
H.~E. Haber, ``{Nonminimal Higgs sectors: The Decoupling limit and its
  phenomenological implications}'' in {\em {Joint U.S.-Polish Workshop on
  Physics from Planck Scale to Electro-Weak Scale (SUSY 94) Warsaw, Poland,
  September 21-24, 1994}}, pp.~1--16.
\newblock 1994.
\newblock
\href{http://arxiv.org/abs/hep-ph/9501320}{{\tt arXiv:hep-ph/9501320
  [hep-ph]}}.
\newblock

\bibitem{Ref:ILCESU1}
H.~Aihara {\em et al.}, ``{The International Linear Collider -- A Global
  Project}''.
  \url{{https://ilchome.web.cern.ch/sites/ilchome.web.cern.ch/files/ILC_Global_Project_Final.pdf}}.

\bibitem{Bambade:2019fyw}
P.~Bambade {\em et al.}, ``{The International Linear Collider: A Global Project
  (long version)}''
\href{http://arxiv.org/abs/1903.01629}{{\tt arXiv:1903.01629 [hep-ex]}}.

\bibitem{Yan:2016xyx}
J.~Yan, S.~Watanuki, K.~Fujii, A.~Ishikawa, D.~Jeans, J.~Strube, J.~Tian, and
  H.~Yamamoto, ``{Measurement of the Higgs boson mass and $e^+e^- \to ZH$ cross
  section using $Z \to \mu^+\mu^-$ and $Z \to e^+ e^-$ at the ILC}''
  \href{http://dx.doi.org/10.1103/PhysRevD.94.113002}{{\em Phys. Rev.} { D94}
  (2016) no.~11, 113002},
\href{http://arxiv.org/abs/1604.07524}{{\tt arXiv:1604.07524 [hep-ex]}}.

\bibitem{Ref:Hinvisible}
A.~Ishikawa, ``{Search for invisible higgs decays at the ILC}''.
  \url{https://agenda.linearcollider.org/event/6389/contributions/30536/}.
  presented at the Linear Collider Workshop, Belgrade, Serbia, October 5-10,
  2014.

\bibitem{Barklow:2017suo}
T.~Barklow, K.~Fujii, S.~Jung, R.~Karl, J.~List, T.~Ogawa, M.~E. Peskin, and
  J.~Tian, ``{Improved Formalism for Precision Higgs Coupling Fits}''
  \href{http://dx.doi.org/10.1103/PhysRevD.97.053003}{{\em Phys. Rev.} { D97}
  (2018) no.~5, 053003},
\href{http://arxiv.org/abs/1708.08912}{{\tt arXiv:1708.08912 [hep-ph]}}.

\bibitem{Barklow:2017awn}
T.~Barklow, K.~Fujii, S.~Jung, M.~E. Peskin, and J.~Tian, ``{Model-Independent
  Determination of the Triple Higgs Coupling at e+e- Colliders}''
  \href{http://dx.doi.org/10.1103/PhysRevD.97.053004}{{\em Phys. Rev.} { D97}
  (2018) no.~5, 053004},
\href{http://arxiv.org/abs/1708.09079}{{\tt arXiv:1708.09079 [hep-ph]}}.

\bibitem{Karl:2019hes}
R.~Karl, \href{http://dx.doi.org/10.3204/PUBDB-2019-03013}{{\em {From the
  Machine-Detector Interface to Electroweak Precision Measurements at the ILC -
  Beam-Gas Background, Beam Polarization and Triple Gauge Couplings.}}}
\newblock PhD thesis, {Hamburg U.}, Hamburg,
2019.
\newblock

\bibitem{Ref:EWBG1}
S.~Kanemura, Y.~Okada, and E.~Senaha, ``{Electroweak baryogenesis and the
  triple Higgs boson coupling}'' {\em eConf} { C050318} (2005)  0704,
\href{http://arxiv.org/abs/hep-ph/0507259}{{\tt arXiv:hep-ph/0507259
  [hep-ph]}}.

\bibitem{Ref:EWBG2}
A.~Noble and M.~Perelstein, ``{Higgs self-coupling as a probe of electroweak
  phase transition}'' \href{http://dx.doi.org/10.1103/PhysRevD.78.063518}{{\em
  Phys. Rev.} { D78} (2008)  063518},
\href{http://arxiv.org/abs/0711.3018}{{\tt arXiv:0711.3018 [hep-ph]}}.

\bibitem{Ref:EWBG3}
D.~E. Morrissey and M.~J. Ramsey-Musolf, ``{Electroweak baryogenesis}''
  \href{http://dx.doi.org/10.1088/1367-2630/14/12/125003}{{\em New J. Phys.} {
  14} (2012)  125003},
\href{http://arxiv.org/abs/1206.2942}{{\tt arXiv:1206.2942 [hep-ph]}}.

\bibitem{Ref:Claude}
{C. D\"urig}, {\em {Measuring the Higgs Self-coupling at the International
  Linear Collider.}}
\newblock PhD thesis, {Hamburg U.}, Hamburg, 2016.
\newblock \url{https://bib-pubdb1.desy.de/record/310520}.

\bibitem{Habermehl:2017dxh}
M.~Habermehl, K.~Fujii, J.~List, S.~Matsumoto, and T.~Tanabe, ``{WIMP Searches
  at the International Linear Collider}''
  \href{http://dx.doi.org/10.22323/1.282.0155}{{\em PoS} { ICHEP2016} (2016)
  155},
\href{http://arxiv.org/abs/1702.05377}{{\tt arXiv:1702.05377 [hep-ex]}}.

\bibitem{Habermehl:2018yul}
M.~Habermehl, \href{http://dx.doi.org/10.3204/PUBDB-2018-05723}{{\em {Dark
  Matter at the International Linear Collider}}}.
\newblock PhD thesis, DESY, Hamburg,
2018.
\newblock

\bibitem{Baer:2016new}
H.~Baer, M.~Berggren, K.~Fujii, S.-L. Lehtinen, J.~List, T.~Tanabe, and J.~Yan,
  ``{Naturalness and light higgsinos: A powerful reason to build the ILC}''
  \href{http://dx.doi.org/10.22323/1.282.0156}{{\em PoS} { ICHEP2016} (2016)
  156},
\href{http://arxiv.org/abs/1611.02846}{{\tt arXiv:1611.02846 [hep-ph]}}.

\bibitem{Horiguchi:2013wra}
T.~Horiguchi, A.~Ishikawa, T.~Suehara, K.~Fujii, Y.~Sumino, Y.~Kiyo, and
  H.~Yamamoto, ``{Study of top quark pair production near threshold at the
  ILC}''
\href{http://arxiv.org/abs/1310.0563}{{\tt arXiv:1310.0563 [hep-ex]}}.

\bibitem{Vos:2016til}
M.~Vos {\em et al.}, ``{Top physics at high-energy lepton colliders}''
\href{http://arxiv.org/abs/1604.08122}{{\tt arXiv:1604.08122 [hep-ex]}}.

\bibitem{Amjad:2015mma}
M.~S. Amjad {\em et al.}, ``{A precise characterisation of the top quark
  electro-weak vertices at the ILC}''
  \href{http://dx.doi.org/10.1140/epjc/s10052-015-3746-5}{{\em Eur. Phys. J.} {
  C75} (2015) no.~10, 512},
\href{http://arxiv.org/abs/1505.06020}{{\tt arXiv:1505.06020 [hep-ex]}}.

\bibitem{Ref:bilokin2017}
S.~Bilokin, {\em {Hadronic showers in a highly granular silicon-tungsten
  calorimeter and production of bottom and top quarks at the ILC}}.
\newblock PhD thesis, Saclay, 2017.
\newblock
\url{https://tel.archives-ouvertes.fr/tel-01946099}.
\newblock

\bibitem{Barklow:2015tja}
T.~Barklow, J.~Brau, K.~Fujii, J.~Gao, J.~List, N.~Walker, and K.~Yokoya,
  ``{ILC Operating Scenarios}''
\href{http://arxiv.org/abs/1506.07830}{{\tt arXiv:1506.07830 [hep-ex]}}.

\bibitem{Behnke:2013xla}
T.~Behnke, J.~E. Brau, B.~Foster, J.~Fuster, M.~Harrison, J.~M. Paterson,
  M.~Peskin, M.~Stanitzki, N.~Walker, and H.~Yamamoto, ``{The International
  Linear Collider Technical Design Report - Volume 1: Executive Summary}''
\href{http://arxiv.org/abs/1306.6327}{{\tt arXiv:1306.6327 [physics.acc-ph]}}.

\bibitem{Harrison:2013nva}
M.~Harrison, M.~Ross, and N.~Walker, ``{Luminosity Upgrades for ILC}'' in {\em
  {Proceedings, 2013 Community Summer Study on the Future of U.S. Particle
  Physics: Snowmass on the Mississippi (CSS2013): Minneapolis, MN, USA, July
  29-August 6, 2013}}.
\newblock 2013.
\newblock
\href{http://arxiv.org/abs/1308.3726}{{\tt arXiv:1308.3726 [physics.acc-ph]}}.
\newblock

\bibitem{Yokoya:2019rhx}
K.~Yokoya, K.~Kubo, and T.~Okugi, ``{Operation of ILC250 at the Z-pole}''
\href{http://arxiv.org/abs/1908.08212}{{\tt arXiv:1908.08212
  [physics.acc-ph]}}.

\bibitem{ild:bib:JAHEP}
JAHEP, ``{A Proposal for a Phased Execution of the International Linear
  Collider Project }''.
  \url{http://www.jahep.org/office/doc/201210_ILC_staging_e.pdf}.

\bibitem{ild:bib:Newsline_Kitakami}
B.~Warmbein, ``{The Road to Kitakami }''.
  \url{http://newsline.linearcollider.org/2014/02/20/the-road-to-kitakami/}.
  ILC Newsline, 20. February 2014.

\bibitem{Keller:2019aak}
L.~Keller and G.~White, ``{Simulation of Muon Background at the ILC}'' in {\em
  {International Workshop on Future Linear Colliders (LCWS 2018) Arlington,
  Texas, USA, October 22-26, 2018}}.
\newblock 2019.
\newblock
\href{http://arxiv.org/abs/1901.06449}{{\tt arXiv:1901.06449
  [physics.acc-ph]}}.
\newblock

\bibitem{Parker:2009zz}
B.~Parker, A.~Mikhailichenko, K.~Buesser, J.~Hauptman, T.~Tauchi, P.~Burrows,
  T.~Markiewicz, M.~Oriunno, and A.~Seryi, ``{Functional Requirements on the
  Design of the Detectors and the Interaction Region of an Electron-Positron
  Linear Collider with a Push-Pull Arrangement of Detectors}'' in {\em
  {Particle accelerator. Proceedings, 23rd Conference, PAC'09, Vancouver,
  Canada, May 4-8, 2009}}, pp.~WE6PFP078, ILC--Note--2009--050.
\newblock 2010.
\newblock
\url{http://www-public.slac.stanford.edu/sciDoc/docMeta.aspx?slacPubNumber=slac-pub-13657}.
\newblock

\bibitem{ild:bib:lstar_cr}
G.~White, ``{Baseline Optics to Provide for a Single FFS L* Optics
  Configuration}''.
  \url{https://edmsdirect.desy.de/item/D00000001082495,A,1,1}. ILC-CR-0002.

\bibitem{ild:bib:beam_gas}
R.~Karl and J.~List, ``{Impact of Beam-Gas Interactions in the IP Region of
  ILD}''.
  \url{https://confluence.desy.de/download/attachments/42357928/ILD-TECH-PUB-2017-001.pdf?version=1&modificationDate=1497399362316&api=v2}.
  ILD-TECH-PUB-2017-001.

\bibitem{Ref:bib:TPCOPT}
M.~Berggren, ``{Overall ILD dimensions and the ILD tracker}'' in {\em
  Proceedings of the ECFA Linear Collider meeting in Santander, Spain}.
\newblock 2016.
\newblock
  \url{https://agenda.linearcollider.org/event/7014/contributions/34672/attachments/30216/45163/berggren-lcecfa-santander-june-2016-opt2.pdf}.

\bibitem{Arominski:2018uuz}
{ CLICdp Collaboration}, D.~Arominski {\em et al.}, ``{A detector for CLIC:
  main parameters and performance}''
\href{http://arxiv.org/abs/1812.07337}{{\tt arXiv:1812.07337
  [physics.ins-det]}}.

\bibitem{Besson:2016ivb}
A.~Besson, A.~Perez~Perez, E.~Spiriti, J.~Baudot, G.~Claus, M.~Goffe, and
  M.~Winter, ``{From vertex detectors to inner trackers with CMOS pixel
  sensors}'' \href{http://dx.doi.org/10.1016/j.nima.2016.04.081}{{\em Nucl.
  Instrum. Meth.} { A845} (2017)  33--37},
\href{http://arxiv.org/abs/1604.02957}{{\tt arXiv:1604.02957
  [physics.ins-det]}}.

\bibitem{Alonso:2012ss}
{ DEPFET Collaboration}, O.~Alonso {\em et al.}, ``{DEPFET active pixel
  detectors for a future linear $e^+e^-$ collider}''
  \href{http://dx.doi.org/10.1109/TNS.2013.2245680}{{\em IEEE Trans. Nucl.
  Sci.} { 60} (2013)  1457},
\href{http://arxiv.org/abs/1212.2160}{{\tt arXiv:1212.2160 [physics.ins-det]}}.

\bibitem{Richter:2003dn}
R.~H. Richter {\em et al.}, ``{Design and technology of DEPFET pixel sensors
  for linear collider applications}''
\href{http://dx.doi.org/10.1016/S0168-9002(03)01802-3}{{\em Nucl. Instrum.
  Meth.} { A511} (2003)  250--256}.

\bibitem{Andricek:2004cj}
L.~Andricek, G.~Lutz, R.~H. Richter, and M.~Reiche, ``{Processing of ultra-thin
  silicon sensors for future e+ e- linear collider experiments}''
\href{http://dx.doi.org/10.1109/TNS.2004.829531}{{\em IEEE Trans. Nucl. Sci.} {
  51} (2004)  1117--1120}.

\bibitem{Paredes:2014kda}
C.~Calancha~Paredes {\em et al.}, ``{Progress in the development of the vertex
  detector with fine pixel CCD at the ILC}''
\href{http://dx.doi.org/10.22323/1.198.0022}{{\em PoS} { Vertex2013} (2013)
  022}.

\bibitem{Denisov:2015jjl}
D.~Denisov, V.~Evdokimov, and S.~Lukic, ``{Time and Position Resolution of the
  Scintillator Strips for a Muon System at Future Colliders}''
  \href{http://dx.doi.org/10.1016/j.nima.2016.03.091}{{\em Nucl. Instrum.
  Meth.} { A823} (2016)  120--125},
\href{http://arxiv.org/abs/1512.06729}{{\tt arXiv:1512.06729
  [physics.ins-det]}}.

\bibitem{ild:bib:TPC_lctpc}
{ LCTPC Collaboration}, ``{The LCTPC collaboration}''.
  \url{https://www.lctpc.org}.

\bibitem{ild:bib:CALICE}
``{}the calice collaboration''.

\bibitem{ild:bib:FCAL}
{ FCAL Collaboration}, ``{The FCAL Collaboration}''.
  \url{http://fcal.desy.de/}.

\bibitem{Contin:2017mck}
G.~Contin {\em et al.}, ``{The STAR MAPS-based PiXeL detector}''
  \href{http://dx.doi.org/10.1016/j.nima.2018.03.003}{{\em Nucl. Instrum.
  Meth.} { A907} (2018)  60--80},
\href{http://arxiv.org/abs/1710.02176}{{\tt arXiv:1710.02176
  [physics.ins-det]}}.

\bibitem{AglieriRinella:2017lym}
{ ALICE Collaboration}, G.~Aglieri~Rinella, ``{The ALPIDE pixel sensor chip for
  the upgrade of the ALICE Inner Tracking System}''
\href{http://dx.doi.org/10.1016/j.nima.2016.05.016}{{\em Nucl. Instrum. Meth.}
  { A845} (2017)  583--587}.

\bibitem{Koziel:2017loo}
M.~Koziel {\em et al.}, ``{Vacuum-compatible, ultra-low material budget
  Micro-Vertex Detector of the compressed baryonic matter experiment at FAIR}''
\href{http://dx.doi.org/10.1016/j.nima.2016.05.093}{{\em Nucl. Instrum. Meth.}
  { A845} (2017)  110--113}.

\bibitem{Nomerotski:2011zz}
{ PLUME}, A.~Nomerotski {\em et al.}, ``{PLUME collaboration: Ultra-light
  ladders for linear collider vertex detector}''
\href{http://dx.doi.org/10.1016/j.nima.2010.12.083}{{\em Nucl. Instrum. Meth.}
  { A650} (2011)  208--212}.

\bibitem{Andricek:2011zza}
L.~Andricek {\em et al.}, ``{Intrinsic resolutions of DEPFET detector
  prototypes measured at beam tests}''
\href{http://dx.doi.org/10.1016/j.nima.2011.02.015}{{\em Nucl. Instrum. Meth.}
  { A638} (2011)  24--32}.

\bibitem{Velthuis:2008zza}
J.~J. Velthuis {\em et al.}, ``{A DEPFET based beam telescope with submicron
  precision capability}''
\href{http://dx.doi.org/10.1109/TNS.2007.914031}{{\em IEEE Trans. Nucl. Sci.} {
  55} (2008)  662--666}.

\bibitem{Marinas:2011zz}
C.~Marinas and M.~Vos, ``{The Belle-II DEPFET pixel detector: A step forward in
  vertexing in the superKEKB flavour factory}''
\href{http://dx.doi.org/10.1016/j.nima.2010.12.116}{{\em Nucl. Instrum. Meth.}
  { A650} (2011)  59--63}.

\bibitem{Abe:2010gxa}
{ Belle-II Collaboration}, T.~Abe {\em et al.}, ``{Belle II Technical Design
  Report}''
\href{http://arxiv.org/abs/1011.0352}{{\tt arXiv:1011.0352 [physics.ins-det]}}.

\bibitem{ild:bib:FPCCD}
H.~Sato, ``{FPCCD Large Prototype Test Status}''.
  \url{https://agenda.linearcollider.org/event/8081/contributions/43021/}.
  Presented at the ILC-JP end-of-year physics and detector meeting, 21 December
  2018.

\bibitem{PELLEGRINI201412}
G.~Pellegrini {\em et al.}, ``Technology developments and first measurements of
  low gain avalanche detectors ({LGAD}) for high energy physics applications''
  {\em Nuclear Instruments and Methods A} { 765} (2014)  12 -- 16.
  \url{http://www.sciencedirect.com/science/article/pii/S0168900214007128}.

\bibitem{Carulla_2016}
M.~Carulla {\em et al.}, ``Technology developments and first measurements on
  inverse low gain avalanche detector ({iLGAD}) for high energy physics
  applications'' {\em Journal of Instrumentation} { 11} (2016) no.~12,
  C12039--C12039. \url{https://doi.org/10.1088%2F1748-0221%2F11%2F12%2Fc12039}.

\bibitem{Curras2019}
E.~Curr\'as {\em et al.}, ``{Inverse Low Gain Avalanche Detectors (iLGADs) for
  precise tracking and timing applications}'' in {\em {15th Vienna Conference
  on Instrumentation (VCI2019) Vienna, Austria, February 18-22, 2019}}.
\newblock 2019.
\newblock
\href{http://arxiv.org/abs/1904.02061}{{\tt arXiv:1904.02061
  [physics.ins-det]}}.
\newblock

\bibitem{ild:bib:TPC_liaison}
{ Linear Collider Collaboration}, ``{The Detector R\&D liaison report}''.
  \url{http://linearcollider.web.cern.ch/physics-detectors/working-group-detector-rd-liaison}.

\bibitem{ild:bib:TPC_desytb}
R.~Diener {\em et al.}, ``{The DESY II Test Beam Facility}''
  \href{http://dx.doi.org/10.1016/j.nima.2018.11.133}{{\em Nucl. Instrum.
  Meth.} { A922} (2019)  265--286},
\href{http://arxiv.org/abs/1807.09328}{{\tt arXiv:1807.09328
  [physics.ins-det]}}.

\bibitem{ild:bib:TPC_lycoris}
M.~Wu {\em et al.}, ``{Development of a large active area beam telescope based
  on the SiD micro-strip sensor}''. \url{http://cds.cern.ch/record/2666439}.
  Poster presented at Vienna Conference of Instrumentation 2019.

\bibitem{ild:bib:TPC_distortions}
P.~Colas, ``{A new scheme for Micromegas TPC readout: the encapsulated
  resistive anode with reverse grounding}'' {\em Talk given at the MPGD
  conference at La Rochelle, May 2019} (2019)  .
  \url{https://indico.cern.ch/event/757322/contributions/3387077}.

\bibitem{Malek:2017xol}
P.~Malek, ``{Development of a GEM based TPC Readout for ILD}'' in {\em
  {Proceedings, International Workshop on Future Linear Colliders 2016
  (LCWS2016): Morioka, Iwate, Japan, December 05-09, 2016}}.
\newblock 2016.
\newblock
\href{http://arxiv.org/abs/1703.05719}{{\tt arXiv:1703.05719
  [physics.ins-det]}}.
\newblock

\bibitem{ild:bib:TPC_quad}
C.~Ligtenberg {\em et al.}, ``{Performance of a GridPix TPC readout based on
  the Timepix3 chip}'' in {\em {International Workshop on Future Linear
  Colliders (LCWS 2018) Arlington, Texas, USA, October 22-26, 2018}}.
\newblock 2019.
\newblock
\href{http://arxiv.org/abs/1902.01987}{{\tt arXiv:1902.01987
  [physics.ins-det]}}.
\newblock

\bibitem{ild:bib:TPC_gatinginbeam}
A.~Shoji, ``{Measurement of dE/dx resolution of TPC prototype with gating GEM
  exposed to an electron beam}'' in {\em {International Workshop on Future
  Linear Collider (LCWS2017) Strasbourg, France, October 23-27, 2017}}.
\newblock 2018.
\newblock
\href{http://arxiv.org/abs/1801.04499}{{\tt arXiv:1801.04499
  [physics.ins-det]}}.
\newblock

\bibitem{ild:bib:TPC_gatingpaper}
M.~Kobayashi {\em et al.}, ``{Measurement of the electron transmission rate of
  the gating foil for the TPC of the ILC experiment}''
  \href{http://dx.doi.org/10.1016/j.nima.2018.11.060}{{\em Nucl. Instrum.
  Meth.} { A918} (2019)  41--53},
\href{http://arxiv.org/abs/1903.01717}{{\tt arXiv:1903.01717
  [physics.ins-det]}}.

\bibitem{Callier:2011zz}
S.~Callier, F.~Dulucq, C.~de~La~Taille, G.~Martin-Chassard, and
  N.~Seguin-Moreau, ``{SKIROC2, front end chip designed to readout the
  Electromagnetic CALorimeter at the ILC}''
\href{http://dx.doi.org/10.1088/1748-0221/6/12/C12040}{{\em JINST} { 6} (2011)
  C12040}.

\bibitem{Suehara:2018mqk}
T.~Suehara {\em et al.}, ``{Performance study of SKIROC2/A ASIC for ILD Si-W
  ECAL}'' \href{http://dx.doi.org/10.1088/1748-0221/13/03/C03015}{{\em JINST} {
  13} (2018) no.~03, C03015},
\href{http://arxiv.org/abs/1801.02024}{{\tt arXiv:1801.02024
  [physics.ins-det]}}.

\bibitem{Boudry:2014bxa}
{ ILD SiW-ECAL Development Group}, V.~Boudry, ``{Development of technological
  prototype of silicon-tungsten electromagnetic calorimeter for ILD.}'' {\em
  PoS} { TIPP2014} (2014)  020.
\url{https://pos.sissa.it/213/020/pdf}.

\bibitem{Poeschl:2015jma}
{ CALICE Collaboration}, R.~P{\"o}schl, ``{R\&D for a highly granular silicon
  tungsten electromagnetic calorimeter}''
\href{http://dx.doi.org/10.1088/1742-6596/587/1/012032}{{\em J. Phys. Conf.
  Ser.} { 587} (2015) no.~1, 012032}.

\bibitem{Kawagoe:2019dzh}
K.~Kawagoe {\em et al.}, ``{Beam test performance of the highly granular
  SiW-ECAL technological prototype for the ILC}''
  \href{http://dx.doi.org/10.1016/j.nima.2019.162969}{{\em Nucl. Instrum.
  Meth.} { A950} (2020)  162969},
\href{http://arxiv.org/abs/1902.00110}{{\tt arXiv:1902.00110
  [physics.ins-det]}}.

\bibitem{Magniette:2019nyg}
F.~Magniette, J.~Nanni, R.~Guillaumat, M.~Louzir, M.~Anduze, E.~Edy,
  O.~Korostyshevskyi, V.~Balagura, V.~Boudry, and J.-C. Brient, ``{ILD Silicon
  Tungsten Electromagnetic Calorimeter First Full Scale Electronic Prototype}''
\href{http://arxiv.org/abs/1909.04329}{{\tt arXiv:1909.04329
  [physics.ins-det]}}.

\bibitem{bib:talk-twepp-jj}
{ CALICE Collaboration}, J.~Jeglot, ``{A new compact electronics for CALICE SIW
  ECAL calorimeter readout}'' in {\em Proceedings of the topical Workshop on
  Electronics for Particle Physics, TWEPP 2019}.
\newblock 2019.
\newblock \url{https://indico.cern.ch/event/799025/contributions/3486491/}.

\bibitem{Collaboration:2293646}
{CMS Collaboration}, ``{The Phase-2 Upgrade of the CMS Endcap Calorimeter}''.
  Tech. Rep. CERN-LHCC-2017-023.

\bibitem{Bouchel:2011zz}
M.~Bouchel, S.~Callier, F.~Dulucq, J.~Fleury, J.~J. Jaeger, C.~de~La~Taille,
  G.~Martin-Chassard, and L.~Raux, ``{SPIROC (SiPM Integrated Read-Out Chip):
  Dedicated very front-end electronics for an ILC prototype hadronic
  calorimeter with SiPM read-out}''
\href{http://dx.doi.org/10.1088/1748-0221/6/01/C01098}{{\em JINST} { 6} (2011)
  C01098}.

\bibitem{ild:bib:hdmppc}
``{Hamamatsu Photonics K.K. MPPC}''.
\newblock
  \url{https://www.hamamatsu.com/eu/en/product/optical-sensors/mppc/mppc_mppc-array/index.html}.

\bibitem{Adloff:2011ha}
{ CALICE Collaboration}, C.~Adloff, J.~Blaha, J.~J. Blaising, C.~Drancourt,
  A.~Espargiliere, R.~Galione, N.~Geffroy, Y.~Karyotakis, J.~Prast, and
  G.~Vouters, ``{Tests of a particle flow algorithm with CALICE test beam
  data}'' \href{http://dx.doi.org/10.1088/1748-0221/6/07/P07005}{{\em JINST} {
  6} (2011)  P07005},
\href{http://arxiv.org/abs/1105.3417}{{\tt arXiv:1105.3417 [physics.ins-det]}}.

\bibitem{Adloff:2012gv}
{ CALICE Collaboration}, C.~Adloff {\em et al.}, ``{Hadronic energy resolution
  of a highly granular scintillator-steel hadron calorimeter using software
  compensation techniques}''
  \href{http://dx.doi.org/10.1088/1748-0221/7/09/P09017}{{\em JINST} { 7}
  (2012)  P09017}, \href{http://arxiv.org/abs/1207.4210}{{\tt arXiv:1207.4210
  [physics.ins-det]}}.
MPP-2012-116.

\bibitem{Adloff:2013vra}
{ CALICE Collaboration}, C.~Adloff {\em et al.}, ``{Track segments in hadronic
  showers in a highly granular scintillator-steel hadron calorimeter}''
  \href{http://dx.doi.org/10.1088/1748-0221/8/09/P09001}{{\em JINST} { 8}
  (2013)  P09001},
\href{http://arxiv.org/abs/1305.7027}{{\tt arXiv:1305.7027 [physics.ins-det]}}.

\bibitem{Adloff:2013kio}
{ CALICE Collaboration}, C.~Adloff {\em et al.}, ``{Validation of GEANT4 Monte
  Carlo Models with a Highly Granular Scintillator-Steel Hadron Calorimeter}''
  \href{http://dx.doi.org/10.1088/1748-0221/8/07/P07005}{{\em JINST} { 8}
  (2013)  07005},
\href{http://arxiv.org/abs/1306.3037}{{\tt arXiv:1306.3037 [physics.ins-det]}}.

\bibitem{Adloff:2013jqa}
{ CALICE Collaboration}, C.~Adloff {\em et al.}, ``{Shower development of
  particles with momenta from 1 to 10 GeV in the CALICE Scintillator-Tungsten
  HCAL}'' \href{http://dx.doi.org/10.1088/1748-0221/9/01/P01004}{{\em JINST} {
  9} (2014) no.~01, P01004},
\href{http://arxiv.org/abs/1311.3505}{{\tt arXiv:1311.3505 [physics.ins-det]}}.

\bibitem{Adloff:2014rya}
{ CALICE Collaboration}, C.~Adloff {\em et al.}, ``{The Time Structure of
  Hadronic Showers in highly granular Calorimeters with Tungsten and Steel
  Absorbers}'' \href{http://dx.doi.org/10.1088/1748-0221/9/07/P07022}{{\em
  JINST} { 9} (2014)  P07022},
\href{http://arxiv.org/abs/1404.6454}{{\tt arXiv:1404.6454 [physics.ins-det]}}.

\bibitem{Bilki:2014bga}
{ CALICE Collaboration}, B.~Bilki {\em et al.}, ``{Pion and proton showers in
  the CALICE scintillator-steel analogue hadron calorimeter}''
  \href{http://dx.doi.org/10.1088/1748-0221/10/04/P04014}{{\em JINST} { 10}
  (2015) no.~04, P04014},
\href{http://arxiv.org/abs/1412.2653}{{\tt arXiv:1412.2653 [physics.ins-det]}}.

\bibitem{Lucaci-Timoce:2013tkf}
{ CALICE Collaboration}, M.~Chefdeville {\em et al.}, ``{Shower development of
  particles with momenta from 15 GeV to 150 GeV in the CALICE
  scintillator-tungsten hadronic calorimeter}''
  \href{http://dx.doi.org/10.1088/1748-0221/10/12/P12006}{{\em JINST} { 10}
  (2015) no.~12, P12006},
\href{http://arxiv.org/abs/1509.00617}{{\tt arXiv:1509.00617
  [physics.ins-det]}}.

\bibitem{Price:2016sce}
{ CALICE Collaboration}, G.~Eigen {\em et al.}, ``{Hadron shower decomposition
  in the highly granular CALICE analogue hadron calorimeter}''
  \href{http://dx.doi.org/10.1088/1748-0221/11/06/P06013}{{\em JINST} { 11}
  (2016) no.~06, P06013},
\href{http://arxiv.org/abs/1602.08578}{{\tt arXiv:1602.08578
  [physics.ins-det]}}.

\bibitem{Repond:2018flg}
{ CALICE Collaboration}, J.~Repond {\em et al.}, ``{Hadronic Energy Resolution
  of a Combined High Granularity Scintillator Calorimeter System}''
  \href{http://dx.doi.org/10.1088/1748-0221/13/12/P12022}{{\em JINST} { 13}
  (2018) no.~12, P12022},
\href{http://arxiv.org/abs/1809.03909}{{\tt arXiv:1809.03909
  [physics.ins-det]}}.

\bibitem{Hartbrich:2016bbz}
O.~Hartbrich, \href{http://dx.doi.org/10.3204/PUBDB-2016-02800}{{\em
  {Scintillator Calorimeters for a Future Linear Collider Experiment}}}.
\newblock PhD thesis, DESY, Hamburg, 2016.
\newblock
\url{https://bib-pubdb1.desy.de/record/301691}.
\newblock

\bibitem{Blazey:2009zz}
G.~Blazey, D.~Chakraborty, A.~Dyshkant, K.~Francis, D.~Hedin, {\em et al.},
  ``{Directly coupled tiles as elements of a scintillator calorimeter with MPPC
  readout}'' \href{http://dx.doi.org/10.1016/j.nima.2009.03.253}{{\em
  Nucl.Instrum.Meth.} { A605} (2009)  277--281}.
FERMILAB-PUB-09-585-E.

\bibitem{Simon:2010hf}
F.~Simon and C.~Soldner, ``{Uniformity Studies of Scintillator Tiles directly
  coupled to SiPMs for Imaging Calorimetry}''
  \href{http://dx.doi.org/10.1016/j.nima.2010.03.142}{{\em Nucl.Instrum.Meth.}
  { A620} (2010)  196--201}, \href{http://arxiv.org/abs/1001.4665}{{\tt
  arXiv:1001.4665 [physics.ins-det]}}.
MPP-2010-10.

\bibitem{Liu:2015cpe}
Y.~Liu, V.~B{\"u}scher, J.~Caudron, P.~Chau, S.~Krause, L.~Masetti,
  U.~Sch{\"a}fer, R.~Spreckels, S.~Tapprogge, and R.~Wanke,
  \href{http://dx.doi.org/10.1109/NSSMIC.2014.7431118}{``{A Design of
  Scintillator Tiles Read Out by Surface-Mounted SiPMs for a Future Hadron
  Calorimeter}''} in {\em {Proceedings, 21st Symposium on Room-Temperature
  Semiconductor X-ray and Gamma-ray Detectors (RTSD 2014): Seattle, WA, USA,
  November 8-15, 2014}}, p.~7431118.
\newblock 2016.
\newblock
\href{http://arxiv.org/abs/1512.05900}{{\tt arXiv:1512.05900
  [physics.ins-det]}}.
\newblock

\bibitem{Sefkow:2018rhp}
{ CALICE Collaboration}, F.~Sefkow and F.~Simon, ``{A highly granular
  SiPM-on-tile calorimeter prototype}''
  \href{http://dx.doi.org/10.1088/1742-6596/1162/1/012012}{{\em J. Phys. Conf.
  Ser.} { 1162} (2019) no.~1, 012012},
\href{http://arxiv.org/abs/1808.09281}{{\tt arXiv:1808.09281
  [physics.ins-det]}}.

\bibitem{Reinecke:2013zua}
{ CALICE Collaboration}, M.~Reinecke,
  \href{http://dx.doi.org/10.1109/NSSMIC.2013.6829522}{``{Performance of the
  large scale prototypes of the CALICE tile hadron calorimeter}''} in {\em
  {Proceedings, 2013 IEEE Nuclear Science Symposium and Medical Imaging
  Conference (NSS/MIC 2013): Seoul, Korea, October 26-November 2, 2013}}.
\newblock
2013.
\newblock

\bibitem{Munwes:2634923}
P.~C. Y.~Munwes and F.~Simon, ``{Performance of test infrastructure for highly
  granular optical readout}''. \url{https://cds.cern.ch/record/2634923}. Tech.
  Rep. AIDA-2020-D14.2.

\bibitem{Kvasnicka:2017bpx}
{ CALICE Collaboration}, J.~Kvasnicka, ``{Data Acquisition System for the
  CALICE AHCAL Calorimeter}''
  \href{http://dx.doi.org/10.1088/1748-0221/12/03/C03043}{{\em JINST} { 12}
  (2017) no.~03, C03043},
\href{http://arxiv.org/abs/1701.02232}{{\tt arXiv:1701.02232
  [physics.ins-det]}}.

\bibitem{Callier:2014uqa}
S.~Callier, J.~B. Cizel, F.~Dulucq, C.~d.~L. Taille, G.~Martin-Chassard, and
  N.~Seguin-Moreau, ``{ROC chips for imaging calorimetry at the International
  Linear Collider}''
\href{http://dx.doi.org/10.1088/1748-0221/9/02/C02022}{{\em JINST} { 9} (2014)
  C02022}.

\bibitem{Buridon:2016ill}
{ CALICE Collaboration}, V.~Buridon {\em et al.}, ``{First results of the
  CALICE SDHCAL technological prototype}''
  \href{http://dx.doi.org/10.1088/1748-0221/11/04/P04001}{{\em JINST} { 11}
  (2016) no.~04, P04001},
\href{http://arxiv.org/abs/1602.02276}{{\tt arXiv:1602.02276
  [physics.ins-det]}}.

\bibitem{Deng:2016obt}
{ CALICE Collaboration}, Z.~Deng {\em et al.}, ``{Resistive Plate Chamber
  Digitization in a Hadronic Shower Environment}''
  \href{http://dx.doi.org/10.1088/1748-0221/11/06/P06014}{{\em JINST} { 11}
  (2016) no.~06, P06014},
\href{http://arxiv.org/abs/1604.04550}{{\tt arXiv:1604.04550
  [physics.ins-det]}}.

\bibitem{Fleury:2014hfa}
J.~Fleury, S.~Callier, C.~de~La~Taille, N.~Seguin, D.~Thienpont, F.~Dulucq,
  S.~Ahmad, and G.~Martin, ``{Petiroc and Citiroc: front-end ASICs for SiPM
  read-out and ToF applications}''
\href{http://dx.doi.org/10.1088/1748-0221/9/01/C01049}{{\em JINST} { 9} (2014)
  C01049}.

\bibitem{Abramowicz:2018vwb}
H.~Abramowicz {\em et al.}, ``{Performance and Moli{\`e}re radius measurements
  using a compact prototype of LumiCal in an electron test beam}''
  \href{http://dx.doi.org/10.1140/epjc/s10052-019-7077-9}{{\em Eur. Phys. J.} {
  C79} (2019) no.~7, 579},
\href{http://arxiv.org/abs/1812.11426}{{\tt arXiv:1812.11426
  [physics.ins-det]}}.

\bibitem{ild:bib:FLAME}
M.~Firlej, ``{Status of FLAME development}''.
  \url{https://indico.cern.ch/event/818956/contributions/3558877/}. presented
  at the 35th FCAL Collaboration Workshop, 19-20 September 2019.

\bibitem{ild:bib:surface_facilities}
H.~Hayano, ``{Kitakami Site-specific CFS Study Update}''.
  \url{https://agenda.linearcollider.org/event/7645/contributions/40035/}.
  International Workshop on Future Linear Colliders LCWS2017, Strasbourg,
  October 2017.

\bibitem{ild:bib:underground_facilities}
M.~Miyahara, ``{Update on Detector Hall and Assembly Hall Layout}''.
  \url{https://agenda.linearcollider.org/event/6910/contributions/33981/}.
  Mini-Workshop on Infrastructure and CFS for Physics and Detectors, KEK, March
  2016.

\bibitem{ild:bib:Access}
M.~Miyahara, ``{Update on Detector Hall and Assembly Hall Design}''.
  \url{https://agenda.linearcollider.org/event/6851/contributions/33640/}.
  Mini-Workshop on Infrastructure and CFS for Physics and Detectors, KEK,
  September 2015.

\bibitem{ild:bib:services_figure}
Y.~Sugimoto, ``{Detector Utility}''.
  \url{https://agenda.linearcollider.org/event/8217/contributions/44784/}.
  International Workshop on Future Linear Colliders, LCWS2019, Sendai, October
  2019.

\bibitem{ild:bib:services}
Y.~Sugimoto, ``{Detector Utility}''.
  \url{https://agenda.linearcollider.org/event/8123/contributions/43421}.
  Mini-Workshop on Infrastructure and CFS for Physics and Detectors, KEK,
  February 2019.

\bibitem{ild:bib:ejade_mdi}
K.~Buesser and T.~Schoerner-Sadenius, ``{MDIPlan, E-JADE Deliverable Report
  22}''.
  \url{https://www.e-jade.eu/sites/sites_custom/site_e-jade/content/e49893/e65922/e84403/D.22.WP3.MDI.v3.pdf}.
  E-Jade.Deliverable.WP3.D22.MDIPlan.v3.

\bibitem{ild:bib:assembly}
Y.~Sugimoto, ``{Detector Assembly Schedule}''.
  \url{https://agenda.linearcollider.org/event/8297/contributions/44286/}.
  Mini-Workshop on Infrastructure and CFS for Physics and Detectors,
  Ichinoseki, October 2019.

\bibitem{ild:bib:gantry_crane}
M.~Oriunno, ``{SiD Underground Assembly}''.
  \url{https://agenda.linearcollider.org/event/6404/contributions/30839/}.
  MDI-CFS Meeting on ILC Interaction Region Issues, Ichinoseki, September 2014.

\bibitem{ild:bib:inner_detector_integration}
A.~Gonnin and C.~Bourgeois, ``{Integration of the inner detector region}''.
  \url{https://edmsdirect.desy.de/item/D00000001003815,A,1,1}. EDMS ID:
  D00000001003815,A,1,1.

\bibitem{ild:bib:VTX_integration}
A.~Besson, ``{VTX and Intermediate Tracking Status}''.
  \url{https://agenda.linearcollider.org/event/8126/contributions/43237/}. ILD
  Integration Meeting, DESY, February 2019.

\bibitem{ild:bib:SI_integration}
A.~Gonnin and R.~Poeschl, ``{ILD Cabling - what we know today}''.
  \url{https://agenda.linearcollider.org/event/8126/contributions/43236/}. ILD
  Integration Meeting, DESY, February 2019.

\bibitem{ild:bib:TPC_ICD}
P.~Colas {\em et al.}, ``{TPC Interface Control Document}''.
  \url{https://edmsdirect.desy.de/item/D00000001162555,A,1,1}. EDMS ID:
  D00000001162555,A,1,1.

\bibitem{ild:bib:SiECAL_ICD}
R.~Poeschl and H.~Videau, ``{SiEcal Interface Control Document}''.
  \url{https://edmsdirect.desy.de/item/D00000001162465,A,1,1}. EDMS-ID:
  D00000001162465,A,1,1.

\bibitem{ild:bib:ScECAL_ICD}
T.~Takeshita, ``{ScECAL Interface Control Document}''.
  \url{https://edmsdirect.desy.de/item/D00000001162515,A,1,1}. EDMS-ID:
  D00000001162515,A,1,1.

\bibitem{ild:bib:VFS_ICD}
S.~Schuwalow and Y.~Benhammou, ``{VFS Interface Control Document}''.
  \url{https://edmsdirect.desy.de/item/D00000001163265,A,1,1}. EDMS-ID:
  D00000001163265,A,1,1.

\bibitem{ild:bib:Magnet_Note}
F.~Kircher {\em et al.}, ``{Conceptual Design of the ILD Detector Magnet
  System}''.
  \url{http://flc.desy.de/lcnotes/notes/localfsExplorer_read?currentPath=/afs/desy.de/group/flc/lcnotes/LC-DET-2012-081.pdf}.
  LC-DET-2012-081.

\bibitem{ild:bib:Cryostat_Note}
R.~Stromhagen, ``{Note on the integration of Cryostat into the Central Barrel -
  ILD Detector }''.
  \url{https://edmsdirect.desy.de/item/D00000001010065,A,1,1}. EDMS-ID:
  D00000001162515,A,1,1.

\bibitem{ild:bib:anti-did}
A.~Seryi, T.~Maruyama, and B.~Parker, ``{IR Optimization, DID and anti-DID}''.
SLAC-PUB-11662.

\bibitem{ild:bib:Solenoid_Manufacturing}
Y.~Makida and T.~Okamura, ``{Status of Magnet Design Studies }''.
  \url{https://agenda.linearcollider.org/event/8126/contributions/43233/}. ILD
  Integration Meeting, Hamburg, February 2019.

\bibitem{ild:bib:anti-did-design}
Y.~Makida {\em et al.}, ``{Interim Report from Toshiba/Hitachi Studies on
  Solenoid and Anti-DID }''.
  \url{https://agenda.linearcollider.org/event/7520}. ILD Software and
  Technical Meeting, Lyon, April 2017.

\bibitem{ild:bib:Magnet_Simulations}
U.~Schneekloth, ``{ILD Magnet Activities}''.
  \url{https://agenda.linearcollider.org/event/7760/contributions/40574/}. ILD
  Meeting, Ichinoseki, February 2018.

\bibitem{ild:bib:Radiation_Hall}
T.~Sanami {\em et al.}, ``{IR Hall Dose Rate Estimates with Detector Concepts
  }''. SLAC Radiation Physics Note, RP-09-08.

\bibitem{ild:bib:Machine_Backgrounds}
D.~Jeans and A.~Miyamoto, ``{Machine-related backgrounds in ILD}''.
  ILD-TECH-PUB-2019-001.

\bibitem{ild:bib:schuetz_thesis}
A.~Schuetz, ``{Optimizing the design of the Final-Focus region for the
  International Linear Collider}''.
SLAC-PUB-11662.

\bibitem{ild:bib:AIDADAQ}
M.~Wing {\em et al.}, ``{Common DAQ system ready for combined tests, AIDA-2020
  report, AIDA-2020-MS80 (2018)}''. \url{http://cds.cern.ch/record/2314260}.

\bibitem{ild:bib:hi-net}
``{High Sensitivity Seismograph Network Japan}''.
\newblock \url{http://www.hinet.bosai.go.jp}.

\bibitem{ild:bib:earthquake}
T.~Tauchi, ``{Standard Reference Earthquake Parameters}''.
  \url{https://edmsdirect.desy.de/item/D00000001164345,A,1,1}. ILC-EDMS
  D*1164345.

\bibitem{ild:bib:Seismic_Damping}
T.~Sanuki {\em et al.}, ``{Seismic Isolation}''.
  \url{https://agenda.linearcollider.org/event/7976/contributions/42723/}.
  Mini-Workshop on ILC Infrastructure and CFS for Physics and Detectors, KEK,
  Tsukuba, November 2018.

\bibitem{ild:bib:edms}
``{ILC Engineering Data Management System}''.
\newblock \url{https://ilc-edms.desy.de/}.

\bibitem{ild:bib:edmsdirect}
``{ILD Technical Documentation}''.
\newblock \url{https://edmsdirect.desy.de/treebrowser/ildtdr/}.

\bibitem{bib:ilcsoft}
``{iLCSoft Project Page}''. \url{https://github.com/iLCSoft}, 2016.

\bibitem{Kilian:2007gr}
W.~Kilian, T.~Ohl, and J.~Reuter, ``{WHIZARD: Simulating Multi-Particle
  Processes at LHC and ILC}''
  \href{http://dx.doi.org/10.1140/epjc/s10052-011-1742-y}{{\em Eur. Phys. J.} {
  C71} (2011)  1742},
\href{http://arxiv.org/abs/0708.4233}{{\tt arXiv:0708.4233 [hep-ph]}}.

\bibitem{Sjostrand:2006za}
T.~Sjostrand, S.~Mrenna, and P.~Z. Skands, ``{PYTHIA 6.4 Physics and Manual}''
  \href{http://dx.doi.org/10.1088/1126-6708/2006/05/026}{{\em JHEP} { 05}
  (2006)  026},
\href{http://arxiv.org/abs/hep-ph/0603175}{{\tt arXiv:hep-ph/0603175
  [hep-ph]}}.

\bibitem{Schulte:1998au}
D.~Schulte, ``{Beam-beam simulations with Guinea-Pig}'' {\em eConf} { C980914}
  (1998)  127--131.
\url{https://cds.cern.ch/record/382453}.

\bibitem{Chen:1993dba}
P.~Chen, T.~L. Barklow, and M.~E. Peskin, ``{Hadron production in gamma gamma
  collisions as a background for e+ e- linear colliders}''
  \href{http://dx.doi.org/10.1103/PhysRevD.49.3209}{{\em Phys. Rev.} { D49}
  (1994)  3209--3227},
\href{http://arxiv.org/abs/hep-ph/9305247}{{\tt arXiv:hep-ph/9305247
  [hep-ph]}}.

\bibitem{Gaede:2003ip}
F.~Gaede, T.~Behnke, N.~Graf, and T.~Johnson, ``{LCIO: A Persistency framework
  for linear collider simulation studies}'' {\em eConf} { C0303241} (2003)
  TUKT001,
\href{http://arxiv.org/abs/physics/0306114}{{\tt arXiv:physics/0306114
  [physics]}}.

\bibitem{Gaede:2006pj}
F.~Gaede, ``{Marlin and LCCD: Software tools for the ILC}''
\href{http://dx.doi.org/10.1016/j.nima.2005.11.138}{{\em Nucl. Instrum. Meth.}
  { A559} (2006)  177--180}.

\bibitem{Frank:2014zya}
M.~Frank, F.~Gaede, C.~Grefe, and P.~Mato, ``{DD4hep: A Detector Description
  Toolkit for High Energy Physics Experiments}''
\href{http://dx.doi.org/10.1088/1742-6596/513/2/022010}{{\em J. Phys. Conf.
  Ser.} { 513} (2014)  022010}.

\bibitem{Frank:2015ivo}
M.~Frank, F.~Gaede, N.~Nikiforou, M.~Petric, and A.~Sailer, ``{DDG4 A
  Simulation Framework based on the DD4hep Detector Description Toolkit}''
\href{http://dx.doi.org/10.1088/1742-6596/664/7/072017}{{\em J. Phys. Conf.
  Ser.} { 664} (2015) no.~7, 072017}.

\bibitem{Agostinelli:2002hh}
{ GEANT4 Collaboration}, S.~Agostinelli {\em et al.}, ``{GEANT4: A Simulation
  toolkit}''
\href{http://dx.doi.org/10.1016/S0168-9002(03)01368-8}{{\em Nucl. Instrum.
  Meth.} { A506} (2003)  250--303}.

\bibitem{bib:lcgeo}
``{lcgeo Project Page}''. \url{https://github.com/iLCSoft/lcgeo}, 2016.

\bibitem{Gaede:2014aza}
F.~Gaede, S.~Aplin, R.~Glattauer, C.~Rosemann, and G.~Voutsinas, ``{Track
  reconstruction at the ILC: the ILD tracking software}''
\href{http://dx.doi.org/10.1088/1742-6596/513/2/022011}{{\em J. Phys. Conf.
  Ser.} { 513} (2014)  022011}.

\bibitem{Marshall:2015rfa}
J.~S. Marshall and M.~A. Thomson, ``{The Pandora Software Development Kit for
  Pattern Recognition}''
  \href{http://dx.doi.org/10.1140/epjc/s10052-015-3659-3}{{\em Eur. Phys. J.} {
  C75} (2015) no.~9, 439},
\href{http://arxiv.org/abs/1506.05348}{{\tt arXiv:1506.05348
  [physics.data-an]}}.

\bibitem{Suehara:2015ura}
T.~Suehara and T.~Tanabe, ``{LCFIPlus: A Framework for Jet Analysis in Linear
  Collider Studies}'' \href{http://dx.doi.org/10.1016/j.nima.2015.11.054}{{\em
  Nucl. Instrum. Meth.} { A808} (2016)  109--116},
\href{http://arxiv.org/abs/1506.08371}{{\tt arXiv:1506.08371
  [physics.ins-det]}}.

\bibitem{Catani:1991hj}
S.~Catani, Y.~L. Dokshitzer, M.~Olsson, G.~Turnock, and B.~Webber, ``{New
  clustering algorithm for multi - jet cross-sections in e+ e- annihilation}''
\href{http://dx.doi.org/10.1016/0370-2693(91)90196-W}{{\em Phys.Lett.} { B269}
  (1991)  432--438}.

\bibitem{Cacciari:2006sm}
M.~Cacciari, ``{FastJet: A Code for fast $k_t$ clustering, and more}'' in {\em
  {Deep inelastic scattering. Proceedings, 14th International Workshop, DIS
  2006, Tsukuba, Japan, April 20-24, 2006}}, pp.~487--490.
\newblock 2006.
\newblock \href{http://arxiv.org/abs/hep-ph/0607071}{{\tt arXiv:hep-ph/0607071
  [hep-ph]}}.
\newblock
[,125(2006)].

\bibitem{Grefe:2014sca}
{ CLIC detector, physics study}, C.~Grefe, S.~Poss, A.~Sailer, and
  A.~Tsaregorodtsev, ``{ILCDIRAC, a DIRAC extension for the Linear Collider
  community}''
\href{http://dx.doi.org/10.1088/1742-6596/513/3/032077}{{\em J. Phys. Conf.
  Ser.} { 513} (2014)  032077}.

\bibitem{Miyamoto:2019xve}
A.~Miyamoto and H.~Ono, ``{ILD MC production for detector optimization}'' in
  {\em {International Workshop on Future Linear Colliders (LCWS 2018)
  Arlington, Texas, USA, October 22-26, 2018}}.
\newblock 2019.
\newblock
\href{http://arxiv.org/abs/1902.02516}{{\tt arXiv:1902.02516
  [physics.ins-det]}}.
\newblock

\bibitem{Berggren:2012ar}
M.~Berggren, ``{SGV 3.0 - a fast detector simulation}'' in {\em {International
  Workshop on Future Linear Colliders (LCWS11) Granada, Spain, September 26-30,
  2011}}.
\newblock 2012.
\newblock
\href{http://arxiv.org/abs/1203.0217}{{\tt arXiv:1203.0217 [physics.ins-det]}}.
\newblock

\bibitem{ild:bib:perfgoal::barklow}
T.~Barklow, ``{Physics Impact of Detector Performance}''.
  \url{http://www-conf.slac.stanford.edu/lcws05/program/talks/18mar2005.ppt}.
  Presentation for LCWS05.

\bibitem{ild:bib:PandoraPFA}
M.~Thomson, ``{Particle Flow Calorimetry and the PandoraPFA Algorithm}''
  \href{http://dx.doi.org/10.1016/j.nima.2009.09.009}{{\em Nucl.Instrum.Meth.}
  { A611} (2009)  25--40}, \href{http://arxiv.org/abs/0907.3577}{{\tt
  arXiv:0907.3577 [physics.ins-det]}}.
CU-HEP-09-11.

\bibitem{Tanabashi:2018oca}
{ Particle Data Group}, M.~Tanabashi {\em et al.}, ``{Review of Particle
  Physics}''
\href{http://dx.doi.org/10.1103/PhysRevD.98.030001}{{\em Phys. Rev.} { D98}
  (2018) no.~3, 030001}.

\bibitem{ILDNote:Hbbccgg}
M.~Kurata and R.~Yonamine, ``{Higgs branching ratio study for new detector
  models as benchmark process in ILD}''. ILD-PHYS-PUB-2019-008,
  \url{https://confluence.desy.de/display/ILD/ILD+notes}, 2019.

\bibitem{Mueller:2016exq}
F.~J. Mueller, \href{http://dx.doi.org/10.3204/PUBDB-2016-02659}{{\em
  {Development of a Triple GEM Readout Module for a Time Projection Chamber \&
  Measurement Accuracies of Hadronic Higgs Branching Fractions in $\nu\nu$H at
  a 350 GeV ILC}}}.
\newblock PhD thesis, DESY, Hamburg, 2016.
\newblock
\url{http://bib-pubdb1.desy.de/search?cc=Publication+Database&of=hd&p=reportnumber:DESY-THESIS-2016-018}.
\newblock

\bibitem{Ono:2013voc}
H.~Ono, ``{Higgs branching ratio study for DBD detector benchmarking in ILD}''
  in {\em {Helmholtz Alliance Linear Collider Forum: Proceedings of the
  Workshops Hamburg, Munich, Hamburg 2010-2012, Germany}}, pp.~203--223, DESY.
\newblock DESY, Hamburg,
2013.
\newblock

\bibitem{Ono:2013sea}
H.~Ono and A.~Miyamoto, ``{A study of measurement precision of the Higgs boson
  branching ratios at the International Linear Collider}''
  \href{http://dx.doi.org/10.1140/epjc/s10052-013-2343-8}{{\em Eur. Phys. J.} {
  C73} (2013) no.~3, 2343},
\href{http://arxiv.org/abs/1207.0300}{{\tt arXiv:1207.0300 [hep-ex]}}.

\bibitem{Fujii:2017vwa}
K.~Fujii {\em et al.}, ``{Physics Case for the 250 GeV Stage of the
  International Linear Collider}''
\href{http://arxiv.org/abs/1710.07621}{{\tt arXiv:1710.07621 [hep-ex]}}.

\bibitem{ILDNote:MH}
J.~Tian, ``{Higgs Mass Measurement at $\sqrt{s}=500$\,GeV as benchmark process
  in ILD}''. ILD-PHYS-PUB-2019-001,
  \url{https://confluence.desy.de/display/ILD/ILD+notes}, 2019.

\bibitem{Kawada:2019isz}
S.-I. Kawada, J.~List, and M.~Berggren, ``{Prospects of measuring Higgs boson
  decays into muon pairs at the ILC}'' in {\em {International Workshop on
  Future Linear Colliders (LCWS 2018) Arlington, Texas, USA, October 22-26,
  2018}}.
\newblock 2019.
\newblock
\href{http://arxiv.org/abs/1902.05021}{{\tt arXiv:1902.05021 [hep-ex]}}.
\newblock

\bibitem{ILDNote:Hmumu}
S.~Kawada, ``{Branching ratio $H \to \mu^+ \mu^-$ at $\sqrt{s}=500$\,GeV as
  benchmark process in ILD}''. ILD-PHYS-PUB-2019-002,
  \url{https://confluence.desy.de/display/ILD/ILD+notes}, 2019.

\bibitem{ILDNote:Hinv}
Y.~Kato, ``{$H \to $ invisible at $\sqrt{s}=500$\,GeV as benchmark process in
  ILD}''. ILD-PHYS-PUB-2019-003,
  \url{https://confluence.desy.de/display/ILD/ILD+notes}, 2019.

\bibitem{Boronat:2014hva}
M.~Boronat, J.~Fuster, I.~Garcia, E.~Ros, and M.~Vos, ``{A robust jet
  reconstruction algorithm for high-energy lepton colliders}''
  \href{http://dx.doi.org/10.1016/j.physletb.2015.08.055}{{\em Phys. Lett.} {
  B750} (2015)  95--99},
\href{http://arxiv.org/abs/1404.4294}{{\tt arXiv:1404.4294 [hep-ex]}}.

\bibitem{ILDNote:tautau}
D.~Jeans and K.~Yumino, ``{ILD benchmark: a study of $e^- e^+ \to \tau^-
  \tau^+$ at 500 GeV}''. ILD-PHYS-PUB--2019-004,
  \url{https://arxiv.org/abs/1912.08403}, 2019.

\bibitem{ILDNote:QGCs}
J.~Beyer, ``{Vector Boson Scatering at $\sqrt{s}=1$\,TeV as benchmark process
  in ILD}''. ILD-PHYS-PUB-2019-005,
  \url{https://confluence.desy.de/display/ILD/ILD+notes}, 2019.

\bibitem{ILDNote:gammaZ}
T.~Mizuno, ``{Calibration of the photon energy scale from $e^+e^- \to \gamma
  \mu^+ \mu^-$ as benchmark process in ILD}''. ILD-PHYS-PUB-2019-006,
  \url{https://confluence.desy.de/display/ILD/ILD+notes}, 2019.

\bibitem{ILDNote:bbtt}
A.~Irles and Y.~Okugawa, ``{$e^+e^- \to b \bar{b}$ and $e^+e^- \to t \bar{t}$
  as benchmark processes in ILD}''. ILD-PHYS-PUB-2019-007,
  \url{https://confluence.desy.de/display/ILD/ILD+notes}, 2019.

\bibitem{Durieux:2019rbz}
{G. Durieux, A. Irles, V. Miralles, A. Penuelas, R. P\"oschl, M. Perell\'o and
  M. Vos}, ``{The electro-weak couplings of the top and bottom quarks - global
  fit and future prospects}''
\href{http://arxiv.org/abs/1907.10619}{{\tt arXiv:1907.10619 [hep-ph]}}.

\bibitem{ILDNote:extraH}
Y.~Wang, ``{Search for additional Higgs bosons as benchmark processes in
  ILD}''. ILD-PHYS-PUB-2019-011,
  \url{https://confluence.desy.de/display/ILD/ILD+notes}, 2019.

\bibitem{Abbiendi:2002qp}
{ OPAL Collaboration}, G.~Abbiendi {\em et al.}, ``{Decay mode independent
  searches for new scalar bosons with the OPAL detector at LEP}''
  \href{http://dx.doi.org/10.1140/epjc/s2002-01115-1}{{\em Eur. Phys. J.} {
  C27} (2003)  311--329},
\href{http://arxiv.org/abs/hep-ex/0206022}{{\tt arXiv:hep-ex/0206022
  [hep-ex]}}.

\bibitem{FIPnote:ESU_BSM}
{M. Berggren, J.List, M. Perelstein and M. Peskin}, ``{{\rm (on behalf of the
  LCC Physics WG)} Response to the ESPP BSM conveners on Feebly Interacting
  Particles}''. {{private communication to the ESPP BSM conveners}}, 2019.

\bibitem{Berggren:2013vfa}
M.~Berggren, F.~Bruemmer, J.~List, G.~Moortgat-Pick, T.~Robens, K.~Rolbiecki,
  and H.~Sert, ``{Tackling light higgsinos at the ILC}''
  \href{http://dx.doi.org/10.1140/epjc/s10052-013-2660-y}{{\em Eur. Phys. J.} {
  C73} (2013) no.~12, 2660},
\href{http://arxiv.org/abs/1307.3566}{{\tt arXiv:1307.3566 [hep-ph]}}.

\bibitem{ILDNote:WIMPs}
R.~Yonamine, ``{WIMP searches in the mono-photon channel as benchmark processes
  in ILD}''. ILD-PHYS-PUB-2019-010,
  \url{https://confluence.desy.de/display/ILD/ILD+notes}, 2019.

\end{thebibliography}\endgroup
%
%
%
%
%

\end{document}